\pdfoutput=1
\documentclass{panda_book}
\markpanda{Strong interaction studies with antiprotons}{Technical Design Report - EMC}
\begin{document}
%
%
\pagenumbering{roman}
\onecolumn
%
%
\thispagestyle{empty}
\vspace*{-0.5cm}
{\it FAIR/PANDA/Technical Design Report - EMC}
\vspace*{1.5cm}
\begin{center}
{\huge Technical Design Report for:\\ \ \\ \Panda{} \\ Electromagnetic Calorimeter (EMC) \\
{\sf\small (Anti\underline{P}roton \underline{An}nihilations at \underline{Da}rmstadt)}\\
\ \\ Strong Interaction Studies with Antiprotons}
\vskip 1cm
{\large \Panda{} Collaboration}
%
\vskip 0.5cm
%
\end{center}
\vskip 1cm
%
%
\vskip 1cm
\begin{center}
\includegraphics[width=0.9\dwidth]{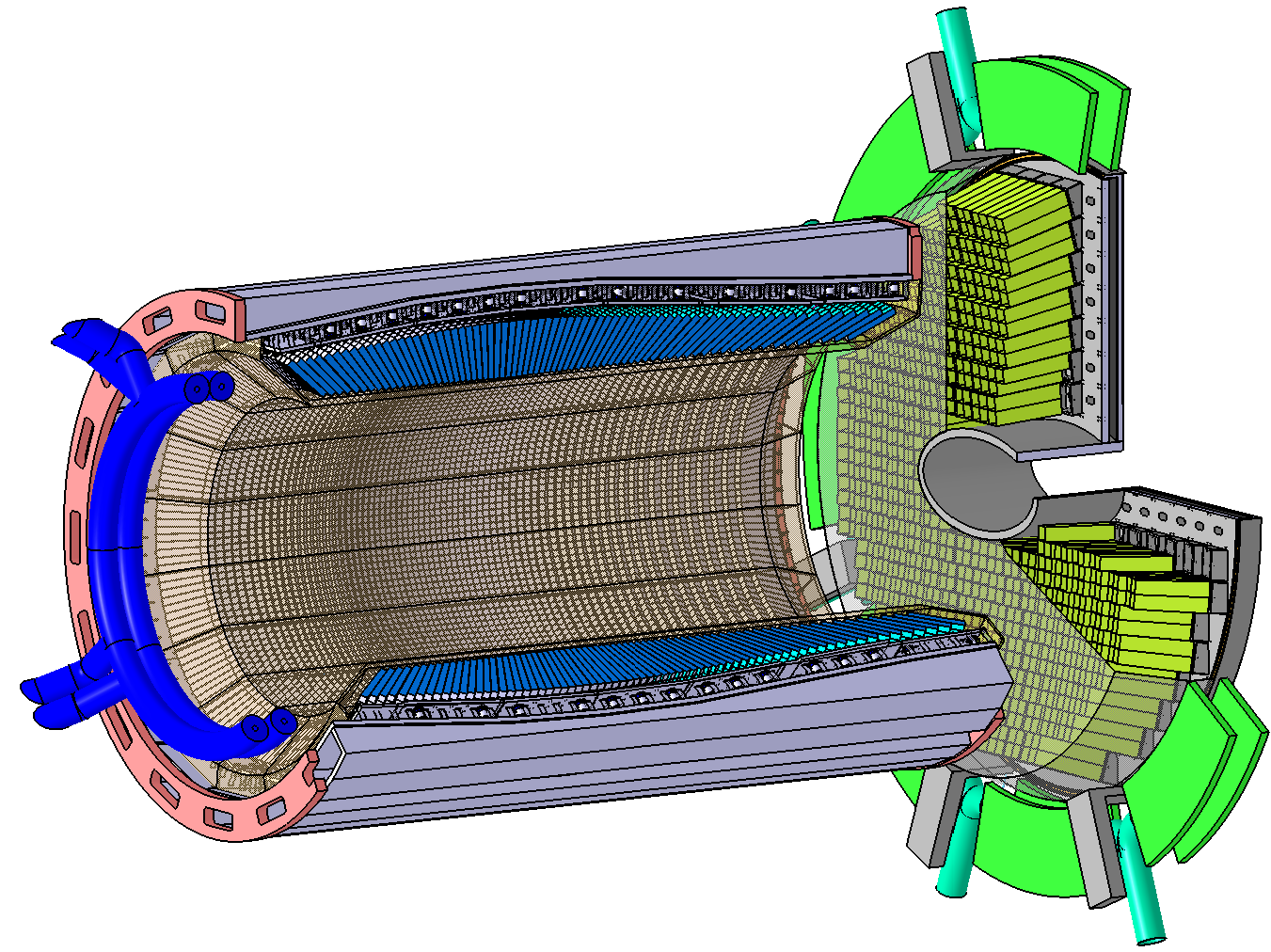}
\end{center}
\vfill
%
%
\newpage

\vspace{16cm}

\vfill

{\bf Cover}: The figure shows the barrel (left) and forward endcap part (right) of the electromagnetic calorimeter.
The parts of the barrel calorimeter are cut to get a better view to the inside.
Displayed are 
the PWO crystals with the carbon fiber alveole packs, the support feet connecting
to the support beam, held by the support rings at front and back.
On the right side the forward endcap crystals inside the carbon fiber packs are shown along with the insulation (shown transparent) and the eight supporting structures on the outside.

\newpage
\begin{center}
\vspace*{3mm }
{\LARGE The \Panda{} Collaboration}
\vskip 7mm
%
%
\institem{Universit\"at {\bf Basel}, Switzerland}
\authitem{W.~Erni},
\authitem{I.~Keshelashvili},
\authitem{B.~Krusche},
\authitem{M.~Steinacher}\lastitem
\institem{Institute of High Energy Physics, Chinese Academy of Sciences, {\bf Beijing}, China}
\authitem{Y.~Heng},
\authitem{Z.~Liu},
\authitem{H.~Liu},
\authitem{X.~Shen},
\authitem{O.~Wang},
\authitem{H.~Xu}\lastitem
\institem{Universit\"at {\bf Bochum}, I. Institut f\"ur Experimentalphysik, Germany}
\authitem{F.-H.~Heinsius},
\authitem{T.~Held},
\authitem{H.~Koch},
\authitem{B.~Kopf},
\authitem{M.~Peliz\"aus},
\authitem{M.~Steinke},
\authitem{U.~Wiedner},
\authitem{J.~Zhong}\lastitem
% ...
%
%

\end{center}
%
%
\vfill
\hrulefill\\
\begin{tabbing}
Editors:\hspace{3cm} \= Fritz-Herbert Heinsius  \hspace{1cm}  \= Email: \verb$heinsius@ep1.rub.de$ \\ \ \\
 \> Bertram Kopf  \hspace{1cm}  \> Email: \verb$bertram@ep1.rub.de$ \\ \ \\
 \> Bernd Lewandowski  \hspace{1cm}  \> Email: \verb$b.lewandowski@gsi.de$ \\ \ \\
 \> Herbert L\"ohner  \hspace{1cm}  \> Email: \verb$h.loehner@kvi.nl$ \\ \ \\
 \> Rainer Novotny  \hspace{1cm}  \> Email: \verb$rainer.novotny@exp2.physik.uni-giessen.de$ \\ \ \\
 \> Klaus Peters  \hspace{1cm}  \> Email: \verb$k.peters@gsi.de$ \\ \ \\
 \> Philippe Rosier \hspace{1cm}  \> Email: \verb$rosierph@ipno.in2p3.fr$ \\ \ \\
 \> Lars Schmitt  \hspace{1cm}  \> Email: \verb$l.schmitt@gsi.de$ \\ \ \\
 \> Alexander Vasiliev  \hspace{1cm}  \> Email: \verb$alexander.vasiliev@ihep.ru$ \\ \ \\
Technical Coordinator: \> Lars Schmitt \> Email: \verb$l.schmitt@gsi.de$ \\ 
Deputy:  \> Bernd Lewandowski  \> Email: \verb$b.lewandowski@gsi.de$ \\ \ \\
Spokesperson: \>  Ulrich Wiedner \> Email: \verb$ulrich.wiedner@tsl.uu.se$ \\
Deputy:  \> Paola Gianotti  \> Email: \verb$paola.gianotti@lnf.infn.it$ \\
\end{tabbing}
\hrulefill\\
\vfill
%
%
\cleardoublepage
%
\begin{center}
\vspace*{2cm}
{\Large\bf Preface}\addcontentsline{toc}{chapter}{Preface}
\vskip 2cm
\begin{minipage}[t]{8cm}
\sloppy\large
This document presents the technical layout and the envisaged performance of
the Electromagnetic Calorimeter (EMC) for the \Panda target spectrometer.
The EMC has been designed to meet the physics goals of the \Panda experiment. 
The performance figures are based on extensive prototype tests and radiation
hardness studies. The document shows that the EMC is ready for construction up to
the front-end electronics interface.
\end{minipage}
\end{center}
\vspace*{2cm}
\centerline{
}
\vfill
\clearpage
\vspace*{18cm}
\hrulefill\\
\vspace*{2cm}\\
\begin{minipage}[t]{10cm}
\sloppy
The use of registered names, trademarks, \etc in this publication does not
imply, even in the absence of specific statement, that such names are exempt
from the relevant laws and regulations and therefore free for general use.
\end{minipage}
\vfill
%

%
%
\cleardoublepage
\tableofcontents
%
%

%
%
%
\cleardoublepage
\pagenumbering{arabic}
\setcounter{page}{1}
\chapter{Executive Summary}
\label{sec:exs}
%
%
\section{The PANDA Experiment}
\label{sec:exs:int}
\PANDA is a next generation hadron physics detector planned to be
operated at the future Facility for Antiproton and Ion Research (FAIR)
at Darmstadt, Germany. It will use cooled antiproton beams with a
momentum between 1.5$\,$GeV$/c$ and 15$\,$GeV$/c$ interacting with
various internal targets.

At FAIR the antiprotons will be injected into HESR, a slow ramping
synchrotron and storage ring with excellent beam energy definition by
means of stochastic and electron cooling. This allows to measure
masses and widths of hadronic resonances with an accuracy of
50--100$\,$keV, which is 10 to 100 times better than achieved in any
$e^+e^-$-collider experiment. In addition states of all quantum
numbers can be directly formed in antiproton-proton annihilations
whereas in $e^+e^-$-collisions states with quantum numbers other than
$J^{PC}=1^{--}$ of a virtual photon can only be accessed by higher
order processes with corresponding lower cross section or production
reactions with much worse mass resolution. In the \PANDA experiment,
the antiprotons will interact with an internal target, either a
hydrogen cluster jet or a high frequency frozen hydrogen pellet
target, to reach a peak luminosity of up to $2\cdot
10^{32}$cm$^{-2}$s$^{-1}$. For reactions with heavy nuclear targets
thin wires or foils are inserted in the beam halo.

The experiment is focusing on hadron spectroscopy, in particular the
search for exotic states in the charmonium mass region, on the interaction
of charmed hadrons with the nuclear medium, on double-hypernuclei to
investigate the nuclear potential and hyperon-hyperon interactions as
well as on electromagnetic processes to study various aspects of nucleon
structure.

From these goals a list of physics benchmarks can be derived defining
the requirements for the \PANDA detector system. For precision
spectroscopy of charmonium states and exotic hadrons in the charmonium
region, full acceptance is required to perform a proper partial wave
analysis. Final states with many photons can occur, leading to a low photon
threshold as a central requirement for the electromagnetic calorimeters. To
reconstruct charmed mesons, a vertex detector and the identification of
kaons is necessary.

Further requirements result from other signals of interest: final states with muons occur in $\jpsi$ decays, semi-leptonic charm
meson decays and the Drell-Yan process. Measuring wide angle Compton
scattering requires detection of high energy photons. Investigating
hypernuclei requires the detection of hyperon cascades. The
measurement of proton formfactors relies on an efficient $e^{\pm}$
identification and discrimination against pions.

For the reconstruction of invariant masses a good momentum resolution
in the order of $\delta p/p \sim 1\%$ is desirable.  Low cross-section
processes and precision measurements lead to a high rate operation at up
to 20 million interactions per second. To perform several measurements in
parallel an efficient event selection is needed.
\begin{figure*}[htb]
\begin{center}
\includegraphics[width=\textwidth]{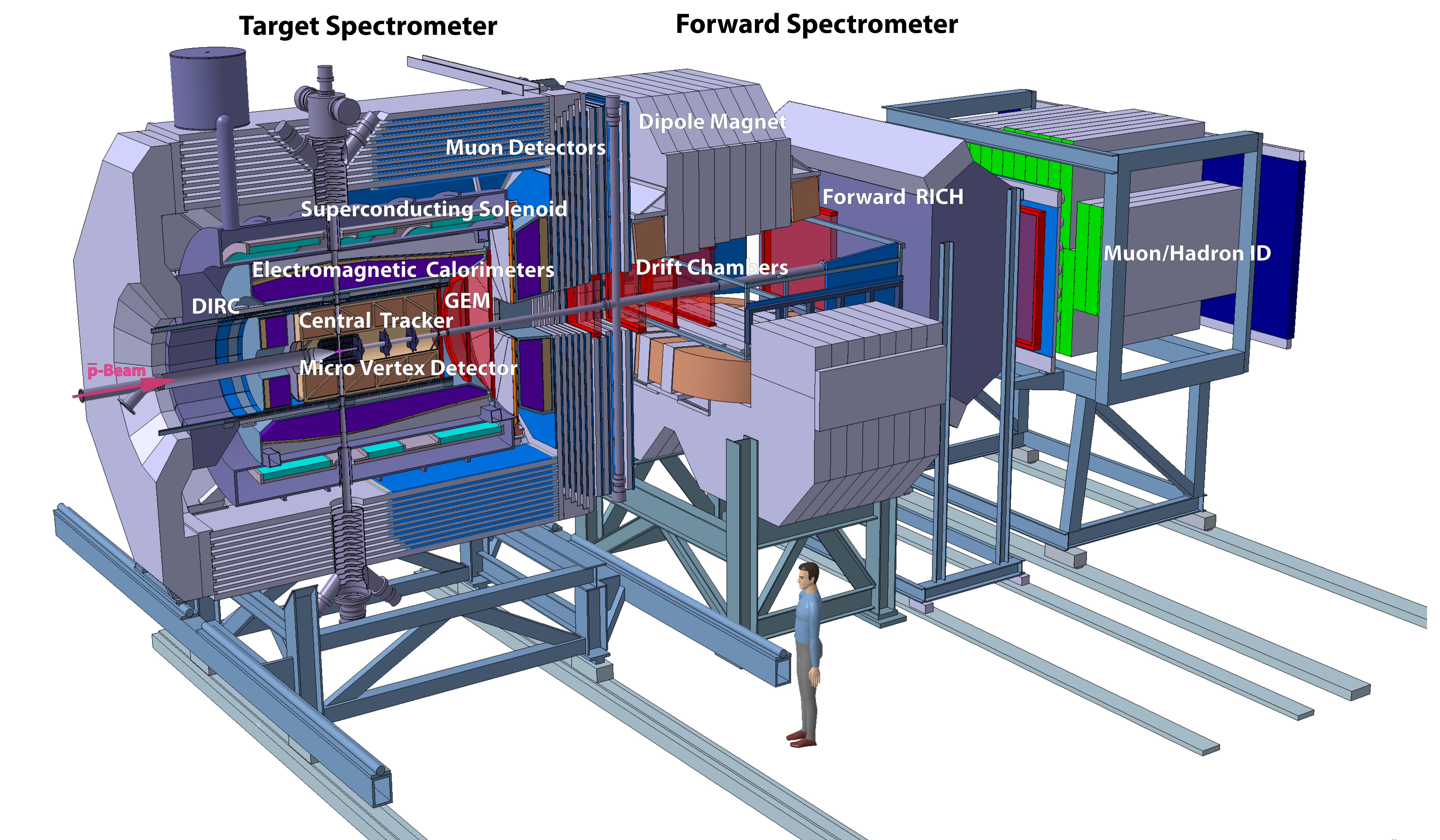}  
\end{center}
\caption[Overview of the \PANDA spectrometer.]{Overview of the \PANDA spectrometer.}
\label{fig:exs:detector}
\end{figure*}

\paragraph*{General Setup.}
To achieve almost 4$\pi$ acceptance and good momentum resolution over
a large range, a solenoid magnet for high $p_T$ tracks ({\em target
  spectrometer}) and a dipole magnet for the forward-going reaction
products ({\em forward spectrometer}) are foreseen
(\Reffig{fig:exs:detector}).  The solenoid magnet is super-conducting, provides a field strength 
of 2$\,$T and has a coil opening of 1.89 m and a coil length of
2.75$\,$m. The target spectrometer is arranged in a barrel part for angles
larger than 22$\degrees$ and an endcap part for the forward range down to
5$\degrees$ in the vertical and 10$\degrees$ in the horizontal plane.

The dipole magnet has a field integral of up to 2$\,$Tm with a 1.4$\,$m wide 
and 70 cm high opening. The forward spectrometer covers the very forward
angles.

Both
spectrometer parts are equipped with tracking, charged particle identification,
electromagnetic calorimetry and muon identification.

To operate the experiment at high rate and with different parallel
physical topologies a self-triggering readout scheme was adopted.  The
frontend electronics continuously digitizes the detector data and
autonomously finds valid hits. Physical signatures like energy
clusters, tracklets or ringlets are extracted on the fly. Compute
nodes make a fast first selection of interesting time slices which
then is refined with further data in subsequent levels. Data logging
happens only after online reconstruction. This allows full flexibility
in applying selection algorithms based on any physics signatures
detectable by the spectrometer. In addition physics topics with
identical target and beam settings can be treated in parallel.

\paragraph*{Tracking detectors.}
The silicon vertex detector consists of two inner layers of hybrid
pixel detectors and two outer layers with silicon strip detectors.
The pixel layers consist of a silicon sensor coupled to a custom-made
self-triggering readout ASIC realized in 0.13$\,\mu \m$ CMOS with
100$\times$100$\,\mu \m^2$ pixel size and cover 0.15$\,\m^2$ with 13
million channels. The strip part is based on double-sided silicon
strips read by a self-triggering 128-channel ASIC with discriminator
and analog readout and covers 0.5 m$^2$ with 70 000 channels.

For the tracking in the solenoid field low-mass straw 
tubes arranged in straight and skewed configuration are foreseen.
The straws have a diameter of 1 cm and a length of 150 cm and are
operated with Ar/CO$_2$ at 1$\,$bar overpressure giving them
rigidity without heavy support frames.

Alternatively an ungated time projection chamber based on GEM
foils as readout stage is being developed. The GEM foils suppress
the ion backflow into the drift volume, minimizing the space charge
build-up. As detector gas Ne/CO$_2$ is used. The readout plane consists
of 100 000 pads of 2$\times$2 mm$^2$. With 500 hits per track and
more than 50$\,\mu$s drift time 500 events are overlapping, leading 
to a very high data rate which has to be handled online.

Tracks at small polar angles ($5\degrees<\theta<22\degrees$) are measured by
large planar GEM detectors. Further downstream, in the forward
spectrometer, straw tube chambers with a tube diameter of 1 cm will be
employed.

\paragraph*{Particle Identification.}
Charged particle identification is required over a large momentum range from
200$\,$ MeV/$c$ up to almost 10$\,$GeV/$c$. Different physical processes are
employed.

The main part of charged particles is identified by various
Cherenkov detectors. In the target spectrometer two detectors based
on the detection of internally reflected Cherenkov light (DIRC) are
under design, one consisting of quartz rods for the barrel region, 
the other one in shape of a disc for the forward endcap. Novel readout
techniques are under study to achieve correction or compensation of
dispersion in the radiator material. In the forward spectrometer a
ring imaging Cherenkov detector with aerogel and C$_4$F$_{10}$ as
radiators is planned.

Time of flight can be partly exploited in \PANDA. Although no
dedicated start detector is available, a scintillator wall after the dipole magnet
can measure the relative timing of charged particles with very good
time resolution in the order of 100$\,$ps. 

The energy loss within the trackers will be employed as well for
particle identification below 1$\,$GeV/$c$ since the individual charge
is obtained by analog readout or time-over-threshold measurement.
Here the TPC option would provide best performance.

The detection system is complemented by muon detectors based on drift
tubes located inside the segmented magnet yoke, between the spectrometer
magnets and at the end of the spectrometer. Muon detection is implemented 
as a range system with interleaved absorbing material and detectors to
best distinguish muons from pions in the low momentum range of \PANDA.

\paragraph*{Calorimetry.}
In the target spectrometer high precision electromagnetic calorimetry
is required over a large energy range from a few MeV up to several GeV.
Lead-tungstate is chosen for the calorimeters in the target spectrometer
due to its good energy resolution, fast response and high density, allowing
a compact setup.

Good identification and reconstruction of multi-photon and lepton-pair
channels are of utmost importance for the success of the \Panda
experiment. Low energy thresholds and good energy and spatial
resolution are important assets to achieve high yield and good
background rejection.  Due to the high luminosity, fast response and
radition hardness are additional requirements which have to be
fulfilled. \Reftbl{tbl:req:summary} shows the detailed list of
requirements for the EMC.
\begin{table*}[hbtp]
\begin{center}
%
%
\begin{tabular}{|l|lll|}
\hline\hline
 & \multicolumn{3}{l|}{Required performance value}\\ \hline
Common properties & & & \\\hline
energy resolution $\sigma_{E}/E$ & \multicolumn{3}{l|}{$\le 1\,\% \oplus \frac{\le 2\,\%}{\sqrt{E/\gev}}$}\\
energy threshold (photons) $E_{thres}$ & \multicolumn{3}{l|}{10\,\mev (20\,\mev tolerable)}\\
energy threshold (single crystal) $E_{xtl}$ & \multicolumn{3}{l|}{3\,\mev}\\
rms noise (energy equiv.) $\sigma_{E,noise}$ & \multicolumn{3}{l|}{1\,\mev}\\
angular coverage \%\,4$\pi$ & \multicolumn{3}{l|}{99\,\%}\\
mean-time-between-failures  $t_{mtbf}$ &  \multicolumn{3}{l|}{ 2000\,y}\\
(for individual channel) & & & \\
\hline
Subdetector specific properties & backward & barrel & forward \\
 & ($\geq 140^\circ$) & ($\geq 22^\circ$) & ($\geq 5^\circ$) \\
\hline
energy range from $E_{thres}$ to & 0.7\,\gev & 7.3\,\gev & 14.6 \,\gev \\
angular equivalent of crystal size $\theta$ & \multicolumn{2}{c}{4$^\circ$} & 1$^\circ$ \\
spatial resolution $\sigma_{\theta}$ & 0.5$^\circ$  & 0.3$^\circ$  & 0.1$^\circ$ \\
maximum signal load $f_{\gamma}$ ($E_\gamma>E_{xtl}$) & \multicolumn{2}{c}{60\,kHz} & 500\,kHz \\
($\ppbar$-events)
maximum signal load $f_{\gamma}$ ($E_\gamma>E_{xtl}$) & \multicolumn{2}{c}{100\,kHz} & 500\,kHz \\
(all events)
shaping time $t_s$ & \multicolumn{2}{c}{400\,ns} & 100\,ns \\
radiation hardness  & 0.15\,Gy & 7\,Gy & 125\,Gy \\
(maximum annual dose $\ppbar$-events) & & & \\
radiation hardness  & \multicolumn{2}{c}{10\,Gy} & 125\,Gy \\
(maximum annual dose from all events) & & & \\
\hline\hline
\end{tabular}
\caption[Main requirements for the \Panda EMC.]
{Main requirements for the \Panda EMC. Rates and doses are based on a luminosity of $L=2\cdot\,10^{32}cm^{-1}s^{-1}$.}

\label{tbl:req:summary}
\end{center}
\end{table*}

To achieve the required very low energy threshold, the light yield has
to be maximized.  Therefore improved lead-tungstate crystals are employed with a
light output twice as high as used in \INST{CMS}. These crystals
are operated at -25 $\degC$ which increases the light output by
another factor of four. In addition, large area APDs are used for
readout, providing high quantum efficiency and an active area four
times larger than used in \INST{CMS}.

The largest sub-detector is the barrel calorimeter with 11360 crystals
of $200\,\mm$ length. In the backward direction 592 crystals provide
hermeticity at worse resolution due to the presence of readout and
supply lines of other detectors. The 3600 crystals in the forward
direction face a much higher range of particle rates across the
acceptance of the calorimeter in the forward endcap. A readout with
vacuum phototriodes is foreseen to be able to safely operate at the
higher particle rates and correspondingly higher radiation load.

The crystal calorimeter is complemented in the forward spectrometer with a shashlik type sampling calorimeter consisting of 1404 modules of $55\times 55\,\mm^2$ cell size
covering 2.97 $\times$ 1.43$\,\m^2$.

This document presents the details of the technical design of the
electromagnetic crystal calorimeter of the \PANDA target spectrometer.

\section{The \Panda Electromagnetic Calorimeter}
\subsection{PWO-II Scintillator Material}
\label{sec:exs:scint}

The concept of \Panda places the \TSEMC inside the super-conducting coil of the solenoid. Therefore, the basic requirements of the appropriate scintillator material are compactness to minimize the radial thickness of the calorimeter layer, fast response to cope with high interaction rates, sufficient energy resolution and efficiency over the wide dynamic range of photon energies given by the physics program, and finally an adequate radiation hardness. In order to fulfill these requirements, even a compact geometrical design must provide a high granularity leading to a large quantity of crystal elements. Their fabrication relies on existing proven technology for mass production to guarantee the necessary homogeneity of the whole calorimeter. Presently and even in the near future there is no alternative material besides lead tungstate available.

The intrinsic parameters of PbWO$_4$ (PWO) as developed for CMS/ECAL are meeting all requirements, except for the light output. Therefore, an extended R$\&$D program was initiated to improve the luminescence combined with an operation at low temperatures such as T=-25$\degC$. Several successful steps have been taken, which have lead to a significantly improved material labelled as \PWOII.
The research was aiming at an increase of the structural perfection of the crystal and the optimized activation with luminescent impurity centers, which have a large cross section to capture electrons from the conduction band, combined with a sufficiently short delay of radiative recombination.
Beyond the improvement of the crystal quality, the reduction of the thermal quenching of the luminescence process by cooling the crystals leads to an additional increase of the light yield without any considerable distortion of the scintillation kinetics. Operating at a temperature of T=-25$\degC$ provides an overall gain factor of about 4 compared to T=+25$\degC$. As a result, full size crystals with $200\,\mm$ length deliver a light yield of 17-20$\,$photoelectrons per MeV (phe/MeV) at 18$\degC$ measured with a photomultiplier with bi-alkali photocathode (quantum efficiency $\sim$20$\percent$). 

The radiation level will be well below typical LHC values by at
least two orders of magnitude even at the most forward direction and
be much lower at larger angles. The radiation hardness and the effective loss of light yield as a function of dose rate and accumulated total dose are well known from the studies for CMS but investigated at room temperature. The effective deterioration of the optical transmission during the experiment is caused by the interplay of damaging and recovering mechanisms. The latter are fast at room temperature and keep the loss of light yield moderate. As shown by detailed investigations, at temperatures well below 0$\degC$ the relaxation times of color centers become extremely slow, reaching values well above 200 hours. As a consequence, there is a continuous and asymptotic reduction of the scintillation output. However, the present quality of \PWOII has reached induced absorption values well below the limits of the CMS crystals. A saturation is reached after an integral dose of 30-50$\,$Gy, leading to a maximum loss of light yield of 30\percent. To emphasize, these effects have to be considered only
in the very forward region of the \FWEMC. Taking the radiation-damage effects in forward-angle detectors into account, the proposed operation of the calorimeter at T=-25$\degC$ can be based on a net increase of the light yield by a factor of 3 compared to room temperature operation in addition to the improvement due to crystal quality and a more effective arrangement of the photosensors. A detector being back at room temperature will recover from radiation damage within 1-2 weeks.

The performance parameters are based on a large quantity of full size crystal samples and the experience of a pre-production run of 710 crystals comprising all geometrical shapes to complete one slice of the barrel. Detailed specifications as well as a quality control program have been elaborated, taking into account the experience of the CMS/ECAL collaboration. 

The requested crystal quality can be provided at the moment by the Russian manufacturer BTCP. The only alternative producer SICCAS in China has not yet delivered full size samples of comparable quality. 

Based on the ongoing developments of \PWOII crystals,
a detailed program has been worked out for the quality
assurance of the crystals. This
includes the preparation of various facilities for irradiation studies. The quality control is planned to be
performed at CERN, taking advantage of the existing infrastructure and
experience developed for CMS. One of the two ACCOS machines,
semi-automatic robots, is presently getting modified for the different
specification limits and geometrical dimensions of the \Panda
crystals.

\subsection{APD and VPT Photo Detectors}
\label{sec:exs:photo}

The low
energy threshold of 10$\,\MeV$ for the \Panda EMC requires the usage of excellent photo detectors. The magnetic field of about $2\,$T precludes the use of conventional
photomultipliers. On the other hand the signal generated by ionization
in a PIN photodiode by a traversing charged particle is too large for
our applications. To solve these problems a photosensor insensitive to
magnetic fields and with a small response to ionizing radiation has to be
used.

Since lead tungstate has a relatively low light yield, the photosensor
is required to have an internal gain in addition.  Due to the
envisaged operation temperature of $T = -25 \degC$ for the EMC, not only the \PWOII crystals but also the
used photo detectors have to be radiation hard in this temperature
region, which implies detailed studies concerning possibly occurring
radiation damages.

For the \BEMC an Avalanche Photodiode (APD) which has an internal
signal amplification (gain) in the silicon structure is chosen as
photo detector.  To maximize the light yield the coverage of the readout 
surface of the crystals has to be as large as possible, leading
to the development of Large Area APDs (\LAAPD) with an active area of
10$\times$10$\,\mm^2$. Their characterisation has been focused on the
given requirements.

Several prototypes of \LAAPDs have been studied for their
characteristics, especially for their radiation hardness: These
devices have been exposed to proton, photon and neutron radiation,
while their dark currents have been monitored to
analyse the damage of the surface and of the bulk structure.

The response to hadrons was
investigated and will provide complementary information to the
particle tracking or even identification of muons based on the energy
deposition or different cluster multiplicity of trespassing hadrons.

In addition to several screenings done using monochromatic light, certain properties of the diodes could only be determined
by scanning the complete spectrum of visible light.  The latter
procedure is not only used for studying the quantum efficiency but
also to detect effects on the structure of an \LAAPD after
irradiation.  These studies enable us to identify the best structure
of the APDs fulfilling the requirements of the \Panda EMC.

The operation of an \LAAPD requires the correct knowledge about its
gain factor.  Since this condition changes with the operation
temperature, the screening of this photo detector is being done at
room temperature as well as at the operating temperature of $T = -25
\degC$ to study its gain-voltage dependence.  During the screening
process the knowledge about this voltage dependence will be used for
grouping \LAAPDs in the detector to enable the usage of one HV line
for one block of sensors. Again we take advantage of the CMS experience for the screening of the photo sensors.

During the development of the readout system for the EMC the typical
size of the crystal readout surface of 27$\times$27$\,\mm^2$ 
lead to the development of an \LAAPD with a rectangular shape (active area: 
14$\times$6.8$\,\mm^2$) to
cover a maximum of this space with two neighboring photo detectors. The new geometry allows to fit two sensors on
each crystal irrespectively of its individual shape.

In contrast to the \BEMC, the \FWEMC has to deal with rates up to
500$\,\kHz$ per crystal and magnetic fields up to 1.2$\,$T.  Therefore
vacuum phototriodes (VPTs) have been chosen for the photo detection in
this EMC part due to the following reasons: rate capability, radiation
hardness, absence of nuclear counter effect and absence of temperature
dependence. Standard photomultipliers are excluded due to the magnetic
field environment. Different to the barrel region, the magnetic field
is oriented in the axial direction of the VPTs and thus makes it
feasible to use them for the endcap readout.  Vacuum phototriodes are
essentially a photomultiplier tube with only one dynode and weak field
dependence. Tests and the parameters presented are based on VPTs produced for the CMS experiment. A new VPT with a significantly higher quantum efficiency combined with a larger internal gain is under development and will be produced by the company Photonis and will be tested as soon as it becomes available to us.

The operation of these photo detectors at different angles relative to
the orientation of a magnetic field as well as the high rate
capability are being studied. These studies are being done both at room
temperature as well as at the envisaged EMC operation temperature of
$T = -25 \degC$.

\subsection{Electronics}
\label{sec:exs:elo}
The \Panda Electromagnetic Calorimeter (EMC) will provide an almost full coverage of the final state phase space for photons and electrons. The low-energy photon threshold will be around 10$\,\MeV$, while the threshold for individual detectors can be as low as 3$\,\MeV$ with correspondingly low noise levels of 1$\,\MeV$. Such settings allow the precision spectroscopy of charmonium states and transitions. With a dynamic range of 12000 for the readout electronics, a maximum photon energy deposition of 12$\,\GeV$ per crystal can be detected which allows the study of neutral decays of charmed mesons at the maximum beam energy of the HESR. Typical event rates of 10$\,\kHz$ and maximum 100$\,\kHz$ are expected for the barrel part of the calorimeter and up to 500$\,\kHz$ in the forward endcap. Two different photo detectors, \LAAPD and vacuum photo triodes (VPTs) will be used. The photo sensors are directly attached to the end faces of the individual crystals and the preamplifier is placed as close as possible inside the calorimeter volume for optimum performance and minimum space requirements. Since every \BEMC crystal is equipped with two APDs, not only redundancy is achieved but also a significantly (max. $\sqrt{2}$) improved signal to noise ratio and a lower effective threshold level.

The readout of small and compact subarrays of crystals requires very small preamplifier geometries. The low-temperature environment of the EMC will improve the noise performance of the analogue circuits. For efficient cooling and stable temperature behavior, low power consumption electronics has been developed in combination with extremely low-noise performance.

Two complementary low-noise and low-power (LNP) charge-sensitive preamplifier-shaper (LNP-P) circuits have been developed.
The LNP design based on discrete components utilizes a low-noise J-FET transistor. The circuit achieves a very good noise performance using signal shaping with a peaking time of 650$\,\ns$.
This preamplifier will be used for the readout of the \FWEMC for which we expect a single-channel rate up to 500$\,\kHz$. Such an approach minimizes the overall power consumption and keeps the probability for pileup events at a moderate level well below 1\percent.
The noise floor of the LNP-P at -25$\degC$, loaded with an input capacitance of 22~pF, has a typical equivalent noise charge (ENC) of 235 e$^-$ (rms) using signal shaping with a peaking-time of 650$\,\ns$. Because the VPT has almost no dark current, the noise is not increased due to the leakage current of that photo detector. 
By applying the quantum efficiency and the internal gain of the VPT, the ENC of 235 e$^-$(rms) corresponds to an energy noise level of 0.8 MeV (rms). 

The second circuit for the \LAAPD readout is a state-of-the-art CMOS ASIC (APFEL), which achieves a similar noise performance with a shorter peaking time of 250$\,\ns$. The advantage of the CMOS ASIC is the very low power consumption.
The noise floor of the APFEL ASIC at -20$\degC$, loaded with an input capacitance of 270$\,\pF$, has a typical equivalent noise charge (ENC) of 4150 e$^-$ (rms). Based on measurements of \PWOII light production and the amplification characteristics of the photon sensor, this corresponds to an energy noise level of 0.9 MeV (rms).
This is about the same level as achieved in the \FWEMC with the VPT readout. Both systems will allow an energy threshold of 3$\,\MeV$ and thus fulfill the requirements. 
The ASIC was designed in a 350$\,\nm$ CMOS technology. The power consumption at -20$\degC$ is 52$\,\mW$ per channel.
Prototypes have been produced on Multi Project Wafer (MPW) runs of the EUROPRACTICE IC prototyping program. 
For the instrumentation of the electromagnetic calorimeter about 23000 pieces of the preamplifier are needed for the \BEMC. These amounts of ASICs can no longer be produced cost effectively with MPW runs, thus a chip production campaign has to be started in due time.

The readout of the electromagnetic calorimeter is based on the digitization of the 
amplified signal-shape response of \LAAPD and VPT photo sensors to the light output of \PWOII crystals. 
The digitizer modules are located at a distance of 20--30$\,\cm$ and 90--100$\,\cm$ for the \BEMC and the \FWEMC, respectively, away from the analogue circuits and outside the cold volume. Signal transfer from the front-end over short distances is achieved by flat cables with low thermal budget.
The digitizers consist of high-frequency, low-power pipelined ADC chips, which continuously sample the amplified and shaped signals. 
With an 80$\,\MHz$ sampling ADC a time resolution better than 1$\,\ns$ has been achieved at energy deposits above 60$\,\MeV$ and 150$\,\ps$ at energies above 500$\,\MeV$. At the lower energies the time resolution is limited by APD- and preamplifier-noise. This time resolution is sufficient for maintaining a good event correlation and rejecting
background hits or random noise. 
The sampling is followed by the digital logic, which processes time-discrete digital values, detects hits and forwards hit-related information to the multiplexer module via optical fibers.
The multiplexer modules will be located in the DAQ hut and they perform advanced signal processing to extract amplitude and signal-time information.

The front-end electronics of the \BEMC is located inside the solenoid magnet where any access for maintenance or repair is limited to shutdown periods of the HESR, expected to occur once a year. Redundancy in the system architecture is achieved by arranging the digitization of the two APDs of every crystal in two blocks, respectively, and providing interconnections at the level of FPGA using high-speed links.
A prototype ADC module has been developed containing 32 channels of 12 bit 65 MSPS ADCs. The total power consumption of the module is 15$\,\W$.

The EMC and its subcomponents will be embedded in the general Detector Control System (DCS) structure of the complete \Panda detector. The aim of a DCS is to ensure the correct and stable operation of an experiment, so that the data taken by the detector are of high quality. The scope of the DCS is therefore very wide and includes all subsystems and other individual elements involved in the control and monitoring of the detector.

\subsection{Mechanics and Integration}
\label{sec:exs:mech}

The calorimeter design in the target region is in full accordance with
the constraints imposed by a fixed target experiment with the strong
focusing of the momenta in forward direction. A nearly full coverage of $\sim$
99$\percent$ solid angle in the center-of-mass system is guaranteed in
combination with the forward electromagnetic calorimeter, which is
located downstream beyond the dipole magnet. The granularity is
adapted to the tolerable maximum count rate of the individual modules
and the optimum shower distribution for energy and position
reconstruction by minimizing energy losses due to dead material. The
front sizes of the crystal elements cover a nearly identical
laboratory solid angle and have absolute cross sections close to the
Moli\`ere radius.

The mechanical design is composed of three parts: the barrel part for a length of 2.5\,m and 0.57\,m of inner radius; 
the forward endcap for a diameter of 2\,m located at 2.1\,m downstream from the target; and the backward endcap of 0.8\,m in diameter located at 1\,m upstream from the target.
The conceptual design of these elements is equivalent. The basic \PWO crystal shape is a truncated pyramid of 200$\,\mm$ length and this principle is based on the ``flat-pack'' configuration used in the \INST{CMS} calorimeter.  Right angle corners are introduced in order to simplify the CAD design and the mechanical manufacturing process to reduce machining costs. The presented geometry foresees that the crystals are not pointing toward the target position. A tilt of a few degrees is added on the focal axis to reduce the dead zone effect and to ensure that particles entering between crystals will always cross a significant part of a crystal. The crystals are wrapped in a foil of a high reflectivity (98\,\%) and inserted in carbon fiber alveoles to hold them by the back. This helps also to avoid any piling-up stress and any heavy material in front of the crystals. The nominal distance between crystals is around 600$\,\mum$ taking into account the thickness of the reflector (2x65$\,\mum$), carbon fiber alveoles (2x200$\,\mum$) and mechanical free gaps.   

All crystals have a common length of $200\,\mm$
corresponding to 22 radiation lengths, which allows optimum shower
containment up to $15\,\GeV$ photon energy and limits the nuclear
counter effect in the subsequent photo sensor to a tolerable level.

The necessary thermal shield for the low-temperature operation of the EMC is made of panels using either thin components with high thermal resistivity and low material budget (in terms of $\,X_0$) in front of the crystals or thicker standard foam in the other areas. The cooling circuit is composed of carbon fiber or copper thermal screens depending on the position. Silicone oil, Syltherm XLT has been chosen as cooling liquid for its low viscosity and high thermal efficiency. To avoid any moisture or ice, dry nitrogen gas is flowing through the detector which therefore has to be made airtight. 
All these elements, crystals, front-end electronics and thermal screens are sustained by a metallic support structure at room temperature connected to the mainframe of the \Panda magnet or yoke, for the barrel or the endcaps, respectively. On these structures boards and various services are implemented as the digitizing, optical fibers for calibration and data transfer, or cables for the supply of electrical power and sensors for the detector control system. The electronics itself consists of a low-noise and low-power consumption charge sensitive preamplifier connected to the digitizer part by flat cables to reduce the heat transfer.

The mechanical designs of the barrel and the endcaps differ in some details. The barrel is divided in 16 slices of 710 crystals each and 11 different shapes of crystals are necessary to fill the volume. The average crystal has approximatively a square front face of 21.3$\,\mm$ and a square rear face (readout) of 27.3$\,\mm$ for an average mass of 0.98$\,\kg$. Each slice is divided in 6 modules for an easier construction and assembled to a stainless steel support beam. Two rings connect all these slices to make a compact cylindrical calorimeter inserted into the magnet by special tools. All the barrel services are going through the yoke on its backward side. 

The endcaps are designed as a wall structure with quadrant symmetry. Each quarter is composed of subunit structures composed of 16 crystals each inserted in carbon fiber alveoles and screwed through an interface insert to a thick aluminum back plate. Contrary to the barrel, only one shape of crystal is necessary for the endcaps and the average dimensions are for the front face 24.4$\,\mm$ and for the rear face 26$\,\mm$.

The feasibility of the construction of such devices has been validated by the construction of several prototypes. The last one and the most representative is a 60-crystal prototype which uses some crystals from the barrel and integrates all the different elements described above. It permits  to check the mechanical design and to focus on the result of the temperature stability which shows a promising result of $\pm$0.05$\degC$.
In a next iteration, a mechanical prototype with 200 crystals is presently under construction. This device is primarily meant for studying stable operation and cooling by simulating two adjacent detector slices of the barrel section of the EMC.

\subsection{Calibration and Monitoring}
\label{sec:exs:cal}

The ultimate performance of the \Emc can only be reached with a precise calibration of the individual crystal channels. The energy resolution of less than 2\% at energies above $1\,\GeV$ demands a precision at the 
sub-percent level.
To reach this goal, three methods are combined: A pre calibration with cosmic muons, in situ calibration with physics events and continuous monitoring with a light-pulser system.

Before the start up of the experiment all crystals will be calibrated in situ with cosmic muons at a level of 10\% accuracy. The correspondence between the light output of a muon and a photon will be determined with test beams. The position reconstruction algorithm will also be optimized at test beams.

The final calibration will be performed with physics events during data taking.
Events with three to four $\piz$ or $\eta$ mesons in the final state serve as an input to apply the calibration by constraining the energy measurements to the invariant mass of the meson. A dedicated software trigger will select the events based on total energy and multiplicity at a rate of about $4\,\kHz$. 
An integrated luminosity of $8\,\pb^{-1}$, accumulated in less than a day will be enough to perform a full calibration. 
With the hardware-trigger free concept of \PANDA the full event selection relies on the online trigger system. An efficient operation requires a quick calibration of the \Emc. This will be achieved by combining the calibration constants from the previous day with the information from the light-pulser system.

The amount of collected light per MeV may change slightly due to radiation damage. Radiation damage affects only the optical transmission and can thus be corrected for by measuring the effect with the light-pulser system.With a light-pulser system operating at several wavelengths ($455\,\nm, 530\,\nm$ and $660\,\nm$) the effects from radiation damage can be disentangled from other effects such as changes of gain in the photodetectors and preamplifiers. The wavelength at $455\,\nm$ is close to the emission wavelength at $420\,\nm$. A dominant
defect center due to Molybdenum is at $530\,\nm$ and far in the red
spectral region one does not expect any radiation damage so that one can control separately the readout chain including the photo sensor. 

Ultra-bright LEDs driven by electronic circuits reproducing the time dependence of the \PWO scintillation light will serve as the light source. The light will be distributed to the crystal back face by radiation hard silica/silica fibers.
The normalizations of the pulses are done with temperature stabilized Si PIN photodiodes. A test system showed stabilities of 0.1 to 0.2\% over a day, which is enough to follow the variations of the light output up to the next full calibration.

A
first prototype of the light-pulser system is already implemented into the PROTO60 array and
operating.

\subsection{Simulation}
\label{sec:exs:sim}
The simulation studies are focused on the expected performance of the planned
EMC with respect to the energy and spatial resolution of reconstructed photons, on
the capability of an electron hadron separation, and also on the feasibility of the 
planned physics program of \Panda.

The software follows an object oriented approach, and most of the code
is written in C++. Several proven software tools and packages from other
HEP experiments have been adapted to the \Panda needs and are in use. It contains
event generators with proper decay models for all particles and resonances involved in
the individual physics channels, particle tracking through the complete \Panda 
detector by using the GEANT4 transport code, a digitization which models the signals 
and the signal processing in the front-end electronics of the individual detectors,
the reconstruction of charged and neutral particles as well as user friendly 
high-level analysis tools.

The digitization of the EMC has been realized with realistic properties of \PWO
crystals at the operational temperature at -25$\degC$. A Gaussian distribution of 
$\sigma$ = 1$\,\mev$ has been used for the constant electronics noise.
The statistical fluctuations were estimated by 80 phe/\mev produced in the LAAPD with
an excess noise factor of 1.38. This results in a photo statistic noise term
of 0.41\% /$\sqrt{E}$. A comparison with a 3x3 crystal array test measurement  
demonstrates that this digitization gives sufficiently realistic results. The simulated 
line shape at discrete photon energies as well as the energy resolution as a function of 
the incident photon energy are in good agreement with the measurements.

Different EMC detector scenarios have been investigated in terms of energy and 
spatial resolutions. 
While the variation of the crystal length (15~cm, 17~cm and 20~cm) show nearly the 
same results for photons below 300~\mev, the performance gets significantly 
better for higher energetic photons for longer crystals. The 20 cm 
setup yields an energy resolution of 1.5\% for 1~\gev photons, and even 
$<$ 1\% for photons above 3$\,\gev$. Another important aspect is the choice of
the individual crystal reconstruction threshold, which is driven by the electronics
noise term. Comparisons of the achievable resolution for the most realistic 
scenario with a noise term of $\sigma$ = 1$\,\mev$
and a individual crystal reconstruction threshold of $\extl=3\,\mev$ and a worse case
($\sigma$ = 3 $\mev$, $\extl=9$ $\mev$) show that the degradation increases by more 
than a factor of 2 for the lowest photon energies. This result 
demonstrates clearly that the single crystal threshold has a strong influence on the 
energy resolution. 

The high granularity of the planned EMC provides an excellent position resolution
for photons. Based on a standard cluster finding and bump splitting algorithm 
a $\sigma$-resolution of less than 3~$\mm$ can be obtained for energies above 1$\,\gev$.
This corresponds to roughly 10\% of the crystal size.

Apart from accurate measurements of photons, the EMC is also the most powerful detector 
for the identification of electrons. Suitable properties for the distinction to hadrons
and muons are the ratio of the energy deposit in the calorimeter to the 
reconstructed track momentum ($E/p$) and shower shape variables derived from the 
cluster. Good electron identification can be achieved with a Multilayer Perceptron. 
For momenta above 1$\,\gevc$ an electron efficiency of greater than 98\% can
be obtained while the contamination by other particles is substantially less than 1\%.

The reconstruction efficiency of the EMC is affected by the interaction of particles with
material in front of the calorimeter. The largest contribution to the material budget
comes from the Cherenkov detectors, which consist of quartz radiators of 1-2$\,\cm$. This
corresponds to a radiation length between 17\% and 50\%, depending on the polar angle.

Four different benchmark channels relevant to the EMC have been investigated in order to demonstrate the 
feasibility of the planned physics program of \Panda. The studies cover charmonium 
spectroscopy, the search of charmed hybrids as well as the measurement of the time-like
electromagnetic formfactors of the proton. All channels have in common that the 
particle detection with the electromagnetic calorimeter plays an essential role.

The first charmonium channel is the production of $h_c$ in the formation mode. One of the 
main decay channels of this singlet P wave state ($1 ^{1}P_1$) is the electromagnetic 
transition to the ground state $\eta_c$ 
($\pbarp \, \to \, h_c \, \to \, \eta_{c} \, \gamma$). The 
analysis is based on the $\eta_c \, \to \, \phi \, \phi$ decay mode with 
$\phi \, \to \, K^{+} \, K^{-}$. The three background channels 
$\pbarp \to K^{+} K^{-} K^{+} K^{-} \pi^{0}$, $\pbarp \to \phi K^{+} K^{-} \pi^{0}$ and
$\pbarp \to \phi \phi \pi^{0}$ are considered as the main contributors, 
having a few orders of magnitude higher cross sections.
With one $\gamma$-ray from a $\pi^{0}$ decay left undetected, these reactions have 
the same list of decay products as the studied $h_c$ decay, and thus the $\gamma$ 
reconstruction threshold plays an essential role for the background suppression. A signal
to background ratio of better than 3 can be obtained with a reasonable photon reconstruction
threshold of 10$\,\mev$. With a 30$\,\mev$ threshold instead the signal to background ratio
decreases by 20\% to 40\%.

The second charmonium channel is the formation of the recently discovered vector-state 
$Y(4260)$ in the reaction $\pbarp \, \to \, Y(4260) \, \to \, \jpsi\piz\piz$. The
challenge of this channel is to achieve an efficient and clean electron identification
for the reconstruction of $\jpsi\,\to\,\ee$, and also an accurate measurement of the 
final state photons originating from the $\piz$ decays. After an event selection by applying
kinematical and vertex fits, the reconstruction efficiency is found to be 14\%, whereas the
suppression rate for the background channels 
$\pbarp\,\to\,\jpsi\,\eta\,\eta$ and $\pbarp\,\to\,\jpsi\,\eta\,\piz$ is better than $10^4$.
Another source of background which has been investigated are
non-resonant $\pbarp\,\to\,\pip\,\pim\,\piz\,\piz$ events. Due to the expected high cross
section of this channel a suppression of at least $10^7$ is required. With the currently 
available amount of MC events a suppression better than $10^7$ is obtained.        

Closely connected with charmonium spectroscopy is the search for charmonium hybrids. The  
ground state $\psi_g$ is generally expected to be a 
spin-exotic $J^{PC}=1^{-+}$ state within the mass range of $4.1\,\gevcc$ and $4.4\,\gevcc$.  
In \pbarp 
annihilations this state can be produced only in association with one or more recoil particles.
In the study, the decay of the charmonium hybrid to $\chi_{c1}\piz\piz$ with the subsequent
radiative $\chi_{c1}\to\jpsi\gamma$ decay is considered. The recoiling meson is reconstructed
from the decay $\eta\to\gamma\gamma$. The reconstruction efficiency after all selection criteria
is $4\%$
and $6\%$ for events with $\jpsi\to\ee$ and $\jpsi\to\mumu$ decays, respectively. One major
background source are events with hidden charm, in particular events including a \jpsi meson.
The suppression is found to be $7\cdot 10^3$ for the channel 
$\pbarp\,\to\,\chi_{c0}\,\piz\,\piz\,\eta$, 
$3\cdot 10^4$ for $\pbarp\,\to\,\chi_{c1}\,\piz\,\eta\,\eta$ and $1\cdot 10^5$ for 
$\pbarp\,\to\,\chi_{c1}\,\piz\,\piz\,\piz\,\eta$ and thus low contamination of the $\psi_g$ 
signal from these
background reactions is expected for the foreseen \Panda EMC. 

The 4th channel is related to the measurement of the time-like electromagnetic formfactors of 
the proton. The electric ($G_E$) and magnetic ($G_M$) form factors can be described by
analytic functions of the four momentum transfer $q^2$ ranging from $q^2=-\infty$ to
$q^2=+\infty$. The \pbarp annihilation process allows to access positive $q^2$ (time like) 
starting from 
the threshold of $q^2 = 4 m_p^2$. In this region $G_E$ and $G_M$ become complex functions and 
their determination for low to intermediate momentum transfers is an open question. At \Panda the
region between 5~$(\gevc)^2$ and 22~$(\gevc)^2$ can be accessed. The factors $|G_E|$ and $|G_M|$ 
can be derived from the angular distribution of $\pbarp\to\ee$ events. For a precise
determination of the form factors a suppression better than $10^8$ of the dominant background 
from $\pbarp\,\to\,\pip\,\pim$ is required. With the currently available amount of 
MC events the background
rejection of $\pbarp\to\pip\pim$ is found to be better than $10^8$. Furthermore it was shown
that the angular distributions can be obtained with a sufficient accuracy 
for the determination of  $G_E$ and $G_M$.   

\subsection{Performance}
\label{sec:exs:perf}

To prove the concept for the EMC, test experiments have concentrated on the
response to photons and charged particles at energies below $1\,\GeV$
since those results are dominated by the photon statistics of the
scintillator, the sensitivity and efficiency of the photo sensor and
the noise contributions of the front-end electronics. The conclusions
are drawn based on crystal arrays comprising up to 25 modules. The
individual crystals have a length of $200\,\mm$ and a rectangular shape
with a cross section of $20\times20\,\mm^{2}$. Only the most recently
assembled array PROTO60 consists of 60 crystals in \Panda
geometry. The tapered shape will improve the light collection due to
the focusing effect of the geometry as known from detailed simulations
at \INST{CMS/ECAL}.

The performance tests completed up to now have been aiming at two
complementary aspects. On one hand, the quality of full size \PWOII
crystals has to be verified in in-beam measurements with energy-tagged
photons covering the most critical energy range up to
$1\,\GeV$. Therefore, the scintillator modules were read out with
standard photomultiplier tubes (Philips XP1911) with a bi-alkali
photocathode, which covers $\sim$35$\percent$ of the crystal endface
with a typical quantum efficiency QE=18$\percent$. The noise
contribution of the sensor can be neglected and the fast response
allows an estimate of the time response. The achieved resolution deduced at various operating temperatures can be
considered as benchmark limits for further studies including
simulations and electronics development. The achieved resolution
represent excellent lower limits of the performance to be
expected. Operation at T=-25$\degC$ using a photomultiplier readout
delivers an energy resolution of $\sigma / E = 0.95\percent
/\sqrt{E/\GeV} + 0.91\percent$ for a 3$\times$3 sub-array accompanied
with time resolutions below $\sigma$=$130\,\ps$.

From the experimental tests one can extrapolate and confirm an energy resolution at an
operating temperature of T$=-25\degC$, which will
be well below 2.5$\percent$ at $1\,\GeV$ photon energy.  The
resolution of 13$\percent$ at the lowest investigated shower energy of
$20\,\MeV$ reflects the excellent statistical term. A threshold of $3\,\MeV$ is expected for an individual detector channel in the final setup. Relevant for the
efficient detection and reconstruction of multi-photon events, an
effective energy threshold of $10\,\MeV$ can be considered for the
whole calorimeter as a starting value for cluster identification,
which will enable us to disentangle from the measured data even
physics channels with extremely low cross section.

The second R$\&$D activity intended to come close to the final readout
concept with large area avalanche photodiodes (\LAAPD), which are
mandatory for the operation within the magnetic field. All the
reported results are obtained by collecting and converting the
scintillation light only with a single quadratic \LAAPD of
$10\times10\,\mm^{2}$ active area with a quantum efficiency above
60$\percent$ using newly developed low-noise preamplifiers but
commercial electronics for the digitization. Not even taking advantage
of an operation at the lowest temperature, an energy resolution of
$\sigma / E = 1.86\percent /\sqrt{E/\GeV} + 0.65\percent$ has been
achieved at T=0$\degC$ for a 3$\times$3 sub-array, which would come
very close to a measurement using photomultiplier readout under
similar conditions. In addition timing information can be expected
with an accuracy well below $1\,\ns$ for energy depositions in a
single crystal above $100\,\MeV$.  The present data document
experimentally only the lower limit of the envisaged performance.

As mentioned above, radiation damage might reduce asymptotically the light output at the
most forward angles, since relaxation processes become very slow
at low temperatures. However, due to the further improved radiation
hardness of the crystals, the loss of light yield will stay below
30$\percent$. Combining all improvements including the foreseen sensor
concept the \TSEMC can rely on a factor of 15-20 higher light yield
compared to \INST{CMS/ECAL}.

In order to study the operation of large arrays, the mechanical
support structures, cooling and temperature stabilization concepts and
long term stabilities, a large prototype comprising 60 tapered
crystals in \Panda-geometry has been designed and brought into
operation. First in-beam tests are scheduled for summer 2008.

\section{Conclusion}
\label{sec:exs:conclusion}
The intrinsic performance parameters of the present quality of \PWOII,
such as luminescence yield, decay kinetics and radiation hardness and
the additional gain in light yield due to cooling down to T=-25$\degC$
fulfill the basic requirements for the electromagnetic calorimeter. The general applicability of \PWO for calorimetry in
High-Energy Physics has been promoted and finally proven by the
successful realization of the \INST{CMS/ECAL} detector as well as the photon
spectrometer \INST{ALICE/PHOS}, both installed at LHC. The necessary
mass production of high-quality crystals has been achieved at least at
BTCP in Russia.

The operation at low temperatures imposes a technological challenge on
the mechanical design, the cooling concept and thermal insulation
under the constraints of a minimum material budget of dead
material. Detailed simulations and prototyping have confirmed the
concept and high accuracy has been achieved in temperature
stabilization taking into account also realistic scenarios of the
power consumption of the front-end electronics. 

Concepts for the photon sensors and the readout electronics have been tested. As the most
sensitive element, a prototype of a custom designed ASIC implementing
pre-amplification and shaping stages has been successfully brought
into operation and will provide a large dynamic range of 12,000 with a
typical noise level corresponding to $\sim1\,\MeV$. 

The overall performance of the calorimeter will be controlled by
injecting light from LED-sources distributed via optical fibers at the
rear face of the crystal. 

Based on the ongoing developments of \PWOII crystals and \LAAPDs,
respectively, a detailed program has been elaborated for quality
assurance of the crystals and screening of the photo sensors. 

The general layout of the mechanical structure is completed including
first estimates of the integration into the \Panda detector. The
concepts for signal- and HV-cables, cooling, slow control, monitoring
as well as the stepwise assembly are worked out to guarantee that the
crystal geometries could be finalized. Prototypes of the individual
crystal containers, based on carbon fiber alveoles, have been
fabricated and tested and are already implemented in the PROTO60
device.

The experimental data together with the elaborate design concepts and
simulations show that the ambitious physics program of \Panda can
be fully explored based on the measurement of
electromagnetic probes, such as photons, electrons/positrons or the
reconstruction of the invariant mass of neutral mesons.

%

%
%

%
%
\cleardoublepage
\chapter{Overview of the PANDA Experiment}
\label{sec:int}

%
%
\PANDA represents the efforts of the world-wide hadron physics
community to attack the most fundamental and burning problems in the
strong interaction. Given the complexity of the strong interaction in
the non-perturbative regime this requires studies of rather diverse
topics. The \PANDA experiment should complement the new antiproton
accelerator complex at the FAIR facility in Darmstadt. The high-energy
storage ring HESR will deliver antiproton beams of unprecedented
precision and intensity. It is a definite challenge for
experimentalists to build an experiment which on one side covers an
``as broad as possible'' range of physics topics and on the other side
has to go beyond the precision reached in the past with specialized
setups.  The proposed detector clearly has to be as modern and
versatile as possible to fulfill the physics needs without
jeopardizing quality.

The concept of \PANDA is based on the experience from previous
experiments in the field like the Crystal Barrel and OBELIX detectors
at LEAR, the E835 experiment at Fermilab, the running FINUDA
experiment at Frascati, and takes into account the concepts, which
have been developed and implemented for the most modern LHC
experiments. \PANDA is a hermetic detector for charged and neutral
particles in the energy range of 10 MeV up to 10 GeV. Precise
micro-vertex tracking is mandatory as is good particle
identification. The possibility to combine all these elements of
information simultaneously for each given event gives the \PANDA
detector its unprecedented power, both in the ability to establish
(rather than just suggest) new unexpected phenomena and in the
redundant identification of interesting processes predicted by present
models. Since the ratio of interesting events to backgrounds is often
rather small we shall need all the capability that we can provide.

Clearly the design choices for a detector represent a well-thought
balance between physics needs and the available resources. The hadron
physics community will not have a second detector besides \PANDA for
this physics available and hence the detector has to be sufficiently
robust, redundant and resistant to radiation damage for an operation
over many years. Superb calibration and monitoring capabilities must
be present for all subsystems. The data rate of 2$\EE{7}$ antiproton
annihilations per second poses not only a challenge for the individual
detectors but also for the data acquisition system and the online data
selection.

In the cost/performance optimization one has to distinguish between
items that at a later stage could be improved by upgrades if the need
arises and items where scope reductions remain forever. To the latter
belong in our opinion the choice of magnets and expensive items like
calorimeters and the overall performance of a central tracker. Even
though savings could be achieved by reducing even their parameters, we
believe that going below the levels proposed in this document would
lead to unacceptable technical and performance degradation. The
current report should justify the parameter choices in view of the
physics that should be achieved.

%

%
%
\section{The Physics Case}
\label{sec:phys}

One of the most challenging and fascinating 
goals of modern physics is the achievement of a fully quantitative
understanding of the strong interaction, which is the subject of hadron physics.
Significant progress has been achieved over the past few years thanks to
considerable advances in experiment and theory. New experimental results
have stimulated a very intense theoretical activity and a refinement of
the theoretical tools. 

Still there are many fundamental questions which remain basically
unanswered.
Phenomena such as the confinement of quarks, the existence of glueballs 
and hybrids, the origin of the masses of hadrons in the context of the
breaking of chiral symmetry are long-standing puzzles and represent
the intellectual challenge in our attempt to understand the nature of the
strong interaction and of hadronic matter.

Experimentally, studies of hadron structure can be performed with different
probes such as electrons, pions, kaons, protons or antiprotons.
In antiproton-proton annihilation particles with gluonic degrees of freedom
as well as particle-antiparticle pairs are copiously produced,
allowing spectroscopic studies with very high statistics and precision.
Therefore, antiprotons are an excellent tool to address the open problems.

The recently approved FAIR facility (Facility for Antiproton and Ion
Research), which will be built as a major upgrade of the existing GSI
laboratory in Germany, will provide antiproton beams of the highest
quality in terms of intensity and resolution, which will provide an
excellent tool to answer the aforementioned fundamental questions.

The PANDA experiment (Pbar ANnihilations at DArmstadt) will use the
antiproton beam from the High-Energy Storage Ring (HESR) colliding
with an internal proton target at a CMS energy between 2.2 GeV and 5.5
GeV within a general purpose spectrometer to carry out a rich and
diversified hadron physics program.

The experiment is being designed to fully exploit the extraordinary 
physics potential arising from the availability of high-intensity, cooled
antiproton beams.
The aim of the rich experimental program is to improve our knowledge of the 
strong interaction and of hadron structure.
Significant progress beyond the present understanding of the field is expected
thanks to improvements in statistics and precision of the data.

Many measurements are foreseen in PANDA, part of which can be carried 
out in parallel. In the following the main topics of the PANDA physics
program are outlined.

\paragraph*{Charmonium spectroscopy.} The $\ccbar$ spectrum
can be computed within the framework of non-relativistic potential models and, more
recently, in Lattice QCD. A precise measurement of all states below and above
open charm threshold is of fundamental importance for a better understanding of QCD.
All charmonium states can be formed directly in $\pbarp$ annihilation.

At full luminosity PANDA will be able to collect several thousand $\ccbar$ states
per day.
By means of fine scans it will be possible to measure masses with accuracies of the
order of 100 keV and widths to 10\% or better.
The entire energy region below and above open charm threshold will be explored.

\paragraph*{Search for gluonic excitations (hybrids and glueballs).}
One of the main challenges of hadron physics
is the search for gluonic excitations, i.e. hadrons in which the
gluons can act as principal components.
These gluonic hadrons fall into two main categories: glueballs, i.e.
states of pure glue, and hybrids, which consist of a $\qqbar$ pair and excited glue.
The additional degrees of freedom carried by gluons allow these hybrids and glueballs
to have $J^{PC}$ exotic quantum numbers: in this case mixing effects with nearby
$\qqbar$ states are excluded and this makes their experimental identification easier.
The properties of glueballs and hybrids are determined by the long-distance features
of QCD and their study will yield fundamental insight into the structure of the QCD
vacuum.
Antiproton-proton annihilations provide a very favourable environment to search for gluonic hadrons. 

\paragraph*{Study of hadrons in nuclear matter.}
The study of medium modifications of hadrons embedded in hadronic matter
is aimed at understanding the origin of hadron masses in the context
of spontaneous chiral symmetry breaking in QCD and its partial restoration
in a hadronic environment. 
So far experiments have been focused on the light quark sector.
The high-intensity $\pbar$ beam of up to 15\ GeV/c will allow an extension of
this program to the charm sector both for hadrons with hidden and open charm.
The in-medium masses of these states are expected to be affected primarily
by the gluon condensate.

Another study which can be carried out in PANDA is the measurement of $\jpsi$
and D meson production cross sections in $\pbar$ annihilation
on a series of nuclear targets. The
comparison of the resonant $\jpsi$ yield obtained from $\pbar$ annihilation
on protons and different nuclear targets allows to deduce the
$\jpsi$-nucleus dissociation cross section, a fundamental parameter to 
understand $\jpsi$ suppression in relativistic heavy ion collisions interpreted
as a signal for quark-gluon plasma formation. 

\paragraph*{Open charm spectroscopy.} 
The HESR running at full luminosity and at $\pbar$
momenta larger than 6.4\ GeV/c would produce a large number
of $D$ meson pairs.
The high yield (e.g. 100 charm pairs per second around the $\psi$(4040)) and
the well defined production kinematics of $D$ meson pairs would allow to
carry out a significant charmed meson spectroscopy program which would include,
for example, the rich $D$ and $D_s$ meson spectra.

\paragraph*{Hypernuclear physics.}
Hypernuclei are systems in which up or down quarks are replaced by
strange quarks. In this way a new quantum number, strangeness, is
introduced into the nucleus.  Although single and double
$\Lambda$-hypernuclei were discovered many decades ago, only 6 events
of double $\Lambda$-hypernuclei were observed up to now. The
availability of $\pbar$ beams at FAIR will allow efficient production
of hypernuclei with more than one strange hadron, making PANDA
competitive with planned dedicated facilities. This will open new
perspectives for nuclear structure spectroscopy and for studying the
forces between hyperons and nucleons.

\paragraph*{Electromagnetic Processes.}
In addition to the spectroscopic studies described above, \PANDA will
be able to investigate the structure of the nucleon using
electromagnetic processes, such as Wide Angle Compton Scattering
(WACS) and the process $\pbarp \to \ee$, which will allow the
determination of the electromagnetic form factors of the proton in the
timelike region over an extended $q^2$ region.
%

%
%



\newcommand{\q}[2]{\ensuremath{#1\ \mathrm{#2}}}
\newcommand{\psip}{\ensuremath{\psi (2S)}}
\newcommand{\psinc}{\ensuremath{\jpsi + X}}
\newcommand{\eeX}{\ensuremath{e^+ e^- X}}
\newcommand{\Gin}{\ensuremath{\Gamma_{\mathrm{in}}}}
\newcommand{\Gout}{\ensuremath{\Gamma_{\mathrm{out}}}}
\newcommand{\Gee}{\ensuremath{\Gamma_{\ee}}}
\newcommand{\Gpp}{\ensuremath{\Gamma_{\pbarp}}}
\newcommand{\GGG}{\ensuremath{\Gee \Gpp / \Gamma}}
\newcommand{\GiGoG}{\ensuremath{\Gin \Gout / \Gamma}}
\newcommand{\Lumn}{\ensuremath{\mathcal{L}}}
\newcommand{\lik}{\ensuremath{\Lambda}}
\newcommand{\sBW}{\ensuremath{\sigma_\mathrm{BW}}}
\newcommand{\sBWr}{\ensuremath{\sigma_\mathrm{BWr}}}
\newcommand{\scf}{\ensuremath{b}}
\newcommand{\sbkg}{\ensuremath{\sigma_\mathrm{bkg}}}
\newcommand{\Bi}{\ensuremath{B_\mathrm{in}}}
\newcommand{\Bo}{\ensuremath{B_\mathrm{out}}}
\newcommand{\fcav}{\ensuremath{f^\mathrm{cav}}}
\newcommand{\frf}{\ensuremath{f^\mathrm{rf}}}
\newcommand{\Vrf}{\ensuremath{V^\mathrm{rf}}}
\newcommand{\vrf}{\ensuremath{v^\mathrm{rf}}}
\newcommand{\betarf}{\ensuremath{\beta^\mathrm{rf}}}
\newcommand{\gammarf}{\ensuremath{\gamma^\mathrm{rf}}}
\newcommand{\wrf}{\ensuremath{w^\mathrm{rf}}}
\newcommand{\frfz}{\ensuremath{f^\mathrm{rf}_0}}
\newcommand{\frfi}{\ensuremath{f^\mathrm{rf}_i}}
\newcommand{\dL}{\ensuremath{\Delta L}}
\newcommand{\reff}{\ensuremath{\varepsilon^X_\mathrm{co}/\varepsilon^X_\mathrm{cf}}}
\newcommand{\muee}{\ensuremath{\mu^{ee}}}
\newcommand{\Nee}{\ensuremath{N^{ee}}}
\newcommand{\effee}{\ensuremath{\varepsilon^{ee}}}
\newcommand{\muX}{\ensuremath{\mu^{X}}}
\newcommand{\NX}{\ensuremath{N^{X}}}
\newcommand{\effXcf}{\ensuremath{\varepsilon^{X}_\mathrm{cf}}}
\newcommand{\resG}{\q{290 \pm 25 \mathrm{(sta)} \pm 4 \mathrm{(sys)}}{keV}}
\newcommand{\resA}{\q{579 \pm 38 \mathrm{(sta)} \pm 36 \mathrm{(sys)}}{meV}}
\newcommand{\eperp}{\ensuremath{\varepsilon_\perp}}
\newcommand{\stot}{\ensuremath{\sigma_\mathrm{tot}}}
\newcommand{\tauloss}{\ensuremath{\tau_\mathrm{loss}}}
\newcommand{\tpbar}{\ensuremath{t_{\bar{p}}}}
\newcommand{\Lmax}{\ensuremath{L_\mathrm{max}}}
\newcommand{\Npbar}{\ensuremath{N_{\bar{p}}}}
\newcommand{\tprep}{\ensuremath{t_\mathrm{prep}}}
\newcommand{\texp}{\ensuremath{t_\mathrm{exp}}}
\newcommand{\tcycle}{\ensuremath{t_\mathrm{cycle}}}

\newcommand{\code}[1]{\textsc{#1}}



\section{The High Energy Storage Ring}
%
%
The FAIR facility at Darmstadt, Germany will provide anti-proton beams
with very high quality. A 30 GeV/c proton beam is used to produce
anti-protons which subsequently are accumulated and cooled at a
momentum of 3.7 GeV/c. A derandomized bunch of anti-protons is then fed
into the High Energy Storage Ring (HESR) which serves as slow
synchrotron to bring the anti-protons to the desired energy and then
as storage ring for internal target experiments. HESR will have both
stochastic and electron cooling to deliver high quality beams. 
\subsection{Overview of the HESR}

An important feature of the new anti-proton facility is the
combination of phase-space cooled beams and dense internal targets,
comprising challenging beam parameter in two operation modes:
high-luminosity mode (HL) with beam intensities up to 10$^{11}$, and
high-resolution mode (HR) with a momentum spread down to a few times 10$^{-5}$,
respectively. Powerful electron and stochastic cooling systems are
necessary to meet the experimental requirements.

\begin{figure*}[bhtp]
  \centering
    \resizebox{\dwidth}{!}{
      \includegraphics{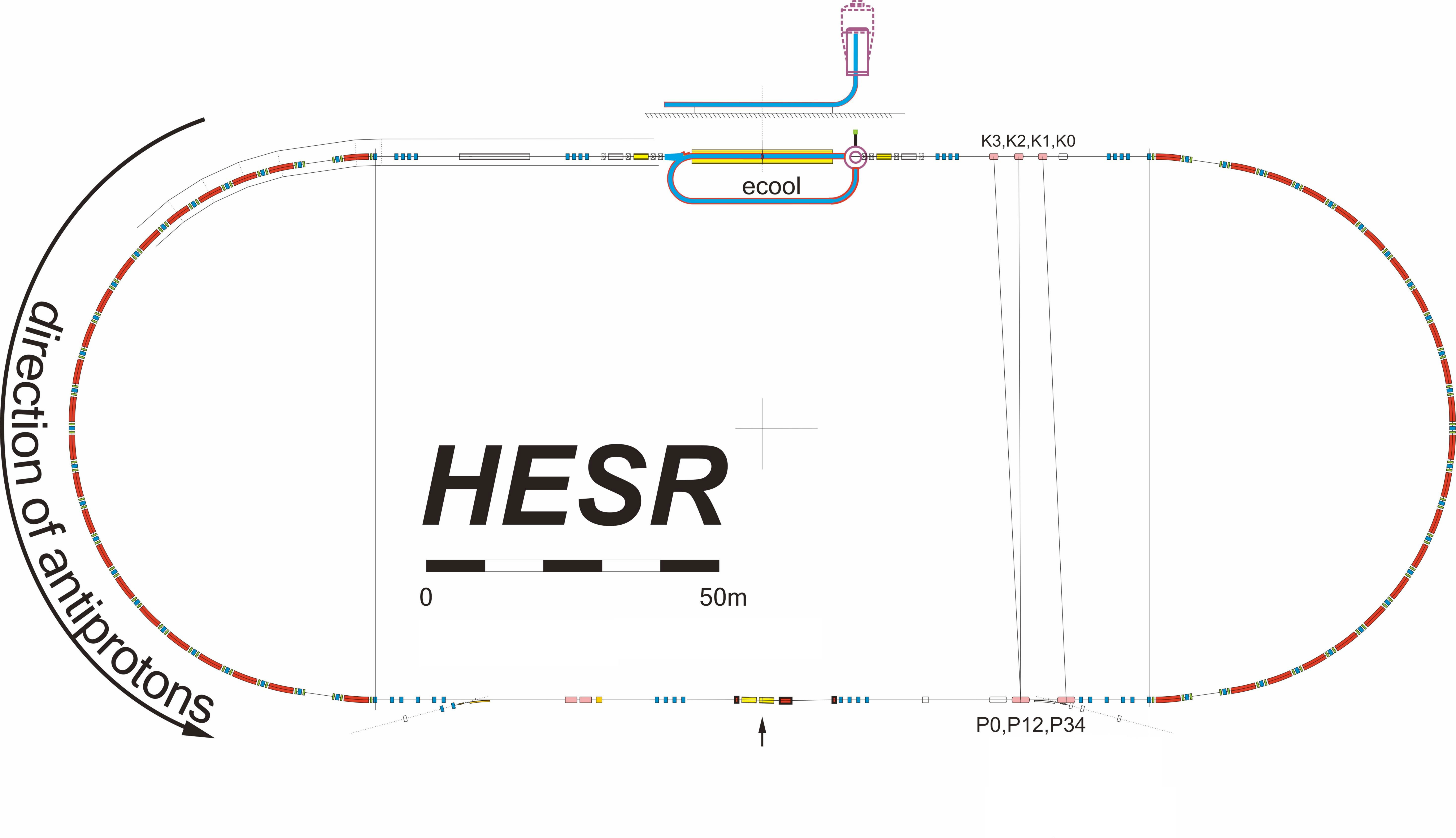}
    }
  \caption[Schematic view of the HESR.]
  {Schematic view of the HESR. Tentative positions for
    injection, cooling devices and experimental installations are
    indicated.}
  \label{fig:HESR}
\end{figure*}

The HESR lattice is designed as a racetrack shaped ring, consisting of
two 180$^\circ$ arc sections connected by two long straight sections. One
straight section will mainly be occupied by the electron cooler. The
other section will host the experimental installation with internal H$_2$
pellet target, RF cavities, injection kickers and septa
(see Fig.~\ref{fig:HESR}).
For stochastic cooling pickup and kicker tanks are also
located in the straight sections, opposite to each other. To improve
longitudinal stochastic cooling a third pickup location in the arc is
presently being investigated.
Special requirements for the lattice are dispersion free straight
sections and small betatron amplitudes in the range of a few meters at
the internal interaction point. In addition the betatron amplitudes at
the electron cooler are adjustable within a large range.

Table~\ref{tab:param} summarizes the specified injection parameters,
experimental requirements and operation modes.

\begin{table*}
\vspace*{0.5cm}
\label{tab:param}
\centering
{\small
\begin{tabular}{ll}
\multicolumn{2}{c}{\textbf{Injection Parameters}}\\ \hline
Transverse emittance &
\q{1}{mm\cdot mrad} (normalized, RMS) for $3.5\cdot 10^{10}$ particles, \\
 & scaling with number of accumulated particles: \(\eperp \sim N^{4/5}\) \\
Relative momentum spread &
$1\cdot 10^{-3}$ (normalized, RMS) for $3.5\cdot10^{10}$ particles, \\
 & scaling with number of accumulated particles: $\sigma_p/p \sim
N^{2/5}$ \\
Bunch length &
Below 200~m \\
Injection Momentum &
\q{3.8}{GeV/c} \\
Injection &
Kicker injection using multi-harmonic RF cavities \\
 & \\
\multicolumn{2}{c}{\textbf{Experimental Requirements}}\\ \hline
Ion species & Antiprotons \\
$\bar{p}$ production rate &
\q{2\cdot 10^7}{/s} ($1.2\cdot 10^{10}$ per 10~min) \\
Momentum / Kinetic energy range &
1.5 to \q{15}{GeV/c} / 0.83 to \q{14.1}{GeV} \\
Number of particles stored in HESR&
$10^{10}$ to $10^{11}$ \\
Target thickness &
\q{4\cdot 10^{15}}{atoms/cm^2} \\
Transverse emittance &
\q{< 1}{mm\cdot mrad} \\
Betatron amplitude E-Cooler &
25--200~m \\
Betatron amplitude at IP &
1--15~m \\
 & \\
\multicolumn{2}{c}{\textbf{Operation Modes}}\\ \hline
High resolution (HR) &
Luminosity of \q{2\cdot 10^{31}}{cm^{-2} s^{-1}} for $10^{10}\
\bar{p}$ \\
 & RMS momentum spread $\sigma_p / p \leq 4\cdot 10^{-5}$, \\
 & 1.5 to \q{9}{GeV/c}, electron cooling  up to \q{9}{GeV/c} \\
High luminosity (HL) &
Luminosity of \q{2\cdot 10^{32}}{cm^{-2} s^{-1}} for $10^{11}\
\bar{p}$ \\
 & RMS momentum spread $\sigma_p / p \sim 10^{-4}$, \\
 & 1.5 to \q{15}{GeV/c}, stochastic cooling  above \q{3.8}{GeV/c}
\end{tabular}
}
\caption[Injection parameters, experimental requirements and operation
  modes.]
{Injection parameters, experimental requirements and operation
  modes.}
\vspace*{0.5cm}
\end{table*}

\subsection{Beam Cooling}

Beam equilibrium is of major concern for the high-resolution
mode. Calculations of beam equilibria for beam cooling, intra-beam
scattering and beam-target interaction are being performed utilizing
different simulation codes like BETACOOL (JINR, Dubna), MOCAC (ITEP,
Moscow), and PTARGET (GSI, Darmstadt). Cooled beam equilibria
calculations including special features of pellet targets have been
carried out with a simulation code based on PTARGET.

An electron cooler is realized by an electron beam with up to 1~A
current, accelerated in special accelerator columns to energies in the
range of 0.4 to 4.5~MeV for the HESR. The 22~m long solenoidal field
in the cooler section has a field range from 0.2 to 0.5~T with a
magnetic field straightness in the order of $10^{-5}$~\cite{FAIR:2006}.  
This arrangement allows beam cooling for beam momenta between
\q{1.5}{GeV/c} and \q{8.9}{GeV/c}.

In the HR mode RMS relative momentum spreads of less than $4\cdot
10^{-5}$ can be achieved with electron cooling.

The main stochastic cooling parameters were determined for a cooling
system utilizing quarter-wave loop pickups and kickers with a
band-width of 2 to 4~GHz. Stochastic cooling is presently specified
above \q{3.8}{GeV/c}~\cite{stockhorst:2006}. Applying stochastic
cooling one can achieve an RMS relative momentum spread of 3 to
$4\cdot 10^{-5}$ for the HR mode. In the HL mode an RMS relative momentum
spread slightly below $10^{-4}$ can be expected. Transverse stochastic
cooling can be adjusted independently to ensure sufficient beam-target
overlap.

\subsection{Luminosity Estimates}

Beam losses are the main restriction for high luminosities, since the
antiproton production rate is limited. Three dominating contributions
of beam-target interaction have been identified: Hadronic interaction,
single Coulomb scattering and energy straggling of the circulating
beam in the target. In addition, single intra-beam scattering due to
the Touschek effect has also to be considered for beam lifetime
estimates. Beam losses due to residual gas scattering can be neglected
compared to beam-target interaction, if the vacuum is better than
\q{10^{-9}}{mbar}. A detailed analysis of all beam loss processes can be found
in~\cite{lehrach:2006,hinterberger:2006}.

\begin{figure}[hbtp]
  \begin{center}
    \begin{tabular}{c}
    \includegraphics[width=0.48\dwidth]{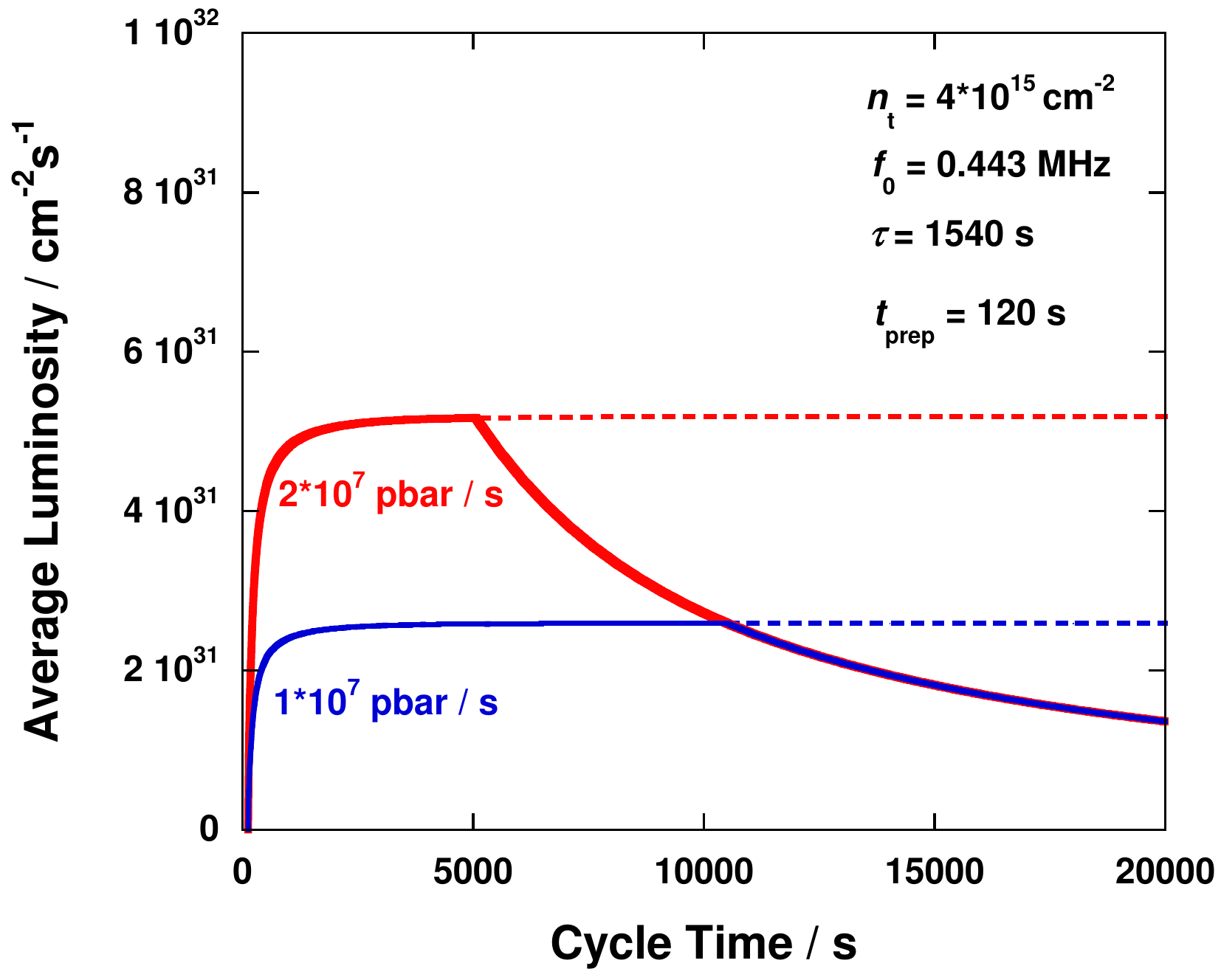} 
    \\
    \includegraphics[width=0.48\dwidth]{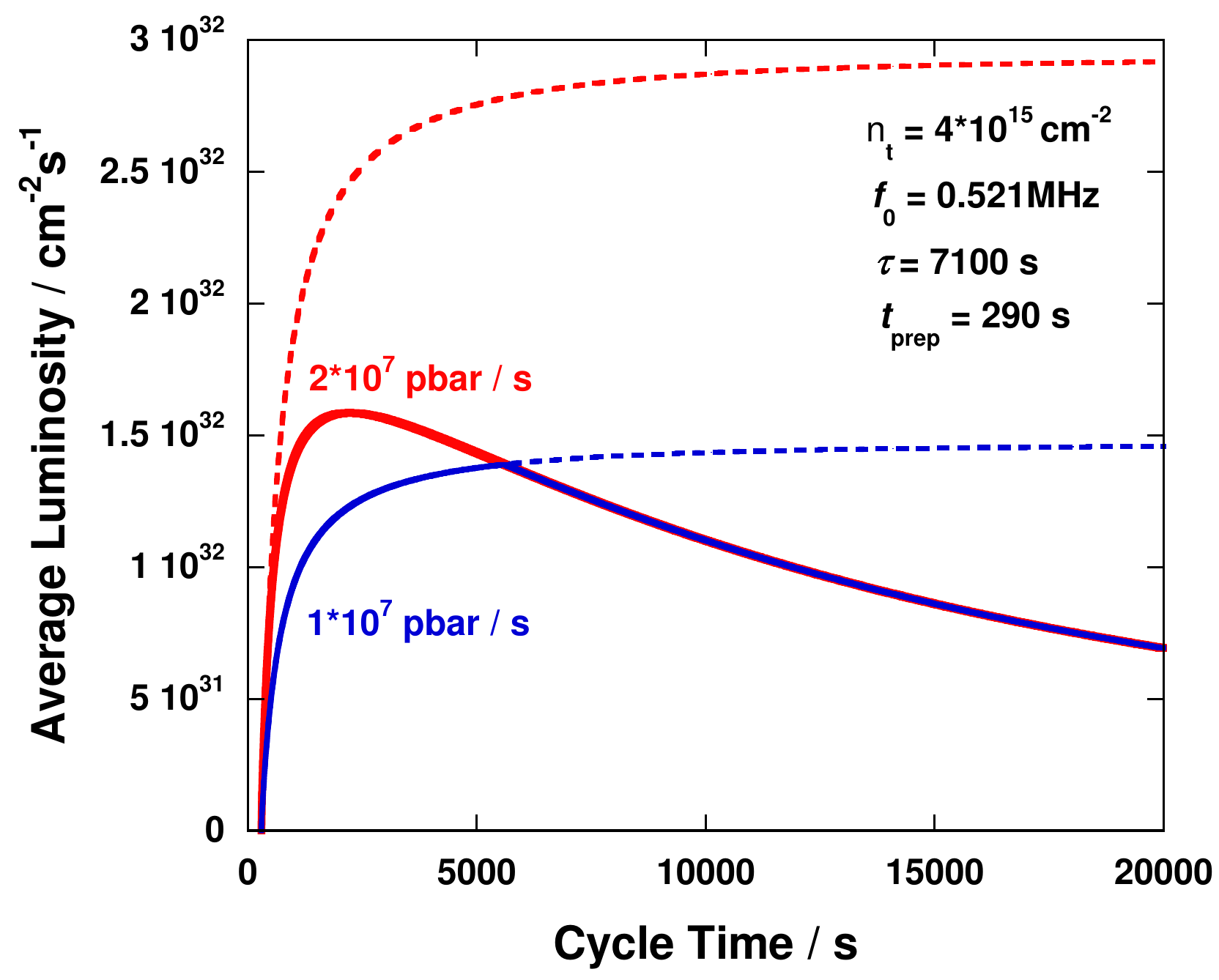}
  \end{tabular}
  \caption[Average luminosity vs. cycle time.]
  {Average luminosity vs. cycle time at 1.5 (top) and
    \q{15}{GeV/c} (bottom). The maximum number of particles is limited
    to $10^{11}$ (solid line), and unlimited (dashed lines).}
  \label{fig:avglum}
\end{center}
\end{figure}

The maximum luminosity depends on the antiproton production rate
\( d\Npbar / dt = \q{2\cdot 10^7}{/s} \) and loss rate
\begin{equation}
\Lmax = \frac{d\Npbar / dt}{\stot}
\end{equation}
and ranges from \q{0.8 \cdot 10^{32}}{cm^{-2} s^{-1}} at \q{1.5}{GeV/c}
to \q{3.9 \cdot 10^{32}}{cm^{-2} s^{-1}} at \q{15}{GeV/c}.

To calculate the average luminosity, machine cycles and beam
preparation times have to be specified. After injection, the beam is
pre-cooled to equilibrium (with target off) at \q{3.8}{GeV/c}. The beam is
then ac-/decelerated to the desired beam momentum. A maximum ramp rate
for the superconducting dipole magnets of 25~mT/s is specified. After
reaching the final momentum beam steering and focusing in the target
and beam cooler region takes place. Total beam preparation time
\tprep\ ranges from 120~s for \q{1.5}{GeV/c} to 290~s for
\q{15}{GeV/c}. 

In the high-luminosity mode, particles should be re-used in the next
cycle. Therefore the used beam is transferred back to the injection
momentum and merged with the newly injected beam. A bucket scheme
utilizing broad-band cavities is foreseen for beam injection and the
refill procedure. During acceleration 1\% and during deceleration 5\%
beam losses are assumed.  The average luminosity reads
\begin{equation}
\bar{L} = f_0 N_{i,0} n_t
  \frac{\tau \left[ 1 - e^{-\texp/\tau} \right]}{\texp + \tprep}
\end{equation}
where~$\tau$ is the $1/e$ beam lifetime, \texp\ the experimental time
(beam on target time), and \tcycle\ the total time of the cycle, with
\( \tcycle = \texp + \tprep \). 
The dependence of the average luminosity on the cycle time is shown
for different antiproton production rates in Fig.~\ref{fig:avglum}.

With limited number of antiprotons of $10^{11}$, as specified for the
high-luminosity mode, average luminosities of up to
\q{1.6\cdot 10^{32}}{cm^{-2} s^{-1}} can be achieved at \q{15}{GeV/c} for cycle
times of less than one beam lifetime. If one does not restrict the
number of available particles, cycle times should be longer to reach
maximum average luminosities close to
\q{3\cdot 10^{32}}{cm^{-2} s^{-1}}.
This is a theoretical upper limit, since the larger momentum spread of
the injected beam would lead to higher beam losses during injection
due to the limited longitudinal ring acceptance. For the lowest
momentum, more than $10^{11}$ particles can not be provided in
average, due to very short beam lifetimes. As expected, average
luminosities are below \q{10^{32}}{cm^{-2} s^{-1}}.

In the operation cycle of HESR the time for re-injection and
acceleration of a bunch of anti-protons can be used by the experiment 
to do pulser calibrations and tests required to monitor noise and
signal quality.

%
%
\begin{figure*}[tp]
\begin{center}
\includegraphics[width=\dwidth]{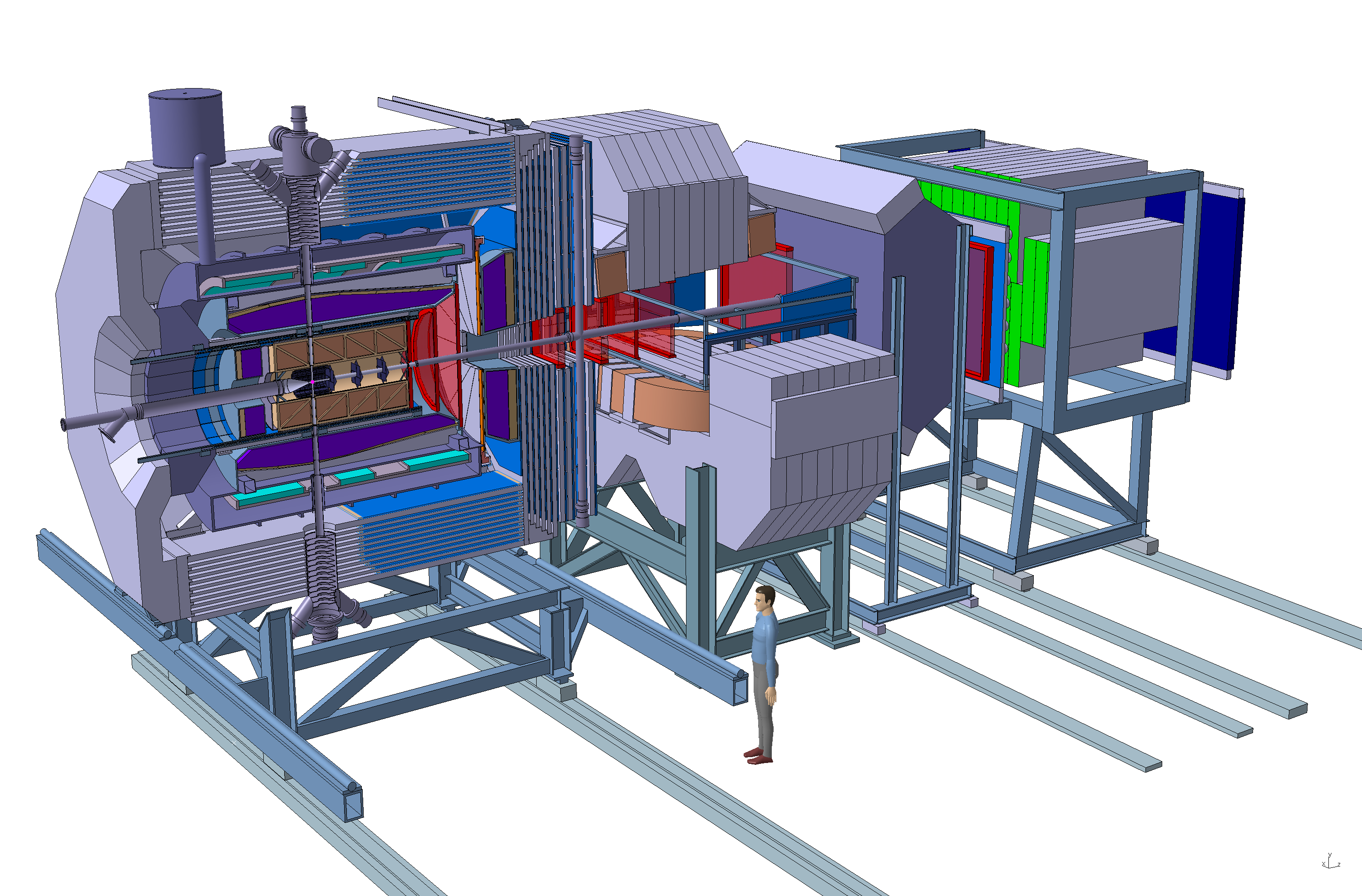}
\caption[Artistic view of the \PANDA Detector.]
{Artistic view of the \PANDA Detector.}
\label{fig:det:exp}
\end{center}
\end{figure*}

\section{The \PANDA Detector}
\label{sec:int:det}
%
%
The main objectives of the design of the \PANDA experiment pictured in
\Reffig{fig:det:exp} are to achieve $4\pi$ acceptance, high resolution
for tracking, particle identification and calorimetry, high rate
capabilities and a versatile readout and event selection. To obtain a
good momentum resolution the detector is split into a {\em target
  spectrometer} based on a superconducting solenoid magnet surrounding
the interaction point and measuring at high angles and a {\em forward
  spectrometer} based on a dipole magnet for small angle tracks. A
silicon vertex detector surrounds the interaction point. In both
spectrometer parts tracking, charged particle identification,
electromagnetic calorimetry and muon identification are available to
allow to detect the complete spectrum of final states relevant for the
\PANDA physics objectives.

In the following paragraphs the components
of all detector subsystems are briefly explained. 

\subsection{Target Spectrometer}
The target spectrometer surrounds the interaction point and
measures charged tracks in a solenoidal field of 2 T. In the manner
of a collider detector it contains detectors in an onion shell like
configuration. Pipes for the injection of target material have
to cross the spectrometer perpendicular to the beam pipe.

\begin{figure*}[hbt]
\begin{center}
\includegraphics[width=0.8\dwidth]{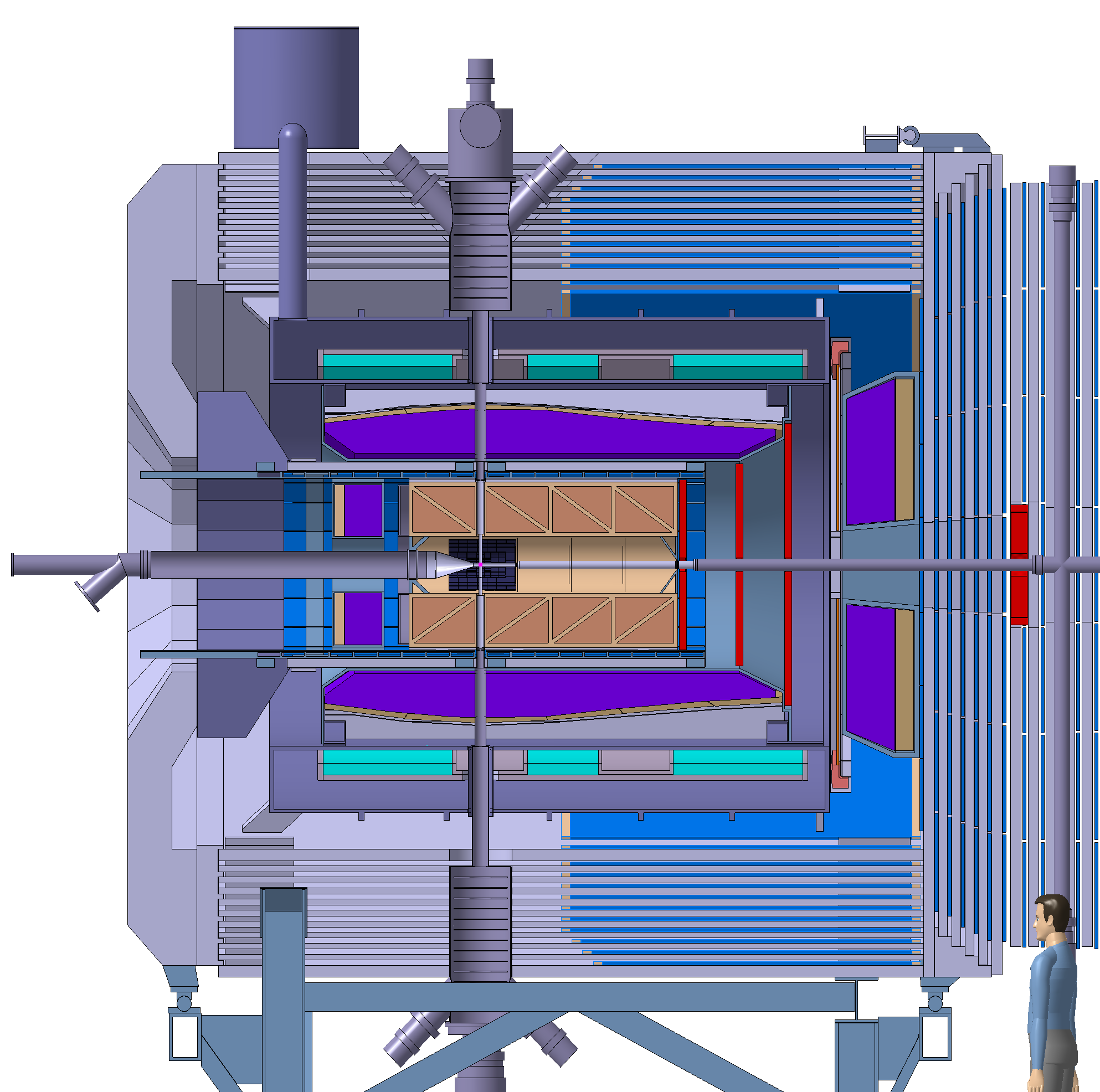}
\caption[Side view of the target spectrometer.]
{Side view of the target spectrometer.}
\label{fig:det:ts}
\end{center}
\end{figure*}

The target spectrometer is arranged in a barrel part for angles
larger than 22$\degrees$ and an endcap part for the forward range down 
to 5$\degrees$ in the vertical and 10$\degrees$ in the horizontal plane.
The target spectrometer is given in a side view in \Reffig{fig:det:ts}.

A main design requirement is compactness to avoid a too large
and a too costly magnet and crystal calorimeter.

\subsubsection{Target}
\label{sec:det:ts:tgt}

\begin{figure*}[hbt]
\begin{center}
\includegraphics[angle=90,width=\dwidth]{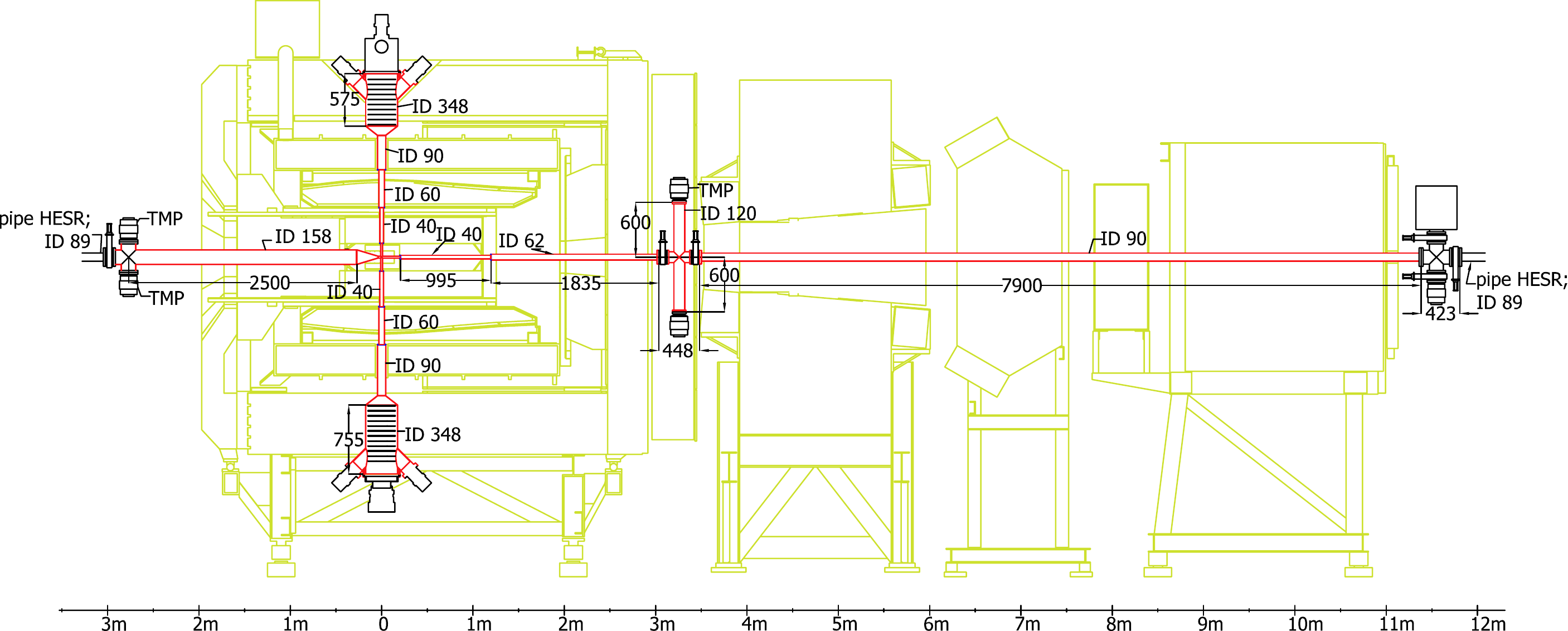}
\caption[Schematic of the target and beam pipe setup with pumps.]
{Schematic of the target and beam pipe setup with pumps.}
\label{fig:det:target}
\end{center}
\end{figure*}

The compact geometry of the detector layers nested inside the
solenoidal magnetic field combined with the request of minimal
distance from the interaction point to the vertex tracker leaves very
restricted space for the target installations.  The situation is
displayed in \Reffig{fig:det:target}, showing the intersection
between the antiproton beam pipe and the target pipe being gauged to
the available space.  In order to reach the design luminosity of
$2\EE{32}$ s$^{-1}$cm$^{-2}$ a target thickness of about
$4\EE{15}$ hydrogen atoms per cm$^2$ is required assuming
$10^{11}$ stored anti-protons in the HESR ring.

These are conditions posing a real challenge for an internal target
inside a storage ring. At present, two different, complementary
techniques for the internal target are developed further: the
cluster-jet target and the pellet target. Both techniques are capable
of providing sufficient densities for hydrogen at the interaction
point, but exhibit different properties concerning their effect
on the beam quality and the definition of the interaction point. In
addition, internal targets also of heavier gases, like deuterium,
nitrogen or argon can be made available.

For non-gaseous nuclear targets the situation is different in
particular in case of the planned hyper-nuclear experiment. In these
studies the whole upstream end cap and part of the inner detector
geometry will be modified.

\paragraph*{Cluster-Jet Target}
The expansion of pressurized cold hydrogen gas into vacuum through a
Laval-type nozzle leads to a condensation of hydrogen molecules
forming a narrow jet of hydrogen clusters. The cluster size varies
from $\EE{3}$ to $\EE{6}$ hydrogen molecules tending to become larger at higher
inlet pressure and lower nozzle temperatures.  Such a cluster-jet
with density of $\EE{15}$ atoms/cm$^3$ acts as a very diluted target since it
may be seen as a localized and homogeneous monolayer of hydrogen atoms
being passed by the antiprotons once per revolution.

Fulfilling the luminosity demand for \PANDA still requires a density
increase compared to current applications. Additionally, due to
detector constraints, the distance between the cluster-jet nozzle and
the target will be larger. The size of the target region will be given
by the lateral spread of hydrogen clusters. This width should stay
smaller than 10 mm when optimized with skimmers and collimators both
for maximum cluster flux as well as for minimum gas load in the
adjacent beam pipes. The great advantage of cluster targets is the
homogeneous density profile and the possibility to focus the
antiproton beam at highest phase space density. Hence, the interaction
point is defined transversely but has to be reconstructed
longitudinally in beam direction. In addition the low $\beta$-function
of the antiproton beam keeps the transverse beam target heating effects
at the minimum. The possibility of adjusting the target density along
with the gradual consumption of antiprotons for running at constant
luminosity will be an important feature.

\paragraph*{Pellet Target} 
The pellet target features a stream of frozen hydrogen micro-spheres,
called pellets, traversing the antiproton beam perpendicularly. A
pellet target presently is in use at the Wasa at COSY experiment.
Typical parameters for pellets at the interaction point are the rate
of 1.0 -1.5 $\EE{4}$ s$^{-1}$, the pellet size of 25 - 40 $\umu$m, and
the velocity of about 60 m/s. At the interaction point the pellet train
has a lateral spread of $\sigma\approx$ 1 mm and an interspacing of
pellets that varies between 0.5 to 5 mm. With proper adjustment of
the $\beta$-function of the coasting antiproton beam at the target
position, the design luminosity for \PANDA can be reached in time
average. The present R\&D is concentrating on minimizing the
luminosity variations such that the instantaneous interaction rate
does not exceed the acceptance of the detector systems.  Since a
single pellet becomes the vertex for more than hundred nuclear
interactions with antiprotons during the time a pellet traverses the
beam, it will be possible to determine the position of individual
pellets with the resolution of the micro-vertex detector averaged over
many events. R\&D is going on to devise an optical pellet tracking
system. Such a device could determine the vertex position to about 50
$\umu$m precision for each individual event independently of the
detector. It remains to be seen if this device can later be
implemented in \PANDA.

The production of deuterium pellets is also well established, the use
of other gases as pellet target material does not pose problems.

\paragraph*{Other Targets} are under consideration for the
hyper-nuclear studies where a separate target station upstream will
comprise primary and secondary target and detectors. Moreover, current
R\&D is undertaken for the development of a liquid helium target and a
polarized $^3$He target. A wire target may be employed to study
antiproton-nucleus interactions.

\subsubsection{Solenoid Magnet}
The magnetic field in the target spectrometer is provided by a
superconducting solenoid coil with an inner radius of 90$\,$cm and a
length of 2.8$\,$m. The maximum magnetic field is 2$\,$T. The field
homogeneity is foreseen to be better than 2 \% over the volume of the
vertex detector and central tracker.  In addition the transverse
component of the solenoid field should be as small as possible, in
order to allow a uniform drift of charges in the time projection
chamber. This is expressed by a limit of $\int B_r/B_z dz < 2$ mm for
the normalized integral of the radial field component.

In order to minimize the amount of material in front of the
electromagnetic calorimeter, the latter is placed inside the magnetic
coil. The tracking devices in the solenoid cover angles down to
5\degrees{}/10\degrees{} where momentum resolution is still
acceptable. The dipole magnet with a gap height of 1.4 m provides a
continuation of the angular coverage to smaller polar angles.

The cryostat for the solenoid coils has two warm bores of 100$\,$mm
diameter, one above and one below the target position, to
allow for insertion of internal targets.

\begin{figure*}[hbt]
\begin{center}
\includegraphics[width=0.8\dwidth]{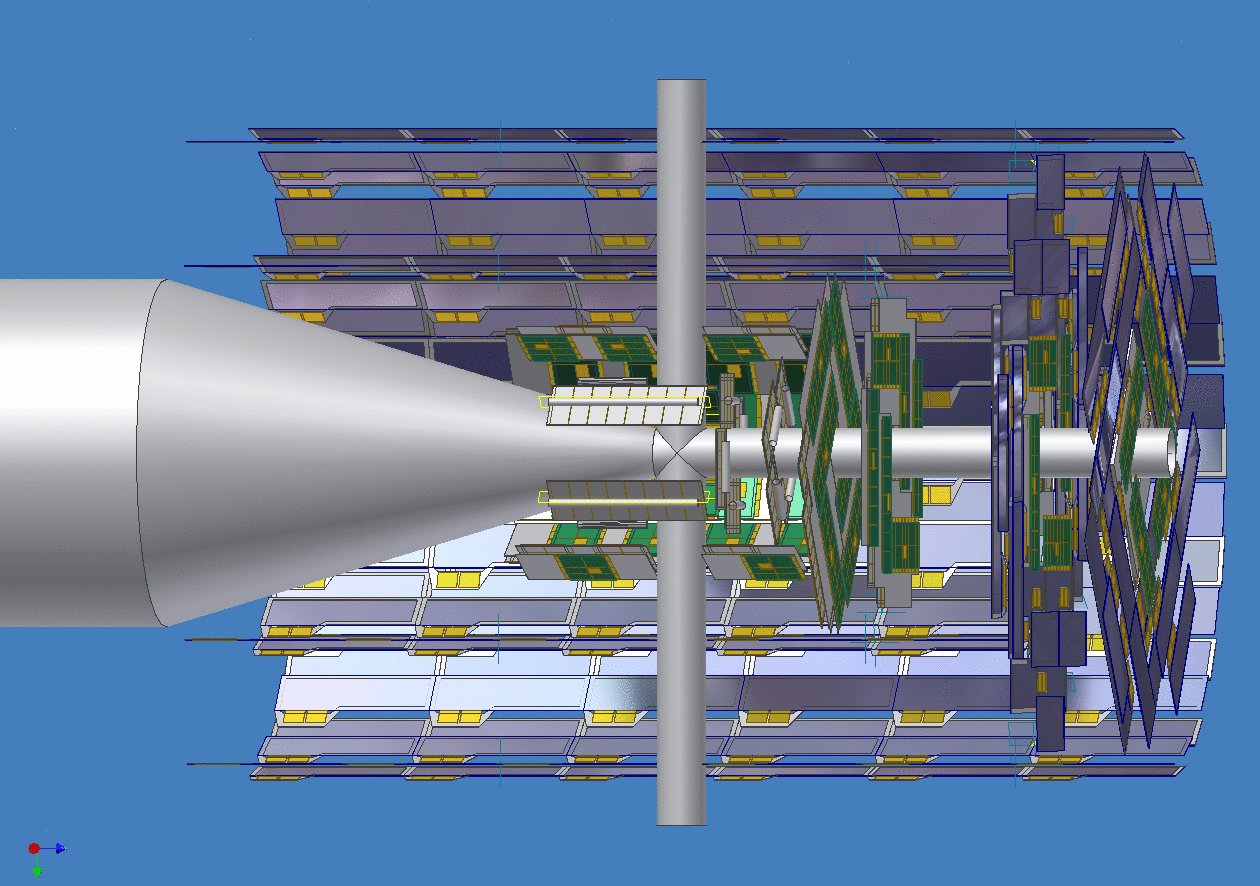}
\caption[The Microvertex detector of \PANDA]
{The Microvertex detector of \PANDA}
\label{fig:det:mvd}
\end{center}
\end{figure*}

\subsubsection{Microvertex Detector}
The design of the micro-vertex detector (\Mvd) for the target
spectrometer is optimized for the detection of secondary vertices from
\D and hyperon decays and maximum acceptance close to the interaction
point. It will also strongly improve the transverse momentum
resolution. The setup is depicted in \Reffig{fig:det:mvd}.
\par
The concept of the \Mvd is based on radiation hard silicon pixel
detectors with fast individual pixel readout circuits and silicon
strip detectors. The layout foresees a four layer barrel detector with
an inner radius of 2.5 cm and an outer radius of 13 cm. The two
innermost layers will consist of pixel detectors while the outer two
layers are considered to consist of double sided silicon strip
detectors.

Eight detector wheels arranged perpendicular to the beam will achieve
the best acceptance for the forward part of the particle spectrum.
Here again, the inner two layers are made entirely of pixel detectors,
the following four are a combination of strip detectors on the outer
radius and pixel detectors closer to the beam pipe. Finally the last 
two wheels, made entirely of silicon strip detectors, are placed
further downstream to achieve a better acceptance of hyperon cascades.

The present design of the pixel detectors comprises detector wafers
which are 200$\,\umu$m thick (0.25\%$\,X_0$).  The readout via
bump-bonded wafers with ASICs as it is used in {\INST{ATLAS}} and
{\INST{CMS}} \cite{Atlas:Tdr:11,Cms:Tdr:5} is foreseen as the default
solution.  It is highly parallelized and allows zero suppression as
well as the transfer of analog information at the same time. The
readout wafer has a thickness of 300 $\umu$m (0.37\%$\,X_0$). A pixel
readout chip based on a 0.13 $\umu$m CMOS technology is under
development for \PANDA. This chip allows smaller pixels, lower power
consumption and a continuously sampling readout without external
trigger.

Another important R\&D activity concerns the minimisation of the
material budget. Here strategies like the thinning of silicon wafers
and the use of ultra-light materials for the construction are
investigated.

\subsubsection{Central Tracker}
The charged particle tracking devices must handle the high particle
fluxes that are anticipated for a luminosity of up to
several $10^{32}\,$cm$^{-2}$s$^{-1}$.
The momentum resolution $\delta p/p$ has to be on the percent level.
The detectors should have good detection efficiency for secondary
vertices which can occur outside the inner vertex detector
(e.g.\  $\Ks$ or $\Lambda$).
This is achieved by the combination of the silicon vertex detectors
close to the interaction point (\Mvd) with two outer systems. One
system is covering a large area and is designed as a barrel around the
\Mvd. This will be either a stack of straw tubes (\Stt) or a
time-projection chamber (\Tpc). The forward angles will be covered using
three sets of GEM trackers similar to those developed for the
{\INST{COMPASS}} experiment at \INST{CERN}. The two options for
the central tracker are explained briefly in the following.

\paragraph*{Straw Tube Tracker (\Stt)}
\label{sec:det:ts:stt}
\begin{figure*}[htb]
\begin{center}
\includegraphics[width=0.6\dwidth]{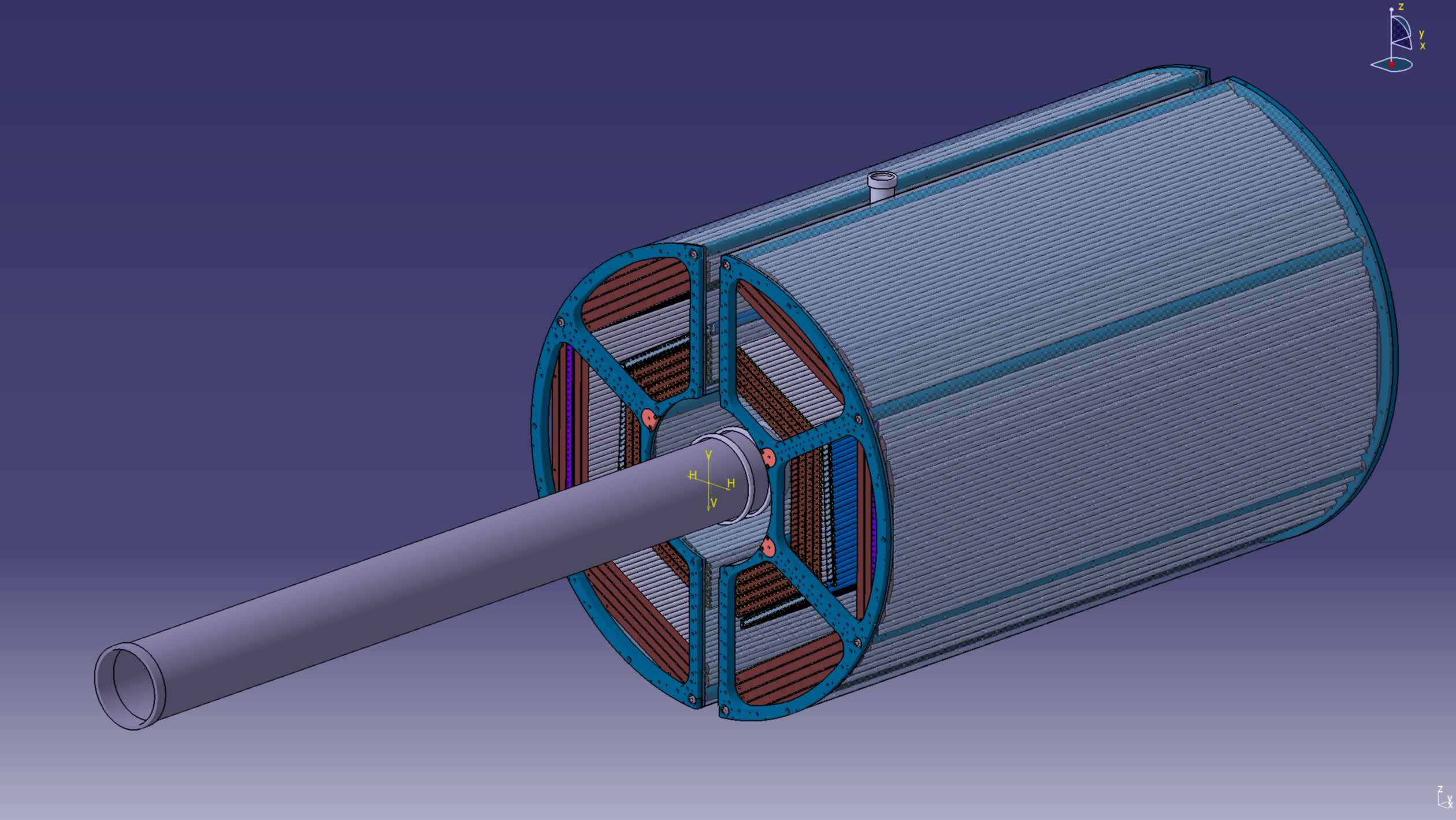} 
\caption[Straw Tube Tracker in the Target Spectrometer]
{Straw Tube Tracker in the Target Spectrometer.}
\label{fig:exp:ts:stt}
\includegraphics[width=0.6\dwidth]{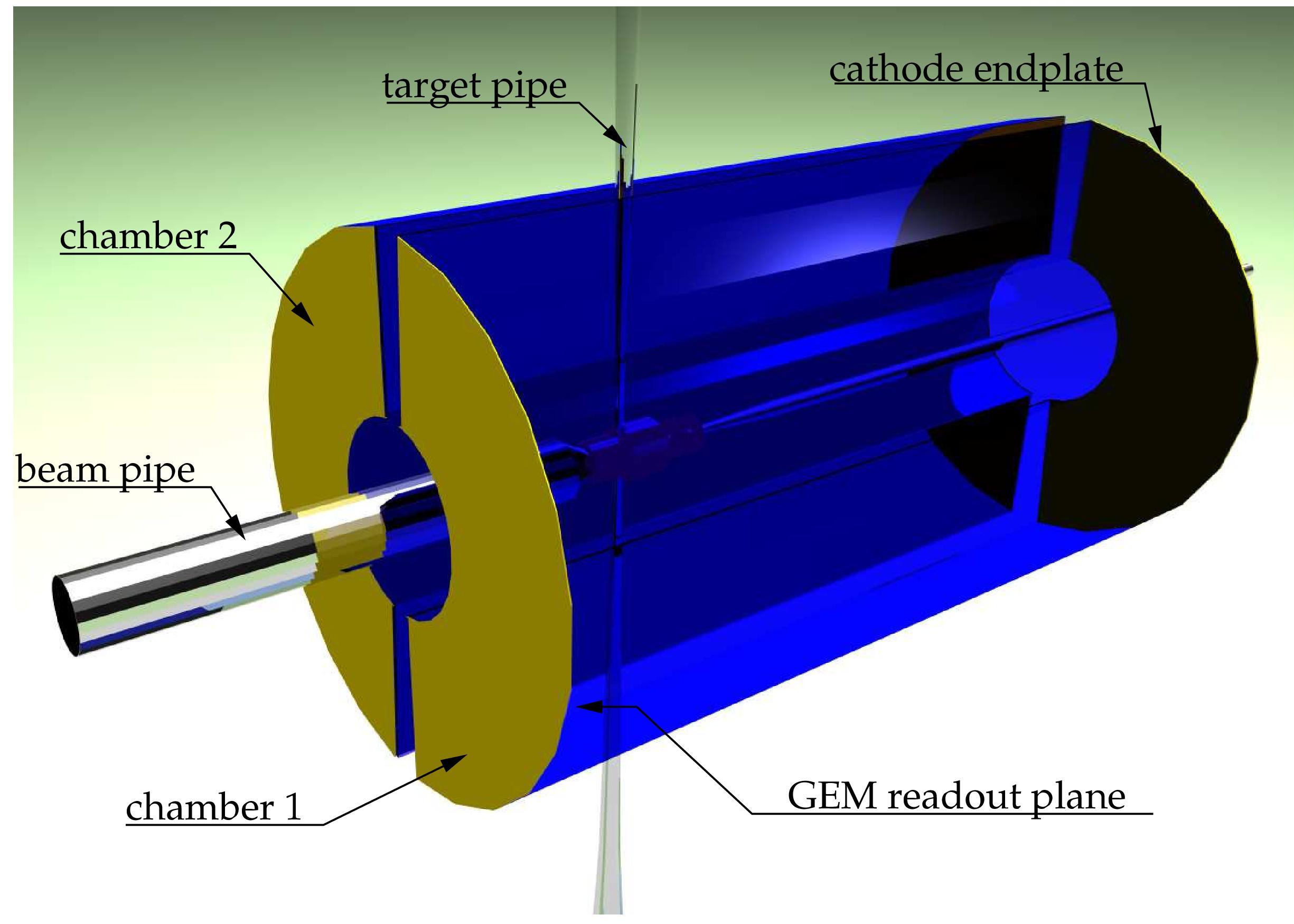}
\caption[GEM Time Projection Chamber in the Target Spectrometer]
{GEM Time Projection Chamber in the Target Spectrometer.}
\label{fig:exp:ts:tpc}
\end{center}
\end{figure*}
This detector consists of aluminized mylar tubes called {\em straws},
which are self supporting by the operation at 1 bar overpressure. The
straws are arranged in planar layers which are mounted in a hexagonal
shape around the \Mvd as shown in \Reffig{fig:exp:ts:stt}. In total
there are 24 layers of which the 8 central ones are tilted to achieve
an acceptable resolution of 3 mm also in z (parallel to the beam). The
gap to the surrounding detectors is filled with further individual
straws. In total there are 4200 straws around the beam pipe at radial
distances between 15$\,$cm and 42$\,$cm with an overall length of
150$\,$cm. All straws have a diameter of 10 mm. A thin and light space
frame will hold the straws in place, the force of the wire however is
kept solely by the straw itself. The mylar foil is 30 $\umu$m thick,
the wire is made of 20$\,\umu$m thick gold plated tungsten. This
design results in a material budget of 1.3 \% of a radiation length.

The gas mixture used will be Argon based with CO$_2$ as quencher. It
is foreseen to have a gas gain no greater than 10$^5$ in order to
warrant long term operation. With these parameters, a resolution in
$x$ and $y$ coordinates of about 150$\,\umu$m is expected.

\paragraph*{Time Projection Chamber (\Tpc)}
\label{sec:det:ts:tpc}
A challenging but advantageous alternative to the \Stt is a \Tpc, which
would combine superior track resolution with a low material budget and
additional particle identification capabilities through energy loss
measurements. 

The \Tpc depicted in a schematic view in \Reffig{fig:exp:ts:tpc}
consists of two large gas-filled half-cylinders enclosing the target
and beam pipe and surrounding the \Mvd. An electric field along the
cylinder axis separates positive gas ions from electrons created by
ionizing particles traversing the gas volume. The electrons drift with
constant velocity towards the anode at the upstream end face and
create an avalanche detected by a pad readout plane yielding
information on two coordinates.  The third coordinate of the track
comes from the measurement of the drift time of each primary electron
cluster. In common TPCs the amplification stage typically occurs in
multi-wire proportional chambers. These are gated by an external
trigger to avoid a continuous backflow of ions in the drift volume
which would distort the electric drift field and jeopardize the
principle of operation.

In \Panda the interaction rate is too high and there is no fast
external trigger to allow such an operation. Therefore a novel readout
scheme is employed which is based on GEM foils as amplification stage.
These foils have a strong suppression of ion backflow, since the ions
produced in the avalanches within the holes are mostly caught on the 
backside of the foil. Nevertheless about two ions per primary
electron are drifting back into the ionisation volume even at moderate
gains. The deformation of the drift field can be measured by a laser
calibration system and the resulting drift can be corrected
accordingly. In addition a very good homogeneity of the solenoid field
with a low radial component is required.

A further challenge is the large number of tracks accumulating in the
drift volume because of the high rate and slow drift. While the TPC is
capable of storing a lot of tracks at the same time, their assignment
to specific interactions has to be done by time correlations with
other detectors in the target spectrometer. To achieve this, first a
tracklet reconstruction has to take place. The tracklets are then
matched against other detector signals or are pointed to the
interaction.  This requires either high computing power close to the
readout electronics or a very high bandwidth at the full interaction
rate.

\paragraph*{Forward GEM Detectors}
Particles emitted at angles below 22\degrees{} which are not covered
fully by the Straw Tube Tracker or \Tpc will be tracked by three
stations of GEM detectors placed 1.1 m, 1.4 m and 1.9 m downstream of
the target. The chambers have to sustain a high counting rate of
particles peaked at the most forward angles due to the relativistic
boost of the reaction products as well as due to the small angle
$\pbar$p elastic scattering. With the envisaged luminosity, the
expected particle flux in the first chamber in the vicinity of the 5
cm diameter beam pipe is about 3$\cdot10^{4}\,$cm$^{-2}$s$^{-1}$. In
addition it is required that the chambers work in the 2$\,$T magnetic
field produced by the solenoid.  Drift chambers cannot fulfill the
requirements here since they would suffer from aging and the occupancy
would be too high. Therefore gaseous micropattern detectors based on
GEM foils as amplification stages are chosen. These detectors have
rate capabilities three orders of magnitude higher than drift
chambers.

In the current layout there are three double planes with two
projections per plane. The readout plane is subdivided in an outer ring
with longer and an inner ring with shorter strips. The strips are
arranged in two orthogonal projections per readout plane. Owing to the
charge sharing between strip layers a strong correlation between the
orthogonal strips can be found giving an almost 2D information rather
than just two projections. 

The readout is performed by the same frontend chips as are used for
the silicon microstrips. The first chamber has a diameter of 90 cm,
the last one of 150 cm. The readout boards carrying the ASICs are
placed at the outer rim of the detectors.
 
\subsubsection{Cherenkov Detectors and Time-of-Flight}
Charged particle identification of hadrons and leptons over a large
range of angles and momenta is an essential requirement for
meeting the physics objectives of \Panda. There will be several
dedicated systems which, complementary to the other detectors, will
provide means to identify particles. The main part of the momentum
spectrum above 1 GeV/$c$ will be covered by Cherenkov detectors. 
Below the Cherenkov threshold of kaons several other processes have 
to be employed for particle identification: The tracking detectors
are able to provide energy loss measurements. Here in particular the
TPC with its large number of measurements along each track excels.
In addition a time-of-flight barrel can identify slow particles.

\paragraph*{Barrel DIRC}
Charged particles in a medium with index of refraction $n$,
propagating with velocity $\beta c\,<\,\mbox{1}/n$, emit radiation at
an angle $\Theta_C\,=\,\arccos(\mbox{1}/n\beta$).  Thus, the mass of
the detected particle can be determined by combining the velocity
information determined from $\Theta_C$ with the momentum information
from the tracking detectors.

A very good choice as radiator material for these detectors is fused
silica (i.e. artificial quartz) with a refractive index of 1.47. This
provides pion-kaon-separation from rather low momenta of 800 MeV/$c$
up to about 5 GeV/$c$ and fits well to the compact design of the
target spectrometer. In this way the loss of photons converting in the
radiator material can be reduced by placing the conversion point as
close as possible to the electromagnetic calorimeter.

At polar angles between 22\degrees{} and 140\degrees{}, particle
identification will be performed by the detection of internally
reflected Cherenkov (\Dirc) light as realized in the {\INST{BaBar}}
detector~\cite{Staengle:1997xp}.  It will consist of 1.7$\,$cm thick
quartz slabs surrounding the beam line at a radial distance of
45 - 54 $\,$cm. At {\INST{BaBar}} the light was imaged across a large
stand-off volume filled with water onto 11\,000 photomultiplier
tubes. At \Panda{}, it is intended to focus the images by lenses onto
micro-channel plate photomultiplier tubes (MCP PMTs) which are
insensitive to magnet fields. This fast light detector type allows a
more compact design and the readout of two spatial coordinates. In
addition MCP PMTs provide good time resolution to measure the time of 
light propagation for dispersion correction and background suppression.

The DIRC design with its compact radiator mounted close to the
\Emc minimises the conversions. Part of these conversions can be
recovered with information from the DIRC detector, as was shown by 
{\INST{BaBar}}\cite{Adametz05}.

\paragraph*{Forward Endcap DIRC} 
A similar concept can be employed in the forward direction for
particles between 5\degrees{} and 22\degrees{}. The same radiator,
fused silica, is to be employed however in shape of a disk. At the rim
around the disk focusing will be done by mirroring quartz elements
reflecting onto MCP PMTs. Once again two spatial coordinates plus the
propagation time for corrections will be read. The disk will be 2 cm
thick and will have a radius of 110 cm. It will be placed directly
upstream of the forward endcap calorimeter.

\paragraph*{Barrel Time-of-Flight}
For slow particles at large polar angles particle identification shall
be provided by a time-of-flight detector. In the target spectrometer
the flight path is only in the order of 50 -- 100 cm. Therefore the
detector must have a very good time resolution between 50 and 100 ps.

Implementing an additional start detector would introduce too much
material close to the interaction point deteriorating considerably the
resolution of the electromagnetic crystal calorimeter. In the absence
of a start-detector relative timing of a minimum of two particles has
to be employed.

As detector candidates scintillator bars and strips or pads of
multi-gap resistive plate chambers are considered. In both cases a
compromise between time resolution and material budget has to be
found. The detectors will cover angles between 22\degrees{} and
140\degrees{} using a barrel arrangement around the \Stt/\Tpc at 
42 - 45$\,$cm radial distance.

\subsubsection{Electromagnetic Calorimeters}
Expected high count rates and a geometrically compact design of the
target spectrometer require a fast scintillator material with a short
radiation length and Moli\`ere radius for the construction of the
electromagnetic calorimeter (\Emc). Lead tungstate (PbWO$_4$) is a
high density inorganic scintillator with sufficient energy and time
resolution for photon, electron, and hadron detection even at
intermediate energies \cite{Mengel:1998si,Novotny:2000zg,Hoek:2002ss}.
For high energy physics PbWO$_4$ has been chosen by the {\INST{CMS}}
and \INST{ALICE} collaborations at \INST{CERN}
\cite{Alice:Tp,Cms:Tp:1994} and optimized for large scale production.
Apart from a short decay time of less than 10$\,$ns good radiation
hardness has been achieved \cite{Auffray:1999}.  Recent developments
indicate a significant increase of light yield due to crystal
perfection and appropriate doping to enable photon detection down to a
few MeV with sufficient resolution. The light yield can be increased
by a factor of about 4 compared to room temperature by cooling the
crystals down to -25$\degC$.

The crystals will be 20 cm long, i.e. approximately 22$\,X_0$, in
order to achieve an energy resolution below 2\percent{} at 1$\,\gev$
\cite{Mengel:1998si,Novotny:2000zg,Hoek:2002ss} at a tolerable energy
loss due to longitudinal leakage of the shower.  Tapered crystals with
a front size of $2.1\times 2.1\,\cm^2$ will be mounted with an inner
radius of 57$\,$cm.  This implies 11360 crystals for the barrel part
of the calorimeter.  The forward endcap calorimeter will have 3600
tapered crystals, the backward endcap calorimeter 592.  The readout of
the crystals will be accomplished by large area avalanche photo diodes
in the barrel and vacuum phototriodes in the forward and backward
endcaps.

The EMC allows to achieve an e/$\pi$ ratio of 10$^3$ for momenta above
0.5$\,\gevc$.  Therefore, e-$\pi$-separation does not require an
additional gas Cherenkov detector in favor of a very compact geometry
of the EMC.

\subsubsection{Muon Detectors}
Muons are an important probe for, among others, $J/\psi$ decays,
semi-leptonic $D$-meson decays and the Drell-Yan process. The
strongest background are pions and their decay daughter muons.
However at the low momenta of \Panda the signature is less clean than
in high energy physics experiments. To allow nevertheless a proper
separation of primary muons from pions and decay muons a range
tracking system will be implemented in the yoke of the solenoid
magnet. Here a fine segmentation of the yoke as absorber with
interleaved tracking detectors allows the distinction of energy loss
processes of muons and pions and kinks from pion decays. Only in this
way a high separation of primary muons from the background can be
achieved.

In the barrel region the yoke is segmented in a first layer of 6 cm
iron followed by 12 layers of 3 cm thickness. The gaps for the
detectors are 3 cm wide. This is enough material for the absorption of
pions in the momentum range in \PANDA at these angles. In the forward
endcap more material is needed. Since the downstream door of the
return yoke has to fulfill constraints for space and accessibility,
the muon system is split in several layers.  Six detection layers are
placed around five iron layers of 6 cm each within the door, and a
removable muon filter with additional five layers of 6 cm iron is
located in the space between the solenoid and the dipole. This filter
has to provide cut-outs for forward detectors and pump lines and has
to be built in a way that it can be removed with few crane operations
to allow easy access to these parts.

As detector within the absorber layers rectangular aluminum drift
tubes are used as they were constructed for the COMPASS muon detection
system.  They are essentially drift tubes with additional capacitively
coupled strips read out on both ends to obtain the longitudinal
coordinate.

\subsubsection{Hypernuclear Detector}
The hypernuclei study will make use of the modular structure of
\Panda{}. Removing the backward endcap calorimeter will allow to
add a dedicated nuclear target station and the required additional
detectors for $\gamma$ spectroscopy close to the entrance of
\Panda{}. While the detection of anti-hyperons and low momentum $K^+$
can be ensured by the universal detector and its PID system, a
specific target system and a $\gamma$-detector are additional
components required for the hypernuclear studies.

\paragraph*{Active Secondary Target}
The production of hypernuclei proceeds as a two-stage process. First
hyperons, in particular $\Xi\bar{\Xi}$, are produced on a nuclear
target. In some cases the $\Xi$ will be slow enough to be captured in
a secondary target, where it reacts in a nucleus to form a double
hypernucleus.

The geometry of this secondary target is determined by the short mean
life of the $\Xi^-$ of only 0.164$\,$ns. This limits the required
thickness of the active secondary target to about 25--30$\,$mm. It
will consist of a compact sandwich structure of silicon micro
strip detectors and absorbing material. In this way the weak decay
cascade of the hypernucleus can be detected in the sandwich structure.

\paragraph*{Germanium Array} 
An existing germanium-array with refurbished readout will be used for
the $\gamma$-spectroscopy of the nuclear decay cascades of
hypernuclei. The main limitation will be the load due to neutral or
charged particles traversing the germanium detectors. Therefore,
readout schemes and tracking algorithms are presently being developed
which will enable high resolution $\gamma$-spectroscopy in an
environment of high particle flux.
  
\subsection{Forward Spectrometer}
\subsubsection{Dipole Magnet}
A dipole magnet with a window frame, a 1$\,$m gap, and more than
2$\,$m aperture will be used for the momentum analysis of charged
particles in the forward spectrometer.  In the current planning, the
magnet yoke will occupy about 2.5$\,$m in beam direction starting from
3.5$\,$m downstream of the target.  Thus, it covers the entire angular
acceptance of the target spectrometer of $\pm$10\degrees{} and
$\pm$5\degrees{} in the horizontal and in the vertical direction,
respectively.  The maximum bending power of the magnet will be 2$\,$Tm
and the resulting deflection of the antiproton beam at the maximum
momentum of 15$\,\gevc$ will be 2.2\degrees{}.  The design acceptance
for charged particles covers a dynamic range of a factor 15 with the
detectors downstream of the magnet.  For particles with lower momenta,
detectors will be placed inside the yoke opening. The beam deflection
will be compensated by two correcting dipole magnets, placed around
the \Panda{} detection system.

\subsubsection{Forward Trackers}
\label{sec:det:fs:trk}
The deflection of particle trajectories in the field of the dipole
magnet will be measured with a set of wire chambers (either small cell
size drift chambers or straw tubes), two placed in front, two within
and two behind the dipole magnet.  This will allow to track particles
with highest momenta as well as very low momentum particles where
tracks will curl up inside the magnetic field.  

The chambers will contain drift cells of 1$\,$cm width. Each chamber
will contain three pairs of detection planes, one pair with vertical
wires and two pairs with wires inclined by +10\degrees{} and
-10\degrees{}.  This configuration will allow to reconstruct tracks in
each chamber separately, also in case of multi-track events. The beam
pipe will pass through central holes in the chambers.  The most
central wires will be separately mounted on insulating rings
surrounding the beam pipe. The expected momentum resolution of the
system for 3$\,\gevc$ protons is $\delta p/p\,=\,0.2$\percent{} and is
limited by the small angle scattering on the chamber wires and gas.

\subsubsection{Forward Particle Identification}
\paragraph*{RICH Detector}
To enable the $\pi$/$K$ and $K$/p separation also at the very highest
momenta a \Rich detector is proposed. The favored design is a dual
radiator \Rich detector similar to the one used at
\INST{Hermes}~\cite{Akopov:2000qi}. Using two radiators, silica
aerogel and C$_4$F$_{10}$ gas, provides $\pi$/$K$/p separation in a
broad momentum range from 2--15$\,\gevc$.  The two different indices
of refraction are 1.0304 and 1.00137, respectively.  The total
thickness of the detector is reduced to the freon gas radiator
\mbox{(5\%$\,X_0$),} the aerogel radiator (2.8\%$\,X_0$), and the
aluminum window (3\%$\,X_0$) by using a lightweight mirror focusing
the Cherenkov light on an array of phototubes placed outside the
active volume. It has been studied to reuse components of the
\INST{HERMES} \Rich{}.

\paragraph*{Time-of-Flight Wall}
A wall of slabs made of plastic scintillator and read out on both ends
by fast phototubes will serve as time-of-flight stop counter placed at
about 7$\,$m from the target. In addition, similar detectors will be
placed inside the dipole magnet opening, to detect low momentum
particles which do not exit the dipole magnet. The relative time of
flight between two charged tracks reaching any of the time-of-flight
detectors in the experiment will be measured. The wall in front of the
forward spectrometer \Emc will consist of vertical strips varying in
width from 5 to 10$\,$cm to account for the differences in count
rate. With the expected time resolution of $\sigma\,=\,50\,$ps
$\pi$-$K$ and $K$/p separation on a 3$\,\sigma$ level will be possible
up to momenta of 2.8$\,\gevc$ and 4.7$\,\gevc$, respectively.

\subsubsection{Forward Electromagnetic Calorimeter}
For the detection of photons and electrons a {Shashlyk}-type
calorimeter with high resolution and efficiency will be employed. The
detection is based on lead-scintillator sandwiches read out with
wave-length shifting fibers passing through the block and coupled to
photomultipliers. The technique has already been successfully used in
the \INST{E865} experiment~\cite{bib:emc:E865}.  It has been adopted
for various other experiments~\cite{bib:emc:PHENIX, bib:emc:HERAB,
  bib:emc:LHCb, bib:emc:KOP99, bib:emc:KOP1, bib:emc:KOP04}.  An
energy resolution of $4\%/\sqrt{E}$~\cite{bib:emc:KOP99} has been
achieved.  To cover the forward acceptance, 26 rows and 54 columns are
required with a cell size of 55 mm, i.e. 1404 modules in total,
which will be placed at a distance of 7--8$\,$m from the target.

\subsubsection{Forward Muon Detectors}
For the very forward part of the muon spectrum a further range
tracking system consisting of interleaved absorber layers and
rectangular aluminium drift-tubes is being designed, similar to the
muon system of the target spectrometer, but laid out for higher
momenta. The system allows discrimination of pions from muons,
detection of pion decays and, with moderate resolution, also the
energy determination of neutrons and anti-neutrons.

\subsection{Luminosity monitor}
In order to determine the cross section for physical processes, it is
essential to determine the time integrated luminosity $L$ for
reactions at the \PANDA interaction point that was available while
collecting a given data sample. Typically the precision for a relative
measurement is higher than for an absolute measurement. For many
observables connected to narrow resonance scans a relative measurement
might be sufficient for \PANDA, but for other observables an absolute
determination of $L$ is required.  The absolute cross section can be
determined from the measured count rate of a specific process with
known cross section. In the following we concentrate on elastic
antiproton-proton scattering as the reference channel.  For most other
hadronic processes that will be measured concurrently in \PANDA the
precision with which the cross section is known is poor.

The optical theorem connects the forward elastic scattering amplitude
to the total cross section.  The total reaction rate and the
differential elastic reaction rate as a function of the
4-momentum transfer {\it t} can be used to determine the total cross
section.

The differential cross section $d\sigma_{el}/dt$ becomes dominated by
Coulomb scattering at very low values of $t$. Since the
electromagnetic amplitude can be precisely calculated, Coulomb elastic
scattering allows both the luminosity and total cross section to be
determined without measuring the inelastic rate
\cite{Armstrong:1996np}.

Due to the 2 T solenoid field and the existence of the MVD it appears
most feasible to measure the forward going antiproton in \PANDA. The
Coulomb-nuclear interference region corresponds to 4-momentum
transfers of $-t\approx 0.001$ GeV$^2$ at the beam momentum range of
interest to \PANDA. At a beam momentum of 6 GeV/$c$ this momentum
transfer corresponds to a scattering angle of the antiproton of about
5 mrad.

The basic concept of the luminosity monitor is to reconstruct the
angle (and thus $t$) of the scattered antiprotons in the polar angle
range of 3-8 mrad with respect to the beam axis.  Due to the large
transverse dimensions of the interaction region when using the pellet
target, there is only a weak correlation of the position of the
antiproton at e.g. z=+10.0 m to the recoil angle. Therefore, it is
necessary to reconstruct the angle of the antiproton at the luminosity
monitor. As a result the luminosity monitor will consist of a sequence
of four planes of double-sided silicon strip detectors located as far
downstream and as close to the beam axis as possible. The planes are
separated by 20 cm along the beam direction. Each plane consists of 4
wafers (e.g. 2 cm $\times$ 5 cm $\times\ 200\ \umu$m, with 50 $\umu$m
pitch) arranged radially to the beam axis. Four planes are required
for sufficient redundancy and background suppression. The use of 4
wafers (up, down, right, left) in each plane allows systematic errors
to be strongly suppressed. 

The silicon wafers are located inside a vacuum chamber to minimize
scattering of the antiprotons before traversing the 4 tracking
planes. The acceptance for the antiproton beam in the HESR is $\pm$3
mrad, corresponding to the 89 mm inner diameter of the beam pipe at
the quadrupoles located at about 15 m downstream of the interaction
point.  The luminosity monitor can be located in the space between the
downstream side of the forward spectrometer hadronic calorimeter 
and the HESR dipole
needed to redirect the antiproton beam out of the \PANDA chicane back
into the direction of the HESR straight stretch (i.e. between z=+10.0
m and z=+12.0 m downstream of the target). At this distance from
the target the luminosity monitor needs to measure particles at a
radial distance of between 3 and 8 cm from the beam axis.

As pilot simulations show, at a beam momentum of 6.2 GeV/$c$ the
proposed detector measures antiprotons elastically scattered in the
range $0.0006 $(GeV)$^2< -t < 0.0035$ (GeV)$^2$, which spans the Coulomb-nuclear
interference region.  Based upon the granularity of the readout the
resolution of $t$ could reach $\sigma_t \approx 0.0001$ (GeV)$^2$. In
reality this value is expected to degrade to $\sigma_t \approx 0.0005$ 
(GeV)$^2$ when taking small-angle scattering into account. At the
nominal \PANDA interaction rate of $2\EE{7}$/s there will be an
average of 10 kHz/cm$^2$ in the sensors. In comparison with
other experiments an absolute precision of about 3\% is considered
feasible for this detector concept at \PANDA, which will be verified
by more detailed simulations.

\subsection{Data Acquisition}
In many contemporary experiments the trigger and data acquisition (DAQ)
system is based on a two layer hierarchical approach. A subset of
specially instrumented detectors is used to evaluate a first level
trigger condition. For the accepted events, the full information of
all detectors is then transported to the next higher trigger level or
to storage. The available time for the first level decision is usually
limited by the buffering capabilities of the front-end electronics.
Furthermore, the hard-wired detector connectivity severely constrains
both the complexity and the flexibility of the possible trigger
schemes.

In \PANDA, a data acquisition concept is being developed which is
better matched to the high data rates, to the complexity of the
experiment and the diversity of physics objectives and the rate
capability of at least $2\EE{7}$ events/s.

\par
In our approach, every sub-detector system is a self-triggering entity.
Signals are detected autonomously by the sub-systems and are preprocessed.
Only the physically relevant information is extracted and transmitted.
This requires hit-detection, noise-suppression and clusterisation at
the readout level. 
The data related to a particle hit, with a substantially reduced rate
in the preprocessing step, is marked by a precise time stamp and
buffered for further processing.
The trigger selection finally occurs in computing nodes which
access the buffers via a high-bandwidth network fabric. The new
concept provides a high degree of flexibility in the choice of trigger
algorithms. It makes trigger conditions available which are
outside the capabilities of the standard approach. One obvious example
is  displaced vertex triggering.

\par
In this scheme, sub-detectors can contribute to the trigger decision
on the same footing without restrictions due to hard-wired
connectivity. Different physics can be accessed either in parallel or
via software reconfiguration of the system.

High speed serial (10$\,$Gb/s per link and beyond) and high-density FPGA
(field programmable gate arrays) with large numbers of programmable
gates as well as more advanced embedded features
are key technologies to be exploited within the DAQ framework.
\par
The basic building blocks of the hardware infrastructure which
  can be combined in a flexible way to cope with varying demands, are
  the following:
  \begin{itemize}
  \item Intelligent front-end modules capable of autonomous hit
    detection and data preprocessing (e.g.\ clustering, hit time
    reconstruction, and pattern recognition) are needed.
  \item A  precise time distribution system is mandatory to
    provide a clock norm from which all time stamps can be derived.
    Without this, data from subsystems cannot be correlated.
  \item Data concentrators provide point-to-point communication,
    typically via optical links, buffering and online data
    manipulation.
  \item Compute nodes aggregate large amounts of computing power in a
    specialized architecture rather than through commodity PC hardware.
    They may employ fast FPGAs (fast programmable gate arrays),
    DSPs (digital signal processors), or other computing units.
    The nodes have to deal with feature extraction, association of data
    fragments to events, and, finally, event selection.
  \end{itemize}
\par
A major component providing the link for all building blocks is the
network fabric. Here, special emphasis is put on embedded switches
which can be cascaded and reconfigured to reroute traffic for
different physics selection topologies. Alternatively, with an even
higher aggregate bandwidth of the network, which according to
projections of network speed evolution will be available by the time
the experiment will start, a flat network topology where all data is
transferred directly to processing nodes may be feasible as well.
This requires a higher total bandwidth but would have a simpler
architecture and allow event selection in a single environment. The
bandwidth required in this case would be at least 200 GB/s. After
event selection in the order of 100-200 MB/s will be saved to mass
storage.

An important requirement for this scheme is that all detectors perform
a continuous online calibration with data. The normal data taking is
interleaved with special calibration runs. For the monitoring of the
quality of data, calibration constants and event selection a small
fraction of unfiltered raw data is transmitted to mass storage.

To facilitate the association of data fragments to events the beam
structure of the accelerator is exploited: Every 1.8 $\umu$s there is
a gap of about 400 ns needed for the compensation of energy loss with
a bucket barrier cavity. This gap provides a clean division between
consistent data blocks which can be processed coherently by one
processing unit.

\subsection{Infrastructure}
\begin{figure*}[htb]
\centerline{
\includegraphics[angle=270,width=1.1\dwidth]{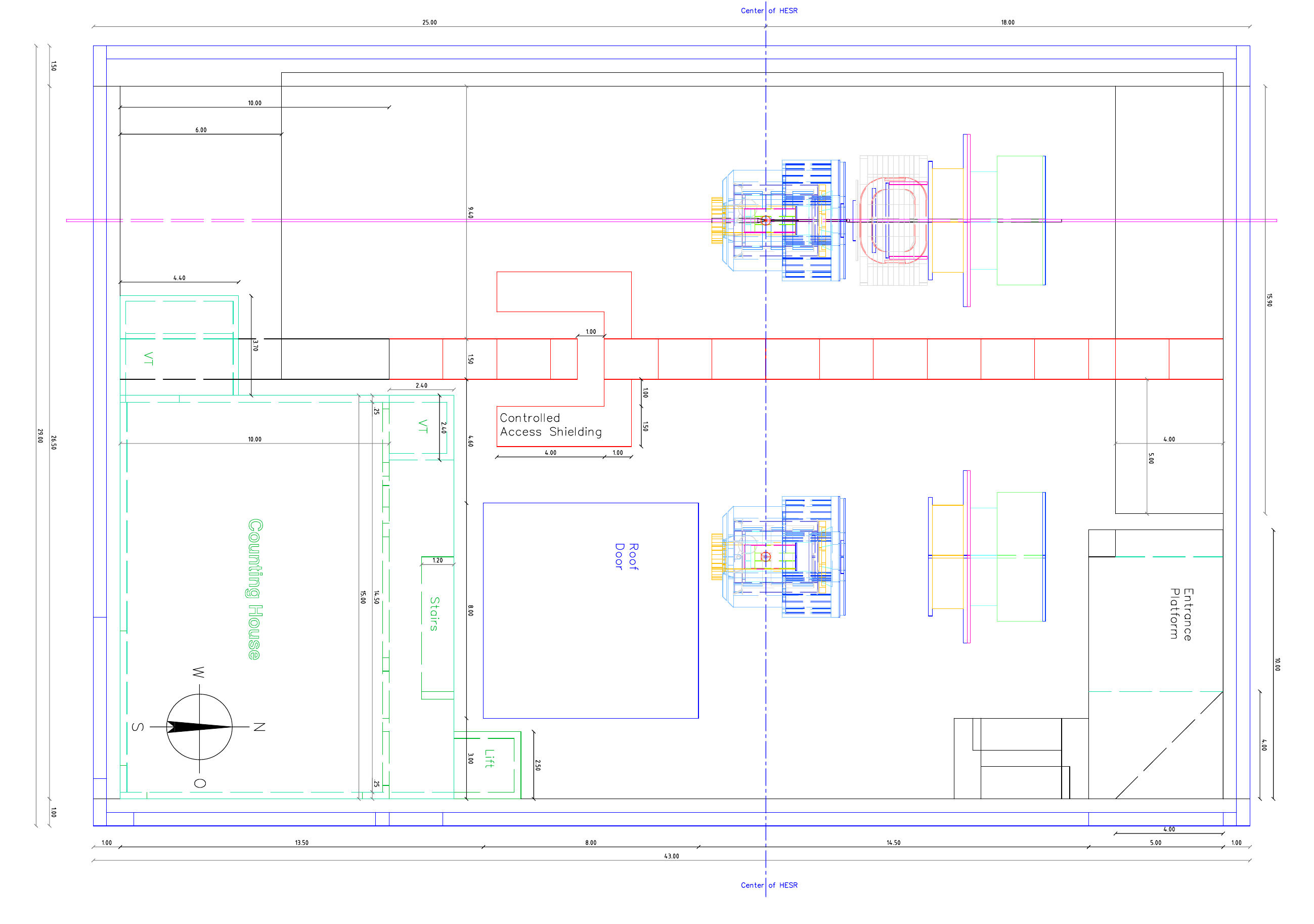}
}
\caption[Top view of the experimental area.]{Top view of the
  experimental area indicating the location of \Panda in the \Hesr
  beam line. The target center is at the center of \Hesr and is
  indicated as vertical dash-dotted line. North is to the right and
  the beam comes in from the left. The roll-out position of the
  detector will be on the east side of the Hall.}
\label{fig:exp:inf:exphall}
\end{figure*}

The target for antiproton physics is located in the straight section
at the east side of the \HESR.  At this location an experimental hall
of 43\,m $\times$ 29\,m floor space and 14.5\,m height is planned (see
\Reffig{fig:exp:inf:exphall}). A concrete radiation shield of 2$\,$m
thickness on both sides along the beam line is covered by concrete
bars of 1$\,$m thickness to suppress the neutron sky shine. Within the
elongated concrete cave the \Panda{} detector together with auxiliary
equipment, beam steering, and focusing elements will be housed. The
roof of the cave can be opened and heavy components hoisted by crane.

The shielded beam line area for the \Panda experiment including
dipoles and focusing elements is foreseen to have 37\,m $\times$
9.4\,m floor space and a height of 8.5\,m with the beam line at a height
of 3.5\,m.  The general floor level of the \HESR is 2\,m higher.
This level will be kept for a length of 4\,m in the north of the
hall (right part in \Reffig{fig:exp:inf:exphall}), to facilitate
transport of heavy equipment into the \HESR tunnel.

The target spectrometer with electronics and supplies will be mounted
on rails which makes it retractable to a parking position outside
the \HESR{} beam line (i.e.\ into the lower part of the hall in
\Reffig{fig:exp:inf:exphall}). The experimental hall provides
additional space for delivery of components and assembly of the
detector parts. In the south corner of the hall, a counting house
complex with five floors is foreseen. The lowest floor will contain
various supplies for power, high voltage, cooling water, gases
etc. The next level is planned for readout electronics including data
concentrators. The third level will house the online computing farm.
The fourth floor is at level with the surrounding ground and will house
the control room, a meeting room and social rooms for the shift crew.
Above this floor, hall electricity supplies and ventilation is placed.
A crane (15$\,$t) spans the whole area with a hook at a height of
about 10$\,$m. Sufficient (300$\,$kW) electric power will be
available.

Liquid helium coolant may come from the main cryogenic liquefier for
the \INST{SIS} rings.  Alternatively, a separate small liquefier
(50$\,$W cooling power at 4$\,$K) would be mounted. The temperature of
the building will be moderately controlled. The more stringent
requirements with respect to temperature and humidity for the
detectors have to be maintained locally. To facilitate cooling and
avoid condensation the target spectrometer will be kept in a tent
with dry air at a controlled temperature.
%

%
%

%

%
\newpage
\bibliographystyle{panda_tdr_lit}
\bibliography{./lit_emc,./int/lit_int_detector,./int/lit_int_hesr}

%

%
%
%
\cleardoublepage
\chapter{Design Considerations}
\label{sec:req}
%
%
%
%
%
The \Panda experiment aims at various physics topics related to the
very nature of large distance strong binding. Although the details and
observables turn out to be different, most channels share one important
feature - many photons and/or electrons/positrons in the final state.
Examples are hidden charm decays of charmonium hybrids with neutral recoils and low-mass
isoscalar S-waves (appearing in $\piz\piz$), radiative charm decays and
the nucleon structure physics. This puts special emphasis on the electromagnetic
calorimeter, and its basic performance parameters have to be tuned to accomplish
the effective detection of these channels in order to succeed in the basic
programme of \Panda.
\par
The basic function of an electromagnetic calorimeter is the efficient reconstruction of
electrons, positrons and photons with high efficiency and low background. This is performed
by measuring the deposited energy ($E$) and the direction via the point of impact.
($\theta$ and $\phi$). High resolution is mandatory for a sufficient
resolving power for final states with multiple electrons, positrons and photons.
\par
Photons in the final state can originate from various sources. The most abundant sources
are $\piz$ and $\eta$ mesons. Important probes are radiative charmonium decays
(like $\chicone\to\jpsi\gamma$), which are suppressed by the charm production yield
or direct photons from rare electromagnetic processes.
To distinguish radiatively decaying  charmonium and direct photons from background
 with undetected photons (from $\piz$ and $\eta$ rich states) it is of utmost
importantance to identify very efficiently $\piz$ and $\eta$ by reducing the number of
undetected photons due to solid angle or energy threshold.
\par
The EMC may also provide timing information. This is needed to accomplish a proper
distinction among different events. The annihilation rate goes up to several $\cdot 10^7$/s
leading to $\sigma_t\approx 10$\,ns. Thus a fast scintillator is required for operation.
\par
\Panda will not have a threshold Cherenkov detector to discriminate pions from
electrons and positrons. Therefore, the EMC has to add complementary
information to the basic $E/p$ information.
Lateral shower shape information is needed to discriminate $e^\pm$ from background.
These informations are deduced from the difference of lateral
shower shapes.
Hadronic showers ($K_L$, $n$, charged hadrons) in an electromagnetic calorimeter differ 
significantly due to the difference in energy loss per interaction and the elementary statistics
of these processes.  The quality of this discrimination does not (to first order) depend
on the  actual choice of crystal geometry and readout, as long as the front face size of the crystal is matched
to the Moli\`ere radius. Therefore, this does not place a strong requirement on the calorimeter design.
Nevertheless, the final design process must incorporate an optimization of the electron-pion separation
power.
\par
One basic aspect of the \Panda EMC is the requirement on compactness to reduce cost.
The price of the scintillator and the surrounding magnet scales with the cube of their dimension
thus, e.g., leading to a 50\,\% increase in price for $\approx 15$\,\% increase in radius.
\section{Electromagnetic Particle Reconstruction}
\subsection{Coverage Requirements}
\subsubsection{Energy Threshold}
\label{sec:ethres}
Apart from energy resolution the minimum photon energy $E_{thres}$ being accessible with the EMC is an important issue since it determines the very acceptance of low energy photons.
\par
\begin{figure}[bth]\begin{center}
\includegraphics[width=\swidth]{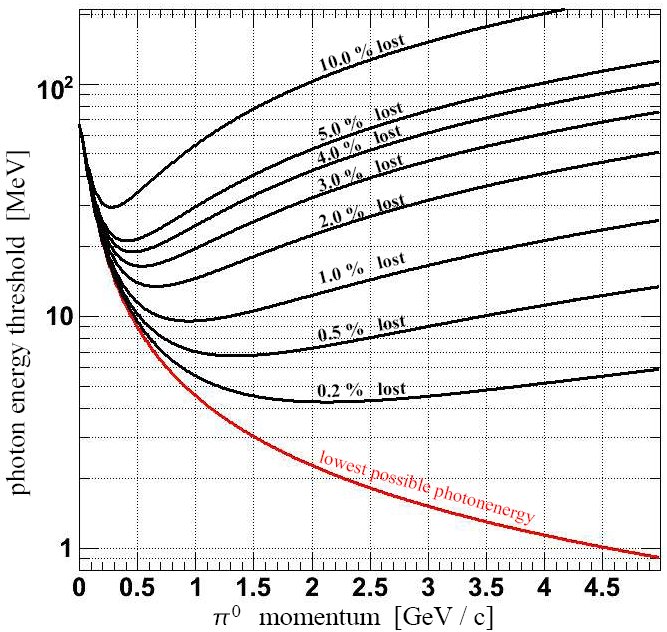}
\caption[$\piz$ loss rate as a function of energy threshold.]
{Percentage of $\piz$ loss as a function of energy threshold.}
\label{fig:pizloss}
\end{center}\end{figure}
It is easily shown that the bare photon and $\piz$ loss rate is not very large even for energy thresholds higher than 20\,\mev as long as a limit below 50\,\mev is maintained (e.g. $\piz$ loss in \Reffig{fig:pizloss}). Also the number of events dropped due to the energy threshold within the limits just mentioned is not dramatic. What is really driving the limit is the fact, that the physics being performed with \Panda requires an effective background rejection to distinguish radiative charmonium decays (as a tag for exotic charmonia) and other electromagnetic probes from background events with at least one undetected photon. So even small losses result in an unacceptable drop of the signal-to-background ratio.
\par
How the energy threshold affects the sensitivity for physics with \Panda can be exemplified by the hybrid production channel $\ppbar\to\eta_{c1}\eta$ with the charmonium hybrid $\eta_{c1}$ decaying to $\chicone\piz\piz$ leading to the final state $\jpsi\gamma\piz\piz\eta$. The production ratio between potential background and signal is expected to have the same order of magnitude. We consider only background with $\ccbar$ content - e.g. $\jpsi3\piz\eta$, since generic light quark background disappears to an undetectable level after the electron-id and charmonium cuts. However the decay branching ratios suppress the signal by one or two orders of magnitude compared to the background. Therefore, every effect of background leaking into the selection has to be minimized. Simulations show that the signal-to-background ratio depends almost quadratically on the minimum photon energy.
\par
This is due to the fact that, if the energies of the undetected photons are small enough, the residual particles may be recognized as an exclusive event and may contribute to the background of a channel with one photon less. Due to imperfect energy resolution, there is a certain probability, that the reduced set of particles of the background event fulfill all selection criteria of the signal channel and even survive a kinematic fit with high probability.
\par
From this discussion it is clear, that the lowest achievable value of  $E_{thres}$ is necessary to get the optimum in terms of photon detection. Although $E_{thres}=10\,\mev$ would be ideal in that context, technical limitations like noise or a reasonable coverage of a low energy shower may increase this value, but at least $E_{thres} \le 20\,\mev$ should be achieved to reach the physics goals of \Panda. More details are given in the simulation part of the report (\Refsec{sec:sim}).
\subsubsection{Geometrical Coverage}
The acceptance due to geometrical cuts is to $1^{st}$ order proportional to $(\Omega/4\pi)^n$ ($n$ being the number of $e^\pm$, $\gamma$. This is illustrated by the example of 6 photons and 90\,\% solid angle coverage where the geometrical acceptance drops to $1/2$. Since final states with many electrons, positrons and/or photons are one of the prime signals, these put, therefore, strong requirements on the angular coverage. As demonstrated in \Refsec{sec:ethres}, undetected photons are an important source of background effects and the loss due to the solid angle coverage should be minimized to the mechanical limit. In the backward region the beampipe is the limiting factor, but a
maximum opening of $\approx 10^\circ-15^\circ$ should be reached. In the forward direction a dipole bends all charged particles. In particular the $\pbar$-beam is inclined by 2$^\circ$, which allows for 0$^\circ$ calorimetry to maximize performance. Additional holes for mechanics, support, pipes and cables have to be considered. Nevertheless, the angular coverage should be maximized and the aim is 99\,\%\,4$\pi$ coverage in the center-of-mass system. In addition to the \TSEMC the forward part down to 0$^\circ$ will be covered by a shashlyk detector. The actual partioning between \FWEMC and forward shashlyk is optimized to allow high momentum tracks to enter the spectrometer dipole. 
\par
\subsubsection{Dynamical Energy Range}
\begin{figure*}[bth]\begin{center}
\includegraphics[width=\dwidth]{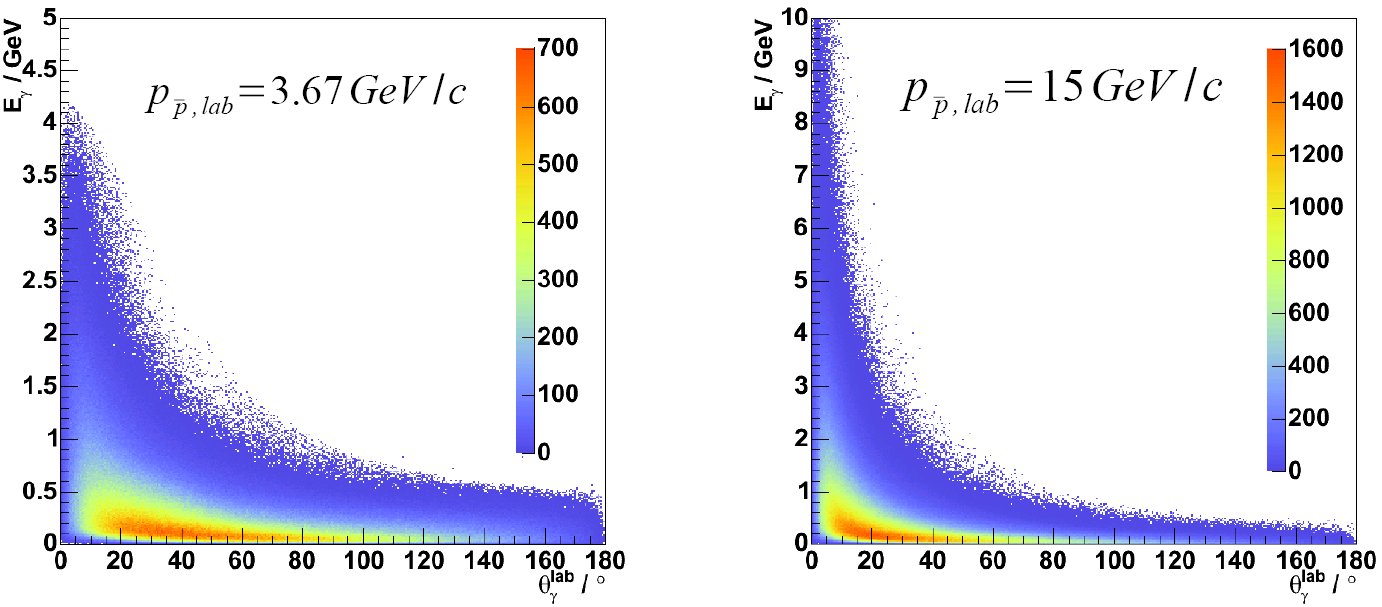}
\caption[Photon energy distribution vs. lab. angle.]
{Photon energy distribution vs. lab. angle for two momentum settings.}
\label{fig:dynrange}
\end{center}\end{figure*}
\Reffig{fig:dynrange} shows the range of energies from the DPM generator for two different momentum settings for the
antiproton beam. The highest energies are in forward direction, while backward particles are relatively low in energy. Since low energy capabilities are mandatory for all regions of the calorimeter the dynamic range is
mainly driven by the highest energy possible. The dynamic range for the various detector parts should
at least cover in 
\begin{itemize}
\item \BWEMC: 10(20)\,\mev - 0.7\,\gev,
\item \BEMC: 10(20)\,\mev - 7.3\,\gev, and
\item \FWEMC: 10(20)\,\mev - 14.6\,\gev.
\end{itemize}
\subsubsection{Vertex Distribution}
The primary vertex  distribution is dictated by the overlap of beam and target stream. The worst case appears for a
cluster-jet target with a spread in the order of a cm. In order to ensure that no photon escapes in the dead area between neighbouring crystals, a non-pointing geometry is needed. Ideally, the distance of closest approach of the pointer normal to the frontface of the crystal to the primary vertex should be at least 4\,cm. Thus the focus of the endcap is off the average vertex position in $z$ by at least 10\,cm. In the barrel part a tilt of 4$^\circ$ is needed to fulfill the same requirement.
\subsection{Resolution Requirements}
\subsubsection{Energy Resolution}
\label{sec:eres}
Apart from obtaining the best resolution to ensure the exclusiveness of events, the choice of the appropriate energy resolution has various additional aspects:
\begin{itemize}
\item Precise measurement of electron and positron energies for
\begin{itemize}
\item very accurate $E/p$ determination, and
\item optimum $\jpsi$ mass resolution
\end{itemize}
\item Efficient recognition of light mesons (e.g. $\piz$ and $\eta$) to reduce potential background.
\end{itemize}
\par
Precise $E/p$ measurement is an important asset to positively identify electrons and positrons against pions.
This is achieved if the error on the electron energy is negligible compared to the momentum error from the tracking detectors ($\approx\,$1\,\%). This puts a limit on the resolution  $\frac{\sigma_E}{E} \le 1.0\,\%$ at high energies.
\par
Another effect of bad energy resolution 
is the bad mass determination of $\piz$ and $\eta$ mesons. At low energies this is due to the $1/\sqrt{E}$ dependence of the energy resolution, while at high energies this is dominated by the constant term. Experiments like Crystal Barrel, CLEO, BaBar and BES, with similarities in the topology and composition of final states have proven, that a $\piz$ width of less than 8\,\mev and $\eta$ width of less than 30\,\mev is necessary for reasonable final state decomposition.
Assuming an energy dependence of the energy resolution of the form
\begin{equation}
\frac{\sigma_E}{E} = a \oplus \frac{b}{\sqrt{E/\gev}}
\end{equation}
leads to the requirement $a\le 1\,\%$ and $b\le 2\,\%$. This balance of values also ensures a $\jpsi$ resolution which is well matched with the resolution of the typical light recoil mesons (like $\eta$ and $\omega$).
\subsubsection{Single Crystal Threshold}
Energy threshold and energy resolution place a requirement on the required minimum single crystal energy $E_{xtl}$. 

\begin{figure}[bth]\begin{center}
\includegraphics[width=\swidth]{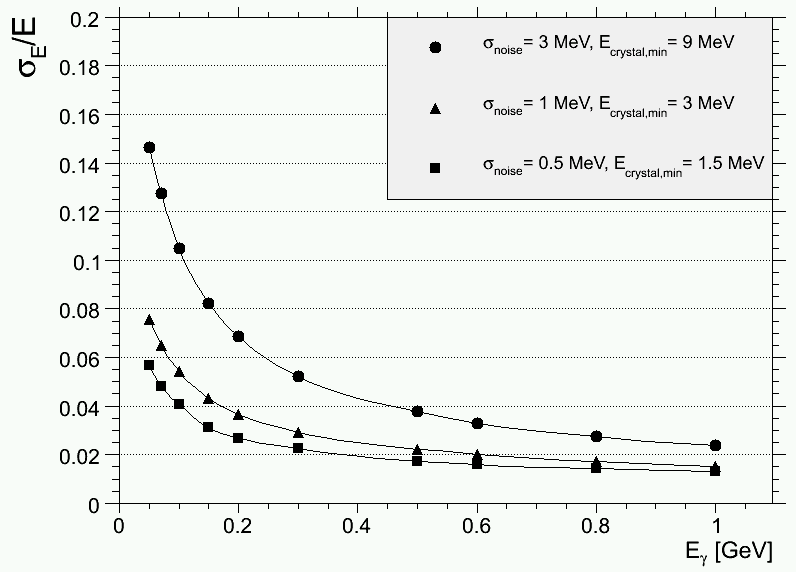}
\caption[Energy resolutions for different single crystal reconstruction 
thresholds.]
{Comparison of the energy resolutions for three different single crystal reconstruction thresholds. 
The most realistic scenario with a noise term of $\sigma \, = \,1 \, \mev$
and a single crystal threshold of $\extl=3\,\mev$ is illustrated by triangles, a worse case
($\sigma \, = \, 3 \mev$, $\extl \, = \, 9 \, \mev$) by circles and the better case ($\sigma \, = \, 0.5 \, \mev$, 
$\extl \,= \, 1.5 \, \mev$) by rectangles.}
\label{fig:scthreshold}
\end{center}\end{figure}

As a consequence this threshold puts a limit on the single crystal noise, since the single crystal cut (several MeV) should be high enough to exclude a random assignment of photons. This requirement can be relaxed by demanding a higher single-crystal energy to identify a bump, i.e. a local maximum in energy deposition (e.g. 10\,\mev). Starting with those seeds, additional crystals are only collected in the vicinity of this central crystal. With typically 10 neighbours and not more than 10 particles the probability of less or equal of one random crystal hit per event for a single crystal cut of $E_{xtl}=3\sigma_{noise}$.
\Reffig{fig:scthreshold} shows that a single crystal threshold of $E_{xtl}=3\,\mev$ is needed to obtain the required energy resolution. Also the containment of low energy photons in more than the central crystal can only be achieved for $E_{xtl}\le3\,\mev$. From this consideration we deduce a limit for the total noise of $\sigma_{E,noise}=1\,\mev$.
\subsubsection{Spatial Resolution}
The spatial resolution is mainly governed by the granularity. 
The reconstruction of the point of impact is achieved by weighted averaging of hits in adjacent crystals. In addition, to identify overlapping photons (e.g. due to $\piz$ with small opening angles) it is mandatory to efficiently split crystal clusters into individual photons. This requires, that the central hits of the involved photons are separated by at least two crystal widths to assure two local maxima in energy deposition.
If this can not be achieved, a cluster moment analysis has to be performed to identify $\piz$ without identifying individual photons. This is possible down to a spatial separation of one crystal.
\par
\begin{figure*}[bth]\begin{center}
\includegraphics[width=\dwidth]{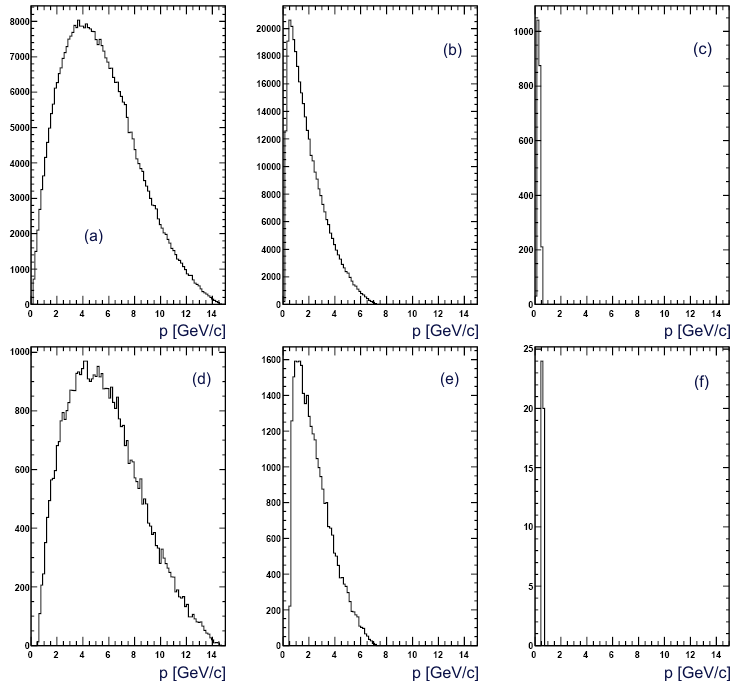}
\caption[$\piz$ and $\eta$ energy spectrum.]
{$\piz$ and $\eta$ energy spectrum (scale in GeV) for $p_{\pbar}=15\,\gevc$ for the \FWEMC (left), \BEMC (middle) and \BWEMC (right), for $\pi^0$ (top) and $\eta$ (bottom).}
\label{fig:pietaspectrum}
\end{center}\end{figure*}
\begin{figure}[bth]\begin{center}
\includegraphics[width=\swidth]{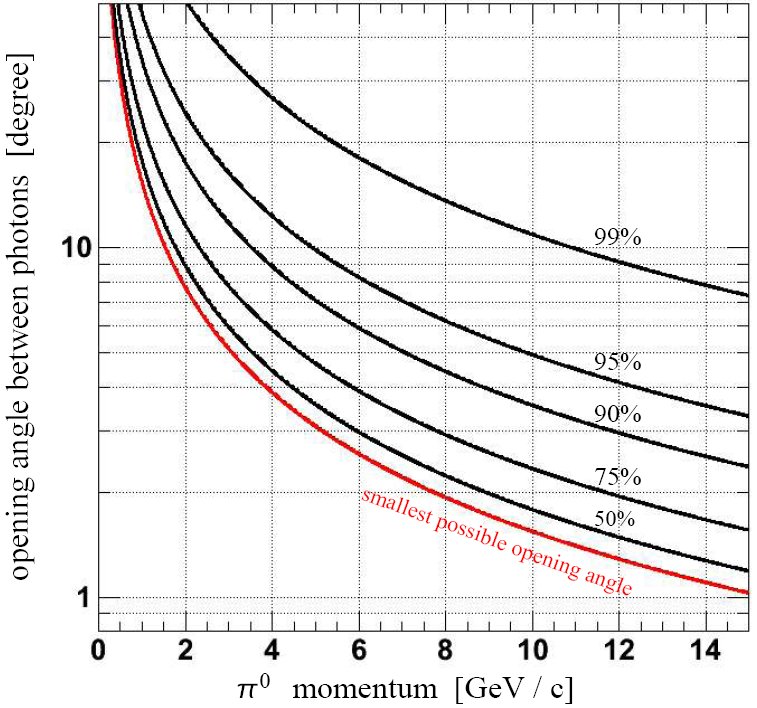}
\caption[Minimum $\piz$ opening angle.]
{Minimum $\piz$ opening angle vs. beam momentum.}
\label{fig:pi0opening}
\end{center}\end{figure}

\Reffig{fig:pietaspectrum} shows the energy distribution of $\piz$ and $\eta$ for a cocktail of events with many photons
for the three detector regions. The average (maximum) $\piz$/$\eta$ momenta for the three detector parts are $<1\,\gev$ ($<1\,\gev$), $\approx 2\,\gev$ ($\approx 7\,\gev$) and $\approx 5\,\gev$ ($\approx 14\,\gev$) for \BWEMC, \BEMC and \FWEMC, respectively, for the highest incident antiproton momentum of 15\,\gevc. \Reffig{fig:pi0opening} shows the minimum opening angle for various $\pi^0$ momenta. The angular coverage of a crystal should be tuned to the smallest $\piz$ opening angle possible for this subdetector and should not exceed 10$^\circ$, 2$^\circ$ and 0.5$^\circ$, respectively, to fully resolve the photons. These angles may be a factor 2 larger when taking cluster moment analysis into account.
\par
\begin{figure}[bth]\begin{center}
\includegraphics[width=\swidth]{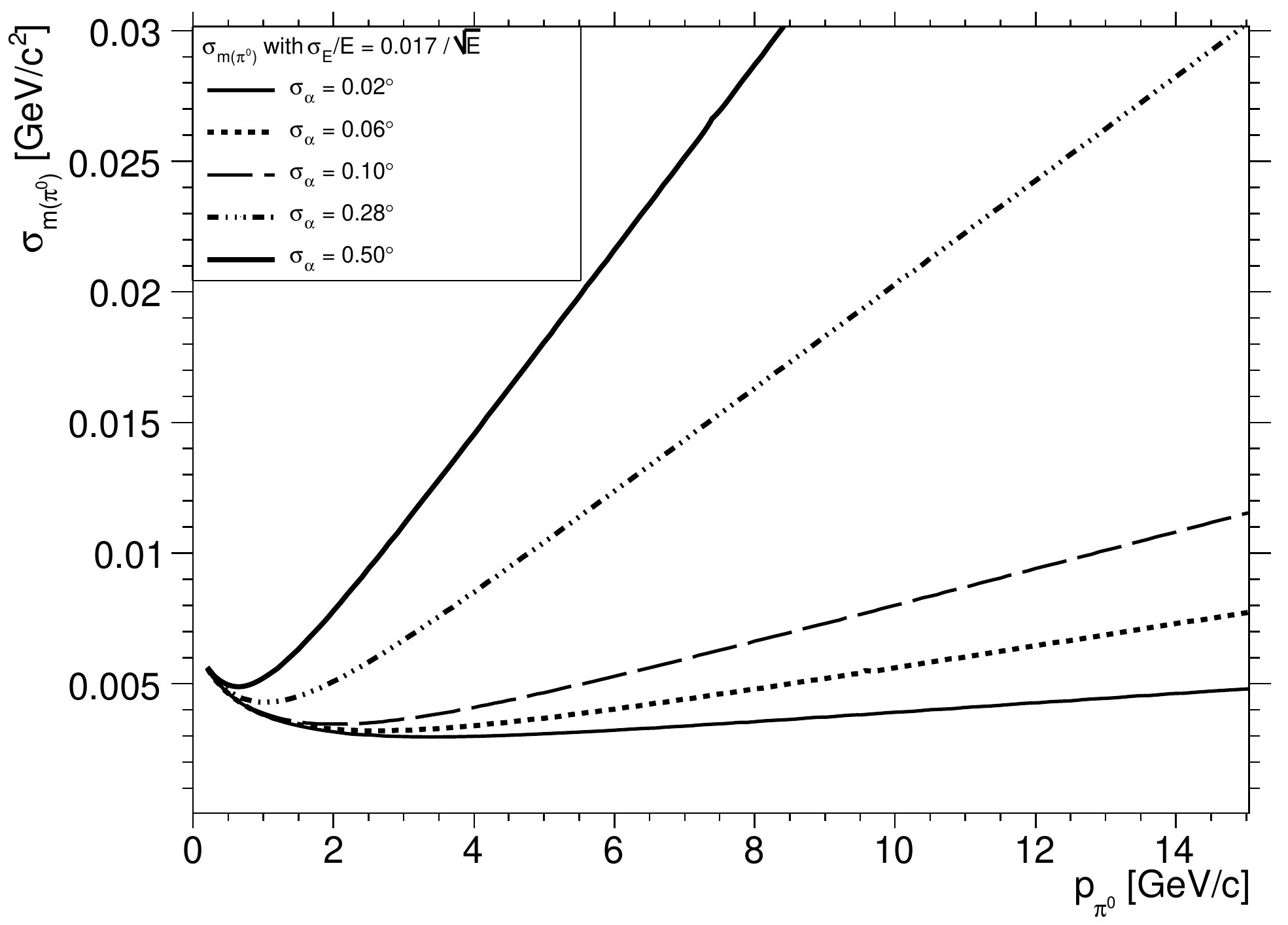}
\caption[Mass resolution for $\piz$ (spatial resolution).]
{$\piz$ mass resolution for various spatial resolution values vs. beam momentum.}
\label{fig:pi0reso:spatial}
\end{center}\end{figure}

The required spatial resolution is mainly governed by the required width of the $\pi^0$ invariant mass peak in order to assure proper final-state decomposition. \Reffig{fig:pi0reso:spatial} shows the effect on the $\piz$ mass resolution as a function of momentum for various spatial resolution values. Taking into account the average $\piz$ energies a narrow $\piz$ width below 8\,\mev is maintained for a $\cos\theta$ dependent spatial resolution of $\le$0.5$^\circ$, $\le$0.3$^\circ$ and $\le$0.1$^\circ$ for \BWEMC, \BEMC and \FWEMC, respectively. These resolutions are in accordance with the granularity requirement. They can be achieved by an
asymmetric geometry where compactness is still maintained for the radial component, while it extends more to the forward direction.
\section{Environment}
\subsection{Surrounding Detectors}
\subsubsection{Magnet System}
The EMC will be operated in a high solenoidal magnetic field (2\,T).
Therefore the sensors perpendicular (barrel part) to the field have to be field insensitive.
In the endcaps the requirements due to the magnetic field are relaxed since the sensors would be
differently oriented. \par
Backscattering in the magnet may take place. These effects are accounted
for in the detector
simulations in subsequent chapters.
\subsubsection{DIRC}
Detectors of Internally Reflected Cherenkov Light (DIRC) are placed in the barrel part and in
the forward endcap for particle identification. They are located near the front face of the
scintillator and add substantially to the material budget before the EMC. It has been shown,
that the preshower information of the Cherenkov light of the shower electrons of such a detector
can be used to repair the distorted calorimeter information.
\subsubsection{Other Systems}
Various different target systems are foreseen for the \Panda experiment. Gas-Jet and Pellet
targets require a target pipe perpendicular to the beam pipe. Since pumps have to be placed outside of the
detector, the target pipes extend to the end of the magnet, thus crossing the EMC. Mechanical
cut-outs for these pipes have to be foreseen.
\subsection{Count-rate and Occupancy}
\subsubsection{Signal Load}
\label{sec:load}
\begin{figure}[bth]\begin{center}
\includegraphics[width=\swidth]{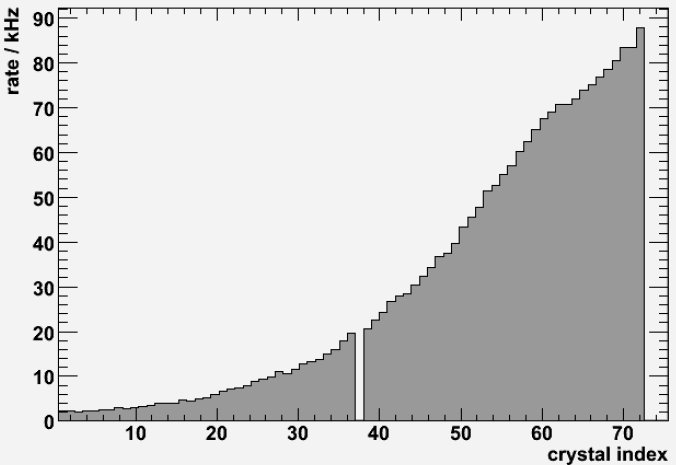}
\caption[Hit rate in the barrel part.]
{Hit rate in the barrel part from the DPM background generator at $p_{\pbar}$=14\,\gevc.}
\label{fig:hitsbarrel}
\end{center}\end{figure}

\begin{figure}[bth]\begin{center}
\includegraphics[width=\swidth,height=\swidth]{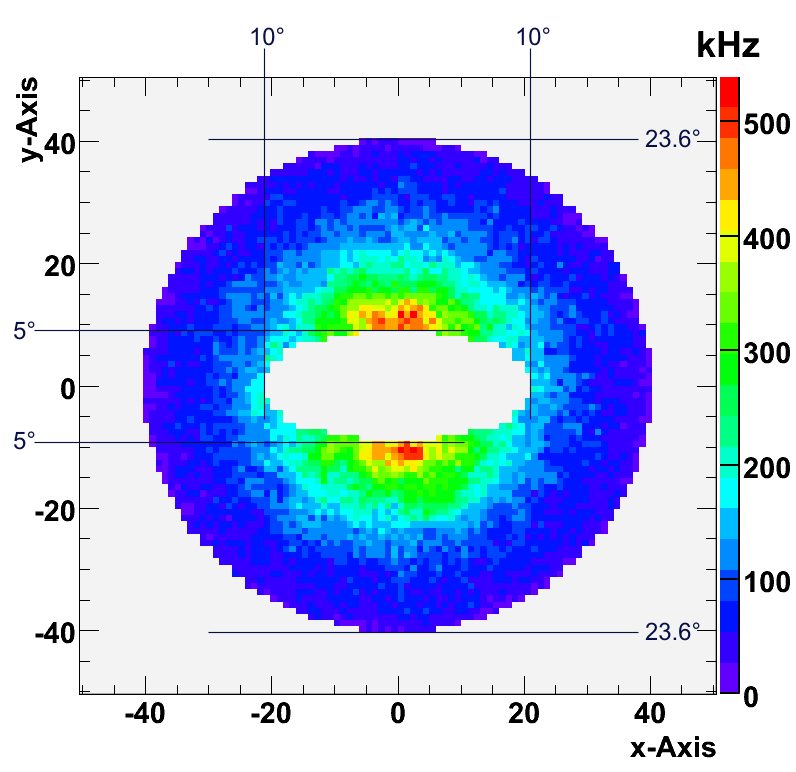}
\caption[Hit rate in the forward part.]
{Hit rate in the forward part from the DPM background generator at $p_{\pbar}$=14\,\gevc.}
\label{fig:hitsfwd}
\end{center}\end{figure}

\Reffig{fig:hitsbarrel} and \ref{fig:hitsfwd} show the hit rates ($E_\gamma>$1\,\mev, DPM background generator) for a geometrical setup which fulfills the spatial requirements already mentioned. The rates are for the barrel part and the forward endcap, respectively. The detector has to be able to digest the maximum hit rate which happens for the highest beam momenta. The maximum rates per crystal are $\approx$60\,kHz and $\approx$500\,kHz for \BEMC and \FWEMC respectively. For heavy targets the single crystal rate for the \BEMC goes up to $\approx$100\,kHz due to the reduced boost compared to $\ppbar$ events.
\subsubsection{Response and Shaping Time}
The response time has to be short enough to allow event identification. Since the $\pbar$ beam will have a time structure,
\Panda has to be prepared to accept and instantaneous rate of up to 50\,MHz on short timescales. This condition leads to a
required time resolution for the relevant hit time $t_0$ of  3\,ns or less. 
The shaping time of the preamplification stage should be longer than the actual
decay time of the scintillation mechanism to collect the whole signal. The shaping time should be maximised to improve the noise level. If the shaping time becomes too long pile-up occurs, since new hits are delivered on top of a preceding hit.
As mentioned previously, undetected photons are a severe cause of background. Therefore a maximum effective pile-up rate of less than 1\% is required. In particular in the forward region this would lead to extremely short (and therefore unrealistic) shaping times. To reduce photon loss, the pile-up hits should be recovered. This can be achieved with a FADC system
with appropriate sample frequency and proper hit detection (in hardware or software). In case of a FADC readout a pile-up rate up to 10\% can be tolerated (including variations of the instantaneous rate) leading to shaping times of not more than $\approx\,100$\,ns for the \FWEMC and $\approx\,400$\,ns for \BEMC and \BWEMC.

Using the hit rates from \Refsec{sec:load} and requiring less than 1\,\% loss due to pile-up the shaping times would drop. 
\subsubsection{Radiation Hardness}
\label{sec:req:rad}
Potential detector setups which fulfill the requirements already mentioned have been used to check for radiation doses.
These simulations show that for the highest beam momenta the innermost crystals (setup as in \Refsec{sec:load}) accumulate an energy dose of 25\,mJ/h for $\ppbar$ generic background (DPM background generator). The maximum dose does not increase with heavier targets. Due to the much reduced boost of the overall system, the hit rate in the innermost crystals is reduced dramatically (even with the same luminosity) and the dose in this case is distributed more evenly among crystal rings. For a crystal weight of 1\,kg (in case of e.g. PWO or BGO) and a typical annual operation of not more than 5\,kh the maximum annual dose would be 125\,Gy (or 12.5\,kRad/year). This strong constraint applies only to the forward crystals. For \BWEMC and \BEMC the maximum energy doses (for $\ppbar$ events) are 30\,$\mu$J/h (0.15\,Gy annually) and 1.4\,mJ/h (7\,Gy annually), respectively. Due to the reduced boost, part of the \BEMC and in particular the whole \BWEMC will have a much higher rate than with $\ppbar$ events. To be on the safe side, all non-forward crystals should be radiation hard to an annual limit of 10\,Gy.
\subsection{Operational Aspects}
\subsubsection{Long- and Medium-term Stability}
\Panda will be operated in a factory mode, thus running for several months (up to 9 months) and will be suspended for the rest of the year for maintenance, repair and upgrade. This requires a long Mean-Time-between-Failure to the level of not more than 10 channels lost per year. 
\subsubsection{Maintenance of Sensors, Electronics and Environment}
\Panda is planned to operate for at least one decade. Therefore it is mandatory to allow access to all
crystals, sensors and electronics for maintenance during down-time periods. It is envisaged to dismount the detector annually on the detector module level for service work. Therefore the overall mounting procedure
has be reversible and detector module oriented.
\subsubsection{Calibration and Monitoring Prerequisites}
The energy calibration has to be performed at a precision much better than the 
energy resolution, thus at the sub percent level.
The \PANDA DAQ relies on online trigger decisions performed on compute nodes.
The use of the electromagnetic calorimeter for the trigger decisions requires
the availability of calibration constants with sufficient precision 
in real time.

The monitoring system should detect any changes in the readout chain starting
from the light transmission in the crystals to the digitizing modules.
\section{Production and Assembly}
\subsection{Production Schemes}
The fabrication of calorimeters has been exercised many times in the past. Many production and assembly procedures
have been used in the past for other large scale projects. Whenever possible existing production schemes,
equipment and personnel should be assimilated to reduce overall cost. Decent quality assurance processes are needed for all mass-produced parts.
\subsection{Assembly Schemes}
Modern large-scale detectors require a large number of channels and many mass-produced parts which have to
be screened and stored for further assembly. Since not all parts can be stored and assembled at a single laboratory, complex logistics has to be arranged to ensure a smooth supply of parts throughout the production process.
\section{Conclusion}
The full list of requirements is compiled in \Reftbl{tbl:req}.
\begin{table*}[hbtp]
\begin{center}
%
%
\begin{tabular}{|l|lll|}
\hline\hline
 & \multicolumn{3}{l|}{Required performance value}\\ \hline
Common properties & & & \\\hline
energy resolution $\sigma_{E}/E$ & \multicolumn{3}{l|}{$\le 1\,\% \oplus \frac{\le 2\,\%}{\sqrt{E/\gev}}$}\\
energy threshold (photons) $E_{thres}$ & \multicolumn{3}{l|}{10\,\mev (20\,\mev tolerable)}\\
energy threshold (single crystal) $E_{xtl}$ & \multicolumn{3}{l|}{3\,\mev}\\
rms noise (energy equiv.) $\sigma_{E,noise}$ & \multicolumn{3}{l|}{1\,\mev}\\
angular coverage \%\,4$\pi$ & \multicolumn{3}{l|}{99\,\%}\\
mean-time-between-failures  $t_{mtbf}$ &  \multicolumn{3}{l|}{ 2000\,y}\\
(for individual channel) & & & \\
\hline
Subdetector specific properties & backward & barrel & forward \\
 & ($\geq 140^\circ$) & ($\geq 22^\circ$) & ($\geq 5^\circ$) \\
\hline
energy range from $E_{thres}$ to & 0.7\,\gev & 7.3\,\gev & 14.6 \,\gev \\
angular equivalent of crystal size $\theta$ & \multicolumn{2}{c}{4$^\circ$} & 1$^\circ$ \\
spatial resolution $\sigma_{\theta}$ & 0.5$^\circ$  & 0.3$^\circ$  & 0.1$^\circ$ \\
maximum signal load $f_{\gamma}$ ($E_\gamma>E_{xtl}$) & \multicolumn{2}{c}{60\,kHz} & 500\,kHz \\
($\ppbar$-events)
maximum signal load $f_{\gamma}$ ($E_\gamma>E_{xtl}$) & \multicolumn{2}{c}{100\,kHz} & 500\,kHz \\
(all events)
shaping time $t_s$ & \multicolumn{2}{c}{400\,ns} & 100\,ns \\
radiation hardness  & 0.15\,Gy & 7\,Gy & 125\,Gy \\
(maximum annual dose $\ppbar$-events) & & & \\
radiation hardness  & \multicolumn{2}{c}{10\,Gy} & 125\,Gy \\
(maximum annual dose from all events) & & & \\
\hline\hline
\end{tabular}
\caption[Main requirements for the \Panda EMC.]
{Main requirements for the \Panda EMC. Rates and doses are based on a luminosity of $L=2\cdot\,10^{32}cm^{-1}s^{-1}$.}

\label{tbl:req}
\end{center}
\end{table*}
\par
In this technical design report of the EMC, we will demonstrate that a compact lead tungstate crystal calorimeter being operated at low temperatures and read out with LAAPDs and VPTs will fulfill the listed requirements on energy resolution, spatial resolution, energy threshold, timing and radiation hardness.
\par
Test equipment of CMS, employed in the past for large-scale production, can be reused for \Panda. Moreover, the expertise of technicians operating these devices can be advantageously exploited.
%

%
%
\newpage
%

%
\cleardoublepage
\chapter{Scintillator Material}
\label{sec:scint}
%
%

\label{sec:scint:pwo}
\section{Inorganic Scintillators}
\label{sec:scint:pwo:scintillator}

The concept of \Panda places the \TSEMC inside the super-conducting
coil of the solenoid. Therefore, the basic requirement of any
appropriate scintillator material has to be its compactness to
minimize the radial thickness of the calorimeter layer
\cite{bib:emc:scint:alt16,bib:emc:scint:alt17}. Of similar importance,
high interaction rates, the ambitiously large dynamic range of photon
energies, sufficient energy resolution and efficiency and finally a
moderate radiation hardness rule out most of the well known
scintillator materials. Finally, even a compact geometrical design
requires due to a minimum granularity a large quantity of crystal
elements, which rely on existing technology for mass production to
guarantee the necessary homogeneity of the whole
calorimeter. Presently and even in the near future, no alternative
materials besides lead tungstate will become available.

\begin{table*}
\begin{center}
\begin{tabular}{lcccccc}
\hline\hline
\multicolumn{2}{c}{Parameter} & CeF$_3$ & LSO/LYSO:Ce & BGO  & PWO & PWO-II\\
\hline
$\rho$            & g/cm$^{3}$             & 6.16 & 7.40 & 7.13  & \multicolumn{2}{c}{8.28}\\
$X_0$             & $\cm$                  & 1.77 & 1.14 & 1.12  & \multicolumn{2}{c}{0.89}\\
$R_M$             & $\cm$                  & 2.60 & 2.30 & 2.30  & \multicolumn{2}{c}{2.00}\\
$\tau$$_{decay}$  & $\ns$                  & 30   & 40   & 300   & \multicolumn{2}{c}{6.5}   \\
$\lambda$$_{max}$ & $\nm$                  & 330  & 420  & 480   & \multicolumn{2}{c}{420}\\
$n$ at $\lambda$$_{max}$&                  & 1.63 & 1.82 & 2.15  & \multicolumn{2}{c}{2.24/2.17}\\
relative LY       & \percent($LY$ NaI)     & 5    & 75   & 9     & 0.3 at RT & 0.6 at RT\\
                  &                        &      &      &       & 0.8 at -25$\degC$ &2.5 at -25$\degC$ \\
hygroscopic       &                        & no   & no   & no    & \multicolumn{2}{c}{no}  \\
dLY/dT            &  \percent/$\degC$      & 0.1  & 0    & -1.6  & - 2.7 at RT & -3.0 at RT\\
dE/dx (MIP)       & MeV/cm                 & 6.2  & 9.6  & 9.0   & \multicolumn{2}{c}{10.2}\\
\hline\hline
\end{tabular}
\caption[Properties of various scintillation crystals.]{Relevant properties of CeF$_{3}$, LSO/LYSO:Ce, BGO, PWO taken from the updated table of the Particle Data Group 2007. The specific parameters of PWO-II are deduced from the presented test measurements.}
\label{tab:scint:alt:prop}
\end{center}
\end{table*}

Of the materials listed in the table below Bismuth Germanate
(Bi$_4$Ge$_3$O$_{12}$, BGO) is a well known scintillator material
since many decades~\cite{bib:emc:alt1}. It has been applied in several
experiments, like \INST{L3} at \INST{CERN} or \INST{GRAAL}at Grenoble,
providing well performing electromagnetic calorimeters
~\cite{bib:emc:alt2,bib:emc:alt3,bib:emc:alt4}. However, presently it
is only used in large quantities but small samples for medical
applications such as Positron-Electron-Tomography (PET) scanners
~\cite{bib:emc:alt5}. The properties of BGO, listed in
\Reftbl{tab:scint:alt:prop}, allow for the design of a very compact
calorimeter. The light yield is comparable to cooled \PWOII
crystals. The emission wavelength is well suited for an efficient
readout with photo- or avalanche diodes. However, the moderate decay
time imposes a strong limitation on the necessary high rate capability
for the envisaged interaction rate of 10$^{7}$/s of \Panda. In
addition, it excludes in general any option for the generation of a
fast timing information on the level of $\leq$1--2 ns or even below for
higher energies. Concerning the radiation hardness, which has not been
a major issue for the former applications, controversial results are
reported in the literature. Some authors report recovery times in the
order of days or online recovery by exposure to intense light from
fluorescent lamps at short wavelengths
\cite{bib:emc:alt6,bib:emc:alt7,bib:emc:alt8,bib:emc:alt9}. Since most
of these studies are rather old, an update with new studies using full
size scintillation crystals produced recently would become necessary.

CeF$_{3}$ has been invented already in 1989 by D. F. Anderson
\cite{bib:emc:scint:alt10} and identified as a fast scintillator with
maximum emission wavelength between 310 and $340\,\nm$. The
luminescence process is dominated by radiative transitions of
Ce$^{3+}$ ions, which explains the fast decay time of $\sim$$30\,\ns$
and the insensitivity to temperature changes. The luminescence yield
is comparable or even higher than the fast component of BaF$_{2}$,
corresponding to approximately 5\percent of NaI(Tl) as a
reference. Since the crystal matrix is extremely radiation hard, it
was considered for the \INST{CMS} calorimeter during a long period of
R$\&$D, supported by the short radiation length and Moli\`ere radius,
respectively. However, homogeneous crystal samples beyond 10X$_{0}$
length, grown by the Bridgeman method, have never been produced with
adequate quality. Detailed studies of the response function
\cite{bib:emc:scint:alt11} to low energy protons and high energy
photons have documented excellent time and energy resolutions, which
are limited only in case of the reconstruction of the electromagnetic
shower energy by the inhomogeneity of the available crystals. Further
improvement relies on a significant optimization of the manufacturing
process. In case of \Panda, the implementation of CeF$_{3}$ has not
been considered. It would on one hand double the radial thickness of
the calorimeter shell. On the other hand, the effective light yield
would be degraded due to the mismatch of the emission wavelength in
the near UV with the optimum quantum efficiency of LAAPDs.

In the last decade cerium doped silicate based heavy crystal
scintillators have been developed for medical applications, which
require high light output for low energy $\gamma$-ray detection and
high density to allow for small crystal units. Mass production of
small crystals of Gd$_{2}$SiO$_{5}$ (GSO), Lu$_{2}$SiO$_{5}$ (LSO) and
Lu$_{2(1-X)}$Y$_{2X}$SiO$_{5}$ (LYSO) has been established in the
meantime. Because of the high stopping power and fast and bright
luminescence, the latter material has also attracted interest in
calorimetry. However, the need for large homogeneous samples and the
presently high costs, partly justified by the high melting point, are
retarding the R$\&$D. LYSO, which has almost identical physical and
scintillation properties as LSO and shows identical emission,
excitation and optical transmittance spectra, has been recently
produced in samples up to $20\,\cm$ length
\cite{bib:emc:scint:alt12,bib:emc:scint:alt13,bib:emc:scint:alt14,bib:emc:scint:alt15}. The
light output of small samples can reach values 8 times of BGO but with
a decay time shorter by one order of magnitude, determined by the
Ce-activation. Detailed studies even within the \PANDA-collaboration
have been performed and response functions have been obtained even for
high-energy photons up to 500 MeV \cite{bib:emc:scint:alt14}. Aiming
at applications for the next generation of homogeneous calorimeters,
detailed studies of full size crystals with respect to homogeneity,
scintillation processes and radiation hardness are part of a research
program promoted by a group at Caltech \cite{bib:emc:scint:alt15}.

\subsection{Specific Requirements for the \TSEMC}
\label{sec:scint:pwo:general}

As pointed out in the previous chapter in selected figures based on
simulations assuming already to some extent parameters of \PWOII, the
\TSEMC requires very ambitious specifications to be adapted to the
physics program. The main features are:

\begin{itemize}
\item high rate capability, which requires a fast scintillation
  kinetics, appropriate granularity to minimize pile-up as well as
  guarantee optimum reconstruction of the center of the
  electromagnetic shower;
\item sufficient luminescence yield to achieve good energy resolution
  in particular at the lowest photon energies in the MeV range, which
  goes in parallel with a minimum energy threshold of the individual
  crystal;
\item timing information, primarily to reduce background and to
  provide an efficient correlation with other detector components for
  particle identification;
\item adapted geometrical dimensions to contain the major part of the
  electromagnetic shower and to minimize the impact of leakage
  fluctuations and direct ionization in the photo sensor;
\item radiation hardness to limit the loss in optical transparency to
  a tolerable level.
\end{itemize}

Based on the its intrinsic parameters, \PWO, as developed for CMS/ECAL,
most of the requirements could already be met except the light
output. Therefore, an extended R$\&$D program was initiated to
improve the luminescence combined with an operation at low
temperatures such as T=-25$\degC$. The different steps to achieve the
quality of \PWOII, as described later, have shown that radiation
hardness becomes a very sensitive parameter, when the operating
temperature of the calorimeter is below T=0$\degC$. Therefore, the
estimated radiation environment is discussed here in more detail.

\subsection{Hit Rates and Absorbed Energy Dose in Single Crystals}
\label{chap:scint:pwo:rates-crystals}

\begin{figure}[htb]
\begin{center}
\includegraphics[width=\swidth]{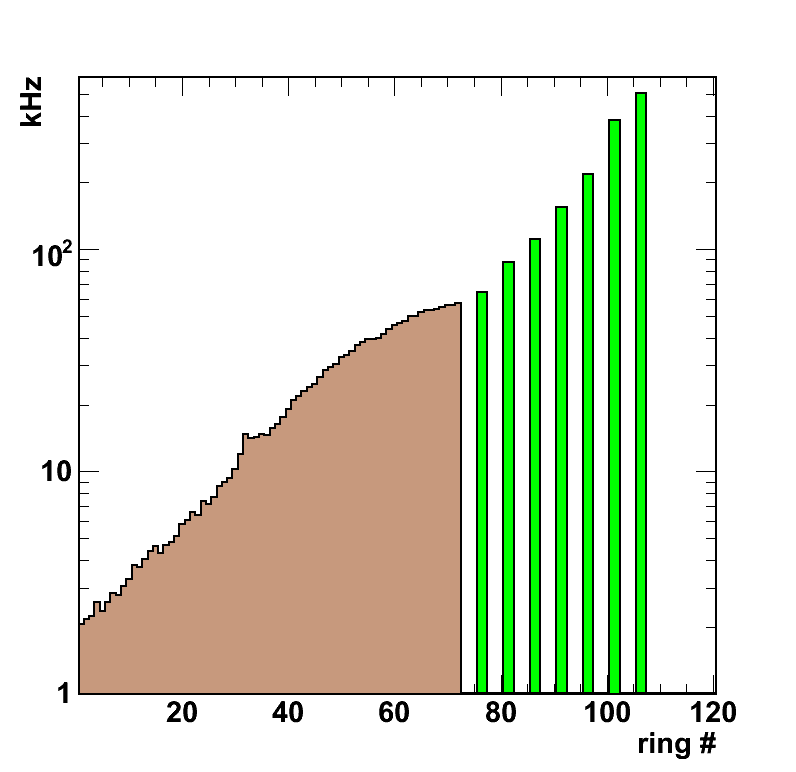}
\caption[Integrated single crystal rate for the barrel section.]
{Integrated single crystal rate for the barrel section for an energy threshold of E$>$$3\,\mev$ using DPM at 15~GeV/c incident
beam momentum.}
\label{fig:perf:Rate}
\end{center}
\end{figure}

\begin{figure}[htb]
\begin{center}
\includegraphics[width=\swidth]{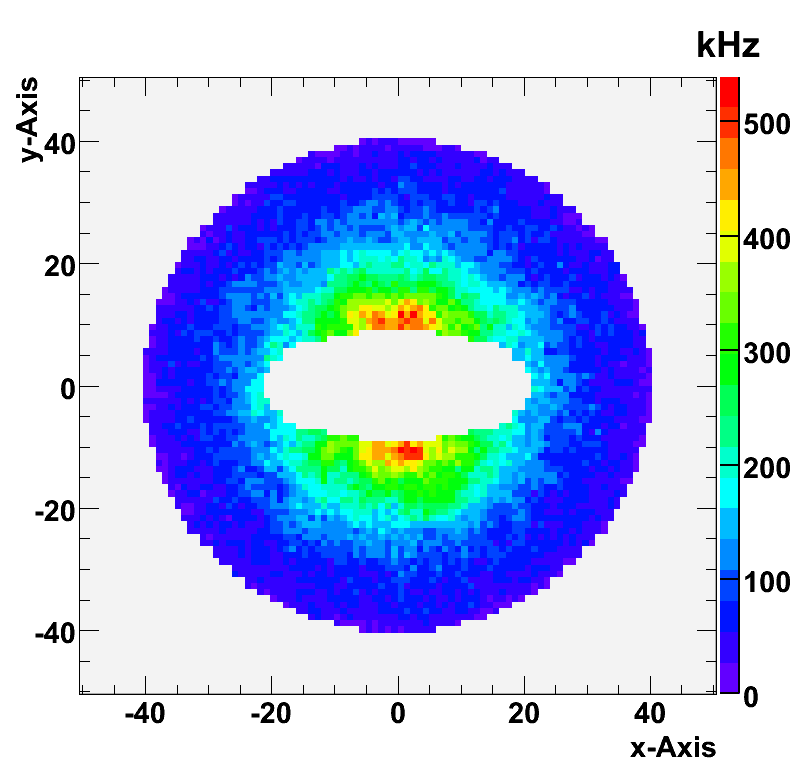}
\caption[Integrated single crystal rate for the forward endcap.]
{Integrated single crystal rate for the forward endcap for an energy threshold of E$>$$3\,\mev$ using DPM at 15~GeV/c incident beam momentum.}
\label{fig:perf:FwdRate}
\end{center}
\end{figure}

The single crystal rates under the proposed running conditions have
been simulated using the Dual Parton Model (DPM) as generator for
primary events. The maximum beam momentum of 15~GeV/c has been used to
evaluate the highest load for $\bar{p} p \rightarrow X$
interactions. A full tracking of the primary generated events through
the whole \PANDA detector using GEANT3/4 was simulated. The results are
scaled to the canonical interaction rate of 10$^{7}$ primary events/s.
Fig.~\ref{fig:perf:Rate} and Fig.~\ref{fig:perf:FwdRate} show the
integral rate for the forward end cap and the barrel part for an
energy threshold E$>$$3\,\mev$.

\begin{figure}[htb]
\begin{center}
\includegraphics[width=\swidth]{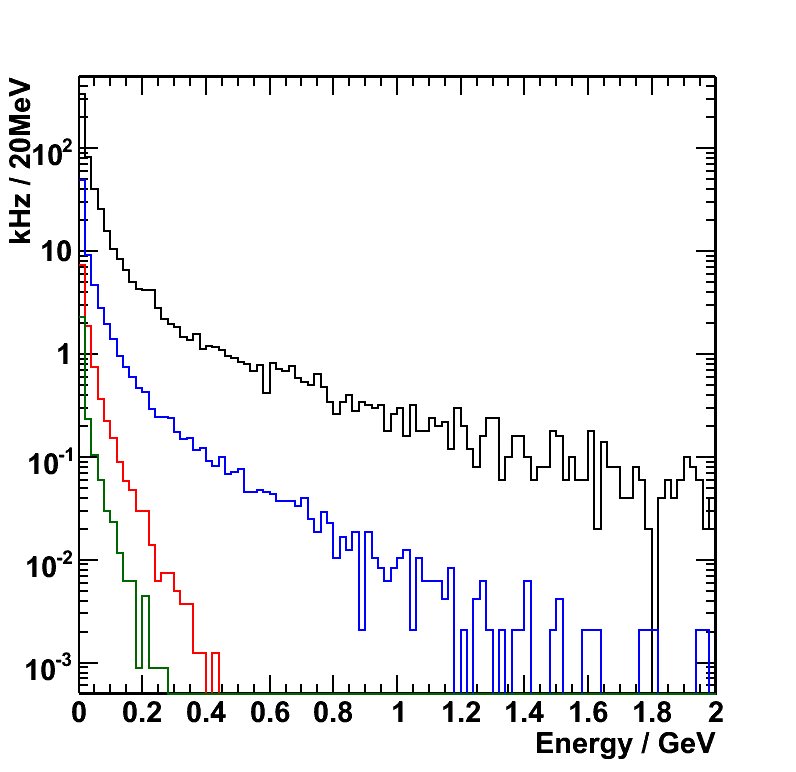}
\caption[Single crystal energy differential rate spectrum.]
{Single crystal energy differential rate spectrum for polar angles of
5$^\circ$ (black), 25$^\circ$ (blue),
 90$^\circ$ (red) and 135$^\circ$ (green) using DPM at 15~GeV/c incident beam momentum.}
\label{fig:perf:Ediff}
\end{center}
\end{figure}

\begin{figure}[htb]
\begin{center}
\includegraphics[width=\swidth]{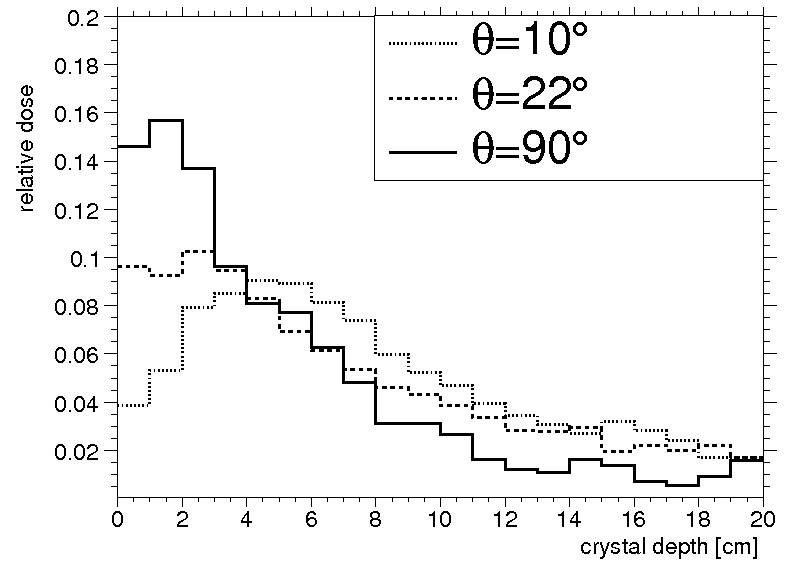}
\caption[Relative energy dose as function of the radial depth.]
{Relative energy dose normalized to 1 as function of the
radial depth in a single crystal using DPM at 15~GeV/c incident beam
momentum.}
\label{fig:perf:Erel}
\end{center}
\end{figure}

The energy spectrum of a single crystal depends strongly on the polar
angle and is represented in Fig.~\ref{fig:perf:Ediff} for four
different polar angles (5$^\circ$, 25$^\circ$, 90$^\circ$ and
135$^\circ$). The summation of the given energy distributions yields
total absorbed energy rates of 27~mGy/h, 1.5~mGy/h, 0.16~mGy/h and
0.03~mGy/h, respectively, for each angle. The energy absorption is
homogeneous in lateral direction of the crystals, but in radial
depth the situation changes. The different interaction mechanisms
of the electromagnetic and hadronic probes with the scintillator
material lead to a strong radial dependence of the energy absorption
as shown in Fig~\ref{fig:perf:Erel} under similar conditions. The
strong variation can bring the stated average numbers up to peak
values, which can be higher by factors of 2-3 within the first few cm
of the crystals. The centers of the electromagnetic shower of photons
up to $10\,\gev$ incident energy will be concentrated within the first
third of the crystal length, which has to be considered for the
optimization of a homogeneous collection of the scintillation light.

Summarizing, the individual scintillation crystals have to be capable
to handle average hit rates above a threshold of $3\,\mev$ of 100 kHz
in the \BEMC and increasing up to 500 kHz in the most forward parts of
the \FWEMC. However, these values might be even exceeded due to
fluctuations in the time structure of the beam. Considering the
complete cocktail of impinging probes, absorbed dose rates up to
20-30 mGy/h have to be expected at most forward regions. The values in
most parts of the \BEMC are two orders of magnitude lower. Therefore,
the absolute values are well below the environment expected for LHC
experiments. Due to the significantly slower recovery processes at
T=-25$\degC$, as outlined in chapter 3.4.1.1, radiation induced
changes of the optical transmittance will reach a final level
asymptotically on a time scale of a typical run period. However, these
limits will only be reached at the most forward angles of the \FWEMC.

\section{Lead Tungstate \PWO}
\label{sec:scint:pwo:pwo}

\subsection{General Aspects}
\label{chap:scint:pwo:aspects}

Lead tungstate, PbWO$_4$ (PWO), crystals meet in general the
requirements to represent an extremely fast, compact, and radiation
hard scintillator material. However, a significant improvement of the
light output was mandatory for the application in
\Panda. \Reftbl{tab:scint:alt:prop} considers already the finally
achieved quality described as PWO-II.

The specifications, which were developed and obtained for the
application in experiments at \INST{LHC} at \INST{CERN} such as the
$\textbf{E}$lectromagnetic $\textbf{CAL}$orimeter (ECAL) of CMS
\cite{bib:emc:pwo6} and the $\textbf{PHO}$ton $\textbf{S}$pectrometer
(PHOS) of ALICE \cite{bib:emc:pwo7} served as a
starting point for the further optimization of lead tungstate crystals
\cite{bib:emc:pwo1}. Radiation hardness, fast scintillation kinetics,
large-scale production and full size crystals of 23$\,\cm$ length
(28$\,X_{0}$) with a light output of 9--11$\,$phe/$\mev$ (measured
with a bi-alkali photocathode at room temperature) became standard.

Looking backwards, in the R{\&}D phase, performed in 1998--2002, the
light output was improved by additional doping
\cite{bib:emc:pwo2,bib:emc:pwo3,bib:emc:pwo4}. Co-doping with
molybdenum (Mo) and lanthanum (La) at concentrations $<$100 ppm
delivered optimum results. However, an additional strong slow
scintillation component ($\tau_{\rm g}= 1-4\,\mus$) appeared limiting
applications at high count rates \cite{bib:emc:pwo17}. Two additional
concepts were pursued:

\begin{enumerate}
\item increase of the structural perfection of the crystal, and
\item activation of the crystal with luminescent impurity
  centers. These have a large cross section to capture electrons from
  the conduction band, combined with a sufficiently short delay of
  radiative recombination.
\end{enumerate}

Beside crystal activation with impurity centers, authors of
Ref.~\cite{bib:emc:pwo5} investigated the possibility to redistribute
the electronic density of states near the bottom of the conduction
band by the change of ligands contained in the crystal, which
unfortunately introduced slow components in the scintillation
kinetics. An obvious possibility to reduce the concentration of point
structure defects is doping with yttrium (Y), lanthanum (La) or
lutetium (Lu) ions, which suppress oxygen and cation vacancies in the
crystal matrix. An activator concentration at a level of 100\,ppm is
needed. This concept has been followed up successfully by the
\INST{CMS} collaboration, and led to mass production technologies of
radiation hard crystals at the Bogoroditsk Technical Chemical Plant
\cite{bib:emc:pwo1}. A further increase of the scintillation yield by
30--50$\percent$ can be achieved by a reduction of the
La-concentration to $\sim$40\,ppm or less, which can compensate
vacancies only if point defects are further suppressed. That was
realized by an improved control of the stoichiometric composition of
the melt.

\subsection{Basic Properties of PbWO$_{4}$ and the Scintillation Mechanism}
\label{chap:scint:pwo:basics}

\begin{figure}[htb]
\begin{center}
\includegraphics[width=\swidth]{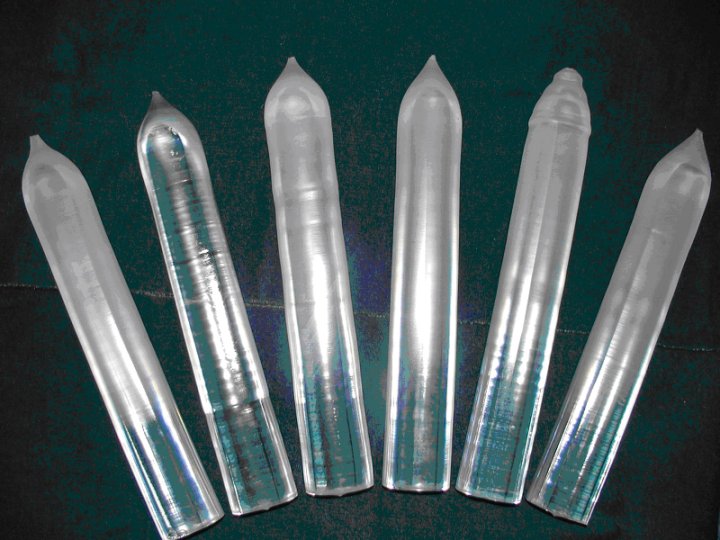}
\caption[Selected \PWOII ingots.]{Selected \PWOII ingots, which have been used to machine crystals in \Panda geometry.}
\label{fig:scint:pwo:fig1}
\end{center}
\end{figure}

The world's largest PWO crystal producer Bogoroditsk Technical Chemical
Plant (BTCP) from Russia produces lead tungstate crystals with high
yield by the Czochralski method, using standard "Crystal 3M" or
"Lazurit" equipment. Crystals are grown from raw material with a
purity level close to 6N specification in Pt crucibles. In order to
grow high quality crystals an additional pre-crystallization is
required. Ingots of $250\,\mm$ length and slightly elliptical
cross-sections with a large axis between 36 and up to $45\,\mm$ in
diameter became available. Crystals can be pulled with identical
quality in the directions close to the a or c axis. Nevertheless, the
production along the direction of the a axis appears to be the best
for producing long samples. The process of pulling is carried out in
an isolated chamber filled with nitrogen. Pulling rate and rotation
speed can be varied over a wide range adapted to the required ingot
diameter. Several crystal ingots to produce modules for the \FWEMC are
shown in \Reffig{fig:scint:pwo:fig1}.

Lead tungstate crystals occur in nature as tetragonal stolzite
\cite{bib:emc:pwo8}, scheelite type or monoclinic raspite
\cite{bib:emc:pwo9}. The structure of the synthetic lead
tungstate crystal is determined by X-ray diffraction as scheelite-type
of tetragonal symmetry with a space group of I4$_{1}$/a and unit-cell
parameters a = b = 5.456(2), c = 12.020(2) {\AA}. The density of the
synthetic crystal can vary by a few percent depending on the
technology.

The electronic structures of the conducting and valence band for pure
and defective lead tungstate crystals are described in detail in
\cite{bib:emc:pwo10}. According to the electron band structure,
luminescence appears in PWO crystals because of charge-transfer
transitions in a regular (WO$_4$)$^{2-}$ anionic molecular
complex. Since the (WO$_4$)$^{2-}$ complex has a T$_{d}$ point
symmetry of the crystalline field, the final configurations of the
energy terms are found to be $^{3}$T$_{1}$, $^{3}$T$_{2}$ and
$^{1}$T$_{1}$, $^{1}$T$_{2}$ with the $^{1}$A$_{1}$ ground state. The
spin and parity forbidden radiative transition $^{1}$T$_{1}$,
$^{1}$T$_{2} \rightarrow ^{1}$A$_{1}$ in the anionic complex is
responsible for the intrinsic blue luminescence of the lead tungstate
crystal. The luminescence has a large Stokes shift of $0.44\,\eV$ and
therefore is characterized by a strong temperature quenching.

\begin{figure}[htb]
\begin{center}
\includegraphics[width=\swidth]{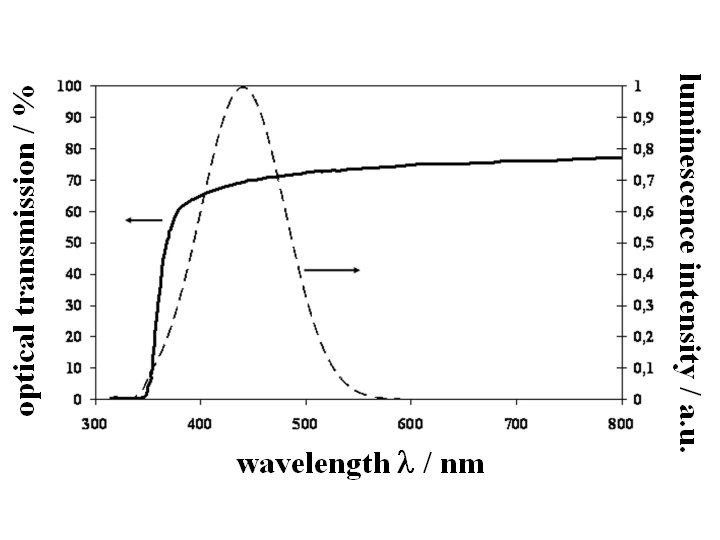}
\caption[The optical transmission and radio-luminescence spectra.]
{The optical transmission and radio-luminescence spectra at room temperature measured for a $200\,\mm$ long PWO crystal.}
\label{fig:scint:pwo:fig2}
\end{center}
\end{figure}

Crystalline material of lead tungstate is transparent in the visible
spectral range and fully colorless. A typical transmission spectrum
measured along the crystallographic axis a of a $200\,\mm$ long sample
and the radio-luminescence spectrum measured at room temperature are
compared in \Reffig{fig:scint:pwo:fig2}. There is an overlap of the
distribution of luminescence with the cut-off of the transmission
spectrum. Therefore, optimizing the light yield relies significantly
on the crystal transmittance in the near UV region.

\begin{figure}[htb]
\begin{center}
\includegraphics[width=\swidth]{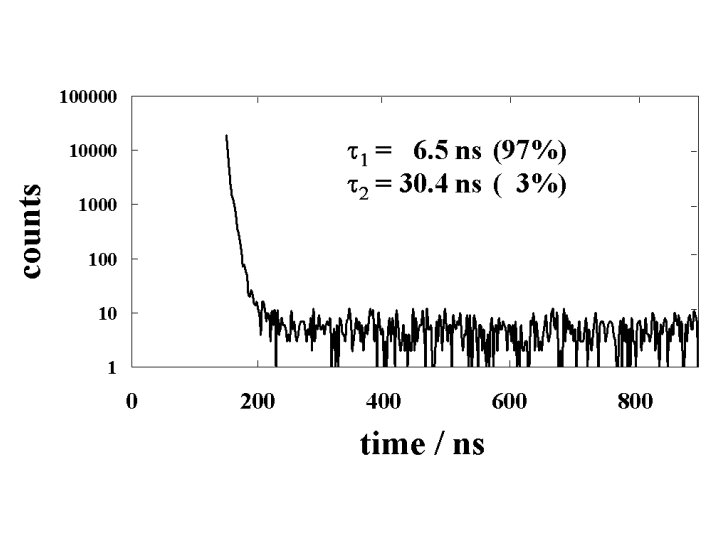}
\caption[The scintillation kinetics of PWO measured at room temperature.]{The scintillation kinetics of PWO measured at room temperature.}
\label{fig:scint:pwo:fig3}
\end{center}
\end{figure}

Due to strong temperature quenching of the luminescence in PbWO$_{4}$,
the light yield shows a strong temperature dependence. As a
consequence, the scintillation kinetics is fast at room temperature
and is dominated by an exponential decay constant of $\tau$=6.5ns. The
temperature dependence of the scintillation light yield in the range
+20$\degC$ to -25$\degC$ was found to be close to linear with a
gradient of -2\percent/$\degC$. The typical kinetics of the
scintillation processes in PWO is shown in
\Reffig{fig:scint:pwo:fig3}.

Both scintillation and optical properties of the currently produced
lead tungstate crystals are rather uniform along the axis of crystal
growth. The variation with respect to the wavelength of the optical
transmission value of T=50\percent does not exceed
$\Delta \lambda = 3\,\nm$ for a $200\,\mm$ long crystal.

\subsection{The Improved Properties of PWO-II}
\label{chap:scint:pwo:improved}

Up to date the CMS Collaboration has the largest comprehensive
experience with the exploration and production of lead tungstate
\cite{bib:emc:pwo11, bib:emc:pwo12}. Recently BTCP has completed the
production of 77,000 scintillation crystals for the CMS/ECAL using the
Czochralski method. A smaller percentage was delivered by the Shanghai
Institute of Ceramics (SICCAS) using a modified Bridgeman method
\cite{bib:emc:pwo13}. As an outcome of seven years of mass production
the following quality parameters of the production technology were
obtained \cite{bib:emc:pwo14}, summarized in
\Reftbl{tab:scint:cms:prop}.

\begin{table*}
\begin{center}
\begin{tabular}{cccc}
\hline\hline
number  & parameter & mean value & standard   \\
of samples        &           &            &  deviation \\
\hline
40607 &  length/mm & 229.95 & 0.02\\
61267 &  LY/phe$\cdot$MeV$^{-1}$& 10.2 & 1.2\\
61267 &  LY non-uniformity /$\percent$X$_{0}$ & -0.13 & 0.02\\
61267 &  opt. transm. at 420nm/$\percent$ & 69.5& 2.2\\
61267 &  rad. ind. absorption at 420nm/m$^{-1}$ & 1.04 & 0.4\\
\hline\hline
\end{tabular}
\caption[Distribution of the crystal properties of accepted PWO crystals.]
{Distribution of the crystal properties of accepted PWO crystals for the CMS/ECAL project.}
\label{tab:scint:cms:prop}
\end{center}
\end{table*}

One can state that both crystal producers reached a high quality of
the crystals. Moreover, the distributions of crystal parameters are
significantly superior to the limits set by the CMS specifications. As
a consequence, only a small fraction at the level of 1.5\percent was
rejected by the certification of scintillation elements repeated by
the customer at CERN.

\begin{figure*}[bth]
\begin{center}
\includegraphics[width=\dwidth]{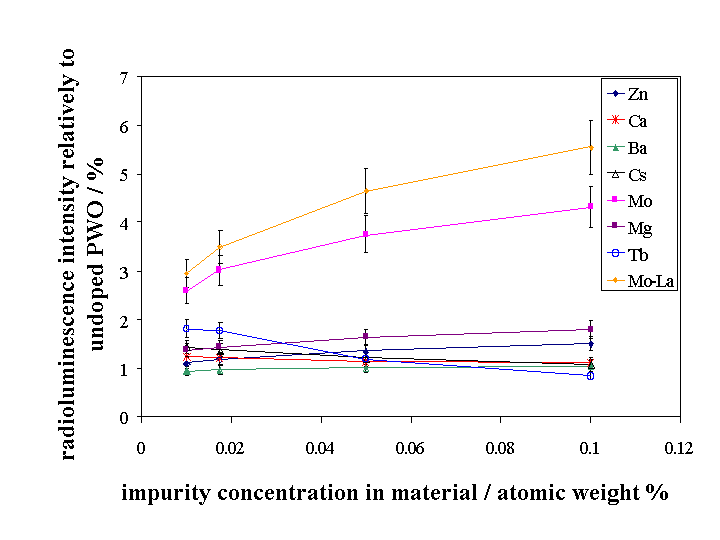}
\caption[Integral dependence of the radioluminescence.]{Integral dependence of the radioluminescence of lead
  tungstate crystals doped with different ions as a
  function of the activator concentration. The measurements have been
  performed at an absolute temperature of T=300\,K.}
\label{fig:scint:pwo:fig4}
\end{center}
\end{figure*}

Experimental applications for photon detection in the MeV range below
200 MeV are predominantly limited by the photon statistics
\cite{bib:emc:pwo15,bib:emc:pwo16}. Obviously, further improvement of
the crystal light yield became mandatory. Detailed studies of the
different dopants in the crystal did not lead to a spectacular
increase of the light yield \cite{bib:emc:pwo2}, as illustrated in
\Reffig{fig:scint:pwo:fig4}.

PWO is very sensitive to the conditions at the synthesis. During the
crystal growth by the Czochralski method from stoichiometric raw
material a dominant leakage of lead from the melt takes place leading
to the creation of cation vacancies V$_{c}$ on the lead site in the
host. This holds also for crystals obtained by the modified Bridgeman
method \cite{bib:emc:pwo13}. The systematic lead deficiency introduces
a superstructure created by cation vacancies as identified by combined
X-ray and neutron diffraction measurements \cite{bib:emc:pwo18}. In
fact, cation vacancies V$_{c}$ are compensated by oxygen vacancies
V$_{O}$ or Frenkel-type defects and their ordering is compensated by
the distortion of the tungstate anionic polyhedra. During the
development, point structure defects, such as electron capturing
centers were identified by Thermo Stimulated Luminescence (TSL) and
Electron Paramagnetic Resonance (EPR) measurements
\cite{bib:emc:pwo19,bib:emc:pwo20,bib:emc:pwo21,bib:emc:pwo22}.  As a
result, the shallowest center occurs in all crystals as an intrinsic
defect: an additional electron auto-localized at an anionic
(WO$_4$)$^{2-}$ complex via a Jahn-Teller distortion creating a
(WO$_4$)$^{3-}$ polaronic center
\cite{bib:emc:pwo23,bib:emc:pwo24,bib:emc:pwo25}. This shallow trap is
emptied near a temperature of 50\,K with an activation energy of
$0.05\,\eV$. Part of the released electrons recombine radiatively or
are caught by deeper traps.  The second one is a Pb$^{1+}$-V$_{0}$
center, which is stable in the crystal up to 175\,K
\cite{bib:emc:pwo26,bib:emc:pwo27}. It is not excluded that instead of
Pb an other ion may create such a center near an anion
vacancy. However, it counts that the electron is trapped by a
hetero-valent cation in the vicinity of an oxygen vacancy. That can be
proved by a relatively large deviation of the g-factor of all such
magnetically non-equivalent species from g$_{e}$.  This center is not
clearly correlated with the detected TSL band, however it is
photo-ionized by IR with the threshold $0.9\,\eV$.  The third one is a
(WO$_4$)$^{3-}$ electron center, which is created on a base of a
regular tungstate anionic complex disturbed by a nearby rare earth
(RE) trivalent impurity ion like La, Lu or Y \cite{bib:emc:pwo24}. It
decays near 97\,K and has an activation energy $0.2\,\eV$. Careful
partial annealing experiments showed that the rate of decrease of the
EPR intensity of (WO$_4$)$^{3-}$-La coincides with the TSL emission in
this temperature region \cite{bib:emc:pwo26}. In addition, the doping
of the crystal with stable trivalent RE ions such as La, Lu, Gd and Y
at concentrations of some tens of ppm redistributes electron trapping
centers and reducing the number of deep ones
\cite{bib:emc:pwo26,bib:emc:pwo28,bib:emc:pwo29}. Such ions localized
at Pb sites introduce in the crystal an additional positive
uncompensated charge and thus will compete with the creation of
vacancies. In addition, the dopants introduced in the crystal above
the stoichiometric ratio occupy empty lead sites in the lattice and
suppress the superstructure fraction with its distorted tetrahedra in
the crystal.

Another trap center with an activation energy of 0.13 eV appears in the
crystal with RE doping. This RE-(WO$_4$)$^{4-}$ center is not
paramagnetic. This hypothesis is in agreement with the fact that
RE-doped PWO crystals show higher TSL intensity in the vicinity of
100 K temperature. An electron release from the 0.13 eV traps causes
simultaneous production of the lowest electron centers by re-trapping
as well as by contributing to an increase of TSL intensity in that
region via the creation of RE-(WO$_4$)$^{3-}$.

The deepest electron trap centers are double vacancy centers having an
activation energy in the range of 0.4-0.5\,eV and Frenkel-type defect
with an activation energy 0.7\,eV, which occurs in the crystal at the
shift of the oxygen ion into the inter-site position. Double vacancies
cause induced absorption in the near IR spectral region in the
crystal. Frenkel-type defects cause an optical absorption band at a
wavelength of 360\,nm, which is converted under ionizing radiation in
the absorption band with a maximum near 410\,nm.

We attribute a molybdenum (Mo) based center to the characteristic
luminescence centers in lead tungstate. It is related to a
(MoO$_4$)$^{2-}$-anion complex impurity, which is a stable electron
trap center. Although the raw material is purified before crystal
growth, especially of Mo, the molybdenum ion is chemically very close
to the tungsten ion and it is rather hard to separate both at the
production quality level of the raw material. The effective molybdenum
impurity in \PWOII is typically at the level of $<$$1\,\ppm$. The
properties of the (MoO$_4$)$^{3-}$-center and its impact on the
scintillation parameters of PWO are described in detail in
\cite{bib:emc:pwo17}.

Following the disposition of electronic centers near the bottom of the
conducting band it appears to be possible to describe the influence of
each kind of defects on the crystal scintillation properties even at
different temperatures. On one hand shallow polaronic and RE-distorted
regular centers contribute to the scintillation via relatively fast
electron exchange with the conducting zone. A reduction occurs due to
loss of electrons into the deep traps. The increase of the
concentration of (WO$_4$)$^{3-}$-RE centers in the crystal leads to a
drop of the light yield. Therefore, an increase of the light yield is in
conflict with optimization of radiation hardness by a higher
concentration of activating ions such as La. To fulfill both,
perfection of the crystal is the only way out.

Decreasing the operating temperature of the calorimeter down to
T=-25$\degC$ slows down the time constants of the spontaneous recovery
of the electron centers by thermal
activation. \Reftbl{tab:scint:pwo:par} shows the parameters of
selected electron capturing centers in PWO crystals
\cite{bib:emc:pwo21} obtained by the TSL method, where E$_{T}$, S,
$\gamma$ are the activation energy, the frequency factor and the order
of the kinetics, respectively. The calculated spontaneous recovery
constants for two temperatures are shown in addition. When the
temperature is decreasing to T=-25$\degC$ the deepest electron center
becomes meta-stable and increases effectively the level of dynamic
saturation of the radiation induced optical absorption in the
crystal. This effect has to be considered and its consequences have to
be investigated under the experimental conditions for \Panda.

\begin{table*}
\begin{center}
\begin{tabular}{ccccc}
\hline\hline
E$_{T}$    &   S   & $\gamma$  & $\tau$$_{rec}$ at T=+25$\degC$  & $\tau$$_{rec}$ at T=-25$^{o}$C\\
 meV       &       &           &    s                            &      s                         \\
\hline
0.05&  4.00E+03& 1.1 & 1.80E-03 & 2.70E-03\\
0.135 &	1.0E+08 & 1.6 & 2.80E-06 & 8.36E-06\\
0.19 & 2.90E+08 & 1.35 & 6.40E-06 & 2.98E-05\\
0.23 & 6.7E+09 & 2.2 & 2.2E-06 & 1.42E-05\\
0.27 & 9.0E+10 & 2.2 & 7.3E-07 & 6.51E-06\\
0.40 & 2.0E+10 & 1.1 & 3.1E-04 & 7.93E-03\\
0.50 & 9.0E+11 & 1.4 & 5.0E-04 & 2.87E-02\\
0.49 & 2.0E+09 & 1,0 & 9.0E-02 & 4.77\\
0.58 & 1.1E+07 & 1.0 & 4.80E+02 & 5.28E+04\\
0.7 & 5.5E+7 & 1.0 & 1.00E+04 & 2.91E+06\\
\hline\hline
\end{tabular}
\caption[Selected parameters of the electron capturing centers in PWO crystals.]{Selected parameters of the electron capturing centers in PWO crystals.}
\label{tab:scint:pwo:par}
\end{center}
\end{table*}

Structural perfection of the crystal has been reached by an improved
control of the stoichiometric composition of the melt. Since the
full-size crystal samples have a length of $200\,\mm$, crystal
activation was pursued at different stages of the raw material
preparation by Y and La ions with distribution coefficients of 0.95
and 1.47, respectively. EPR studies have confirmed that only single
RE-(WO$_4$)$^{3-}$ (RE=La,Y) centers are formed, if the concentration
of the dopants remains below 20\,ppm for each ion species. These
crystals contain finally at least two times less Frenkel type defects
leading to an increase of light by $\>$ 80\percent compared to CMS
quality. The kinetics of the enhanced material remains fast and allows
to integrate $\>$97\percent of the light within a 100 ns wide time gate
at room temperature.

\section{Radiation Induced Absorption in PWO-II Crystals}

\begin{figure*}[bth]
\begin{center}
\includegraphics[width=\dwidth]{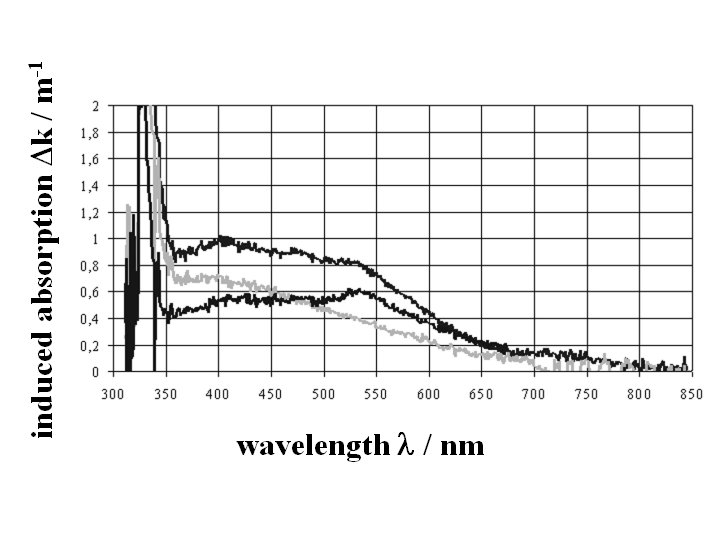}
\caption[The measured induced absorption of various \PWOII samples.]
{The measured change of the induced absorption coefficient of various \PWOII samples
  after irradiation with a dose of 0.2 kGy ($^{60}$Co) over the relevant range of wavelength.}
\label{fig:scint:pwo:fig13}
\end{center}
\end{figure*}

The stability of the parameters of the scintillation material in a
radiation environment is one of the most important properties required
to build a precise electromagnetic calorimeter. As common for most of
the scintillators, the scintillation mechanism in PWO itself is not
damaged when grown under optimized conditions \cite{bib:emc:pwo11}.
Irradiation affects only the optical transmission due to the creation
of color centers with absorption bands in the visible spectral
region. The induced absorption evolves with time, depending
on the dose rate and the origin of the participating centers and their
mechanism of annihilation and relaxation. The absorption is quantified by the
absorption coefficient k given in m$^{-1}$ assuming a pure exponential change
of the transmission loss. \Reffig{fig:scint:pwo:fig13}
shows typical spectra of the induced absorption loss measured for \PWOII
samples at room temperature after a total absorbed dose of 0.2 kGy
using a $^{60}$Co source expressed by the difference of the coefficients
before and after irradiation at given wavelengths.

The obtained values stay below the considered quality limit
$\Delta$k=1m$^{-1}$ in the relevant spectral range of the
luminescence. Nevertheless, enhanced absorption is visible near the
maximum of scintillation emission at $420\,\nm$ or at a broad
structure near $\lambda$=$550\,\nm$.


In contrast to previous applications, the calorimeter will be operated
at T=-25$\degC$. Therefore, a detailed research program had to be
started to investigate possibly new and unexpected effects due to
irradiation at crystal temperatures below 0$\degC$. Besides the
facilities performing quality control during the phases of development
and pre-production at the manufacturer locations BTCP or INP (Minsk,
Belarus) two additional and complementary facilities at IHEP
(Protvino, Russia) and JLU (Giessen, Germany) have been adapted for
that purpose.

\subsection{The Irradiation Facility at IHEP, Protvino}

\begin{figure}[htb]
\begin{center}
\includegraphics[width=\swidth]{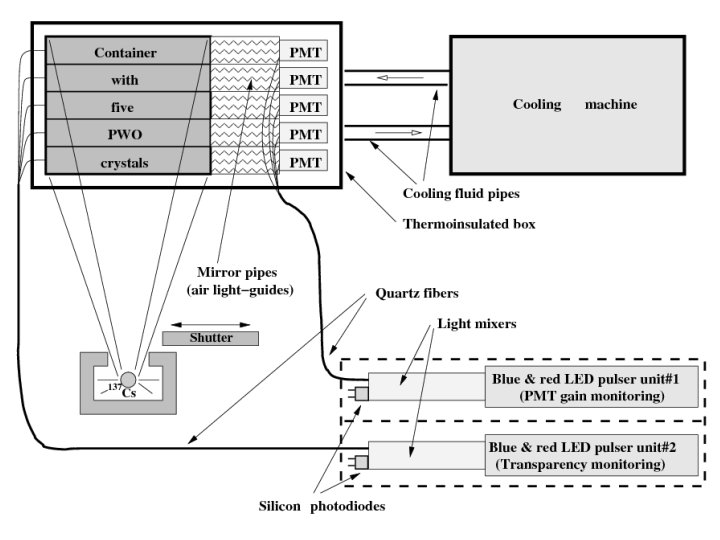}
\caption[Schematic layout of the irradiation facility at IHEP.]
{Schematic layout of the irradiation facility at IHEP, Protvino.}
\label{fig:scint:pwo:fig25}
\end{center}
\end{figure}

\begin{figure}[htb]
\begin{center}
\includegraphics[width=\swidth]{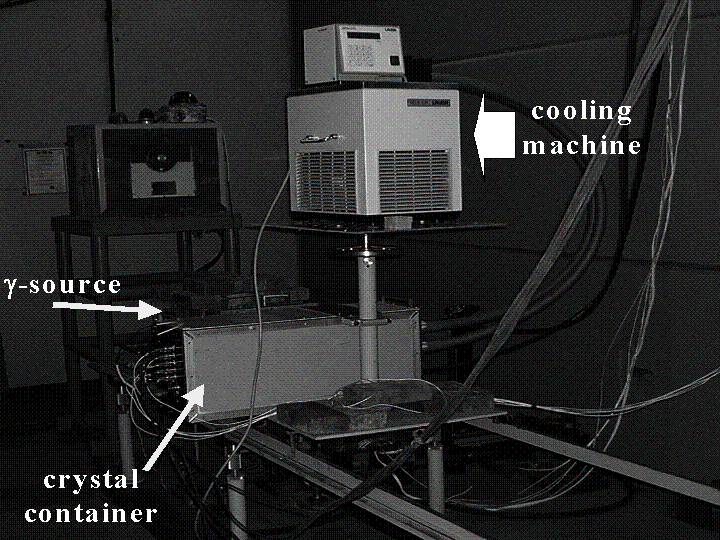}
\caption[Photograph of the major components of the irradiation setup at IHEP.]
  {Photograph of the major components of the irradiation setup
  at IHEP, Protvino.}
\label{fig:scint:pwo:fig26}
\end{center}
\end{figure}

An irradiation facility, dedicated to the investigation of the
radiation damage of cooled and temperature stabilized scintillation
crystals, was designed and installed at IHEP (Protvino, Russia). A
$^{137}$Cs radioactive source, emitting $\gamma$-rays of 662\,keV energy
with an activity of 5$\cdot$10$^{12}$Bq, is used for the irradiation
measurements. The dose rate can be continuously adjusted by changing
the distance to the exposed crystals. The dose rate in air is measured
close to the crystals with a commercial dosimeter (DKS-AT1123) with a
typical accuracy of $<$20$\percent$ \cite{bib:emc:pwo32}.
\Reffig{fig:scint:pwo:fig25} shows the schematic layout of the
facility and \Reffig{fig:scint:pwo:fig26} the major components of the
installed setup. All major components are remote controlled and the
relevant parameters are recorded for the off-line analysis. The up to
five full size crystals are mounted in an insulated container, which
can be cooled and temperature stabilized by a cryo-thermostat (LAUDA
RC6CP) with an absolute accuracy of 0.1$\degC$ even during a test
cycle of several days. The crystals are attached via air light-guides
to photomultiplier tubes (PMT). A gas flow of dry nitrogen helps to
equalize the temperature and to avoid condensation of
moisture. Several thermo-sensors measure and control the temperature.
Any change of the optical transmittance of the crystals due to
irradiation as well as possible gain instabilities of the PMTs are
controlled by a LED-based monitoring system distributed by optical
quartz fibers. Two spectral regions near the wavelengths of 450\,nm and
640\,nm, respectively, are inspected using blue and red LEDs. The highly
stabilized monitor system is similar to the one described in reference
\cite{bib:emc:pwo33,bib:emc:pwo34}.

The DC current of the PMT is directly digitized via a 10 k$\Omega$
resistor. That signal, which is proportional to the product of light
yield and light collection efficiency in the crystal, provides general
information on any change of the light output due to irradiation or
temperature change. Complementary, the signal triggered by the LED
pulses is monitoring any variation of the light transmittance of the
crystal.

The facility will be used during final production to investigate
selected crystal samples with respect to the radiation damage of
cooled crystals, complementary to the standard procedures at the
manufacturer location performed at room temperature. The measurements
will confirm the linear relation between the radiation hardness at
both temperature regions. Finally, tests will be performed for all
crystal geometries in order to determine the correlation between the
loss in transmission and the reduction of the integrated scintillation
light collected after multiple reflections inside the crystal.

\subsection{The Irradiation Facility at the Justus-Liebig-University Giessen}

\begin{figure}[htb]
\begin{center}
\includegraphics[width=\swidth]{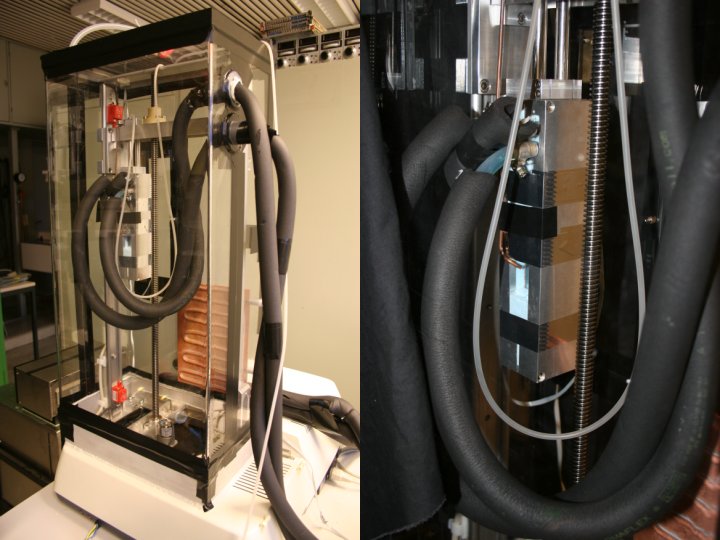}
\caption[The experimental setup for irradiation studies at Giessen.]
{The experimental setup arranged on top of a photon
  spectrometer for irradiation studies at JLU, Giessen (left). The
  right part of the figure shows the movable crystal container with
  the open region for the measurement of the transmission.}
\label{fig:scint:pwo:fig14}
\end{center}
\end{figure}

The irradiation facility at Giessen at the so called Strahlenzentrum
is based on a set of five $^{60}$Co sources, which can be used in any
combination. The maximum dose rate amounts to $\sim$30\,Gy/h measured at
a distance of 1m.  Complementary to the possibilities at Protvino an
experimental setup including a double beam photospectrometer has been
adapted to irradiate cooled crystals and measure immediately after
irradiation or hours and days later any changes of the spectral
distribution of the optical transmission. That arrangement allows to
determine at various temperatures the induced absorption as a function
of integral dose as well as the recovery as a function of
time. Present experiments were performed with a typical dose rate of
12\,Gy/h at a distance of 1m (see \Reffig{fig:scint:pwo:fig14}).

For the future quality control, the crystal samples can be either
exposed in a radiation protected laboratory room or placed into a
sufficiently large container, which can be loaded outside and moved in
front of the array of sources. The equipment includes a \emph{Varian}
photospectrometer to measure afterwards the radiation induced
absorption via the longitudinal optical transmission. All measurements
are meant to be performed at room temperature and will be the final
acceptance test for a significant percentage of all the crystals as a
cross check of the measurements performed at the manufacturer.

\subsection{The Irradiation Facilities at INP (Minsk) and BTCP}

\begin{figure}[htb]
\begin{center}
\includegraphics[width=\swidth]{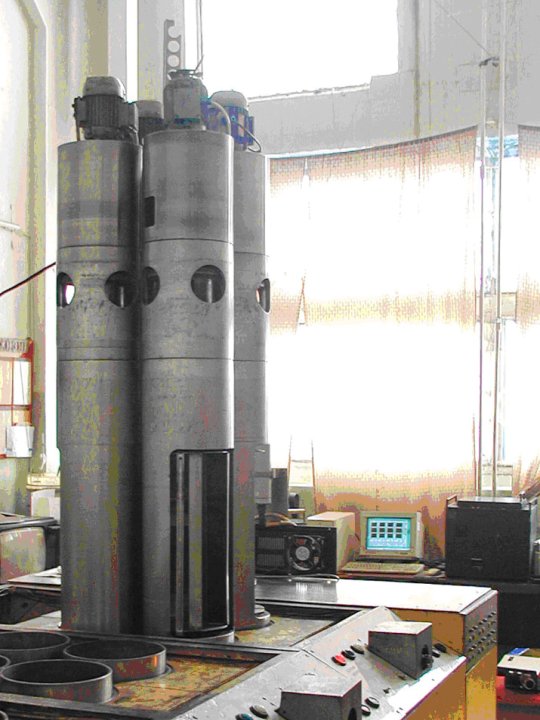}
\caption[The irradiation facility operating at BTCP in Russia.]
{The irradiation facility operating at BTCP in Russia.}
\label{fig:scint:pwo:fig22}
\end{center}
\end{figure}

\begin{figure}[htb]
\begin{center}
\includegraphics[width=\swidth]{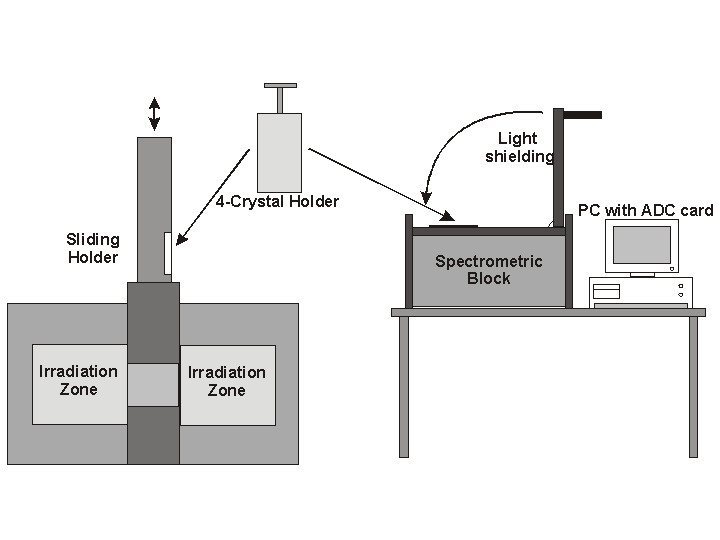}
\caption[Schematic layout of the spectrometric setup at BTCP.]
{Schematic layout of the spectrometric setup at BTCP.}
\label{fig:scint:pwo:fig23}
\end{center}
\end{figure}

For the mass production for the CMS detector, an irradiation facility was installed
at BTCP with technical assistance of INP (Minsk). A $^{60}$Co source of round geometry is placed in a
well for radiation protection, see \Reffig{fig:scint:pwo:fig22}. Up to four
crystals, placed in cylindrical light-tight containers, are irradiated
laterally from all sides at a dose rate of 1200\,Gy/h. In addition, an
inner container equipped with optical windows allows to determine
simultaneously the longitudinal induced absorption at three
characteristic wavelengths in an optical spectrometer with four
independent channels. The procedure of measurement is illustrated in
\Reffig{fig:scint:pwo:fig23}.

\begin{figure}[htb]
\begin{center}
\includegraphics[width=\swidth]{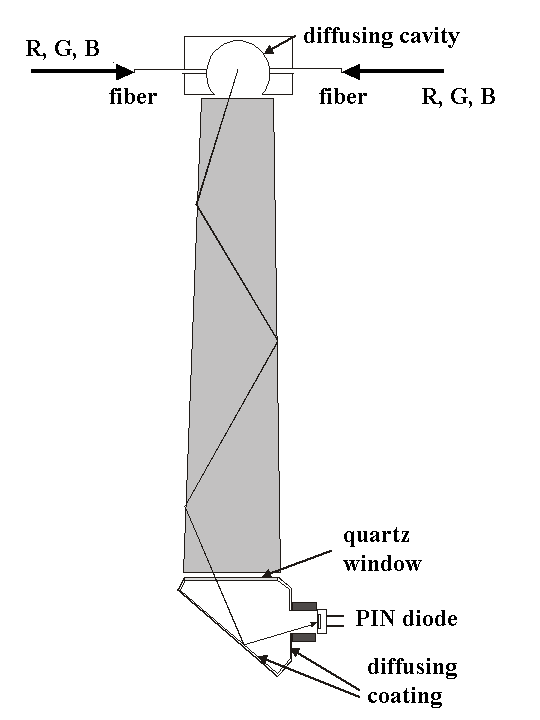}
\caption[Schematic layout of a single spectrometric channel of the setup at BTCP.]
{Schematic layout of a single spectrometric channel of the setup at BTCP.}
\label{fig:scint:pwo:fig24}
\end{center}
\end{figure}

The change of longitudinal transmission after irradiation is inspected
at three wavelengths: 650\,nm (RED), 570\,nm (GREEN) and 420\,nm (BLUE)
\cite{bib:emc:pwo30}. According to these three colors the
spectrometric setup is called RGB System. Three light emitting diodes
emit alternately short (tens of nanoseconds), stable light pulses. The
light emitted from each LED is mixed by four diffusion cavities and
illuminates the samples in longitudinal direction as illustrated in
\Reffig{fig:scint:pwo:fig24}. After traversing the crystal and the
diffusion of the light on a receiving reflector, it impinges on
PIN-photodiodes. The amplified signals are shaped and sent to a
PC-based ADC card.

Some fraction of the light from the LEDs is directed to a reference
channel based on a thermo-stabilized PIN-diode to ensure the long-term
stability at the level of 10$^{-4}$. This concept of the automated
control of the light pulse intensity is similar to one described in
details in \cite{bib:emc:pwo31}.  The measurement cycle of four PWO
crystals includes 2.5 minutes of irradiation and optical transmission
measurements before and after irradiation (taking about 2 minutes
each). Due to short irradiation time and short time of the crystal
transmission measurement after irradiation, unwanted fast crystal
recovery components can also be measured along with induced optical
absorption. Per working day at least 100 crystals can be characterized
with respect to their radiation hardness. This equipment was adopted
to measure \PANDA barrel type crystals of $200\,\mm$ length. The
radiation hardness of the crystals of the pre-production lot has been
characterized already with the adapted equipment.

The irradiation facility at INP Minsk comprises a well shaped
$^{60}$Co source with a constant dose rate of 2300\,Gy/h. Variations of the
crystal transmission are measured 60\,min after irradiation by a Varian
\emph{Cary1E} photospectrometer in the wavelength range of
300-900\,nm. The control of other scintillation parameters can also be
carried out at INP.

BTCP and INP irradiation facilities are cross calibrated two to three
times per year on the occasion of change of the crystal dimensions. It
is agreed that a set of reference crystals including all types of
\PANDA scintillation elements will be produced before crystal mass
production. It will be used for examination of the equipment before
each measurement session and to verify the cross calibration of the
irradiation facilities within the project.

\subsection{Irradiation Studies with Neutrons, Protons and High Energy $\gamma$-rays}

It is foreseen to perform selective irradiation studies with hadrons
and electromagnetic probes. Within the collaboration there are several
accessible facilities to perform such studies. The following options
are available and have been already used for the characterization
presented in the previous section.

\begin{itemize}
\item The irradiation facility at IHEP, Protvino provides a
  Pu-$\alpha$-Be source with a fluence of 3$\cdot$10$^{4}$
  neutrons/(cm$^{2}$$\cdot$s). The emitted neutrons have a mean
  kinetic energy of E$_{n}$=3.5\,MeV. However, there is an accompanying
  background of $\gamma$-rays with a mean energy of
  E$_{\gamma}$=$300\,\keV$ corresponding to a dose of 7\,mGy/h at
  the location of irradiation of samples.
\item The cyclotron at the KVI at Groningen, (The Netherlands)
  provides a dedicated experimental area for irradiation studies with
  protons up to an energy of $150\,\MeV$. The facility allows to
  irradiate locally as well as homogeneously over a typical surface of
  $15\times15\,\cm^{2}$ with an accurate control over the overall dose rate
  and integral dose.
\item Besides the opportunity to measure response functions to
  high-energy photons over the range up to $3.5\,\GeV$, both tagging
  facilities at the electron accelerators MAMI (Mainz, Germany) and
  ELSA (Bonn, Germany) allow to study the radiation damage in a
  controlled way. The absolute flux of the energy marked photons is
  known with high accuracy and can be used to test pure crystals as
  well as assembled detectors or arrays. High-energetic electrons can
  be simultaneously used with well known energies and rates directly
  behind the focal plane of the magnetic dipole
\end{itemize}

\section{The \PWOII Crystals for \Panda}

The achievements in the understanding of the scintillation mechanism,
the defect structures and the implications for the production
technology have provided the quality of lead tungstate fully adequate
for the \Panda application. \PWOII, when operated at T=-25$\degC$
becomes a sufficiently bright scintillator, in particular when nearly
50$\percent$ of the crystal endface are covered and readout with
LAAPDs of high quantum efficiency. The specific temperature keeps the
decay time fast to cope with high count rates. However, the processes
recovering the irradiation damage become extremely slow at low
temperatures and cause an asymptotic reduction of the light yield due
to the changing optical transparency. High quality crystals with a
minimized concentration of defects limit the loss to a value below
30$\percent$. Therefore, for the operating scenario of the \TSEMC a
net increase of the light yield of a factor 3 compared to room
temperature can be considered even for the most forward located
modules in the \FWEMC.  The next chapters are describing in detail the
overall parameters of the scintillator material for \Panda.

\subsection{Light Yield and Decay Kinetics}

\begin{figure*}[htb]
\begin{center}
\includegraphics[width=\dwidth]{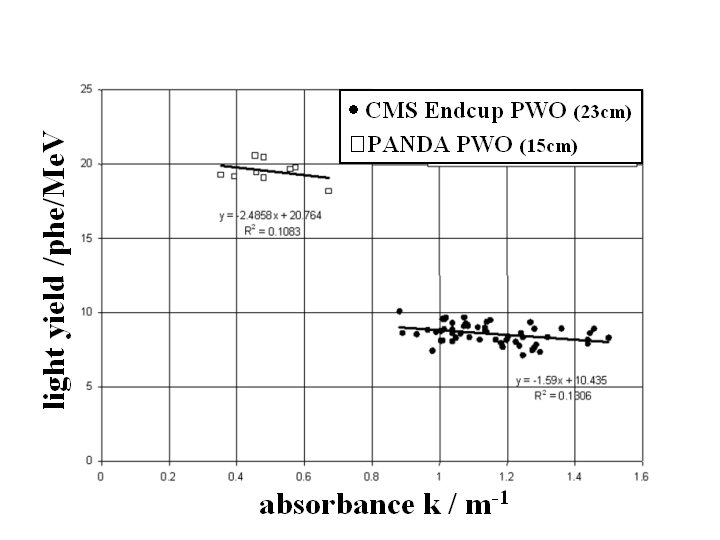}
\caption[Light yield of full size polished scintillation crystals.]
{Light yield of full size polished scintillation crystals (28$\times$28$\times$230mm$^{3}$) used for CMS endcap
  construction (right distribution) and of \Panda prototypes of slightly smaller size (20$\times$20$\times$150mm$^{3}$) shown as a function of the absorbance to compensate for different geometries.}
\label{fig:scint:pwo:fig5}
\end{center}
\end{figure*}

As a result of the significant improvement of the technology of
crystal production, full size crystals with $200\,\mm$ length deliver
today a light yield of 17-20\,phe/MeV at 18$\degC$ measured with a
photomultiplier with bi-alkali photocathode (quantum efficiency
$\sim$20$\percent$.  Documenting the breakthrough years ago,
\Reffig{fig:scint:pwo:fig5} shows for only $150\,\mm$ long \PWOII
samples the light yield in comparison to CMS Endcap type crystals of
$230\,\mm$ length. The light yield is shown as a function of the
optical absorption at $360\,\nm$ wavelength, which is sensitive to the
concentration of Frenkel-type defects in the crystal, in order to take
in to account the different crystal dimensions.

\begin{figure}[htb]
\begin{center}
\includegraphics[width=\swidth]{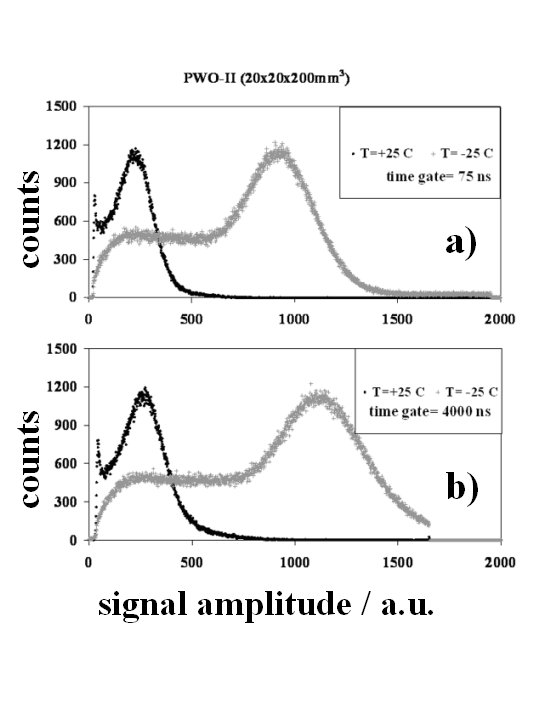}
\caption[Pulse height spectra measured for a \PWOII crystal.]
{Pulse height spectra measured for a \PWOII crystal
  (20$\times$20$\times$200mm$^{3}$) for $\gamma$-rays of a
  $^{60}$Co-source (E$_{\gamma}\sim$1.2MeV). The scintillation light
  has been converted with a photomultiplier tube (Hamamatsu R2059) at
  two temperatures and integrated over time windows of $75\,\ns$
  (part a) or $4000\,\ns$ (part b), respectively.}
\label{fig:scint:pwo:fig6}
\end{center}
\end{figure}

Reducing the thermal quenching of the luminescence process by cooling
the crystals leads to the expected increase of light yield. Obviously,
the low concentration of deep traps in \PWOII does not distort the
scintillation kinetics at least down to a temperature of
T=-25$\degC$. \Reffig{fig:scint:pwo:fig6} a) and b) show typical pulse
height spectra, which were measured for a full size crystal
(20$\times$20$\times$200mm$^{3}$) at two operating temperatures of
T=$\pm$25$\degC$ and two different integration gates of $75\,\ns$ and
$4000\,\ns$, respectively, using a photomultiplier as sensor
(Hamamatsu R2059). The overall gain factor of the light yield at
T=-25$\degC$ compared to T=+25$\degC$ amounts to more than 4 for both
time gates.

Compared to the standard material, \PWOII crystals follow  a different
temperature dependence of the scintillation yield, as shown in
\Reffig{fig:scint:pwo:fig6} in the relevant temperature region between
T=+25$\degC$ and T=-25$\degC$, respectively. The average temperature
gradient of the light yield amounts to $\sim-3\percent/\degC$, which
is 50$\percent$ larger than for standard PWO. The obviously steeper
increase of the light yield with lower temperature can be addressed to
a lower concentration of capturing centers.

\begin{figure}[htb]
\begin{center}
\includegraphics[width=\swidth]{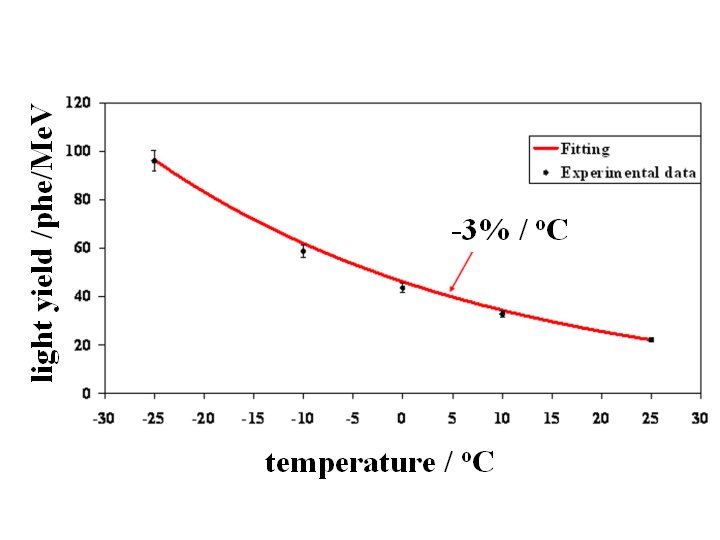}
\caption[Temperature dependence of the yield of \PWOII scintillation light.]
{Temperature dependence of the yield of scintillation light of
  \PWOII in the temperature range between T=+25$\degC$ and
  T=-25$\degC$, respectively.}
\label{fig:scint:pwo:fig8}
\end{center}
\end{figure}

\begin{figure*}[htb]
\begin{center}
\includegraphics[width=\dwidth]{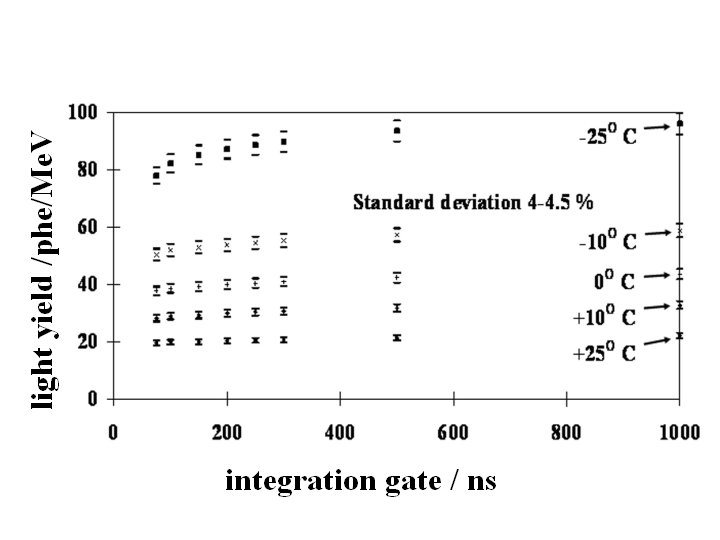}
\caption[Average \PWOII light yield vs. integration gate.]
{Dependence of the average light yield on the temperature and the
  integration gate for 10 \PWOII crystals (20$\times$20$\times$200mm$^{3}$) produced at BTCP. The variations are given as error bars.}
\label{fig:scint:pwo:fig9}
\end{center}
\end{figure*}

\Reffig{fig:scint:pwo:fig9} shows for different temperatures the mean
value of the scintillation yield per MeV deposited energy including
standard deviations obtained for low energy $\gamma$-rays. The signals
obtained with photomultiplier tubes equipped with bi-alkali
photocathodes (Hamamatsu R2059) were integrated over a gate width
between $75\,\ns$ up to $1\,\mus$, respectively. The figure confirms a
mean gain factor up to 4 between the two extreme temperatures and a
sufficiently fast decay time, which allows to collect 95\percent of
the scintillation light within a time window of $300-400\,\ns$.

\begin{figure}[htb]
\begin{center}
\includegraphics[width=\swidth]{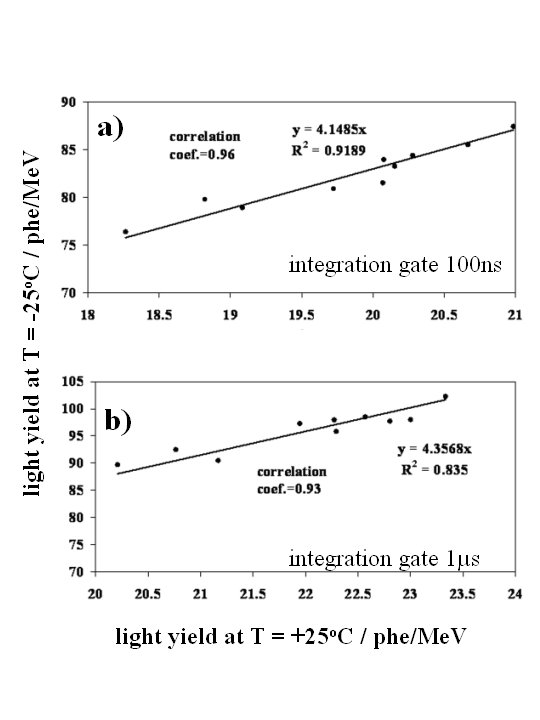}
\caption[Correlation of light yield at low and high temperature.]
{Correlation between the light yield induced by low energy
  $\gamma$-rays at temperatures of T=-25$\degC$ and T=+25$\degC$ for
  10 \PWOII crystals (20$\times$20$\times$200mm$^{3}$). The
  photomultiplier response has been integrated over a time gate of
  $100\,\ns$ (part a) and $1000\,\ns$ (part b).}
\label{fig:scint:pwo:fig11}
\end{center}
\end{figure}

Mass production requires sensitive criteria for the certification of
crystals. As shown in \Reffig{fig:scint:pwo:fig11}, a strong correlation has been
verified  with a coefficient above 4 for the
light yield observed at T=+25$\degC$ and the corresponding value at
T=-25$\degC$. Experimental data are shown in part a) and b) for ten \PWOII crystals
after integrating the response within $100\,\ns$ and $1000\,\ns$,
respectively. Consequently, the linear correlation allows quality
control to be performed in the more convenient environment at room
temperature.

\begin{figure}[htb]
\begin{center}
\includegraphics[width=\swidth]{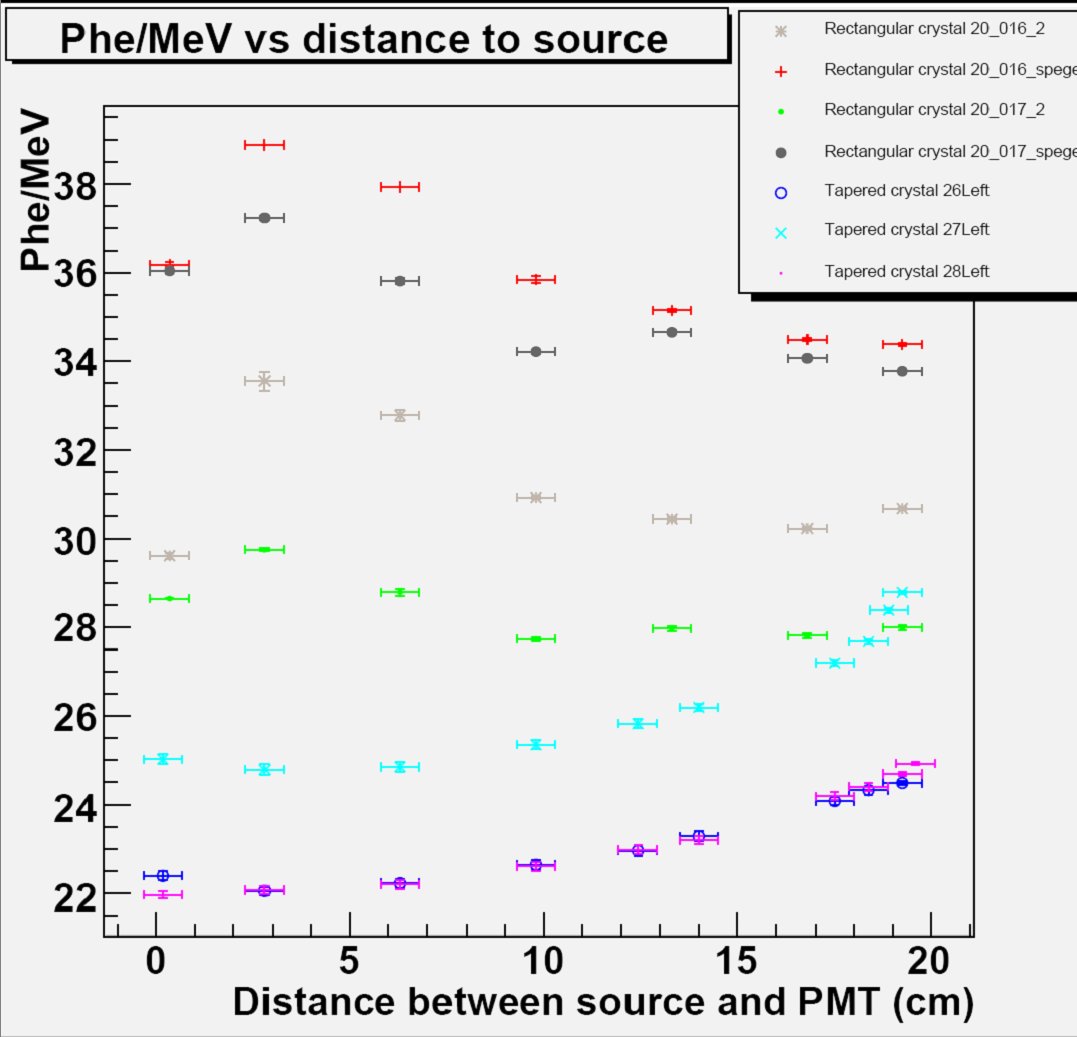}
\caption
{Variation of the light output measured for different crystal geometries and reflector materials when a collimated $\gamma$-source is moved relative to the sensor position along the crystal axis.}
\label{fig:scint:pwo:fig99}
\end{center}
\end{figure}

The overall light collection depends on the optical quality, the geometrical shape, surface treatment and reflector material. The distribution of the emission of the scintillation light and the variation of the light path cause a position dependence of the light yield collected at the photo sensor located at the rear end of the crystal. These effects cause a typical non-linearity, which can have an impact on the final energy resolution, explicitly the constant term. The effect has to be corrected in particular, when the maximum of energy deposition within the crystal is spread over a large region, which is not the case as illustrated in \Reffig{fig:perf:Erel}. Nevertheless, investigations have been started to optimize the additional surface treatment by partly de-polishing or selection of various reflector materials. \Reffig{fig:scint:pwo:fig99} illustrates the change of light output for different geometries and reflector materials when a collimated source is moved relative to the sensor position along the crystal axis.

\subsection{Radiation Hardness of \PWOII}

\begin{figure}[htb]
\begin{center}
\includegraphics[width=\swidth]{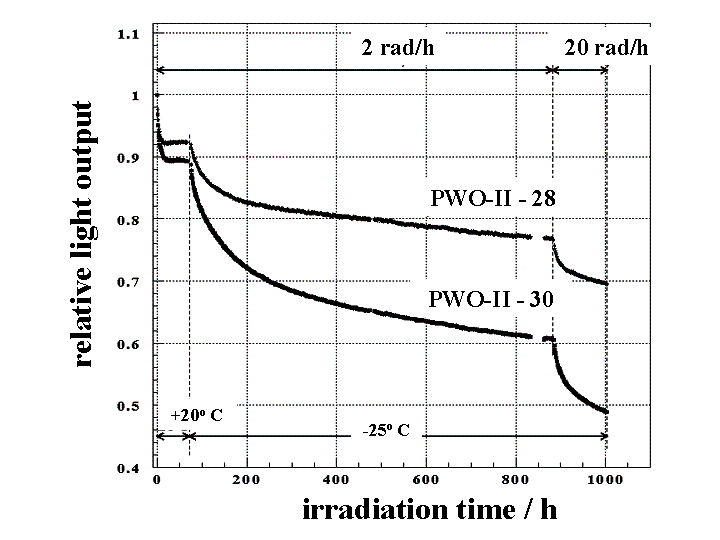}
\caption[Relative change of the light yield of two \PWOII samples.]
{Relative change of the light yield of two \PWOII samples
  with irradiation time for two different dose rates and two
  temperatures. The curves are normalized to the initial light yield
  at T=+20$\degC$.}
\label{fig:scint:pwo:fig15}
\end{center}
\end{figure}

\begin{figure}[htb]
\begin{center}
\includegraphics[width=\swidth]{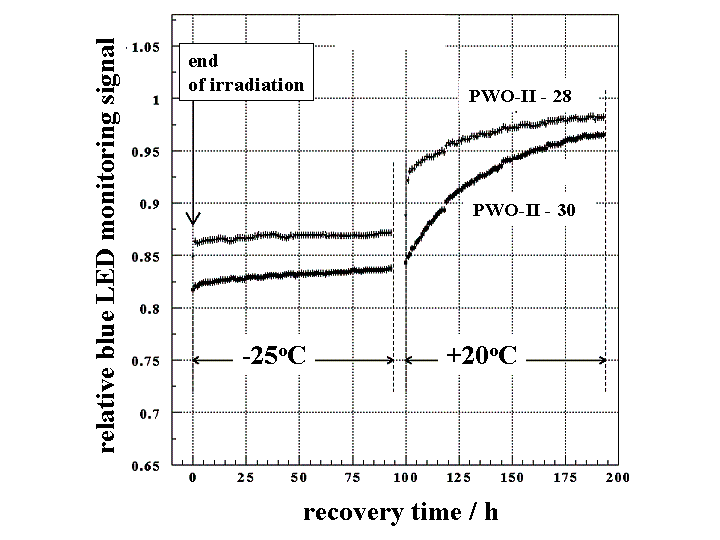}
\caption[Relative change of the monitoring signal.]
{Relative change of the monitoring signal provided by the blue
  emitting LED illustrating the recovery of two \PWOII crystals
  after $\gamma$-irradiation at two different temperatures.}
\label{fig:scint:pwo:fig16}
\end{center}
\end{figure}

The radiation hardness and the effective loss of light yield as a
functions of dose rate and accumulated dose are well known from the
studies for CMS but investigated at room temperature. The
deterioration of the optical transmission in the experiment is caused
by the interplay of damaging and recovering mechanisms. The latter are
fast at room temperature and keep the loss of light yield moderate.
First exploratory experiments at IHEP (Protvino) have shown a
completely different scenario if the irradiated crystals are cooled
down to T=-25$\degC$ and are kept cool even after the irradiation has
been stopped. \Reffig{fig:scint:pwo:fig15} and
\Reffig{fig:scint:pwo:fig16} illustrate for two crystal samples the
typical behavior. At a starting dose rate of 20\,mGy/h the scintillation
response drops continuously and does not even reach saturation after
1000 hours. In a similar way, almost no recovery can be noticed after
irradiation for the two samples kept at T=-25$\degC$.

\begin{figure}[htb]
\begin{center}
\includegraphics[width=\swidth]{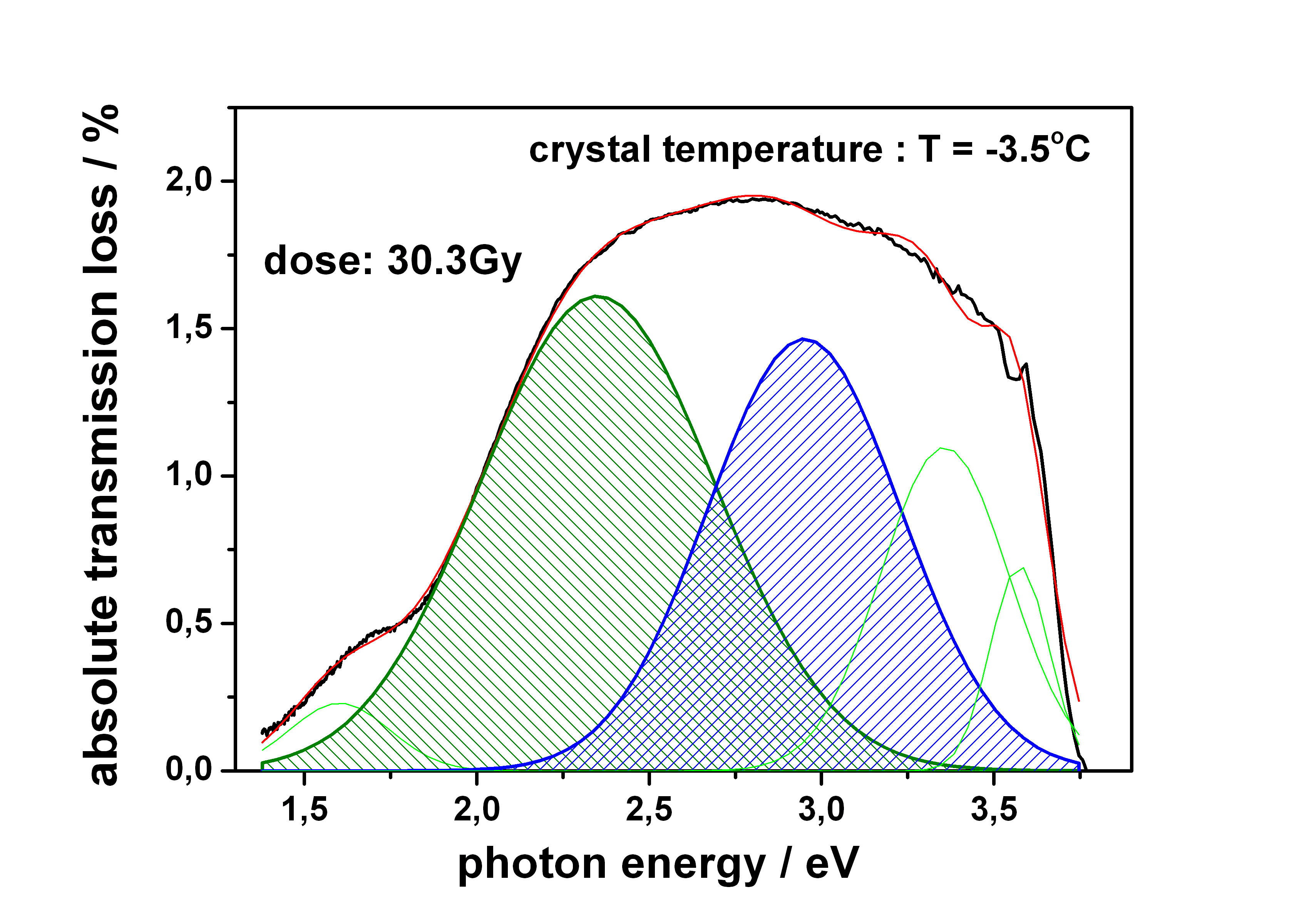}
\caption[The absolute loss in transmission due to irradiation.]
{The absolute loss in transmission due to the irradiation of a
  \PWOII crystal with a dose of 30.3\,Gy at a temperature of
  T=-3.5$\degC$ shown as a function of photon energy. The
  contributions of two absorption bands at photon energies of
  $2.35\,\eV$ and $2.95\,\eV$, respectively, are marked for detailed
  analysis.}
\label{fig:scint:pwo:fig17}
\end{center}
\end{figure}

\begin{figure}[htb]
\begin{center}
\includegraphics[width=\swidth]{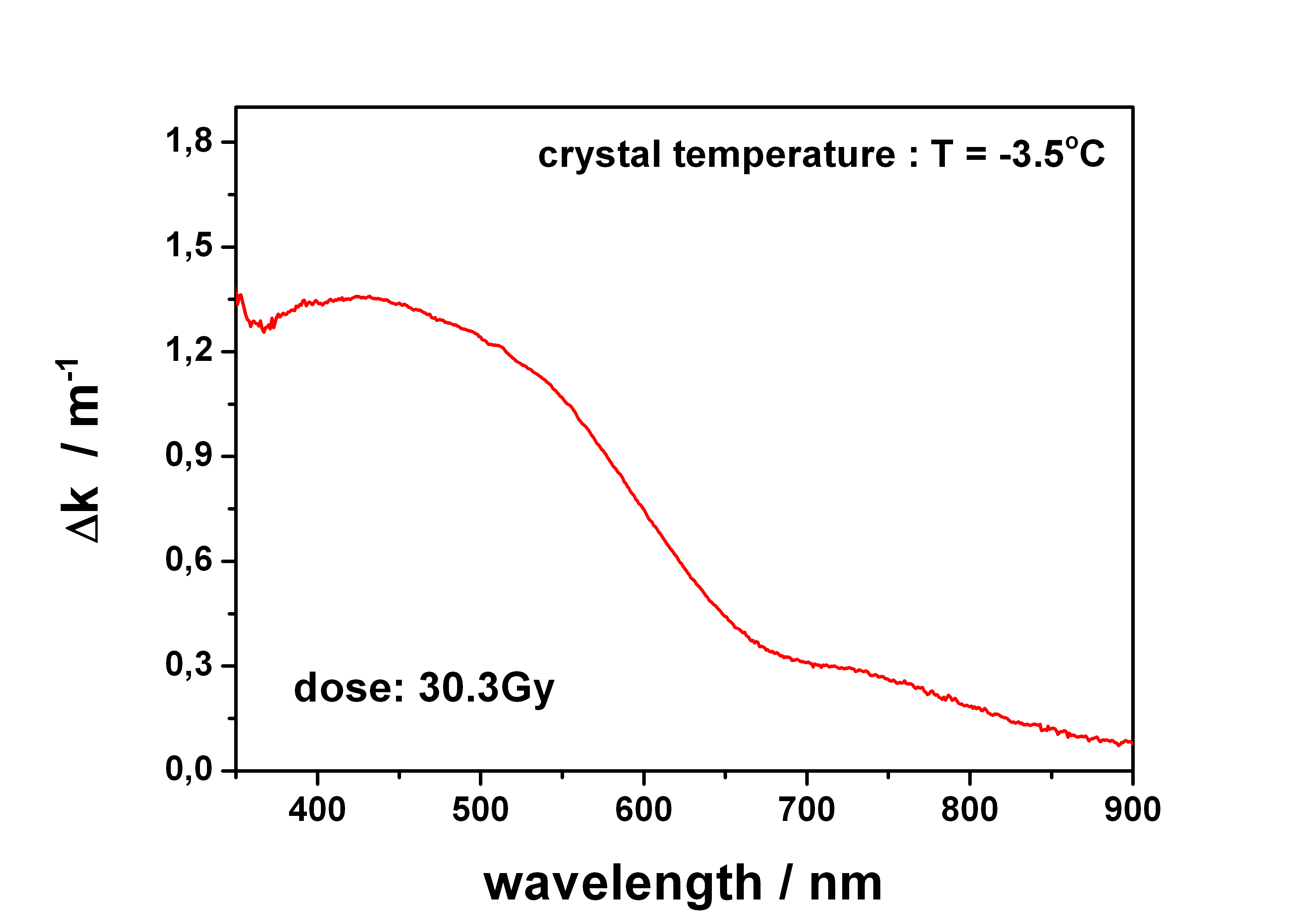}
\caption[The induced absorption due to irradiation.]
{The induced absorption due to the irradiation of a \PWOII
  crystal with a dose of 30.3\,Gy at a temperature of T=-3.5$\degC$
  given as a function of wavelength.}
\label{fig:scint:pwo:fig40}
\end{center}
\end{figure}

Therefore, more specific studies became necessary in order to
characterize the nature and properties of defects with slow decay
times. The experimental installation at Giessen allows to measure in a
wavelength specific manner the deterioration of the optical
transmission expressed as the change of the induced absorption
coefficient for crystals, while they are kept cool after irradiation
for a longer time. Investigations of full size crystals were
performed at temperatures of T=-3.5$\degC$, T=-10.0$\degC$ and
T=-22.5$\degC$. \Reffig{fig:scint:pwo:fig17} shows the
representative result obtained for the absolute loss in transmission
at T=-3.5$\degC$ right after irradiation as a function of photon
energy. The contributions of two specific absorption bands are marked
for further analysis. The shape of the spectrum is similar to the
results at the two lower temperatures. \Reffig{fig:scint:pwo:fig40}
repeats the above data obtained at T=-3.5$\degC$ but re-scaled as
induced absorption as a function of wavelength. The previous
measurement at room temperature (see \Reffig{fig:scint:pwo:fig13})
shows in a similar manner the induced absorption primarily in the same
wavelength range between 400-600 nm and two broad absorption bands
located at wavelengths of $528\,\nm$ and $420\,\nm$, corresponding to
photon energies of $2.35\,\eV$ and $2.95\,\eV$, respectively.

Performing the irradiations of cooled crystals with significantly
higher dose rates a saturation of the damage can be observed at an
integral dose of 30-60 Gy. This value does not depend on the
temperature and might only reflect the individual concentration of
defects in the crystal. In the final \Panda experiment such an
integral dose will be accumulated only in the most forward part of the
\FWEMC at maximum beam intensity after $\>$ 3000 hours of operation
and might be never reached in the barrel region over a typical
experimental cycle lasting 6 months.

\begin{figure}[htb]
\begin{center}
\includegraphics[width=\swidth]{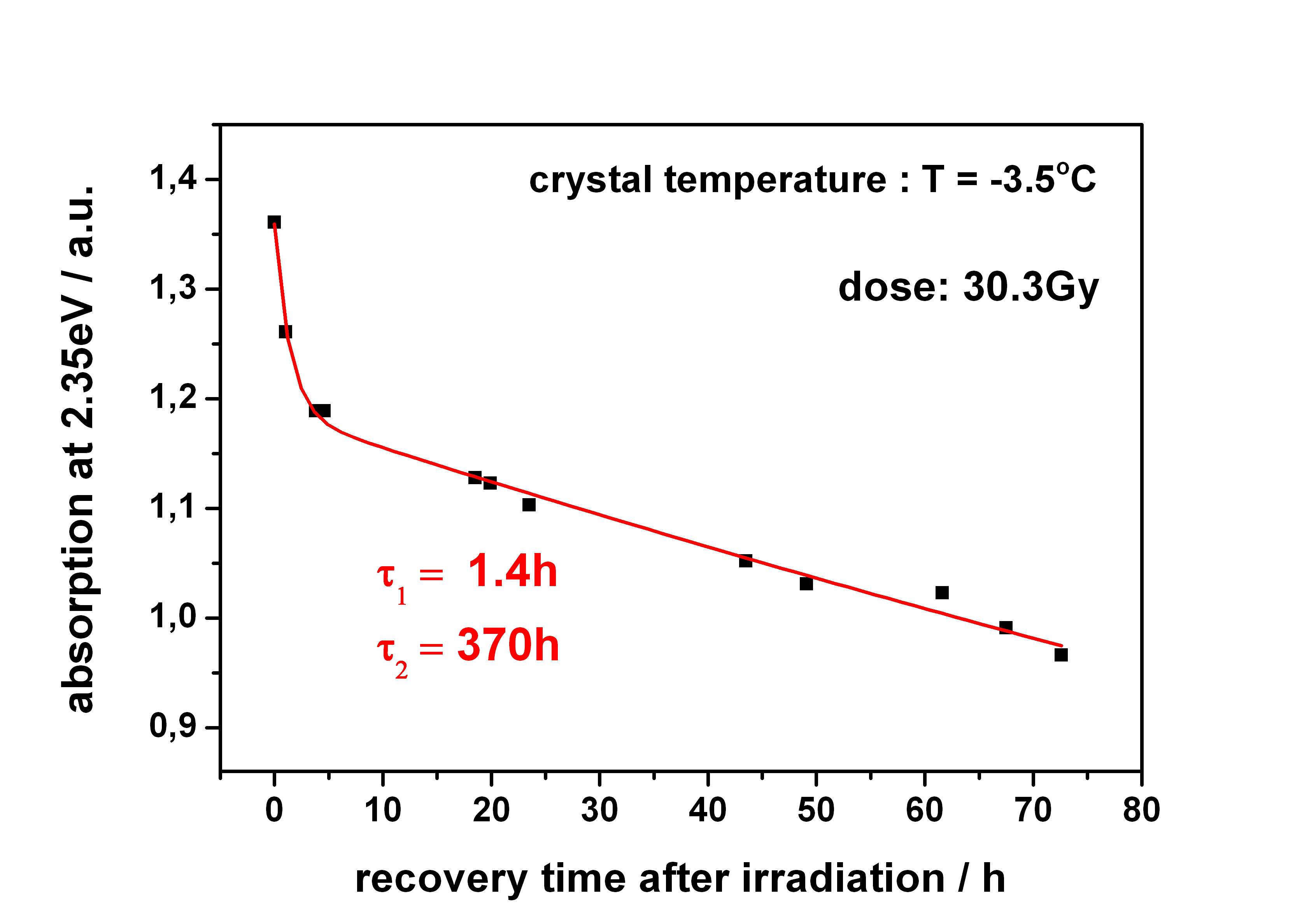}
\caption[The decay kinetics of an absorption region after irradiation.]
{The decay kinetics of an absorption region near a photon
  energy of $2.35\,\eV$ ($\lambda$=$528\,\nm$) after the irradiation
  of a \PWOII crystal with a dose of 30.3\,Gy at a temperature of
  T=-3.5$\degC$.}
\label{fig:scint:pwo:fig41}
\end{center}
\end{figure}

\begin{figure}[htb]
\begin{center}
\includegraphics[width=\swidth]{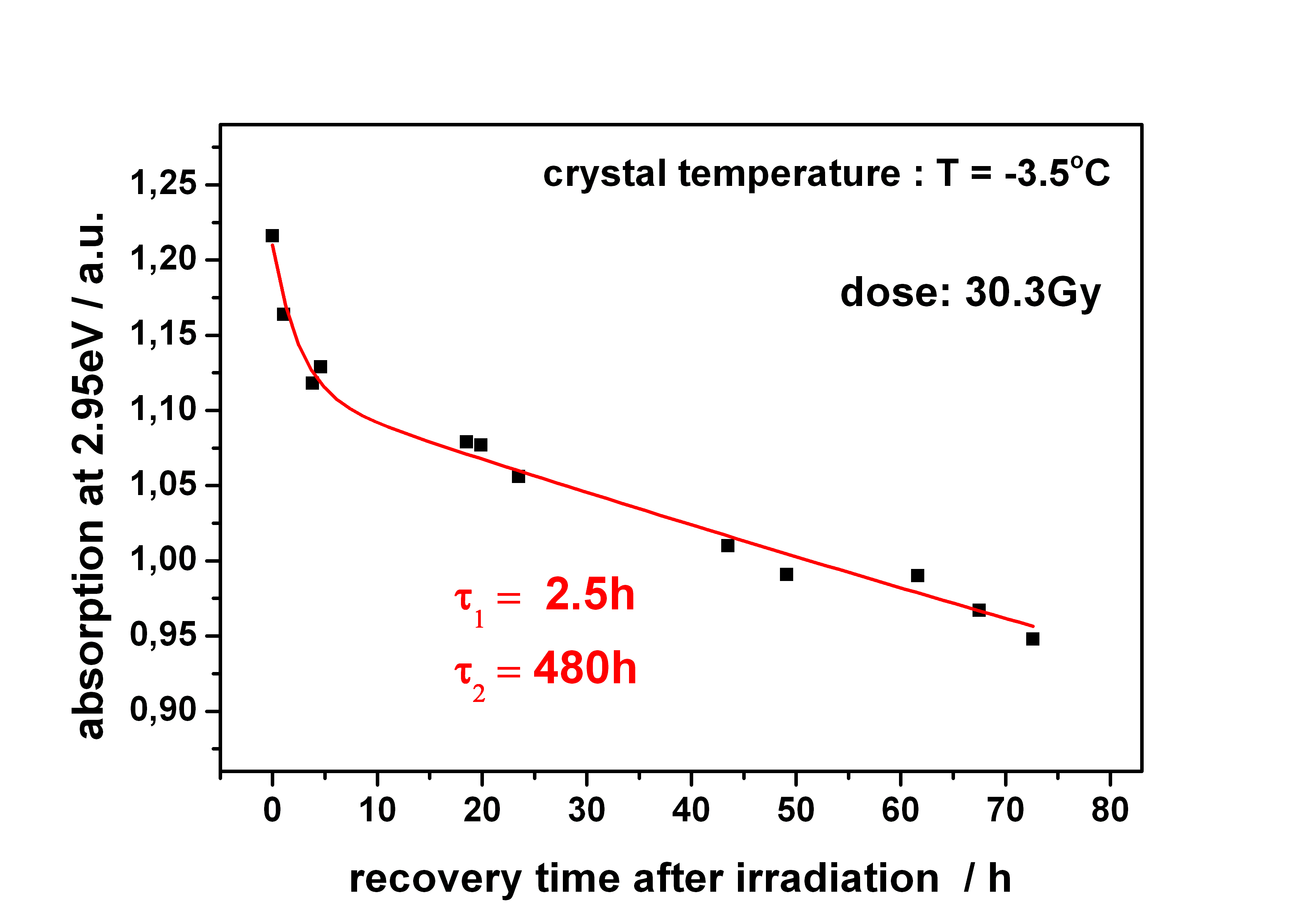}
\caption[The decay kinetics of an absorption region after irradiation.]
{The decay kinetics of an absorption region near a photon
  energy of $2.95\,\eV$ ($\lambda$=$420\,\nm$) after the irradiation
  of a \PWOII crystal with a dose of 30.3\,Gy at a temperature of
  T=-3.5$\degC$.}
\label{fig:scint:pwo:fig42}
\end{center}
\end{figure}

The recovery of the optical transmission has been studied in a time
range up to 5 days and at different temperatures.  From the
measurements similar to \Reffig{fig:scint:pwo:fig17} but performed at
different times after the end of irradiation an estimate of the
lifetimes of the color centers can be
deduced. \Reffig{fig:scint:pwo:fig41} and \Reffig{fig:scint:pwo:fig42}
illustrate the kinetics of the two selected absorption
regions. Besides a fast component the recovery times are in the order
of hundreds of hours in both cases. At all investigated temperatures
below 0$\degC$ the relaxation of the color centers remains slow and
does not compensate even partially the damage in parallel.

\begin{figure}[htb]
\begin{center}
\includegraphics[width=\swidth]{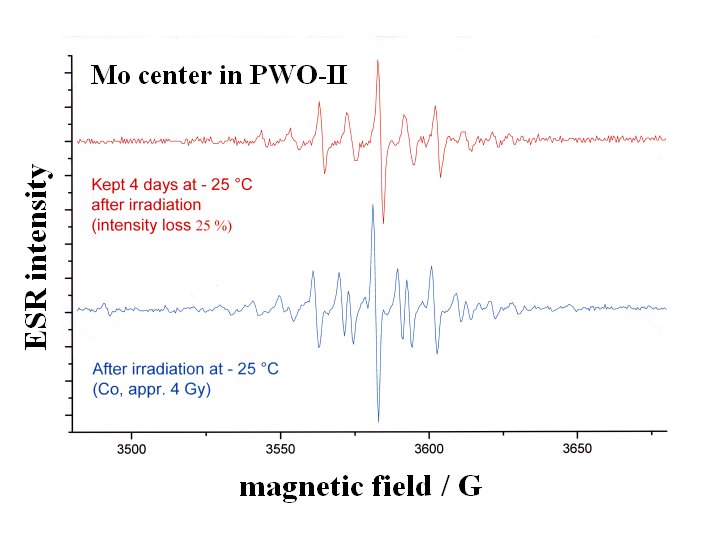}
\caption[EPR signal measured after irradiation.]
{EPR signal of the (MoO$_{4}$)$^{3-}$ center in a \PWOII
  crystal measured at T=-25$\degC$ immediately after irradiation
  (upper part) and 96 hours later. Both probes were kept continuously
  at T=-25$\degC$.}
\label{fig:scint:pwo:fig18}
\end{center}
\end{figure}

Detailed studies of the impact of Mo impurity on the scintillation
performance of the crystal allowed also to clarify its contribution to
the radiation damage. It is well known that a contamination with
MoO$_{4}$-complexes gives rise to green luminescence and extremely
slow decay times in PWO crystals \cite{bib:emc:pwo17}. Therefore, the
present generation of \PWOII crystals maintains the impurity on the
level of several ppm. Our recent studies have identified the presence
of (MoO$_{4}$)$^{3-}$ centers exploiting the EPR technique. Spectral
distributions of the induced absorption of various samples measured at
room temperature gave first indications based on the broad absorption
band peaked near 535\,nm as already shown in
\Reffig{fig:scint:pwo:fig13}. Systematic studies of paramagnetic
resonance and thermoluminescence of PbWO$_{4}$/PbMoO$_{4}$ mixed
crystals \cite{bib:emc:pwo19} have investigated the properties of
(MoO$_{4}$)$^{3-}$ centers with respect to the molybdenum
concentration. The temperature of maximum thermal decay depends
strongly on the Mo concentration and reaches values near 250\,K at
concentrations below 1\percent. TL spectra of mixed crystals after
$50\,\keV$ X-ray excitation show a pronounced glow peak slightly above
250\,K for mixed crystals with low Mo concentration. Measurements of the
concentration of (MoO$_{4}$)$^{3-}$ and (WO$_{4}$)$^{3-}$ paramagnetic
centers in CaWO$_{4}$:Pb after X-ray irradiation and subsequent
stepwise heating allow deducing the radiation induced optical
absorption due to (MoO$_{4}$)$^{3-}$ centers \cite{bib:emc:pwo20}. The
spectral distribution indicates a broad distribution peaked at a
photon energy of 2.35\,eV (FWHM=0.8\,eV) corresponding to a wavelength of
528\,nm.  At T=-25$\degC$, which is below the activation temperature of
(MoO$_{4}$)$^{3-}$, one observes as a consequence of the
thermo-activation a slow decay time. \Reffig{fig:scint:pwo:fig18}
shows the EPR signal identifying the (MoO$_{4}$)$^{3-}$ center in the
\PWOII crystal, which was stored directly after the irradiation in
liquid nitrogen. The drop of the signal intensity after a recovery of
96 hours was estimated to 25\percent. Similar measurements at different
temperatures are in preparation.

Therefore, even a small concentration of Mo impurities can contribute
to the reduced transmittance within that range of wavelengths. In order
to ensure sufficient radiation hardness one has to put even a stronger
limit on the Mo contamination in the crystal, even at a level below
$1\,\ppm$. This limit of Mo concentration has been so far specified
for the crystal mass production.

\subsection{Radiation Hardness Required for \Panda}

The radiation damage reduces the output of the scintillation light and
consequently the number of photoelectrons per unit of deposited
energy. Therefore, the expectable light output has to be correlated
with the quantified radiation resistivity. The latter is specified by
the radiation induced absorption coefficient $\Delta$k at the
wavelength of maximum scintillation emission at $\lambda$=$420\,\nm$,
which can be easily measured and cross-checked, if it can be
determined at the relevant facilities at a temperature of
T=+20$\degC$.  The correlation depends on the individual crystal
shape, light collection and efficiency of the photo sensor and has not
been experimentally determined for the \TSEMC up to date. We intend to
follow the approach used by CMS \cite{bib:emc:pwo35}. The information
is obtained at small dose rates, which are similar to the experimental
conditions of \Panda. The procedure determines the ratio of the
changes of the scintillator and monitoring response, respectively, and
models the relationship by the expression
S/S$_{0}$=(R/R$_{0}$)$^{\alpha}$, where S/S$_{0}$ and R/R$_{0}$ are
the relative variations of the response to the scintillation and
monitoring signal, respectively. Since the geometrical shape of the
crystals of CMS and \Panda are comparable, a typical value of 1.5 can
be assumed for the parameter $\alpha$.

\begin{figure}[htb]
\begin{center}
\includegraphics[width=\swidth]{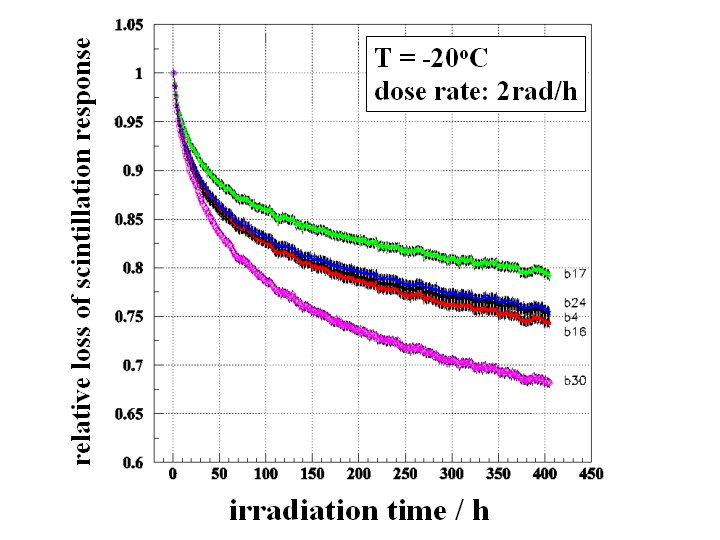}
\caption[Change of the scintillation response due to a continuous irradiation.]
  {Change of the scintillation response due to a continuous
  irradiation with low-energy $\gamma$-rays ($^{137}$Cs) for 5 samples
  of the pre-production run deduced from the DC-signal of the
  photomultiplier response. The measurement has been performed at
  IHEP, Protvino.}
\label{fig:scint:pwo:fig27}
\end{center}
\end{figure}

The measurement at the irradiation facility at IHEP (Protvino)
estimates the change of optical transmission, deduced from the
response to the transmitted light of two LEDs of different color, and
the reduction of the scintillation response via the luminescence
initiated by the absorbed $\gamma$-rays emitted from the $^{137}$Cs
source. This measurement is not fully compatible with the response to
electromagnetic probes hitting the crystal from the front face, but
can give a first hint. \Reffig{fig:scint:pwo:fig27} shows the dropping
DC-current of the photomultiplier tube as a function of the
irradiation time. The five investigated \PWOII samples were
characterized before by the value of the induced absorption at
$\lambda$=$420\,\nm$ ($\Delta$k(420 nm)) determined at T=+20$\degC$ at
BTCP and afterwards annealed at T=200$\degC$ for 2
hours. \Reftbl{tab:scint:rad:prop} summarizes the experimental data
including the relative loss of scintillation response measured at
T=-20$\degC$.

\begin{table*}
\begin{center}
\begin{tabular}{cccc}
\hline\hline
crystal ID& $\Delta$k/m$^{-1}$ & R/R$_{0}$    & estimated S/S$_{0}$\\
          &  T=+20$\degC$      & T=-20$\degC$ & T=-20$\degC$       \\
\hline
b4  &  0.402& 0.905 & 0.86\\
b16 &  0.420& 0.910 & 0.87\\
b17 &  0.289& 0.915 & 0.88\\
b24 &  0.405& 0.920 & 0.88\\
b30 &  0.753& 0.890 & 0.84\\
\hline\hline
\end{tabular}
\caption[Comparison of the induced absorption.]
{Comparison of the induced absorption $\Delta$k(420 nm)
  measured at BTCP, the value R/R$_{0}$ at $\lambda$=450nm measured at
  IHEP (Protvino) and the estimated change of scintillation response.}
\label{tab:scint:rad:prop}
\end{center}
\end{table*}

\begin{figure}[htb]
\begin{center}
\includegraphics[width=\swidth]{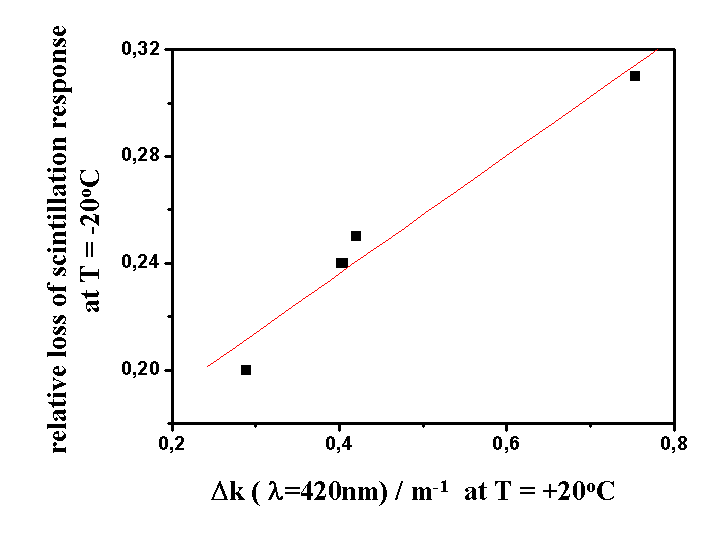}
\caption[Maximum change of the scintillation
  response vs. induced absorption.]
  {Correlation between the maximum change of the scintillation
  response, as shown in \Reffig{fig:scint:pwo:fig27}, and the induced
  absorption coefficient $\Delta$k at $\lambda$=420\,nm measured at BTCP
  at room temperature.}
\label{fig:scint:pwo:fig28}
\end{center}
\end{figure}

Both specification parameters are correlated in
\Reffig{fig:scint:pwo:fig28}. In spite of the lack of sufficient
statistics, a linear correlation can be proposed, which supports that
the quality limits defined for control measurements at room
temperature are sufficiently selective for the final operation at low
temperatures. The data are consistent with the assumption of
$\alpha$=1.5 as discussed above. Therefore, due to the slow relaxation
of color centers in cooled crystals one has to cope with a typical
loss in scintillation response between 20 and 30\percent as an
asymptotic value after a deposited dose of $\>$30-50\,Gy for typical
\PWOII crystals.

\begin{figure}[htb]
\begin{center}
\includegraphics[width=\swidth]{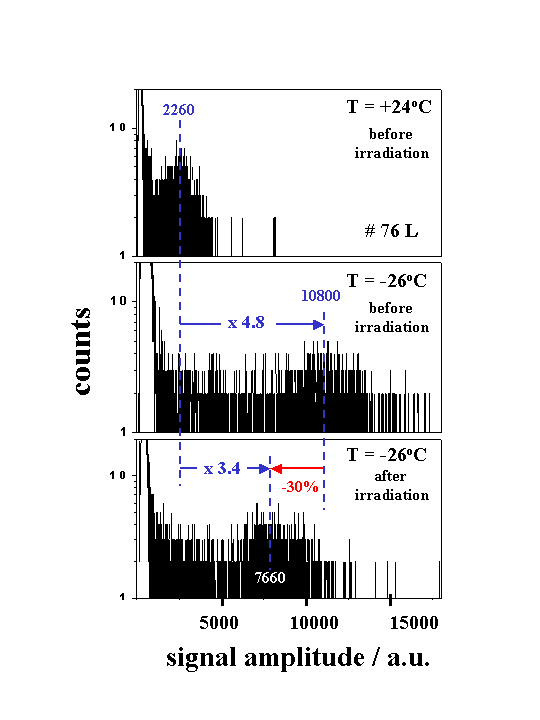}
\caption[Change of the scintillation response before and after irradiation.]
{Change of the scintillation response expressed by the
  measured energy deposition of cosmic muons in an assembled \PWOII
  detector. The figure shows the measurement at T=+24$\degC$ and
  T=-26$\degC$, respectively. The spectra at the lower temperature
  are obtained before and after irradiation with an integral dose of 53Gy. The measurement was
  carried out at Giessen.}
\label{fig:scint:pwo:fig43}
\end{center}
\end{figure}

These results have been further confirmed by a direct measurement of
the pulse height spectrum of the energy deposition of minimum ionizing
cosmic muons measured with a cooled crystal detector
assembly. \Reffig{fig:scint:pwo:fig43} compares the spectra measured
at T=+24$\degC$ and T=-26$\degC$, before and after irradiation at the
lower temperature with an integral dose of 53Gy. In spite of a reduction of the light output by
31\percent the photon statistics remains a factor 3 above the value at
room temperature. In spite of the significant asymptotic degradation
of the response, the slow change can be very well monitored with high
precision and stability.

\subsection{Pre-Production Run of \PWOII Crystals}

\begin{figure}[htb]
\begin{center}
\includegraphics[width=\swidth]{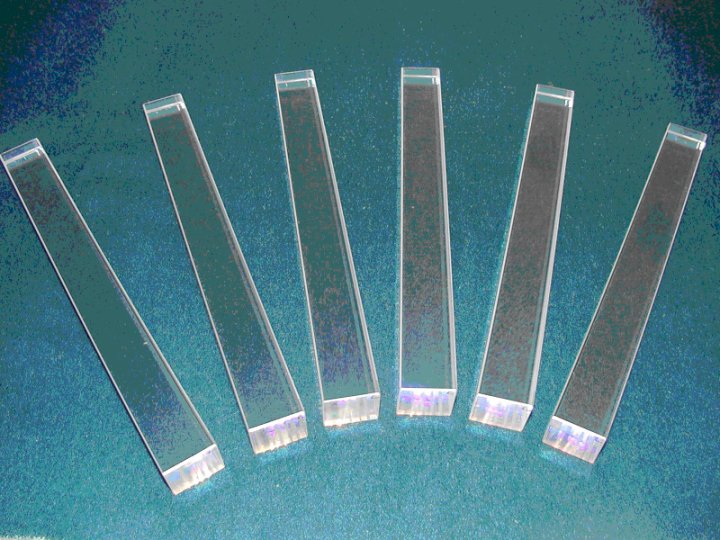}
\caption[Machined and optically polished barrel type \PWOII crystals.]
{Machined and optically polished barrel type \PWOII crystals for the \Panda electromagnetic calorimeter.}
\label{fig:scint:pwo:fig19}
\end{center}
\end{figure}

A subgroup of scintillator modules of the \BEMC (40 crystals of each
type: 8,9,10) has been produced by BTCP in December 2007/February
2008. \Reffig{fig:scint:pwo:fig19} shows a few optically polished
barrel type crystals produced at BTCP. The remaining crystals to
complete one slice of the \BEMC have been delivered in May 2008 and
are presently inspected for a detailed quality control and to adjust
and calibrate the ACCOS machine and test installations at CERN and
JLU, respectively. The pre-production run was aiming for:

\begin{enumerate}
  \item check of the reproducibility of the production technology
  \item confirmation of the correspondence of the crystal quality to
    the specification
  \item production of several reference crystals to be used for
    adaptation of the certification tools to \PANDA type crystals
  \item estimation of the technological yields of the crystals at
    different modes of operation of the production.
\end{enumerate}

The performed tests of the first 120 units have confirmed, that all
mechanical dimensions are within the specified tolerances. Some of the
relevant quality parameters are summarized in the next three figures.

\begin{figure}[htb]
\begin{center}
\includegraphics[width=\swidth]{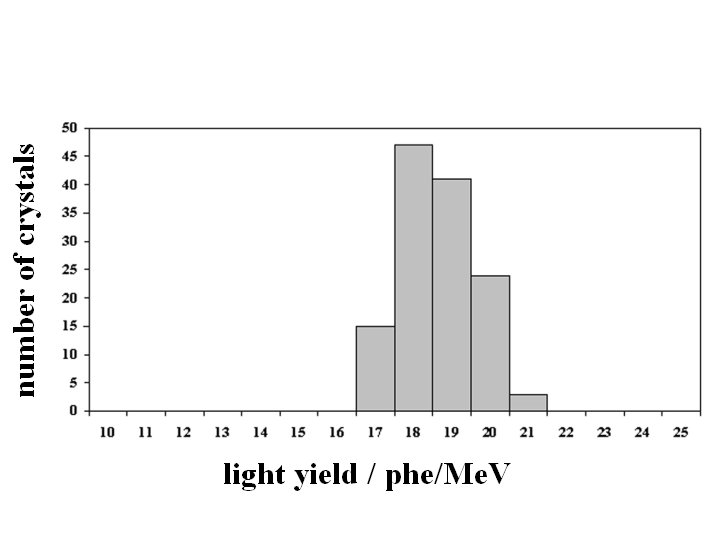}
\caption[Distribution of the light yield of barrel type \PWOII
  crystals.]
  {Distribution of the light yield of barrel type \PWOII
  crystals of the pre-production lot measured at room temperature
  using a photomultiplier tube with bialkali photocathode.}
\label{fig:scint:pwo:fig20}
\end{center}
\end{figure}

\begin{figure}[htb]
\begin{center}
\includegraphics[width=\swidth]{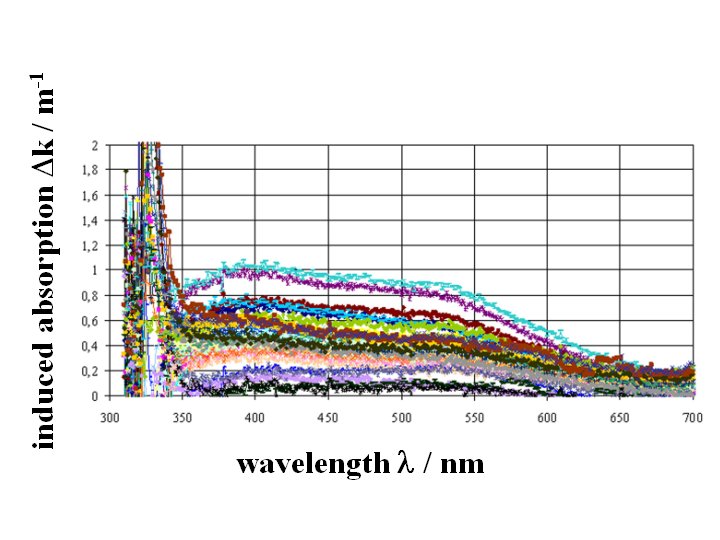}
\caption[Radiation induced absorption spectra.]{Radiation induced absorption spectra of 1cm thick \PWOII
  samples cut from the ingots of a pre-production lot measured at room
  temperature. A $^{60}$Co source has deposited a total dose of
  1 kGy at a dose rate of 2 kGy/h.}
\label{fig:scint:pwo:fig12}
\end{center}
\end{figure}

\begin{figure}[htb]
\begin{center}
\includegraphics[width=\swidth]{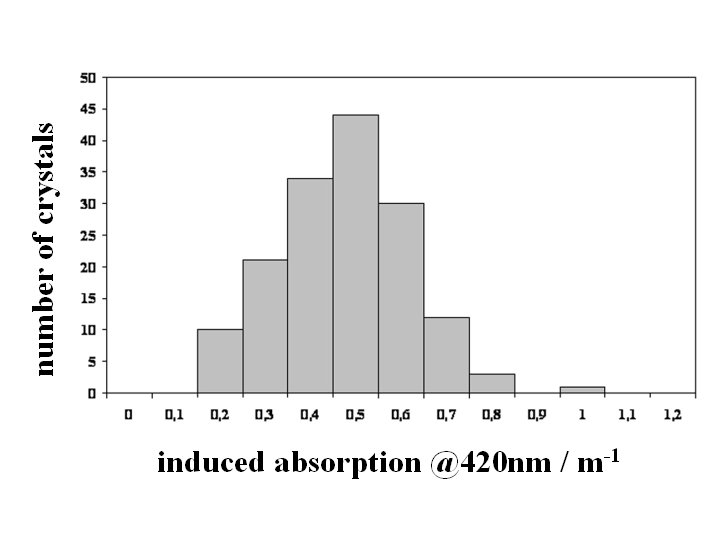}
\caption[Distribution of the induced absorption after an absorbed dose of 50 Gy.]
 {Distribution of the induced absorption measured at
  $\lambda$=420nm after an absorbed dose of 50 Gy at room temperature
  for \PWOII crystals of the pre-production run.}
\label{fig:scint:pwo:fig21}
\end{center}
\end{figure}

\Reffig{fig:scint:pwo:fig20} shows the range of the achieved light
yield measured with a standard photomultiplier with bi-alkali
photocathode at room temperature. The next two figures document the
variation of the induced absorption. \Reffig{fig:scint:pwo:fig12}
shows for $1\,\cm$ thick samples cut from the ingots of a
pre-production lot the induced absorption as a function of
wavelength. A $^{60}$Co source was used to deposit a total dose of
1 kGy at a rate of 2 kGy/h. \Reffig{fig:scint:pwo:fig21} projects
the distribution of the induced absorption value at the wavelength of
$420\,\nm$. These measurements were performed at the irradiation
facility at BTCP.

\section{Quality Requirements and Control}

The electromagnetic calorimeter of \PANDA requires the production of
nearly 16,000 scintillating crystals. The envisaged high
performance of the detector imposes strict requirements on the crystal
parameters, such as sufficient and uniform light yield, short decay
time free of slow components, very limited light yield loss under
irradiation and finally precise geometrical dimensions. To ensure a
high and efficient production rate, the certifying procedure for such a
large scale project has been elaborated for both the producer and the
customer site taking advantage of the many years of experience for
CMS/ECAL \cite{bib:emc:pwo36}. The equipment for automatic measurement
of the optical and scintillation properties, the associated software
tools as well as the methods and facilities for the radiation hardness
tests are described in the upcoming sections.

The \TSEMC requires the production of PbWO$_{4}$ crystal elements of
$200\,\mm$ length of one shape for both endcaps and 11 different shapes for the barrel. The latter ones
are used in two symmetric versions. Each crystal has to pass two certifying procedures,
one before delivery at the manufacturer site and a final one performed by the
customer. Presently, the specification limits, which are listed below,
are more stringent mainly due to the requested higher light yield and
the operation at low temperature down to T=-25$^{o}$C.

\begin{enumerate}
  \item \textbf{General Properties}
  \begin{itemize}
  \item Crystal dimensions have to be conform to the drawings given
    in \Reffig{fig:mech:mech-crystal_definition}, \Reffig{fig:ecap:FwEndCapCrystal},
    and \Reffig{fig:mech:ecap:BwEndCapCrystal}.
  \item Crystals have to be oriented with their front face to the seed
    part of the ingot.
  \item The appearances of crystals with respect to color, visible
    spatial and surface defects have to conform to the coordinated
    reference crystal.
  \item The polished surfaces are polished with a roughness of Ra
    $\leq$0.02$\mu$m, the lapped face has a roughness Ra =
    0.40$\pm$0.05$\mu$m. This parameter may be subject of modification
    under mutual agreement with the producer.
  \item The surface finish of chamfers should be made at a roughness
    of not more than 0.5$\mu$m (lapping).
  \item No cracks prolonging deeper than $0.5\,\mm$ into the crystal are allowed on the chamfers.  
  \end{itemize}
\item \textbf{Optical Properties:} These properties of all crystals
  should be measured with cross-checked equipment.
  \begin{enumerate}
    \item  Longitudinal transmission (absolute values)\\
            $\geq$35\percent at $\lambda$=360nm\\
            $\geq$60\percent at $\lambda$=420nm\\
            $\geq$70\percent at $\lambda$=620nm
          \item The non-uniformity of the transversal transmission
            $\delta$$\lambda$ at the transmission value of
            T=50\percent has to be $\delta$$\lambda$$\leq$=3 nm, for 6
            measurements every 4 cm; the first one is at 1.5 cm from the
            front face.
          \item Light yield $\geq$15 phe/MeV at T=18$\degC$ (for
            crystals with one lateral side de-polished).
          \item Light yield $\geq$16 phe/MeV at T=18$\degC$ (for
            crystals with all sides polished).
          \item Decay time: LY(100 ns)/LY(1000 ns)$>$90\percent at
            T=18$\degC$.
  \end{enumerate}
  \item \textbf{Radiation Hardness}
  \begin{enumerate}
  \item Radiation hardness is evaluated by the measurement of the
    induced optical absorption in the crystal along its axis. The crystals shall be kept at all times at a temperature T=$20\pm5\degC$ between the irradiation and the completion of the measurements. The value of the
    induced absorption calculated from the optical transmission of the crystal caused by the irradiation is limited
    to $\Delta$k$\leq$1m$^{-1}$ at $\lambda$=420 nm due to lateral
    $^{60}$Co irradiation. The totally accumulated dose shall be 30 Gy
    at a rate of 50-500 Gy/h. The mean value of the $\Delta$k distribution defined for each lot and based on the measurements of not less that 50\% of the lot shall be $< \Delta k > \ge 0.75\,\m^{-1}$.
  \end{enumerate}
\end{enumerate}

\begin{figure}[htb]
\begin{center}
\includegraphics[width=\swidth]{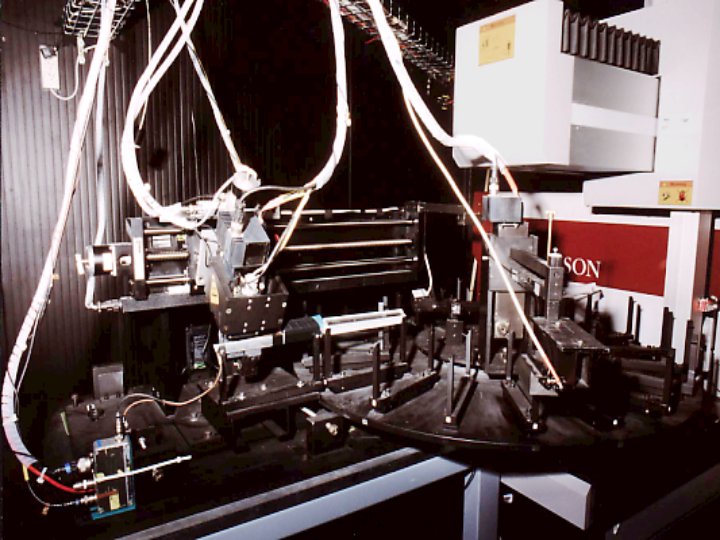}
\caption[The ACCOS machine installed at the CMS regional center at CERN.]
{The ACCOS machine installed at the CMS regional center at CERN to perform the quality inspection of \PWOII crystals in a semi-automatic procedure.}
\label{fig:scint:pwo:fig30}
\end{center}
\end{figure}

The conventional laboratory methods used for crystal measurements
cannot provide the necessary throughput. Therefore, a
high-productivity industrial system - Automatic Crystal Control System
(ACCOS) has been constructed \cite{bib:emc:pwo37} and installed at
both the crystal production facility and at CERN during implementation
of the CMS project (see \Reffig{fig:scint:pwo:fig30}). There is an
agreement with the CMS management on the use of this equipment and the
directly related infrastructure. The equipment is highly automated in
order to reduce the influence of the human factor and to provide an
output of at least 30 crystals per day. The PWO crystal is
mechanically fragile and can easily be damaged by contact with hard
equipment parts. Therefore, unnecessary frequent handling of crystals
should be strictly avoided. The adopted solution is to keep crystals
lying on special supports during all measurements and to move the
spectrometers along them. Consequently, special devices of a very
compact layout were designed. The overall setup including the
associated 3D-machine is installed in a temperature-controlled and
light-tight room. A bar-code reader identifies each crystal from the
bar-coded label glued on the small end-face at the production stage.
Two identical ACCOS machines are installed both at the production site
and at CERN to provide non-stop operation even in case of
malfunctioning.

\subsection{Measurements of the Light Yield and the Light Yield Non-uniformity}

A \emph{start-stop} or delayed coincidence method is used both for the
measurement of the scintillation decay time and light yield (LY). The
proportionality between the count rate of the stop signals and the
light yield of the scintillator for small quantities of light is used
to determine the LY by integration of the decay time spectrum. It
means that both parameters can be measured simultaneously. In this
case, neither optical coupling between scintillator and PM nor
scintillator wrapping are needed in contrast to the very widely used
method of pulse-height spectrum measurement by using radioactive
$\gamma$-sources (most commonly $^{60}$Co) with consequent total
absorption peak position determination. The \emph{start-stop} method
puts two constraints on the event rates, which limits the
productivity. One is the limitation on the annihilation source
activity. It should not exceed ~10$^{5}$\,Bq. At higher activity the
probability of random coincidences of two $\gamma$-quanta from
different decays becomes too high and leads to background. The second
is the limitation on the probability of detecting scintillation
photons in the stop channel per single excitation. A conventional TDC
requires that the average number of detected photons to be not more
than 0.1 per single excitation. Otherwise, the measured scintillation
decay curve will be distorted, especially in the slow component
region. The use of a multi-hit TDC would avoid this problem. The dead
time of the stop-channel detector can also cause distortion, in
particular in the region of the fast component, when the fastest
component in scintillation is less than or comparable to the stop
channel dead time, which is the case for PWO. A single-hit TDC (or TDC
with multi-stop rejection \cite{bib:emc:pwo38}) was implemented to
overcome the distortion problem. The most comprehensive analysis of
TDC with multi-stop rejection is made in \cite{bib:emc:pwo39}. The
start-channel detector consists of a Hamamatsu UV-extended
photomultiplier (R5900) coupled to a $20\times20\times20$mm$^{3}$
BaF$_{2}$ scintillator. The $^{22}$Na annihilation source with an
activity of 200\,kBq is mounted as close as possible to the scintillator
and delivers about $2\cdot 10^{4}\,/\s$ starts. The start
photomultiplier with scintillator and source is mounted on a moving
platform together with the transversal transmission photospectrometer
for scanning along the crystal. The stop-channel is based on another
Hamamatsu R5900 selected with a low count rate of dark pulses. The
average number of stop signals (detected scintillation photons)
amounts to $\sim$0.3 per single excitation for the most luminous PWO
crystals. The overall count rate of good events (scintillation photons
produced by scintillation decay in PWO and collected in the decay time
spectrum) is typically 1\,kHz, and a spectrum with acceptable statistics
can be acquired within one minute.

\subsection{Inspection of the Optical Properties}

The longitudinal transmission can be reconstructed using transmission
data taken in transverse direction. However, it is impossible to
obtain the required precision in case of PWO scintillation elements
because of uncertainties in the Fresnel reflection level from crystal
to crystal. These uncertainties are caused by the deviation of the
crystal growth axis from the crystallographic axis a by
$\pm$ 5\degrees. Therefore, both types of transmission measurements
should be performed. Since PWO is a birefringent crystal and the shape
of the scintillation elements varies, a wide-aperture photo detector
is required. This prevents from the use of spectrometers, which are
based on photodiode arrays with a light dispersing prism. On the other
hand, the PWO transmission spectrum has no narrow absorption bands, so
it is not necessary to measure the spectrum at many wavelengths. It
can be measured only at several well-chosen wavelengths with
subsequent reconstruction of the total spectrum to speed up the
measurements. For this purpose, a set of 11 interference filters was
used. Two dedicated compact photospectrometers for optical
transmission measurement have been designed and built. The optical
part of each photospectrometer includes a 20\,W halogen lamp, a
four-lens collimator, three objective lenses, a rotating changer with
interference filters, collecting mirrors, a $20\times20\,$mm$^{2}$
UV-extended photodiode (Hamamatsu S6337) and readout electronics. The
spectrometer dimensions are $70\times80\times220\,$mm$^{3}$. The
compact design allows the fast movement of the spectrometer from the
measurement zone to the calibration position (air measurement). During
operation, the rotating wheel with interference filters crosses the
light beam sequentially. Such a system provides a fast switching
(0.5\,s) of the wavelength. The time necessary to measure a transmission
spectrum in the range from 330\,nm to 700\,nm takes less than 10\,s for one
spatial point.

\subsection{Measurement of the Geometrical Dimensions}

All dimension measurements are carried out with the standard
three-dimensional machine \emph{TOPAZ 7} supplied by Johansson AB
(Sweden), which provides three coordinates of several points on each
crystal face with a precision of $\pm$5$\,\mu$m and subsequent
calculation of the crystal dimensions and the planarity of all faces.

\subsection{Analysis and Documentation}

Software is provided for a user interface, acquisition, data
processing, results storage and presentation. It also supports
communication with the C.R.I.S.T.A.L. database software
\cite{bib:emc:pwo40} and the 3D-machine. Software components are
implemented in LabView$^{TM}$ graphical programming system and
Microsoft$^{TM}$ Visual C++ and runs on an IBM-compatible PC under
Microsoft$^{TM}$ Windows operating system. A special algorithm was
developed for the processing of the scintillation kinetics data
\cite{bib:emc:pwo41}. Instead of a non-linear iterative fitting
procedure usually used for multi-exponential functions and giving
sometimes unstable results due to inherent convergence problems, a new
approach is based on non-iterative 3-step integrating/subtraction
procedure working without user intervention.

\subsection{Control Measurements of Production Stability}

To provide the stability of the technology and minimize the number of
rejections the following set of control measurements should be carried
out in addition to the certification measurements along the total
manufacturing process.

\begin{itemize}
\item \textbf{Quality check} of the raw material provided by pilot
  crystal growth and measurements (1 per 100 crystals) - the only
  acceptable rate in terms of time and expenses;
\item \textbf{Technology discipline check:}\\
  analysis of average crystals parameters per lot;\\
  analysis of average crystals parameters during some period for each
  growing machine;\\
  analysis of rejected crystals
\item \textbf{Radiation hardness characterization on the sampling basis:}\\
  measurement of radiation induced absorption in top parts of crystals
  (5-10 top parts per 100 crystals);\\
  measurement of radiation induced absorption in full-size crystals
  devoted to the specific oven in case a previous check gives
  marginal results
\end{itemize}

Pilot crystals along with top parts of regular crystals will be
analyzed in detail at the INP (Minsk) using the same methods and
laboratory equipment as were exploited during the development phase of
PWO crystals. All crystals rejected by the certification measurements
have to be investigated by independent laboratory methods with full
confirmation of the results. Even accepted crystals, which have passed
the certification procedure, will be checked by laboratory methods on
the sampling basis.  All required measurements during PWO
mass-production can be summarized in a 3-level scheme to make
decisions. On the first level, the manufacturer and INP provide pilot
crystal growth and tests to prove the specific raw material batch to
be acceptable for crystal production. On the second level, the
produced crystals pass through the certification procedure on the
ACCOS system and radiation hardness checks at the manufacturer
plant. The data analysis is performed at INP including the decision on
the delivery of the crystals to the customer. On the third level, the
customer carries out the visual inspection, the certification on the
ACCOS machine and radiation hardness tests. If all requirements are
fulfilled the crystals will be accepted for the assembly into the
calorimeter. The first two steps of the final acceptance will be
performed at the CERN facilities. Final radiation tests will be done
at the facility at Giessen. Depending on the overall productivity and
stability of the manufacturing process some tests at the customer
facilities might be done either on all crystals or on selected
samples.

\subsection{Manufacturer for the Final Production}

All presently installed or operating calorimeters, such as ECAL (CMS),
PHOS (ALICE) and HYCAL of the PrimEx experiment at JLAB
\cite{bib:emc:pwo46} are composed of PWO crystals manufactured by
$\INST{Shanghai Institute of Ceramics}$ (SICCAS, China) or the two
Russian producers $\INST{Bogoroditsk Technical Chemical Plant}$ or
$\INST{North Crystals}$. The latter one, the major
producer for PHOS, has moved out of the scintillator business.  As
outlined in the previous chapters, the specification limits for the
calorimeter crystals have to be identical to the quality of \PWOII,
which was developed in close collaboration with BTCP and the CMS/ECAL
group.  Up to now a first pre-production run performed at BTCP was
completed in May 2008 comprising more than 710 crystals of all 23
different shapes with an overall length of $200\,\mm$. The first lot
allows to complete one out of the 16 slices of the \BEMC and a first
subarray of the \FWEMC. Some of the achieved quality parameters were
shown in the previous chapter.

There exists a long collaboration with the Shanghai Institute of
Ceramics and many state-of-the-art test crystals of various sizes were
investigated. Finally, a full matrix comprising 25 modules of
$200\,\mm$ length with identical cross section of
20$\times$20\,mm$^{2}$ were delivered in 2005 for comparison.

\begin{figure*}[htb]
\begin{center}
\includegraphics[width=\dwidth]{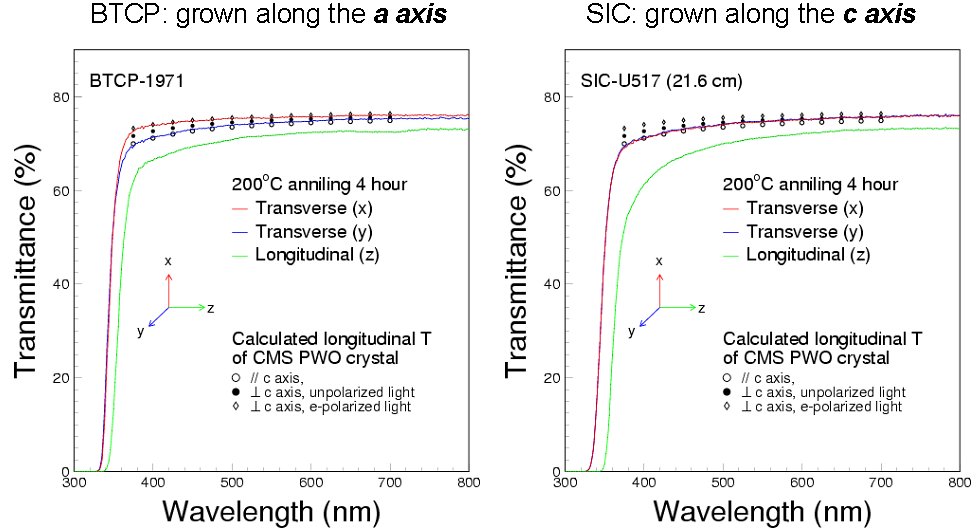}
\caption[Optical transmission of crystal samples from BTCP and SICCAS.]
 {Comparison of the optical transmission of two crystal samples from
 the manufacturers BTCP and SICCAS. The experimental
 data are compared to theoretical calculations described in the
 text. The figures are taken from Ref.~\cite{bib:emc:pwo44}.}
\label{fig:scint:pwo:fig32}
\end{center}
\end{figure*}

These crystals are grown along the $c$-axis by a modified Bridgeman
method~\cite{bib:emc:pwo42} and consequently differ in the longitudinal
and transverse optical transmittance. \Reffig{fig:scint:pwo:fig32}
shows a comparison of measured transverse and longitudinal
transmission. The measurements performed for a sample from both
manufacturers are compared to theoretical limits assuming no internal
absorption \cite{bib:emc:pwo43,bib:emc:pwo44}. The chosen crystal from
BTCP corresponds to \INST{CMS} quality. The three different
calculations are using the refractive index of ordinary light, which
propagates along the $c$-axis and has a polarization perpendicular to
the $c$-axis. Extraordinary light propagates perpendicular to the
$c$-axis and has a polarization along the $c$-axis. Finally,
unpolarized light is assumed to propagate perpendicular to the
$c$-axis. Because of the birefringence, the theoretical limit of the
extraordinary light is about 3\percent higher than that of the
ordinary light. Since the crystals produced in China are grown along
the $c$-axis, the experimental longitudinal transmission should be
compared to the theoretical limit for propagation along the $c$-axis
independent of polarization. Consistent with the theoretical
calculations, the samples from BTCP show a better longitudinal
transmission.  However, the measurement of the absolute yield of the
detected scintillation light is slightly superior for the Chinese
products, which are doped exclusively with yttrium ions. Compared to
\INST{CMS} standards, these crystals show approximately 30--40\percent
more light.

\begin{figure}[htb]
\begin{center}
\includegraphics[width=\swidth]{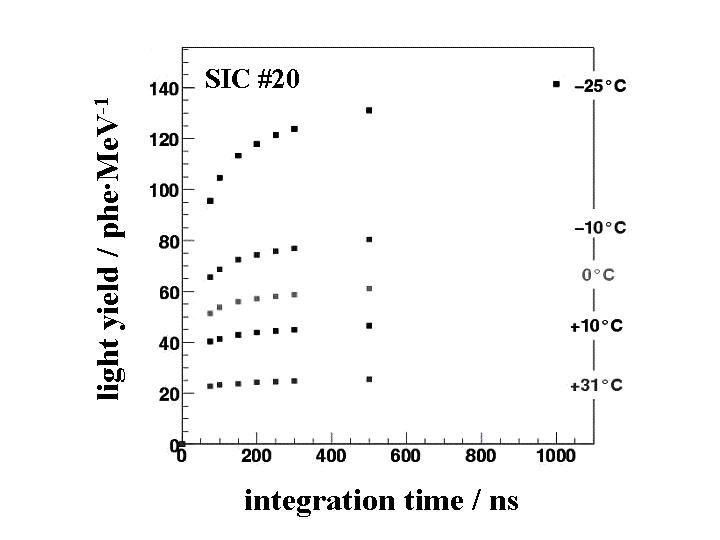}
\caption[Absolute light yield.]{Absolute light yield of a $200\,\mm$
  long rectangular crystal manufactured by SICCAS. The scintillation
  light has been measured with a photomultiplier tube with bialkali
  photo cathode. The response is shown at different temperatures as a
  function of integration gates varying between $75\,\ns$ and
  $1\,\mus$, respectively.}
\label{fig:scint:pwo:fig33}
\end{center}
\end{figure}

Small crystal samples (16$\times$16$\times$30\,mm$^{3}$), which were
readout with photomultiplier tubes, confirm the good photon statistics
expressed by the excellent energy resolution of
$\sigma_{E}$/E=14.5\percent obtained for $662\,\kev$ $\gamma$-rays at
a temperature of \mbox{T=-26$\degC$.} The 25 full size crystals
at room temperature deliver a mean light yield of $\sim$18.0 phe/MeV
and are comparable to \PWOII. Both values are determined with
photomultiplier tubes. \Reffig{fig:scint:pwo:fig33} summarizes for one
of the brightest samples the change of light yield at reduced
temperatures. The values are obtained with a calibrated
photomultiplier tube (Hamamatsu R2059, quantum efficiency
QE($420\,\nm$)=20.5$\percent$) by integrating the response over time
gates ranging between 75\,ns and 1$\,\mu$s, respectively. The figure shows
a high and nearly constant light yield at T=+31$\degC$. There is a
strong increase of the emitted light when the temperature is decreased
down to T=-25$\degC$. However, in strong contrast to crystals of
similar size from BTCP, there is a significant and rising contribution
of slow scintillation components. Therefore, in order to collect
$\geq$90$\percent$ of the light output, integration times between 0.5
and $\gg$$1.0\,\mus$ become mandatory leading to a severe impact on
the count rate capabilities of a calorimeter \cite{bib:emc:pwo45}.

\begin{figure}[htb]
\begin{center}
\includegraphics[width=\swidth]{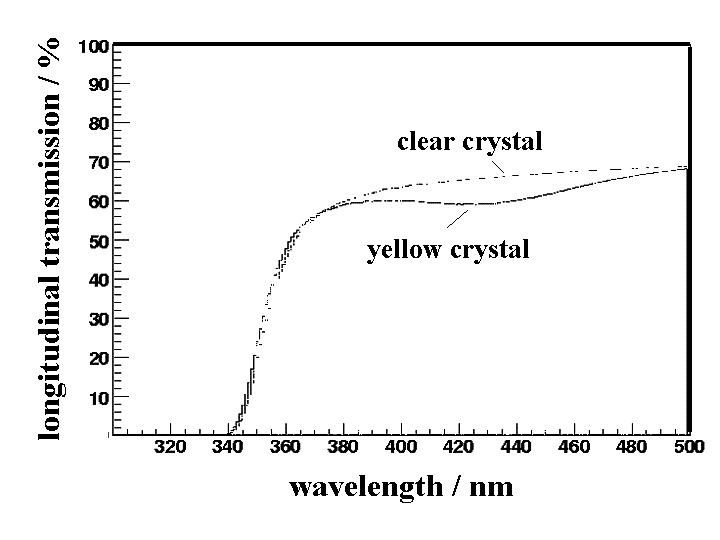}
\caption[Longitudinal optical transmission of two $200\,\mm$ long crystals.]
{Longitudinal optical transmission of two $200\,\mm$ long
  crystals manufactured by SICCAS. The slightly colored sample shows a
  strong absorption band peaking at $\lambda$$\sim$$420\,\nm$.}
\label{fig:scint:pwo:fig34}
\end{center}
\end{figure}

\begin{figure}[htb]
\begin{center}
\includegraphics[width=\swidth]{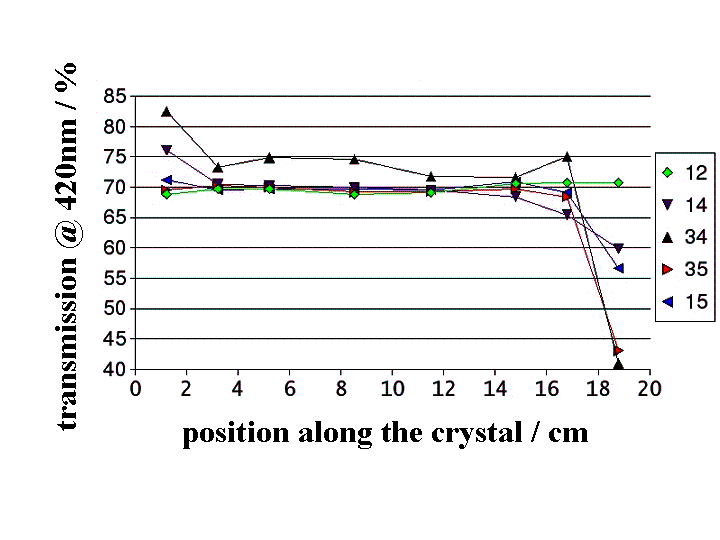}
\caption[Inhomogeneity in the transversal optical transmission.]
{Inhomogeneity in the transversal optical transmission of
  several $200\,\mm$ long crystals manufactured by SICCAS and measured at the wavelength $\lambda$$\sim$$420\,\nm$. The cross section of the rectangular crystals is 20$\times$20\,mm$^{2}$. Most of the samples indicate deviations even at both end
  sections of the crystal.}
\label{fig:scint:pwo:fig35}
\end{center}
\end{figure}

In addition, detailed investigations have observed significant
inhomogeneities of the crystals. In particular, about half of the
delivered samples show a yellowish color, which leads in the optical
transmission spectrum to a strong absorption region peaked at
$\lambda$$\sim$420nm as shown in \Reffig{fig:scint:pwo:fig34}. A complementary study of the transversal optical transmission, determined at several positions along the crystal axis, documents an
insufficient homogeneity of several
crystals. \Reffig{fig:scint:pwo:fig35} shows strong deviations of the
transmission values appearing typically at both ends of the crystal
when measured at the critical wavelength of $\lambda$$\sim$$420\,\nm$.

As a consequence, the manufacturer SICCAS has started further
developments in order to adapt the scintillation kinetics at lower
temperatures to the \Panda needs. Some of the samples with a volume of
a few cm$^{3}$ show faster decay kinetics but there has been up to
date not yet the delivery of full size scintillation crystals for a realistic
comparison.

%
%
%

%
%
\newpage
\bibliographystyle{panda_tdr_lit}
\bibliography{./lit_emc}
%

%
\cleardoublepage
\chapter{Photo Detectors}
\label{sec:photo}
%
%
\label{sec:photo:intro}
The detection of scintillation light of lead tungstate under the given
conditions for the \Panda EMC requires excellent photo detectors.

The magnetic field of about $2\,$T precludes the use of conventional
photomultipliers. On the other hand the signal generated by ionization
in a PIN photodiode by a traversing charged particle is too large for
our applications. To solve these problems a photosensor insensitive to
magnetic fields and with a small response to ionizing radiation has to be
used.

Since lead tungstate ($PbWO_4$) has a relatively low light yield, the
photosensor is required to have internal gain in addition.  Due to the
increase of the crystal light yield accomplished by cooling the
scintillator down to a temperature of $T = -25 \degC$, the used photo
detectors have to be radiation hard in this temperature regime, which
implies detailed studies concerning possibly occurring radiation
damages.

For the \BEMC an Avalanche Photodiode (APD) which has an internal
signal amplification in the silicon structure is chosen as photo
detector.  The low-energy photon threshold in the order of 10$\,\MeV$
requires to maximise the coverage of the readout surface of the
crystals, leading to the development of Large Area APDs (\LAAPD) with
an active area of 10$\times$10$\,\mm^2$, which have been tested for
the given requirements.  During the development of the readout system
for the EMC the typical size of the crystal readout surface of 27 x
27$\,\mm^2$ was leading to the development of an \LAAPD with a
rectangular shape to cover a maximum of this space with two
neighboring photo detectors.

The photon detection in the \FWEMC has to deal with rates up to
500$\,\kHz$ and magnetic fields up to 1.2$\,$T.  Therefore we have
chosen vacuum phototriodes (VPT) with a diameter of $22\,\mm$ as
photon detectors.  The main 
reasons for this choice are rate capability, radiation hardness,
absence of nuclear counter effect and absence of temperature
dependence. Standard photomultipliers are excluded due to the magnetic
field environment. In contrast to the barrel region, the magnetic
field is oriented in the axial direction of the VPTs and thus makes it
feasible to use VPTs for the endcap.  Vacuum phototriodes are
essentially a photomultiplier tube with only one dynode and weak field
dependence.

\section{Avalanche Photodiodes (APD)}
\label{sec:photo:APD}
\subsection{Introduction}
\label{sec:photo:APD:Intro}
%
%
\begin{figure}[b]
  \begin{center}
    \includegraphics[width=\swidth]{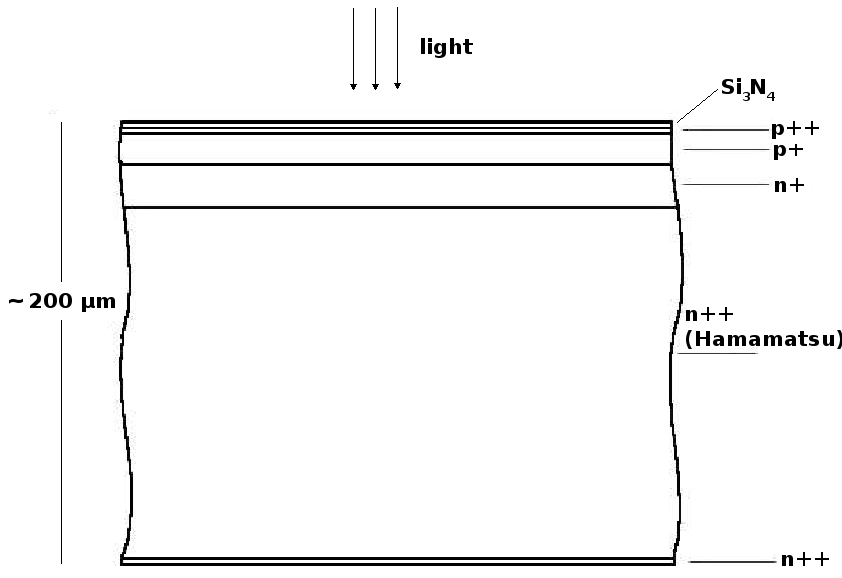}
    \caption[APD reverse structure.]{Schematic view of an APD with reverse
    structure \cite{bib:emc:photo:APD:CMS_TDR}.}
    \label{fig:photo:APD:Intro:APD_scheme}
  \end{center}
\end{figure}
\begin{figure*}
  \begin{center}
  \includegraphics[width=\dwidth]{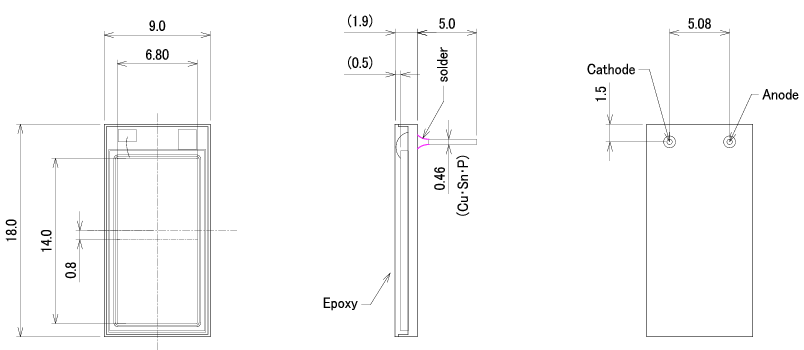}
    \caption[Technical drawing of the rectangular APD type.]{Technical drawing of the developed rectangular APD type.
    }
    \label{fig:photo:APD:Intro:APD_rectangular}
  \end{center}
\end{figure*}

%
The requirements to detect the scintillation light of lead tungstate in
the barrel part of the EMC can be satisfied by silicon avalanche photodiodes
(APDs). Avalanche photodiodes are reverse biased diodes with an internal
electric field used for avalanche multiplication of charge carriers.
The \INST{CMS} collaboration has developed APDs in
collaboration with Hamamatsu Photonics optimized to detect the
scintillation light from lead tungstate crystals. Those APDs have
several advantages: compactness with an overall thickness of $\approx
200\,\mum$ (\Reffig{fig:photo:APD:Intro:APD_scheme}), high quantum
efficiency of 70--80\percent at the
wavelength of maximum emission intensity of lead tungstate, insensitivity to magnetic
fields and low cost in mass production.
A disadvantage of this photosensor is its relatively small active area of
$5\times 5\,\mm^2$ compared to the area of the crystal end faces. 
Therefore we started the \mbox{R\&D} of large area APDs (LAAPDs) 
with an active area of ($10 \times 10$)$\,\mm^2$ in collaboration with Hamamatsu Photonics.
Our R\&D of LAAPDs is based on the same internal structure
that has already been tested by the \INST{CMS} collaboration
(see \Reffig{fig:photo:APD:Intro:APD_scheme}):
Light enters the APD via the $p^{++}$ layer and is absorbed
in the following $p^{+}$ layer, where electron-hole pairs are generated.
The electrons drift in the electric field towards the p-n junction where they are amplified by
impact ionization, yielding avalanche gain, and drift through the n-material to the $n^{++}$
electrode where the charge collection takes place.
In front of the $p^{++}$ layer a passivation layer made of silicon nitride Si$_3$N$_4$ is used 
which reduces the decrease of quantum efficiency caused by reflection
losses from the surface of the Si wafer.\\
Due to the large capacitance of the ($10 \times 10$)$\,\mm^2$ LAAPDs 
(see \Reftbl{tab:photo:APD:Char:summprop:character}) an additional prototype 
of same dimensions but lower capacitance has also been tested. As shown in the Technical
Progress Report \cite{bib:emc:photo:APD:Panda_TPR} the first prototype of the ($10 \times 10$)$\,\mm^2$ APDs showed a nonsatisfying 
radiation hardness concerning proton irradiation. Therefore
the APD internal structure was modified by adding a groove to reduce the occuring surface
current and in the present status of R\&D work 
mainly two different APD types are under investigation. Those are LAAPDs with an active area of ($10 \times 10$)$\,\mm^2$ with
different capacitance values developed in cooperation with Hamamatsu Photonics: a so called 'normal C'
type LAAPD ($C = 270\,pF$) and a 'low C' type LAAPD with a capacitance of $C = 180\,pF$. 
The foreseen rear side dimensions of the lead tungstate crystals (typ. \mbox{($21.4 \times 21.4$)$\,\mm^2$}) make it impossible to use two LAAPDs 
with quadratic active area of our dimensions (package size: \mbox{($14.5 \times 13.7$)$\,\mm^2$}). Therefore LAAPDs with rectangular 
package/active area have to be used to cover the main part of the crystal readout area. Those diodes with
rectangular shape are 
presently under development and first prototypes
will be available mid of 2008. A preliminary drawing of the LAAPD dimensions of the
rectangular LAAPD type is shown in \Reffig{fig:photo:APD:Intro:APD_rectangular}. This avalanche photodiode will
have an active area of \mbox{($7 \times 14$)$\,\mm^2$}.    
Two of these rectangular LAAPDs will be used to detect the scintillation light of one crystal in the barrel
part of the electromagnetic calorimeter. 
\subsection{Characteristics}
\label{sec:photo:APD:Char}
To ensure a stable operation of an avalanche photodiode several device properties have to be measured
during screening or rather 
recorded during operation. The main parameter to be reported is the temperature of the APD during
operation, because the dark current as well as the internal gain are strongly depending on temperature. Therefore
the temperature has to be held stable down to an uncertainty of $\Delta T = \pm 0.1\,\degC$. Another essential 
parameter is the value of the applied bias voltage, which has to be kept constant down to 
$\Delta U_R = \pm 0.1\,V$ to ensure a proper determination of the internal gain as well as a proper measurement
of the device dark current depending on bias voltage. Special attention will be devoted to the main APD 
parameters in the following paragraphs.
\subsubsection{Dark Current and Gain}
\label{sec:photo:APD:Char:Dark}
\begin{figure}
  \begin{center}
    \includegraphics[width=\swidth]{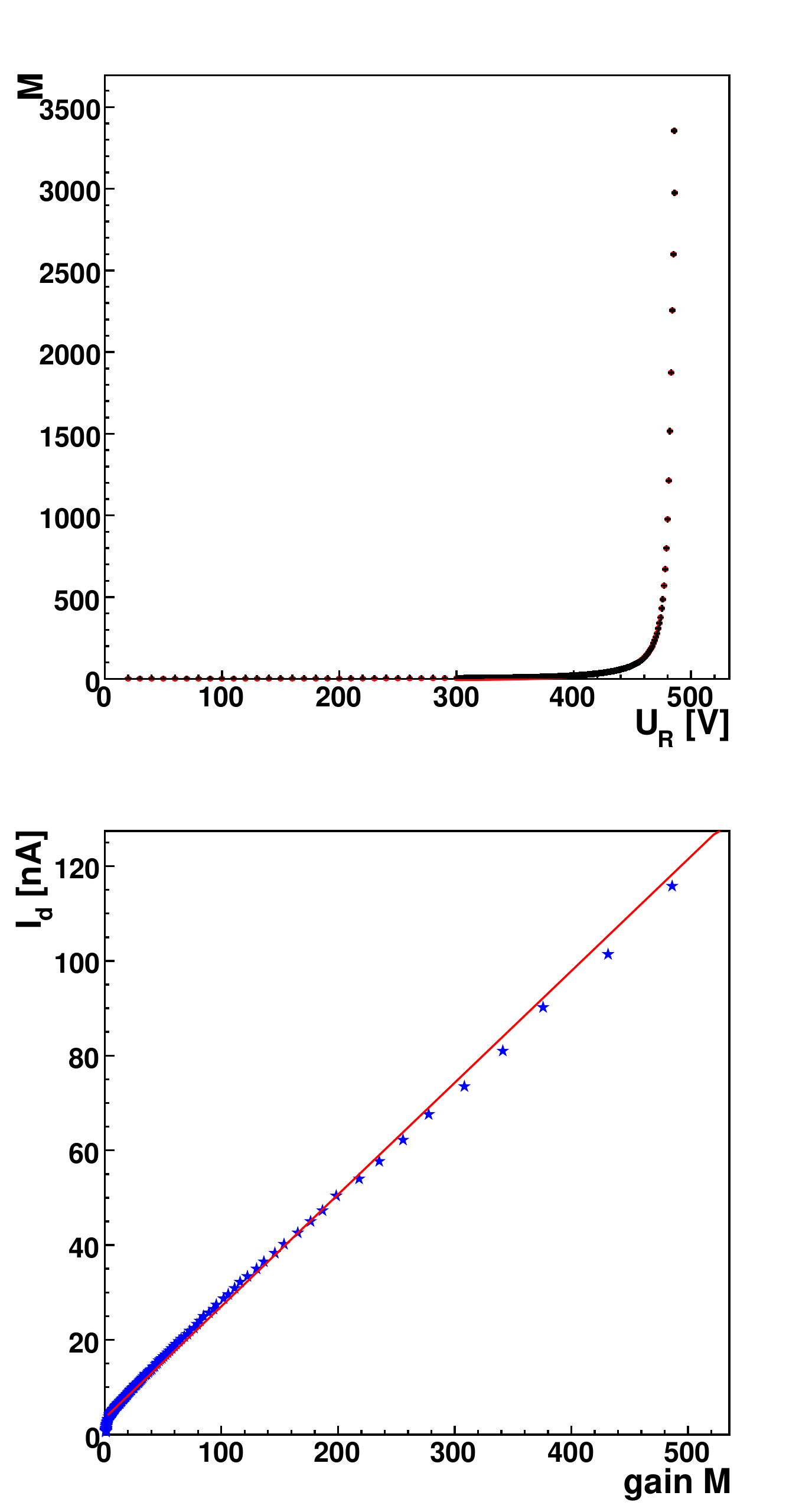}
    \caption[Gain and dark current of 'normal C' version LAAPD.]{Determined gain $M$ of the 'normal C' version LAAPD
    (above) and the corresponding dark current dependence of $M$ in the range of linear dark current behaviour (below).}
    \label{fig:photo:APD:Char:Dark:M_Id_normal}
  \end{center}
\end{figure}

\begin{figure}
  \begin{center}
    \includegraphics[width=\swidth]{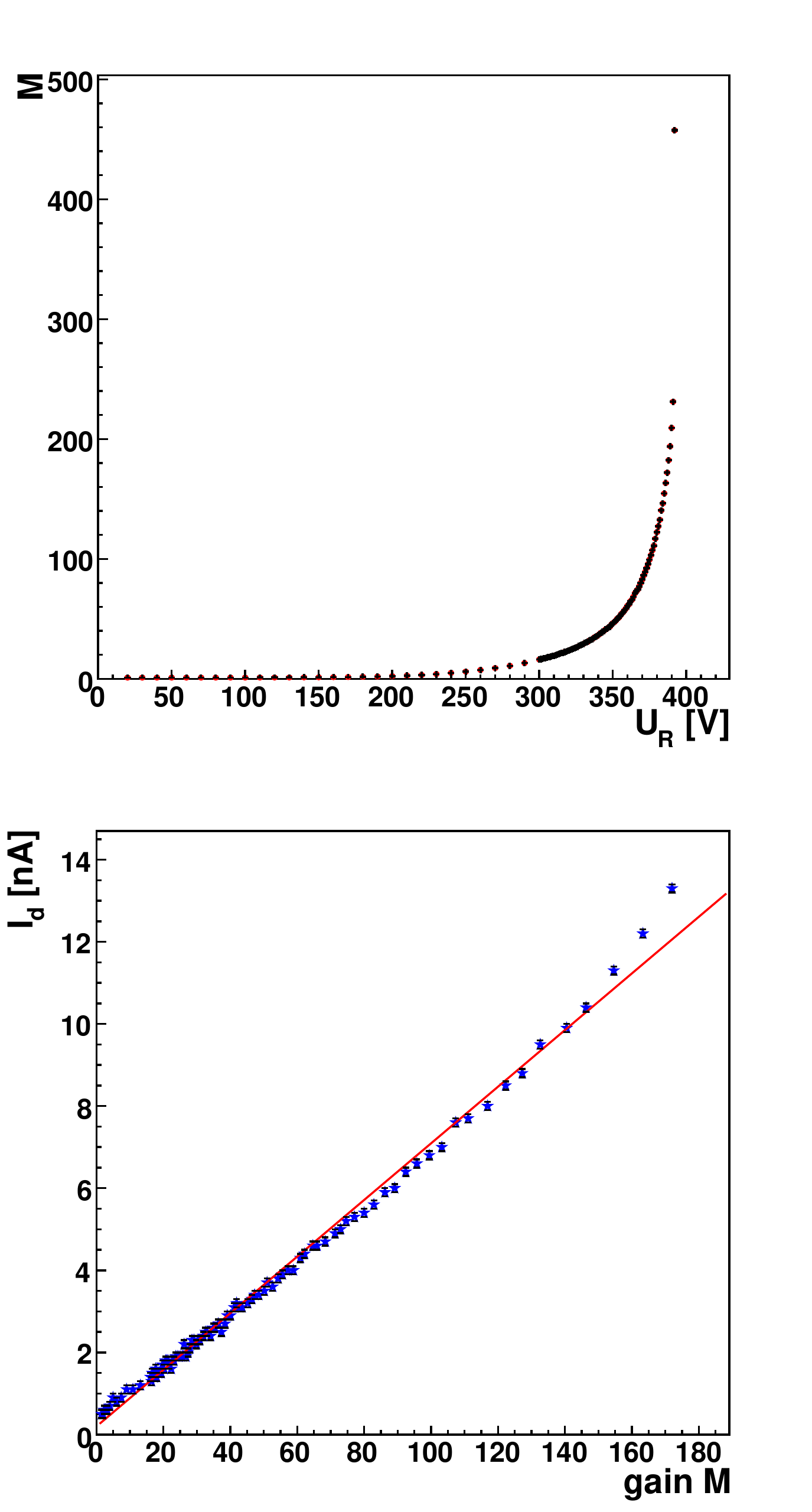}
    \caption[Gain and dark current of 'low C' version LAAPD.]{Determined gain $M$ of the 'low C' version LAAPD
    (above) and the corresponding dark current dependence of $M$ in the range of linear dark current behaviour (below).}
    \label{fig:photo:APD:Char:Dark:M_Id_low}
  \end{center}
\end{figure}

\begin{figure}
  \begin{center}
    \includegraphics[angle=90,width=\swidth]{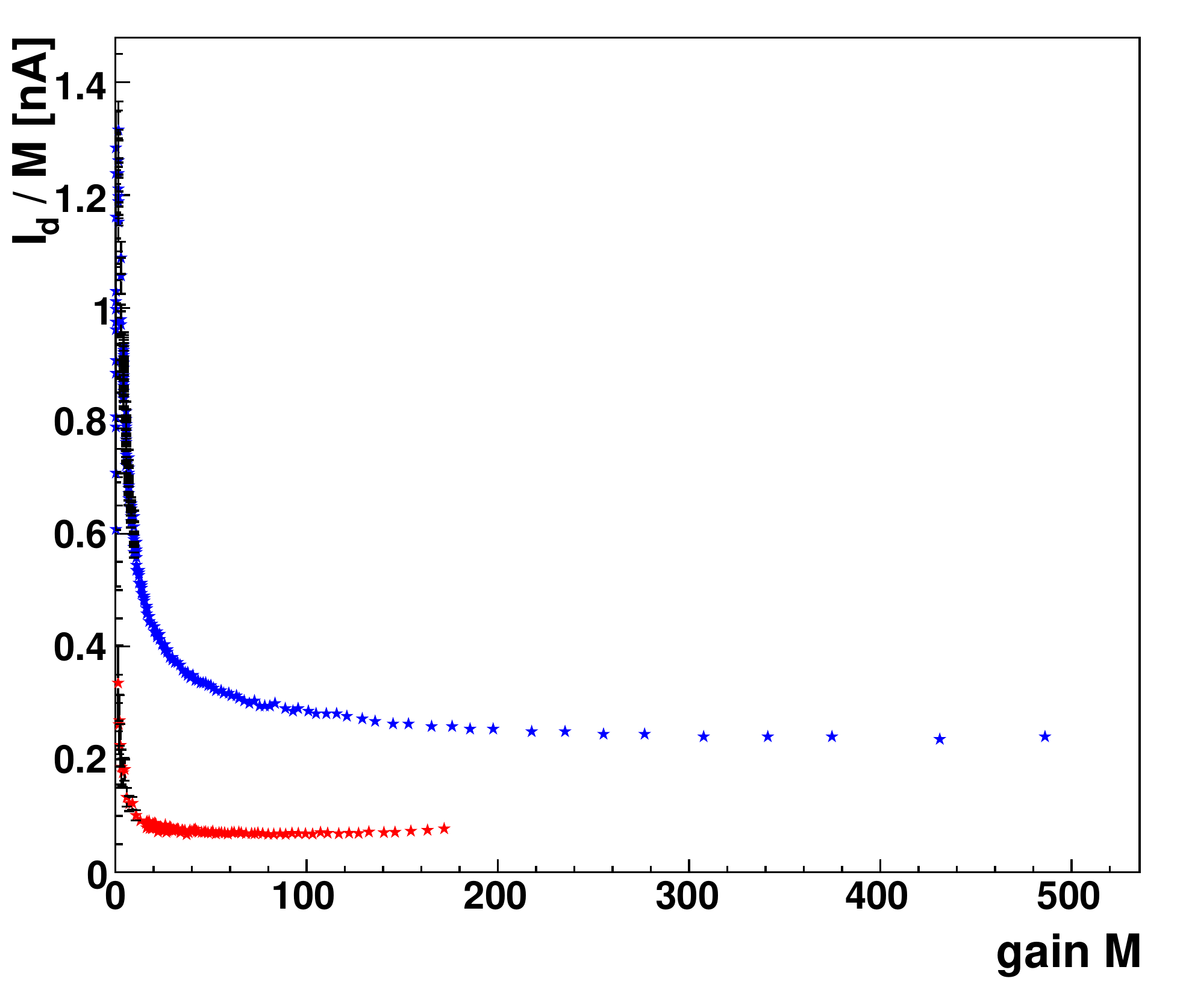}
    \caption[Ratio Id/M for different APD types.]{$I_d/M$ depending on $M$ of the 'normal C' version 
    (blue) and the corresponding contribution for the 'low C' type APD
    (red).}
    \label{fig:photo:APD:Char:Dark:IdM_M}
  \end{center}
\end{figure}

Due to interactions of charge carriers with phonons the gain of an APD
decreases with increasing temperature. On the other hand the free mean path and the 
band gap of semiconductor devices are temperature dependent. In our case these two parameters
can be neglected because of the envisaged operation temperature of $T = -25 \degC$. The degree of sensitivity to temperature changes can be determined by the
proximity of the applied bias/reverse voltage ($U_R$) to the value of breakdown voltage ($U_{Br}$)
and by the device characteristics as a result of internal structure:\\
A decrease of capacitance of the APD results in an increasing sensitivity
to temperature changes. The temperature dependence of gain has a value of
-2.2\percent/$\degC$ at an internal gain of $M\,=\,50$ in case of a
\INST{CMS}-APD \cite{bib:emc:photo:APD:Renker_Calor2000}.\\ 
The dark current $I_d$ of an APD can be devided into the bulk current $I_{b}$ and the surface current 
$I_{s}$. While the surface current is independent from the applied gain $M$ the value of the bulk current 
increases with increasing gain. The contributions of those two can be calculated separately by using the
following relation if the gain $M$ and the overall dark current $I_d$ have been determined:
\begin{equation}
I_d = I_b \cdot M + I_s.
\label{eq:photo:APD:Char:Dark:I_d} 
\end{equation}
The internal gain $M$ of the APDs at a fixed temperature is measured by using the following method: The dark current ($I_d$)
and the current under continous illumination ($I_{ill}$) at a fixed wavelength
of $\lambda\,=\,420\,\nm$, according to the maximum emission wavelength of $PbWO_4$, is recorded for 
several bias voltages up to breakdown. 
From the ratio of photo currents ($I_{ill}-I_d$) of high bias voltages and the voltage value where no
amplification occurs (equivalent $M\,=\,1$) the gain $M$ is calculated by:
\begin{equation}
M = \frac{I_{ill}(U_R)-I_d(U_R)}{I_{ill}(M=1)-I_d(M=1)}.
\end{equation}
The results of these measurements done at room temperature are shown for a 'normal C' type APD 
in \Reffig{fig:photo:APD:Char:Dark:M_Id_normal} and for a 'low C' type version in 
\Reffig{fig:photo:APD:Char:Dark:M_Id_low}.
For the evaluation of the maximum gain of the device up to which a stable operation of the diode can be ensured,
the dark current dependence of gain has to be regarded in parallel. For the investigated 'normal C' LAAPD gains
up to a value of $M = 3300$ could be reached before the breakdown voltage of the diode was reached. A closer look
on the dark current reduces this value down to $M = 500$, because at higher gains the linearity of the dark current
given by \Refeq{eq:photo:APD:Char:Dark:I_d} is no longer guaranteed (see \Reffig{fig:photo:APD:Char:Dark:M_Id_normal}). 
Similar behaviour can be observed in case of the 'low C' version, where the maximum gain decreases from $M = 460$
to a value of $M = 160$ due to the observed dark current unlinearity.
The discrepancy in the maximum usable gain of the two APD types may originate from their
different dark current behaviour depending on the internal APD gain and is caused by their
different internal structure. This circumstance could be clarified by considering the ratio
$I_d/M$ depending on gain which is shown in \Reffig{fig:photo:APD:Char:Dark:IdM_M}.  
As it can be seen the two APD types behave completely different for gain values larger than
$M = 20$: In case of the 'normal C' LAAPD the ratio is decreasing with increasing gain in contrast
to the $I_d/M$ behaviour of the 'low C' version diode where the ratio is slightly
increasing.\\ 
The values of the gain variation depending on bias voltage changes $1/M \cdot dM/dV $ were 
also determined. At gain $M=50$ a
value of \mbox{$1/M \cdot dM/dV = 3.4$\percent/V} has been evaluated for the 'normal C' version and
a similar value for the 'low C' version of \mbox{$1/M \cdot dM/dV = 3.2$\percent/V}. These values
are in good agreement with the value given for \INST{CMS}-APDs in
\cite{bib:emc:photo:APD:Renker_Calor2000}.

\subsubsection{Excess Noise Factor}
\label{sec:photo:APD:Char:ENF}
The APD internal charge carrier multiplication via avalanche is a
statistical process. A characterization of these statistical fluctuations of the APD
gain $\sigma_M$ is given by the excess noise factor $F$ and has its origin in
inhomogeneities in the avalanche region and in hole multiplication. 
The value of the excess noise factor is determined by the internal
structure of the APD and is related to the amplification of electrons and
holes at a given value of the internal gain $M$:
\begin{equation}
F \approx k\times M +\left(2-\frac{1}{M}\right) \times (1-k),
\label{eq:photo:APD:Char:ENF:EqNF} 
\end{equation}
with $k$ defined as the ratio of the ionization coefficients for electrons to
holes. Its value is determined by the shape of the electric field near
the p-n junction. Its r.m.s. broadening of a signal
from $N_{pe}$ photoelectrons is given by $\sqrt{F/N_{pe}}$.
If $N_\gamma$ photons per MeV are emitted by the crystal the number of photoelectrons created
inside the conversion layer of the diode with quantum efficiency QE can be calculated by
$N_{pe} = N_\gamma \cdot QE$ (assuming full coverage of the crystal rear side by the APD). 
Therefore a shower of energy E(MeV) creates $EN_{pe}$
photoelectrons.\\
The processes mentioned above result in an additional contribution to the energy resolution 
of \cite{bib:emc:photo:APD:CMS_CR97}
\begin{eqnarray}
\frac{\sigma_E}{E} &=& \frac{1}{\sqrt{EN_{pe}}} \cdot
\frac{\sqrt{M^2+{\sigma_{M}}^2}}{M} \nonumber \\ 
&=& \frac{1}{\sqrt{E}} \cdot \sqrt{\frac{F}{N_{pe}}}.
\label{eq:photo:APD:Char:ENF:sigma_ENF} 
\end{eqnarray}
From \Refeq{eq:photo:APD:Char:ENF:sigma_ENF} it can be clearly seen that the excess noise
factor should be as small as possible to maintain an excellent energy resolution.  
\begin{figure}
  \begin{center}
    \includegraphics[width=\swidth]{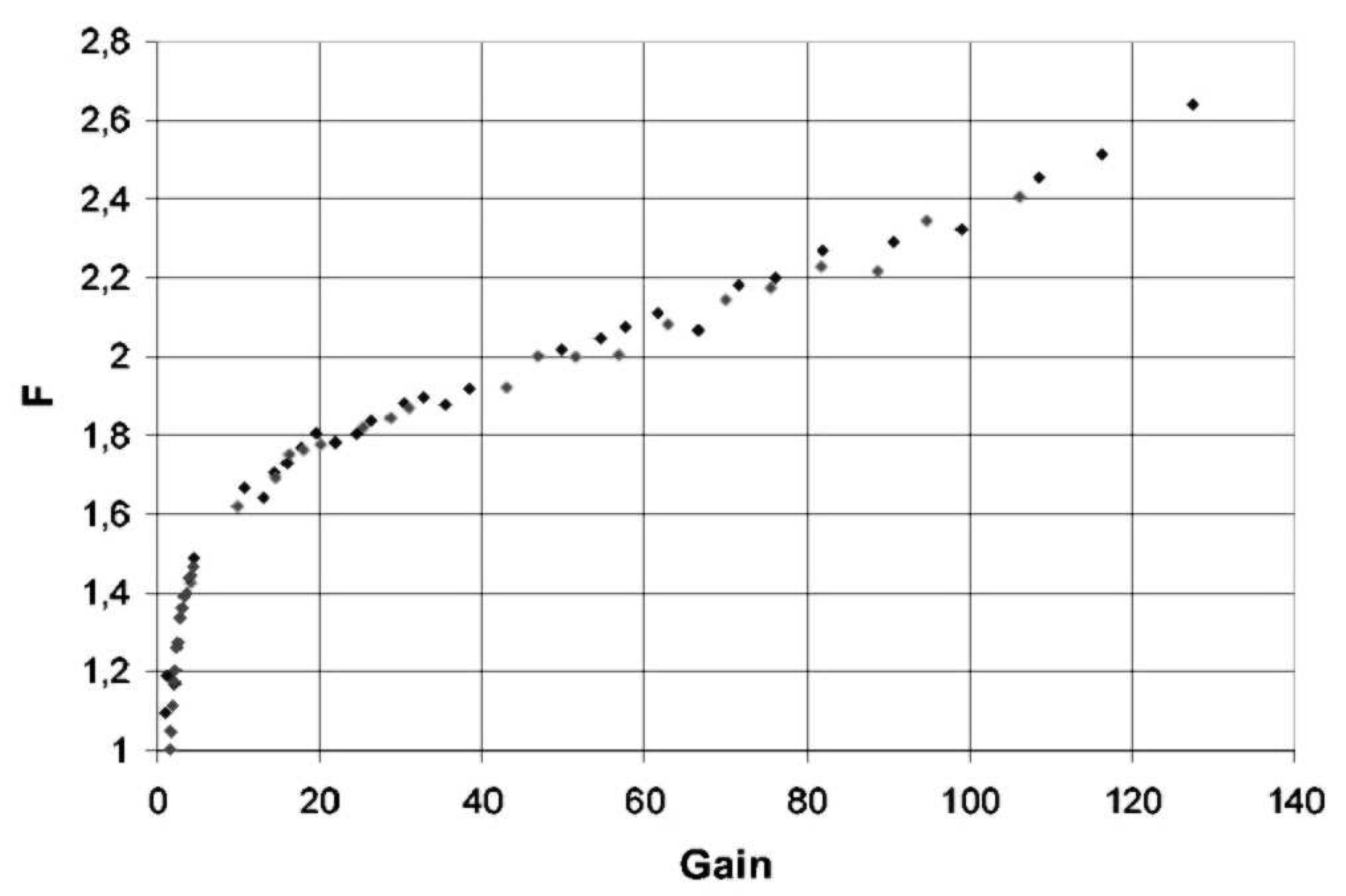}
    \caption[Excess noise factor of \INST{CMS}-APDs.]{Measured excess noise factor $F$ as a function of the internal gain
    of a \INST{CMS} APD \cite{bib:emc:photo:APD:Patel}.}
    \label{fig:photo:APD:Char:ENF:CMS_excess}
  \end{center}
\end{figure}
The value of the excess noise factor at an internal gain of $M\,=\,50$ of the 
\INST{CMS}-APDs is
$F\,=\,2.0$ (\Reffig{fig:photo:APD:Char:ENF:CMS_excess}), at gain $M\,=\,100$
the excess noise factor is increasing to $F=2.33$.
\begin{figure}
  \begin{center}
    \includegraphics[width=\swidth]{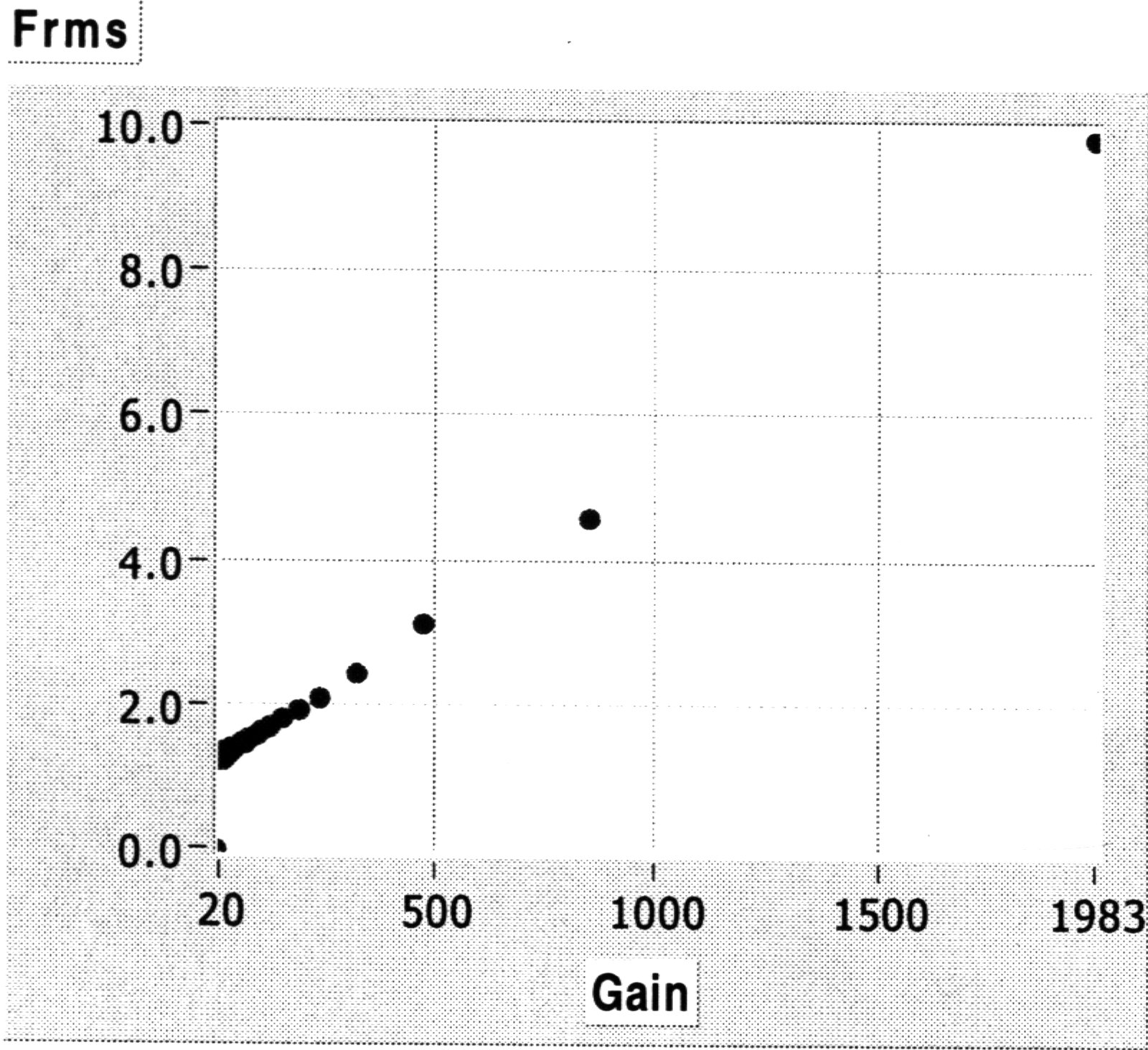}
    \caption[Excess noise factor of 'normal C' version LAAPD from Hamamatsu.]{Measured excess noise factor $F$
     as a function of the internal gain of the first prototype of the LAAPD S8664-1010SPL
     with 'normal C' manufactured by Hamamatsu.}
    \label{fig:photo:APD:Char:ENF:LAAPD_excess}
 \end{center}
\end{figure}
The excess noise factor of the first LAAPD 'normal C' prototype was measured at the
APD laboratory of the \INST{CMS} collaboration at \INST{CERN} in April 2004. The result of this
measurement is shown in \Reffig{fig:photo:APD:Char:ENF:LAAPD_excess}.
At an internal gain of $M=50$ the excess noise factor of these
LAAPDs has a value of $F=1.38$, at gain $M=100$ its value is $F=1.57$.

\subsubsection{Nuclear Counter Effect (NCE)}
\label{sec:photo:APD:Char:NCE}
The Nuclear Counter Effect (NCE) describes an extra amount of charge produced inside a photodiode by a charged particle 
directly hitting it, in addition to the charge produced by the scintillation light of the used crystal
\cite{bib:emc:photo:APD:BGO_NCE}.
Therefore the NCE can cause a decrease of resolution in the energy measurement of an electromagnetic shower due to 
the leakage (secondary produced electron/hole pairs) of charged particles from the crystal.\\ 
Charged particles crossing a silicon layer of given thickness create a number of electron/hole pairs calculated by:
\begin{equation}
\frac{dn}{dx} = \frac{dE}{dx} \cdot \varrho \cdot \frac{1}{E_{e/h}} \approx 100\,e/h\,pairs\,/\,\mu m.
\end{equation}
In an APD only the photoelectrons created in front of the avalanche region, inside the conversion layer, always experience 
full amplification. Are the electrons created inside the avalanche region their reachable amplification depends on their
place of creation (see details \Refsec{sec:photo:APD:irradiation}). 
To give a quantitative measurement of this effect, an effective thickness $d_{eff}$ of the APD can be defined.
Its value can be determined by exposing the APD directly to a $^{90}Sr$ source, emitting beta electrons up 
to an energy of $2\,\mev$. The charge Q collected in the APD, operating at gain $M$, will be compared to the value measured 
with a PIN diode of well-known thickness $d_{PIN}$ which results in the following relation \cite{bib:emc:photo:APD:CMS_CR97}:
\begin{equation}
d_{eff} = \frac{d_{PIN}}{Q(PIN)} \cdot \frac{Q(APD)}{M}.
\end{equation}
The calculation of the NCE includes the determination of the number of photoelectrons $N_{pe}$ detected by the 
crystal-APD system. Its value depends on the light yield (LY) of the crystal, the APD quantum efficiency QE and the
fraction of the crystal rear area covered by the APD (f). The asumption of a $22\,\percent$
coverage using a ($10 \times 10$)$\,\mm^2$ large APD mounted on a ($2.14 \times 2.14$)$\,\cm^2$
crystal endface ($f = 0.22$) results in:
\begin{eqnarray}
N_{pe}  =  LY \cdot QE \cdot f & \approx & 500 \times 0.7 \times 0.22 \nonumber \\ 
  & = & 77.0 \,\frac{pe}{\mev}
\end{eqnarray}
at an operation temperature of $T = -25\degC$.
Based on these values one MIP creates a signal inside the APD (operating at $T = -25\degC$) equivalent to the 
light signal released in the crystal by a photon of energy:
\begin{equation}
E_{MIP} = \frac{dn}{dx} \times \frac{d_{eff}}{N_{pe}} \approx 1.3\,\mev \times d_{eff}(\mu m).
\label{eq:photo:APD:Char:NCE:E_MIP} 
\end{equation}
From \Refeq{eq:photo:APD:Char:NCE:E_MIP} it can clearly be seen, that the
effective thickness of the diode should be as small as possible to minimize the
influence of the Nuclear Counter Effect (assuming $d_{eff}= 5.6\,\mu m$ (\INST{CMS}) $E_{MIP}$ would 
have a value of $7.28\,\mev$). 
\begin{figure}
  \begin{center}
    \includegraphics[width=\swidth]{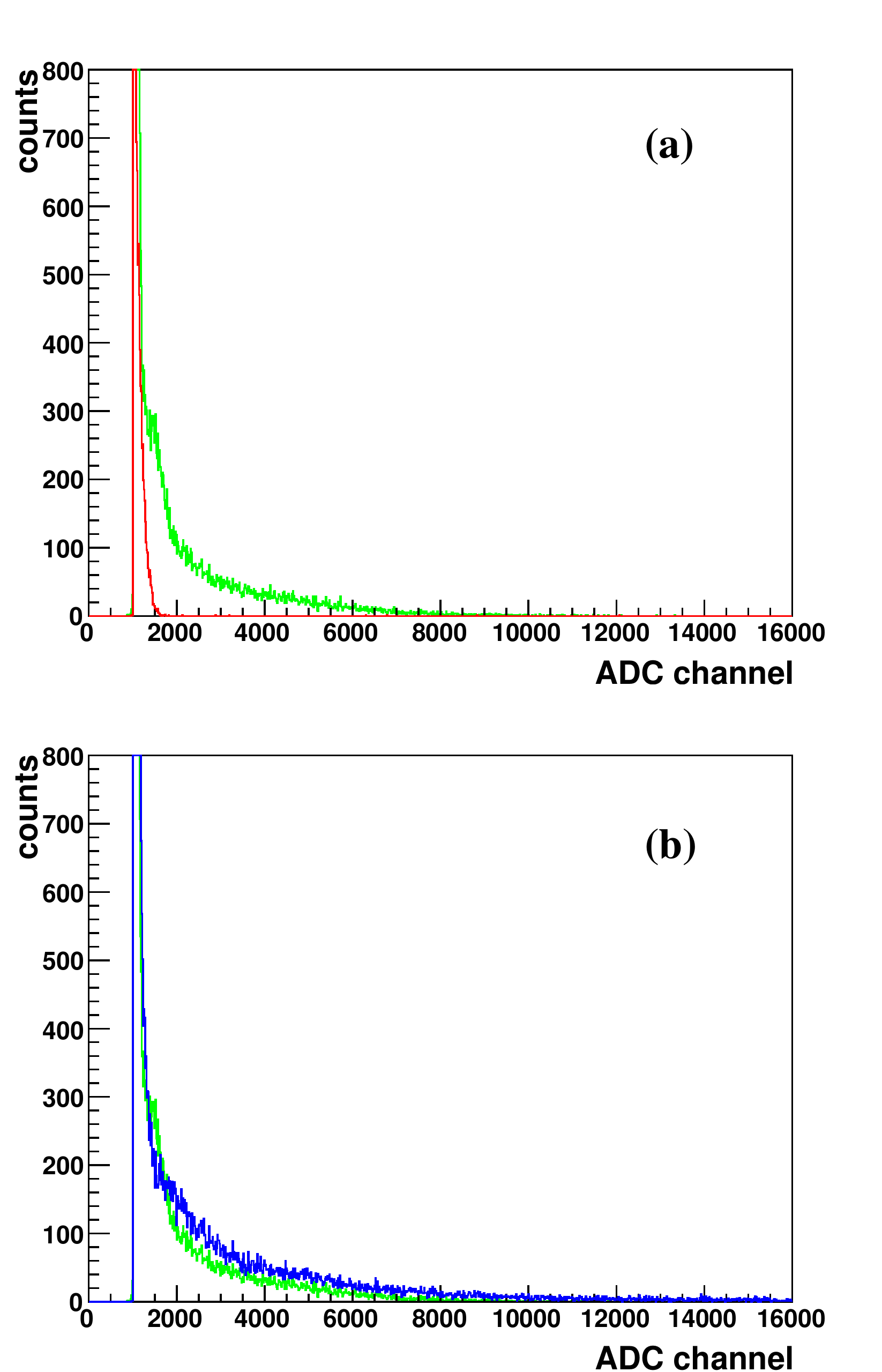}
    \caption[NCE of the investigated LAAPD versions.]{Measured Nuclear Counter Effect of a PIN
    diode of $300\,\mu m$ thickness (green), of
    the 'normal C' version LAAPD (red) and of the 'low C' version LAAPD (blue). In part a) the comparison between 
    PIN diode and 'normal C' version APD is shown and b) shows the corresponding
    comparison between PIN and the 'low C' version LAAPD.}
    \label{fig:photo:APD:Char:NCE:LAAPD_NCE}
 \end{center}
\end{figure}
On the other hand a reduction of
$d_{eff}$ increases the capacitance of the APD. Therefore a compromise has to be
found and in fact as shown in \Reffig{fig:photo:APD:Char:NCE:LAAPD_NCE} one of the tested APD 
types appears to fulfill these
requirements. While the exposure with electrons seems to have no measureable effect on the 'normal C' version
LAAPD (\Reffig{fig:photo:APD:Char:NCE:LAAPD_NCE}b), the 'low C' version shows a clear signal created by the 
electrons directly hitting the APD (\Reffig{fig:photo:APD:Char:NCE:LAAPD_NCE}c). The determination of the effective thickness of the
used LAAPDs is hindered compared to the measurement done by \INST{CMS} because of the much larger device capacitance. 
This value gives a major impact on the diodes noise behaviour and has to be known to conceive 
the layout of a low noise preamplifier which will be a major part of the APD readout chain.  
For \PANDA the measurement of $d_{eff}$ requires the minimization of any possible noise source and is currently under development.    

\subsubsection{Electrical and Optical Properties of the Tested APD Types}
\label{sec:photo:APD:Char:summprop}
\begin{figure}
  \begin{center}
    \includegraphics[width=\swidth]{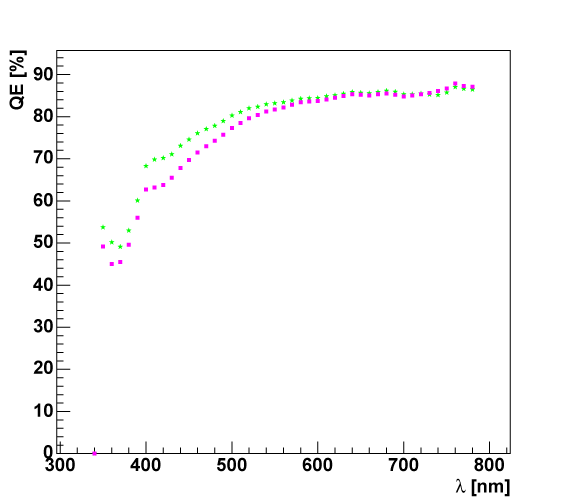}
    \caption[QE of a \PANDA-LAAPD.]{QE of a PIN diode (squares/pink) compared to the QE measured for a LAAPD
    (stars/green).}
    \label{fig:photo:APD:Char:summprop:QE}
 \end{center}
\end{figure}

The optical and electrical properties of the tested two LAAPDs mentioned above are 
summarized in \Reftbl{tab:photo:APD:Char:summprop:character} and compared to the APDs to be
used in the electromagnetic calorimeter ECAL of the \INST{CMS} experiment. Most of the measured
parameters of the LAAPDs are
in good agreement with the properties of the APDs developed by the \INST{CMS}
group in collaboration with Hamamatsu Photonics.
The measured quantum efficiency of the \PANDA-LAAPDs is shown in \Reffig{fig:photo:APD:Char:summprop:QE}
compared to values measured with a PIN diode.
Due to the four times larger active area of the LAAPD prototypes their
capacitance is much larger than those of the tested ($5 \times 5$)$\,\mm^2$
APD type of \INST{CMS}. 
\begin{table*}
\begin{center}
\begin{tabular}{llccc}
\hline\hline
Property & Condition & \INST{CMS} APD  & S8664-1010SPL & S8664-1010SPL \\
& & & 'normal C' & 'low C'\\
\hline
Active area [$\mm^2$]& & $5 \times 5$& $10 \times 10$ & $10 \times 10$\\
Quantum efficiency $QE$ [\percent]   & $M=1$,$\lambda=420\,\nm$ & 70 & 70 & 70\\
Breakdown voltage $U_{br}$ [$V$]& $I_d=100\,\mu A$ & 400 & 400 & 500-600 \\
Dark current $I_d$ [$nA$]& $M=50$& 5 & 10-40 & 20-50\\
Capacitance $C$ [$pF$]& $M=50$ & 80 & 270 & 180\\
Excess noise factor F& $M=50$& 2.0  & 1.38 & not measured\\
Excess noise factor F& $M=100$& 2.33 & 1.57 & not measured\\
\hline\hline
\end{tabular}
\caption[Properties of tested APDs.]{Electrical and optical properties of the tested APD types.}
\label{tab:photo:APD:Char:summprop:character}
\end{center}
\end{table*}
\subsection{Radiation Damage due to Different Kinds of Radiation}
\label{sec:photo:APD:irradiation}
To analyze the influence of radiation on a semiconductor device the knowledge of the internal device structure 
has to be assumed. Therefore several measurements to gain inside on the internal APD structure are inevitable.
The possibility of calculating the effective thickness $d_{eff}$ of the tested devices was already mentioned in 
\Refsec{sec:photo:APD:Char:NCE}. Another important parameter of the diode which was not discussed in detail so far and has
also influence on the NCE is the conversion 
layer thickness $d_{conv}$.
\begin{figure}[b]
  \begin{center}
   \includegraphics[width=\swidth,height=5.5cm]{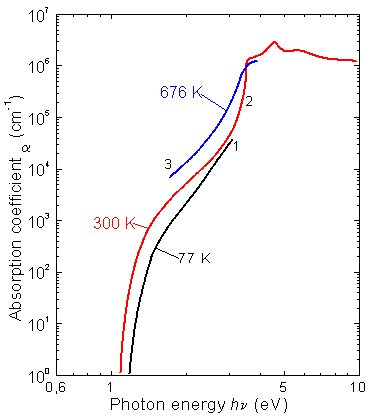}
    \caption[Absorption coefficient $\alpha(\lambda,T)$ of Si.]{Absorption coefficient $\alpha(\lambda,T)$ of Si 
    depending on the photon energy for different temperatures \cite{bib:emc:photo:APD:Sze}.}
    \label{fig:photo:APD:irradiation:alpha_Si}
 \end{center}
\end{figure}
In \Refsec{sec:photo:APD:Char:NCE} it was already mentioned that the longitudinal position $x$ where the electron
hole creation inside the diode takes place has influence on the reachable amplification of the photoelectrons passing the
device structure: For light absorbed before reaching the avalanche region the gain is constant. The absorption of light
inside the avalanche region leads to a decrease of the internal gain and therefore to additional gain fluctuations.
Assuming a constant electric field in the avalanche region of width $W$, the gain $M$ could be calculated by
using the relation
\begin{equation}
M = e^{\alpha_{i} \cdot (W - x)}
\end{equation}
in which $\alpha_{i}$ defines the ionization coefficient.
Additionally the light absorption inside the avalanche region follows the exponential law 
\begin{equation}
N(x) = N_0 \cdot e^{-(\alpha(\lambda,T) \cdot x)}.
\end{equation} 
\begin{figure}[b]
  \begin{center}
   \includegraphics[angle = 90,width=\swidth,height=5.5cm]{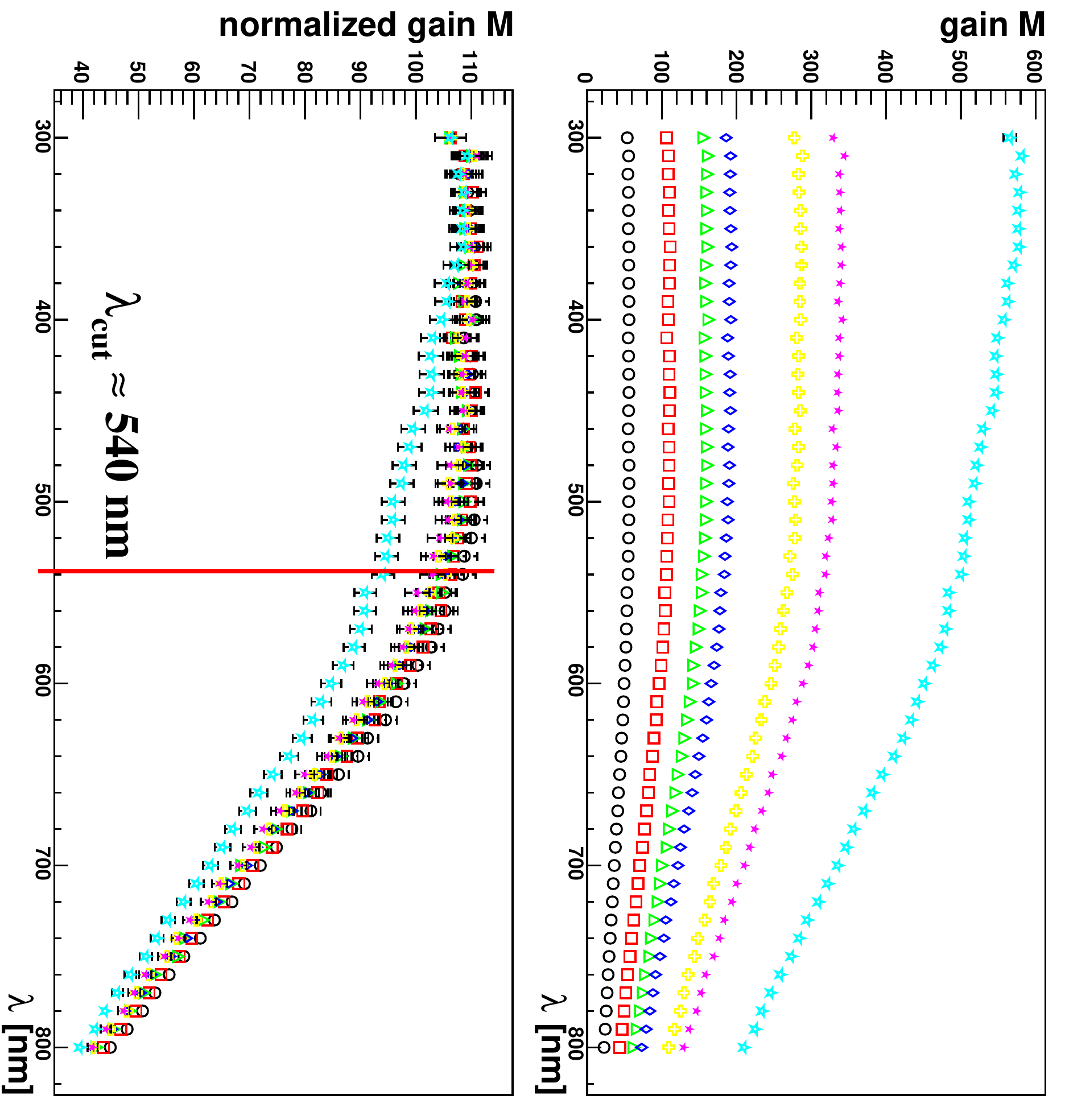}
    \caption[Normalized gain of 'normal C' type APD.]{$M(\lambda)$ for the 'normal C' version APD
    for different $U_R$ values (above). Determination of $\lambda_{cut}$ by normalizing the data (below).}
    \label{fig:photo:APD:irradiation:normgain_normalC}
 \end{center}
\end{figure}
\begin{figure}
  \begin{center}
   \includegraphics[angle = 90,width=\swidth,height=5.5cm]{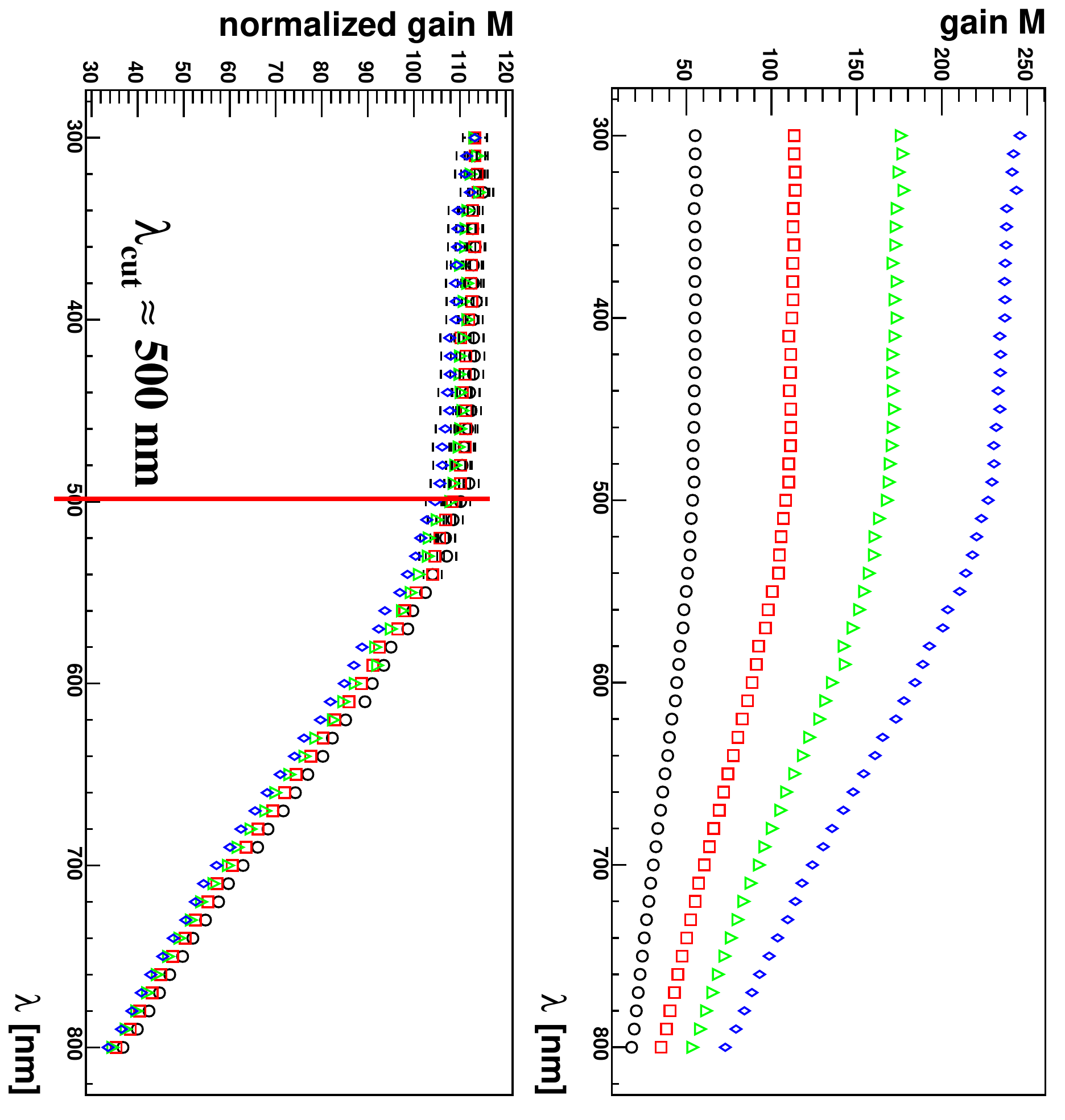}
    \caption[Normalized gain of 'low C' type APD.]{$M(\lambda)$ for the 'low C' version APD
    for different bias voltages (above). Determination of $\lambda_{cut}$ by normalizing the data (below).}
    \label{fig:photo:APD:irradiation:normgain_lowC}
 \end{center}
\end{figure}
The absorption coeffcient $\alpha(\lambda,T)$ respectively $\alpha(E_\gamma,T)$ of silicon is shown in 
\Reffig{fig:photo:APD:irradiation:alpha_Si} depending on the incident photon energy. 
The conversion layer thickness of the APD could be determined by applying the definition of the optical mean depth of 
penetration \cite{bib:emc:photo:APD:Kneifel}:
\begin{equation}
d_{conv} = \frac{1}{\alpha(\lambda,T)}.
\end{equation}
By normalizing the measured gain values depending on wavelength, as shown in 
the lower plot of \Reffig{fig:photo:APD:irradiation:normgain_normalC} for the 'normal C' type APD 
(\Reffig{fig:photo:APD:irradiation:normgain_lowC} for the 'low C' type respectively), it is possible to define a 'cut-off'
wavelength $\lambda_{cut}$ above which light is absorbed inside the avalanche region \cite{bib:emc:photo:APD:Kirn}.
The evaluated conversion layer thicknesses of the two different APD types and their
corresponding 'cut-off' wavelenghts are assorted in \Reftbl{tab:photo:APD:irradiation:d_conv} 
using the $\alpha(\lambda)$-values given for a temperature of $300\,K$. 
\begin{table}
\begin{center}
\begin{tabular}[width=\swidth]{|l|c|c|c|}
\hline
 & wavelength &$\alpha(\lambda)$ & $d_{conv}$ \\
APD version & $\lambda_{cut}$ [$nm$] &  [$1/cm$] & [$\mu m$]\\
\hline
'low C' type & 500 & $1.1 \cdot 10^4 $& 0.9\\
\hline
'normal C' type & 540 &$ 7.05 \cdot 10^3$ & 1.4 \\
\hline
\end{tabular}
\caption[Conversion layer thicknesses of the tested APD types.]{Determined conversion layer thicknesses $d_{conv}$ of the tested
APD types.}
\label{tab:photo:APD:irradiation:d_conv}
\end{center}
\end{table}
It is quite obvious that any kind of irradiation of the APD has an impact on the $M(\lambda)$ behaviour of the device.
Therefore measurements as e.g. shown in \Reffig{fig:photo:APD:irradiation:normgain_normalC} will lead to a more consolidated
understanding of the caused radiation damages and their locations inside the APD structure.
\par
The results of different irradiation experiments will be reported in the following section. All irradiations have been done
at $T = -25\,\degC$ to account for the real operation conditions of the \Panda electromagnetic calorimeter.

\subsubsection{Proton Irradiation}
\label{sec:photo:APD:irradiation:proton}
Irradiation of an APD with protons causes two different kinds of
radiation damage:
\begin{itemize}
\item ionization effects at the surface reflected in the surface current $I_s$,
and 
\item atom displacements inside the silicon bulk increasing the bulk current $I_b$.
\end{itemize}
The proton irradiation of the diodes has been done at the \INST{KVI} Groningen (the Netherlands) using a $90\,\mev$ proton beam 
providing a homogenous beam profile. The diodes were irradiated until an integrated fluence of $1.1 \cdot 10^{13}\,p$ was reached
which is comparable with the dose expected for 10 years of \Panda operation. Using the NIEL theory this proton fluence
corresponds to an equivalent neutron fluence of $1.42 \cdot 10^{13}\,n$ of an energy of $1\,\mev$
\cite{bib:emc:photo:APD:Mol}. 
\par
\begin{figure}
  \begin{center}
   \includegraphics[angle=90,width=\swidth,height=5.5cm]{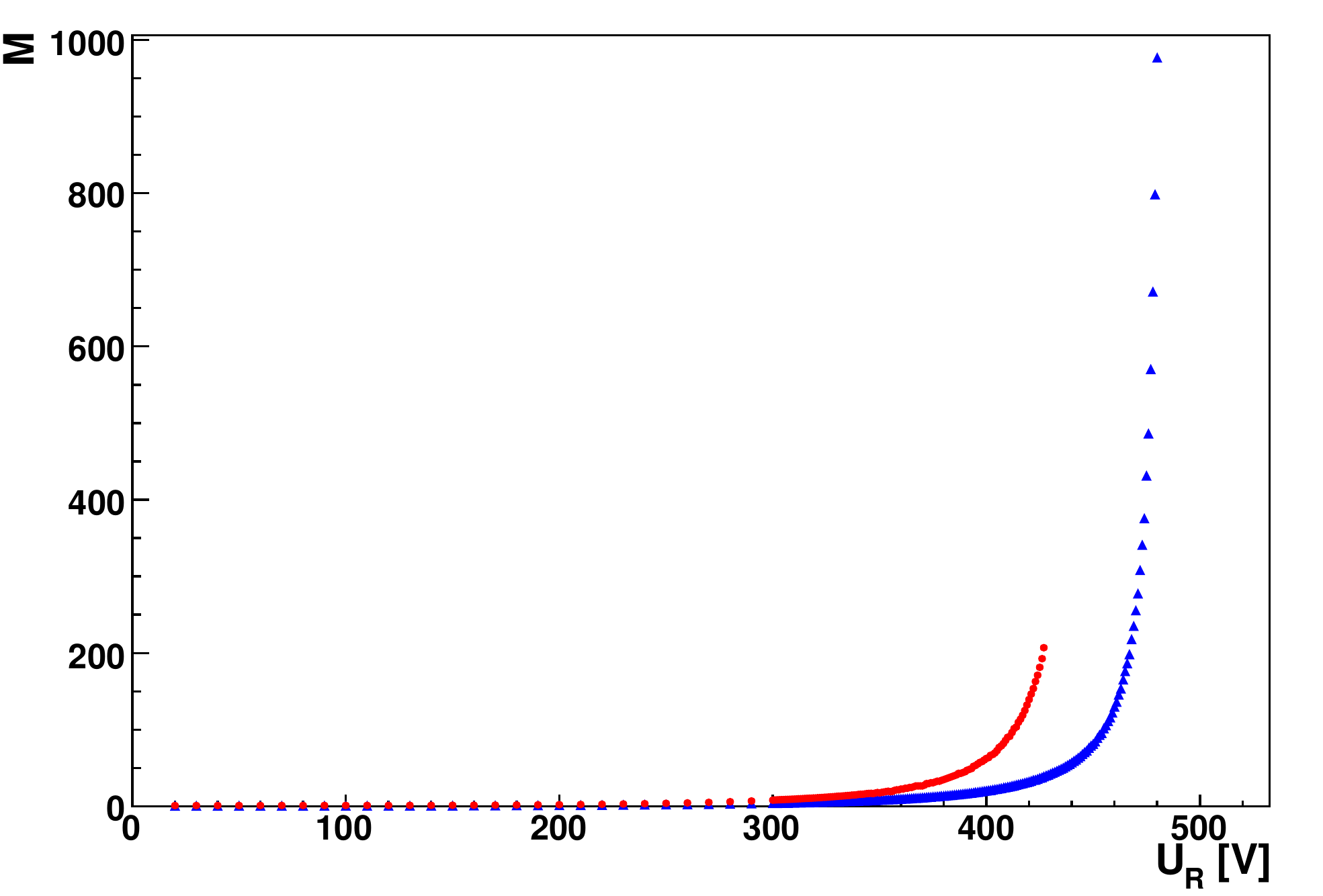}
    \caption[Gain $M$ of 'normal C' type after and before protonirradiation.]{Determined gain values $M$ of the 
    \mbox{'normal C'} type APD
    before (blue) and after (red) exposure with $1.1 \cdot 10^{13}\,p$ of $E_p = 90\,\mev$.}
    \label{fig:photo:APD:irradiation:proton:normal_M_change_KVI}
 \end{center}
\end{figure}
\begin{figure}
  \begin{center}
  \includegraphics[angle=90,width=\swidth,height=5.5cm]{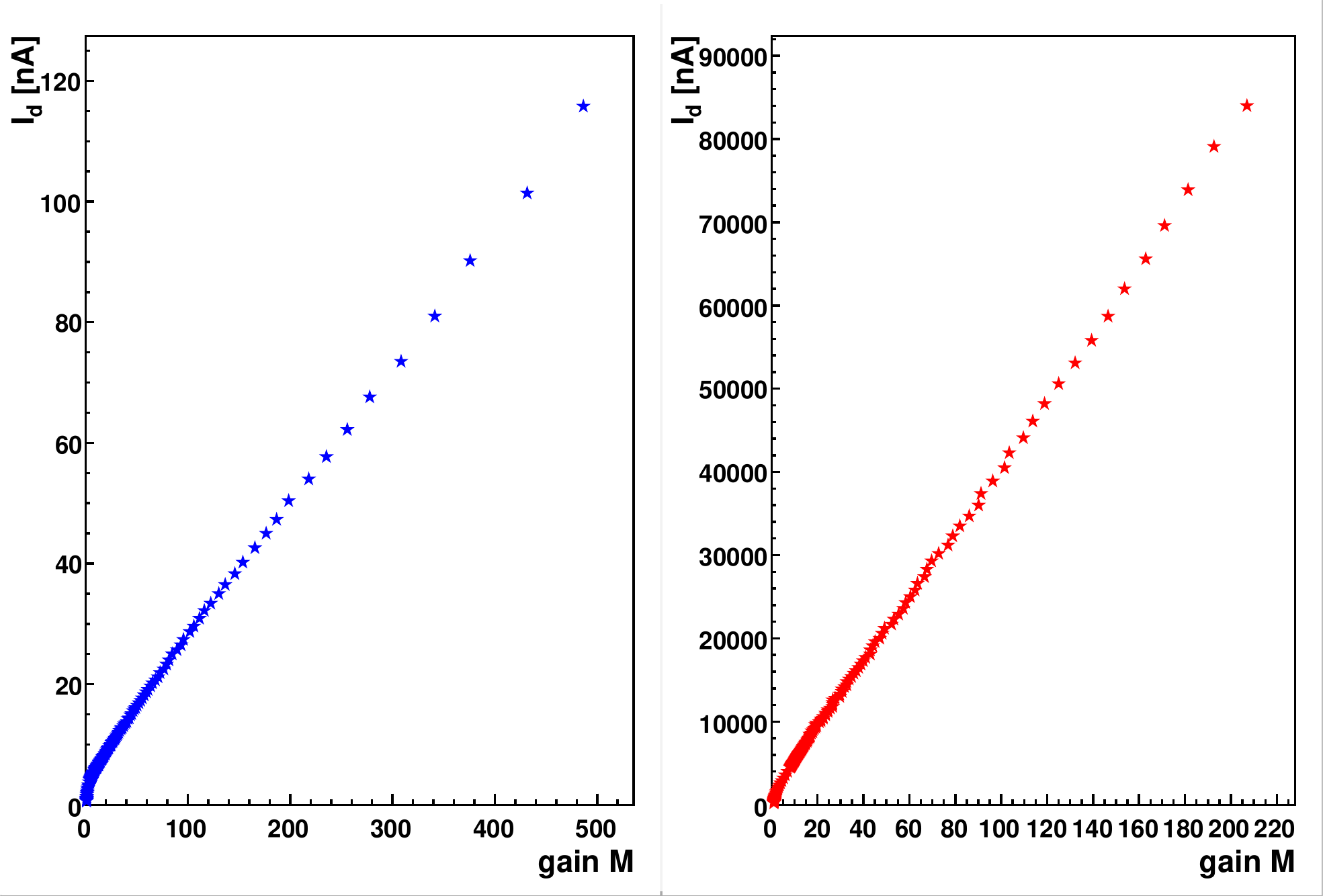}
    \caption[Dark current rise of 'normal C'diode caused by proton irradiation.]{Comparison of the measured dark current
    values of the 'normal C' type APD before (left/blue) and after (right/red) proton irradiation.}
    \label{fig:photo:APD:irradiation:proton:normal_Id_M_change_KVI}
 \end{center}
\end{figure}
\begin{figure}
  \begin{center}
  \includegraphics[angle=90,width=\swidth,height=5.5cm]{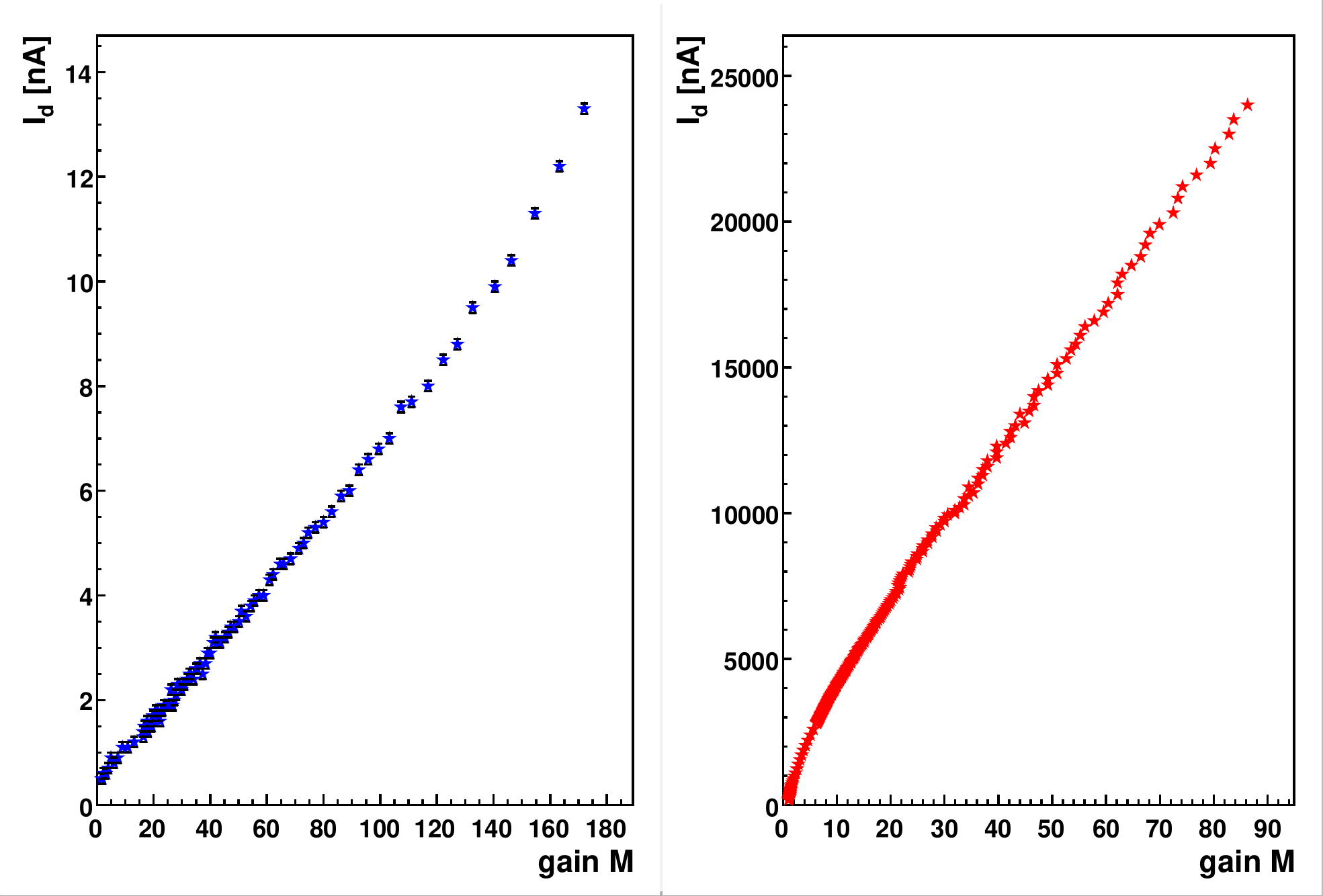}
    \caption[Dark current increase of 'low C'diode caused by proton irradiation.]{Comparison of the measured dark current
    values of the 'low C' type APD before (left/blue) and after (right/red) proton irradiation.}
    \label{fig:photo:APD:irradiation:proton:low_Id_M_change_KVI}
 \end{center}
\end{figure}
\begin{figure}
  \begin{center}
  \includegraphics[angle=90,width=\swidth,height=4.0cm]{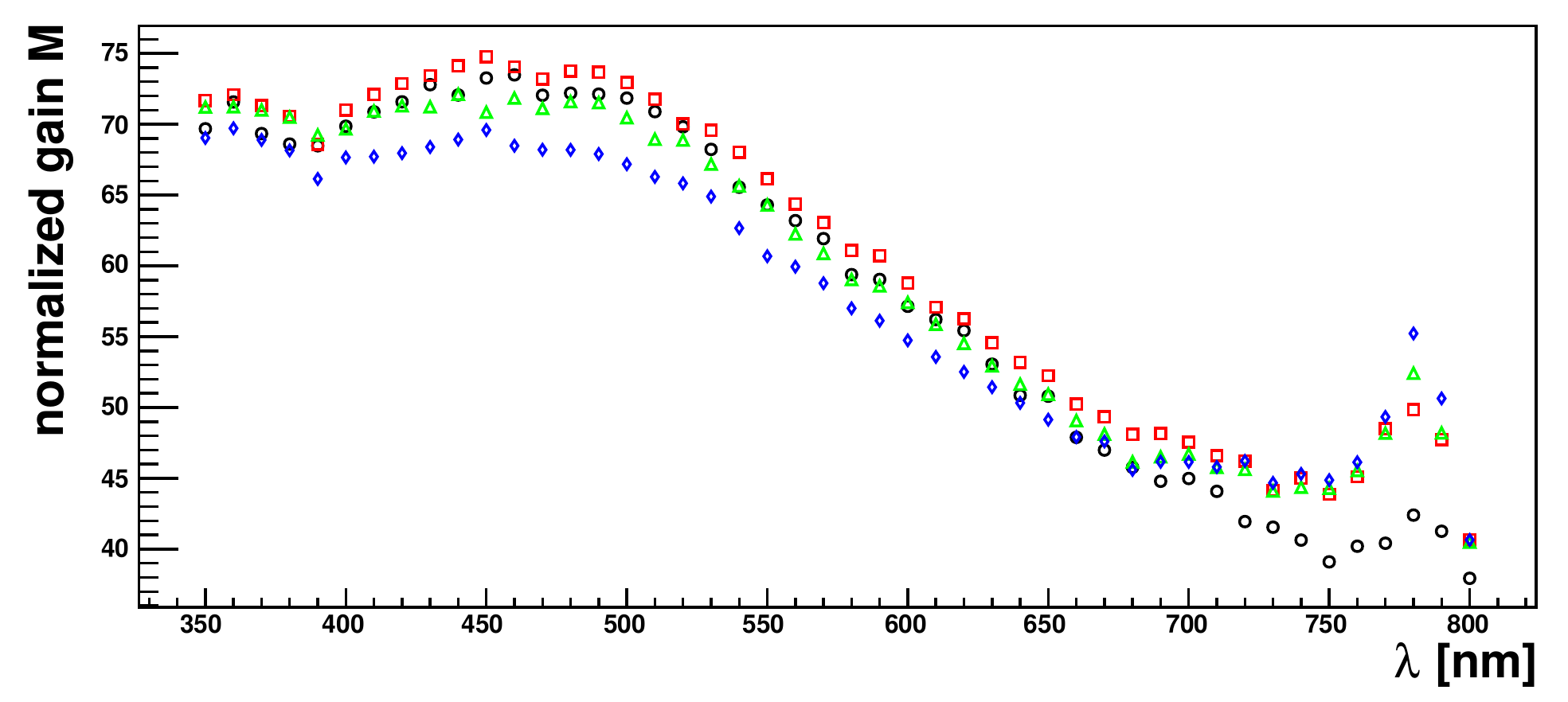}
    \caption[$M(\lambda)$ of 'normal C' type after proton exposure.]{$M(\lambda)$ of 'normal C' type after 
    proton exposure with $1.1 \cdot 10^{13}\,p$ at \INST{KVI} Groningen.}
    \label{fig:photo:APD:irradiation:proton:normalC_Mlambda_afterp}
 \end{center}
\end{figure}
\begin{figure}
  \begin{center}
  \includegraphics[angle=90,width=\swidth,height=4.0cm]{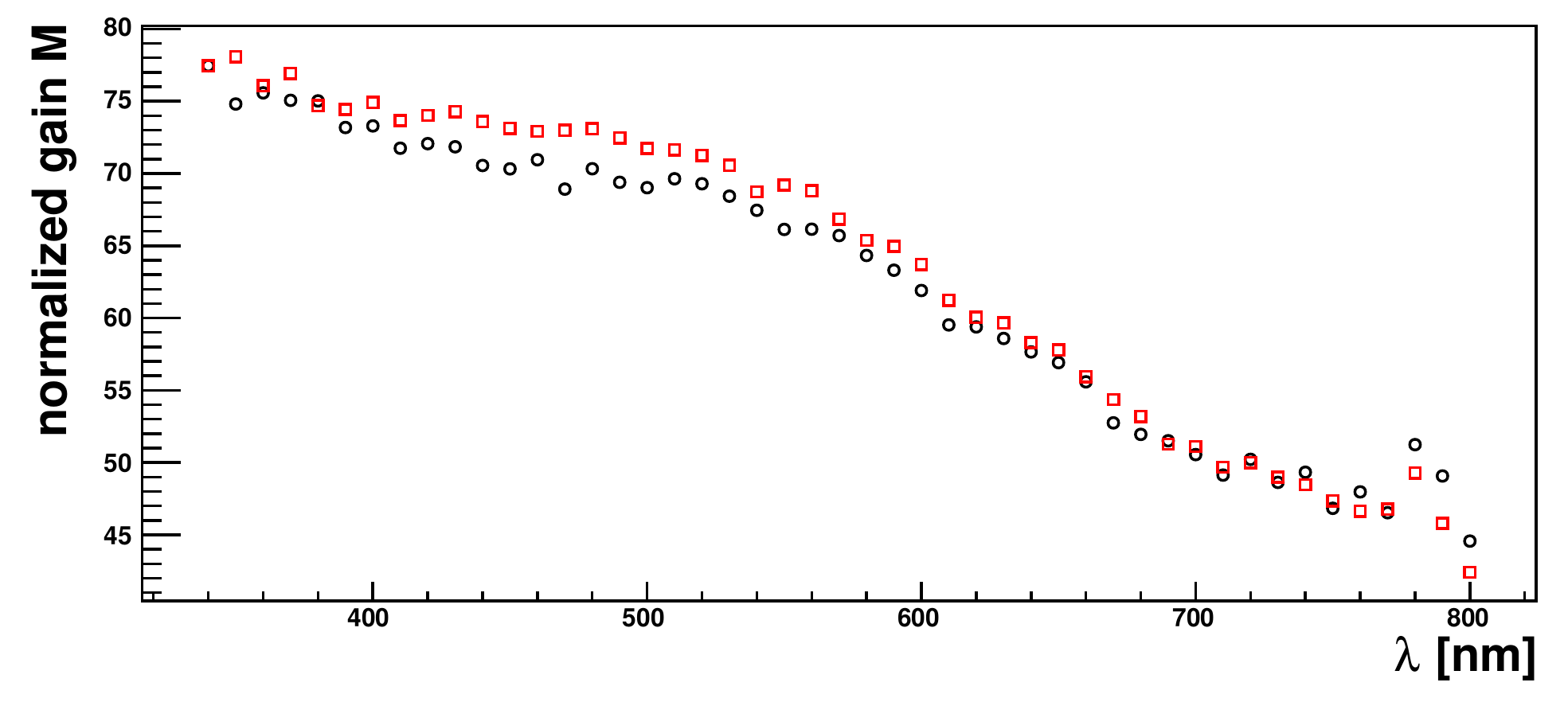}
    \caption[$M(\lambda)$ of 'low C' type after proton exposure.]{$M(\lambda)$ of 'low C' type after 
    proton exposure with $1.1 \cdot 10^{13}\,p$ at \INST{KVI} Groningen.}
    \label{fig:photo:APD:irradiation:proton:lowC_Mlambda_afterp}
 \end{center}
\end{figure}

The first step in evaluating the influence of irradiation on the APD characteristics is the remeasurement of the
internal gain of the diode. The typical change of $M(U_R)$ due to proton irradiation is shown for instance in 
\Reffig{fig:photo:APD:irradiation:proton:normal_M_change_KVI} for the 'normal C'
version. It can be clearly seen that the gain of this diode type has dramatically decreased. Taking into account the
assumed linearity of the dark current depending on $M$ (see \Reffig{fig:photo:APD:irradiation:proton:normal_Id_M_change_KVI}), 
the usable maximum gain decrease of the 'normal C' type APD 
after proton exposure is in the order of 60\%. The decrease in case of the 'low C' version diode (see x-axes of 
\Reffig{fig:photo:APD:irradiation:proton:low_Id_M_change_KVI}) has a similar value about 50-60\%. 
The influence of proton irradiation on the internal APD structure and the associated characteristic changes could be 
concretized by having a closer look on the dark current behaviour. For the two tested APD versions the dark current was
measured in dependence of the internal gain before and after proton exposure. The results are shown in 
\Reffig{fig:photo:APD:irradiation:proton:normal_Id_M_change_KVI} and
\Reffig{fig:photo:APD:irradiation:proton:low_Id_M_change_KVI}.
As already mentioned above this kind of measurement makes it possible to evaluate the different 
rates of the surface current and the bulk current to the overall dark current.
\begin{table*}
\begin{center}
\begin{tabular}[width=\dwidth]{|l|c|c|c|c|c|c|}
\hline
APD version & $I_{b1}$ [nA]& $I_{b2}$ [nA]& $I_{s1}$ [nA]& $I_{s2}$ [nA] & $I_{b2}/I_{b1}$ & $I_{s2}/I_{s1}$\\
\hline
'low C' type & $ 0.07$ & $ 268$ & $ 0.2$ & $\approx 1554$ & $\approx 3829$ & $\approx 7770$\\
\hline
'normal C' type & $ 0.24$ & $ 389$ & $ 3.47$ & $\approx 1770$ & $\approx 1621$ & $\approx 510$\\
\hline
\end{tabular}
\caption[Dark current parameters after proton irradiation of the tested APDs.]{Influence of proton irradiation on the APD dark current. The
values $I_{s1}$ and $I_{b1}$ have been measured before irradiation, $I_{s2}$ and $I_{b2}$ afterwards.}
\label{tab:photo:APD:irradiation:proton:Dark_currents}
\end{center}
\end{table*}
The values of the single dark current contributions shown in \Reftbl{tab:photo:APD:irradiation:proton:Dark_currents} were
evaluated by fitting \Refeq{eq:photo:APD:Char:Dark:I_d} to the datapoints in 
\Reffig{fig:photo:APD:irradiation:proton:normal_Id_M_change_KVI} and
\Reffig{fig:photo:APD:irradiation:proton:low_Id_M_change_KVI} respectively.
The enormous increase of the 'low C' bulk current compared to the increase factor $I_{b2}/I_{b1}$ measured for the 'normal C' type APD has
its origin in the increased layer thickness of this diode type used for the required reduction of the device capacitance.\\  
To get an impression where these observed proton induced damages are mainly located the measurement of the gain dependence on
wavelength has to be reconsidered after proton exposure for the two different kinds of tested avalanche photodiodes.
The results of this measurement are shown in \Reffig{fig:photo:APD:irradiation:proton:normalC_Mlambda_afterp} and 
\Reffig{fig:photo:APD:irradiation:proton:lowC_Mlambda_afterp}.
Obviously in case of the 'normal C' type APD the proton induced damages are mainly located inside the conversion layer 
and at the end of the avalanche region. To clarify the corresponding situation in case of the 'low C' version APD
more APDs of that type have to be irradiated. Nevertheless \Reffig{fig:photo:APD:irradiation:proton:lowC_Mlambda_afterp}
provides an indication of the locations of the damages similar to those found in case of the normal capacitance APDs.

\subsubsection{Photon Irradiation}
\label{sec:photo:APD:irradiation:photon}
The $\gamma$-irradiation of the diodes took place at the \INST{Strahlenzentrum} Giessen, Germany. The APDs were irradiated 
using $^{60}Co$ sources of different dose rates to reach the envisaged integrated dose of $10^{12}\,\gamma$s per APD. To determine the influence of $\gamma$ exposure on the APD characteristics
basically the same procedure was used as described in the former paragraph in case of proton exposure. The
irradiation has been accomplished for the 'normal C' version only, due to the fact that only 5 pieces of the 'low C' type APDs 
were available at the specific date of measurement. Similar to the proton irradiation done at \INST{KVI} the photon exposure of the
diodes has been done at $T = -25\,\degC$. 
\begin{figure}
  \begin{center}
   \includegraphics[angle=90,width=\swidth,height=5.5cm]{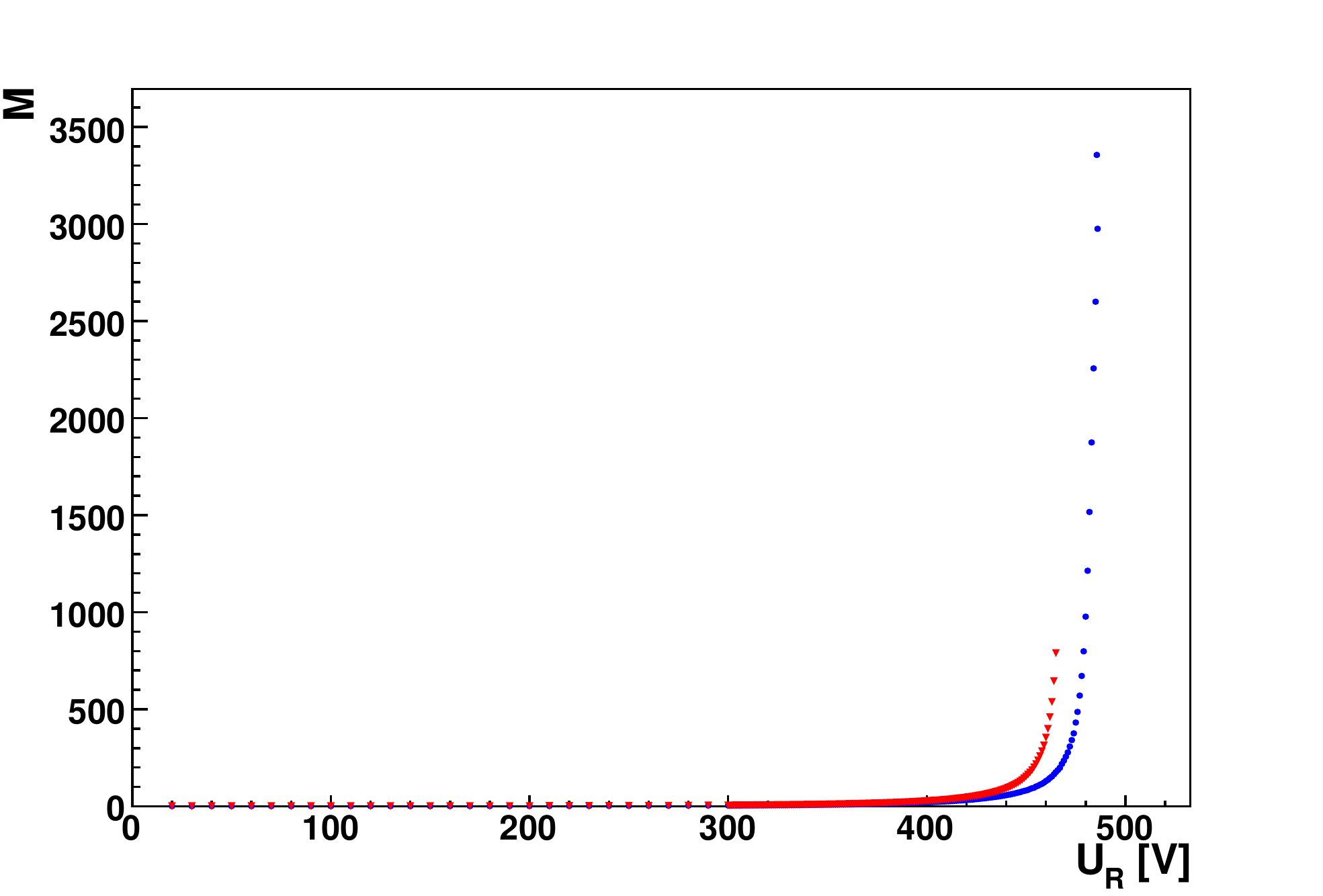}
    \caption[Gain $M$ of 'normal C' type after and before photon irradiation.]{Determined gain values $M$ of the 
    \mbox{'normal C'} type APD
    before (blue) and after (red) exposure with a $^{60}Co$ source.}
    \label{fig:photo:APD:irradiation:photon:normal_M_change_Giessen}
 \end{center}
\end{figure}
The remeasurements of the APD characteristics have been done after a storage of the diodes 
at room temperature in the order of one month. 
Similar to the results presented for the proton irradiation mesurement the gain values after $\gamma$ exposure are
compared to those measured using an unirradiated diode of the same internal structure 
(\Reffig{fig:photo:APD:irradiation:photon:normal_M_change_Giessen}). Taking into account the determination of the
maximum usable gain of the 'normal C' type APD no obvious gain decrease could be observed 
(see x-axes of \Reffig{fig:photo:APD:irradiation:photon:normal_Id_M_change_Giessen}), which is in contrast to the
dramatical gain decrease of the diode in terms of proton exposure. Comparing the measured dark current after $\gamma$ irradiation and one
month of storage with
the values determined for an unirradiated diode (\Reffig{fig:photo:APD:irradiation:photon:normal_Id_M_change_Giessen})
a nearly complete annealing of the dark current could be observed (see y-axes values). Looking at the gain range between 180 and 300
the dark current behaviour is dramatically changing so that the required linearity of the dark current depending on the internal 
gain seems to be no longer ensured. Detailed studies have to be done in the near future to determine the influence of surface 
defects caused by photon irradiation in this gain region.
\begin{figure}
  \begin{center}
  \includegraphics[angle=90,width=\swidth,height=5.5cm]{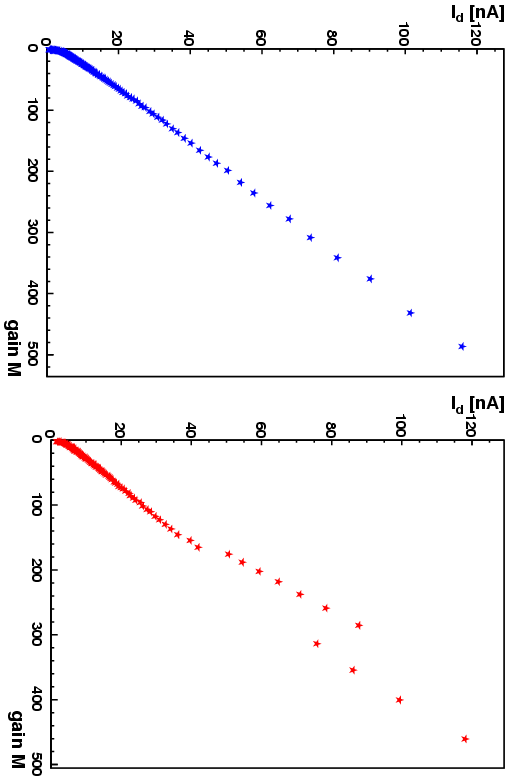}
    \caption[Dark current $I_d$ of 'normal C' type after and before photon irradiation.]{Dark current dependence on
    gain of the \mbox{'normal C'} type APD
    before (blue/left) and after (red/right) exposure with $10^{12}\,\gamma$s.}
    \label{fig:photo:APD:irradiation:photon:normal_Id_M_change_Giessen}
 \end{center}
\end{figure}
\begin{figure}
  \begin{center}
  \includegraphics[angle=90,width=\swidth,height=4.0cm]{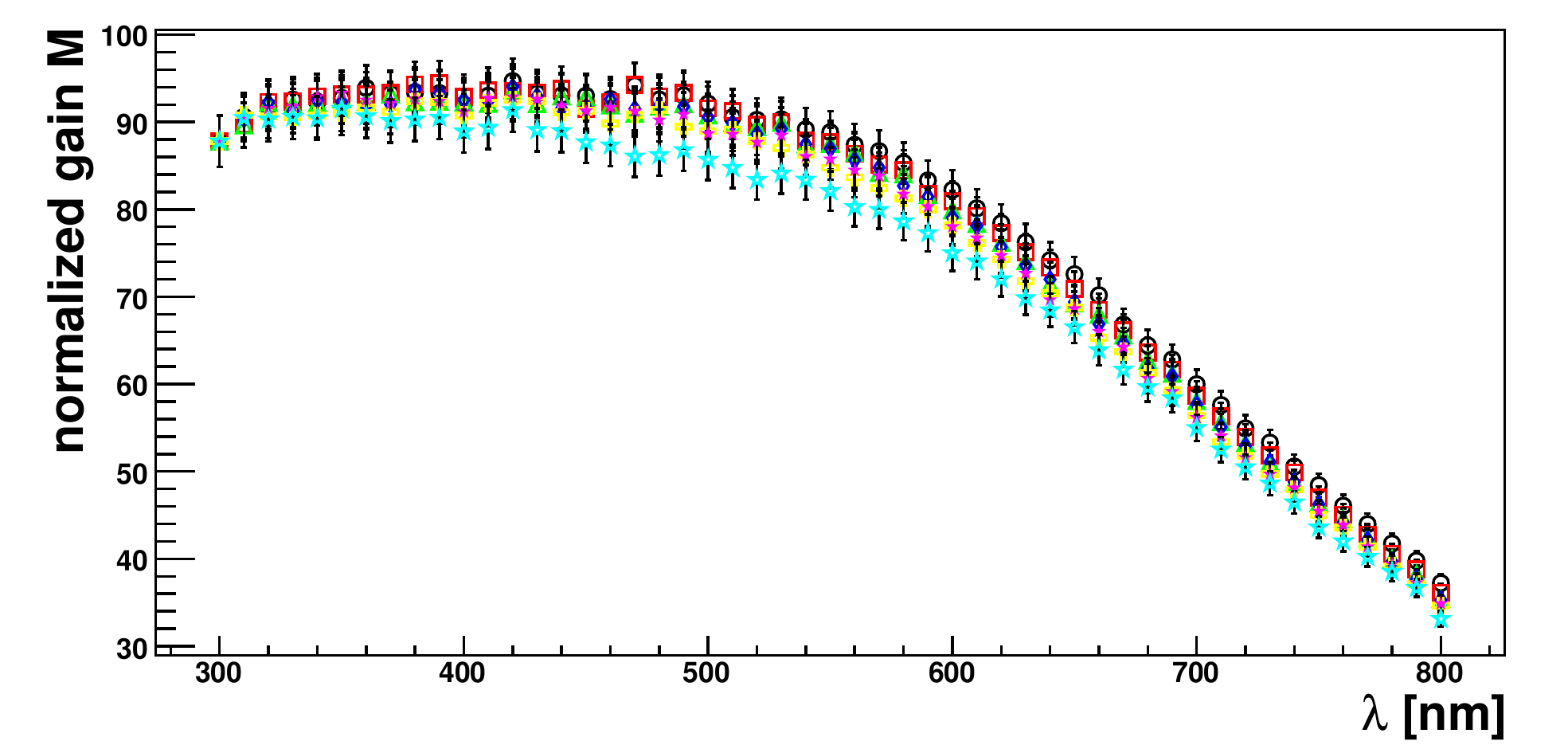}
    \caption[$M(\lambda)$ of 'normal C' type after $\gamma$ exposure.]{$M(\lambda)$ of 'normal C' type after 
    photon exposure with an integrated dose of $10^{12}\,\gamma$s at \INST{Strahlenzentrum} Giessen.}
    \label{fig:photo:APD:irradiation:photon:normalC_Mlambda_aftergamma}
 \end{center}
\end{figure}
On closer examination of \Reffig{fig:photo:APD:irradiation:photon:normalC_Mlambda_aftergamma} no indication for any
kind of bulk damage induced by the accomplished photon exposure could be observed. This result suggests some kind of
surface damage to be cause of the atypical behaviour of the dark current shown in 
\Reffig{fig:photo:APD:irradiation:photon:normal_Id_M_change_Giessen}. 

\subsubsection{Neutron Irradiation}
\label{sec:photo:APD:irradiation:neutron}
\begin{figure}
  \begin{center}
  \includegraphics[width=\swidth,height=6.0cm]{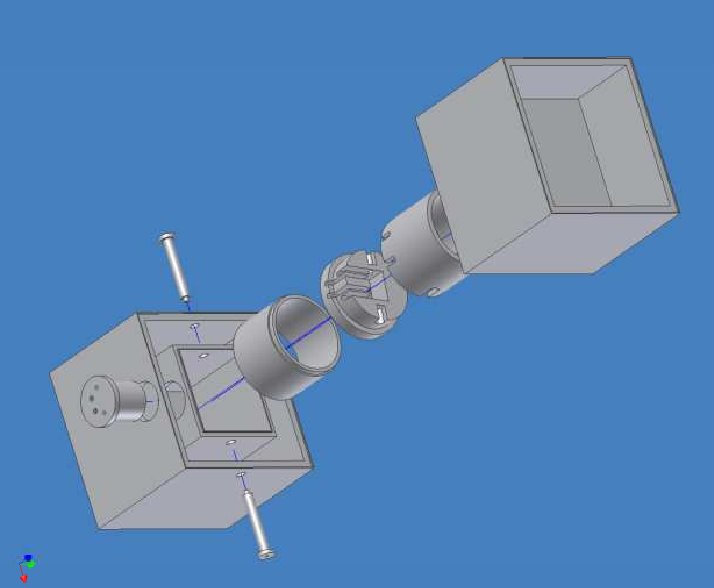}
    \caption[Apparatus used for neutron irradiation.]{Schematic of the apparatus used for neutron irradiation
    of the avalanche diodes including APD housing und holder for the used neutron source.}
    \label{fig:photo:APD:irradiation:neutron:Noras_kiste}
 \end{center}
\end{figure}

The neutron irradiation of the quadratic shaped APDs which has been done until now 
was executed at the \INST{APDlab} Frankfurt using
a $^{241}Am-\alpha-Be$ source. This source generates approximately \mbox{$0.65 \cdot 10^{-4}$} neutrons per
emitted $\alpha$ particle of the alpha decay of americium via the reaction 
\begin{equation}
^4He + ^9Be \rightarrow ^{12}C + n.
\end{equation}
Taking into account the activity of the
employed source ($1.1\,GBq$) a neutron rate of \mbox{$7.15 \cdot 10^4 n/s$} could be reached.\\
In contrast to the exposure procedures described in the former paragraphs the irradiation results
described in this section are based on measurements done at $T = 20\degC$.
The measuring equipment providing a save handling of the neutron source and which has to be completely 
filled with boron silicate due to radiation protection reasons is shown in 
\Reffig{fig:photo:APD:irradiation:neutron:Noras_kiste}.
\begin{figure}
  \begin{center}
  \includegraphics[angle=90,width=\swidth,height=5.5cm]{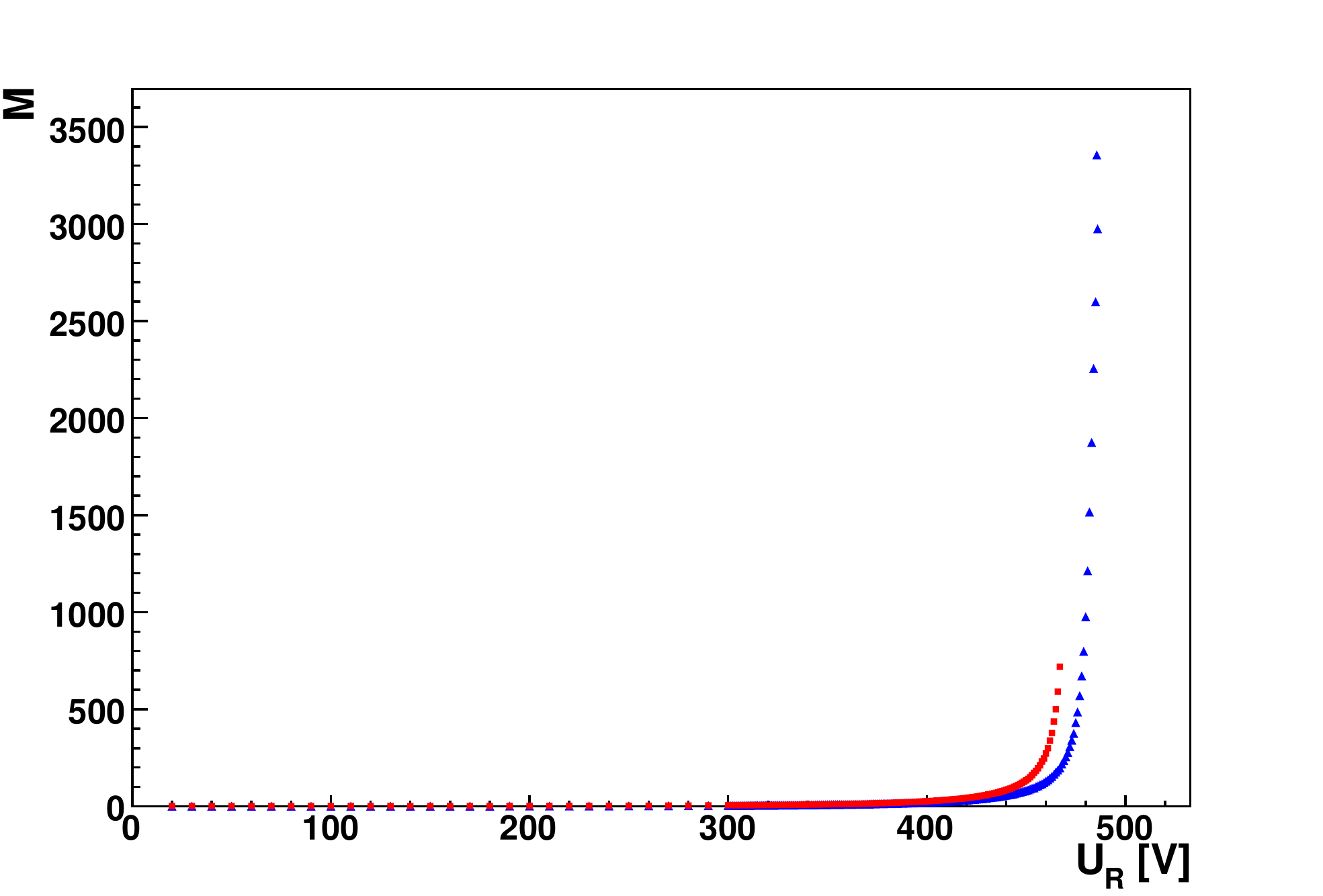}
    \caption[Gain $M$ of 'normal C' type after and before neutron irradiation.]{Determined gain values $M$ of the 
    \mbox{'normal C'} type APD
    before (blue) and after (red) exposure using a $^{241}AmBe$ source.}
    \label{fig:photo:APD:irradiation:neutron:normal_M_change_Neutrons}
 \end{center}
\end{figure}
As already discussed in cases of proton and photon exposure of the diodes the gain of the two tested 
APD types has decreased after irradiation. The curve progression of the measured gain dependence on the
applied bias voltage is shown exemplarily for the 'normal C' type in \Reffig{fig:photo:APD:irradiation:neutron:normal_M_change_Neutrons} in
comparison to the behaviour observed before exposure.
The dark current dependence on the determined gain values is shown for both APD types in 
\Reffig{fig:photo:APD:irradiation:neutron:normal_Id_M_change_Neutron} and in 
\Reffig{fig:photo:APD:irradiation:neutron:low_Id_M_change_Neutron} respectively. In case of the 'normal
C' type the dark current shows nearly no increase and its linearity depending on $M$ is still ensured.
The converse situation could be observed in case of the 'low C' type APD where the maximum useable gain
has dramatically decreased to a value 50\% below the determined one for the unirradiated device. 
\begin{figure}
  \begin{center}
  \includegraphics[angle=90,width=\swidth,height=5.5cm]{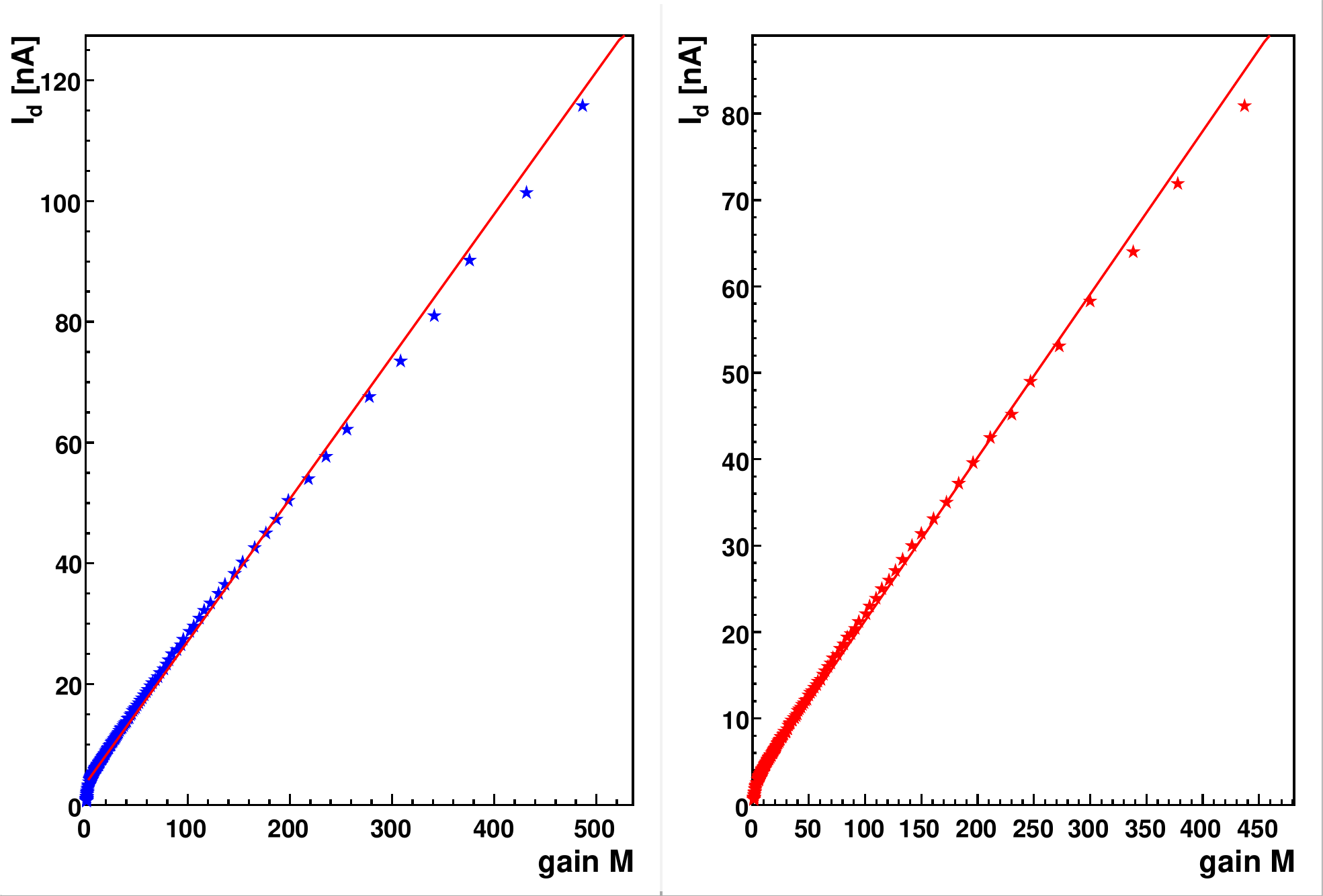}
    \caption[Dark current $I_d$ of 'normal C' type after and before neutron irradiation.]{Dark current dependence on
    gain of the \mbox{'normal C'} type APD
    before (blue/left) and after (red/right) exposure using a $^{241}AmBe$ source (linear fits are added).}
    \label{fig:photo:APD:irradiation:neutron:normal_Id_M_change_Neutron}
 \end{center}
\end{figure}
\begin{figure}
  \begin{center}
  \includegraphics[angle=90,width=\swidth,height=5.5cm]{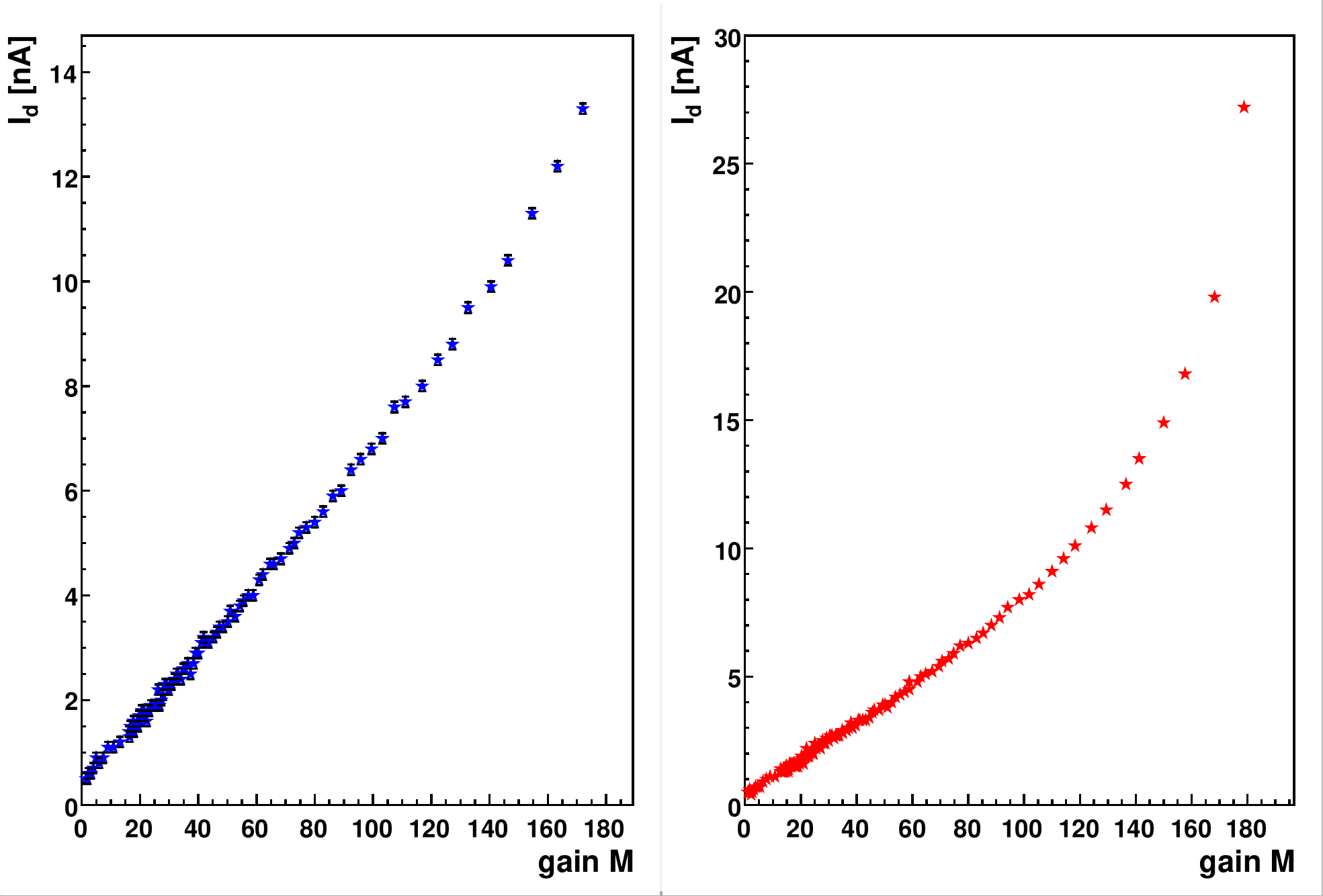}
    \caption[Dark current $I_d$ of 'low C' type after and before neutron irradiation.]{Dark current dependence on
    gain of the \mbox{'low C'} type APD
    before (blue/left) and after (red/right) exposure using a $^{241}AmBe$ source.}
    \label{fig:photo:APD:irradiation:neutron:low_Id_M_change_Neutron}
 \end{center}
\end{figure}
Having a closer look on the wavelength dependence of the gain measured after neutron exposure lead to the
situations shown in \Reffig{fig:photo:APD:irradiation:neutron:normalC_Mlambda_afterneutron} 
and \Reffig{fig:photo:APD:irradiation:neutron:lowC_Mlambda_afterneutron}. 
\begin{figure}
  \begin{center}
  \includegraphics[angle=90,width=\swidth,height=4.0cm]{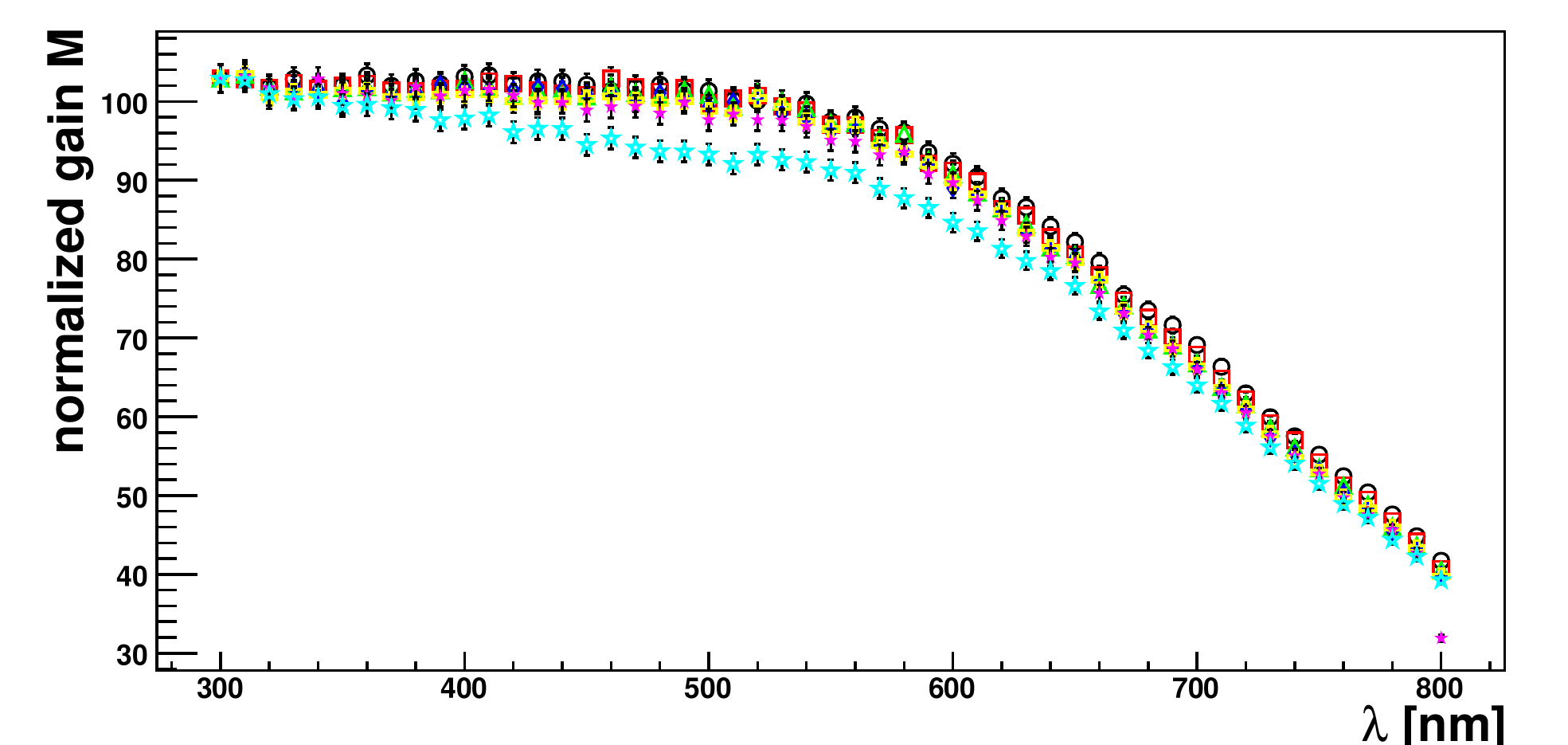}
    \caption[$M(\lambda)$ of 'normal C' type after neutron exposure.]{$M(\lambda)$ of 'normal C' type after 
    neutron exposure using a $^{241}AmBe$ source.}
    \label{fig:photo:APD:irradiation:neutron:normalC_Mlambda_afterneutron}
 \end{center}
\end{figure}
\begin{figure}
  \begin{center}
  \includegraphics[angle=90,width=\swidth,height=4.0cm]{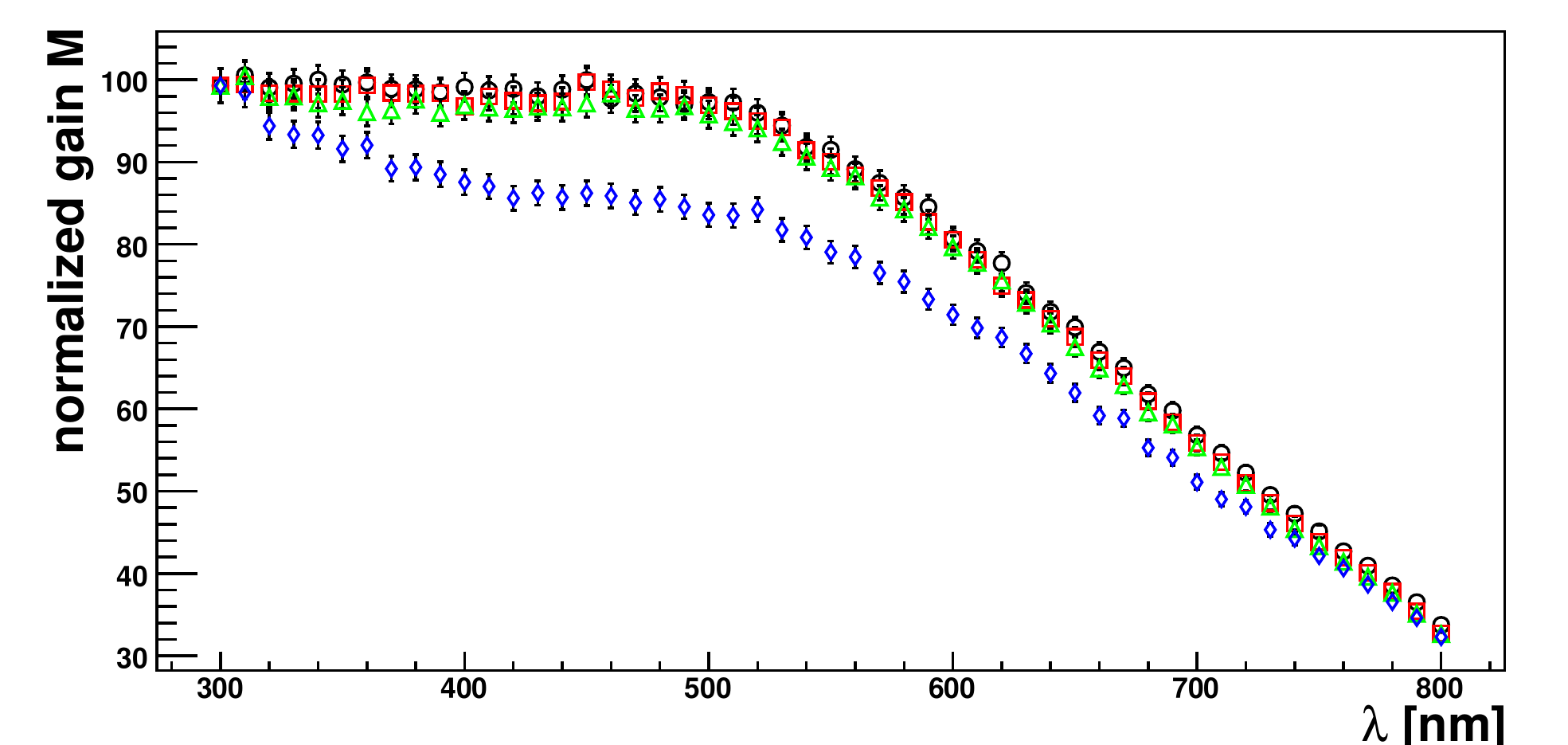}
    \caption[$M(\lambda)$ of 'low C' type after neutron exposure.]{$M(\lambda)$ of 'low C' type after 
    neutron exposure using a $^{241}AmBe$ source.}
    \label{fig:photo:APD:irradiation:neutron:lowC_Mlambda_afterneutron}
 \end{center}
\end{figure}
For both APD types, 'normal C' and 'low C', no indication of the existence of any kind of bulk damages
could be found using the $^{241}Am-\alpha-Be$ source described above.
\par
To complete the radiation hardness studies of the APDs an irradiation of the devices 
using a neutron generator located at the \INST{FZ Rossendorf}
providing neutrons with an energy of $14\,\mev$ is foreseen. This beamtime will take place during first
half of 2008 and will clarify open questions concerning the device radiation hardness due to neutron
exposure.

\subsection{APD Screening Procedure}
\label{sec:photo:APD:Screening}
To ensure an optimum operation of the calorimeter all components have to be tested before
their mounting as a part of the EMC. In case of the readout devices of the EMC barrel several
stations of the APD testing have to be passed before the APDs will be definitively
glued on the rear side of the scintillator crystals. The screening procedure for the LAAPDs
described in this section contains the timeline from delivery of the diodes until their mounting to the crystals.\\
Due to the stability of one APD production run provided by Hamamatsu Photonics some 
properties of the APDs will only be measured for control samples and are not foreseen as a part of
the APD mass screening:
\begin{itemize}
\item the quantum efficiency QE,
\item the Excess Noise Factor at an internal gain of $M = 50$ and $M = 100$ at room temperature and at
\mbox{$T = -25\,\degC$},
\item the capacitance of the APD depending on bias voltage/gain and
\item the gain uniformity of the APD active area surface.
\end{itemize}
To ensure a stable operation of the large amount of LAAPDs the stability of several parameters during
operation have to be guaranteed in addition. Therefore a complex screening procedure for the diodes
used for the EMC crystal readout has to be conceived.
The measurements of the APD parameters have to be done for each single diode at room 
temperature (to be comparable
to the parameter values given by Hamamatsu) and at the envisaged operation temperature of
\mbox{$T = -25\,\degC$}.
For each APD the gain-bias voltage characteristic has to be measured as well as the dark
current dependence of $U_R$. From these measurements several APD parameters, which will be stored in a 
database, will be extracted:
\begin{itemize}
\item Bias voltage values for the corresponding envisaged operation gain (in case of
\INST{CMS}: $M = 50$) and other gain values,
\item Dark current $I_d$ for different fixed gain values (e.g. $M = 50$ and $M = 100$) and
\item the gain variation with varying bias voltage $U_R$: \mbox{$1/M \cdot dM/dV$ [\percent/V]} for different gain
values.  
\end{itemize}
To accommodate the influence of radiation on all these parameters an irradiation of all APDs
using a high dose $^{60}Co$ source will be part of the screening process. After irradiation
all APDs have to be annealed (under bias voltage) using an oven operating at a temperature 
of $T = 80\,\degC$ until the monitored dark current of the APDs will reach an equilibrium.
After this procedure all APD parameters listed above have to be measured again and have to
be compared to the values determined before irradiation. 
\par 
Due to the application of one preamplifier board (including two channels) per crystal one additional 
step, as part of the screening procedure, 
has to be done: Based on the results of the remeasured APD characteristics two of them have to be assorted
in terms of similar dark current and bias voltage values at a given gain value for the readout of one lead
tungstate crystal of the EMC barrel part.

\subsection{Mounting Procedure}
\label{sec:photo:APD:Mounting}
\begin{figure}
  \begin{center}
    \includegraphics[width=\swidth]{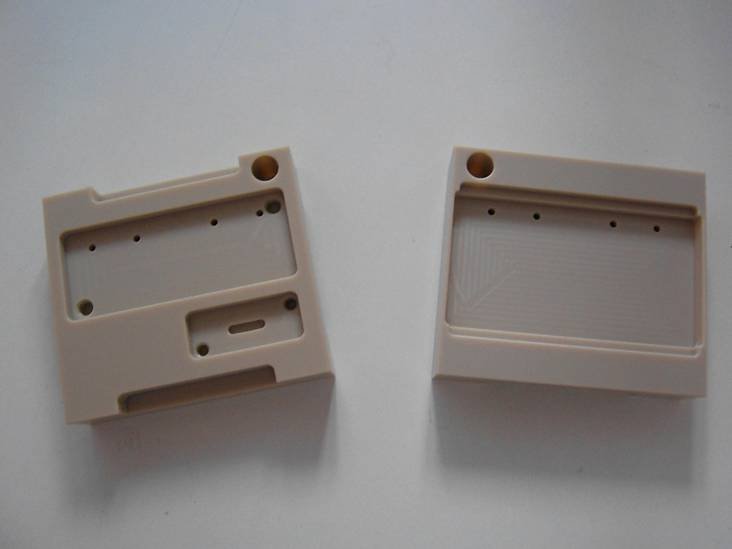}
    \caption[PEEK APD capsule.]{PEEK capsule designed to carry two ($10 \times
    10$)$\,\mm^2$ LAAPDs.}
    \label{fig:photo:APD:Mounting:Capsule_with}
 \end{center}
\end{figure}

After all APDs have passed the screening procedure explained above they have to be mounted on
the rear side of the lead tungstate crystals. The mounting technique has to fulfill three main
requirements:
\begin{enumerate}
\item The thermal contact between the APD and the crystal has to be as good as possible to
guarantee an overall temperature stability of the crystal-APD system down to $\Delta T = \pm 0.1\,\degC$
at \mbox{$T = -25\,\degC$}.
\item The optical coupling between crystal and APD has to be done as proper as possible to avoid 
any kind of light/signal loss in this region. Additional emphasize should lie on the optical temperature
dependent properties of
the used coupling material especially in view of the envisaged operation temperature of the
calorimeter.  
\item The readout diodes have to be as good as possible electrically insulated to 
avoid the appearance of additional noise sources (increase of dark current, etc.$\cdots$).
\end{enumerate}  
Most of these specifications could be fulfilled by using a special kind of plastic (PEEK) produced by
injection-moulding as carrier
for the APDs. During the ongoing R\&D work a so called 'capsule' built out of PEEK was
designed and manufactured for the housing of 2 APDs of quadratic shape. A picture of one of
the first prototypes is shown in \Reffig{fig:photo:APD:Mounting:Capsule_with}. In addition to
its function as APD housing a hole for one fibre of the envisaged light pulser system (see \Refsec{sec:cal:monitoring}) used for
radiation damage monitoring during the \Panda operation was implemented. On the rear side of
the capsule space for the ASIC preamplifier and for a possibly needed temperature
sensor is allocated (see right side of \Reffig{fig:photo:APD:Mounting:Capsule_with}). 
To avoid scintillation light losses due to a low reflectivity of the capsule material the admixture of 
titanium oxide of
different granularity to the moulding process is under investigation. 
The assembled capsule (including light fibre, preamplifier, thermistor and cables) offers a proper adjustment of the APDs on each crystal of the
EMC barrel part and will be coupled to the scintillator using 
a special kind of optical glue between the APD entrance window and the rear side of the
crystal. 
%
%
%

%
\section{Vacuum Phototriodes (VPT)}
\label{sec:photo:VPT}
\subsection{Introduction}
\label{sec:photo:VPT:intro}

\begin{figure}[htb]
\begin{center}
\includegraphics[width=\swidth]{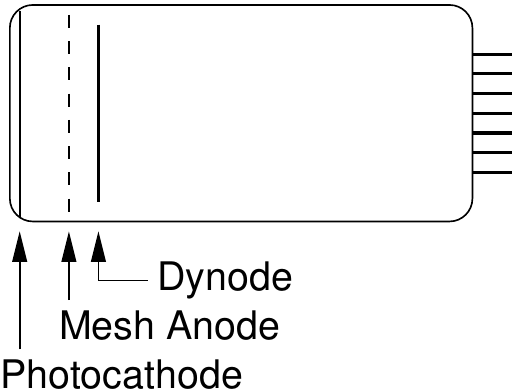}
\caption[Schematic of VPT.]{Schematic diagram of a 22$\,\mm$ diameter VPT. Typical distances between the photocathode and the mesh anode are 4.5$\,\mm$. The distance between mesh anode and dynode is 4.5$\,\mm$. The total length is around $46\,\mm$ and is needed for connections in and outside the tube and for production purposes.}
\label{fig:photo:vpt}
\end{center}
\end{figure}

For the photon detection in the \FWEMC we have chosen vacuum photo triodes (VPT).
A schematic view of a VPT is shown in \Reffig{fig:photo:vpt}. Typical high voltage settings are: cathode at $-1000\,$V, the anode on ground and the dynode at $-200\,$V. Photoelectrons emitted from the photocathode are accelerated towards the anode consisting of a fine metal mesh. The electrons passing the anode strike the solid metal dynode, located behind the anode. The secondary electrons ejected from the dynode are accelerated back towards the anode and are collected there.
The VPTs have a high rate capability and can be produced radiation hard, thus they are matching the requirements of the \FWEMC.

Vacuum phototriodes were previously in operation at the electromagnetic calorimeter endcap of the OPAL~\cite{bib:emc:vpt-opal}, DELPHI~\cite{bib:emc:vpt-delphi} and CMD-2~\cite{bib:emc:vpt-cmd-2} experiments. Recently radiation hard vacuum phototriodes were developed for the CMS endcap electromagnetic calorimeter for operation in a magnetic field of up to 4$\,$T.  

\subsection{Available Types}
\label{sec:photo:VPT:types}

Vacuum phototriodes produced by three different companies are
investigated for \PANDA. Samples of the R2148 from
Hamamatsu\footnote{Hamamatsu Photonics, Electron Tube Division Export
Sales Dept. 314-5, Shimokanzo, Iwata City, Shizuoka Pref., 438-0193, Japan} provide a typical gain of 10 and were tested in the lab. The Research Institute Electron\footnote{National Research Institute Electron (RIE), Ave M.~Toreza 68, 194223 St. Petersburg, Russia} produced 15000 radiation hard VPTs for the CMS endcap. 
Presently Photonis\footnote{PHOTONIS France SAS, Avenue Roger Roncier, 19100 Brive La Gaillarde, France} is developing phototriodes with high quantum efficiency photocathodes and high gain ($>$40 at 1000 V). The latter VPTs may produce only about a factor of two less charge per deposited energy in PWO as $1\,\cm^2$ sized APDs.
Currently we assume only the availability of VPTs with a standard
quantum efficiency and gain of about 10 as already available from RIE
and Hamamatsu.

\subsection{Characteristics and Requirements}
\label{sec:photo:VPT:reqs}

For the \FWEMC 4000 phototriodes are needed. To match the size of
crystals a maximum diameter of $22\,\mm$ and to stay within the space
assigned to the detector an overall length of $46\,\mm$ is
available. The specifications are listed in
\Reftbl{tab:photo:VPT:specs}. Parameters of the RIE FEU-189 VPT are
listed in \Reftbl{tab:photo:VPT:feu189}. It matches most of the
specifications, though further development is needed.

\begin{table*}
\begin{center}
\begin{tabular}{lc}
\hline\hline
Parameter & Value\\ \hline
External diameter & max. 22 mm \\ 
Overall length & about 46 mm\\
Dynode number & 1 or 2\\
Gain & 10 to 30 or more\\
Region of max. spectral response & 420 nm (PbWO$_4$)\\
Magnetic field & max 1.2 T \\
 & in 0-17 degrees in axial direction of VPT\\
Photocathode useful diameter & 16-20 mm \\
Quantum efficiency at 430 nm& 20\% or higher \\
Radiation hardness & 10 Gy per annum\\
Operational temperature range & -30$\degC$ to 35$\degC$ \\
Rate capability & above $500\,\kHz$\\
\hline\hline
\end{tabular}
\caption[Specification for \PANDA VPT.]{Specifications for the \PANDA VPTs.}
\label{tab:photo:VPT:specs}
\end{center}
\end{table*}

\begin{table*}
\begin{center}
\begin{tabular}{lc}
\hline\hline
Parameter & Value\\ \hline
External diameter & 21 mm \\
Photocathode useful diameter & 15 mm \\
Overall length & 41 mm\\
Operating bias voltage: $V_u, V_d (V_c=0$V$)$ & 1000 V, 800 V\\
Dark current & 1 - 10 nA\\
$(dM/dV)/M$ & $<0.1$\%/V \\
$(dM/dT)/M$ & $<0.1$\%/C \\
Quantum efficiency at 430 nm & $>15$\% \\
Range of spectral response & 300 -- 620 nm \\
Effective gain ($B$=0\,T) & 12 \\
Effective gain ($B$=4\,T, $\Theta=0^\circ$) & 6 \\
Effective gain ($B$=4\,T) $\Theta=20^\circ$)& 8 \\
Anode pulse rise time & 1.5 ns \\
Excess noise factor $F$ at $B=0$ & 2.0 -- 2.5 \\
Excess noise factor $F$ at $B=$1 -- 4 T, $\Theta=20^\circ$  & 2.2 -- 2.6 \\
\hline\hline
\end{tabular}
\caption[Parameters of the RIE FEU-189 VPT.]{Parameters of the RIE FEU-189 VPT~\cite{bib:emc:CMS-note-80}.}
\label{tab:photo:VPT:feu189}
\end{center}
\end{table*}

\begin{figure}[htb]
\begin{center}
\includegraphics[width=\swidth]{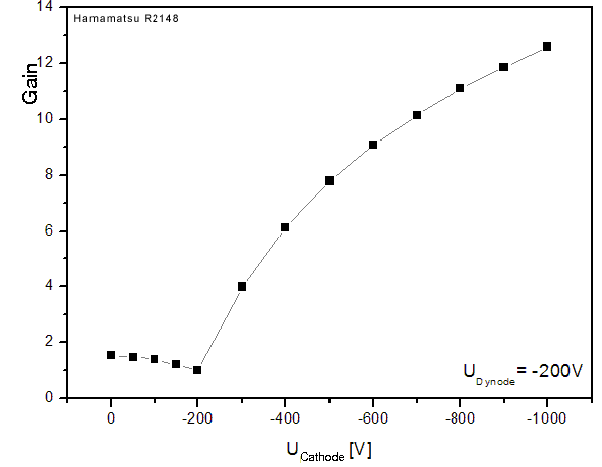}
\caption[VPT gain vs. cathode HV for R2148.]{Variation of VPT gain as a function of cathode voltage at a dynode voltage of -200 V for R2148 VPT.}
\label{fig:photo:vpt_gain_hamamatsu}
\end{center}
\end{figure}

\begin{figure}[htb]
\begin{center}
\includegraphics[width=\swidth]{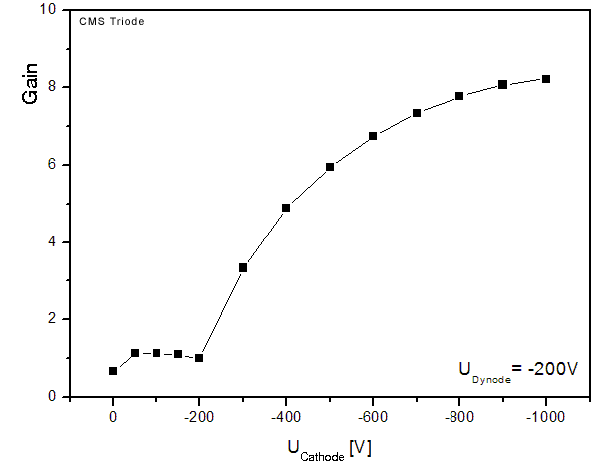}
\caption[VPT gain vs. cathode HV for FEU-188.]{Variation of VPT gain as a function of cathode voltage at a dynode voltage of -200 V for CMS VPT.}
\label{fig:photo:vpt_gain_cms}
\end{center}
\end{figure}

In \Reffig{fig:photo:vpt_gain_hamamatsu} and \Reffig{fig:photo:vpt_gain_cms} the gain is shown as a function of anode voltage for Hamamatsu R2148 and CMS FEU-188 VPT, respectively. Compared to APDs they exhibit only a small dependence of the gain on the high voltage. Typical values are less than 0.1$\,$\% per volt.
New developments in the material design of the dynodes allow the production of VPTs with gain above 40. Prototypes from Photonis are currently investigated. The high gain is favourable since it reduces the noise contribution for low energy calorimeter signals.

\begin{figure}[htb]
\begin{center}
\includegraphics[width=\swidth]{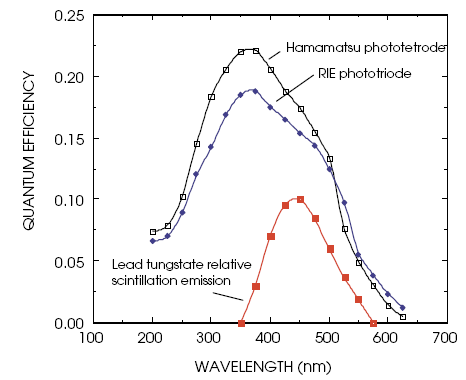}
\caption[Quantum efficiency VPT.]{Quantum efficiencies of a caesium antimony photocathode in a Hamamatsu R5189 tetrode and a RIE triode compared with the emission spectrum of \PWO \cite{bib:emc:photo:APD:CMS_TDR}.}
\label{fig:photo:vpt:qe}
\end{center}
\end{figure}

The spectral response of the photocathodes should cover a region from $350\,$nm  to $650\,$nm.
Typical quantum efficiencies of bialkali photocathodes are above
15$\,$\%, much less than the quantum efficiencies of APDs, which are
above 65\%. In \Reffig{fig:photo:vpt:qe} the quantum efficiency of
VPTs are compared to the \PWO emission spectrum. An increase of the
quantum efficiency would be desirable to improve the signal over noise ratio.
Recently photocathodes with quantum efficiencies above 30\% were produced by Photonis and Hamamatsu~\cite{bib:emc:vpt:highQE-photocathode,bib:emc:vpt:highQE-Hamamatsu}. These would be favourable. However, in the current design and simulations we assume only the availability of standard photocathodes.

\begin{figure}[htb]
\begin{center}
\includegraphics[width=\swidth]{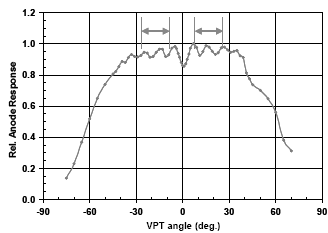}
\caption[VPT response to magnetic field vs. angle.]{Variation of anode response for constant pulsed LED illumination as a function of the VPT angle to the axial field of 1.8~T for the CMS type VPT \cite{bib:emc:vpt-nima504}.}
\label{fig:photo:vptmag}
\end{center}
\end{figure}

The endcap is located close to the door of the solenoid magnet. At the position of the photodetectors the magnetic field varies between 0.5$\,$T and 1.2$\,$T and has a direction between 0 and 17 degrees with respect to the axial direction of the VPTs. The gain variation under a magnetic field depends on the mesh size of the anode. In \Reffig{fig:photo:vptmag} the variation of the anode response is shown as a function of angle to the axial field for the VPTs produced for the CMS experiment. 


The dependence of the VPT quantum efficiency and gain on the temperature has been measured to be small and can be completely neglected compared to the temperature dependence of the light yield of PWO. VPTs can be operated like photomultipliers in temperature ranges down to $-30\,\degC$.


Unlike PIN diodes, where energetic charged particles may produce electron-hole pairs in the silicon (nuclear counter effect), photomultipliers and VPTs are not susceptible to charged particles, due to the thinness of the entrance window and photocathode.

The capacitance of the VPT is small compared to the cables connecting to the preamplifiers, such that low noise values can be reached. Typical values are about $22\,$pF or less. Therefore the cables to the preamplifiers must be kept as short as possible. In the design it is foreseen to place the preamplifiers within less than 10$\,$cm to the VPTs (see  \Refchap{sec:elo}). Also the dark current of $1\,$nA is very small compared to the maximum accepted $50\,$nA of the LAAPD.

\subsection{Testing}
\subsubsection{Radiation Hardness}
\label{sec:photo:VPT:radhard}

\begin{figure}[htb]
\begin{center}
\includegraphics[width=\swidth]{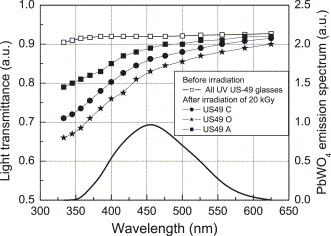}
\caption[Light transmission of VPT window after gamma irradiation.]{Light transmittance spectra of different UV glasses produced in Russia before and after $20\,$kGy $^{60}$Co gamma irradiation and the emission spectrum of PWO. Dose rate is $0.24],$kGy/h \cite{bib:emc:vpt-nima535}.}
\label{fig:photo:vpt:radiation1}
\end{center}
\end{figure}

\begin{figure}[htb]
\begin{center}
\includegraphics[width=\swidth]{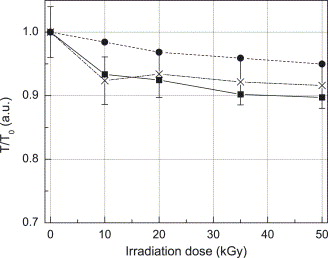}
\caption[Light transmission of VPT window as function of gamma irradiation.]{Relative US-49C (squares) and US-49A (circles) faceplate light transmission in the range of the PWO emission spectrum and relative anode response of the VPT as a function of the gamma dose at $B=0\,\T$ (crosses) \cite{bib:emc:vpt-nima535}.}
\label{fig:photo:vpt:radiation2}
\end{center}
\end{figure}

Radiation may change the response of the anode and the excess noise factor.
Damages resulting in severe modifications of the response would deteriorate the energy resolution. 
Extensive studies of the radiation hardness were performed by the CMS experiment for the FEU-188 VPTs \cite{bib:emc:vpt-nima535}. 
It was shown that under gamma irradiation the VPT anode response is fully determined by the loss of the VPT faceplate transmittance (\Reffig{fig:photo:vpt:radiation1}). The CMS VPTs use the UV glass type US-49A. The VPTs were irradiated in Russia with $20\,$kGy at a dose rate of $0.24\,$kGy/h. The decrease of the VPT anode signal did not exceed 4\% at 20$\,$kGy (\Reffig{fig:photo:vpt:radiation2}) \cite{bib:emc:vpt-nima535}. 
At \PANDA we do not expect gamma dose rates above 0.2$\,$kGy. 

\begin{figure}[htb]
\begin{center}
\includegraphics[width=\swidth]{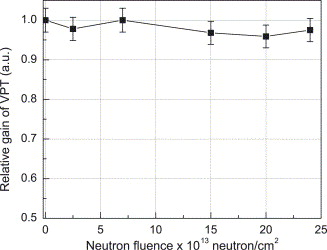}
\caption[Relative anode response of VPT versus the neutron fluence.]{Relative anode response of VPT FEU-188 with faceplates from cerium-doped glass C1-96 versus the neutron fluence ($E_n=14\,\MeV$) \cite{bib:emc:vpt-nima535}.}
\label{fig:photo:vpt:radiation5}
\end{center}
\end{figure}

Radiation hardness tests with reactor neutrons have a high level of gamma background and activate the VPT material by a thermal neutron flux. Therefore they are difficult to evaluate. To circumvent this problem the FEU-188 VPTs were irradiated at a neutron generator providing $E_{\mathrm n}=14\,\MeV$.
The anode response of VPTs with cerium-doped glass was found independent of the neutron fluence up to $2.4\EE{15}\,$n/cm$^2$ within the experimental error of $\pm5\,$\% (\Reffig{fig:photo:vpt:radiation5}).

\subsubsection{Rate Studies}
\label{sec:photo:VPT:rates}

The particle rate in the \FWEMC can be as high as $500\,\kHz$. Signals of the VPT are as short as the \PWO scintillation pulses. 
To check the rate capabilities of the VPTs a LED pulser producing 445$\,\nm$ pulses not shorter than the \PWO scintillation pulses was built. 
The VPTs of the different producers will be tested with the light pulser at varying rates up to $1\,\MHz$. 

\subsection{Screening Procedure}
\label{sec:photo:VPT:screening}

After delivery of the VPTs they will undergo a screening procedure to check relevant parameters. The results of the test of each of the 4000 VPTs will be stored in a database. Basic dimensions of the VPTs and the connectors will be checked first after the arrival. Following a burn-in procedure the anode leakage current and the gain times the quantum efficiency will be tested with an LED pulser system at wavelengths of $455\,\nm$ and $470\,\nm$.
The stability of the gain will be checked by varying the frequency of the LED pulser system between 500 kHz and 0 kHz and back to 500 kHz.
 Finally the variation of gain between $0\,$T and $1\,$T axial magnetic field and the excess noise factor will be measured. 
  
%

%
%
%

%
%
\newpage
\bibliographystyle{panda_tdr_lit}
\bibliography{./lit_emc}
%

%
\cleardoublepage
\chapter{Electronics}
\label{sec:elo}
%
%
%
\label{sec:elo:Intro} 
The \Panda Electromagnetic Calorimeter (EMC) will consist of PbWO$_4$
(\PWOII) crystals arranged in the cylindrical \BEMC with 11360
crystals, the \FWEMC with 3600 crystals and the \BWEMC with 592
crystals. The purpose of this detector is an almost full coverage, as
far as the acceptance of the forward spectrometer allows, of the final
state phase space for photons and electrons. Since one of the physics
goals is e.g. precision spectroscopy of the charmonium spectrum, the
low-energy photon threshold should be around 10$\,\MeV$, which
requires the threshold for individual crystals to be about 3$\,\MeV$
and correspondingly low noise levels of 1$\,\MeV$. Neutral decays of
charmed mesons require the detection of a maximum photon energy
deposition of 12$\,\GeV$ per crystal at the given maximum beam energy
of the HESR. These requirements dictate a dynamic range of 12000 for
the readout electronics.

The placement of the calorimeter inside the 2$\,\T$ solenoidal
magnetic field requires photo sensors which provide a stable gain in
strong magnetic fields. Therefore a Large Area Avalanche PhotoDiode
(\LAAPD) has been developed for the \Panda EMC and will be employed in
the \BEMC part where typical event rates of 10$\,\kHz$ and maximum
100$\,\kHz$ are expected (see
\Refsec{chap:scint:pwo:rates-crystals}). Because of higher rates (up
to 500$\,\kHz$) vacuum photo triodes (VPTs) have been chosen for the
\FWEMC. The photo sensors are directly attached to the end faces of
the individual crystals and the preamplifier has to be placed as close
as possible inside the calorimeter volume for optimum performance and
minimum space requirements. The readout of small and compact subarrays
of crystals requires very small preamplifier geometries. In order to
gain maximum light output from the \PWOII crystals, the calorimeter
volume will be cooled to -25$\degC$. Efficient cooling thus requires
low power consumption electronics to be employed in combination with
extremely low-noise performance.
 
To minimize the input capacitance and pickup noise, the analogue
front-end electronics is placed near the APD and kept at the same
temperature as envisaged for the \PWOII crystals, namely at
-25$\degC$. The low-temperature environment improves the noise
performance of the analogue circuits and, at the same time, constrains
the power consumption of the analogue front-end electronics. The
\PANDA collaboration has developed two complementary low-noise and
low-power (LNP) charge-sensitive preamplifier-shaper (LNP-P) circuits.

First, a LNP design was developed based on discrete components,
utilizing a low-noise J-FET transistor. The circuit achieves a very
good noise performance using signal shaping with a peaking time of
650$\,\ns$.  Second, a state-of-the-art CMOS ASIC was developed, which
achieves a similar noise performance with a shorter peaking time of
250$\,\ns$. The advantage of the CMOS ASIC is the very low power
consumption.  Both designs are complementary since the preamplifiers,
based on discrete components, will be used for the readout of the
\FWEMC for which we expect a maximum rate per crystal of
500$\,\kHz$. Such an approach minimizes the overall power consumption
and keeps the probability for pileup events at a moderate level well
below 1\percent.

The subsequent digitization stage will be placed as close as possible
to the calorimeter volume but outside the low-temperature area. This
allows signal transfer from the front-end over short distances by flat
cables with low thermal budget. Optical links will be employed to
transfer digitized data via a multiplexing stage to the compute node
outside the experimental setup.

In the following paragraphs the requirements and performance of the
various stages of the readout chain will be discussed: the general
readout scheme, the preamplifiers, the digitizer modules, the
multiplexer stage and, finally, the detector control system
supervising the performance of the whole detection chain.

\section{General EMC Readout Scheme}
\label{sec:elo:General} 

\begin{figure*}[tb]
\includegraphics[width=\linewidth]{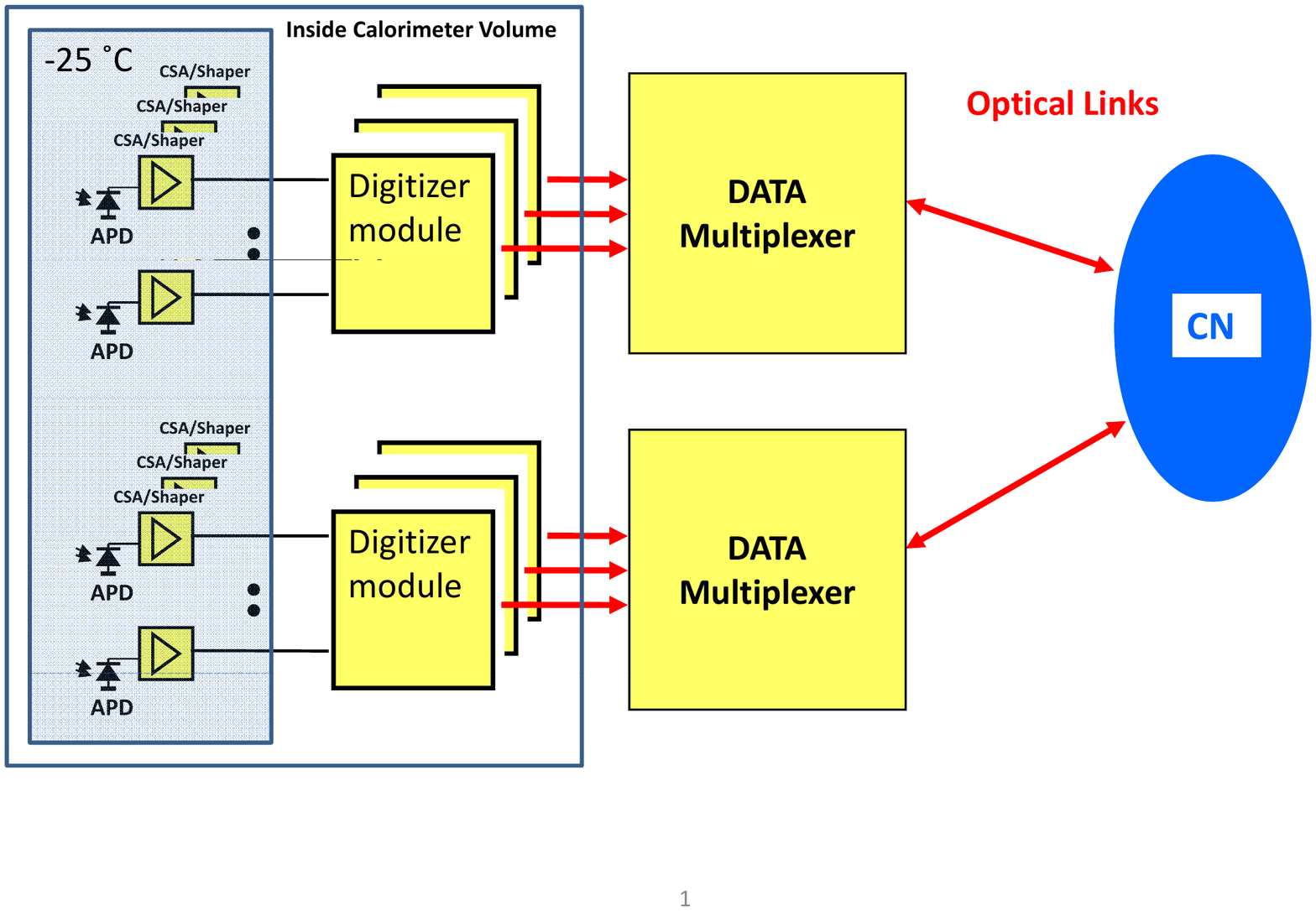}
\caption[The readout chain of the Electromagnetic Calorimeter.]
{The readout chain of the Electromagnetic Calorimeter.}
\label{fig:elo:ro_chain}
\end{figure*}

The readout of the electromagnetic calorimeter is based on the fast,
continuous digitization of the amplified signal-shape response of
\LAAPD and VPT photo sensors to the light output of \PWOII
crystals. The readout chain will consist of extremely low-noise
front-end electronics, digitizer modules and data multiplexer (see
\Reffig{fig:elo:ro_chain} ).

The digitizer modules are located at a distance of 20--30$\,\cm$ and
90--100$\,\cm$ for the \BEMC and the \FWEMC, respectively, away from
the analogue circuits and outside the cold volume. The digitizers
consist of high-frequency, low-power pipelined ADC chips, which
continuously sample the amplified and shaped signals. The sampling is
followed by the digital logic, which processes time-discrete digital
values, detects hits and forwards hit-related information to the
multiplexer module via optical fibers. At this step the detection of
clusters of energy deposition can be efficiently implemented. A
cluster seed requires an energy deposition of typically at least 10
MeV with a number of neighboring crystals surpassing the
single-crystal threshold of typically 3 MeV.  The multiplexer modules
will be located in the DAQ hut and they perform advanced signal
processing to extract amplitude and signal-time information.

The front-end electronics of the \BEMC is located inside the solenoid
magnet where any access for maintenance or repair is limited to
shutdown periods of the HESR, expected to occur once a year. This
condition requires to implement a redundancy in the system
architecture. One of the most important decisions, that has been taken
by the collaboration, is to equip every EMC crystal with two
APDs. Apart from redundancy, the system with two independent readout
channels offers a significantly (max. $\sqrt{2}$) improved signal to
noise ratio and a lower effective threshold level.

\section{Preamplifier and Shaper for \BEMC APD-readout}
\label{sec:elo:Barrel} 
A low-noise and low-power charge preamplifier ASIC (APFEL) was
designed and developed for the readout of the \LAAPD for the \Panda
EMC. Two LAAPDs with an active area of $7\times14\,\mm^2$ each are
attached to the end face of the lead tungstate scintillating crystals
(\PWOII) which have a typical geometry of
$(200\times27\times27)\,\mm^3$. Contrary to a photomultiplier, the
LAAPDs can also be operated in a strong magnetic field. In the \BEMC
the LAAPDs act as photo detectors converting the scintillating light
to an electrical charge signal. The preamplifier linearly converts the
charge signal from the LAAPD to a voltage pulse which is transmitted
to the subsequent electronics.

\subsection{Requirements and Specifications}
\label{sec:elo:Barrel:Require} 

\subsubsection{Power Consumption}
\label{sec:elo:Barrel:Require:Power} 
Since the complete \BEMC, together with the APDs and the
preamplifiers, will be cooled to low temperatures (-25$\degC$) to
increase the light-yield of the \PWOII crystals, the power dissipation
of the preamplifier has to be minimized. Low power dissipation leads
to a smaller cooling unit and thinner cooling tubes; it also helps to
achieve a uniform temperature distribution over the length of the
crystals.

\subsubsection{Noise}
\label{sec:elo:Barrel:Require:Noise}       
To reach the required low detection threshold of about 1 MeV, the
noise performance of the preamplifier is crucial. The newly developed
rectangular LAAPD from Hamamatsu (Type S8664-1010) has an active area
of $14\times7\,\mm^2$ resulting in a quite high detector capacitance
which requires a low-noise charge preamplifier. The total output noise
is a combination of the preamplifier noise and the noise generated by
the dark current flowing through the APD. By cooling the APD to
-25$\degC$ the dark current is reduced by a factor of about ten, with
respect to room temperature. Using a low-leakage LAAPD at low
temperature, the charge preamplifier is the dominating noise source
due to the relatively high detector capacitance of around
270$\,\pF$. The noise floor of the APFEL ASIC at \mbox{-20$\degC$},
loaded with an input capacitance of 270$\,\pF$, has a typical
equivalent noise charge (ENC) of 4150 e$^-$ (rms) (see
\Refsec{sec:elo:Barrel:IC:Noise}).

Investigations of \PWOII light production (see
\Reffig{fig:scint:pwo:fig9}) yield on average 90 photoelectrons per
MeV, measured at -25$\degC$ with a photomultiplier tube of 18\%
quantum efficiency and an integration gate width of 300$\,\ns$. This
value results in 500 photons/MeV at the end face of the cooled
(-25$\degC$) \PWOII crystal. With the average back face cross section
of barrel crystals of 745$\,\mm^2$ (see
\Reffig{fig:mech:mech-crystal_definition}) we obtain 66 photons/MeV on
the active area of $7\times14\,\mm^2$ of a single rectangular
LAAPD. The quantum-efficiency of the LAAPD is around 70\% for the
scintillating light of the \PWOII crystals and the voltage biased
LAAPD will be operated at an internal gain M~=~100. Applying these
numbers, a primary photon with the energy of 1 MeV induces an input
charge of 0.74 fC (4620 e$^-$) to the preamplifier. Thus, an ENC of
4150 e$^-$ (rms) corresponds to an energy noise level of about 0.9 MeV
(rms).

\subsubsection{Event Rate}
\label{sec:elo:Barrel:Require:Rate} 
To cope with the expected event rates in the \BEMC of maximum
100$\,\kHz$ per crystal, a feedback time constant has to be chosen
which is a trade-off between noise performance and pile-up
problematic.  The preamplifier is designed for an event rate up to
350$\,\kHz$.
 
\subsubsection{Bias Voltage}
\label{sec:elo:Barrel:Require:Bias} 
At the maximum event rate of 100$\,\kHz$ with the maximum expected
photon energy deposition per crystal of 12$\,\GeV$ (8.9$\,\pC$ from
the LAAPD) a mean current of 890$\,\nA$ is flowing through the
LAAPD. Under these extreme conditions, the voltage drop over the
low-pass filter for the APD bias voltage gets relevant, since the
internal gain (M~=~100) of the LAAPD varies by about 3\%/V.  This means,
that the measured energy will be dependent on the event rate. To
minimize this effect one can design the low-pass filter with low
series resistance resulting in a degraded noise performance of the
preamplifier and the need for larger filter capacitors. Eventually,
the various regions of the \BEMC with different event rates could be
equipped with adapted low-pass filters to obtain an optimum noise
performance combined with low rate/energy dependence.  Another
solution is that the APD bias-voltage supply sources a voltage which
is corrected on the output current (the more current the more
voltage). This means, that the output resistance of the bias-voltage supply
is negative and therefore compensates the series resistance of the
low-pass filter. The advantage is a better noise performance of the
preamplifier due to the highly resistive APD bias-voltage filter. This
solution would demand a more complex bias-voltage supply system with a
sophisticated overvoltage and overload control. Each LAAPD has its own
measured bias voltage where it reaches the nominal internal gain of
M~=~100. To reduce the number of APD bias-voltage channels, it is
foreseen to group LAAPDs with similar bias-voltages. Since this
grouping will be very local, in regions where similar event rates are
expected, the solution with the negative resistance bias-supply could
still be feasible.

\subsubsection{FADC Readout}
\label{sec:elo:Barrel:Require:FADC} 
Readout with a flash ADC (FADC) does not need a dedicated timing
amplifier. Only a good anti aliasing low-pass filter/amplifier has to
be implemented in front of the FADC. The low-pass filter/amplifier can
be safely installed at distances below 70$\,\cm$ from the preamplifier
directly at the FADC in the room-temperature environment. The cut-off
frequency of that filter is given by the sampling frequency of the
FADC divided by 2.5; this prevents aliasing effects due to the
sampling.  To determine the energy with a high signal to noise ratio,
the digitized values are processed by a digital shaping filter
followed by a digital peak determination. It could be possible to
choose different digital energy shaping filters for the different
regions of the \BEMC: A short peaking-time for the small angles in
forward direction and the region orthogonal to the target position,
where high event rates are expected. For other regions of the
detector, where lower event rates are estimated, the peaking-time
could be larger and therefore result in a better energy
resolution. Even a rate-dependent automatic adaptation of the digital
energy shaping filter could be imagined.  The timing information is
extracted by processing the digitized values similarly to the
traditional signal chain with a fast digital timing filter followed by
a digitally implemented CFD. Measurements have shown, that a timing
resolution well below 1$\,\ns$ can be reached with that regime.  One
has to point out, that the processing for the energy- and timing
information extraction must be performed in real time. This can be
implemented by using programmable digital signal processors (DSPs) or
completely in hardware by using field programmable gate arrays
(FPGAs). The needed signal throughput at the high sampling frequency
(80$\,\MHz$) combined with the complex algorithm will result in a
remarkable power dissipation.

\subsection{Integrated Circuit Development}
\label{sec:elo:Barrel:IC} 
The high compactness of the electromagnetic calorimeter in connection
with the required temperature homogeneity at low operation temperature
puts high demands on space and power consumption. It appears almost
mandatory to integrate the preamplifier and shaper on a single chip to
fulfill these demands.  For this reason an Application Specific
Integrated Circuit (ASIC "APFEL") for the readout of Large Area APDs
as foreseen at the \PANDA EMC was developed.

\subsubsection{Circuit Description}
\label{sec:elo:Barrel:IC:Circuit} 
\begin{figure}[tb]
\includegraphics[width=\linewidth]{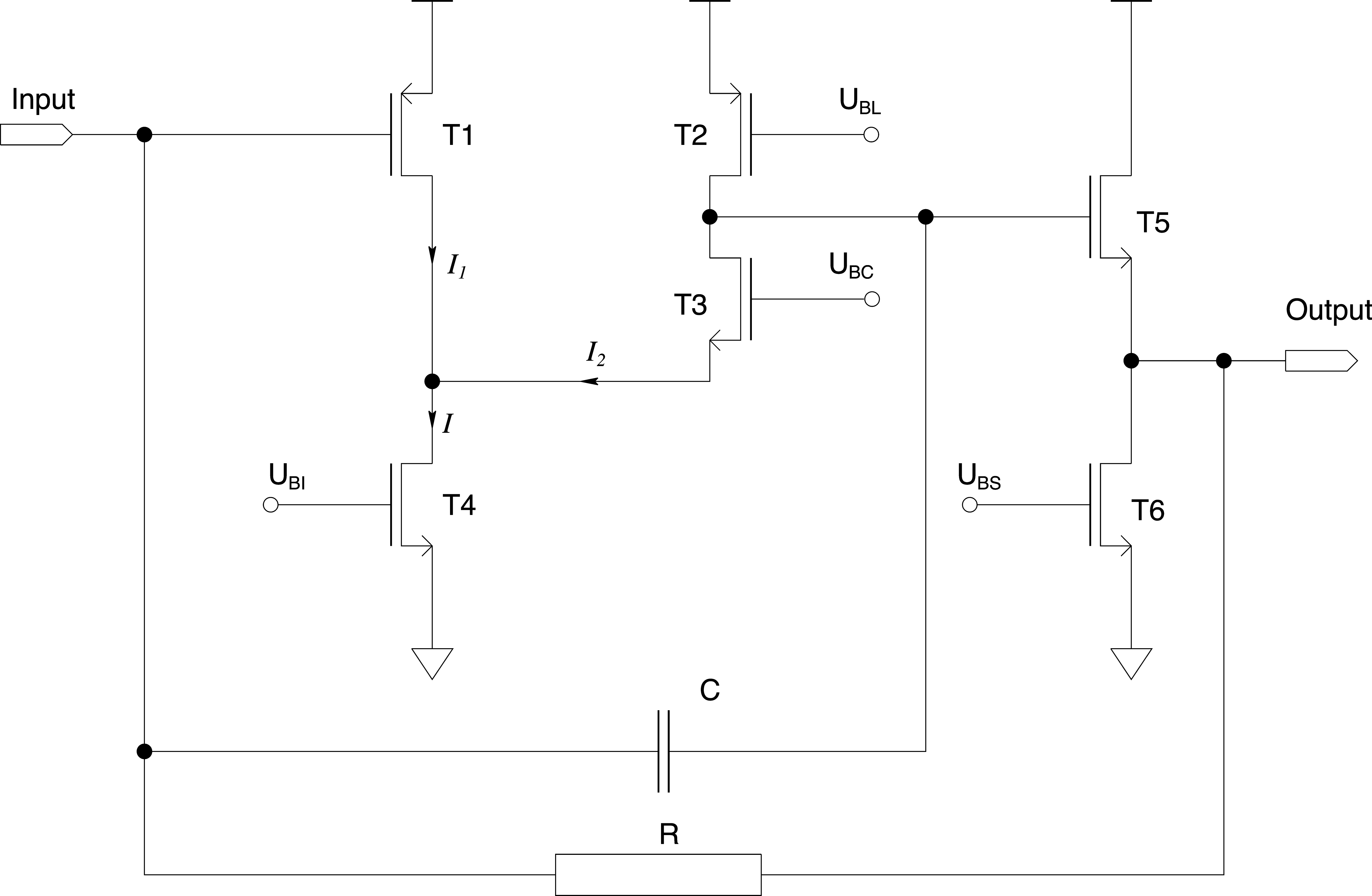}
\caption[Schematic diagram of the folded cascode front end.]{Schematic
  diagram of the folded cascode front end.}
\label{fig:elo:fcfront}
\end{figure}
For the front-end amplifier design a single ended folded cascode architecture was chosen as it is frequently described in literature \cite{bib:emc:elo:ieee1,bib:emc:elo:ieee2,bib:emc:elo:ieee3} and 
shown in \Reffig{fig:elo:fcfront}. This architecture combines a high open loop gain with a large output swing.

A signal at the input transistor T1 effects a current change of $I_1$. As the current $I = I_1 + I_2$ is kept constant by the current sink T4 also $I_2$ changes 
by the same value. This variation creates a voltage drop at the output node of T2. The cascode transistor T3 separates the input transistor from the output node. 
To get the best gain and stability the ratio of the  currents $I_1$ and $I_2$ in \Reffig{fig:elo:fcfront} should be in the order of $I_1 / I_2 = 9$.\\
  
The signal charge is integrated on a capacitance C which has to be discharged by a resistor R to prevent the preamplifier from saturation. The integration capacitance 
is not across the feedback resistor as usual but between the input and the dominant pole node realizing a Miller compensation. This results in a reduction of wideband 
output noise \cite{bib:emc:elo:ieee1} and minimizes the sensitivity to variations in detector capacitance \cite{bib:emc:elo:ieee4}.

For the preamplifier design the noise consideration played the leading part. The parts in equivalent input noise of transistors T2 and T4 scale with their transconductances $g_{m2} / g_{m1}$ and 
 $g_{m4} / g_{m1}$, respectively, so they can be minimized by a dedicated choice of parameters \cite{bib:emc:elo:designofan}. 
The main noise contributor is the input transistor T1. From noise theory one can see, that the drain source current $I_{DS}$ and the transistor width $W$ are the 
free parameters to control the transistor noise which decreases with increasing $I_{DS}$ and $W$. Since at $W \approx 10^4$~$\mu$m the noise reaches a 
minimum, $W = 12.8$~mm was chosen. A tradeoff between noise performance and the power consumption led to a current of $I_{DS} = 2$$\,\mA$.

Another tradeoff is the choice of the feedback resistor to balance low 
parallel noise (large R) and the hit rate the preamplifier has to cope with 
(small R).
The capacitance $C$ is given by the maximum of the input charge the preamplifier has to deal with.

\begin{figure}[tb]
\includegraphics[width=\linewidth]{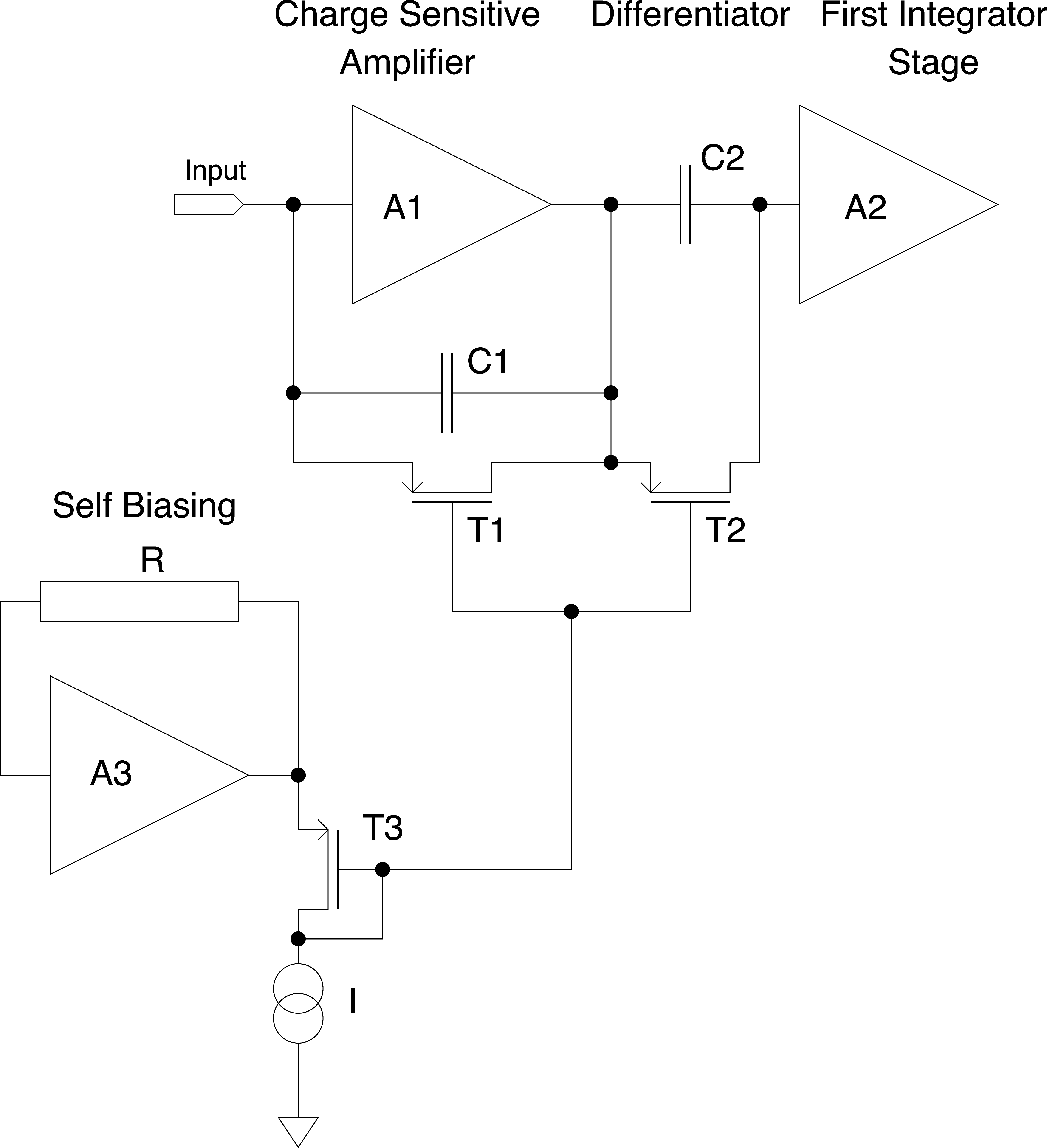}
\caption[Principle diagram of the self biasing feedback network.]
{Principle diagram of the self biasing feedback network.}
\label{fig:elo:selfbias} 
\end{figure}
To realize the mandatory high resistivity for the feedback resistor R1, a transistor operating in the subthreshold region is used. Concerning the temperature and 
process independence, a self biasing technology as described by O'Connor et al. in \cite{bib:emc:elo:ieee3} is  realized. As shown in \Reffig{fig:elo:selfbias} a MOS transistor T3 in 
diode connection is used together with a current sink to generate the gate source voltage $V_{GS}$ of the feedback transistor T1. The source potential of this MOS diode 
is fixed by a downscaled version A3 of the preamplifier circuit A1.

The pole which is introduced by the output resistance $r_{oT1}$ of T1 and the capacitance C1 is compensated by transistor T2 in parallel connection to the 
differentiation capacitance C2. This way a zero is introduced into the transfer function. By choosing the time constants $\tau_2= \tau_1$ with $\tau_1 = C1 \cdot r_{oT1}$ 
and $\tau_2 = C2 \cdot r_{oT2}$ any undershoot in the pulse shape is eliminated.

The gate of T2 is connected to the same potential as the gate of T1. To ensure that the drain and source potentials of T1, T2 and T3 are equal, the amplifiers A2 and A3 are downscaled versions of the input amplifier A1.

The amplifier A2 is used as the first stage of a 3rd order integrator. With a capacitive feedback a 1st order low pass filter with a time constant 
of $\tau = 90\ {\rm ns}$ is realized. At this point the signal is split into two paths. To add two more poles to the transfer function, a 2nd order integrator stage based on a fully differential operational amplifier follows on each path.

\begin{figure*}[t]
\begin{minipage}[b]{0.45\textwidth}
\makebox[0cm]{}\\
\includegraphics[width=\linewidth]{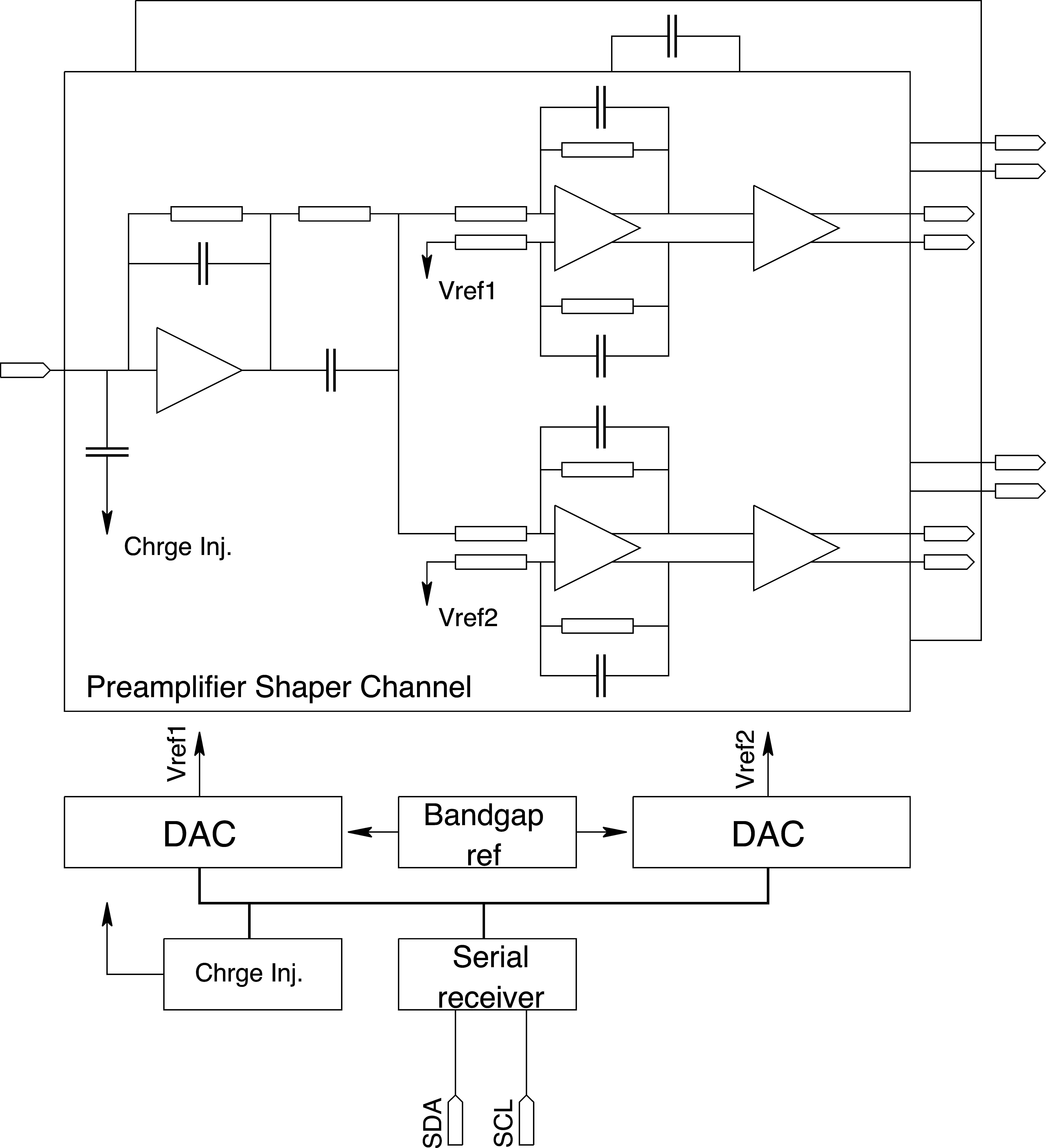}
\end{minipage}
\begin{minipage}[b]{0.55\textwidth}
\makebox[0cm]{}\\
\includegraphics[width=\linewidth]{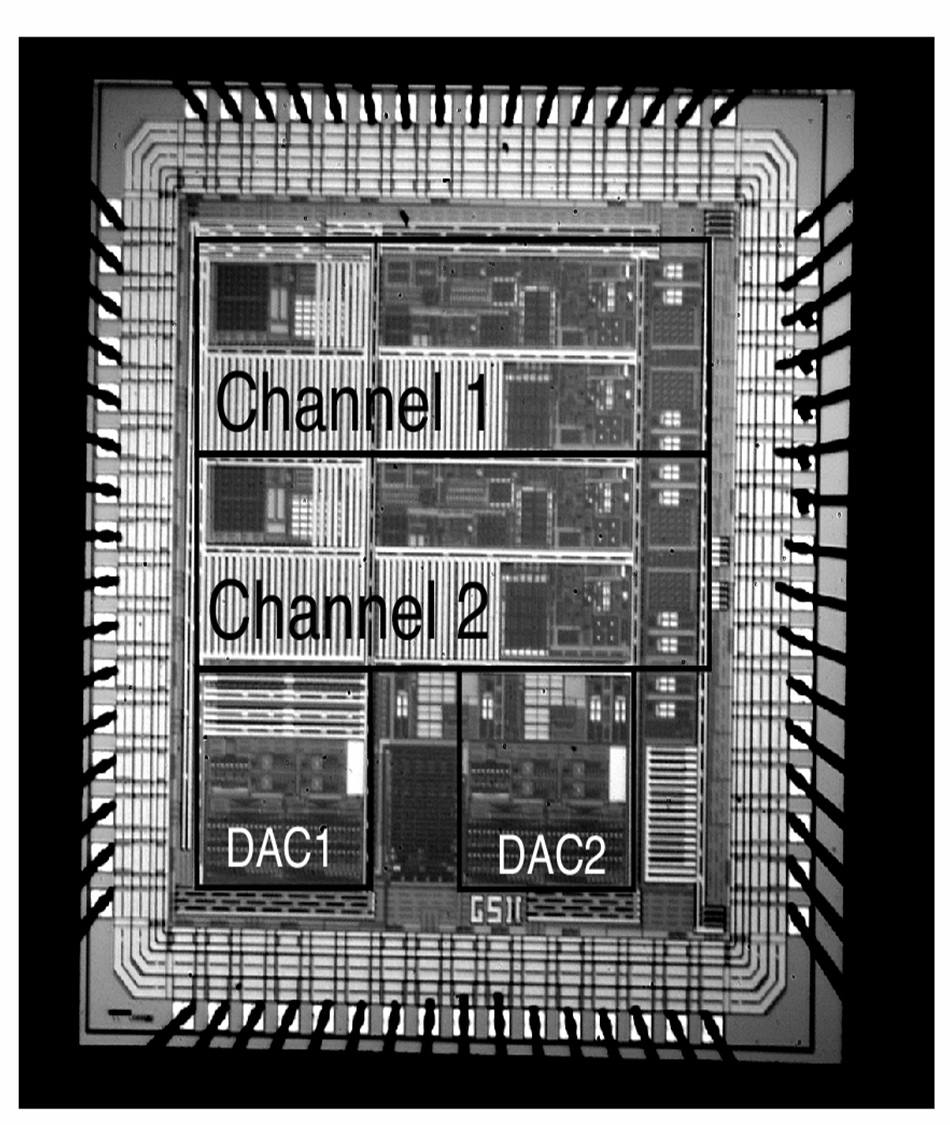}
\end{minipage}
\caption[Block diagram and photograph of the preamplifier and shaper ASIC.]
{Left side: Overall block diagram of the preamplifier and shaper ASIC. Right side: Photograph of the prototype preamplifier ASIC.}
\label{fig:elo:apfel}
\end{figure*}

On one of these paths an amplification factor of 16 is realised so this signal path is optimised to measure at the low energy part of the dynamic range with a minimized 
influence of pick up noise on the connection between preamplifier and ADC. The other path has no additional amplification so it covers the whole upper part of the dynamic range.

After the shaper build up by the differentiator stage and the integrators, detector pulses have a semi Gaussian pulse shape with a peaking time of 250~ns. 
The shapers are followed by output drivers with a driving capability of 10~pF parallel to 20~k$\Omega$.
For shaper operation two reference voltages are needed. To guarantee the full dynamic range over a temperature range from $-30\degC$ to $+30\degC$ these 
reference voltages have to be adjusted. An adjustable voltage reference based on two 10 bit digital to analogue converters is implemented on the ASIC to provide this functionality.

A charge injection unit for each channel is implemented on the chip which gives the possibility to inject a defined amount of charge into the preamplifier input for 
readout electronics monitoring. The amount of charge can be programmed in four steps.
For programming of the adjustable voltage references as well as for programming and triggering of the charge injection a serial interface to a two-wire serial bus 
is implemented on the integrated circuit. Data transfer has to follow a specific bus protocol which was defined for this circuit. Each transfer consists of 20 bits. 
After a start signature the first eight bits are used as an address to select a readout chip. The next two bits are used to select one of four internal data 
registers into which the following 10 data bits are written.

To avoid introduction of additional noise by substrate coupling from a running digital logic, there is no continuous clock signal for the digital logic. 
Receiving data and latching into the internal registers is triggered by the external serial clock signal. After data transmission the clock line stays 
on high level and the on-chip digital logic is inactive. So it neither produces any additional noise nor it consumes power.

An overall block diagram of the integrated preamplifier circuit is shown in \Reffig{fig:elo:apfel}
on the left side. The choice of technology was driven by the fact that larger feature size technologies have the advantage of higher core voltages, which affects directly the dynamic range. 
On the other hand, the noise performance benefits from a smaller feature size only on a minor level. 
Therefore, for the preamplifier design a 350$\,\nm$ technology from AMS\footnote{Austria Mikrosysteme AG} was 
chosen, which provides a sufficient high core voltage and which is widely 
used e.g. in automotive industry. Thus, we may expect that this technology 
is available with a long term perspective.
The prototype shown on the right side of \Reffig{fig:elo:apfel} was produced in 2007. Since 
summer 2007 several measurements have been done to specify the ASIC 
prototype.

\subsubsection{Prototype Performance}
\label{sec:elo:Barrel:IC:Noise} 
\begin{figure}[tb]
\includegraphics[width=\linewidth]{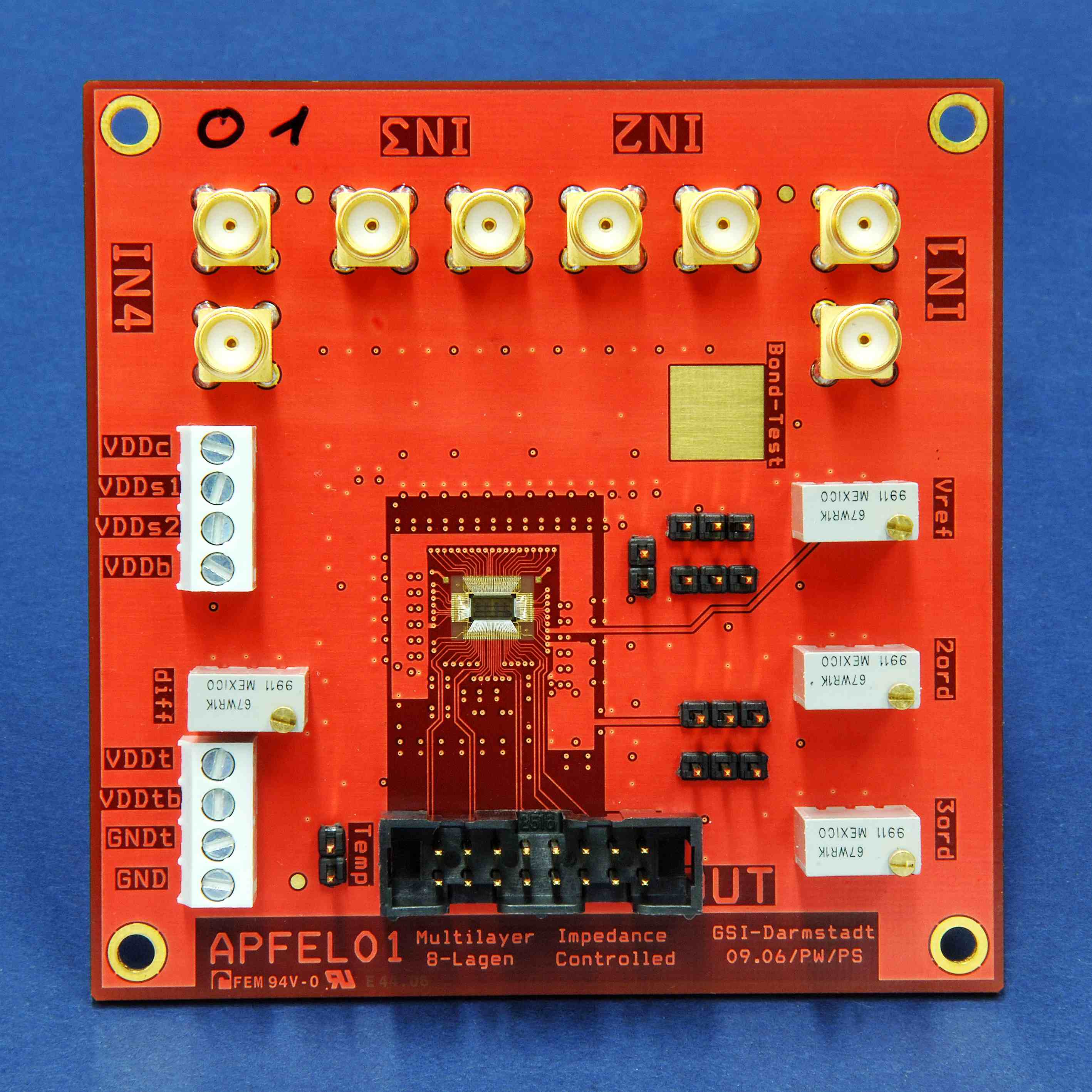}
\caption[Test PCB with glued and bonded preamplifier ASIC.]
{Test PCB with glued and bonded preamplifier ASIC.}
\label{fig:elo:testpcb}
\end{figure}

For specifying the preamplifier prototype over a temperature range from $-20\degC$ to $+20\degC$ the ASIC was mounted on a printed circuit board (see \Reffig{fig:elo:testpcb}) which can be 
cooled by a Peltier element. To avoid condensation of air humidity the setup is placed in an evacuated chamber.  For measurements a voltage step generated by an AWG 510 
on a well defined capacitor is used as charge injection. A second well defined capacitor connected parallel to the amplifier input simulates the detector capacitance. 
The output signals are monitored with a digital oscilloscope type DPO 7254. Serial programming is done with a data timing generator DTG 5078.
\begin{figure}[tb]
\includegraphics[width=\linewidth]{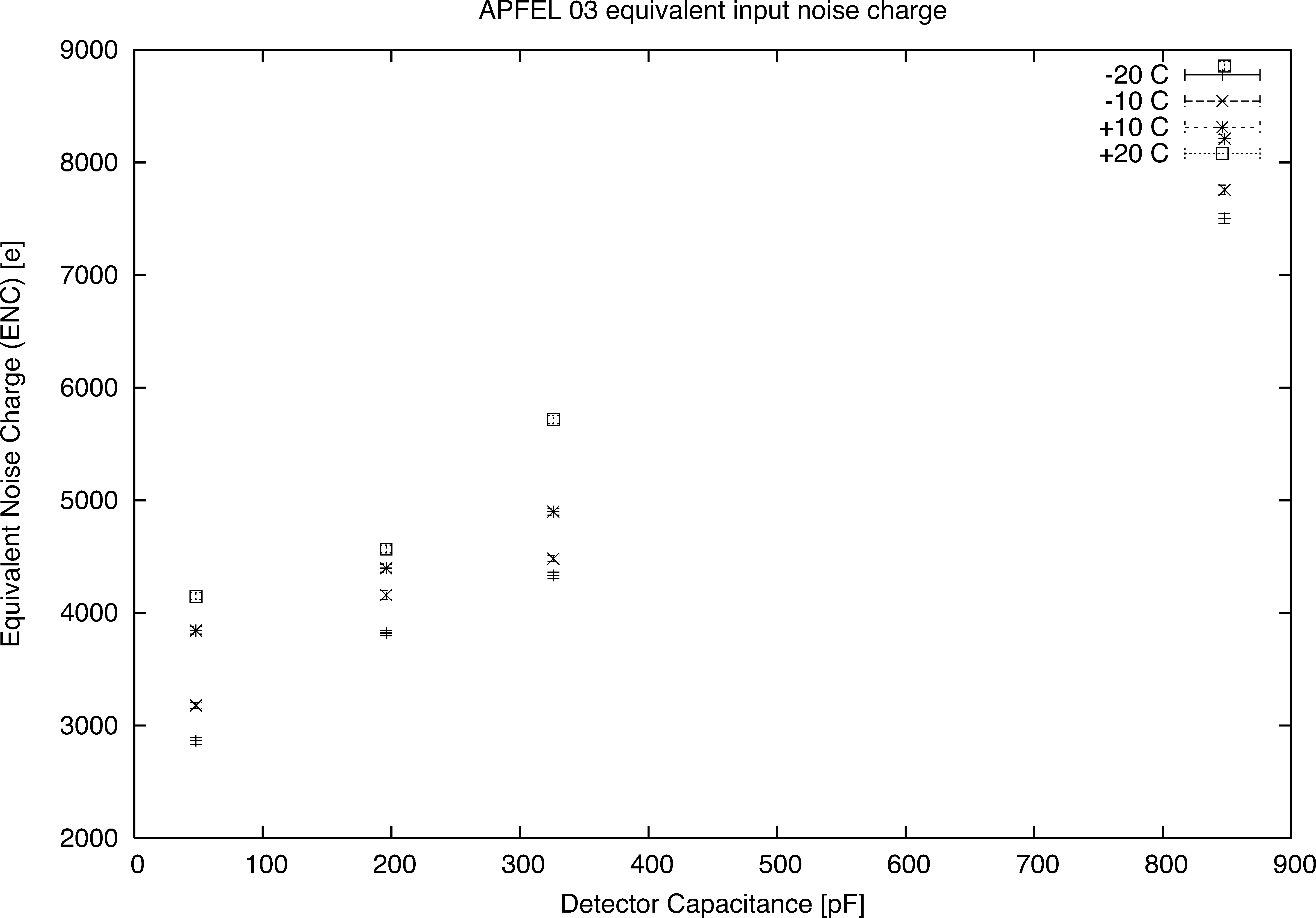}
\caption[Measured noise values of the preamplifier prototype.]
{Measured noise values of the preamplifier prototype.}
\label{fig:elo:noise}
\end{figure}
For the temperatures $-20\degC$, $-10\degC$, $+10\degC$ and $+20\degC$ dynamic range, gain and output noise voltage were measured so that the equivalent input noise charge could be calculated. The results for four different detector capacitances are plotted in \Reffig{fig:elo:noise}. As expected, the noise increases linearly with the detector capacitance. 
Also the temperature dependence is clearly visible.

A linear fit indicates that the slope $S_{ENC} = ( 5.73 \pm 0.5 )$~e$^-$/pF of the equivalent noise charge is almost independent of temperature. 
The constant term increases linearly with temperature with a temperature coefficient of $S_T = 26.15$~e$^-/\degC$.

For an operating temperature of $-20\degC$ the measured noise of the preamplifier prototype is 
\begin{equation*}
ENC = \left[2610 \pm 103 + ( 5.69 \pm 0.3 ) \cdot C_{Det}\right]\ {\rm e}^-
\end{equation*}
so with a detector capacitance of $C_{Det} = 270$~pF a noise of $ENC = ( 4146 \pm 131)\ {\rm e}^-$ (rms) can be expected. This value corresponds (see \Refsec{sec:elo:Barrel:Require:Noise}) to an energy of $0.9 \pm 0.02\,\MeV$.

\begin{figure}[tb]
\includegraphics[width=\linewidth]{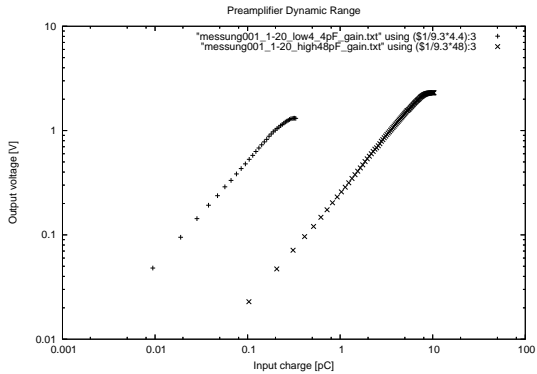}
\caption[Amplification characteristics of the preamplifier at $-20\degC$.]
{Amplification characteristics of the preamplifier at $-20\degC$.}
\label{fig:elo:dynamic}
\end{figure}

\Reffig{fig:elo:dynamic} shows the measured characteristics of the integrated preamplifier for the high and the low amplification path. 
The measurement covers a range from 10 fC to 10 pC input charge. Both paths show an excellent linear behavior with an overlapping range up to 200 fC. 
The -1 dB compression point of the low amplification path can be determined at 7.84 pC. Together with the measured noise of 0.66 fC this leads to a covered dynamic range of 11900.

\begin{figure}[tb]
\includegraphics[width=\linewidth]{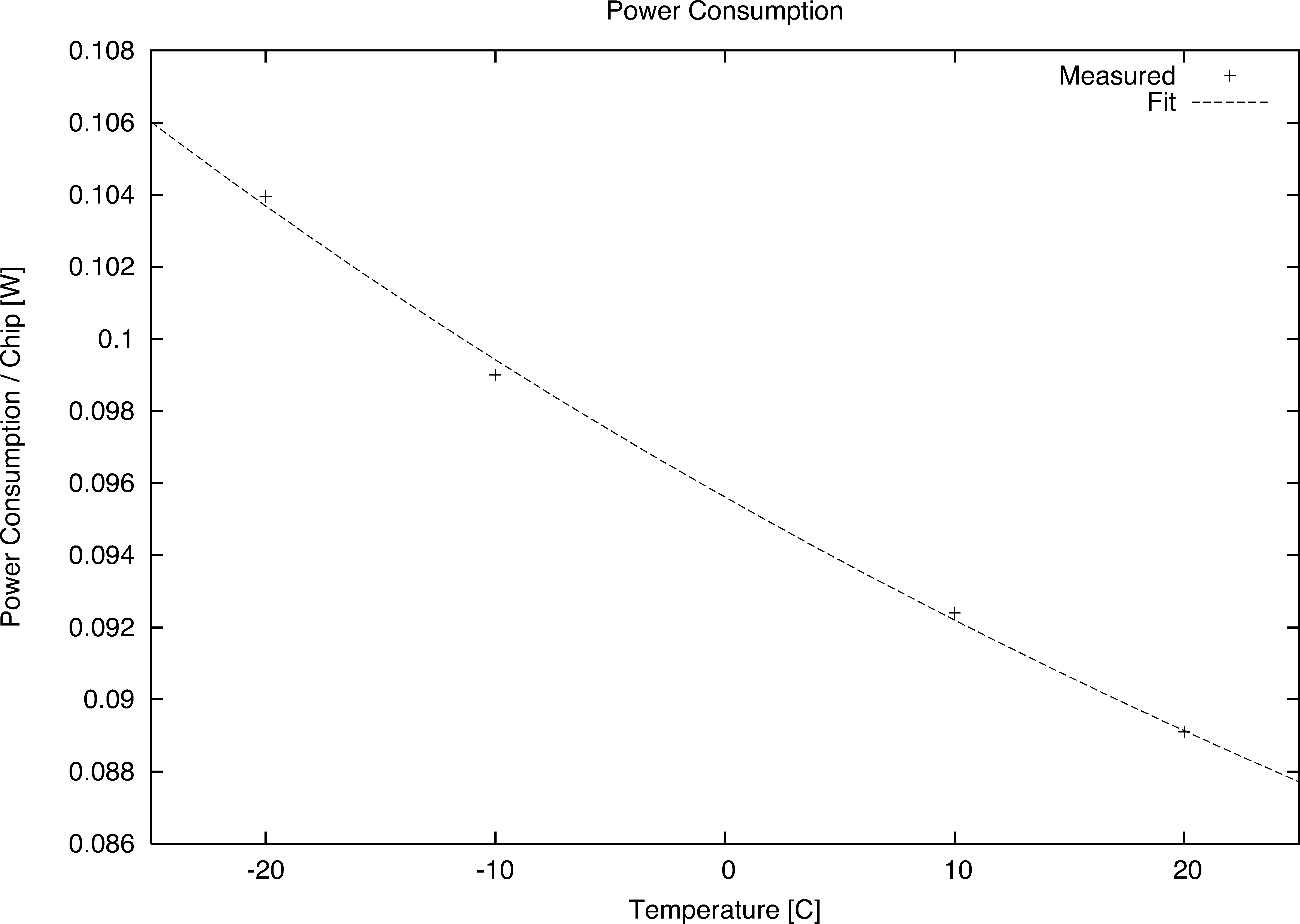}
\caption[Power consumption for one chip in dependence of temperature.]{Power consumption for one chip with two channels in dependence of temperature.}
\label{fig:elo:power}
\end{figure}

In \Reffig{fig:elo:power} the dependence of power consumption on the temperature is shown. For a temperature of $T=$ $+20\degC$ the power consumption for one channel amounts to 10 mW for the charge sensitive amplifier,
to 15$\,\mW$ for the shaper stages and to 17 mW for the buffers. In addition there are about 3$\,\mW$ for the bias circuit and the digital part. So the overall power consumption 
for one chip with two channels is 90$\,\mW$ at $+20\degC$.
At the temperature of $-20\degC$ the power consumption slightly increases to 104 mW for one chip with two channels.
Neither simulations nor measurements revealed any significant dependence of the power consumption on the event rate
for an event rate up to 350$\,\kHz$.

A compilation of the measured results is given in \Reftbl{tab:elo:parameter}.

\begin{table}[htb]
\begin{center}
\begin{small}
\begin{tabular}{lr@{\ }l}
Parameter&& \\ \hline
ENC (-20$\degC$, 270$\,\pF$)& 4146 $\pm$ 131&$e^-$ \\
Max. input charge& 7.84&pC \\
Dynamic Range&11900& \\
Max. Event rate&350&kHz \\
Peaking Time&250&ns\\
Power Consumption&104&mW \\  
(2 channel, -20$\degC$)&& \\ \hline
\end{tabular}
\end{small}
\end{center}
\caption[Measured preamplifier parameters.]{Measured preamplifier parameters.}
\label{tab:elo:parameter}
\end{table}

\section{Preamplifier and Shaper for \FWEMC VPT-readout}
\label{sec:elo:VPT} 
\subsection{Requirements and Specifications}
\label{sec:elo:VPT:Require} 
A discrete charge preamplifier, the Low Noise / Low Power Charge Preamplifier (LNP-P) has been developed in first instance for the LAAPD readout of the \BEMC and was implemented in the \BEMC prototype detector. It has an excellent noise performance in combination with low power consumption. This preamplifier has been further developed and adapted for the readout of Vacuum Photo Triodes (VPTs). VPTs will be used as photo detectors in the \FWEMC. The VPT is attached to the end face ($26\times26\,\mm^2$) of the lead tungstate scintillating crystals (\PWOII) which have a length of 200~mm. Due to the higher event rates and large photon energies in the \FWEMC with respect to the \BEMC, the APDs would suffer from radiation damage (increased noise) and from the nuclear counter effect. Therefore, APDs are not suitable at that position of the EMC. The VPT is a single-stage photomultiplier with only one dynode which can also be operated in a strong magnetic field without loosing significantly in gain. The VPT translates the scintillating light of the \PWOII crystals into an electrical charge which is linearly converted by the LNP-P to a positive voltage pulse; this output pulse is then transmitted via a 50~$\Omega$ line to the subsequent electronics. The low capacitance of the VPT favors the development of a faster low-noise preamplifier with discrete components which is suitable for the high-rate environment. 
For the following design considerations, we adopt the parameters of the VPT type RIE-FEU-188 used in the CMS ECAL. A new VPT with a significantly higher quantum efficiency ($>$~30\%) combined with a larger internal gain ($>$~40) is under development and will be produced by the company Photonis. By using this new VPT, the energy noise level will be reduced remarkably.

\subsubsection{Power Consumption}
\label{sec:elo:VPT:Require:Power} 
\begin{figure*}
\begin{center}
\includegraphics[width=\linewidth]{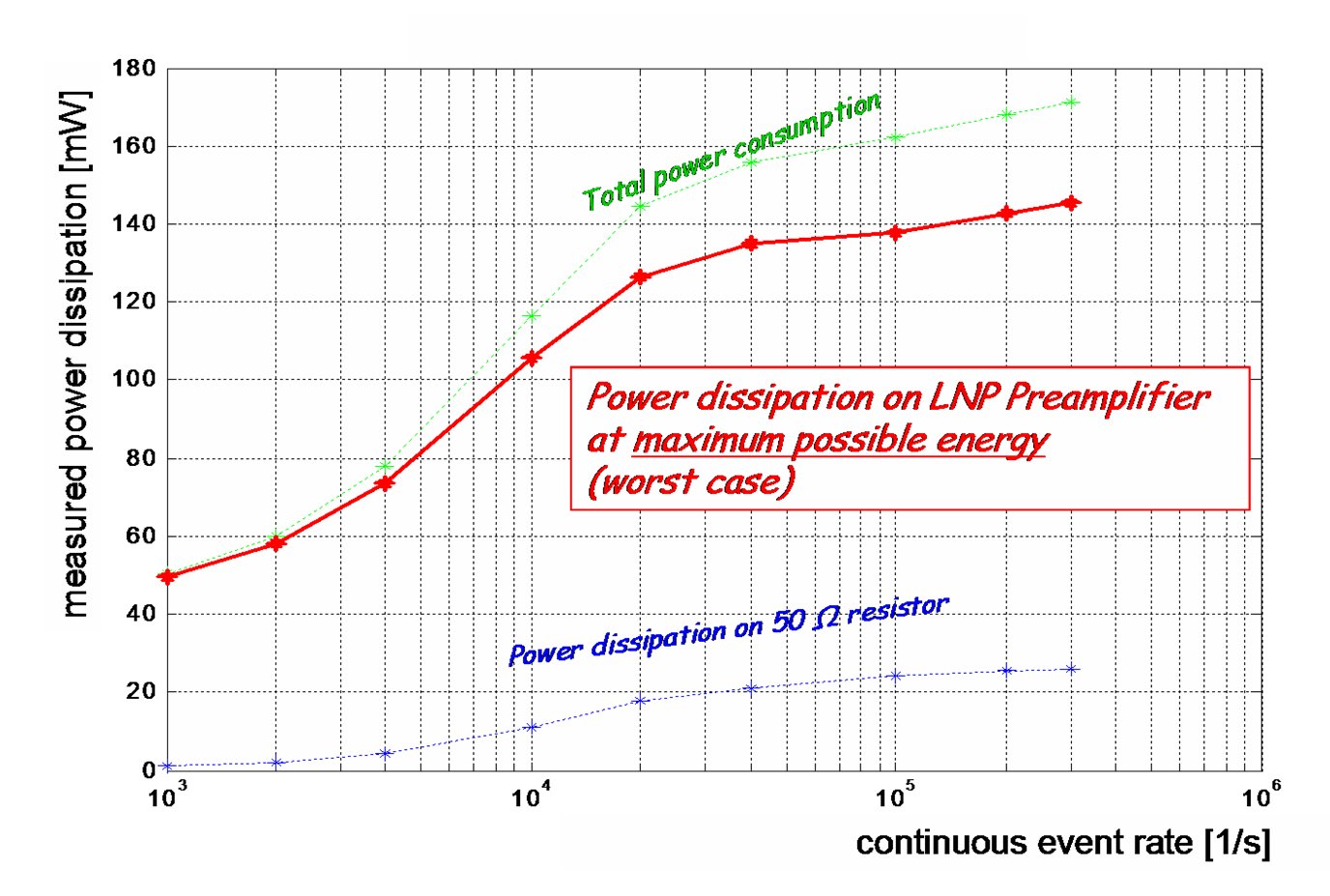}
\caption[Power dissipation as function of continuous event rate.]
{Power dissipation as function of continuous event rate for the LNP preamplifier for the VPT readout.}
\label{fig:elo:LNPpower-dissipation}
\end{center}
\end{figure*}

Since the complete \FWEMC, including VPTs and preamplifiers, will be cooled to low temperatures (-25$\degC$) to increase the light-yield of the \PWOII crystals, the power dissipation of the preamplifier has to be minimized. Low power dissipation leads to a smaller cooling unit and thinner cooling tubes; it also helps to achieve a uniform temperature distribution over the length of the crystals. The LNP-P has a quiescent power consumption of 45 mW. 
The power dissipation is dependent on the event rate and the photon energy; \Reffig{fig:elo:LNPpower-dissipation} shows the measured power dissipation on the LNP-P versus count-rate in the worst case of maximum output amplitude. Since the high rates are predominantly occurring at lower energies, a reasonable maximum power consumption of 90 mW can be presumed.

\subsubsection{Noise}
 \label{sec:elo:VPT:Require:Noise} 
To reach the required low detection threshold of only several MeV, the noise performance of the preamplifier is crucial. The VPT has an outside diameter of 22 mm and a minimum photocathode diameter of 16 mm (see \Reftbl{tab:photo:VPT:specs}), resulting in an active area of ca. 200~mm$^2$. This area is the same as the combined active area of two rectangular 
$7\times14\,\mm^2$ LAAPD. The VPT anode capacitance is around 22~pF which is more than 10 times lower than the capacitance of the LAAPD; this results in a much lower noise from the LNP-P. Thus, the shielded cable between the VPT and the LNP-P has a significant impact on the total detector capacitance; it must be kept as short as possible. 
The dark current of the VPT (1~nA) is significantly lower than the one from the LAAPD (50~nA), both measured at room temperature. The quantum efficiency of the standard available VPT type RIE-FEU-188 is about 20\%, compared to 70\% of the LAAPD. Further, the internal gain of the VPT is around ten, which is five times lower than that of the LAAPD. 
The noise floor of the LNP-P at -25$\degC$ loaded with an input capacitance of 22~pF, has a typical equivalent noise charge (ENC) of 235 e$^-$ (rms). This is measured with an ORTEC450 shaping filter/amplifier with a peaking-time of 650~ns. Because the VPT has almost no dark current, the noise is not increased due to the leakage current of that photo detector. 

As already discussed in \Refsec{sec:elo:Barrel:Require:Noise}, measurements of PWO light production yield 500 photons/MeV at the end face ($26\times26\,\mm^2$ for the \FWEMC) of the cooled (-25$\degC$) \PWOII crystal. This results in 150 photons/MeV on the active area of the VPT. By applying the quantum efficiency and the internal gain of the VPT, a primary photon with the energy of 1~MeV induces an input charge of 48~aC (300 e$^-$) to the preamplifier. So an ENC of 235 e$^-$(rms) corresponds to an energy noise level of 0.78 MeV (rms). This is about the same level as achieved under the same conditions in the \BEMC with the LAAPD readout (0.9~MeV (rms)). Therefore, the signal to noise is in the same order when using a VPT or an LAAPD for the readout of a \PWOII crystal. 
The noise level is increased as the shaping time is decreased. Shorter shaping times are mandatory to cope with the expected high event rates in the \FWEMC. By decreasing the peaking-time from 650~ns (reference values) to 200~ns, the noise level is raised by around 25\%. So, the noise floor with the more realistic shaping using a peaking-time of 200~ns corresponds to about 1$\,\MeV$ (rms) for the presently available VPT.
As already mentioned in \Refsec{sec:elo:VPT:Require}, a new VPT with a significantly higher quantum efficiency (ca. 30\%) combined with a larger internal gain (ca. 40) is under development and will be produced by Photonis. 
By applying these values for quantum efficiency and the internal gain of the VPT, a primary photon with the energy of 1~MeV induces an input charge of 290$\,\aC$ (1800 e$^-$) to the preamplifier. In this case the ENC of 235 e$^-$(rms) corresponds to a significantly reduced energy noise level of 160$\,\keV$ (rms) for a peaking-time of 200$\,\ns$.

\subsubsection{Event Rate}
\label{sec:elo:VPT:Require:Rate} 
The expected event rate in the \FWEMC is maximum 500$\,\kHz$ per crystal. The LNP-P has a feedback time constant of 25~$\mu$s. This feedback time constant is a trade-off between noise performance and pile-up problematic. Reducing the feedback time constant by a factor of two will increase the noise by about 10\%.  
For a single pulse (or very low rates) the LNP-P accepts an input charge of up to 4 pC; for a continuous event rate of 500$\,\kHz$ an input charge of up to 8~pC is allowed. This discrepancy is due to the following reason: A single output pulse starts from zero output voltage and is limited by the positive supply voltage (+6 V) of the LNP-P. At high continuous event rates the output pulses will swing between the negative (-6 V) and the positive (+6 V) supply voltage; therefore the maximum input charge is doubled. If a 500$\,\kHz$ event rate is applied abruptly (burst) to the LNP-P it takes around one second until a continuous input charge of up to 8~pC is allowed. During that transition period, a maximum input charge of 0.3~pC can be handled. With this charge restriction, the output voltage of the preamplifier stays always in the linear range and is never limited by the power supply voltages. Nevertheless, the electronics after the preamplifier has to perform a good base-line correction, because at higher rates it is likely that one pulse sits on the trailing edge of the previous one.
If a charge of 48$\,\aC / \MeV$ (290$\,\aC / \MeV$) is assumed from the VPT (see \Refsec{sec:elo:VPT:Require:Noise}) the maximum expected photon energy deposition of 12~GeV per crystal results in an input charge of 0.58$\,\pC$ (3.5$\,\pC$). 
Under the expected operational conditions with high rates predominantly occurring at low energy, the LNP-P will not be restricted by pile-up even with the relatively long feedback time constant of 25$\,\mu\s$.

\subsubsection{Bias Voltage}
\label{sec:elo:VPT:Require:Bias} 
Because the anode of the VPT will be referenced to ground by the LNP-P, the photocathode (PC) and the dynode (DY) must be biased with negative high voltages (HV). To have the input of the preamplifier referenced to ground has the advantage that any noise on the HV supply is not directly coupled into the charge sensitive input. The typical bias voltages for the VPT type RIE-FEU-188 are: Photocathode: V$_{PC}$ = -1000 V; Dynode: V$_{DY}$ = -250 V. 
Even if these bias voltages do not directly couple to the charge input of the preamplifier, they have to be cleaned from external noise by an efficient low pass (LP) filter before they are wired to the VPT. Also, if all the VPT are biased with the same two high voltages, each VPT must have its own LP filter to prevent crosstalk. These LP filters for the two negative bias voltages (V$_{PC}$, V$_{DY}$) are not integrated on the preamplifier printed circuit board (PCB). A separate LP filter board has to be used; to minimize the noise level it is important that the ground of this LP filter board is tightly connected to the ground of the LNP-P. During the prototyping phase it is reasonable that both bias voltages of the VPT can be adjusted independently. In the final realization a passive voltage divider can eventually be used to generate the two bias voltages.     
At the maximum event rate of 500$\,\kHz$ with the maximum expected photon energy deposition per crystal of 12~GeV (0.58 pC from the VPT) a mean current of 290~nA is flowing through the VPT. This current is mainly drawn from the dynode bias supply. Since the internal gain of the VPT varies only by about 0.1\%/V, the voltage drop over the LP filter for the VPT dynode bias is not critical.  

\subsubsection{Dynamic Range}
\label{sec:elo:VPT:Require:Range} 
As explained in the \Refsec{sec:elo:VPT:Require:Rate}, the LNP-P is designed for a single pulse charge input of maximum 4~pC. With an input charge of 48$\,\aC / \MeV$ (290$\,\aC / \MeV$) coming from the VPT (see \Refsec{sec:elo:VPT:Require:Noise}) this corresponds to a maximum photon energy of 83$\,\GeV$ (14$\,\GeV$). Therefore the dynamic range of the LNP-P is restricted by the noise floor only. Thus, the specification of a dynamic range is strongly dependent on the applied shaping filter. 
In principle, the energy range of the LNP-P spans from the noise floor of 1$\,\MeV$ (rms) (presently available VPT, peaking-time of 200~ns, see \Refsec{sec:elo:VPT:Require:Noise}) up to the maximum input charge corresponding to an energy of 83 GeV; this corresponds to a theoretical dynamic range of 83000. In practice, the typical energy range will start at 2$\,\MeV$ (2$\cdot \sigma_{noise}$) and end at 12~GeV which corresponds to an effective dynamic range of 6000. 

\subsubsection{FADC Readout}
\label{sec:elo:VPT:Require:FADC} 
For the readout with a flash ADC (FADC) the same arguments apply as discussed in \Refsec{sec:elo:Barrel:Require:FADC}.
The distance from the preamplifier in the cold volume to the digitizing electronics in the warm environment is maximally 110$\,\cm$ which can be bridged with flat cables and differential signal lines. Therefore also in this case the  
anti-aliasing low-pass filter/amplifier will be placed right in front of the FADC.

\subsection{Circuit Description}
\label{sec:elo:VPT:Circuit} 
\begin{figure*}
\begin{center}
\includegraphics[width=\linewidth]{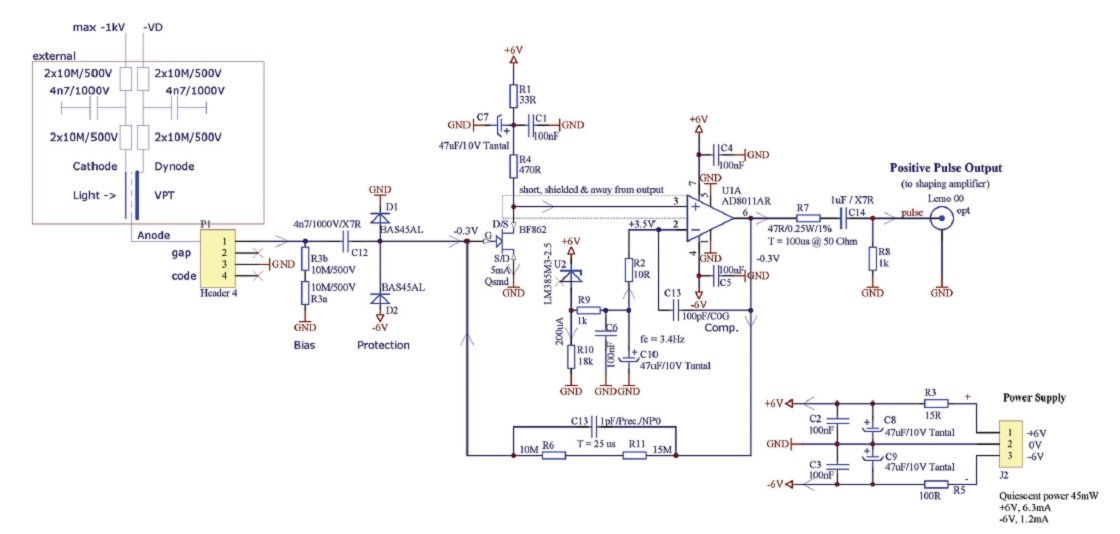}
\caption[Circuit diagram of the LNP-P prototype for the VPT readout.]
{Circuit diagram of the LNP-P prototype for the VPT readout. The flexibility of the discrete design allows easy modifications in the future development process. The HV filter indicated in the top left is not integrated on the preamplifier PCB. This is the revision 1 of the LNP-P and it has the identification number SP 883a01.}
\label{fig:elo:VPTcircuit-diagram}
\end{center}
\end{figure*}

The LNP-P (Version SP 883a01) for the VPT is a further development of the charge preamplifier described in \cite{bib:emc:elo:TDRpreamp}. Some modifications on the circuit are made and a couple of components are changed to SMD types.
The circuit diagram of the LNP-P is shown in \Reffig{fig:elo:VPTcircuit-diagram}. The AC-coupled input stage consists of a low-noise J-FET of the type BF862 from the company NXP Semiconductors (former Philips). This industrial standard J-FET is often used in preamplifiers of car radio receivers. It is specified with a typical input voltage noise density of 0.8$\,\nV/\sqrt{\Hz}$ at 100$\,\kHz$ and at room temperature. The J-FET input capacitance is 10~pF and the forward transductance is typically 30~mS at a drain-source current (IDS) of 5~mA. Along with the 470~$\Omega$ AC-dominant drain resistor this transductance results in a typical AC-voltage gain of 14 for the J-FET input stage. The gate of the J-FET is protected against over-voltages by two low-leakage silicon diodes of the type BAS45AL.
The input stage is followed by a broadband (300$\,\MHz$), fast (2000$\,\V/\s$) and low power (1$\,\mA$) current feedback operational amplifier of the type AD8011AR from the company Analog Devices. With its typical input voltage noise density of only 2$\,\nV/\sqrt{\Hz}$ at 10$\,\kHz$, this amplifier suits well for such a low noise design.
The proper frequency compensation is performed by the capacitor C13 (100~pF), in combination with R2 (10~$\Omega$); this leads to high-frequency feedback to the inverting input of the operational amplifier. Overshoot and ringing can be efficiently suppressed and this compensation also prevents from oscillations when no VPT is connected. 

The output of the operational amplifier is DC-coupled via the feedback network (1~pF $||$ 25~M$\Omega$) to gate of the J-FET. In parallel the output is AC-coupled via a 1~$\mu$F capacitor and a 47~$\Omega$ series resistor to the output of the LNP-P. Therefore, the output voltage is divided by a factor of two if it is terminated with 50~$\Omega$. 

With a symmetrical supply voltage of $\pm$6~V the output voltage can swing symmetrically between the positive and negative supply when high continuous event rates at high energies occur. The LNP-P draws a typical quiescent current of 6.3~mA from the +6~V supply and 1.2 mA from the -6~V supply; this leads to a total power consumption of only 45~mW. 

To set the 5~mA operating point of drain-source current through the J-FET, a gate voltage in a range of -0.2~V to -0.6~V (typically -0.3~V, depending on the DC characteristics of the individual J-FET) has to be applied. This negative DC voltage is fed from the output of the operational amplifier via the 25~M$\Omega$ resistor to the gate of the J-FET. The operating point (I$_{DS}$ = 5~mA) is fixed by the well filtered DC voltage applied to the inverting input of the operational amplifier. This set-point voltage is obtained by subtracting 2.5~V from the positive supply voltage (+6~V) by using a 2.5~V reference diode. So, the same voltage drop of 2.5~V must also be present over the total drain resistor of 503~$\Omega$ (470~$\Omega$ + 33~$\Omega$); this results in a stabilized DC drain current of 5~mA. 

As shown in \Reffig{fig:elo:VPTcircuit-diagram} the anode of the VPT is referenced to ground by a 20~M$\Omega$ resistor and the gate input of the J-FET is decoupled by a 4.7~nF high voltage capacitor.

As already discussed in the \Refsec{sec:elo:VPT:Require:Bias}, the voltage drop over the LP filter for the VPT bias voltage has to be proven. At high rates in combination with high energies, a maximum current of 240~nA is flowing through the VPT, mainly drawn from the dynode bias voltage supply. The planned series resistance of the LP filter is 40~M$\Omega$, resulting in a maximum voltage drop of about 10$\,\V$. By using the typical gain sensitivity of 0.1\%/V of the VPT, this voltage drop corresponds to a maximum energy/rate error of 1\%. By reducing the series resistance of the LP filter, this energy/rate error can be kept at an acceptable level.

\subsection{Performance Parameters}
\label{sec:elo:VPT:Perform} 

\begin{figure}[tb]
\includegraphics[width=\linewidth]{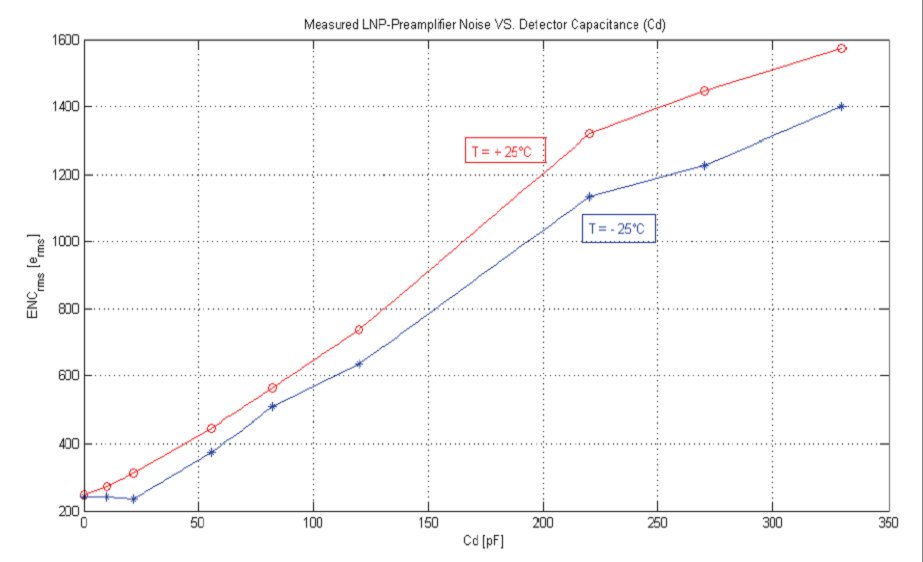}
\caption[The measured noise performance of the LNP-P.]
{The measured noise performance of the LNP-P versus the detector capacitance (C$_d$) at room temperature and at -25$\degC$. Measurements are performed by using an ORTEC450 Research Amplifier with T$_{int}$ = 250~ns and T$_{diff}$ = 2~$\mu$s which corresponds to a peaking-time of 650~ns. One can notice the strong decrease of the noise if the detector capacitance drops from 270~pF for a LAAPD, 
(1250 e$^-$ (rms)) to 22~pF for the VPT (235 e$^-$ (rms)). At high event rates, a more adequate shaping filter with a peaking-time of 200~ns must be used; in that case the noise for a detector capacitance of 22~pF is increased by 25\%, which results in an ENC of around 300 e$^-$ (rms). }
\label{fig:elo:VPTnoise-performance}
\end{figure}

A summary of the LNP-P (Version SP 883a01) performance and specifications is given below:
\begin{itemize}
\item J-FET  (BF862,  NXP Semiconductors) in combination with a low power, high-speed current-feedback operational amplifier (AD8011AR, Analog Devices)

\item Supply +6$\,\V$ at 6.3$\,\mA$, -6$\,\V$ at 1.2$\,\mA$ leading to 45$\,\mW$ quiescent power consumption

\item Rise-time 13$\,\ns$ at C$_d$ = 22$\,\pF$ 

\item	Feedback time-constant 25$\,\mu \s$

\item	Gain: 0.5$\,\V/\pC$ at 50 $\Omega$ termination
\item	Maximum single pulse input charge: 4$\,\pC$ 
\item	Maximum 500$\,\kHz$ burst input charge: 0.3$\,\pC$ 
\item Maximum continuous 500$\,\kHz$ input charge: 8$\,\pC$ 

\item	Single channel LNP-P version: PCB size $48\times18\,\mm^2$

\item	Typical noise performance at -25 $\degC$ and C$_d$ = 22$\,\pF$ (see also \Reffig{fig:elo:VPTnoise-performance})
\begin{itemize}
\item	ENC = 235 e$^-$ (rms) (shaping with a peaking-time of 650$\,\ns$)
\item	ENC = 300 e$^-$ (rms) (shaping with a peaking-time of 200$\,\ns$)
\end{itemize}
\end{itemize}
The single-ended output of the LNP-P is designed to drive a 50~$\Omega$ transmission line. The charge sensitivity is 0.5$\,\V/\pC$ and so the maximum input charge of 4$\,\pC$ corresponds to a positive output pulse with a peak voltage of 2$\,\V$ at 50$\,\Omega$. 
As an incident photon energy of 100$\,\MeV$ corresponds to a pulse peak of only 2.4$\,\mV$, the subsequent electronics has also to be designed with low-noise performance. If the following electronics is located at distances of several 10$\,\cm$ from the preamplifier, it may be necessary to add an additional amplifier onto the LNP-P printed circuit board. Advantageously an amplifier with a differential output driver should be integrated because differential signals are less sensitive to noise-pickup caused by an improper ground system. The differential driver AD8137YR from the company Analog Devices seems to be applicable for an extra gain of 5, while capable to drive a 120$\,\Omega$ terminated differential line. It has a typical voltage noise density of only 9$\,\nV/\sqrt{\Hz}$ at 10$\,\kHz$. By using such an additional amplifier/driver, the quiescent power consumption of the preamplifier would increase to around 85$\,\mW$ and a larger rise time of about 20$\,\ns$ is expected at a detector capacitance of 22$\,\pF$. Also more space on the printed circuit board of the LNP-P would be needed for such an additional amplifier.

The LNP-P (Version SP 883a01) can handle detector capacitances in a range from 0~pF to 250~pF. To reach an optimal rise time, the frequency compensation of the amplifier can be tuned by a capacitor. For different ranges of detector capacitances the frequency compensation must be matched. The actual frequency compensation is suitable for detector capacitances in a range of 0~pF to 100~pF. It results in a short rise-time of only 13$\,\ns$ at a detector capacitance of 22~pF; this allows precise timing measurements. The anode of the VPT is connected to the LNP-P via a short and shielded cable. 

\subsection{SPICE Simulations}
\label{sec:elo:VPT:Spice} 
\begin{figure}[tb]
\includegraphics[width=\linewidth]{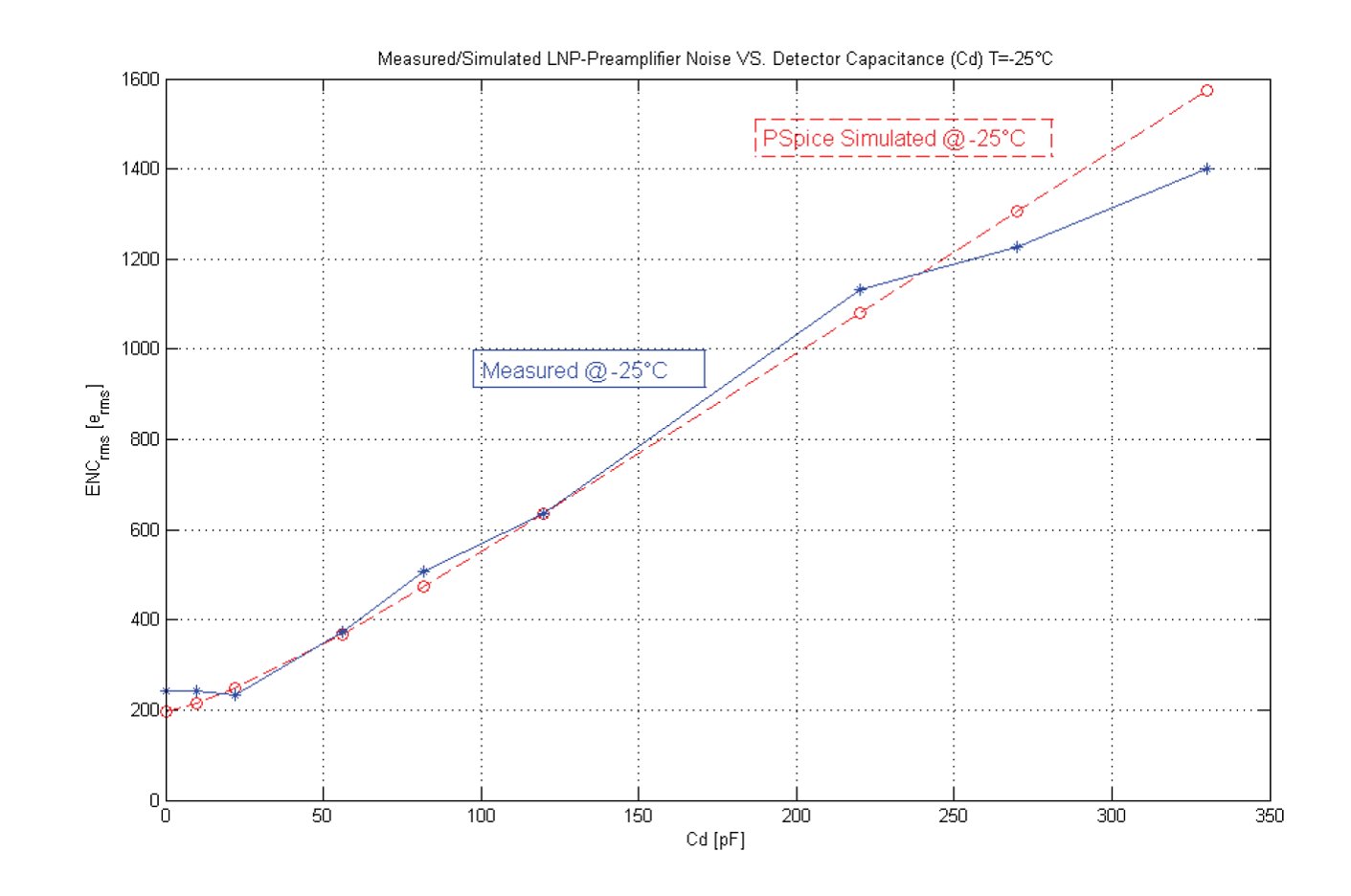}
\caption[PSPICE simulation of ENC versus the detector capacitance.]
{PSPICE simulation of ENC versus the detector capacitance (dashed red) together with the measured ENC (blue line), both at -25$\degC$. The simulation and the measurement are in very good agreement over the entire capacitance range.}
\label{fig:elo:VPTsimulationENC}
\end{figure}

A precise SPICE model of the LNP-P including the shaping filter (peaking-time 650~ns) has been developed. 
The LNP-P circuit is based on the SPICE models of the BF862 (March 2007, NXP Semiconductors) and the model of the AD8011 (Rev. A 1997, Analog Devices). The shaping filter is modeled noiseless by using the Laplace block from the analog behavioral modeling (ABM) library. All simulations are made with PSPICE version 16.0 from the company Orcad/Cadence. An example of the good agreement between simulation and measurement is given in \Reffig{fig:elo:VPTsimulationENC}. 

%

\section{APD Timing Performance with FADC Readout}
\label{sec:elo:Timing} 
\begin{figure}[tb]
\includegraphics[width=\linewidth]{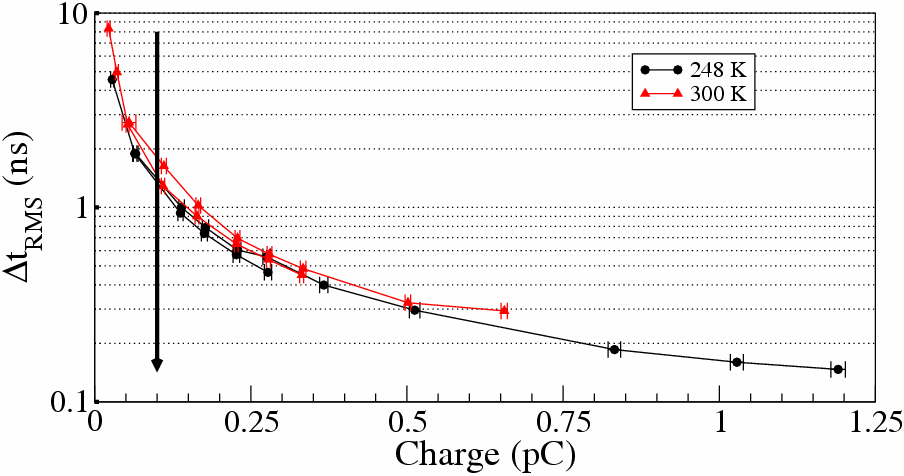}
\caption[Time resolution as function of the corresponding pulse energy.]
{Time resolution as function of the corresponding pulse energy.}
\label{fig:elo:SADCtimeResolution}
\end{figure}

The EMC readout electronics is being designed to provide the best
possible energy resolution and highest dynamic range. However,
a time resolution in the order of at least 1 ns is desirable to reject
background hits or random noise. To investigate the timing performance
of the APD readout a series of test measurements was performed. The experimental
setup consisted of two Hamamatsu APD S8664-1010 mounted on
opposite sides of a 150$\,\mm$ long \PWOII crystal. The crystal was
mounted in an alcohol-cooled aluminium case placed in a
dry-nitrogen flooded dark box. Light pulses of 3 ns rise time
and variable intensity were supplied by a LED pulser using quartz
fibers, coupled perpendicularly to either crystal end face.
The APD signals after the LNP preamplifier were shaped using a newly
developed two-channel two-stage shaper unit and were digitized
by 10 bit 80$\,\MHz$ sampling ADC. The full pulse shape was digitized in typically 60 time samples with 7 time samples in the leading edge of the pulse. Time information for each
pulser event was determined using the method of constant
fraction timing.
\Reffig{fig:elo:SADCtimeResolution} shows the time resolution as function of the collected charge.
The measurements were performed at room temperature and at -25$\degC$. An arrow indicates the position of the cosmic-ray peak with an energy deposit of roughly 22 MeV, which was used to establish the charge-energy calibration. At the lower energies the time resolution is limited by APD- and preamplifier-noise.
These measurements show that it is possible to achieve a time resolution better than 1 ns at energy deposits above 60 MeV and 150 ps at energies above 500 MeV using the proposed sampling ADC readout scheme
of the APD signals.

\section{Digitizer Module}
\label{sec:elo:Digit} 
\begin{figure}[tb]
\includegraphics[width=\linewidth]{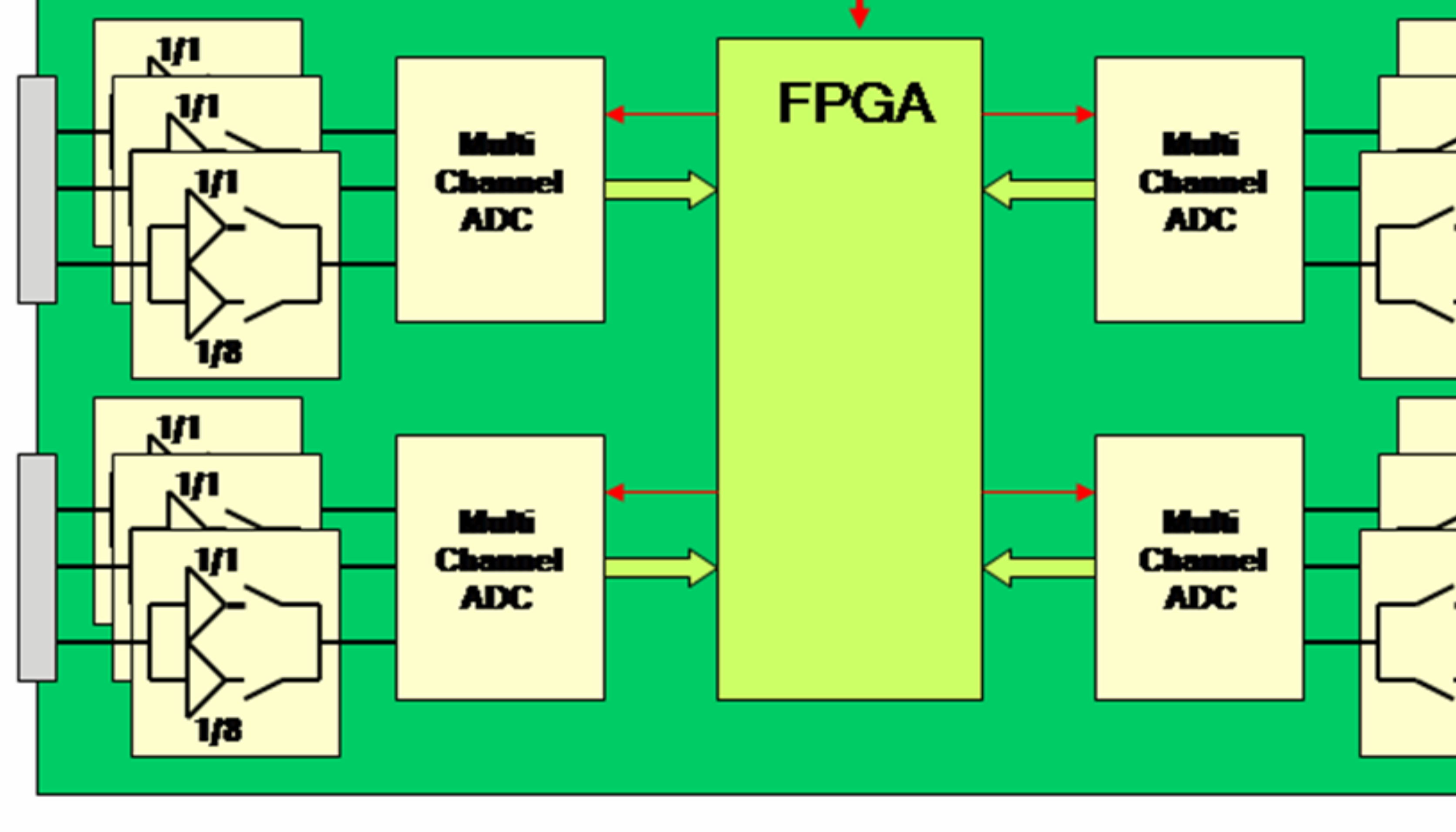}
\caption[The functional diagram of the digitizer module.]
{The functional diagram of the digitizer module.}
\label{fig:elo:adc_card}
\end{figure}

A functional diagram of the digitizer module is shown in \Reffig{fig:elo:adc_card}. The module employs commercial multi-channel 12 bit ADC chips. One module houses up to 120 ADC channels, FPGAs and two fiber optic links.

There will be two versions of the digitizer module with Low and High digitization frequency for the ASIC and for the LNP preamplifier, respectively. In both cases the digitization frequency is a factor of three higher than the frequency of the highest harmonic. The digitization frequency range will be arranged between 
40$\,\MHz$ for Low and 80$\,\MHz$ for High frequencies. These values will be implemented for test experiments with the Proto60 prototype of the \BEMC equipped with a prototype version of the ADC module and the LNP preamplifier.

The ADC chips have a resolution of 12 bit so that with two overlapping amplification ranges (see \Reffig{fig:elo:dynamic})
digitized in two ADC channels the full dynamic range of 12000 can be covered. A special range selection circuit, suitable for operation in conjunction with the LNP preamplifier, is introduced in front of the ADC chip. The circuit multiplexes direct or attenuated signals depending on the signal amplitude. The range selection circuit consists of a comparator, an attenuator and an analogue switch. The switch is synchronized with the ADC clock. Since the switching time is less than one clock period, one sample can be distorted during switching. The two endcaps, equipped with VPT and LNP preamplifier, thus require in total 4192 ADC channels.
The APFEL preamplifier ASIC provides two outputs with different gain and does not require the range selection circuit.
For the independent readout of two LAAPD per crystal we thus require 4$\times$11360 ADC channels for the barrel part.
 
The FPGAs perform the following tasks:
\begin{itemize}
\item time adjustment and distribution of the global clock signal;
\item noise calibration;
\item common mode noise suppression;
\item pedestal subtraction;
\item autonomous hit detection;
\item conversion of ADC data and linearization of the full data range;
\item transporting the hit information together with the time stamp to the data multiplexer;
\item slow control.
\end{itemize}

The architecture of the digitizer module preserves the redundancy policy, introduced by equipping every crystal with two APDs. The digitizer consists of two blocks of 60 channels each. The blocks have interconnections at the level of FPGAs but may function independently. The EMC channels are mapped in a way that the first APD of the crystal is connected to block one and the second APD to block two.
In case of failure of any component at most 60 channels out of 120
will not provide data. However, during normal operation the data of two APDs of the crystal are merged inside one of the FPGAs by using high-speed links between FPGAs.

An important parameter for the construction of the detector system is the power consumption and the channel density of the readout system.
If one would start the development of the digitizer module today, using presently available commercial components, the power consumption of the digitizer would be below 400$\,\mW$ per channel and the channel density would be about 3$\,\cm^2$/channel.

\begin{figure*}
\begin{center}
\includegraphics[width=0.8\dwidth]{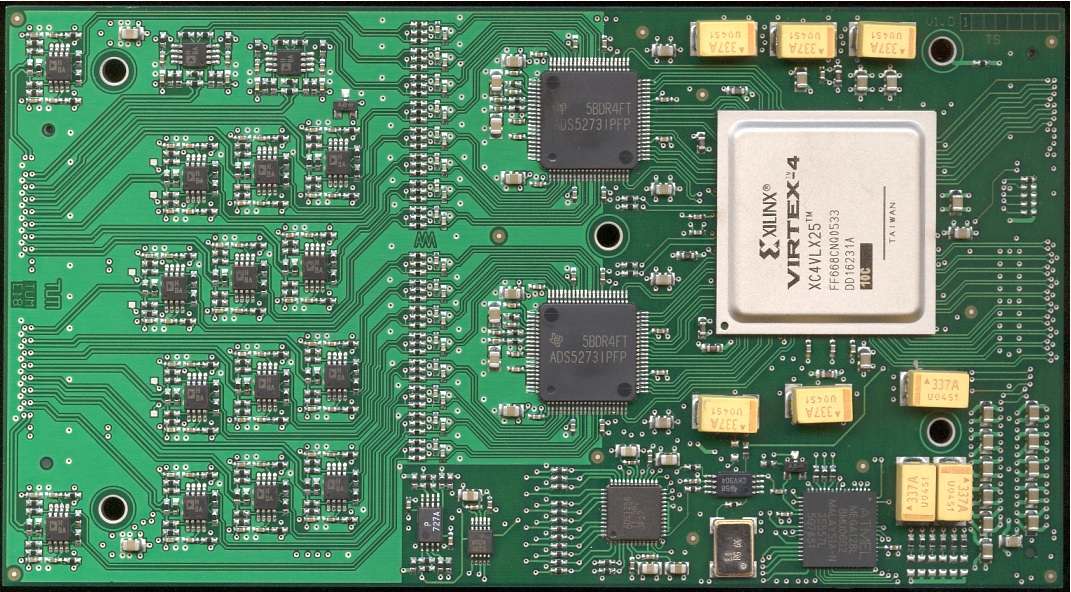}
\caption[The layout of the prototype ADC module.]
{The layout of the prototype ADC module.}
\label{fig:elo:msadc}
\end{center}
\end{figure*}

A prototype ADC module, shown in \Reffig{fig:elo:msadc} 
\cite{bib:emc:elo:msadc}, has a size of $70\times130\,\mm^2$ and contains 32 channels of 12 bit 65 MSPS ADCs. The total power consumption of the module is 15~W. The parameters of this prototype module
are used as a reference for the design of the final readout system. 
The module is part of the setup for experiments with the Proto60 prototype of the \BEMC.

\section{Data Multiplexer}
\label{sec:elo:Multiplex} 
The data multiplexer provides the interfaces between the user program and the front ends, and between the front-end and the DAQ system. The foreseen data multiplexer module will comply to the new Advanced Telecommunication Architecture standard. The following Physical Interfaces will be included:

\begin{itemize}
\item 1 bidirectional 1~Gbit/s optical link to/from the Time Distribution System;
\item 10 bidirectional 1~Gbit/s optical links to/from the front-end electronics (the digitizer modules);
\item 2 bidirectional 2~Gbit/s copper links to/from the backplane to the neighboring multiplexers for  
\item 2 bidirectional 2~Gbit/s links to the DAQ system;
\item 1 Ethernet link to a general purpose network for configuration and slow control.
\end{itemize}

The data multiplexer performs advanced data processing by extracting the signal amplitude and time, combining single hits into clusters, and sorting the clusters in a time-ordered sequence. 
A data flow through the module of up to 200~MBytes/s seems feasible.

For transmission to the data acquisition system a maximum data rate of 4 GByte/s is estimated. This estimate is based on an event rate of 2$\cdot 10^7$ antiproton annihilations per second, a multiplicity of 5 detected clusters with a typical cluster size of 10 crystals, 8 Bytes per hit crystal using two independent readout channels per crystal (as foreseen in the barrel). Further is assumed, that a clusterization process is applied before sending data to the data acquisition system. If clusterization would not be included, the data rate would increase by a factor 2 to 3. 
For 4 GByte/s a number of 40 optical links is required, assuming an optical link capacity of 100 MByte/s, and an equivalent number of data multiplexers. 
%


\section{Signal Routing and Cabling}
\label{sec:elo:Cabling} 

\subsection{Requirements}
\label{sec:elo:Cabling:Require} 
The detector system cooled to -25$\degC$ requires a minimum amount of heat conducting copper into the system. However, to provide shielded, controlled impedance of 50~$\Omega$ signal lines and high voltage insulation, we can not use standard solutions. In Proto60 we designed a rigid multilayer-back-PCB 
(\Reffig{fig:elo:CableStack}). The drawback was the reduction of flexibility in mechanical design.
Possible solutions are, apart from round cables, flat ribbon cable, Flat Foil Cable (FFC), Flat Laminated Cable (FLC), exFC and Flexible Printed Circuit (FPC).
Because of the heterogeneous types of conductors it is planned to implement flexible flat multilayer cables (FPC) instead of conventional bulky flexible solutions with coax cables. 
The signal, $\pm$6~V power supply and the bias voltage (HV) is designed in one 4-Layer cable per channel. 
These flexible cables are usually custom made in different lengths (for \Panda EMC about 350$\,\mm$ for the \BEMC and 1000$\,\mm$ for the \FWEMC) and not available as a standard product.
Long cables are increasing the noise, especially between APD/VPT and the preamplifier, but also between the preamplifier and the Shaper/ADC. Therefore, good shielding is essential.
The temperature sensors (0.2$\,\mm$ thermocouples) and the light guides for calibration are not discussed here.

\subsection{Cable Performance and Specifications for Proto60 Assembly}
\label{sec:elo:Cabling:Proto60} 
\begin{figure*}
\begin{center}
\includegraphics[width=0.7\dwidth]{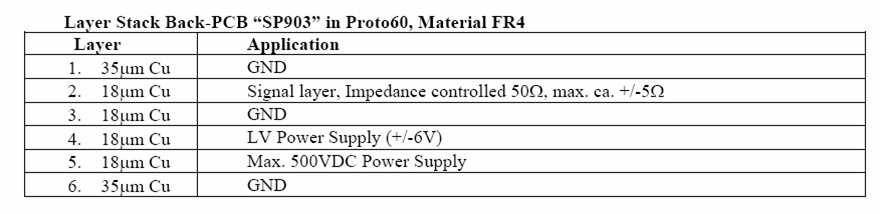}
\caption[The layout of the Layer Stack Back-PCB.]
{The layout of the Layer Stack Back-PCB "SP903" used in the \BEMC prototype Proto60}
\label{fig:elo:CableStack}
\end{center}
\end{figure*}

The distance between the preamplifier and the APD acts as a thermal decoupling to the APD and the crystals but must not be too long because of noise induced into the cable.
A four fold device reduces the amount of cables from the warm to the cooled zone and saves space.
Serviceability is provided through the use of a removable multilayer backplane-PCB 
(\Reffig{fig:elo:CableStack}, which is a cheap and technically suitable solution to distribute supply voltages to the preamplifiers with a minimum amount of copper and to break out to ambient with maximum signal integrity through impedance controlled signal lines. Further connections are then made through Lemo00-connectors and RG174 coax cables. At the moment, the disadvantage of this solution is the limitations imposed on the mechanical construction in the rigid PCB version.
 
From APD to the preamplifier a 70$\,\mm$ long shielded twisted pair cable (Krophon Liff2Y-DY $2\times0.073\,\mm^2$) with an outer diameter of 2.2$\,\mm$ and specified operating voltage of 500 VDC. The flexibility of the cable is sufficient to mount the parts using 2.54$\,\mm$ connectors.

\subsection{Cable Performance and Specifications for \BEMC and \FWEMC}
\label{sec:elo:Cabling:EMC} 
\subsubsection{Barrel}
\label{sec:elo:Cabling:EMC:Barrel} 
\begin{figure*}
\begin{center}
\includegraphics[width=0.7\dwidth]{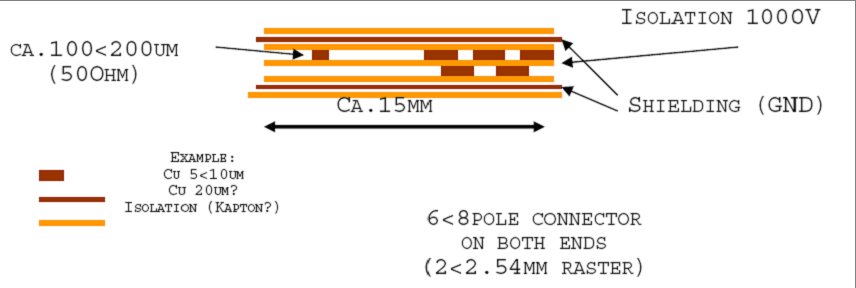}
\caption[Scheme of the layer stack of flat cables.]
{Scheme of the layer stack of flat cables.} 
\label{fig:elo:CableLayerStack}
\end{center}
\end{figure*}

The cabling between preamplifier and ADC for each channel is planned to be provided by a $350\,\mm \times\, ca.\, 15\,\mm$ Flex-PCB flat cable (see scheme in \Reffig{fig:elo:CableLayerStack}). The bias voltage is maximum 500 VDC for the \BEMC.

\subsubsection{Forward Endcap}
\label{sec:elo:Cabling:EMC:FwEMC} 
A new version of the single-channel LNP preamplifier (SP883a01) was designed for use in the \FWEMC with improved stability also for the small capacitance of the VPT's. This preamplifier will be implemented in a "Proto16" VPT-subunit. The cabling between preamplifier and ADC for each channel is planned to be provided by $1000\,\mm \times\, ca.\, 15\,\mm$ Flex-PCB flat cable.
The preamplifier works also with a 50~$\Omega$ line over 1000$\,\mm$ to the ADC or one of the eight patch panels from where RG178 (diameter 1.8$\,\mm$) or other coax cables can be connected.
The maximum bias voltage is 1000~VDC and two high voltages might be supplied to the \FWEMC VPTs. 

From the HV-Filter to the VPT and from the VPT to the preamplifier the distance must be as short as possible. The signal must be shielded and hold the 1000 VDC bias voltage.

The distance between the preamplifier and the Patchpanel/Shaper/ADC is about 1~m. The VPT gives about 10 times lower pulse height (with VPT gain of 10) than the APD at the moment. A new VPT with gain up to 40 is under development.

\subsection{Circuit Description}
\label{sec:elo:Cabling:Circuit} 
The AC coupled signal is transmitted over a 50~$\Omega$ line. A suitable conductor width for an impedance controlled microstrip design is for example 0.2$\,\mm$ on an isolation layer of 0.1$\,\mm$ to ground, but depends on the material specifications.
Crosstalk is minimized by separate shielding of every channel.

\subsection{Manufacturing, Operation, and Safety}
\label{sec:elo:Cabling:Manufact}
The materials are polyimide (e.g. Kapton), copper, tin-lead and are not flammable.
Connectors have only local relevance but care has to be taken also (UL94-V0 approval).
Polyimide (e.g. Kapton) is widely used in accelerator environments.
Soldertype Sn60Pb40 is used to avoid "tin pest" of leadfree solder points at long time temperature exposure of $<$~13$\degC$  ($\beta$-Sn to $\alpha$-Sn transformation).

\subsubsection{Manufacturing and Connectors}
\label{sec:elo:Cabling:Manufact:Connect} 
Economic automatic processing of connectors without soldering can be provided with the use of crimping contacts (e.g. with Schleuniger HFC) or/and press-fit connectors.

\subsection{Alternatives}
\label{sec:elo:Cabling:Alternativ} 
There are alternatives to custom made cables with separated signal line, power supply and HV-cables. This is probably cheaper but needs also more space and has a significantly higher thermal impact, which also leads to higher costs.
An additional differential amplifier stage on the preamplifier can reduce the sensitivity of the signal lines but causes higher power consumption in the cooled stage.
The solution with the ADC directly mounted behind the preamplifier is not recommended because of higher cooling power required at that point.
A resistive voltage divider near the VPT consumes roughly 1$\,\mW$ ($1000\,\V \,\times\, 1\,\mu \A$) but can save a separate bias voltage line.

\section{Detector Control System}
\label{sec:elo:DCS} 
\subsection{Goals}
\label{sec:elo:DCS:Goals} 
The aim of a Detector Control System (DCS) is to ensure the correct and stable operation of an experiment, so that the data taken by the detector are of high quality. The scope of the DCS is therefore very wide and includes all subsystems and other individual elements involved in the control and monitoring of the detector.
The EMC and its subcomponents will therefore be embedded in the more general DCS structure of the complete \Panda detector. Here we will focus primarily on DCS aspects relevant for design and operation of the electromagnetic calorimeter.

The DCS extends from the active elements of the complete setup of the experiment, the electronics at the detector and in the control room, to the communications with the accelerator. The DCS also plays a major role in the protection of the experiment from any adverse occurrences. Many of the functions provided by the DCS are needed at all times, and as a result some parts of the DCS must function continuously, on a 24-hour basis, during the entire year.
The primary function of the DCS will be the overall control of the detector status and its environment. In addition, the DCS has to communicate with external entities, in particular with the run control and monitoring system, which is in charge of the overall control and monitoring of the data-taking process and of the accelerator.
System wide we require the DCS to be:
\begin{itemize}
\item reliable, with respect to safe power, as well as redundant and reliable hardware in numerous places;
\item modular and hierarchical;
\item partitionable in order to allow independent control of individual subdetectors or parts of them;
\item automated to speed up the execution of commonly requested actions and also to avoid human errors in highly repetitive actions;
\item	easily operated such that a few non-experts are able to control the routine operations of the experiment;
\item	scalable, such that new subdetectors or subdetector components can be integrated;
\item	generic: it must provide generic interfaces to other systems, e.g. the accelerator or the run control and monitoring system;
\item	easily maintainable;
\item	homogeneous, which will greatly facilitate its integration, maintenance and possible upgrades, and displays a uniform 'look and feel' throughout all of its parts.
\end{itemize}

\begin{figure}[tb]
\includegraphics[width=\linewidth]{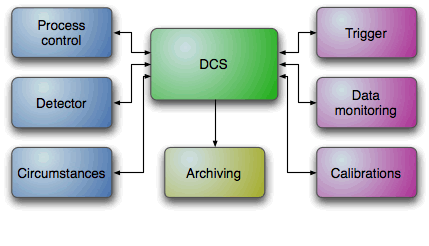}
\caption[The various functions of the Detector Control System.]
{The various functions of the Detector Control System.}
\label{fig:elo:DCStasks}
\end{figure}

The DCS has to fulfill the following functions (see \Reffig{fig:elo:DCStasks}):
\begin{itemize}	
\item process control
\item detector control
\item detector monitoring
\item ambient circumstances control
\item trigger control
\item data monitoring
\item calibrations
\item archiving
\end{itemize}

\begin{figure}[tb]
\includegraphics[width=\linewidth]{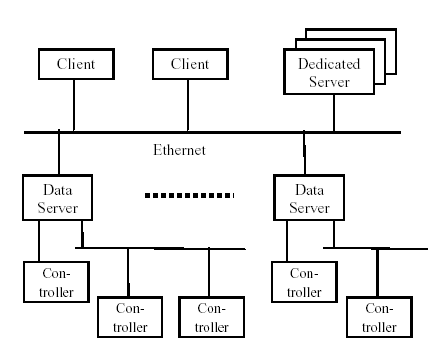}
\caption[The DCS network structure.]
{The DCS network structure.}
\label{fig:elo:DCSnetwork}
\end{figure}

\subsection{Process Control}
\label{sec:elo:DCS:Control} 
The Electromagnetic Calorimeter (EMC) is one part of a complex detector system and it is called a subdetector. The subsystem must be compatible with the requirements imposed on the Detector Control System as stated above. One important requirement is the process control,
realized by the Supervisory Control and Data Acquisition (SCADA) software. The process control supervises the following items:
\begin{itemize}
\item	system status
\item	data flow
\item	run start/stop status
\item critical conditions
\end{itemize}
The definite knowledge of the system status is the heart of each control system. It enables a high level of abstraction and a simplified representation of detector control systems. A finite set of well-defined states is introduced, in which each of its subsystems can be, and rules are defined, that govern transitions between these states. The system status of each subsystem depends on the current state of the underlying hardware. At the same time, the system status enables a logical grouping of DCS subsystems into a hierarchical tree-like structure, where "parent" states are uniquely determined by states of its children and system-specific logic. Each parent in such system status tree can issue an action command to its children. Action commands at the lowest level imply appropriate commands issued to the controlled hardware. The \PANDA EMC control software will be implemented in this way. The software granularity is driven by the EMC subsystem structure. The High Voltage (HV), Low Voltage (LV), cooling, temperature, humidity and safety systems are controlled by independent applications. On top of these applications there is the EMC supervisory application that implements an hierarchical structuring of the whole EMC control software. In addition the EMC DCS applications include numerous other functionalities, such as e.g. full parametrization and visualization of each subsystem, loading from and storing to the \PANDA configuration database the start-up and operational parameters for EMC DCS subsystems.
Among the characteristics of the SCADA system we can distinguish two of them as most important: its capability to collect data from any kind of installation and its ability to control these facilities by sending (limited) control instructions.
The standard SCADA functionality can be summarized as follows:
\begin{itemize}
\item	data acquisition
\item	data logging and archiving
\item	alarm and alert handling
\item	access control mechanism
\item	human-machine interface, including many standard features such as alarm display.
\end{itemize}
Two basic layers can be distinguish in a SCADA system: the 'client layer', which caters for the man-machine interaction, and the 'data-server layer', which handles most of the process data control activities (see \Reffig{fig:elo:DCSnetwork}). The data servers communicate with devices in the field through process controllers.
The latter, e.g. programmable logic controllers (PLCs), are connected to the data servers either directly or by networks or fieldbuses. Data servers are connected to each other and to client stations via an Ethernet local area network (LAN).
There should be also a detector database fully integrated into the SCADA framework. This database should be accessible from the SCADA so that the SCADA features (e.g. alarming or logging) can be fully exploited. This database can incorporate:
\begin{itemize}
\item	firmware
\item	readout settings (thresholds etc.)
\item	hardware settings (e.g. HV, LV)
\item	alignment values
\item	calibration constants.
\end{itemize}

\subsection{Detector Control and Monitoring}
\label{sec:elo:DCS:Monitor} 
The EMC Detector Control System should provide the monitoring of the detector conditions of the on-detector electronics as well as of all EMC subsystems (HV, LV, cooling system, gas system, status of laser monitoring system). All these monitored data should be recorded and archived as part of the common \PANDA 'conditions database'. The DCS also has to provide early warnings about abnormal conditions, issue alarms, execute control actions and trigger hardwired interlocks to protect the detector and its electronics from severe damage. Regarding control functions, the DCS will switch on/off and ramp up/down the HV and LV, as well as set up their operational parameters.
Overall \PANDA will have a hierarchical DCS tree. This tree is a software layer built on top of the experiment controls. Every detector integrates its controls in this tree-like structure. The \PANDA DCS supervisor, which is connected to the \PANDA run control, will sit on top of the tree. In this way the EMC DCS will be directly controlled by the \PANDA supervisor. However, when EMC runs separately (i.e. commissioning) its DCS is under control of an EMC run control. The EMC DCS also has connections to the \PANDA detector safety system. All of the DCS functionalities must be constantly available during the EMC (\PANDA) run time and some functionality practically non-stop (24h/365d) during the whole \PANDA detector life time.
Parts of the DCS functionalities are going to be implemented through software applications running on dedicated DCS computers, as is the case with the HV, LV, cooling and laser monitoring systems. These applications communicate to hardware or to embedded computers using standard network or field-bus protocols.
The other part of the functionalities will be implemented via dedicated DCS applications whose readout systems are completely independent of the EMC DAQ. These are the EMC monitoring system for temperature and humidity, and the monitoring  system for the air temperature of the EMC electronics environment, the water leakage sensors, the proper functioning of the EMC cooling and LV cooling systems, and the control system to automatically perform predefined safety actions and generate interlocks in case of any alarm situation.
The EMC DCS must have an interface to the data acquisition system (DAQ). A communication mechanism between DAQ and the DCS is needed for several purposes. During normal physics data taking, the DCS will act as a slave to the run control and monitoring system and will therefore have to receive commands and send back status information. Partitions will also have to be synchronized between the DAQ and the DCS: the run control and monitoring system will instruct the DCS to set up specific partitions, which will be put under the control of the run control and monitoring system. Furthermore, alarms will be sent from one system to the others.

\section{Production and Assembly}
\label{sec:elo:prod}
\subsection{ASIC preamplifier}
\label{sec:elo:prod:ASIC} 
Since the beginning of the integrated preamplifier development in 2005 two prototypes have been designed. These prototypes have been produced on Multi Project Wafer (MPW) runs of the 
EUROPRACTICE IC prototyping program. At least one more prototype iteration is expected to be needed before a final ASIC design is reached for production.

Typically the designer gets some 20 chips by each MPW prototyping run which is sufficient for testing and device characterisation but may be not sufficient for a detector test 
setup with a medium scale detector array like the Proto60. For such cases the EUROPRACTICE program offers the possibility to process additional wafers on a MPW run. Each additional wafer yields about 50 pieces.
If the decision will be made to adapt the existing integrated preamplifier for the VPT readout of both endcaps of the EMC as well, additional prototypes have to be designed, produced and tested for this versions. 
Also these prototypes can be produced very cost effective by MPW runs.

For instrumentation of the electromagnetic calorimeter about 23000 pieces of the preamplifier are needed for the \BEMC part (2 ASICs for one crystal with 2 LAAPD) and about 5000 pieces for both endcaps (1 ASIC for one crystal with VPT readout). 
These amounts of ASICs can no longer be produced cost effectively with MPW runs, so one has to start a chip production campaign.
Such a campaign starts with the production of a set of photomasks for the lithography steps during chip production. With this mask set an engineering run is started 
to optimize the production parameter for this design. During this engineering run 6 wafers are produced but only 2 wafers are guaranteed to be within the specifications.

\begin{figure}[tb]
\includegraphics[width=\linewidth]{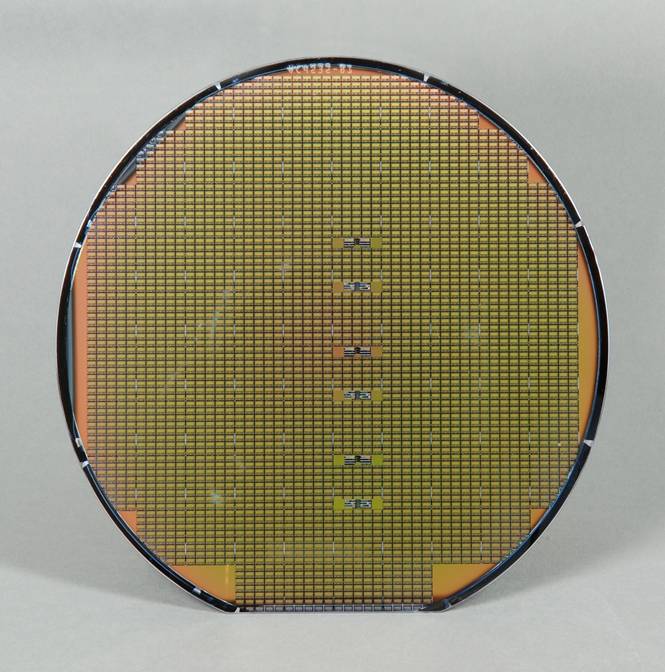}
\caption[Microchip wafer.]
{Microchip wafer.}
\label{fig:elo:wafer}
\end{figure}

With a die size of 10~mm$^2$ and a wafer diameter of 8 inches one will get about 3000 pieces on each wafer, see \Reffig{fig:elo:wafer}, so two wafers do not suffice to get enough preamplifier ASICs 
for the whole detector. That means, after testing the dies produced on the engineering run, one has to order the production of 1 wafer lot in addition which, in case 
of a production run at Austria Mikrosysteme, is 25 wafers.

\begin{figure}[tb]
\includegraphics[width=\linewidth]{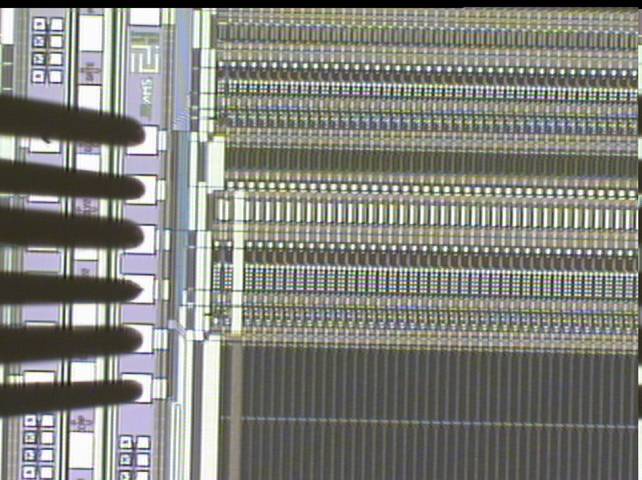}
\caption[An integrated circuit connected by needles during a wafertest.]
{An integrated circuit connected by needles during a wafertest.}
\label{fig:elo:wafertest}
\end{figure}
\begin{figure*}[tb]
\begin{center}
\includegraphics[width=0.7\dwidth]{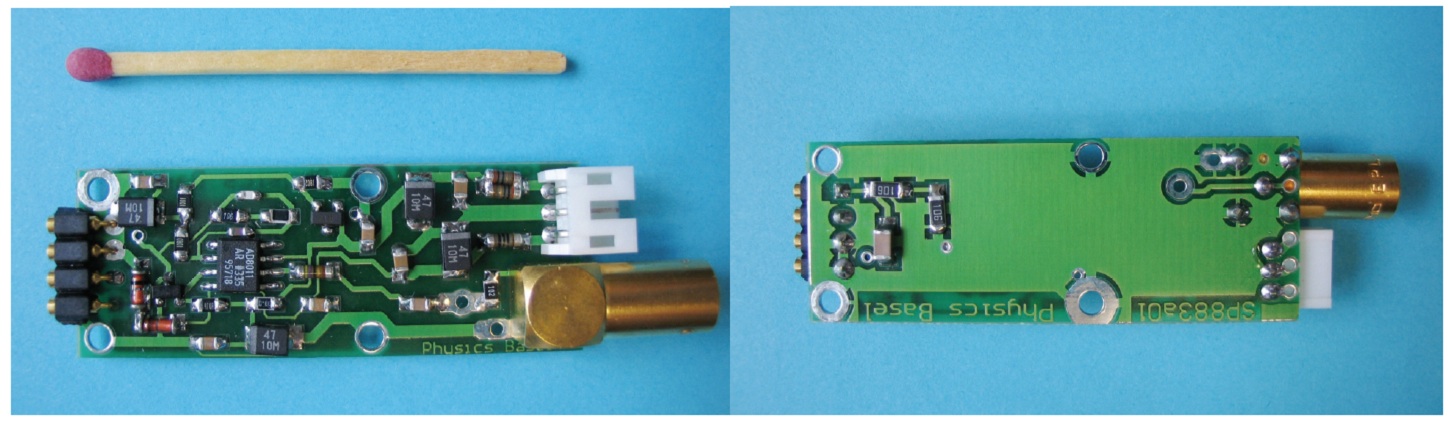}
\caption[The top- and the bottom side of the single-channel LNP-P prototype.]
{The top- and the bottom side of the single-channel LNP-P prototype for the VPT readout. 
It has a PCB size of $48\times18\,\mm^2$ and four holes (connected to ground) with a diameter of 
2.3~mm are foreseen for mounting. The connector for the VPT is on the left side and the supply voltage 
($\pm$6~V) is connected via the white socket on the right. For the testing phase a Lemo-00 connector is equipped at the signal output. On the bottom side the two 10~M$\Omega$ HV resistors and the HV gate decoupling capacitor are located.}
\label{fig:elo:VPTfotoLNP}
\end{center}
\end{figure*}
After wafer production an intensive test phase has to follow. For wafer tests so called needle cards are used to connect the circuits on the wafer temporary. 
These needle cards are special printed circuit boards with very fine needles which are placed to fit the bonding pads of the integrated circuit. On a wafer prober this needle card is 
lowered on the wafer so a single circuit on the wafer can be tested electrically. After the test is finished the needle card is left off and the wafer has to be moved 
by one chip to start the next test. In \Reffig{fig:elo:wafertest} one can see 6 needles connecting an integrated circuit.

For mass tests this procedure has to be done with a semiautomatic prober which can do the lowering, lifting and wafer stepping in an automatic manner. Nevertheless, wafer changing 
and frequently cleaning of the needles has to be done manually.
As all of these tests have to be done 3000 times for each wafer and 81000 times in total, test-algorithms have to be very efficient and fast.
After testing, the chip assembly has to be started by an external assembly company. Assembly for Chip On Board (COB) technology consists of several steps:
\begin{itemize}
\item Sawing wafers into single dies
\item Picking up good dies controlled by a wafermap which was compiled from wafer test results
\item Placing and gluing dies on a PCB
\item Wire bonding the dies
\item Globtop the dies for mechanical protection
\item Placing additional components
\item Soldering discrete components
\end{itemize}

The preamplifier printed circuit board modules produced this way have to be tested once again before they are ready to be mounted on the detector.
\subsection{Discrete preamplifier}
\label{sec:elo:prod:LNP} 
The LNP-P is a simple, robust and low-cost combination of a standard J-FET with a fast integrated operational amplifier. The single-channel version of the LNP-P prototype for the VPT readout is implemented on a small-size double layer printed circuit board (PCB) with the mechanical dimensions of $48\times18\,\mm^2$ (see \Reffig{fig:elo:VPTfotoLNP}).

All components, except the connectors, are surface-mount devices (SMD). Therefore the LNP-P is well suited for automated mass production. 
Due to the discrete design of the LNP-P for the VPT readout, adaptations and modifications can be smoothly made in the future. For example, the power consumption can be easily reduced by changing only the value of two resistors. Of course, lowering the power consumption would also increase the noise level of the preamplifier. 


%
%
\newpage
\bibliographystyle{panda_tdr_lit}
\bibliography{./lit_emc}
%

%
\cleardoublepage
\chapter{Mechanics and Integration}
\label{sec:mech}
%
%
\label{sec:emc:mech:int}

\begin{figure*}[h!]
\begin{center}
\includegraphics[width=0.75\dwidth]{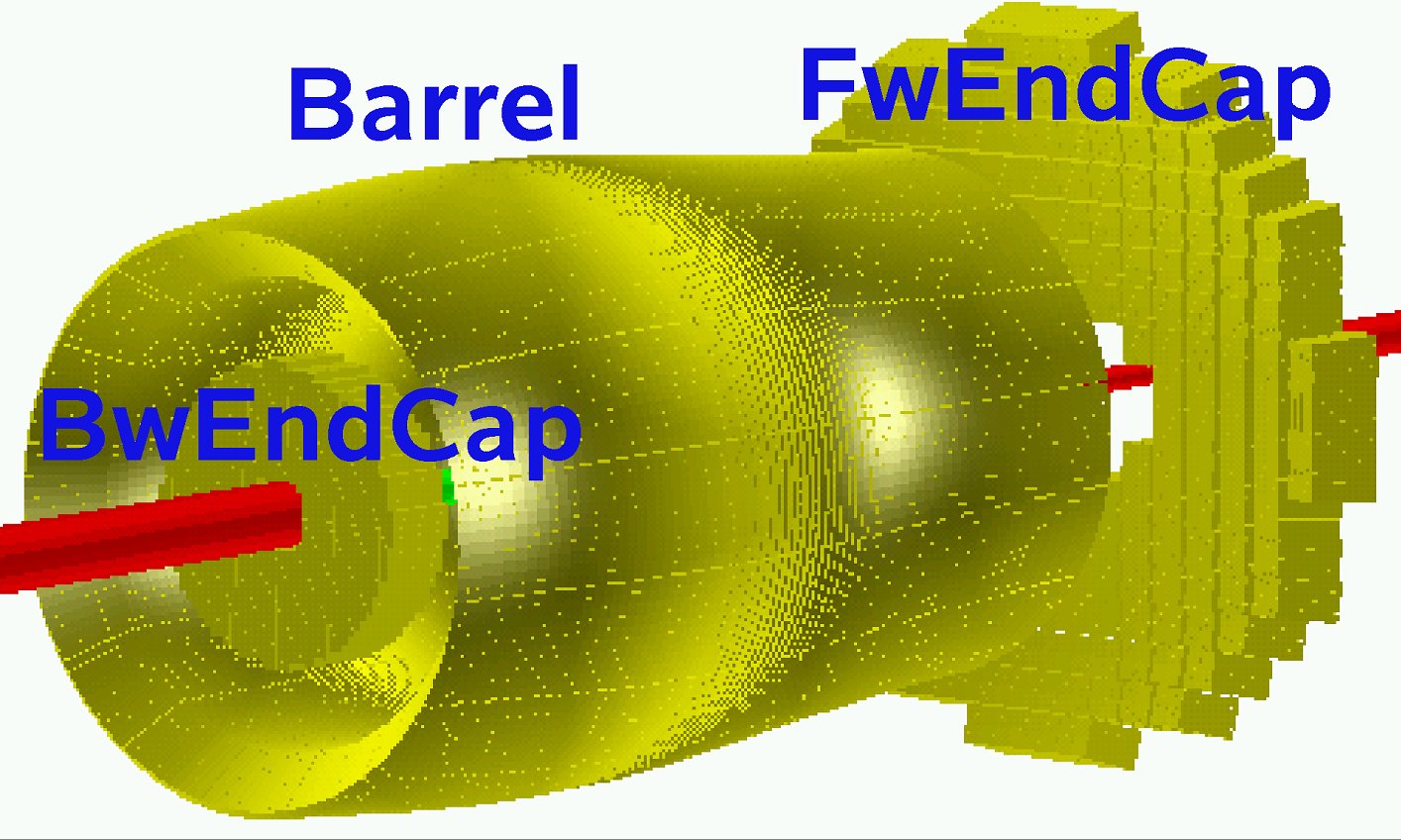}
\caption[The components of the \PANDA electromagnetic calorimeter.]{The components of the \PANDA electromagnetic calorimeter and their acronyms as used in the PandaRoot simulation framework: \BEMC (Barrel), \FWEMC (FwEndCap) and \BWEMC (BwEndCap), with the beam going from left to right.}
\label{fig:ecap:EMCGraph}
\end{center}
\end{figure*}

The electromagnetic calorimeter of \Panda comprises two main parts:  the
central target calorimeter covers in cylindrical geometry almost
completely the target area. A second planar arrangement located further
downstream behind the dipole magnet serves the most forward range up to an
azimuthal angle of 5\degrees with respect to the beam axis. 
The target calorimeter is illustrated in \Reffig{fig:ecap:EMCGraph} and comprises three major parts as summarized in \Reftbl{tab:mech:Geometrical_parameter}.
\begin{table}[b]
\begin{center}
\begin{tabular}{lccc}
\hline\hline
{\bfseries Parts} & Barrel & Forward & Backward \\
 & & downstream & upstream \\
\hline\hline
Crystals & 11360 & 3600 & 592 \\
Axial depth & 2.5\,m & - & - \\
Distance &-& 2.05\,m & 0.55\,m \\
from target & &&\\
Inner radius & 0.57\,m & 0.18\,m & 0.1\,m\\
Outer radius & 0.94\,m & 0.92\,m & 0.3\,m \\
Inner angle & 22\degrees &5\degrees vert.& 169.7\degrees \\
 & &10\degrees horiz.& \\
Outer angle & 140\degrees & 23.6\degrees & 151.4\degrees \\ 
Solid angle  &84.7 &3.2 &5.5 \\
(\%4$\pi$) & & & \\
\hline
\hline\hline
\end{tabular}
\caption[Geometrical parameters of the \Emc.]{Geometrical parameters of the \Emc referring to the front face of the crystal arrangement of 200 mm long crystals.}
\label{tab:mech:Geometrical_parameter}
\end{center}
\end{table}

The present design concept is based on a homogeneous electromagnetic
calorimeter composed of fast and compact scintillator crystals as active
absorber material to be operated within the solenoidal magnetic field of maximum 2$\,T$ field strength. Large area avalanche photodiodes (\LAAPD) are considered
as photosensors in the barrel part. To achieve the envisaged large dynamic
range in energy reaching from 10$\,\gev$ down to 10 $\mev$, the
proposed scintillator PbWO$_4$ (\PWO) has to be operated at low temperatures
down to -25$\degC$ to guarantee sufficient luminescence yield. The limited size
of the photosensor and, consequently, the restricted coverage of the crystal
endface can be compensated by a significantly higher quantum
efficiency of the sensor compared to a standard bialkali photocathode
and an improved scintillator performance. 

The operation conditions impose additional, but still feasible, 
requirements on the mechanical construction, the insulation and the temperature
stability in order to control the strong temperature dependence of the
luminescence yield and keep the \LAAPD gain at a tolerable level. 

\begin{figure}[h!]
\begin{center}
\includegraphics[width=1\swidth]{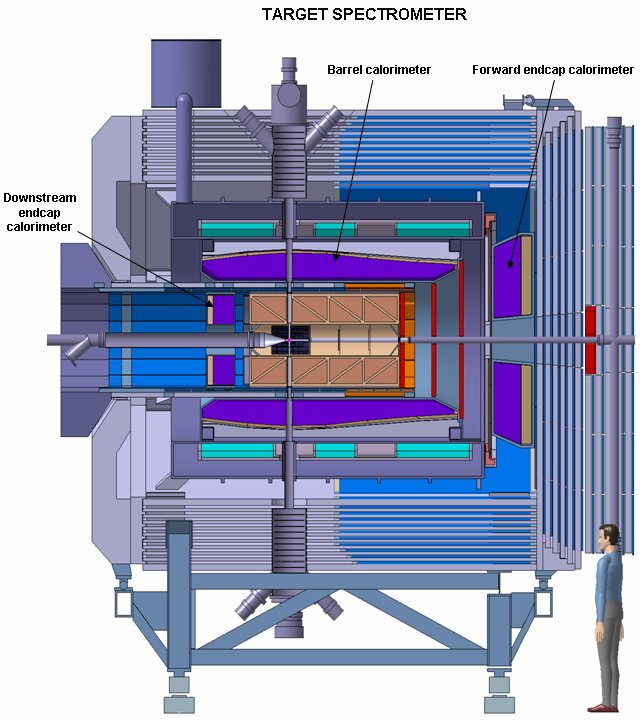}
\caption[Overall view of the target spectrometer.]{Overall view of the integration of the electromagnetic calorimeter into \PANDA.}
\label{fig:mech:mech-Target_spectrometer}
\end{center}
\end{figure}

This report presents the principles of the calorimeter design focusing 
on: 
\begin{itemize} 
\item{the definition of crystal geometry}
\item{the mechanical housing and support structure}
\item{the thermal aspects with respect to cooling \mbox{(-25$\degC$)} and its fine regulation ($\pm$0.1$\degC$)}
\item{the integration with respect to the other detector components and the overall geometry as shown in \Reffig{fig:mech:mech-Target_spectrometer}.} 
\end{itemize}
The calculations are based on \PWO as detector material. The proposed crystal depth is chosen in order to obtain 22 radiation length ($\,X_0$). The overall granularity of the calorimeter is related to the
Moli\`ere radius, which describes the radial shower profile. The granularity has to guarantee the reconstruction of the electromagnetic showers with an adequate energy and position resolution and to limit the
occupancy even for the highest event multiplicities.

The presented concept is based on experience within the collaboration and
on similar calorimeter concepts for \INST{BaBar} \cite{bib:emc:TDR:BaBar}, \INST{CMS} \cite{bib:emc:overview:ref1},
\INST{ALICE} \cite{bib:emc:TDR:AlicePhos} or \INST{CLAS-DVCS} \cite{bib:emc:mech1}.

This design relies on the construction of prototypes primarily to study the technology of extremely
light crystal containers as well as the cooling and temperature control. The prototype results are also presented  in this report.

\section{The Barrel Calorimeter} 
\label{sec:emc:mech:bar}

The barrel calorimeter including the mechanical structure covers the polar angular region between 22\degrees and 140\degrees with an inner radius of 570$\,mm$ and an outer radius of 950$\,\mm$.

\begin{figure*} 
\begin{center}
\includegraphics[width=0.85\dwidth]{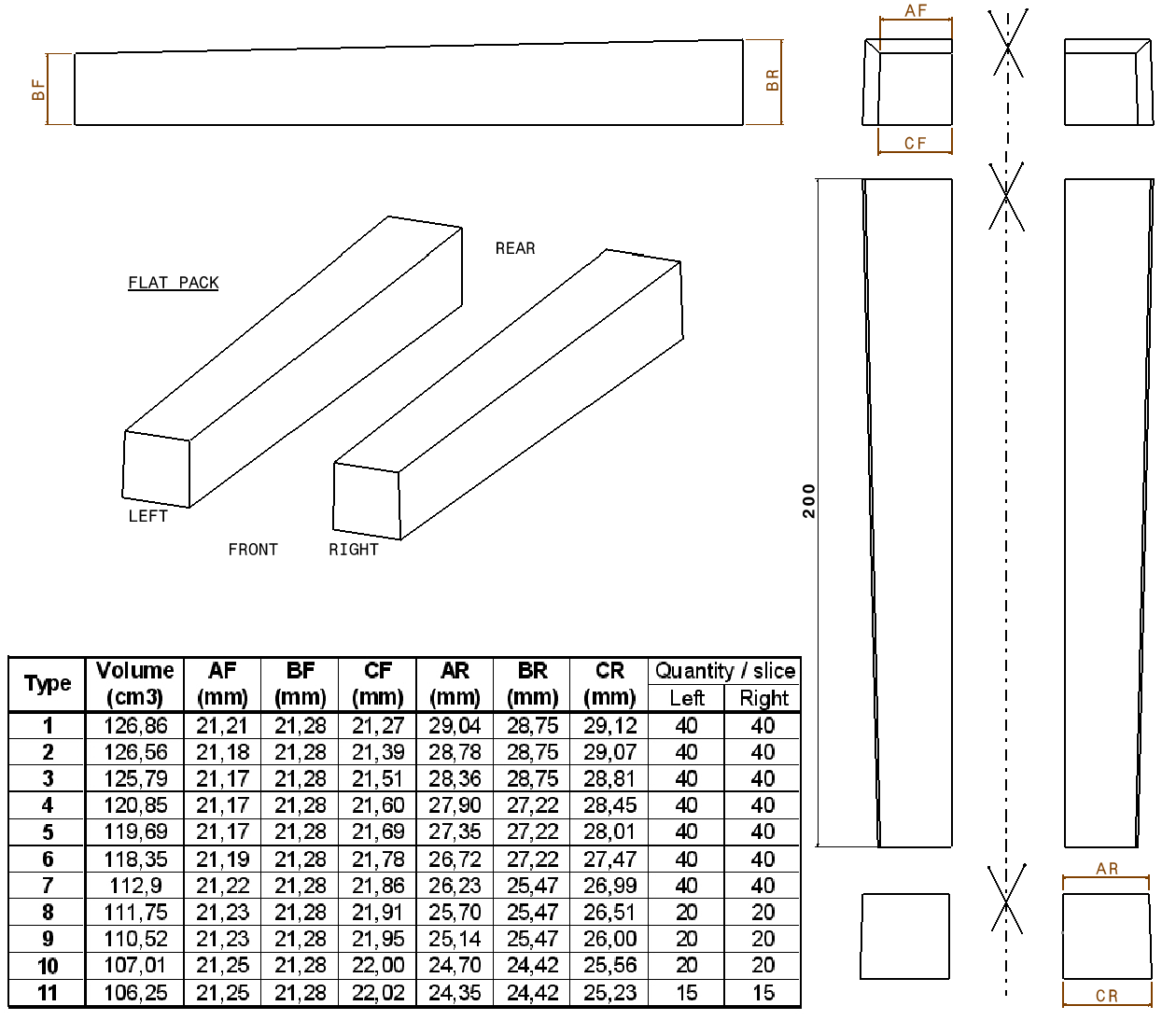}
\caption[The shape of the scintillator crystal.]{The shape 
of the scintillator crystal and the definition of geometrical parameters. The average crystal corresponds to squares of 21.3$\,\mm$ for the front face and 27.3$\,\mm$ for the back face and its mass is 0.98$\,\kg$.}
\label{fig:mech:mech-crystal_definition}
\end{center}
\end{figure*}

\begin{figure}
\begin{center}
\includegraphics[width=0.95\swidth]{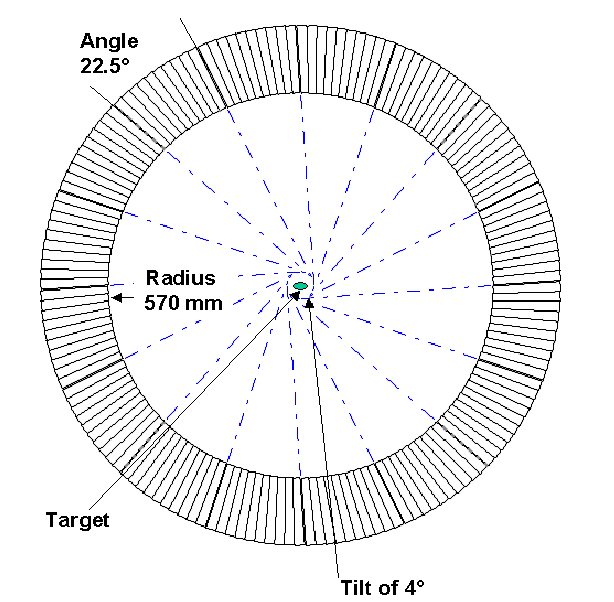}
\caption[Segmentation of the calorimeter along the circumference of the barrel.]{The segmentation of the calorimeter along the circumference of
the barrel part. The 160 crystals are grouped into 16 subunits named
slices.}
\label{fig:mech:mech-Circumference_definition}
\end{center}
\end{figure}

\subsection{The crystal geometry and housing}
\label{sec:emc:mech:bar:cry_geo}

The basic crystal shape is a tapered parallelepiped, shown in
\Reffig{fig:mech:mech-crystal_definition}, and is kept fixed for all
calorimeter elements.  It is based on the ``flat-pack'' configuration
used in the \INST{CMS} calorimeter.  Right angle corners are
introduced in order to simplify the CAD design and the mechanical
manufacturing process to reduce machining costs.  The average mass of
one crystal is 0.98$\,\kg$ (from 0.88 to 1.05$\,\kg$). All given
dimensions are nominal and the tolerances of the crystals are
0/-100$\,\mum$ (the achievable tolerance is based on the present
delivery for \INST{CMS} from the Bogoroditsk plant as well as for the
\INST{DVCS} calorimeter at \INST{JLAB}).  The dimensions of the individual
crystals are related to the global shape and to the discretization of
the calorimeter, defined circumferentially and longitudinally in
\Refsec{sec:emc:mech:bar:cry_geo:Circumference_arrangement} and
\Refsec{sec:emc:mech:bar:cry_geo:Longitudinal_arrangement}, respectively.

\subsubsection{Crystal Arrangements along the Circumference}
\label{sec:emc:mech:bar:cry_geo:Circumference_arrangement}

\Reffig{fig:mech:mech-Circumference_definition} shows the crystal
arrangement on the ring based on the gap dimensions defined in
\Refsec{sec:emc:mech:bar:cry_geo:Distances_between_crystals}. Choosing
the front size of an individual crystal close to 20$\,\mm$
(21.28$\,\mm$ exactly) at a radius of 570$\,\mm$, which corresponds to
the Moli\`ere radius, the ring is divided into 160 crystals. The
crystals are grouped into packs of 4 $\times$ 10 (one alveole pack)
leading to 16 sectors of 22.5\degrees coverage, which are termed
slices.

The presented geometry foresees that the crystals are not pointing
towards the target position. A tilt of 4\degrees is added on the focal
axis of the slice to reduce the dead zone effect. This means, that tracks originating at the target never pass through gaps between crystals, but always cross a significant part of a crystal.

\begin{figure*} 
\begin{center}
\includegraphics[width=1\dwidth]{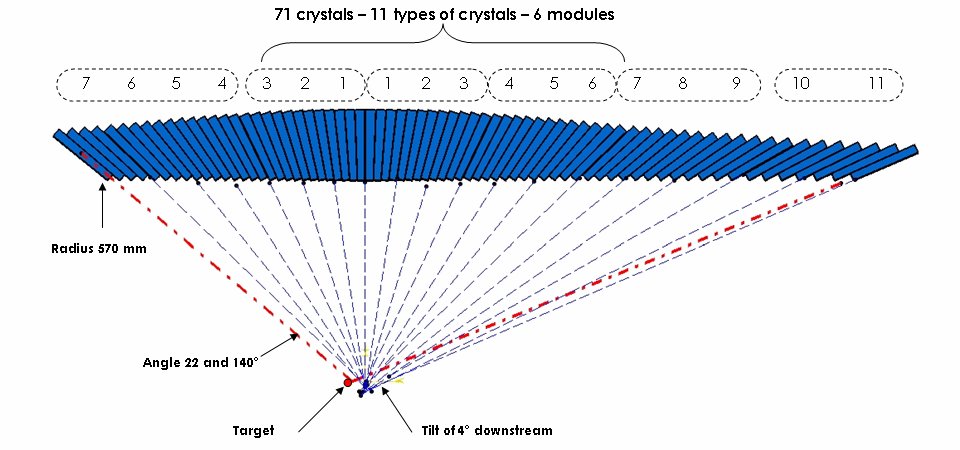}
\caption[Geometrical arrangement of the crystals of the
barrel.]{Geometrical arrangement of the crystals of the barrel in a cut
  along the beam axis. The definition of subgroups by pack of 4 and by
  module is indicated.  The use of the mirror symmetry decreases from
  18 to 11 different types of shapes according to the definition in
  \Reffig{fig:mech:mech-crystal_definition}.}
\label{fig:mech:mech-Barrel-71_crystals}
\end{center}
\end{figure*}

\subsubsection{Longitudinal Crystal Positioning} 
\label{sec:emc:mech:bar:cry_geo:Longitudinal_arrangement}

\begin{figure} 
\begin{center}
\includegraphics[width=1\swidth]{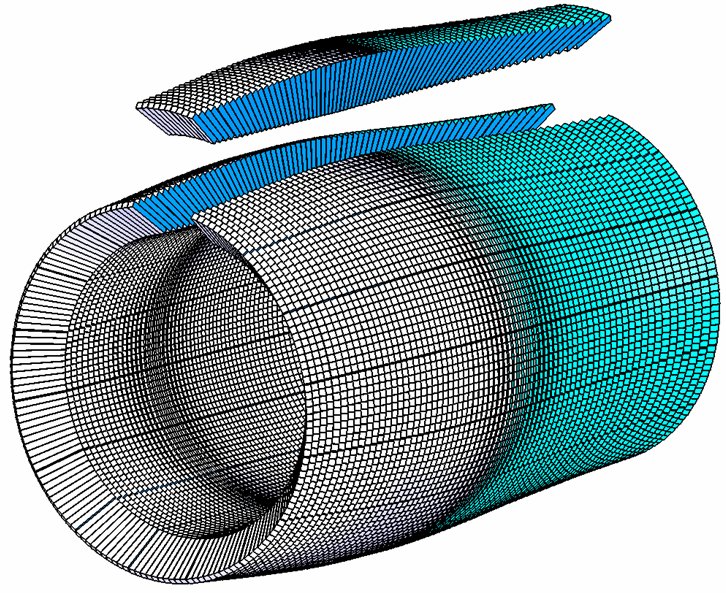}
\caption[View of the barrel of 11360 crystals with a separated slice.]
{View of the total barrel volume composed of 11360 crystals and, separated, a single slice of 710 crystals covering 1/16 of the barrel volume.}
\label{fig:mech:mech-Barrel-11360_crystals}
\end{center}
\end{figure}

Along the length of the barrel (parallel to the beam axis) the crystal
positions and individual geometries are shown in
\Reffig{fig:mech:mech-Barrel-71_crystals}. The mirror symmetry with respect to the vertical axis reduces the number of different crystal shapes in the arrangement from 18 to 11. The lateral sizes of the rear (readout) 
faces vary between 24.35$\,\mm$ and 29.04$\,\mm$ and the average area is equivalent to a square of 27.3$\,\mm$ ($\pm$15\percent area variation between the 11 types of crystals). For the front face, the lateral sizes vary between 21.18$\,\mm$ and 22.02$\,\mm$ and the average area is equivalent to a square with lateral size of 21.3$\,\mm$. In total 71 crystals
are aligned at the radius of 570$\,\mm$.  A tilt angle of 4\degrees is
introduced to reduce the dead zone effect and this angle corresponds
to a shift of the focus by $\approx\,37\,\mm$ downstream.  In one
slice of 710 crystals, 3 or 5 alveole packs are grouped together into
6 modules: 4 modules of 120 crystals each, 1 module with 70 and 1 with 160
crystals. The entire barrel contains 11360 crystals for a total length
of 2466$\,\mm$ and a volume of 1.3$\,\m^3$. \Reffig{fig:mech:mech-Barrel-11360_crystals} highlights one slice out of
the total barrel volume.

\begin{figure} 
\begin{center}
\includegraphics[width=0.9\swidth]{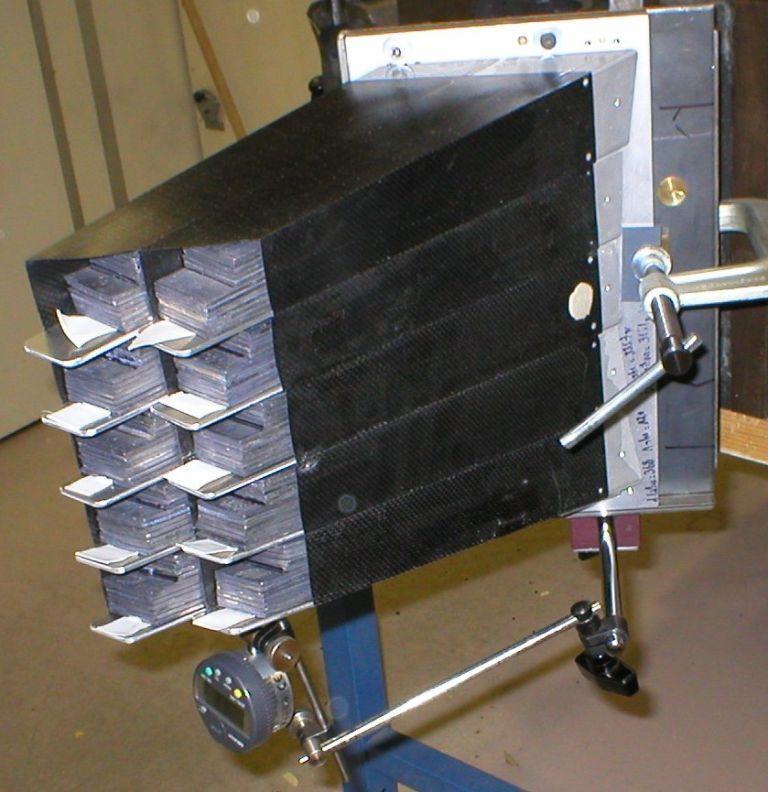}
\caption[Deformation test of carbon alveoles.]{Deformation test of
  carbon alveoles in vertical position. Here the measure is around
  10$\,\mum$ compared to the 80$\,\mum$ in the horizontal position.}
\label{fig:mech:mech-Deformation_test_alveoles}
\end{center}
\end{figure}

\begin{figure*} 
\begin{center}
\includegraphics[width=1\dwidth]{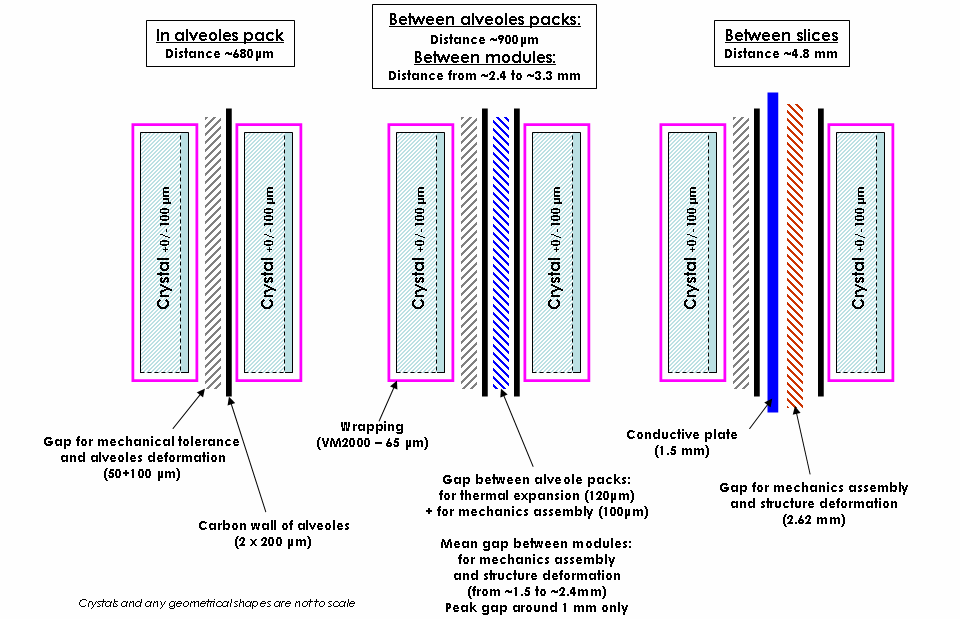}
\caption[Summary of the expected dead space between calorimeter 
elements.]{Summary of the expected dead space between calorimeter elements.}
\label{fig:mech:mech-Gaps_crystals}
\end{center}
\end{figure*}

\subsubsection{Crystal Light Collection}
\label{sec:emc:mech:bar:cry_geo:crystal_light_collection}

The crystals are wrapped with a reflective material in order to
optimize light collection as well as to reduce optical cross talk. Considered
material is Radiant Mirror Film ESR from 3M, commonly called "VM2000"
in the past, accounting for a thickness of 63.5$\,\mum$. This wrapping material, a non-metallic
multilayer polymer, was employed for the crystals of the \INST{DVCS} calorimeter at \INST{JLAB})
and the \INST{GLAST} calorimeter. In order to foresee a small air gap between the reflector and the
crystal face, the foil must be shaped in a mold heated at 80$\degC$ to
sharpen the corners.  In order to compensate for the longitudinal dependence of the light
collection, the structure of the crystal surface (optically polished,
roughed, lapped) can be modified or an inhomogeneous reflector can be selected. This is still under study.

\subsubsection{Carbon Fiber Alveoles} 
\label{sec:emc:mech:bar:cry_geo:carbon_fiber_alveoles}

In the present design, 4 crystals will be contained in one carbon
fiber alveole in order to avoid any load transfer to the fragile \PWO
while the crystals are held in place by their rear end. The expected wall thickness
of the alveoles is 200$\,\mum$ and they are grouped to compose an
alveole pack of 40 crystals. Each alveole is epoxy-glued to an
aluminum insert which is the interface with the support
elements. Temperature cycling tests of the gluing between -40$\degC$
and 30$\degC$ are performed to check the reliability of this support
stressed by the differential thermal expansions. In front of the
alveoles, a carbon plate is added to avoid the movement of crystals.
Epoxy pre-impregnated carbon plain weave fabric is precisely moulded
in complex tools to fabricate the alveoles. Each type of crystal
corresponds to one mold composed, for technical reasons, of 2 alveoles to
overlap the 2 wrapping joints. Real size alveole prototypes have been produced to check the feasibility, to optimize the final thickness and to perform mechanical
tests. \Reffig{fig:mech:mech-Deformation_test_alveoles} shows one of
the deformation tests, here in the vertical position. The results
80$\,\mum$ horizontally and 10$\,\mum$ vertically are tolerable and
are in agreement with the analytical calculation.

\subsubsection{Distances between Crystals}
\label{sec:emc:mech:bar:cry_geo:Distances_between_crystals}

The distance between two crystals is calculated from the thickness of
materials, structure deformation and mechanical
tolerances. \Reffig{fig:mech:mech-Gaps_crystals} presents drawings of
the different gaps which are explained in detail below. A conservative
concept has been chosen, which can be considered as an upper limit of
the expected dead space.

The basic distance between crystals inside a pack is defined to 0.68$\,\mm$ and
represents the sum of:
\begin{itemize}
\item{400$\,\mum$, the double thickness of the carbon alveoles;}
\item{130$\,\mum$, the double thickness of the wrapping material;}
\item{100$\,\mum$, the free distance left for the alveole deformation;}
\item{50$\,\mum$, the approximate manufacturing tolerance.}
\end{itemize}
This calculation is based on the nominal dimension of the crystal. There might appear
an additional distance of  $< 0.2\,\mm$ between two adjacent crystals due to
polishing tolerances. Gaps between the identical shapes of 4 crystals in one alveole pack might introduce an
additional lateral spacing up to 350$\,\mum$. 

Between alveole packs, the thermal expansion of the mounting plate amounts to about 120$\,\mum$ and the mounting tolerance is assumed 100$\,\mum$. Together with the basic distance between crystals of 0.68$\,\mm$ this adds up to a total value of 0.9$\,\mm$.
 
Between modules, a distance varying between 2.4 and 3.3$\,\mm$ has to
be considered to take into account the mounting feasibility and
tolerances in the mechanical assembly, and in the structure
deformations.

Between adjacent slices, a first gap up to 2.62$\,\mm$ due to
mechanical mounting tolerances caused by the deformation of the whole
structure and a second gap of 1.5$\,\mm$ have to be assumed.  The
distance between crystals in this area amounts to $< 4.8\,\mm$.

\begin{figure*} 
\begin{center}
\includegraphics[width=1\dwidth]{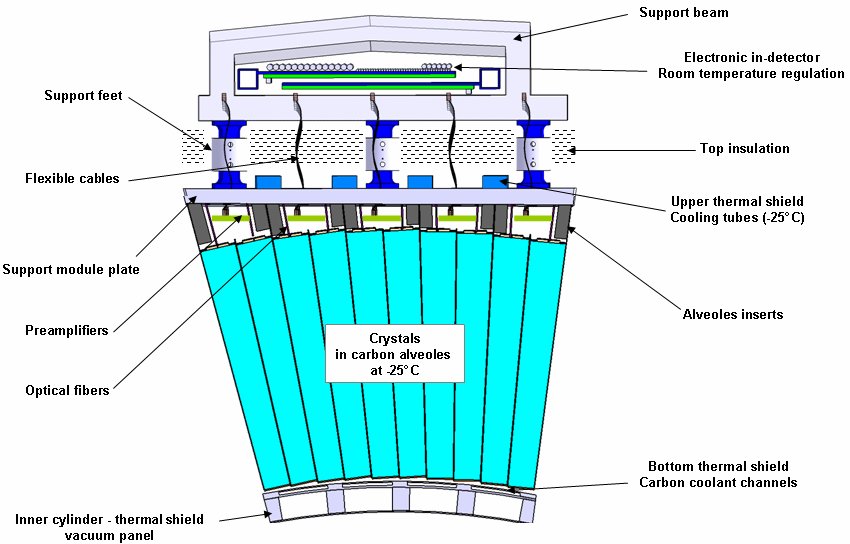}
\caption[Schematic view of the concept and of the major components of a slice.]{Schematic view of the concept and of the major components of a slice.}
\label{fig:mech:mech-Barrel_principle}
\end{center}
\end{figure*}

\subsection{Mechanics around Crystals - Slice Definition}
\label{sec:emc:mech:bar:slice_def}

\Reffig{fig:mech:mech-Barrel_principle} presents the design principle
of one slice coping with all constraints: thermal, mechanical and
electronical integration.

\subsubsection{Insert and Module Definition} 
\label{sec:emc:mech:bar:slice_def:insert_module}

\begin{figure} 
\begin{center}
  \includegraphics[width=1\swidth]{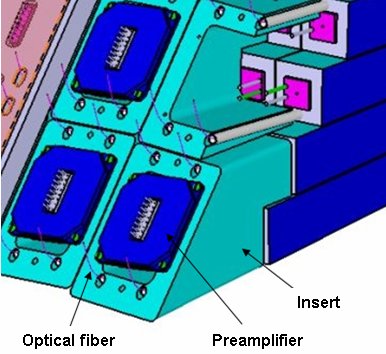}
  \caption[CAD integration of the insert with the preamplifier and
  optical fiber.]{CAD integration of the insert
    with the preamplifier and optical fiber at the back of crystals. First design uses single square LAAPD but 2 rectangular LAAPDs fit as well.}
\label{fig:mech:mech-vue_CAD_insert}
\end{center}
\end{figure}

\begin{figure*} 
\begin{center}
\includegraphics[width=1\dwidth]{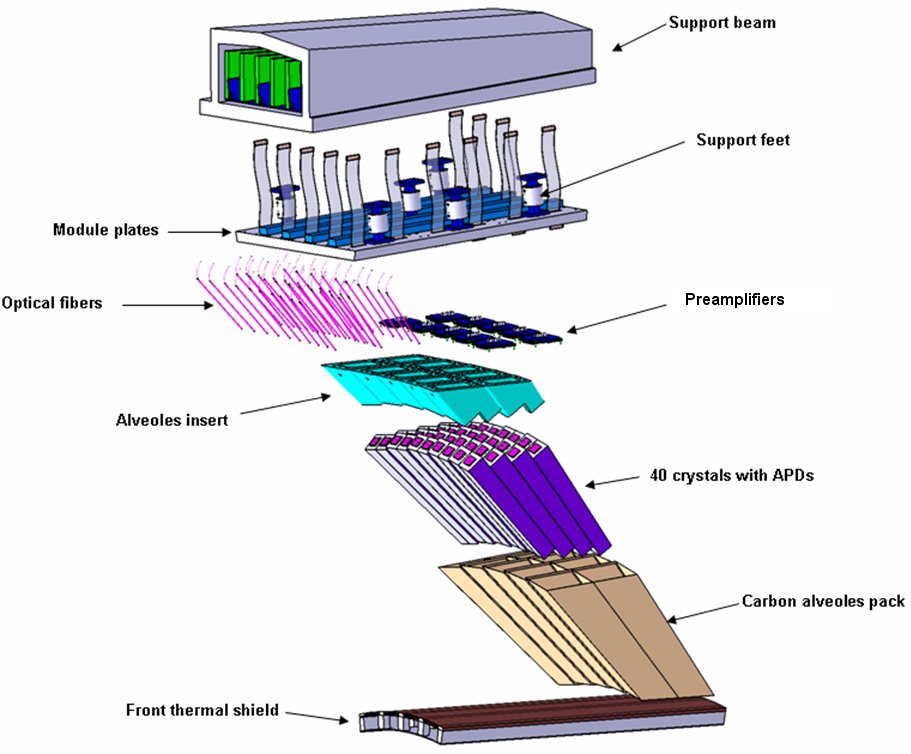}
\caption[Exploded view of a slice.]{Exploded view of one slice showing all the individual components.}
\label{fig:mech:mech-Exploded_view_barrel}
\end{center}
\end{figure*}

The inserts, glued to the carbon alveoles, make the precise connection
to the back module plates and also hold the preamplifiers and the
optical fibers which are shown in 
\Reffig{fig:mech:mech-vue_CAD_insert}. The shape of these inserts are
all different and complex to fabricate due to the tapered slopes of
crystals and due to the facetized shape of the barrel
calorimeter. Corresponding to the 6 modules defined in the
\Refsec{sec:emc:mech:bar:cry_geo:Longitudinal_arrangement}, the back
module plates have a thickness of 14 mm machined with a good flatness in order
to stay a reference plane even under the weight of the crystals.  Each
plate is linked to a support beam through 6 support feet
designed for low thermal transfer and low deformation. In addition, these
feet have motional freedom in particular directions in order to allow
translations due to the thermal expansion. The back module plates are
cooled down by the upper thermal screen. Further thermal
details are given in \Refsec{sec:emc:mech:bar:therm}.
\Reffig{fig:mech:mech-Exploded_view_barrel} shows an exploded view of
one module with all individual components.

\subsubsection{Support Beam Definition} 
\label{sec:emc:mech:bar:slice_def:support_beam}

\begin{figure} 
\begin{center}
\includegraphics[width=1\swidth]{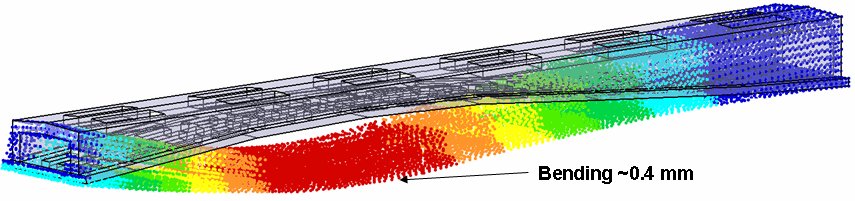}
\caption[Deformation of the support beam.]{Deformation of 0.4$\,\mm$ of the support beam in the upper position (horizontal beam).}
\label{fig:mech:mech-deform_support_beam}
\end{center}
\end{figure}

\begin{figure*} 
\begin{center}
\includegraphics[width=1\dwidth]{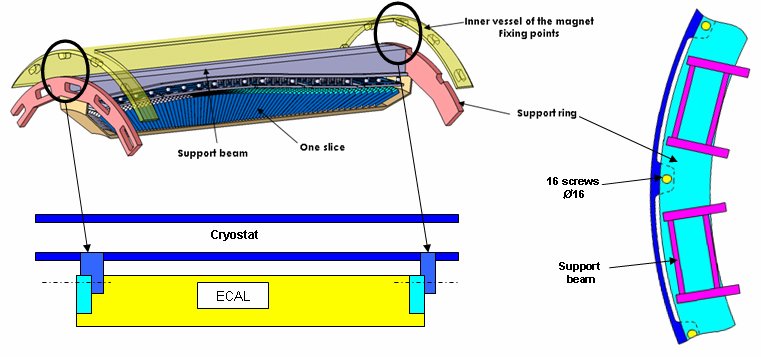}
\caption[Barrel supported on the magnet.]{Barrel supported on the magnet.}
\label{fig:mech:mech-Barrel_support}
\end{center}
\end{figure*}

The 710 crystals of one slice are supported by a stainless steel
support beam whose shape is a rectangular tube of 2.7$\,\m$
length. The bending of this beam is calculated to be between 0.1 and
0.4$\,\mm$ for the horizontal and the vertical position of
the slice, respectively, as shown in
\Reffig{fig:mech:mech-deform_support_beam}. These values show its
rigid behavior but the position of the modules will have to be
corrected in order to align all the crystals. The magnetic field of the
solenoid requires to verify that the applied stainless steel material is indeed of sufficient non-magnetic quality.
The magnetic quality may have been altered by the machining or the welding process of this part and a magnet-relieving anneal is foreseen in an oven.  The internal part of this tube is used for
storing all the services as electronics boards and power supply
cables. The details about the integration of services are provided in
\Refsec{sec:emc:mech:bar:elect_int}. These services
are located at room temperature and access must be
possible simply by opening the top external cover when the barrel is
in maintenance position.  The support beam is fixed on 2 support
rings at its both extremities, where a possibility to adjust
the alignment of the slices is foreseen. These rings (shown in
\Reffig{fig:mech:mech-Barrel_support}), rest on support points added to the inner vessel of the coil cryostat.

\subsubsection{Adapted Design for the Target} 
\label{sec:emc:mech:bar:slice_def:target}

The target system is passing through the calorimeter barrel on the
vertical axis. It is foreseen to place two slices, an upper and a lower one,
specially designed with a central hole. Some crystals are removed, the
mechanics and the thermal shields are modified and a hollow cylinder
of insulation is added. In any case, the target tube will not be in
contact with the cold area and will be let free to move.

\begin{figure*} 
\begin{center}
\includegraphics[width=1\dwidth]{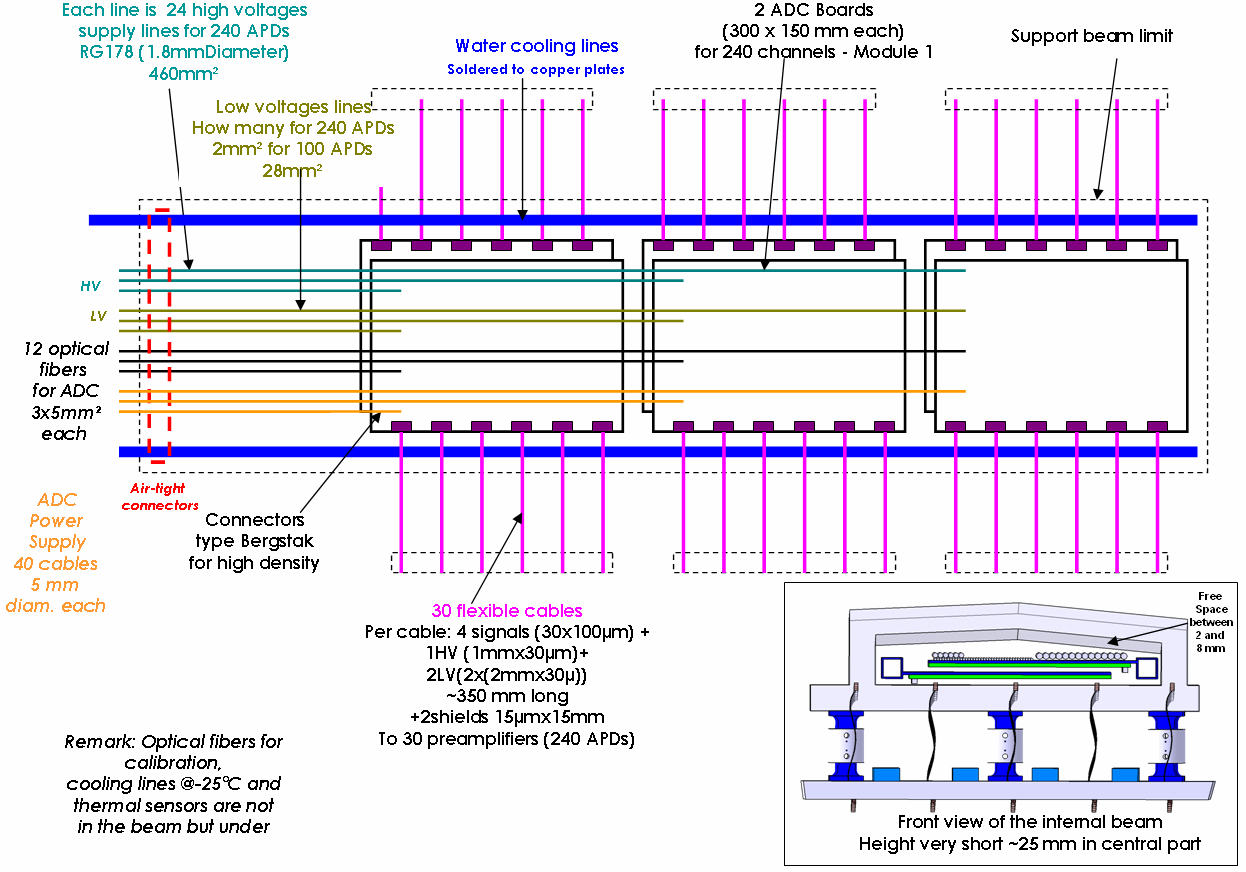}
\caption[Integration of the services in the support beam.]{Integration
  of the services in the support beam.}
\label{fig:mech:mech-Support_beam_integration}
\end{center}
\end{figure*}

\subsection{Electronics Integration}  
\label{sec:emc:mech:bar:elect_int}

\subsubsection{Photosensor, Preamplifier, Flat Flexible Cables} 
\label{sec:emc:mech:bar:elect_int:preamp}

Due to the magnetic field, Large Area Avalanche Photo Diodes (\LAAPD) are employed. 
Two LAAPD of $7\times14\,\mm^2$ each are used and glued on the back face of the crystal. Each \LAAPD is connected
to a charge preamplifier with a 40$\,\mm$ long twisted wire. The
length is discussed in the \Refsec{sec:emc:mech:therm:intAPD} as it
plays a role in the heat transfer to the \LAAPD. The preamplifier has a power consumption of
52\,mW/channel and is fixed on the inserts. It is connected to the
read-out electronics through a 350$\,\mm$ long flat flexible cable
which drives: the 8 signals, 1 high and 2 low voltages, stacked
between 2 shielding layers. The sections are resumed in
\Reffig{fig:mech:mech-Support_beam_integration}. The lack of space
requires high density connectors. The length of the flat flexible
cables is calculated in order to be able to group them and thus reduce
the number of holes in the bottom face of the support beam. Besides, a increased length and a smaller cross section decrease the conductive heat transfer.

\subsubsection{Read-out Electronics} 
\label{sec:emc:mech:bar:elect_int:read-out}

\Refsec{sec:emc:mech:bar:slice_def:support_beam} indicated that the
read-out electronics is integrated into the support beam. This
electronics comprises the digitizing ADC boards. In addition to this equipment, the
support beam also contains services as the high and low voltage
power supply, the ADC read-out optical fibers, the ADC power supply
and the water cooling tubes for the room temperature regulation
(discussed in
\Refsec{sec:emc:mech:therm:room_temp}). \Reffig{fig:mech:mech-Support_beam_integration}
illustrates this arrangement and resumes the sections of the
services. The size of the boards is $300\times 150\times 10\,\mm^3$. Using high-density connectors and a stacking in pairs, 2 boards can fit even in the small central region of the beam and could perform the readout of up to 240
channels, equivalent to a module equipped with 2 \LAAPD per crystal.

\subsubsection{Laser Calibration System} 
\label{sec:emc:mech:bar:elect_int:laser}

The \LAAPD is enclosed in a light tight plastic box into which an optical
fiber for light injection is inserted in one corner. This LED- or
LASER-light is injected from the rear side of the crystals due to the
limited space in front of the calorimeter.  The light pulser system is primarily intended for stability
control of the complete readout chain including the \LAAPD. A first
prototype is under construction based on a LED light source with
optical fiber distribution (see \Refsec{sec:cal:monitoring:lps}).
The routing of the optical fibers is taken care of in the early stage of
the design in order to respect the minimum bending radius and to integrate special
guiding tubes inside the mechanical construction. The fiber installation must be finished
before the upper thermal insulation is closed. The use of a
non-rigid insulator, as vermiculite granulate mentioned in the
\Refsec{sec:emc:mech:therm:thermal_shields}, is a reasonable solution due to
the high number of fibers in random positions.

\subsection{Thermal Cooling} 
\label{sec:emc:mech:bar:therm}

\subsubsection{Requirements and Method}
\label{sec:emc:mech:bar:therm:requirements}

The crystals are to be cooled down to a nominal operating temperature
of -25$\degC$. The temperature gradients of the \LAAPD gain
and of the crystal light yield of -2.2\percent/$\degC$
and -1.9\percent/$\degC$, respectively, require a stable temperature with peak to peak variation over time 
of at most $\pm$0.1$\degC$ in order to keep the initial calibration.  
The read-out electronics is stabilized at room temperature and this
regulation is described in the \Refsec{sec:emc:mech:therm:room_temp}.

\subsubsection{Thermal Properties of the PbWO$_{4}$}
\label{sec:emc:mech:bar:therm:properties_PWO}

Based on information from the \INST{CMS}/ECAL Collaboration, the thermal properties of
\PWO are listed in \Reftbl{tab:scint:pwo:th_prop} and are used in the
thermal analysis discussed below.

\begin{table}
\begin{center}
\begin{tabular}{lcc}
\hline\hline
Parameter & \multicolumn{2}{c}{\PWO} \\
\hline
$\rho$           &  8.28 & $\g/\cm^3$       \\
Specific heat    &  262 & $\J/\kg.\degC$    \\
Conductivity     &  3.22 & $\W/\m.\degC$    \\
\hline\hline
\end{tabular}
\caption[Thermal properties of \PWO.]{The relevant properties of PbWO$_{4}$ (\PWO).}
\label{tab:scint:pwo:th_prop}
\end{center}
\end{table}
 
\subsubsection{Thermal Shields} 
\label{sec:emc:mech:therm:thermal_shields}

\begin{figure} 
\begin{center}
\includegraphics[width=1\swidth]{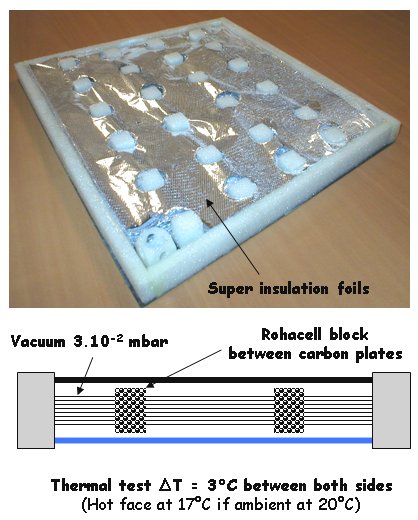}
\caption[Front thermal screen.]{Front thermal screen.}
\label{fig:mech:mech-vaccum_panel}
\end{center}
\end{figure}

The crystals are surrounded by thermal shields basically made of
panels, cooled by serpentines filled with a coolant, and of foam to
insulate from the ambient air. As shown in
\Reffig{fig:mech:mech-Barrel_principle}, two thermal shields are
introduced:
\begin{itemize}
\item{Above the crystals, the module plate is cooled via a thin copper 
    plate and serpentine tubes brazed on it. These tubes have a square cross section of $10\times20\,\mm^2$ to limit the height and keep enough insulating foam. This upper area, containing lots of services like
    flexible read-out cables and optical fibers, is filled with 50 mm
    of vermiculite granulate. The module plate and the insert are made of
    aluminum because of its good thermal conductivity and thus keep a good
    thermal transfer for the preamplifier and for the area up to the
    rear end of the crystal.}
\item{For the front side of the crystals, a special thin thermal screen was
    developed, in order to reduce the distance to the \Dirc for achieving a
    better resolution. Carbon fiber material is used for its
    low radiation length, in order to improve the transparency for
    particles, and for its negligible thermal expansion
    coefficient. This shield has a thickness of only 25 mm distributed in 4$\,\mm$ of carbon 
    coolant channels and in 21$\,\mm$ of a vacuum super-insulating panel, respectively. 
    Super-insulation is
    used in the cryogenic technology and is in our case 2.5 times more
    efficient in terms of thermal conductivity versus total
    thickness. The vacuum (0.03\,Torr), however, is absolutely necessary and
    is contained between two skins linked with Rohacell blocks, a
    structural and vacuum tight foam. The warm side is aluminum, in
    order to homogenize the temperature and avoid local cold points in
    front of each Rohacell block, while the cold side is a skin of carbon
    material. Its low thermal expansion reduces the differential
    constraints between the two faces of the sandwich, which thus
    keeps a good flatness in any case. \Reffig{fig:mech:mech-vaccum_panel} shows the first sandwich
    structure constructed for the thermal and mechanical tests. Installed on prototype, it shows on its external face a satisfactory temperature variation of 3$\degC$ below room temperature.}
\end{itemize}

The temperature on the sides of the slice is assumed to be constant at -25$\degC$, because of the adiabatic boundary condition with the neighboring slice at the same temperature. For reasons of protection, a 1.5$\,\mm$ thick
aluminum plate is inserted between two slices, which additionally improves the thermal cooling homogeneity. 
In fact, the insulation between slices is achieved at the level of the thermal shields.

The heat transfer through the thermal shields is the first external
heat source of the cooling system. The definition of their thickness
minimizes the heat transfer and the limit is to have their external
faces at approximatively room temperature (above the dew point). This criterium ensures that the internal 
temperature stability is nearly independent of the ambient air temperature variation. A ratio of 25 has
been found from measurements performed with crystals in the prototype detector Proto60,
which means that for an ambient temperature change of 1$\degC$ the internal temperature could vary by as much as 0.04$\degC$ without chiller regulation.

\subsubsection{Thermal Bridges}
\label{sec:emc:mech:therm:thermal_bridges}

The second external heat source is introduced by the heat transfer
through the mechanical supports and the metallic conductors of cables for the
readout. In this conductive process, the section and the number of
these thermal bridges have to be minimized. Design studies and simulations
have been performed on the support feet in order to reduce the thermal
transfer. Flexible long cables are preferred.

\subsubsection{Internal Heating of \LAAPD and Crystal} 
\label{sec:emc:mech:therm:intAPD}

\begin{figure} 
\begin{center}
\includegraphics[width=1\swidth]{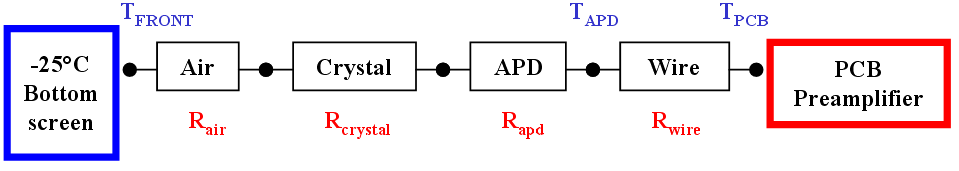}
\caption[Equivalent chain for the analytical analysis of the \LAAPD
temperature.]{Equivalent chain for the analytical computation of the \LAAPD temperature.}
\label{fig:mech:mech-Therm-apd-analytique}
\end{center}
\end{figure}

The preamplifiers introduce an internal heat source in the
calorimeter. Their total power consumption is 50\,mW/channel, and due to the
thermal impedance of their components (200$\degC$/\,W maximum), the hottest point is +4$\degC$ higher. 
This heat produced is partially transferred by direct contact to the metallic support
through its fixing screws or through the conductive silicone interface
(Bergquist gap pad) put on top of the printed circuit board of the
preamplifiers. But the heat propagates overall to the \LAAPD connector,
and therefore to the \LAAPD itself through the twisted pair wire. The \LAAPD temperature
is calculated and settles at an equilibrium under the influence of the
bottom thermal screen in front of the crystals. The analytical formula
is given in \Refeq{eq:thermal:APDtemp:1} and is deduced from the crystal/LAAPD/preamplifier
model illustrated in \Reffig{fig:mech:mech-Therm-apd-analytique}.

\begin{center}
\begin{equation}
T_{apd}=T_{pcb}+R^{th}_{wire}*\frac{(T_{front}-T_{pcb})}{R^{th}_{air}+R^{th}_{crystal}+R^{th}_{apd}+R^{th}_{wire}}
\label{eq:thermal:APDtemp:1}
\end{equation}
\end{center}

\begin{figure}
\begin{center}
\includegraphics[width=1\swidth]{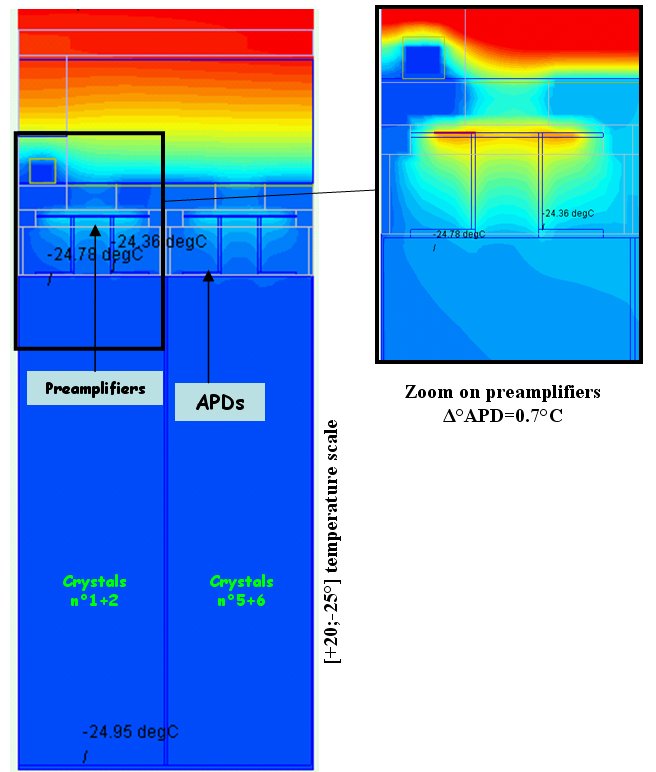}
\caption[Thermal simulation.]{Thermal simulation with a preamplifier of
  50\,mW linked to the \LAAPD with a $40\,\mm$ wire.}
\label{fig:mech:mech-Temperature_preamplifier}
\end{center}
\end{figure}

From this equation, the calculated length of the twisted pair wire must be
at least 150 mm in order to have less than 0.1$\degC$ of temperature
variation on the \LAAPD. This length, however, is not compatible with the
available space on top of the crystals and with the good electronics
functioning where the preamplifiers have to be placed as close as
possible. In the present design, the length is fixed to 40 mm, and the
\LAAPD temperature rises up to 0.7$\degC$. Unfortunately, the temperature
inside the barrel will not be uniform and the \LAAPD stability will be
dependent of the preamplifier power consumption which has to be stable in
the order of 10\%.  The low thermal conductivity of \PWO (see
\Reftbl{tab:scint:pwo:th_prop}) attenuates any crystal temperature
changes on a short timescale ($\approx$ 10$\,\sec$). The crystals
are affected linearly, however, by their longitudinal temperature non-uniformity
which, in fact as for the \LAAPD, is corrected during the pre-calibration. \Reffig{fig:mech:mech-Temperature_preamplifier}
shows a simulation of the design model.

\begin{figure}[b] 
\begin{center}
\includegraphics[width=0.9\swidth]{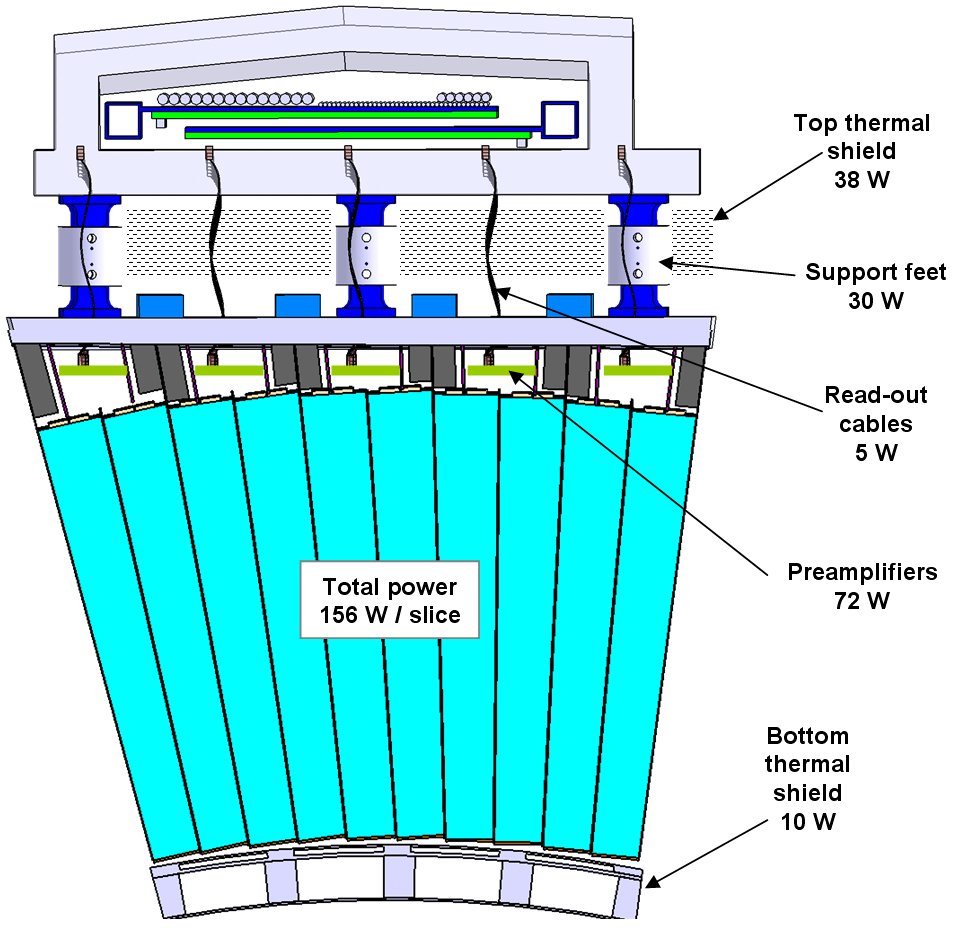}
\caption[Cooling power summary.]{Cooling power summary.}
\label{fig:mech:mech-Total power}
\end{center}
\end{figure}

\subsubsection{Definition of the Cooling System} 
\label{sec:emc:mech:therm:cooling_system}

Three types of heat sources have been defined previously and are
resumed in \Reffig{fig:mech:mech-Total power}. The total power consumption for 16
slices is $\approx$ 2\,600\,W. The external cooling circuit is
subject to heat transfer, too, estimated to be $\approx$
700\,W. Finally, the cooling machine must have a minimum effective
cooling capacity of 3\,300\,W.

The coolant is SYLTHERM XLT from Dow Corning. This liquid is a high
performance silicone polymer designed for use at low
temperature. Compared with water/alcohol or with hydrofluoroether
(used in \INST{ALICE/PHOS)} fluids, it offers the best ratio based on
the heat transfer versus pumpability due to its high specific heat and
low viscosity. It has essentially no odor, and is not corrosive for
long term use.

The design of the cooling circuit is taking in account the pump
capacities, flow rate and pressure in order to optimize the flow
and thus minimize the temperature variation along the longitudinal axis of the crystal.  Inside
a slice, the nominal flow is around 15 liters/min which gives a
non-uniformity of 1.1$\degC$ between the inlet and the outlet. Two
cooling machines, one for each half-barrel, will supply the barrel circuit, basically split into 4 parallel
sectors. 

The stability of the coolant temperature at the entrance of the barrel must be
much better than the required stability in the barrel. A starting
point for the specifications in the machine design is $\pm$0.05$\degC$ and
a time reactivity of the order of tens of seconds. This value depends
mainly of the quality of the cooling machine. Further studies will be
performed and industrial solutions are foreseen to get the optimal reliability versus price ratio.
The low thermal conductivity of the 20\,tons of \PWO material creates a very long time
constant for the barrel. The expected time to reach the final temperature is several tens of hours.

\subsubsection{Read-out of the Temperatures}
\label{sec:emc:mech:therm:readout_temp}

A slice is equipped with up to 50 thermal sensors placed in
representative positions: along the length of the crystals, in the
center or in the extremities of the slice, near the \LAAPD, close to the
thermal shields and in contact with inlet and outlet cooling tubes. Sensors are also placed
externally, in the support beam to check the water regulation. In
fact, the temperature measurement controls the stability of the
calibration but also can return information about possible problems in
the cooling system.  All the thermal sensors will be cross-calibrated at
the nominal temperature. Two types of thermal sensors are used: a) type T
thermocouples working at low temperatures which give correct results
for relative measurements and are thin enough to be inserted into carbon fiber
alveoles, or b) a few Pt100 sensors installed in parallel to give a better
absolute value. The data acquisition is performed reliably and cheap with a commercial system
at a frequency of one readout per several minutes.
  
\subsubsection{Air-tightness Studies and Risk Analysis for Low-temperature Running} 
\label{sec:emc:mech:therm:airtightness}

\begin{figure*} 
\begin{center}
\includegraphics[width=1\dwidth]{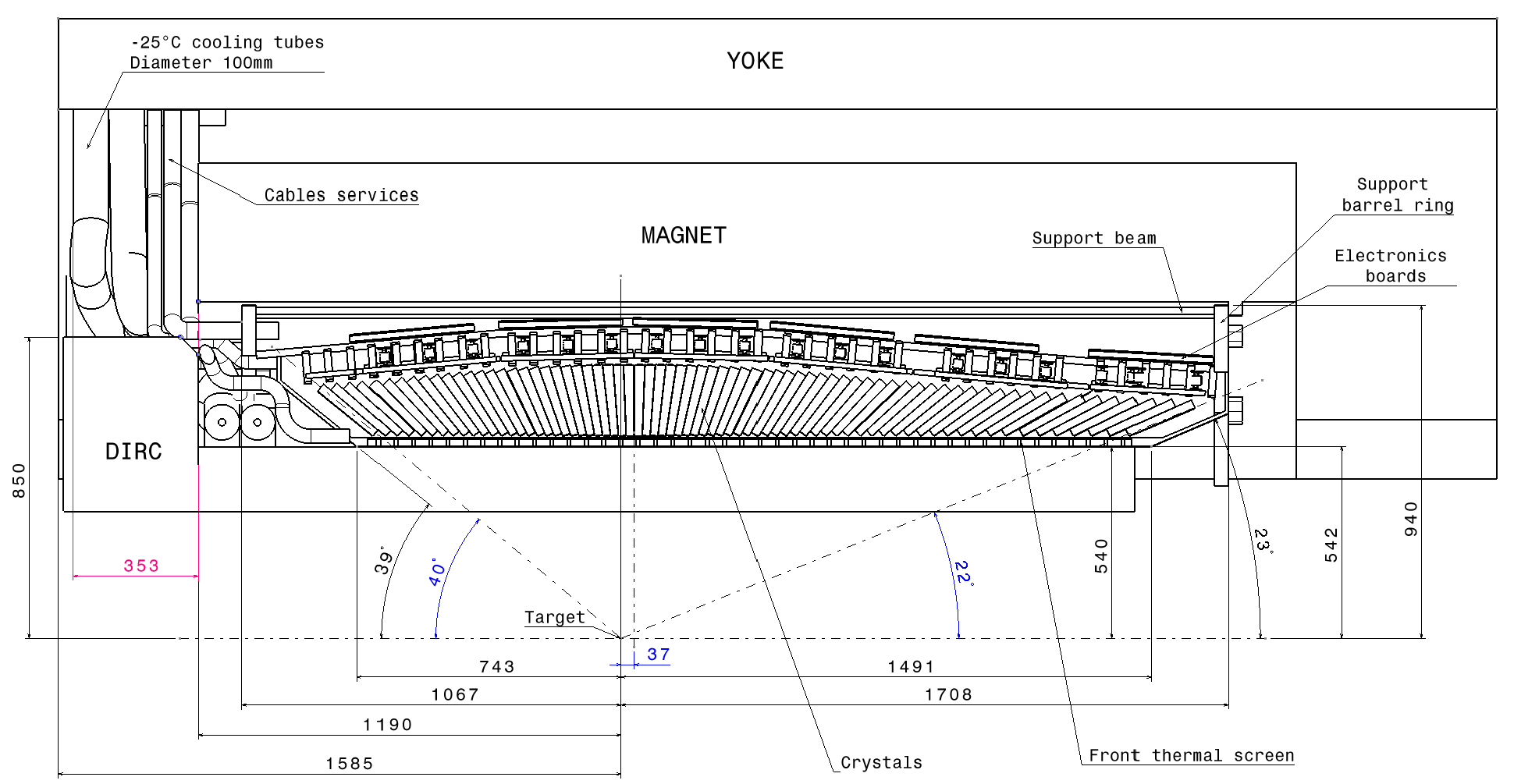}
\caption[Dimensions of a slice.]{Dimensions and volume of one slice.}
\label{fig:mech:mech-vue_kiv3}
\end{center}
\end{figure*}

\begin{figure}
\begin{center}
\includegraphics[width=1\swidth]{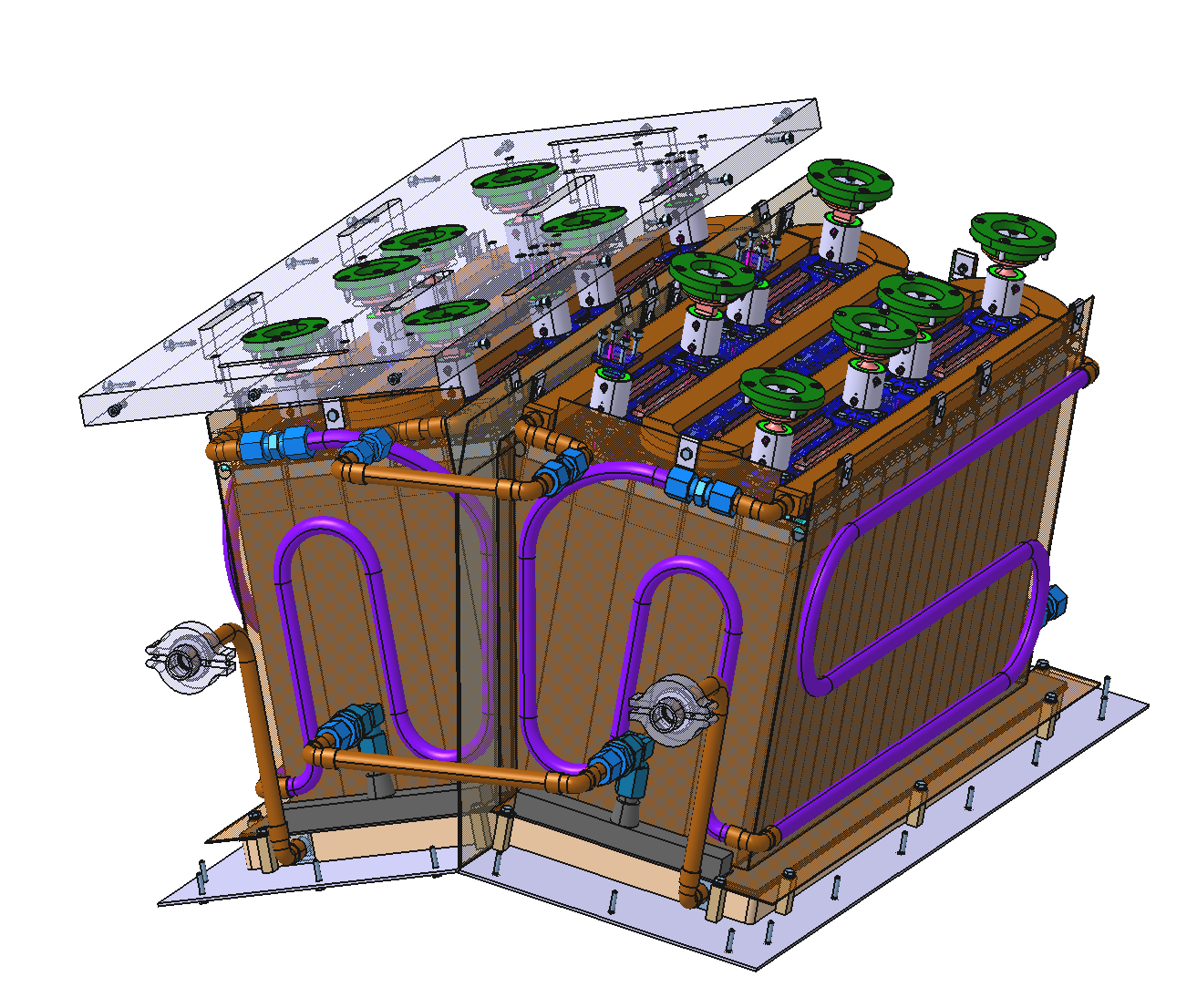}
\caption[Next thermal prototype 480.]{Next thermal prototype 480.}
\label{fig:mech:mech-air-tightness_studies_proto480}
\end{center}
\end{figure}

The condition for a feasible operation at low temperature is to avoid
any ice generation as well inside as outside of the calorimeter. If this would not be achieved, the ice
could introduce water at reheating of the detector or could break the crystals
by ice pressure between the crystals. The method is to remove the humidity
of the air by circulating a dry gas like nitrogen inside an air-tight
envelop surrounding all the calorimeter. The continuity of this
sealing is kept at the boundaries with a feed-through like those used for
electrical or supply connectors. This envelop is mainly made with
plastic covers wrapped with a special industrial thick adhesive
developed for long term domestic gas sealing. For the outside
part, the method is to avoid any temperature lower than the dew point
(around 12$\degC$) on the structure like on the external cooling tubes
or on the mechanical supports for instance. Thermal studies are done
on every critical part and massive parts at room temperature are used
in order to equilibrate the low temperature transfered by conduction.
From a risk analysis the following 2 difficulties arise:
\begin{itemize} 
\item{Danger for the calorimeter in case of ice and moisture: if the
    dry gas is not completely well circulating or if humidity remains
    after any switch off of the dry gas circulation. Humidity sensors
    can provide some information on the quality of the atmosphere. A
    break of the air-tightness envelop can lead to ice, too, and even
    alter the quality of the thermal shields.}
\item{Danger for the other detectors if the thermal shields break. The
    cold can reach the outside sides and put moisture and water in the
    target spectrometer.}
\end{itemize} 
The Proto60 crystals have shown the feasibility of this
air-tightness envelop.  Thermal studies and risk analysis will
continue and be performed with the next prototype of 480 crystals to
improve the reliability and define the emergency procedures.
\Reffig{fig:mech:mech-air-tightness_studies_proto480} introduces the
CAD design of this set-up which will be available in summer 2008. It
integrates all the components of a slice in their final shape (e.g. support
feet, front thermal shield). It is equipped with stainless steel
dummy crystals for cost reasons but should exhibit the same thermal
behavior.

\subsubsection{Room Temperature Stabilization for the Electronics}
\label{sec:emc:mech:therm:room_temp}

The electronics is dissipating 560\,W/slice and is mounted on copper
plates brazed to cooling tubes (see
\Reffig{fig:mech:mech-Support_beam_integration}). Water flows inside the tubes
to regulate the support beam at approximatively room temperature
($\pm$2$\degC$). The machine must have a minimum effective cooling
capacity of 9\,000\,W, and a total flow of 65 liters/min. This level of
requirements can be found in standard industrial systems.

\begin{figure*} 
\begin{center}
\includegraphics[width=0.7\dwidth]{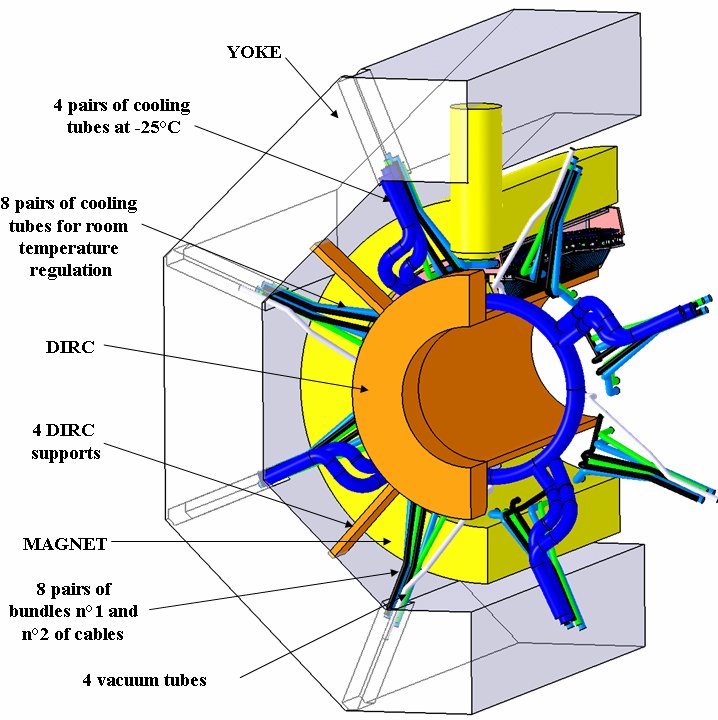}
\caption[Services out of barrel.]{Services going outside the barrel into the corners of the octagonal yoke.}
\label{fig:mech:mech-vue_services barrel}
\end{center}
\end{figure*}

\subsection{Integration in the \PANDA Target Spectrometer} 
\label{sec:emc:mech:bar:integ}

\Reffig{fig:mech:mech-vue_kiv3} defines the complete volume
of the barrel. The overall dimensions are required for the integration of the neighboring
detectors. The total mass is $\approx$ 20\,tons composed of 11\,tons of crystals and 7\,tons of support structure.  The services are going out of the slices in the backward area and
afterwards through the corners of the octagonal shape of the yoke as
shown in \Reffig{fig:mech:mech-vue_services barrel}. A list of services and respective details is given in
\Reftbl{tab:mech:int_serv}.

\begin{table}
\begin{center}
\begin{tabular}{lccc}
\hline\hline
{\bfseries Service} & Detail & $\oslash (\mm)$ & Qty. \\
\hline\hline
Bundle 1 & High/Low volt. & 50 & 16   \\ 
 & ADC supply & & \\
 & optical fibers & & \\
\hline
Bundle 2 & Gas, humidity/  & 40 & 16  \\
 & temperatur sensors & & \\
 & calibr. opt. fibers& & \\
\hline
Tube 1 & Water 20$\degC$ & 50 & 16   \\
\hline
Tube 2 & Coolant -25$\degC$ & 105 & 8   \\
\hline
Vacuum & & 50 & 4   \\
\hline\hline
\end{tabular}
\caption[List of services in the backward area.]{List of services
in the backward area.}
\label{tab:mech:int_serv}
\end{center}
\end{table}

\subsection{Construction of the Slices and Assembly of the Barrel} 
\label{sec:emc:mech:bar:Assembly}

This report presents the design of a slice. A preliminary mounting
sequence is proposed here:
\begin{enumerate}
	\item{Gluing of \LAAPD onto crystals (after control and reference measurements).}
	\item{Wrapping of crystals.}
	\item{Insertion into alveoles. Installation of thermal sensors
            between crystals. Gluing of the insert after having
            controlled the good functioning of the \LAAPD with twisted pair
            wire.}
	\item{Assembly of the alveole pack on the module
            plate. Installation of the guiding tube for the optical
            fiber. Fixing the preamplifier. Control of the
            alignment.}
	\item{Mounting of the modules plates on the mainframe tool (as in
            \INST{CMS}). Adding top and bottom cooling circuit. Adding
            calibration optical fibers. Checking of the keep in volume
            and alignment. Mounting the support feet and spreading the
            insulation.}
	\item{Put the support beam. Connecting flexible cables to the
            boards and adding all electronics. First-level air sealing.}
	\item{Transfer the slice to its support for storage and
            shipment and for the support with insulation sides to cool it
            down. This prepares the test with cosmic muons.}
	\item{Mount the rings on the rolling mounting tool. Then mount
            the slices one by one.}
\end{enumerate}

To perform all these tasks, special tools need to be designed and
manufactured.  The construction of the first prototype of a slice is
foreseen and will validate this sequence.
 
The construction of a single slice and the assembly of the complete barrel require a large amount of manpower, installation space and time for testing. The production time is of
the order of 3 or 4 years and will need a good coordination between the different laboratories.

\subsubsection{Final Assembly of the Barrel} 
\label{sec:emc:mech:bar:Assembly:final}

\begin{figure} 
\begin{center}
\includegraphics[width=1\swidth]{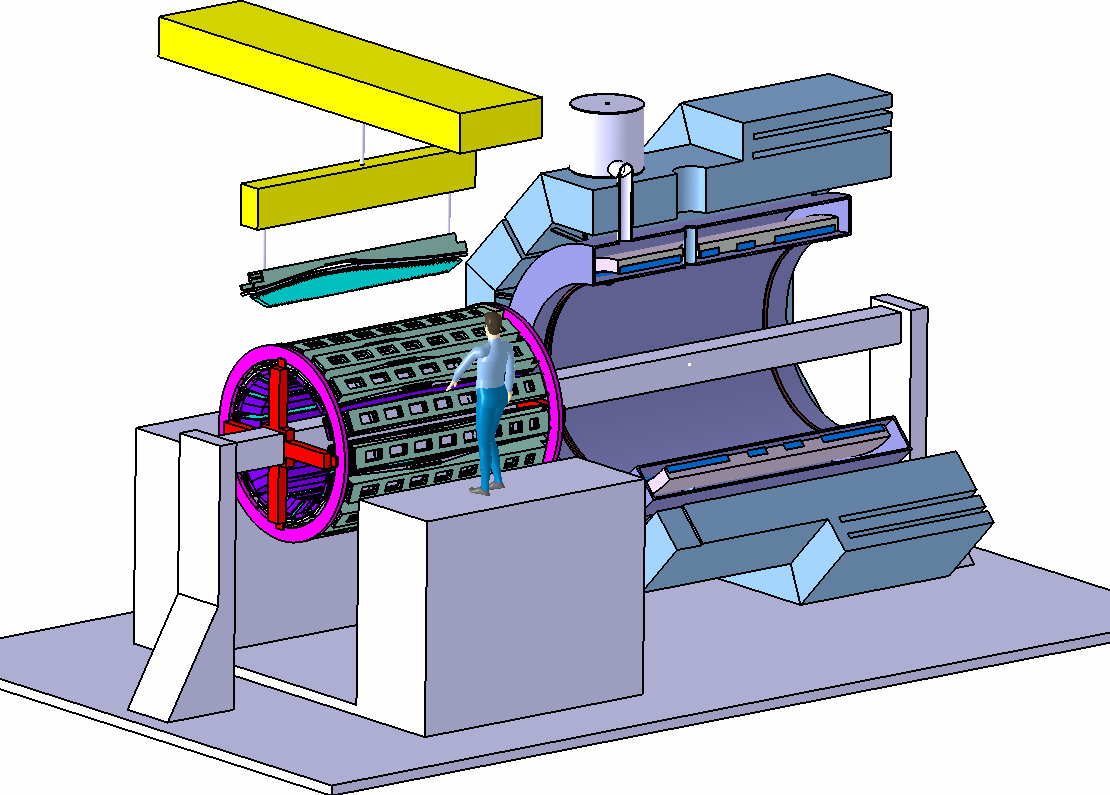}
\caption[Barrel final mounting.]{Barrel final mounting.}
\label{fig:mech:mech-Barrel_mounting}
\end{center}
\end{figure}

Each slice is installed one by one on the two support rings and a
special rolling system (as in \INST{CMS}) can be used.
\Reffig{fig:mech:mech-Barrel_mounting} presents a sketch of the
mounting sequence of the barrel.  Once the barrel is complete, tested and
air-tight insulated, it can slide to its final position on a stable central beam going through
the \PANDA detector.


\section{Forward Endcap}
\label{sec:emc:mech:fwemc}

\subsection{Requirements}
\label{sec:emc:mech:fwemc:requirements}

The envisaged physics program of \PANDA requires measurements of photons and charged particles with good position and timing resolution over a wide dynamic range from a few MeV up to several GeV energy. The electromagnetic calorimeter of \PANDA comprises the \BEMC, the \FWEMC, and the \BWEMC, see \Reffig{fig:ecap:EMCGraph}.

The \FWEMC is designed as a wall structure with off-pointing projective geometry, i.e. the crystals are oriented to a point on the beam axis which is located at a certain distance (in this case 890 mm) away from the target. This arrangement guarantees that particles originating from the target will never pass exactly along the boundaries between two neighbouring crystals where they could remain undetected. Since at the same time quadrant symmetry is required, this condition can not be maintained for the boundaries between the four quadrants of the \FWEMC which will be oriented along the lines (x,y=0) and (x=0, y).

In order to catch electromagnetic showers completely at the boundary between the Barrel and the \FWEMC, the acceptance of the \FWEMC must foresee one full crystal overlap with the Barrel. The acceptance is limited at the most forward angles by the space required for the forward magnetic spectrometer. This defines the outer opening angle to be $<$~23.6\degrees and the inner opening angle to be $>$~5\degrees in vertical and $>$~10\degrees in horizontal direction. 

Simulations performed at 15$\,\GeV$ (see \Refsec{chap:scint:pwo:rates-crystals}) indicate that the particle rates per calorimeter cell with a detection threshold of 3$\,\MeV$ reach 500$\,\kHz$ at the smallest angles and still amount to 100$\,\kHz$ at the largest angles. Because of these high rates and the increased risk of radiation damage the large-area avalanche photodiodes (\LAAPD), foreseen for the Barrel, will be replaced by vacuum phototriode (\VPT) photosensors in the \FWEMC. From two production sites \VPT's are available with an outer diameter of 22$\,\mm$. This size can be accommodated on a crystal with a rear cross section of $26\times26\,\mm^2$, while still maintaining the condition for optimum position resolution, namely an average crystal width of about one Moli\`ere radius (20$\,\mm$ for \PWO).    
The \FWEMC will contain \PWO crystals (PWO-II) of 200$\,\mm$ length, which is equivalent to 22 radiation length and sufficient for 95\% containment of the maximum expected photon energy of 15$\,\GeV$. 

\subsection{Crystal shape}
\label{sec:emc:mech:fwemc:crystal_shape}

\begin{figure*}
\begin{center}
\includegraphics[width=0.65\dwidth]{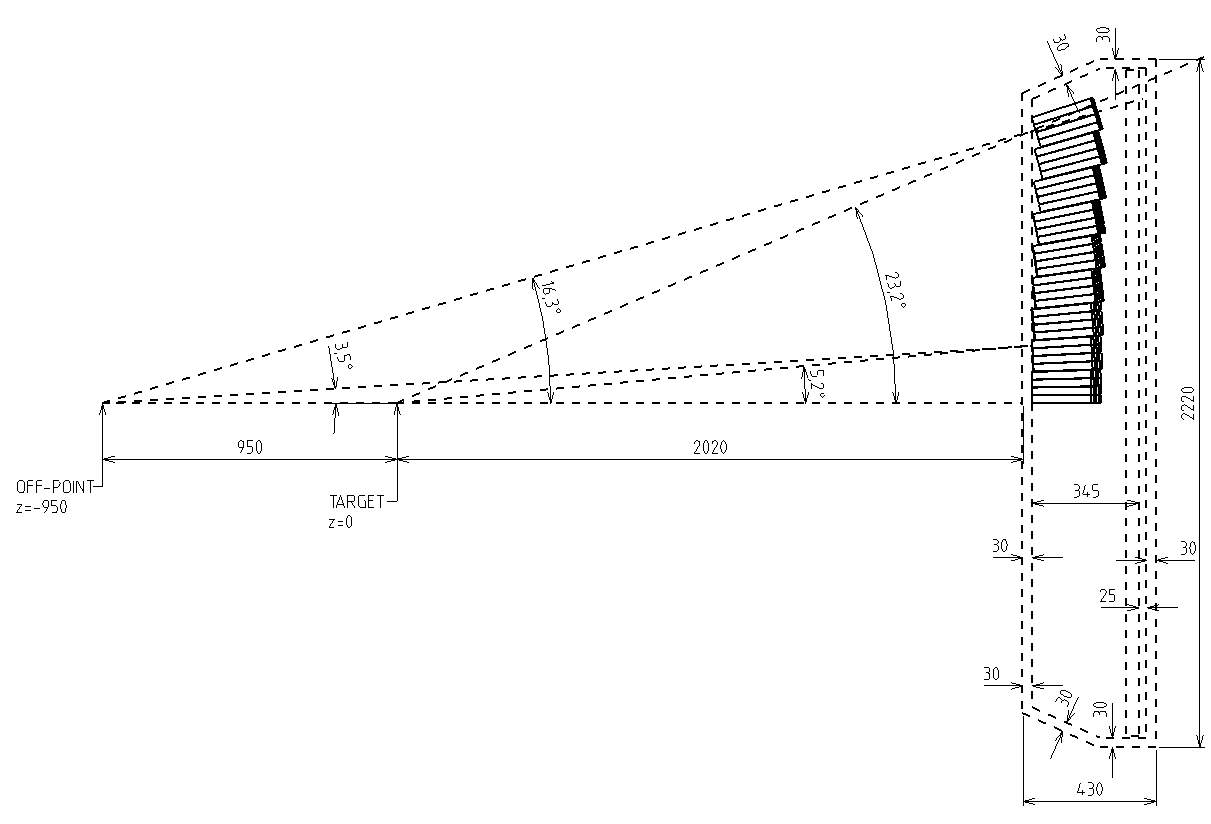}
\caption[The position of the \FWEMC with respect to the target.]{The position of the \FWEMC with respect to the target.}
\label{fig:mech:ecap:FwEndCapDistance}
\end{center}
\end{figure*}

\begin{figure*}
\begin{center}
\includegraphics[width=0.65\dwidth]{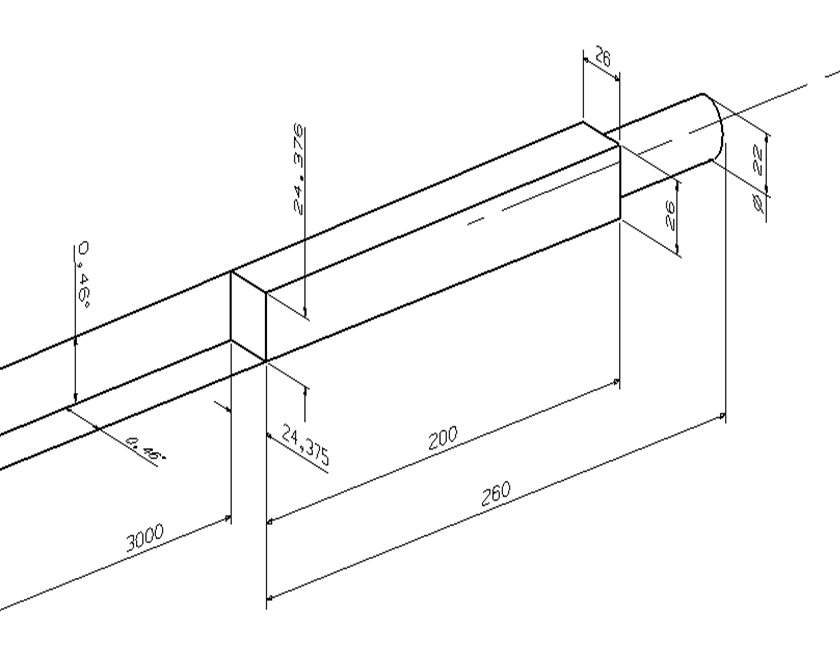}
\caption[The geometry of a single \FWEMC crystal.]
{The geometry of a single \FWEMC crystal.}
\label{fig:ecap:FwEndCapCrystal}
\end{center}
\end{figure*}

In the \FWEMC the crystals are closely packed in "off-pointing" geometry, i.e. oriented towards a point on the beam axis 950$\,\mm$ farther than the target and 3000$\,\mm$ away from the front face of the crystal plane. 
The off-pointing geometry is illustrated in \Reffig{fig:mech:ecap:FwEndCapDistance}. The ratio of target distance to off-pointing distance has been chosen such that the angle of incidence of particles on the crystal front face is minimally 1.6\degrees with respect to normal incidence. This arrangement guarantees that particles originating from the target will never pass more than 15\% of the crystal length in the gap between two neighbouring crystals.
Including photo sensors, front-end electronics, cooling and insulation, the overall depth of the \FWEMC will amount to 430 mm. 
The off-pointing geometry and the \VPT diameter determine that each crystal has a front-face of $24.4\times24.4\,\mm^2$, see \Reffig{fig:ecap:FwEndCapCrystal}. In order to save costs for cutting and polishing crystals, each crystal will be shaped with two tapered and two right-angled sides. A cluster of 4 crystals, touching at the right-angled sides, thus forms a mini-unit of trapezoidal cross section, mounted in a single carbon-fiber alveole with 0.18$\,\mm$ wall thickness. 
The crystals should be manufactured with tolerance of +0/-0.1$\,\mm$ in all transversal and longitudinal dimensions. An angular precision of $<$~0.01\degrees will be requested while a planarity within 0.02$\,\mm$ should be maintained for all faces with chamfers between 0.5 and 0.7$\,\mm$. All surfaces will be polished with roughness Ra $<$~0.2$\,\mm$. 
 
\subsection{Subunit Structure}
\label{sec:emc:mech:fwemc:subunit}

\begin{figure*}
\begin{center}
\includegraphics[width=0.7\dwidth]{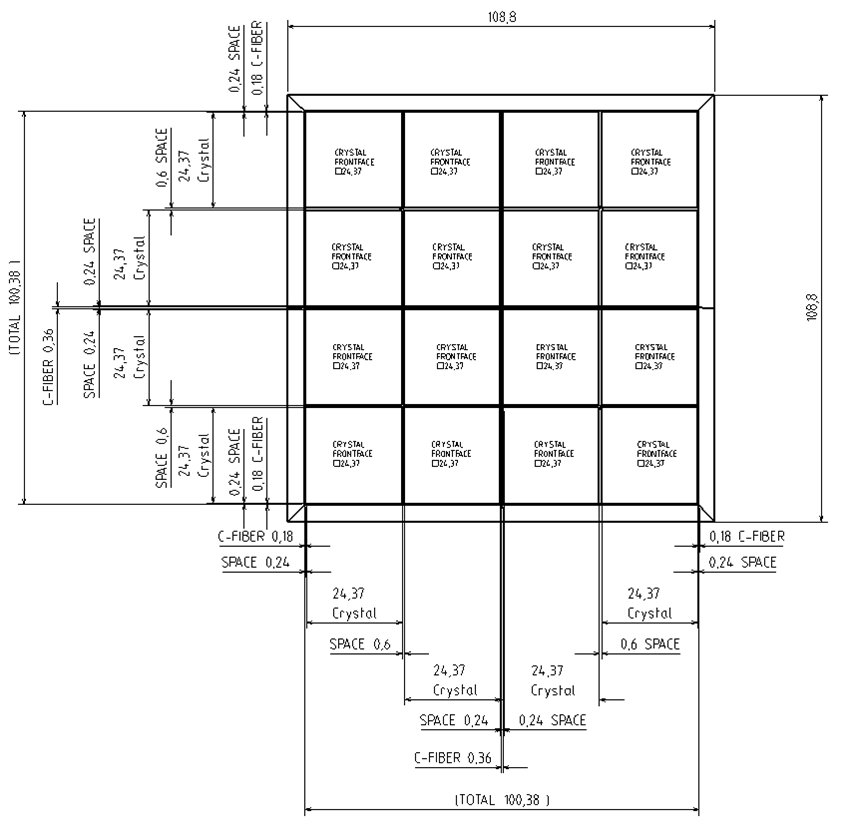}
\caption[The geometry of a C-fiber alveole for 16 crystals.]
{The geometry of a C-fiber alveole for 16 crystals.}
\label{fig:ecap:FwEndCapAlveole}
\end{center}
\end{figure*}

\begin{figure*}
\begin{center}
\includegraphics[width=0.7\dwidth]{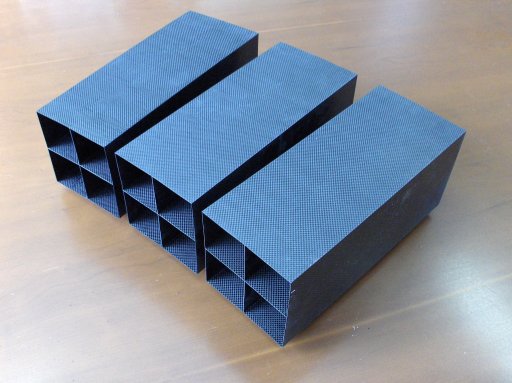}
\caption[Photograph of three produced C-fiber alveoles for 16 crystals each.]
{Photograph of three produced C-fiber alveoles for 16 crystals each.}
\label{fig:ecap:FwEndCapAlveolePhoto}
\end{center}
\end{figure*}

Since PWO crystals are very fragile, they must not be exposed to bending- or shear forces. Therefore crystals are mounted in frames made from composite material (carbon fiber alveoles) which are designed to absorb tolerances in crystal dimensions and accommodate the thickness of the light-reflecting foils (100$\,\mu\m$). Four mini-units of crystals will be combined to form a 16-crystal subunit of ca. 19$\,\kg$ that can be attached individually to the 30~mm thick aluminum mounting plate. 
\Reffig{fig:ecap:FwEndCapAlveole} shows the alveole wall thickness and the space foreseen between crystals. The front face of the alveole will be covered with 1$\,\mm$ thick composite C-fiber/epoxy material. The company FiberWorx B.V. (Netherlands) has designed, engineered and prototyped a C-fiber alveole for 16 crystals. Loaded with the weight of the crystals the alveole material is requested to stretch less than 1.5\percent. The proposed and prototyped design results in a maximum stretch of 0.04\percent, which results in a comfortable safety factor of 30.  
\Reffig{fig:ecap:FwEndCapAlveolePhoto} shows three C-fiber alveoles produced according to the above given specifications.

\begin{figure*}
\begin{center}
\includegraphics[width=0.7\dwidth]{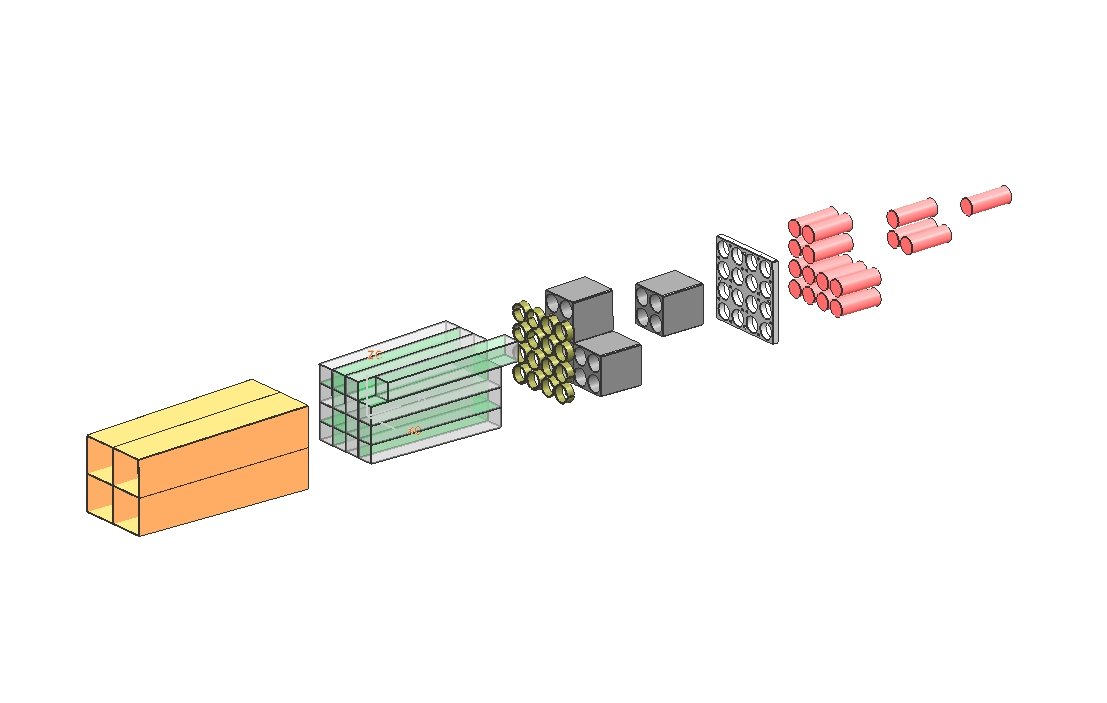}
\caption[The explosion view of a 16-crystal subunit.]
{The explosion view of a 16-crystal subunit.}
\label{fig:ecap:FwEndCapExplAlveole}
\end{center}
\end{figure*}

\begin{figure*}
\begin{center}
\includegraphics[width=0.7\dwidth]{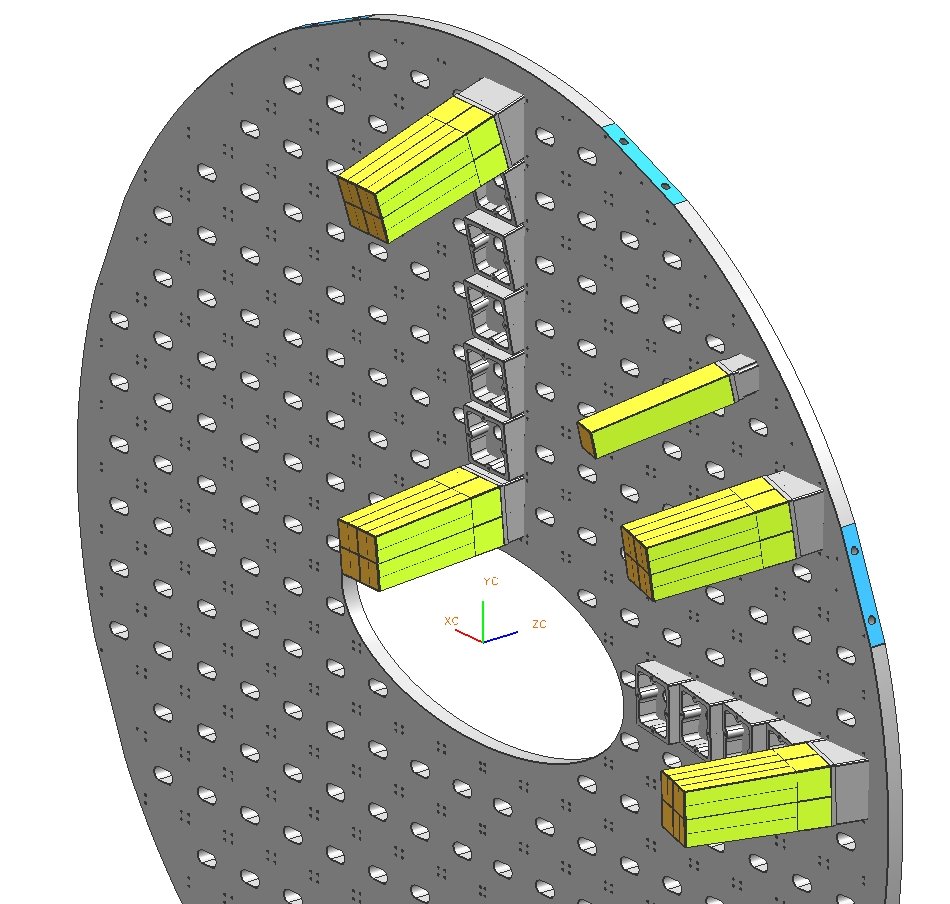}
\caption[The attachement of the C-fiber alveoles in front of the mounting plate.]
{The attachement of the C-fiber alveoles in front of the mounting plate
using angled interface pieces.}
\label{fig:ecap:FwEndCapMountAlveole}
\end{center}
\end{figure*}

The alveoles carrying 16 crystals will be mounted individually to the aluminum backplane without touching the neighboring alveole. For this purpose aluminum inserts will be glued into the rear part of the alveole downstream of the crystal, thus allowing a good thermal contact, providing a rigid support for the \VPT photosensor, and creating a mounting structure for the backplane. \Reffig{fig:ecap:FwEndCapExplAlveole} shows the explosion view of a subunit housing 16 crystals, \VPT and inserts. In \Reffig{fig:ecap:FwEndCapMountAlveole} is demonstrated how the subunits will be attached to the mounting plate using angled interface-pieces which allow a precise arrangement. In total 3520 crystals will be mounted in this way in $4\times4$ subunits, and 80 crystals in smaller $2\times2$ subunits in order to approach as much as possible a homogeneous coverage between minimum and maximum acceptance angle of the \FWEMC. Optionally, for even larger coverage at the expense of more specific alveole development, an additional number of ca. 40 subunits housing only 1 or 2 crystals is being considered.
   
Preliminary thermal calculations indicate a temperature gradient in the crystal of ca. 4$\degC$ if only the mounting plate will be cooled. Presently a prototype setup for 16 crystals is being constructed in order to test the mechanical accuracy, the thermal properties and photon response in photon-beam experiments.

\subsubsection{Mounting Structure and Implementation in the Solenoid}
\label{sec:emc:mech:fwemc:subunit:mounting_struct}

\begin{figure*}
\begin{center}
\includegraphics[width=0.7\dwidth]{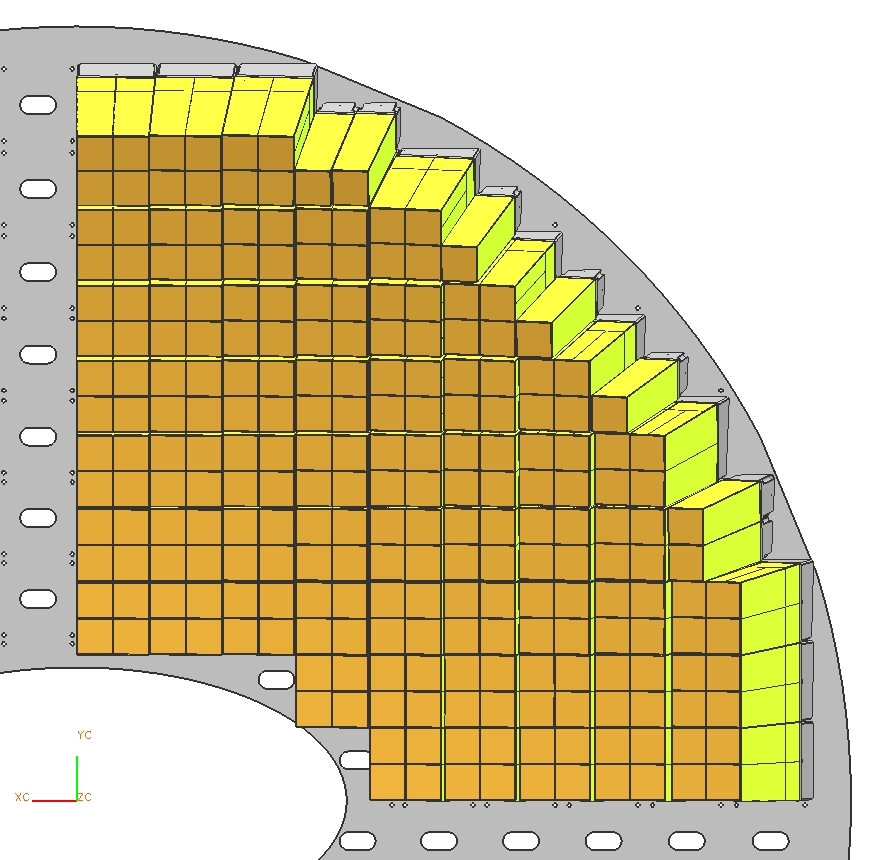}
\caption[One quadrant of the \Panda forward endcap calorimeter.]
{One quadrant of the \Panda forward endcap calorimeter.}
\label{fig:ecap:FwEndCapQuadrant}
\end{center}
\end{figure*}

One quadrant of the \FWEMC houses 900 crystals in subunits of mostly 16 and occasionally 4 crystals. The arrangement as seen from the target is shown in \Reffig{fig:ecap:FwEndCapQuadrant}. The backplane is constructed from 30~mm thick aluminum with elliptic holes at the center of every alveole in order to feed the cables from the front-end electronics to the downstream part of the mounting plate and from there to the circumference of the \FWEMC structure.  

The total weight of the \FWEMC will amount to ca. 5100~kg. For the design of the mounting plate the stiffness is the most important criterium. The loads acting on the mounting plate are the gravity and the moment caused by the center of gravity of the crystals attached in front of the plate. In its final vertical position the maximal Von Mises stress is ca. 20~MPa which results in a maximum deflection of 0.23~mm. 

The endcap mounting plate will shrink by 0.16~mm per row or column of subunits at the operation temperature of -25\degrees C. With 18 rows and columns of subunits, there are 17 gaps in between in  horizontal and vertical direction. Per gap a space has to be reserved of approximately 0.16~mm in order to absorb shrinking or expansion. In addition we have to take into account that the magnetic field of the solenoid will cause deflections of the mounting structure. It is foreseen that the \FWEMC will be  completely constructed outside the \PANDA area, inserted into the downstream part of the solenoid magnet yoke, and held in place by 8 holding arms inside an octagonal frame structure. This structure must be able to absorb a shrinking or expansion of maximally 3~mm.

\subsubsection{Insulation and Cooling}
\label{sec:emc:mech:fwemc:subunit:insulation_cooling}

\begin{figure*}
\begin{center}
\includegraphics[width=0.7\dwidth]{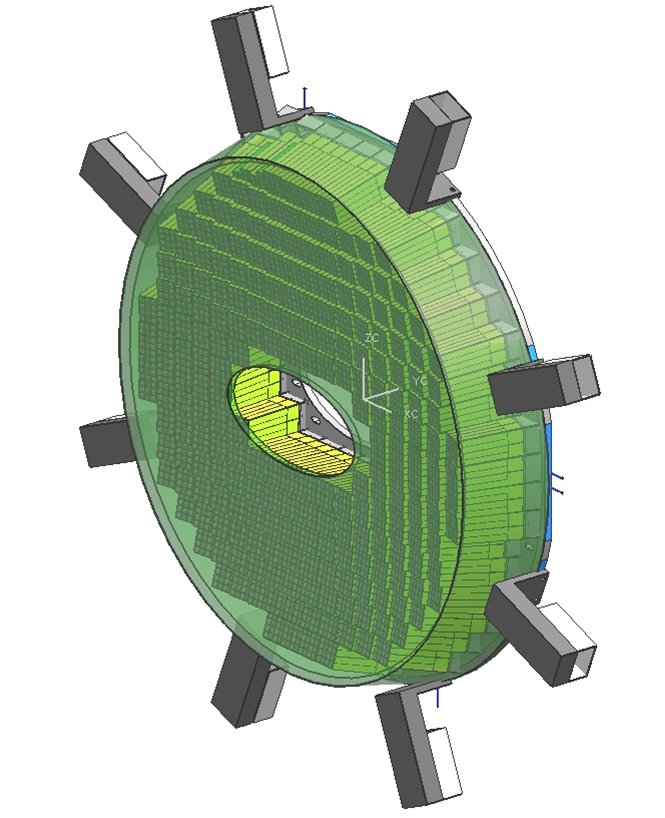}
\caption[The complete endcap.]
{The complete endcap with thermal insulation cover and holding structures.}
\label{fig:ecap:FwEndCapInsulation}
\end{center}
\end{figure*}

For the thermal insulation of the crystal area at -25$\degC$ from room temperature, a space of 30~mm is reserved for a layer thermal-insulation foam. Detailed thermal calculations are needed and have been started in order to determine the required cooling power and shielding material. Preliminary calculations and experience with the Barrel prototype indicate, that the crystals also need cooling from the front side in order to avoid a temperature gradient in the crystal. However, maintaining a stationary temperature gradient will be investigated, since the corresponding gradient in light production could compensate inhomogeneities due to light collection in a tapered crystal. This method would avoid an expensive and time consuming "depolishing" treatment of the crystal surface. Cooling at the front side of the crystal wall could be achieved by cooling a thin (ca. 2~mm) carbon plate in front of the crystals. A flow of cooled dry N$_2$ gas inside the insulated volume would support an homogeneous cooling of the crystal volume. To prevent ice forming around the crystals, dry N$_2$ gas will be supplied to the individual alveoles by gas pipes inserted through the cable-feedthrough holes in the mounting plate. 
\Reffig{fig:ecap:FwEndCapInsulation} shows part of the thermal insulation cover and the holding structures to facilitate the mounting inside the solenoid.

It is foreseen to embed four separate cooling circuits into grooves in the downstream part of the mounting plate. The total cooling circuit will consist of a meander of 10~mm diameter cooling pipes with a length of 56~m.  

\subsubsection{Sensors and Front-End Electronics}
\label{sec:emc:mech:fwemc:subunit:sensors_elect}

The crystals of the \FWEMC will be equipped with \VPT's as photosensors for which a gain of 30 to 50 is expected. Development of a highly sensitive \VPT with super-photocathode and quantum efficiency above 40\percent is in progress. A Low Noise / Low Power (LNP) Charge Preamplifier has been developed on basis of the equivalent LNP preamplifier for \LAAPD readout. The output pulse is transmitted via a 50~$\Omega$ line to the subsequent digitizer module electronics at the circumference of the \FWEMC. The LNP-Preamp has a quiescent power consumption of 45~mW and is suited for operation in the cooled volume. It is foreseen to attach the preamplifier directly to the \VPT on the upstream side of the mounting plate.

\subsubsection{Cabling and Supplies for Monitoring}
\label{sec:emc:mech:fwemc:subunit:cabling}

\begin{figure*}
\begin{center}
\includegraphics[width=0.7\dwidth]{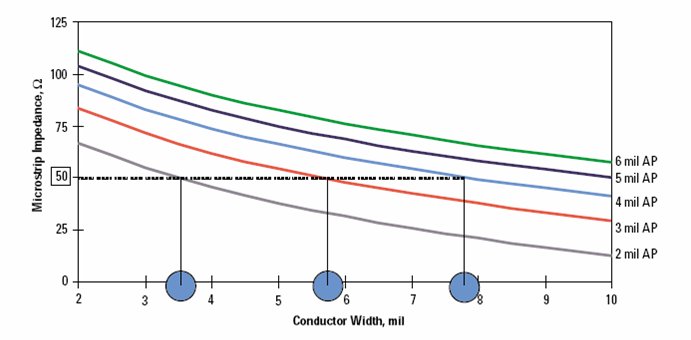}
\caption[Impedance of laminate cable.]
{Impedance of laminate cable as function of conductor width.}
\label{fig:ecap:FwEndCapFlatcable}
\end{center}
\end{figure*}

For signal transmission from the \VPT preamplifier to the digitizer electronics a multi-pin flexible laminate cable of .15~mm thickness is foreseen. Q.P.I. BV Netherlands is able to produce a double-sided, copper-clad all-polyimide composite (PyraluxAP). This is a polyimide film bonded to copper foil.   
\Reffig{fig:ecap:FwEndCapFlatcable} gives the relation between impedance and conductor width and allows to make a suitable choice for 50~$\Omega$ signal transmission.

\subsubsection{Arrangement of Readout Electronics}
\label{sec:emc:mech:fwemc:subunit:readout_elect}

\begin{figure}
\begin{center}
\includegraphics[width=1\swidth]{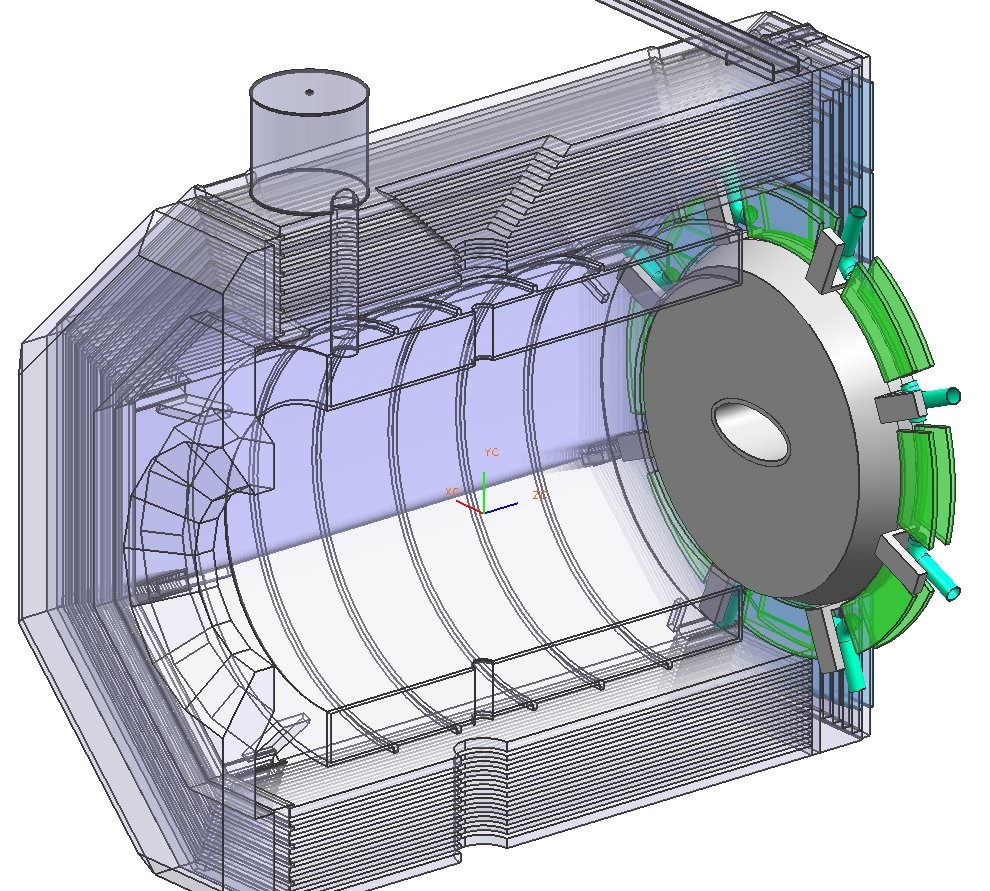}
\caption[Arrangement of digitizer modules for the \FWEMC.]
{Arrangement of Digitizer modules at the circumference of the \FWEMC.}
\label{fig:ecap:FwEndCapADCelectronics}
\end{center}
\end{figure}

The digitizing electronics will use sampling ADC's for signal-shape analysis and timing determination and will be located near the detector in the warm volume inside the solenoid magnet. From there signals can be transmitted via optical link to multiplexer units and compute nodes outside the \PANDA experimental area. With commercial components an ADC channel density of about 300~mm$^2$/channel can be reached. This means that per octant of the \FWEMC an area of 3200~cm$^2$ must be available. Between the circumference of the \FWEMC mounting plate (radius = 1050 mm) and the inner boundary of the solenoid (radius = 1450 mm) a surface area of maximally 2000~cm$^2$ is available. Thus with a sandwich of two ADC boards there is sufficient space available for accommodating the digitizing electronics. \Reffig{fig:ecap:FwEndCapADCelectronics} shows the position of the endcap inside the solenoid structure and the space available at the \FWEMC circumference for arranging the electronics boards in a sandwich layer. In addition, the 8-fold mounting structure is indicated and the routing of cables and cooling pipes. 

\section{Backward Endcap of the Calorimeter}
\label{sec:emc:mech:bwemc}

The \BWEMC is required in order to complete the hermetic coverage of the \TSEMC for electromagnetic energy, with exception of the beam entrance and the acceptance of the forward spectrometer. As shown in \Reffig{fig:perf:Ediff}, at polar angles above 135\degrees the differential rates per crystal are about 1$\,\kHz$ per 20$\,\MeV$ at energies of 50$\,\MeV$ and the maxium energy deposition is about 200 MeV. Since both endcaps require an almost planar arrangement of crystals, the basic design concept of the \FWEMC has been chosen also for the \BWEMC. This approach simplifies the mechanical construction of crystals and submodules and creates synergy in the application of photosensors and readout electronics. In addition,  
the readout with \VPT is superior in timing performance at the low energies expected in the backward region which allows efficient reduction of background. Since design and development work on other detectors in the surrounding of the \BWEMC is still in progress, the mechanical constraints are not yet  defined wellenough, so that the integration in \Panda could not yet be specified in detail. 
\begin{figure}
\begin{center}
\includegraphics[width=\linewidth]{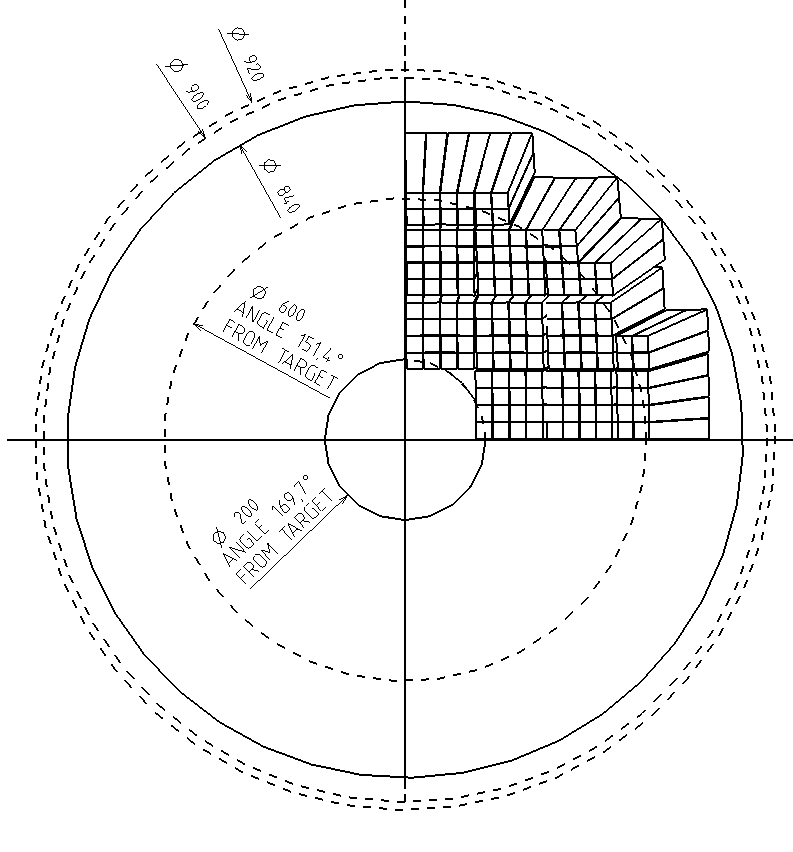}
\caption[Acceptance of the \BWEMC]{Acceptance of the \BWEMC.}
\label{fig:mech:ecap:BwEndCapAcceptance}
\end{center}
\end{figure}
In \Reffig{fig:mech:ecap:BwEndCapAcceptance} the angular acceptance is presented which is allowed by the radial space available inside the \TSEMC and the distance of 550 mm of the crystal front face to the target (see \Reffig{fig:mech:mech-Target_spectrometer}). The maximum opening diameter is 920 mm. Allowing 20 mm for tolerances and mounting space  and 60 mm for thermal insulation, we arrive at 840 mm maximum diameter of the mounting plate to which the readout end of the crystals is attached. The inner hole of 200 mm diameter of the \BWEMC is determined by the beam pipe and a safe distance to prevent disturbance from beam halo. This defines the maximum polar angle of 169.7\degrees. The tilting of the crystals due to the off-pointing projective geometry, oriented towards a point 200 mm farther than the target, determines the minimum polar angle of 151.4\degrees. 
The off-pointing geometry is illustrated in \Reffig{fig:mech:ecap:BwEndCapDistance}. The same ratio of target distance to off-pointing distance has been chosen as for the \FWEMC. Including photo sensors, front-end electronics and insulation, the overall depth of the \BWEMC will amount to 430 mm. 
\begin{figure*}
\begin{center}
\includegraphics[width=0.65\dwidth]{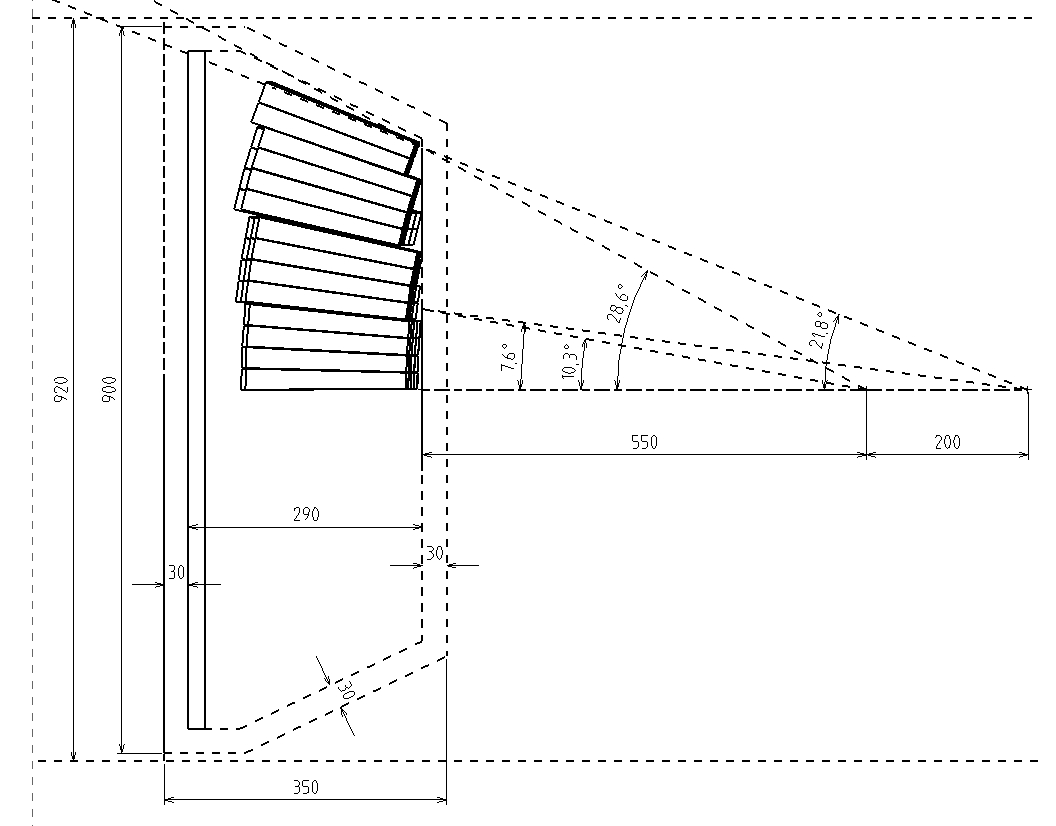}
\caption[The position of the \BWEMC with respect to the target.]{The position of the \BWEMC with respect to the target.}
\label{fig:mech:ecap:BwEndCapDistance}
\end{center}
\end{figure*}
\begin{figure}
\begin{center}
\includegraphics[width=\linewidth]{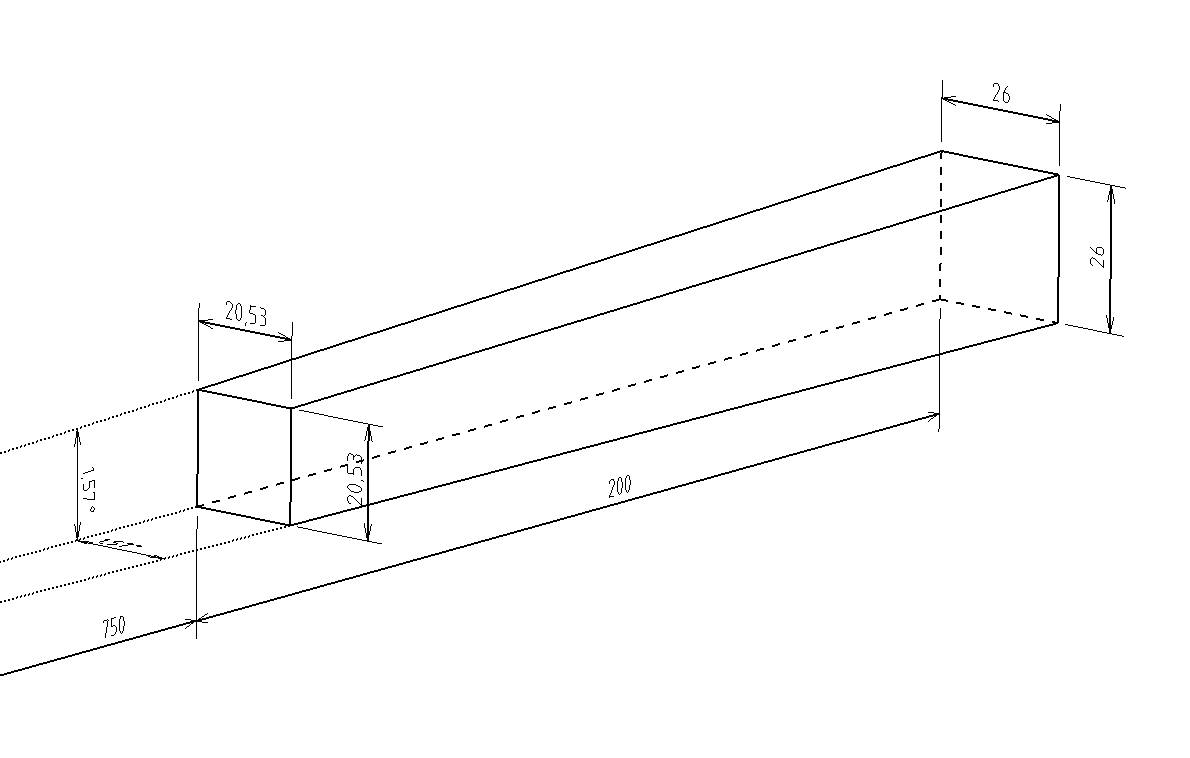}
\caption[The dimensions of a single crystal of the \BWEMC.]{The dimensions of a single crystal of the \BWEMC.}
\label{fig:mech:ecap:BwEndCapCrystal}
\end{center}
\end{figure}

The geometry of a single crystal is shown in \Reffig{fig:mech:ecap:BwEndCapCrystal}. Given the cross section of $26\times26\,\mm^2$ of the readout face to accommodate the VPT, the front face of each individual crystal will result under the given geometrical conditions in $20.5\times20.5\,\mm^2$. The overall dimensions are close to the average dimensions of the barrel crystals.  
\begin{figure}
\begin{center}
\includegraphics[width=\linewidth]{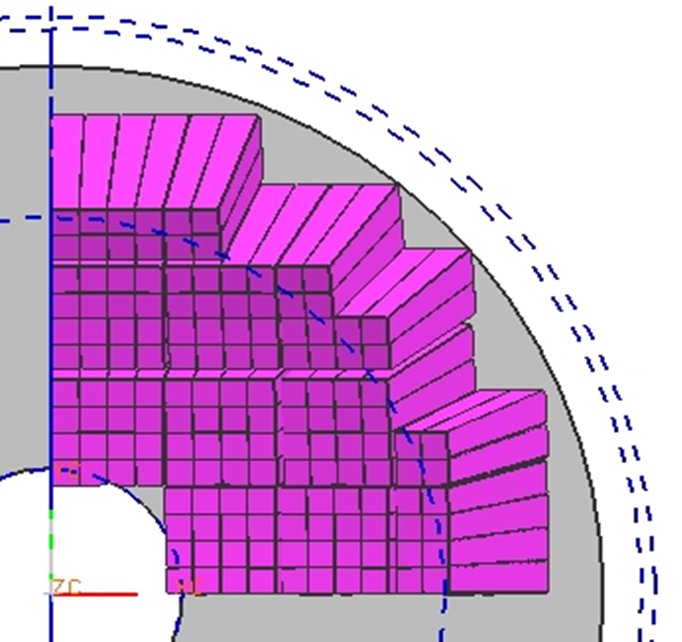}
\caption[The geometry of one quadrant of the \BWEMC.]{The geometry of one quadrant of the \BWEMC.}
\label{fig:mech:ecap:BwEndCapQuadrant}
\end{center}
\end{figure}
The arrangement of individual crystals in one quadrant of the \BWEMC is shown in \Reffig{fig:mech:ecap:BwEndCapQuadrant}.
As for the \FWEMC, the size of the carbon fiber alveoles, the tolerances and the configuration in subunits of $4\times4$ or $2\times2$ crystals has been chosen. The optimum coverage of the available geometrical acceptance is achieved with 7 subunits of 16 crystals and 9 subunits of 4 crystals. This results in 148 crystals per quadrant and 592 crystals for the whole \BWEMC. 
The provisions for nitrogen gas flow and cooling will be arranged as in the \FWEMC. Due to space restrictions, the Digitizer modules can not be attached to the mounting plate, but will be positioned further upstream between the upstream barrel part and the solenoid yoke. 

%
%
\newpage
\bibliographystyle{panda_tdr_lit}
\bibliography{./lit_emc}
%

%
\cleardoublepage
\chapter{Calibration and Monitoring}
\label{sec:cal}
%
%
\label{sec:cal:intro}
To achieve the required energy and spatial resolution of the electromagnetic calorimeter it is essential to precisely calibrate the individual crystal channels. Time dependent variations of the calibration factors are expected due to several effects: change in light transmission, change in the coupling between the crystal and the photodetectors and variations in the photodetectors itself and the following electronics. Both the scintillation of the crystals and the amplification of the APDs depend on the temperature. Therefore stable cooling at the level of 0.1$\degC$ is required. Long term variations in the temperature need to be tracked by the calibration. 
Given an energy resolution between 1\% and 2\% at energies above $1\,\GeV$, the precision of the calibration needs to be at the sub-percent level.

For the calibration several methods performed in stages are planned:
\begin{itemize}
\item Precalibration at test beams and with cosmic muons at the level of 10\%
\item In situ calibration with physics events (neutral mesons and electrons)
\item Continuous monitoring with a light-pulser system
\end{itemize}

They are discussed in the following subsections.

\section{Calibration}
\label{sec:cal:calibration}

\subsection{Calibration with Physics Events}
\label{sec:cal:physics}
Based on the present experience with the temperature sensitivity of the complete detector elements including the crystal and the photosensor, a final in-beam calibration of the whole calorimeter does not appear appropriate.  All calibrations have to be performed using the complete setup operating at the final temperature. 

Therefore energy calibration of each crystal channel will be performed in situ with physics events. Low multiplicity events with \piz and $\eta$ decaying into two photons will be selected. The constraint on the invariant mass $m^2_{\mathrm{meson}} = E_{\gamma_1} E_{\gamma_2} (1-\cos \theta_{12})$ gives corrections to the reconstructed photon energy in the crystal channels of the well separated clusters $\gamma_1$ and $\gamma_2$. After a few iterations all calibration constants are available with a sufficient precision. The precalibration of the crystals with cosmic muons will provide the initial seed to start the procedure for the first time. The method was applied successfully at the Crystal Barrel experiment. There 1.2 million $\pbarp$ events with up to 8 photons in the final state (mainly $\pbarp \to 3\piz$) were used to calibrate 1380 crystals~\cite{bib:emc:cal:augustin,bib:emc:cal:cb}.

For a full calibration of the \PANDA electromagnetic calorimeter a total number of about $5\EE7$ events are required. It is essential to select low occupancy channels to avoid the overlap of clusters in the electromagnetic calorimeter.
The channels $\pbarp \to \piz\piz\piz$ and $\pbarp \to \piz\piz\eta$ have a cross section of about $30\,\mu$b, thus producing 4200 events per second at a luminosity of $10^{32}\,\cm^{-2}\s^{-1}$ ($\eta\to\gamma\gamma$ only). Having a dedicated software trigger, calorimeter calibration data will be produced at a rate of about $4\,\kHz$. The selection of the events with completely neutral final states makes the calibration independent of any other detectors and thus quickly to perform. 

With the available statistics it is expected that a calibration can be performed once a day.

Variations of the calibration for shorter timescales than required for the calibration will be tracked by the monitoring system.


\subsection{Precalibration with Cosmic Muons and at Test Beams}
\label{sec:cal:cosmics}
The absolute geometrical position of the individual calorimeter elements has to be deduced from the mechanical design. The point of impact of the photon relies on the reconstruction of the electromagnetic shower and primarily on the appropriate mathematical algorithm, which will be optimized for the prototype arrays. The optimal position reconstruction algorithm depends on the point of impact and will be evaluated with full size prototypes at test beams.

The precise calibration of the calorimeter by using the constraint on
the $\piz$ and $\eta$ mass relies on a precalibration at a 10\% level.
Before the final assembly of the calorimeters all submodules will
undergo a check and precalibration with cosmic muons. Part of the
submodules will be subject to beam tests to verify the precalibration
and the position reconstruction algorithms.
The final precalibration of all crystal channels will be performed in situ with cosmic muons.
This will be done before the startup with physics beam and at the final and stable temperature.
Depending on the position relative to the earth surface different energies are deposited by the minimum ionizing particles. Even for upward pointing crystals enough statistics will be reached within one day of dedicated running. Due to the hardware-trigger-free concept of the \PANDA DAQ system it is easy to set up algorithms to store all relevant events by requiring certain cuts on energy depositions in neighboring crystals.
The precision of the calibration will be verified at test beams by
comparing the calibration obtained with cosmic muons to the
calibration with defined photon energies. The \INST{CMS} experiment has
obtained a precision of 2.5\% with the calibration of \PWO with
muons. It is expected that the
resulting precision at \PANDA will be better than 10\%, enough for the final
calibration with physics events. 

\subsection{Online Calibration}
\label{sec:cal:online}
The hardware-trigger-free concept of the \PANDA DAQ system requires the availability of calibration data at real time. The trigger decision is taken by software operated on compute nodes where full event information is available. For efficient triggering with the EMC it is therefore essential to have reasonable calibration constants available, to perform a quick analysis of the event. 

The calibration constants can be determined once a day by the algorithm described in \Refsec{sec:cal:physics}. The compute nodes filter the events where only calorimeter information and no charged track is seen. In addition a cut on 5 to 8 clusters and a minimum cut on the total energy are performed.
These filtered events are stored for the calibration.
Variations on the timescale of minutes and hours are detected by the light monitoring system (\Refsec{sec:cal:monitoring}). Each crystal is flashed 100 times per minute by the light pulser. The data is evaluated on the compute nodes online and variations in the light transmission are measured every minute. These are applied to the calibration constants determined before. 

\section{Monitoring}
\label{sec:cal:monitoring}

\begin{figure}[htb]
\begin{center}
\includegraphics[width=\swidth]{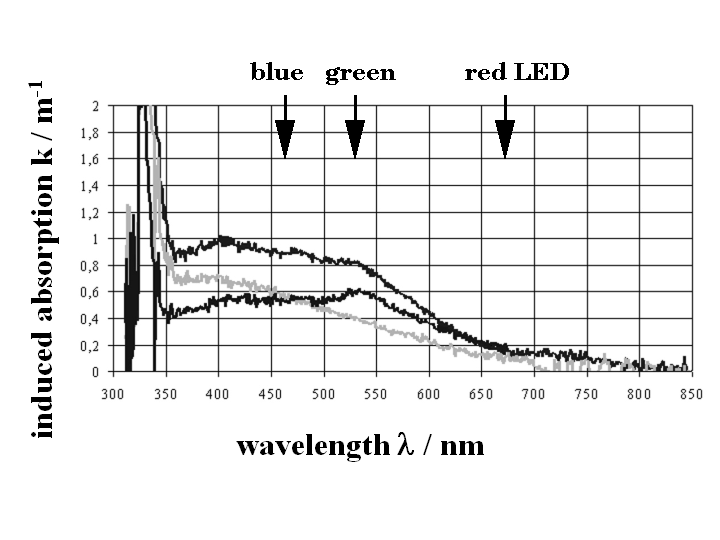}
\caption[Induced absorption of various \PWO samples after irradiation.]
{Induced absorption of various \PWO samples after irradiation with a
  dose of 20$\,$krad ($^{60}$Co) shown with the indication of the
  available wavelengths of LEDs, which are used for monitoring purposes.}
\label{fig:cal:radiation}
\end{center}
\end{figure}

The monitoring system provides a reference to identify and record any
changes of the coefficients for the energy calibration of all
calorimeter cells. The response can shift due to changes of the
luminescence process and the optical quality of the PWO crystals, the
quantum efficiency and gain of the APD, and of the preamplifier/ADC
conversion gain, for example. These effects can originate from
temporary and permanent radiation damages and/or temperature
changes. Due to extensive research on the radiation damage of PWO
carried out during the last decade it was established that the
luminescence yield of PWO crystals is not affected by irradiation,
only the optical transmission. Any change of transparency can be
monitored by the use of light injection from a constant and stabilized
light source. The system will be based on light sources at multiple
wavelengths. The monitoring of the radiation damages will be performed
at low wavelengths (UV/blue to green, $455\,$nm and $530\,$nm). With
light in the red region at $660\,$nm, a wavelength where radiation
damages play no role, the chain from the light coupling to the
photosensitive detector and the amplifier digitization can be
monitored. The wavelengths, which are used for monitoring purposes,
are indicated in \Reffig{fig:cal:radiation} together with the induced
absorption of radiation damaged \PWO crystals. The value of $455\,$nm
corresponds to the closest available LED with respect to the peak
of the \PWO emission.
The concept to be used depends on the time scale when these deteriorations appear. As outlined in \Refsec{sec:req:rad} the expected dose rate will stay well below the values expected for \INST{CMS} operation. Therefore, one can assume that the possible degradation of the optical performance will evolve gradually and very slowly in time. In that case, trends and corrections can be deduced from kinematic parameters such as invariant masses calculated off-line. 
To avoid photon conversion in front of the calorimeter, the overall thickness of dead material has to be minimized. Mechanical structures for support or cooling should be installed as close as possible to the crystal front face. A system of optical fibers for light injection from the front side is excluded due to the space needed to cope with the large bending radius of fibers. Injecting light from the rear with reflection at the front side back to the photosensor is possible. However, the effects due to enhanced multiple scattering must be studied.
The temperature gradient of the luminescence yield of PWO, which varies due to temperature quenching between 2 and 3\percent/K requires a temperature stabilization of the whole system with a precision of $\sim\,0.1\degrees$ including a finely distributed temperature measurement. In addition the intrinsic gain and the noise level of the APD photosensors show similar temperature sensitivity. The thermal contact to the crystals should be sufficiently reliable. 
Checks of the linearity of the electronics at the sub-percent level require to cover a dynamic range of $10\,000$. This will be achieved by neutral density filters. 

\subsection{Concept of a Light Source and Light Distribution System}
\label{sec:cal:monitoring:lps}

To maintain exact correspondence between observed changes of the measured amplitude of the light monitoring signal and the detector signal initiated by high energy photons, which is influenced by the variation of the optical transmission of PWO crystals, it is necessary to meet two requirements: the optical emission spectrum of the light source should be similar to the radio-luminescence spectrum of PWO; the effective optical path length for monitoring light in the crystal should be identical to the average path length of the scintillation light created by an electromagnetic shower in the crystal.  Further technical requirements for the monitoring light source are: the number of  photons per pulse should generate the output signal equivalent to about $2\,\gev$ in each cell; the duration of the pulse should be similar to the decay time of the scintillation of PWO; a long-term pulse stability better than 0.1\percent has to be achieved; minimal pulse height variation from pulse to pulse (determined only by photon statistics without any additional line broadening to keep the required number of monitoring events at the minimum).  Nowadays available ultra-bright LEDs
(brightness $>\,10\,$cd) emitting at various wavelengths give the opportunity to create a system to meet most of the above requirements. In case of blue-violet LEDs the emission spectrum (centered at 420--430$\,$nm) as well as the spectral width (approx. $70\,$nm FWHM) correspond exactly to those for PWO radio-luminescence. The number of photons per $20\,$ns pulse can be as large as 10$^9$ and can be further increased by more than one order of magnitude by combining optically many LEDs in one emitting block. Moreover, combining LEDs of different color allows fine-tuning to the entire scintillator emission spectrum. Even taking into account the light losses in the distribution system one might illuminate 1\,000--1\,200 calorimeter cells with one large LED block. Consequently 10 to 15 such stabilized blocks can provide the monitoring of the whole electromagnetic calorimeter. Such a concept allows the selective monitoring of different parts of the calorimeter by electronic triggering of the appropriate LED blocks without the necessity of optical switching. 
The distribution system should provide the light transfer to each cell with the option to fire selectively only $\sim\,10$\percent of all cells simultaneously to avoid interference problems in the front-end electronics and an overload of the data acquisition system. This can be implemented with optical fibers grouped into 10 bunches attached to the outputs of multi LED blocks. Each fiber bunch will contain additional fibers controlling possible losses in fiber transparency. 

\subsection{Concept of the Light Monitoring System} 

\begin{figure*}[htb]
\begin{center}
\includegraphics[width=\dwidth]{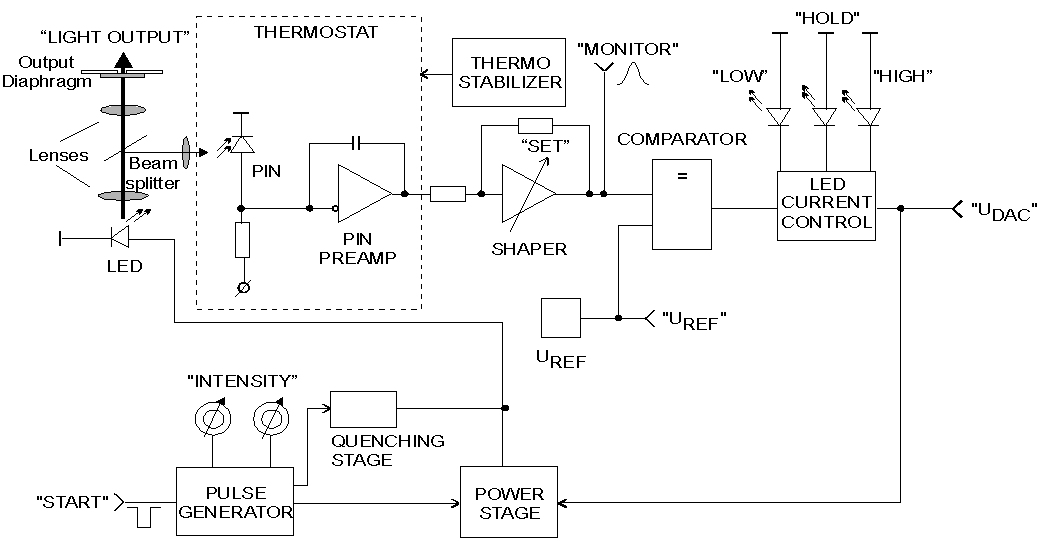}
\caption[Schematic layout of the stabilized light pulser system.]
{Schematic layout of the stabilized light pulser system.}
\label{fig:scint:pwo:fig_mon1}
\end{center}
\end{figure*}

\begin{figure}[htb]
\begin{center}
\includegraphics[width=\swidth]{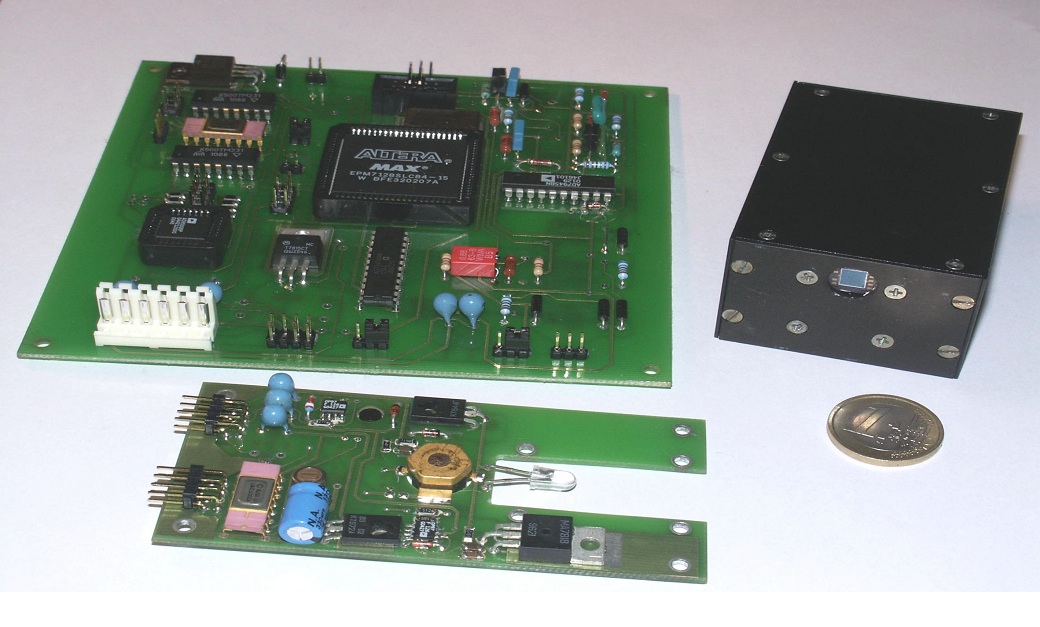}
\caption[Major components of a first prototype of the monitoring system.]
{Major components of a first prototype of the monitoring system.}
\label{fig:scint:pwo:fig_mon2}
\end{center}
\end{figure}

The concept of the system should provide flexible control and redundancy in terms of measured parameters to distinguish between different sources of instability. Each emitting block will be equipped with an optical feedback and a thermo-stabilized PIN-photodiode as reference. Such a device has been successfully implemented for \INST{CMS} ECAL~\cite{bib:emc:cal:mon1}. A central sequencer will trigger all blocks with maximum frequency up to a few kHz. A separate unit is foreseen to measure the light amplitude of the reference fiber in each bundle. The system does not require regular maintenance. \Reffig{fig:scint:pwo:fig_mon1} and \ref{fig:scint:pwo:fig_mon2} show the schematic layout and the major components of a first functioning prototype.

\subsection{Light Pulser Prototype Studies}\label{sec:lightpulser}

\subsubsection{Experience at Test Beam}\label{sec:lightpulser_tb}
A LED-based light pulser system with four LEDs of different
wavelengths was made for the test beam studies at Protvino to
monitor gain variations of the PMTs and transmission variations of
the PWO crystals. The LEDs emit at red (660~nm), yellow (580~nm),
green (530~nm), and blue (470~nm) wavelengths. For the actual
analysis of data, the red and the blue LEDs were most useful.

Between two accelerator spills, 10 light pulses of one color were
sent to the crystals. Then in the following interval, light pulses
of another color were injected in the crystals. This way, four
spills were needed to collect data for all four colors. The light
from all the LEDs was fed into the same set of optical fibers and
they delivered the light to individual crystals. 

In the test beam setup, the LED light was injected at the front
end of the crystals. So the typical path length of LED light in
the crystal approximately equals the length of the crystal. Since
the light comes out of the optical fiber with a characteristic
full angle spread of 25$^\circ$, and this angle is reduced to
11$^\circ$ as the light enters the crystal from air, the path
length of light in the crystal should be increased by
$1/\cos{11^\circ}$. As for the scintillation light from incident
particles, half of the light travels directly to the PMT while the
other half will travel towards the front of the crystal and gets
reflected before it is detected by the PMT. Averaging these two cases, 
the mean path length of scintillation light to
the PMT also equals the crystal length in 0th approximation.
In order to estimate the 1st order correction, we need to know how
much the light zigzags on its way to the PMT. The maximum angle
that the light makes with respect to the crystal axis is
determined by the reflection angle on the side
surfaces due to the total internal reflection angle, which is about
64$^\circ$. This leads to many more zigzag paths than the paths
for LED light. Taking into account that the scintillation light is
emitted isotropically, the average $<1/\cos{\theta}>$ factor
arising from the zigzag paths is about 1.4.

The LED system monitors the transparency of the crystal at a
specific wavelength (in our case, 470 nm was chosen partially due
to the availability of blue LEDs) and thus does not sample the
entire spectrum of scintillation light.  The radiation damage
effect is less severe at 470~nm than at 430~nm, the center of the
PWO scintillation emission peak. From these considerations, we
expect that the ratio, $R$, of the light loss factors for the LED
signal and the particle signal is about 1/1.4 = 0.7 to 1/1.6 =
0.6.

One of our goals in the test beam studies of the calibration
system is to measure this ratio, $R$, experimentally, and to observe how it
varies from crystal to crystal.  Naively, since this ratio only
depends on the geometrical lengths of light paths for the LED and
scintillation light, it should not vary from one crystal to the
next. If there are variations in the shape of the absorption as a
function of wavelength among crystals, the ratio, $R$, may vary
among crystals. In addition, since the crystals will not be
polished to optical flatness, actual reflections of light by the
side surfaces do not follow the simple law of geometrical light
reflection.  This may also lead to variations of the ratio, $R$,
among crystals.  Thus we feel that it is very important to measure
the variation of $R$ values experimentally.

Since it is not practical to measure this ratio for all produced
crystals at a test beam facility (it would take too much time), we
need to know if the variation, if there is any, is small enough so
that we will not spoil the resolution even if we assume and use an
average value of the ratio for all crystals.

\subsubsection{Monitoring Systems for the Light Pulser System and Stability
of Light Pulser}\label{sec:pulserdesign}

We built two monitoring systems to check the stability of the
magnitude of the light pulses.  ({\em i.e.} monitoring systems of
the monitoring system.) One was based on a PIN photodiode, which
is considered very stable even when the temperature varies.
According to the literature, the temperature variation of PIN
photodiodes is less than 0.01\%/$^{\circ}$C. However, since we needed an
amplifier to detect the PIN photodiode signal, the amplifier gain
needed to be stabilized by housing it in the crystal box where the
temperature was stable to $\pm$ 0.1$\degC$.

\begin{figure}
\begin{center}
\includegraphics[width=\swidth]{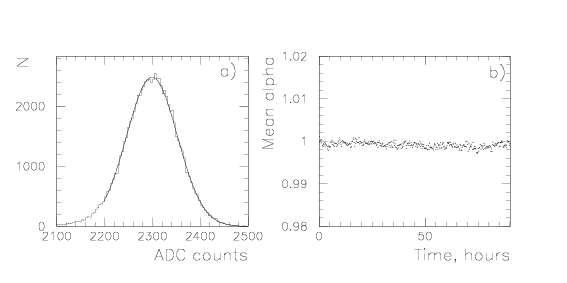}
\caption[$\alpha$ signal stability.]
{(a) $\alpha$ energy spectrum accumulated over 1.5 hours.
 (b) $\alpha$ spectrum peak position as a function of time over 85 hours. Each
point corresponds to a 15-minute measurement duration. }
\label{fig:alpha}
\end{center}
\end{figure}

The second system used a PMT, a scintillation crystal and a
radioactive source. The PMT (Hamamatsu R5900) monitored the LED
pulser while the PMT was monitored using the stable scintillation
light produced when a $YAlO_3:Ce$ crystal was irradiated with an
$^{238}Pu$ alpha source (YAP)~\cite{bib:emc:cal:yap}. The $\alpha$ energy
spectrum measured by the PMT and the peak position of this
spectrum as a function of time is presented in
Fig.~\ref{fig:alpha}. The width of the peak is 2.3\% r.m.s. as
determined by a fit to a Gaussian. The peak position was stable
over 85~hours to better than 0.2\%.

The measured variations (drifts) of the magnitudes of light pulses
(averaged over 120 pulses) over different time periods were
measured for the periods of test beam studies.
They were:

\begin{itemize}
\item{0.1 to 0.2\% over a day;}
\item{0.5\% over a week;}
\item{1\% over a few months.}
\end{itemize}

Temperature variations were the main cause for the variations in
the size of pulses. When corrections based on the temperature were
made in the pulse-height analysis, the long term variation
significantly decreased to 0.4\% over a few months and down to 
0.3\% over a week. No LED ageing effects were observed after 3000
hours of operation.

The stability of this system was better than we needed to monitor
the gain variations of the PMTs and the transparency variations of
the crystals over the relevant time periods. For example, we were
able to track the crystal transparency change with an accuracy of
better than 1\% over a week when we measured how much radiation
damage the crystals suffered. Additionally, we were able to track
the change in crystal transparency with an accuracy of better than
1\% over a few months when we measured the recovery process of the
radiation-damaged crystals. Finally, we were able to
track the PMT gain variations over a day well enough so that it
did not contribute appreciably to the energy resolution.
This last accomplishment implies that we already have a good
enough system for \PANDA except that we need to have a much larger
system, and temperature stabilization must be considered.

\subsubsection{Crystal Light Output Monitoring}\label{sec:R}

\begin{figure}
\begin{center}
\includegraphics[width=\swidth]{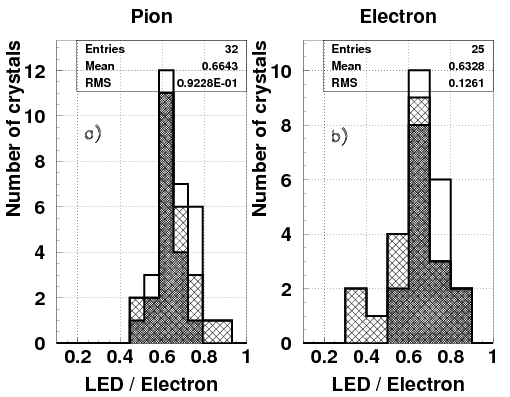}
\caption[Linear fit coefficients for pion and electron irradiation.]
{Linear fit coefficients calculated from the correlation plots of relative changes for the
blue LED vs. electron signal under (a) pion and (b) electron irradiation.}
\label{fig:fit_coeff}
\end{center}
\end{figure}

As was shown by radiation hardness studies, PWO crystals behave in a similar way
in radiation environments of different nature;
clear correlations between electron and LED signal changes were observed.
The dedicated study has been carried out to confirm that these correlations
are not dependent on the type of irradiation using a particular
optical monitoring scheme.
To be more specific, the same crystals were calibrated with a low intensity 
electron beam first, then they were exposed to the highly intense electron 
radiation.
The irradiation of crystals continued with a pion beam.  
Both electron and pion irradiations alternated with calibration runs
using a low intensity electron beam.
Changes in the crystal transparency were monitored continuously with 
the use of the LED monitoring system.
A linear fit of the distributions of signal change of the relative blue LED signal 
vs. the electron signal was calculated for both the electron and pion irradiations.
Coefficients of the linear fit are presented in Fig.~\ref{fig:fit_coeff}

As it was expected, on average the measured coefficients are not different 
and are in a good agreement with the calculations for the current optical 
scheme.
The fact that the linear approximation works well, significantly simplifies
the procedure of the EMC intercalibration that will be performed using
an LED-based light monitoring system over the time intervals between two 
subsequent in-situ calibrations.

\subsection{Light Monitoring System for \PANDA Calorimeter}
\label{sec:cal:monitoring:panda}

The light monitoring system
is designed to inject light pulses
into each PWO crystal in order to measure optical transmission
near the scintillation spectrum peak (430~nm). The red light
pulses are used to monitor a photodetector gain stability. The system includes
both blue and red LEDs, their driver circuits, and optical fibers
to deliver light pulses to each of the PWO crystals.

We plan to use very powerful LEDs, assisted by a reflector and a
light mixer so that each light pulsing system produces enough
light for $\sim$3000 crystals, each receiving light pulses
equivalent to scintillation light from 2 GeV photons.

The distribution of light among the 3000 fibers should be very
uniform. This is accomplished by designing a good light mixer
which will distribute light uniformly across an area of 38
$\times$38~mm$^2$. Each bunch of fibers (containing about 3000 fibers, out of
which 400 are spares) and two reference PIN silicon photodiodes
will be contained in this area. Several light pulsers will serve
the whole \PANDA calorimeter.

The time dependence of pulse heights from the pulser is monitored
by the two reference PIN photodiodes. One of the pulser
systems will be activated at any given time to limit the power
requirements of the light source, the size of data transfers, as
well as high and low voltages current demands.

The principal goal of the system is to monitor short-term
variation in the photodetector gains and the light transmission of the
crystals.
The system will also be used to check out the entire
crystal-readout chain during the assembly of the calorimeter. It
will also permit a rapid survey of the full calorimeter
during the installation or after long shutdowns. Furthermore, the
light monitoring system can be used to measure the response
linearity of the PWO crystal's photodetector and its readout
chain.
This should complement measurements with electronic charge injection at the preamplifier level which does not test the photodetector.

\begin{figure*}
\centering
\includegraphics[width=\dwidth]{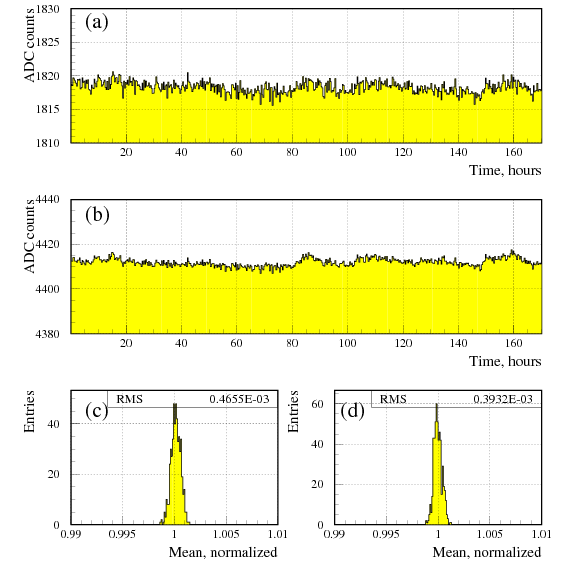}
\caption[Stability of the LED pulser prototype.]
{Stability of the LED pulser prototype: (a),(b) - behaviour in time 
of blue and red LED signals correspondingly detected by one of the photodiodes
over one week of measurements. Each entry is a mean value of amplitude 
distribution collected over 20 min;
(c),(d) - normalized projections on the vertical axis of the diagrams (a) and 
(b). R.m.s. characterizes instability of the system over the period of 
measurements.} 
\label{fig:mon_stab}
\end{figure*}

Some results obtained with a prototype system are presented in \Refsec{sec:pulser_test} and in \Reffig{fig:mon_stab}. In more detail, the monitoring 
system prototype design and performance are discussed 
in \cite{bib:emc:cal:ledperf}.

\begin{figure}
\includegraphics[width=\swidth]{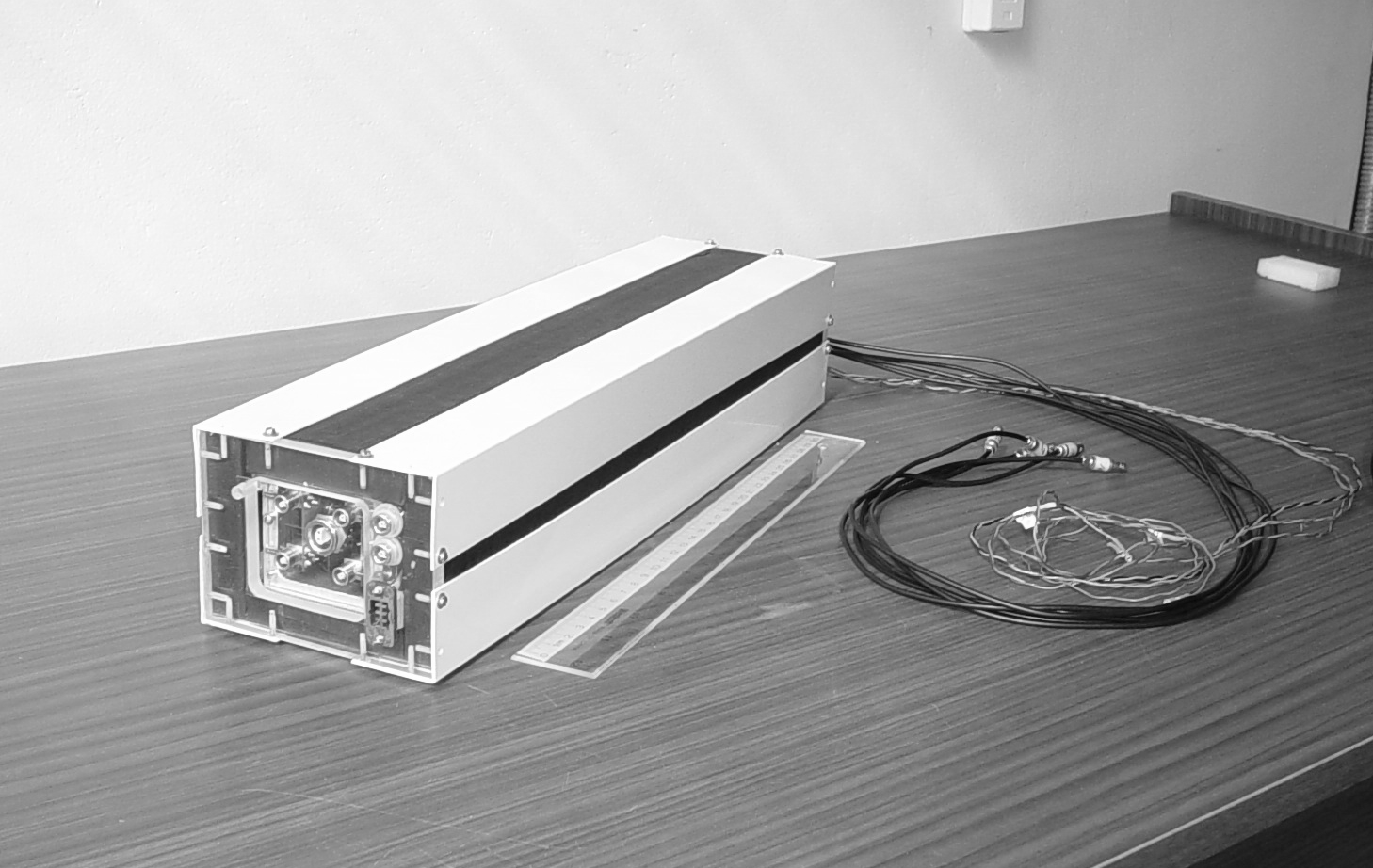}
\caption[Monitoring system prototype.]
{View of the prototype monitoring system. Inside the box
there is a LED driver, blue and red LEDs, a light mixer, a
temperature stabilization system, and a referenced PIN-diode
system. In use we have a fiber bunch coming out the far side of
the box instead of the cables which are pictured; the cables are
for tests only and will not go to the calorimeter. The size
of the box is 370 mm x 70 mm x 60 mm.} \label{fig:mon_general}
\end{figure}

\begin{figure}
\includegraphics[width=\swidth]{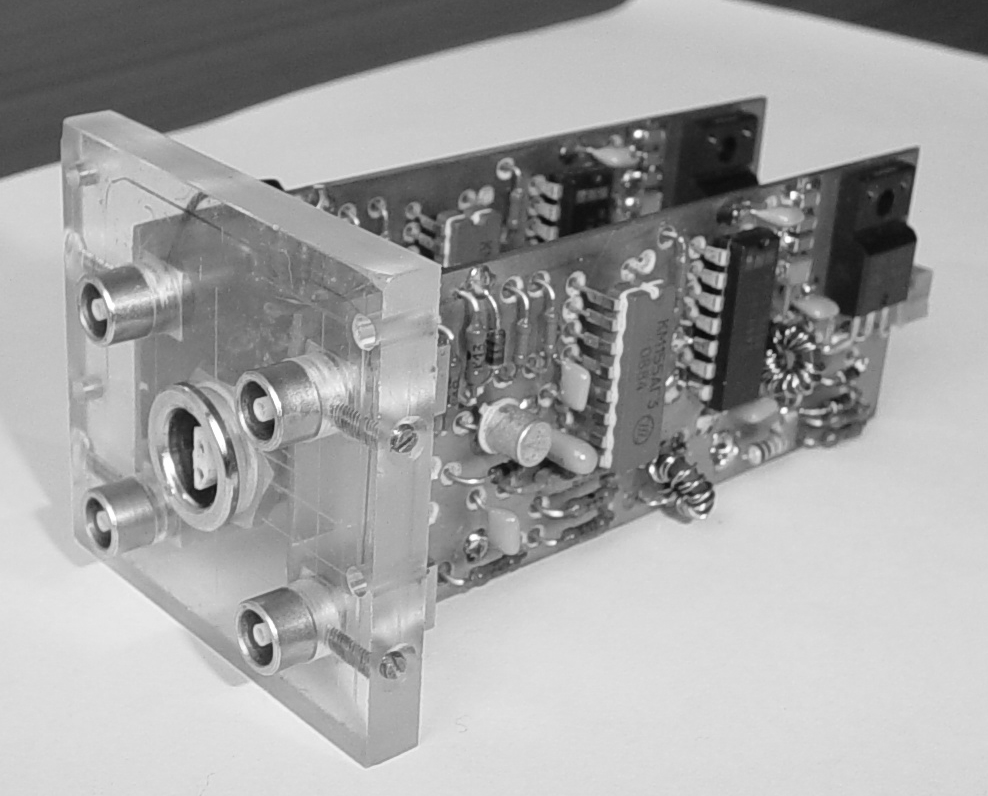}
\caption[Prototype LED driver.]
{ LED driver of the Prototype of the monitoring system. It will
be inside the box (see the previous Figure) near the side opposite to
the one with a bunch of fibers.}
\label{fig:mon_driver}
\end{figure}

A picture of the whole prototype system is presented in
\Reffig{fig:mon_general}. A LED
driver is presented in \Reffig{fig:mon_driver}.

\subsubsection{Monitoring System Components}

The monitoring system will be located directly at the outer radius of the calorimeter support structure and an optical-fiber light distribution system connects the pulser to the crystals.
Since the light pulser is located in a low radiation zone, its components and electronics are not required to be radiation hard.
In contrast, many fibers are routed through a high radiation zone and they must be made of radiation-hard materials.

The characteristics of the LEDs we plan to use in the \PANDA calorimeter monitoring system are given in \Reftbl{tab:LED}.
\begin{table*}
    \label{tab:LED}
    \begin{center}
      \begin{tabular}{ccc} \hline \hline
Property    & blue (royal blue) LED & red LED \\ \hline
Brand & Luxeon 5-W emitter & Luxeon 1-W emitter \\
Typical Luminous flux & 30 lm (@700 mA) & 45 lm (@350 mA) \\
Radiation Pattern & Lambertian & Lambertian \\
Viewing Angle & 150$\,\degrees$ & 140$\,\degrees$ \\
Size of Light Emission Surface & 5 $\times$ 5~mm$^2$ & 1.5$\times$1.5~mm$^2$ \\
Peak Wavelength & 470~nm (455 nm) & 627 nm \\
Spectral Half-width & 25 nm (20 nm) & 20 nm \\
Average Forward Current & 700 mA & 350 mA \\ \hline \hline
      \end{tabular}
    \end{center}
\caption[Properties of LEDs.]{Properties of LEDs.}  
  \end{table*}

Besides the exceptional luminous fluxes, we find that two additional features of the Luxeon technology are very important for our monitoring system: very long operating lifetime (up to 100,000 hours in DC mode), and small temperature dependence of the light output ($\approx 0.1\%/\degC$).
The producer is Lumileds Lighting, USA.
The reflector which was made at IHEP has a trapezoidal shape and is made of
aluminum plated Mylar or Tyvek.

The optical fibers we plan to use are produced by Polymicro Technologies, USA.
Their properties are:
\begin{itemize}
\item Silica /  Silica  optical  fiber
\item High - OH  Core
\item Aluminum  Buffer
\item Core   Diameter   270  $\mu \m$
\item Outer  Diameter   400  $\mu \m$
\item Numerical  Aperture  0.22
\item Full  Acceptance  Cone  25.4$\,\degrees$ 
\end{itemize}

This fiber has very good radiation hardness.
According to the tests made by the \INST{CMS} ECAL group, this fiber has shown no signal degradation under gamma irradiation with an absorbed dose of up to 12 Mrad.

\subsubsection{Reference PIN Silicon Photodiodes}

An essential element of the light monitoring system is a stable reference
photodetector with
good sensitivity at short wavelengths.
PIN silicon photodiodes with a sensitive area of about 6~mm$^2$ are well suited for this task.
In particular, such low leakage currents are achieved with PIN diodes, due to their very narrow depletion zone resulting from heavy (p and n) doping, which is less sensitive to the type inversion than typical PIN diodes.
The rather large sensitive area of this photodiode allows us to work without preamplifiers and improve the stability of the reference system itself.
A PIN silicon photodiode S1226-5BQ (Hamamatsu) was  used in our test measurements.
It has an active area of 2.4$\times$2.4~mm$^2$ and a dark current less
than 50~pA (at 5 V reverse-bias voltage).

\begin{figure}
\begin{center}
\includegraphics[width=\swidth]{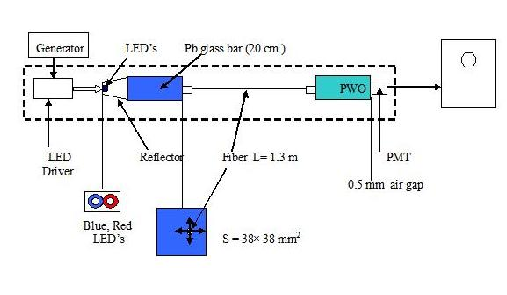}
\caption[Schematic view of the light pulser prototype.]
{Schematic view of the light pulser prototype.}
\label{fig:pulser}
\end{center}
\end{figure}

\subsubsection{Tests of the Light Pulser Prototype}\label{sec:pulser_test}

\begin{figure}
\includegraphics[width=\swidth]{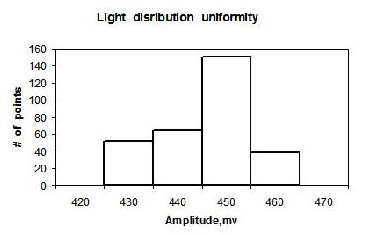} 
\caption[Light distribution uniformity.]
{Distribution of pulse heights (in mV) measured over 
an area of 34 $\times$ 34 mm$^2$ in 2~mm steps.} 
\label{fig:uniformity}
\end{figure}

A schematic view of the light pulser prototype is shown in
\Reffig{fig:pulser}. The light distribution uniformity was
measured with a single fiber scanner. All the measurements were
made with a scope and a manual scan with a step size of 2 mm. The
scan area was 34$\times$34 mm$^2$. The results are shown in
\Reffig{fig:uniformity}.
The FWHM of this pulse height distribution is 2\%, and the full
width is 8\%. The energy equivalent is 20 GeV for the whole scan
area. The average forward current in the tests was 20 mA.
The maximum forward current is 700 mA. So we have a large safety
factor for the amount of light. This light pulser can illuminate
more than 3000 fibers.

    The short-term stability of light uniformity over the area of 34$\times$34~mm$^2$ for a day has been measured to be 0.05\% and a long-term stability was 0.1\% over 20 days.
In spite of these encouraging results, thermostabilisation of the light pulser by means of the Peltier cell is foreseen in the design of the whole system.

%

%
%
\newpage
\bibliographystyle{panda_tdr_lit}
\bibliography{./lit_emc}
%

%
\cleardoublepage
\chapter{Simulations}
\label{sec:sim}
%
%

The simulation of the EMC and the facilitated software is described in this chapter.
The goal of the studies described in the following is the expected performance of 
the planned EMC. We focus here on the energy and spatial resolution of reconstructed photons, 
the capability of an electron hadron separation, and also the feasibility of the \Panda 
physics program. A couple of benchmark studies will be 
presented. These accurate simulations highlight the necessity of the planned EMC. Due to the
fact that most of the physics channels have very low cross sections \-- typically 
between pb and nb \-- , a background
rejection power up to $10^9$ has to be achieved. This requires an electromagnetic 
calorimeter which allows an accurate photon reconstruction within the energy 
range between 10$\,\mev$ and 15$\,\gev$, and an effective and an almost clean electron hadron 
separation. 
         
\section{Offline Software}
The offline software has been devised for detailed design studies of the \Panda detector
and for the preparation of the \Panda Physics Book, which will be 
published by September 2008. It follows an object oriented approach, and most of the code
is written in C++. Several proofed software tools and packages  have been adapted 
from other HEP experiments to the \Panda needs. The software contains
\begin{itemize}
\item event generators with proper decay models for all particles and resonances involved in 
the individual physics channels as well as in the relevant background channels,
\item particle tracking through the complete \Panda detector by using the GEANT4 transport
code \cite{bib:sim:Allison, bib:sim:Agostinelli},
\item a digitization which models the signals and the signal processing in the 
front-end-electronics of the individual detectors,
\item the reconstruction of charged and neutral particles, comprising a particle identification
that provides lists of particle candidates for the final physics analysis, and
\item user friendly high level analysis tools with the purpose to make use of vertex
and kinematical fits and to reconstruct even complicate decay trees in an easy way.
\end{itemize}  

The simulations have been done with the complete setup which was already described in detail
in \Refsec{sec:int:det}. The still not finally established Time Of Flight and the Forward RICH
detectors have not been considered and the Straw Tube option has been used for the central tracker 
device. 

\subsection{Photon Reconstruction}
\subsubsection{Reconstruction Algorithm} \label{sec:sim:recoalgo}
A photon entering one crystal of the EMC develops an electromagnetic shower which, 
in general, extends over several crystals. A contiguous area of such crystals is called a cluster.

The energy deposits and the positions of all   crystals hit in a cluster allow a 
determination of the four vector of the initial photon. Most of the EMC reconstruction
code used in the offline software is based on the cluster finding and bump splitting 
algorithms which were developed and successfully applied by the BaBar experiment 
\cite{bib:sim:BaBarDetector, bib:sim:Strother}.

The first step of the cluster reconstruction is the finding of a contiguous area of
crystals with energy deposit. The algorithm starts at the crystal exhibiting 
the largest energy deposit.
Its neighbors are then added to the list of crystals  if the
energy deposit is above a certain threshold $\extl$. The same procedure is continued on
the neighbors of newly added crystals  until no more crystal fulfills the threshold criterion. 
Finally a cluster gets accepted if the total energy deposit in the contiguous area is above
a second threshold $\ecl$.

A cluster can be formed by more than one particle if the angular 
distances of the particles are small. In this case the cluster has to be subdivided
into regions which can be associated with the individual particles. This procedure 
is called the {\it{bump splitting}}. A bump is defined by a local maximum inside the cluster:  
The energy deposit of one crystal $E_{\mbox{LocalMax}}$ must be above $\emax$, while all neighbor
crystals have smaller energies. In addition the highest energy $E_{\mbox{NMax}}$ of any of the 
$N$ neighboring crystals must fulfill the following requirement:
\begin{eqnarray}
 0.5\,(N-2.5) \, > \, E_{\mbox{NMax}} \, / \,  E_{\mbox{LocalMax}}
\end{eqnarray} 
The total cluster energy is then shared between the bumps, taking into account the shower shape of the 
cluster. For this step an iterative algorithm is used, which assigns a weight $w_i$
to each crystal, so that the bump energy is defined as 
$E_{\mbox{bump}}= \sum_{i} \, w_i \, E_i$. $E_i$ represents the energy deposit in the ith crystal 
and the sum runs over all crystals within the cluster. 
The crystal weight for each bump is calculated by  
\begin{eqnarray} 
w_i = \frac{E_i \, exp(-2.5 \, r_i\, / \,r_m)}
            {\sum_{j} E_j \, exp(-2.5 \, r_j\, / \,r_m)}
\end{eqnarray}
, with
\begin{itemize}
\item $r_m$ = Moli\`ere radius of the crystal material,
\item  $r_i$,$r_j$ = distance of the ith and jth crystal to the center of the bump and
\item  index $j$ runs over all crystals. 
\end{itemize} 
The procedure is iterated until convergence. The center position is always determined from the weights
of the previous iteration and convergence is reached when the bump center stays stable within a tolerance 
of 1$\,\mm$.
       
The spatial position of a bump is calculated via a center-of-gravity method. Due to the fact 
that the radial energy distribution originating from a photon decreases mainly exponentially, 
a logarithmic weighting with
\begin{eqnarray}
  W_i \, = \, max(0, A(E_{\mbox{bump}}) \, + \, ln\,(E_i / E_{\mbox{bump}}))
\end{eqnarray}
 was chosen, where only crystals with positive weights are used. The energy dependent factor 
$A(E_{\mbox{bump}})$ varies between 2.1 for the lowest and 3.6 for the highest photon energies.           
   
\subsubsection{Digitization of the Crystal Readout} 
\label{sec:sim:digi}
Fast FADC will digitize the analog response of the first amplification and shaping stage.
Therefore signal waveforms are important to be considered in the simulation. For performance reasons  
a simplified method has been applied in the simulation studies. It is based on an effective 
smearing of the extracted Monte Carlo energy deposits. 
Reasonable properties of PbWO$_4$ at the operational temperature of -25$^\circ$C have been considered. 
A Gaussian distribution of $\sigma$ = 1$\,\mev$ has been used for the constant electronics noise. 
The statistical fluctuations were estimated by 80 phe/\mev produced in the LAAPD.
An excess noise factor of 1.38 has been used which corresponds to the measurements with 
the first LAAPD 'normal C' prototype at an internal gain of $M=50$ (see 
\Refsec{sec:photo:APD:Char:ENF}). This results in a photo statistic noise term 
of 0.41\% /$\sqrt{E (\gev)}$.  

\paragraph{Comparison with 3$\times$3 Crystal Array Measurements}

\begin{figure}[htb]
\begin{center}
\includegraphics[width=\swidth]{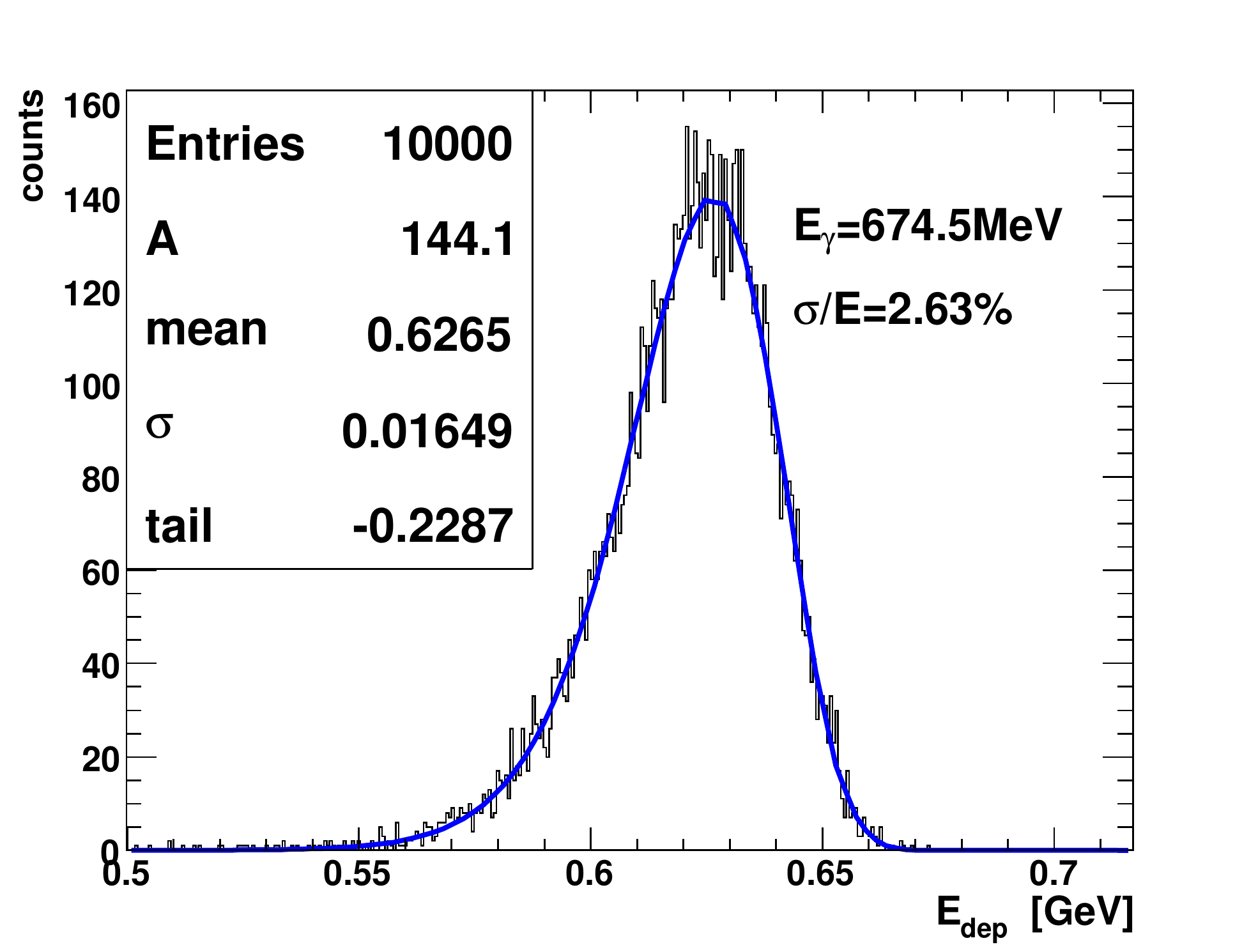}
\caption[Simulated line shape at the photon energy of 674.5$\,\mev$.]
{Simulated line shape of the 3$\times$3 crystal matrix at the photon energy of 674.5$\,\mev$. The fit with
a Novosibirsk function gives an energy resolution of $\sigma/E\,=\,2.63\,$\% }
\label{fig:sim:reso3x3_674}
\end{center}
\end{figure}

\begin{figure}[htb]
\begin{center}
\includegraphics[width=\swidth]{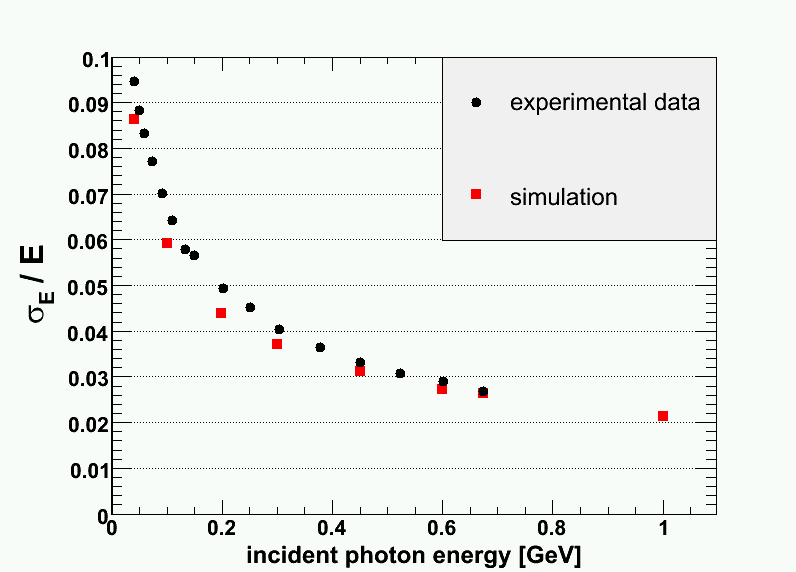}
\caption[Measured and simulated energy resolution vs. incident photon energy.]
{Energy resolution as a function of incident photon energy between 40.9$\,\mev$
and 1$\,\gev$ for a 3$\times$3 crystal array. The measurements are represented by black circles and
the simulation results are illustrated with red rectangles.}
\label{fig:sim:reso3x3_all}
\end{center}
\end{figure}

In order to demonstrate that the digitization is sufficiently described by the simplified
method, the simulation was compared with the results of a measurement with a 3$\times$3 crystal array
at an operational temperature of 0$^\circ$C (see \Refsec{subchap:perform:energy-apd}).
\Reffig{fig:perf:perf16} and \Reffig{fig:perf:perf17} show the obtained line shapes, and
\Reffig{fig:perf:perf19} the measured energy resolution as a function of incident photon energy.
The simulations have been done for a 3$\times$3 array of crystals of 2\,x\,2\,x\,20$\,$cm$^3$. Photons 
with discrete energies between 40.9$\,\mev$ and 1$\,\gev$ were shot to the center of the innermost 
crystal. To consider the crystal temperature of 0$^0$C the digitization has been done with just 
40\,phe/\mev instead of 80\,phe/\mev produced in the LAAPD. The resulting line shape at the photon 
energy of 674.5$\,\mev$ is illustrated in \Reffig{fig:sim:reso3x3_674}. A fit with a Novosibirsk 
function defined by
\begin{eqnarray}
f(E) = A_S \exp ( -0.5 { \ln^2 [1 + \Lambda \tau \cdot (E - E_0) ] / \tau^2 + \tau^2 } ) \nonumber
\end{eqnarray}
,with
\begin{itemize}
\item $\Lambda = \, \sinh ( \, \tau \, \sqrt{\, \ln 4} ) / ( \, \sigma \, \tau \, \sqrt{\ln 4} )$,
\item $E_0\,$= peak position,
\item $\sigma\,$= width, and
\item $\tau\,$= tail parameter
\end{itemize}
yields to a resoultion of $\sigma/E\;=\;2.63\,$\%, which is in good agreement
with the measurements (see \Reffig{fig:perf:perf17}). Moreover the energy resolution as a function of
the photon energy can be reproduced very well. A comparison between the measurements and the simulations
are shown in \Reffig{fig:sim:reso3x3_all}, and the results are consistent within a tolerance of 10\%.
The resolutions obtained with the simulation are systematically better, which may be caused by the
uncertainties of the estimated electronic noise term and the number of  phe/\mev, as well as by
inhomogeneities of the light yield of the crystals, which were not taken into account.

\subsubsection{Reconstruction Thresholds}\label{sec:sim:recothresh}
In order to detect low energetic photons and to 
achieve a good energy resolution, the three photon reconstruction thresholds should be set as 
low as possible. On the other hand the thresholds must be sufficiently high to suppress the 
misleading
reconstruction of photons from the noise of the crystal readout and from statistical fluctuations 
of the 
electromagnetic showers. Based on the properties of the \Panda EMC
(see \Refsec{sec:sim:digi}) the following thresholds were chosen:

\begin{itemize}
\item $\extl\,=\,3\,\mev$
\item $\ecl\,=\,10\,\mev$
\item $\emax\,=\,20\,\mev$  .
\end{itemize}

As already discussed in \Refsec{sec:ethres} the ability to identify photons down to 
approximately $10\,\mev$ is extremely important for 
\Panda. A lot of channels \--- especially in the charmonium sector (exotic and conventional)  like 
$\pbarp\,\to\,\eta_c\,\to \,\gamma\,\gamma$,
$\pbarp\,\to\,h_c \,\to\,\eta_c\,\gamma$ or  $\pbarp\,\to\,\jpsi\,\gamma$ 
\-- require an accurate and clean reconstruction of isolated photons. The main background channels 
here have the same final states with just the isolated photon being replaced by a $\piz$. 
If one low energetic photon from a $\piz$ decay gets lost, the background event will
be misidentified. The cross sections for the 
background channels are expected to be orders of magnitudes higher than for the channels of interest.
Therefore an efficient identification of $\piz$ is mandatory for this important part of the 
physics program of \Panda. \Reffig{fig:pizloss} in \Refsec{sec:ethres} shows the upper limit for 
the 
identification of $\piz$'s with respect to different photon thresholds.
While roughly 1\% of the $\piz$'s get lost for a threshold of 10$\,\mev$, the 
misidentification increases by one order of magnitude for the scenario where photons below 
30$\,\mev$ can not be detected.          
    

\subsubsection{Leakage Corrections}
\begin{figure}[htb]
\begin{center}
\includegraphics[width=\swidth]{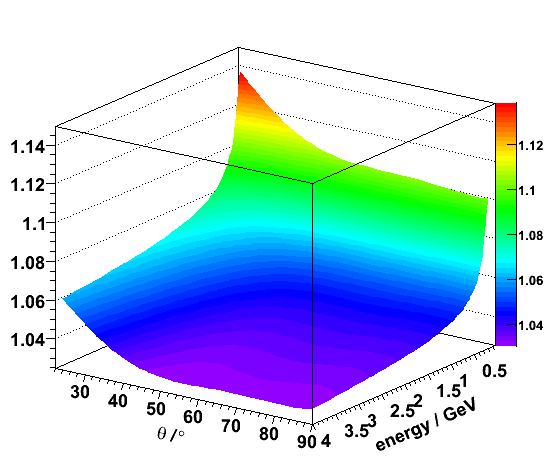}
\caption[Leakage correction function for the barrel EMC.]
{Leakage correction function for the barrel EMC in the $\Theta$ range 
between 22$^\circ$ and 90$^\circ$.}
\label{fig:sim:leakCorr}
\end{center}
\end{figure}

The sum of the 
energy deposited in the crystals in general is a few percent less than the energy of the
incident photon. This is caused by shower leakages particularly in the crystal gaps.
The reconstructed energy of the photon is expressed as a product
of the total measured energy deposit and a correction function which depends 
logarithmically on the energy and \--- due to the layout of the \Panda EMC \--- also on 
the polar angle. Single photon Monte Carlo simulations have been carried out to determine the
parameters of the correction function $E_{\gamma,\mbox{cor}} = E * f(\ln E,\Theta)$ with
\begin{eqnarray}
f(\ln E,\Theta) = exp( a_0 + a_1\,ln E + a_2 \,ln^2 E + a_3 \,ln^3 E \nonumber\\
                   + a_4 \, cos(\Theta)+ a_5 \, cos^2(\Theta) + a_6 \, cos^3(\Theta) \nonumber\\
                  + a_7 \,cos^4(\Theta) + a_8 \, cos^5(\Theta) \nonumber\\
                  + a_9 \, ln E \, cos(\Theta) ) \nonumber 
\end{eqnarray} 
\Reffig{fig:sim:leakCorr} shows the result for the barrel part in the $\Theta$ range 
between 22$^\circ$ and 90$^\circ$.  

\subsubsection{Energy and Spatial Resolution}

\begin{figure}[htb]
\begin{center}
\includegraphics[width=\swidth]{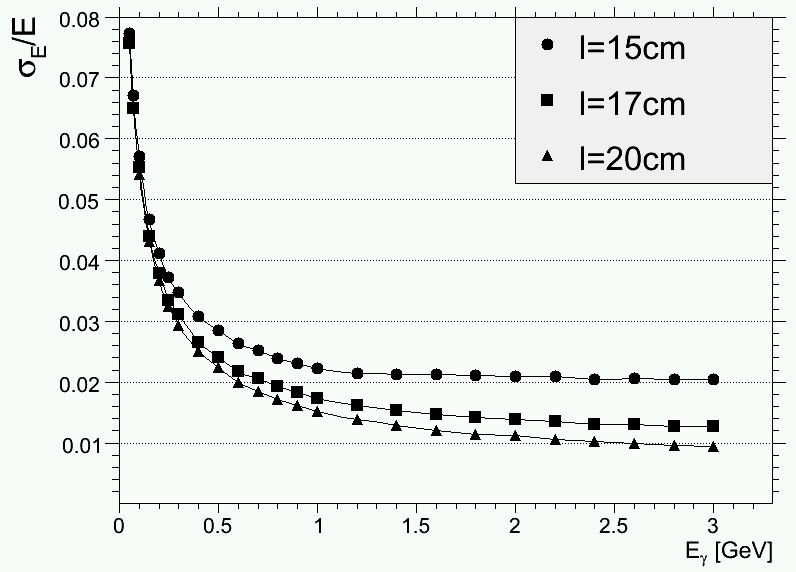}
\caption[Energy resolution for three different crystal length scenarios.]
{Comparison of the photon energy resolution for three different crystal
lengths. The resolution for 15\,cm crystals is illustrated by circles, for 17\,cm 
by rectangles, and the 20\,cm scenario is shown with triangles.}
\label{fig:sim:eResol}
\end{center}
\end{figure}

The energy resolution of the EMC strongly depends on the length of the crystals. 
A sufficient containment of the electromagnetic shower within the detector and
marginal fluctuations of the shower leakages must be guaranteed for the detection 
of photons with energies up to 15$\,\gev$. \Reffig{fig:sim:eResol} compares the simulation
results for the foreseen \Panda EMC equipped with crystals of 15\,cm, 17\,cm and
20\,cm length. While the resolution for low energetic photons \--- below 300$\,\mev$ \--- is 
nearly 
the same for all three scenarios, the performance becomes significantly better for higher 
energetic photons with an increasing crystal length. The 20 cm setup yields in an energy 
resolution of 1.5\% for 1$\,\gev$ photons, and of below 1\% for 
photons above 3$\,\gev$. Another aspect which favors the 20 cm scenario is the tolerable level of the 
nuclear counter effect in the photo sensors (\Refsec{sec:int:det}).   

\begin{figure}[htb]
\begin{center}
\includegraphics[width=\swidth]{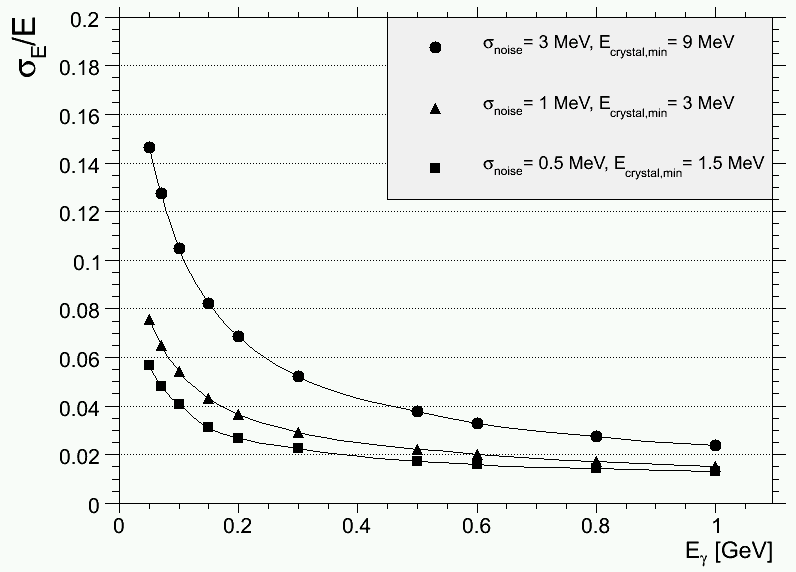}
\caption[Energy resolutions for different single crystal reconstruction thresholds.]
{Comparison of the energy resolutions for three different single crystal reconstruction thresholds. 
The most realistic scenario with a noise term of $\sigma \, = \,1 \, \mev$
and a single crystal threshold of $\extl=3\,\mev$ is illustrated by triangles, a worse case
($\sigma \, = \, 3 \mev$, $\extl \, = \, 9 \, \mev$) by circles and the better case ($\sigma \, = \, 0.5 \, \mev$, 
$\extl \,= \, 1.5 \, \mev$) by rectangles.}
\label{fig:sim:eResolThreshold}
\end{center}
\end{figure}

As already described in \ref{sec:sim:digi} and \ref{sec:sim:recothresh} the choice of the
single crystal threshold $\extl$, which is driven by the electronics noise term, strongly effects
the resolution. Three different scenarios have been investigated: \Reffig{fig:sim:eResolThreshold}
compares the achievable resolution for the most realistic scenario with a noise term of $\sigma$ = 1$\,\mev$
and a single crystal reconstruction threshold of $\extl=3\,\mev$ with a worse case 
($\sigma$ = 3 $\mev$, $\extl=9$ $\mev$) and a better case ($\sigma\,= \,0.5 \,\mev$, $\extl=1.5\,\mev$). 
While the maximal 
improvement for the better case is just 20\% for the lowest photon energies, the degradation
in the worse case increases by more than a factor of 2. This result demonstrates clearly
that the single crystal threshold has a strong influence on the energy resolution.

\begin{figure}[htb]
\begin{center}
\includegraphics[width=\swidth]{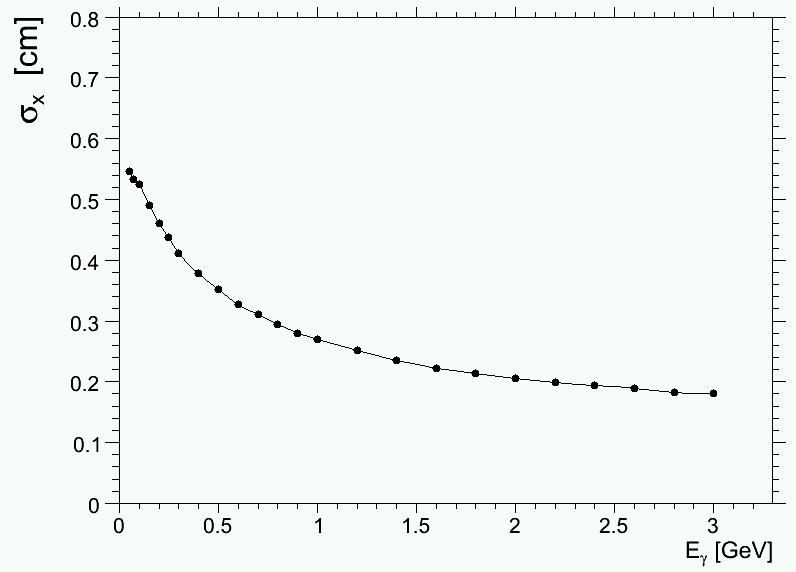}
\caption[Position resolution in $x$-direction for photons below 3$\,\gev$. ]
{Position resolution in $x$-direction for photons below 3$\,\gev$.}
\label{fig:sim:posResol}
\end{center}
\end{figure}

The high granularity of the planned EMC provides an excellent position reconstruction 
of the detected photons. The accuracy of the spatial coordinates is
mainly determined by the dimension of the crystal size with respect 
to the Moli\`ere radius. \Reffig{fig:sim:posResol} shows the resolution in $x$ direction 
for photons up to 3$\,\gev$. A $\sigma_x$-resolution of less than 0.3\,cm can be obtained 
for energies above 1$\,\gev$. This corresponds to roughly 10\% of the crystal size.
For lower energies the position becomes worse due to the fact that
the electromagnetic shower is contained in just a few crystals. In the worst case
of only one contributing crystal the x-position can be reconstructed within an 
imprecision of 0.5\,-\,0.6\,cm.

\subsection{Electron Identification}
\label{sec:sim:electronpid}
Electron identification will play an essential role for most of the physics program of 
\Panda. An 
accurate and clean measurement of the \jpsi decay in $e^+ \, e^-$ is needed for many  
channels in the charmonium sector as well as for the study of the 
$\bar{p}$ annihilation in nuclear matter like the reaction $\pbarA \, \to \, \jpsi X$.
In addition the determination of electromagnetic form factors of the proton via 
$\pbarp \, \to \,e^+ \,e^- $ requires a suppression of the main background channel
$\pbarp \, \to \, \pi^+ \, \pi^-$ in the order of $10^8$.

\begin{figure}[htb]
\begin{center}
\includegraphics[width=\swidth]{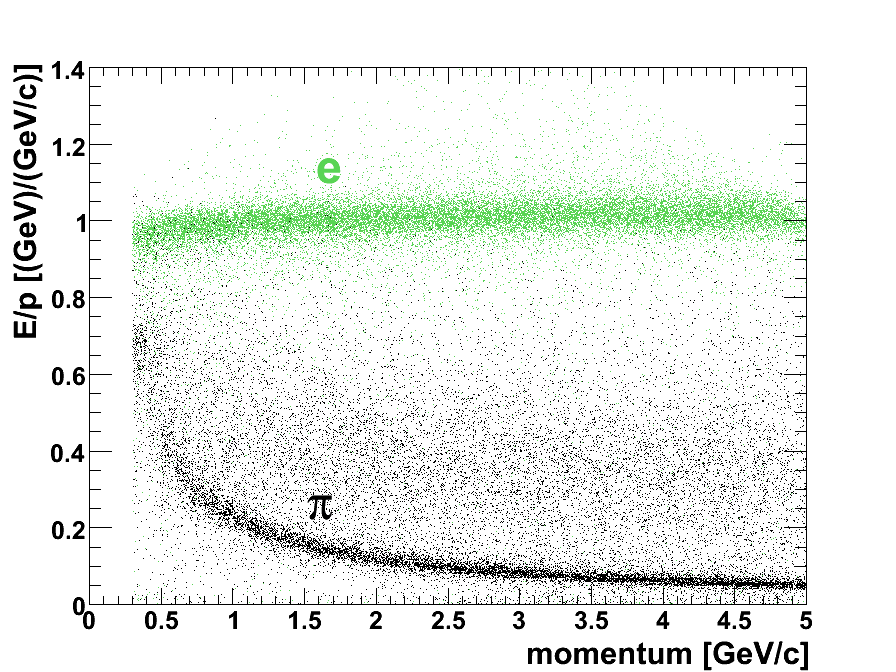}
\caption[E/p versus track momentum for electrons and pions.]
{E/p versus track momentum for electrons (green) and pions (black) in the momentum 
range between 0.3$\,\gevc$ and 5$\,\gevc$.}
\label{fig:sim:eOpElectronsPions}
\end{center}
\end{figure}

The EMC is designed for the detection of photons. Nevertheless it is also the most powerful
detector for an efficient and clean identification of electrons. 
The character of an electromagnetic shower is distinctive for electrons, muons and hadrons. The most suitable
property is the deposited energy in the calorimeter. While muons and hadrons in 
general loose only
a certain fraction of their kinetic energy by ionization processes, electrons deposit
their complete energy in an electromagnetic shower. The 
ratio of the measured energy deposit in the calorimeter to the reconstructed track momentum 
($E/p$) will be approximately unity. Due to the fact that hadronic 
interactions within the crystals 
can take place, hadrons can also have a higher $E/p$ ratio than expected from ionization.
Figure \Reffig{fig:sim:eOpElectronsPions} shows the reconstructed $E/p$ fraction for electrons and pions
as a function of momentum.

\begin{figure}[htb]
\begin{center}
\includegraphics[width=\swidth]{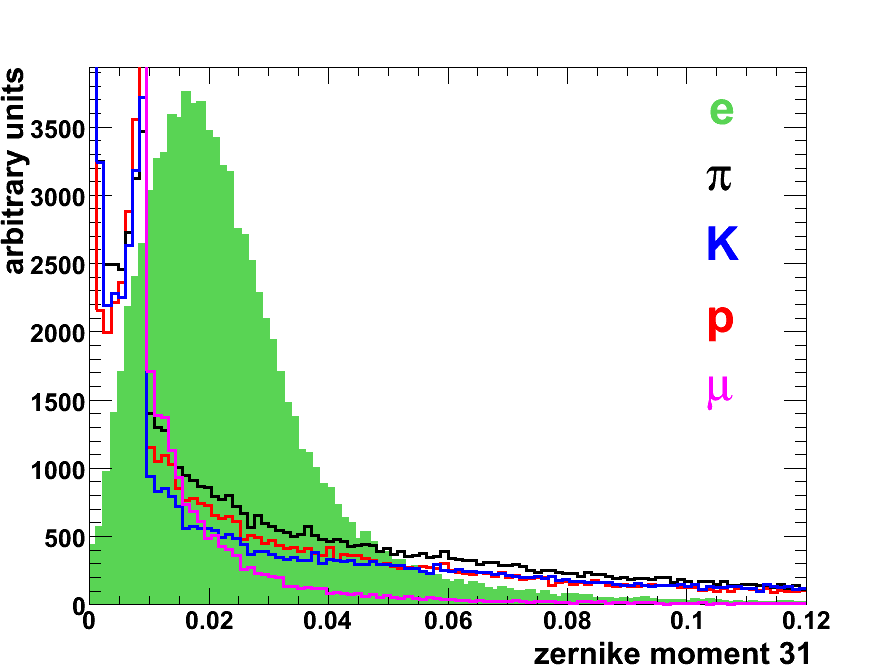}
\caption[Zernike moment $z_{31}$ for electrons, muons and hadrons.]
{Zernike moment $z_{31}$ for electrons, muons and hadrons.}
\label{fig:sim:zernike31}
\end{center}
\end{figure}

Furthermore the shower shape of a cluster is helpful to 
distinguish between electrons, muons and hadrons. Since the chosen size of the crystals 
corresponds to the Moli\`ere radius of lead tungstate, the largest fraction of
an electromagnetic shower originating from an electron is contained in just a few 
crystals. Instead, an hadronic shower with a similar energy deposit is less concentrated.
These differences are reflected in the shower shape of the cluster, which can be 
characterized by the following properties:
\begin{itemize}
\item $E_1/E_9$ which is the ratio of the energy deposited in the central crystal and in 
the 3$\times$3 crystal array containing the central crystal and the first innermost ring.
Also the ratio of $E_9$ and the energy deposit in the 
5$\times$5 crystal array  $E_{25}$ is useful for electron identification.    
 \item The lateral moment of the cluster defined by
      $mom_{\mbox{LAT}} = \frac{\sum_{i=3}^n E_i r_i^2}{\sum_{i=3}^n E_i r_i^2 + E_1 r_0^2 +  E_2 r_0^2}$ with
\begin{itemize}
\item n: number of crystals associated to the shower.
\item $E_i$: deposited energy in the i-th crystal with $E_1 \geq E_2 \geq ... \geq E_n$.
\item $r_i$: lateral distance between the central and the i-th crystal.
\item $r_0$: the average distance between two crystals. 
\end{itemize}
\item A set of zernike moments \cite{bib:sim:zernike} which describe the energy distribution within a 
cluster by
radial and angular dependent polynomials. An example is given in \Reffig{fig:sim:zernike31}, 
where the zernike moment $z_{31}$ is illustrated for all particle types.  
\end{itemize}

Due to the fact that a lot of partially correlated EMC properties are suitable for electron 
identification, a Multilayer 
Perceptron (MLP) has been applied. The advantage of a neural network is that it can 
provide a correlation between a set of input variables and one or several output 
variables without any knowledge of how the output formally depends on the input. Such
techniques are also successfully used by other HEP experiments \cite{bib:sim:Abramowicz, bib:sim:Breton:1995rg}. 

\begin{figure}[htb]
\begin{center}
\includegraphics[width=\swidth]{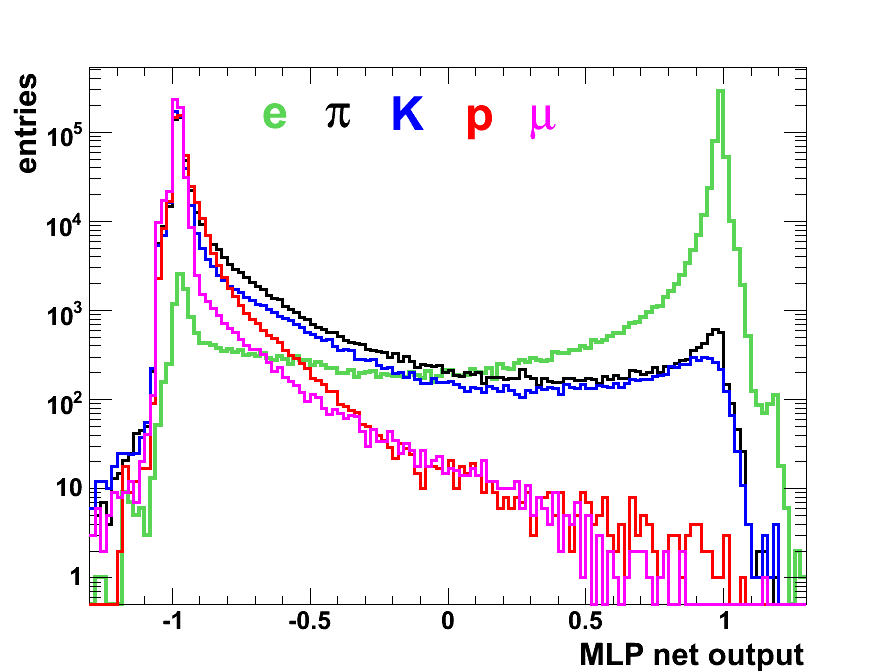}
\caption[MLP output for electrons and other particle species.]
{MLP output for electrons and the other particle species 
in the momentum range between 300$\,\mevc$ and 5$\,\gevc$.}
\label{fig:sim:nno}
\end{center}
\end{figure}

The
training of the MLP has been done with a data set of 850.000 single tracks for each 
particle species (e, $\mu$, $\pi$, K and p) in the momentum range between 200$\,\mevc$ 
and 10$\,\gevc$ in such a way  that the output values are constrained to be 1 for 
electrons and -1 for all other particle types. 10 input variables in total have been
used, namely $E/p$, $p$, the polar angle $\Theta$ of the cluster, and 7 shower 
shape parameters
($E_1/E_9$, $E_9/E_{25}$, the lateral moment of the shower and 4 zernike moments). The
response of the trained network to a test data set of single particles in the 
momentum rage between 300$\,\mevc$ and 5$\,\gevc$ is illustrated in \Reffig{fig:sim:nno}. 
The logarithmically scaled histogram shows that an almost clean electron recognition with a
quite small contamination of muons and hadrons can be obtained by applying a cut on the network output. 

\begin{figure}[htb]
\begin{center}
\includegraphics[width=\swidth]{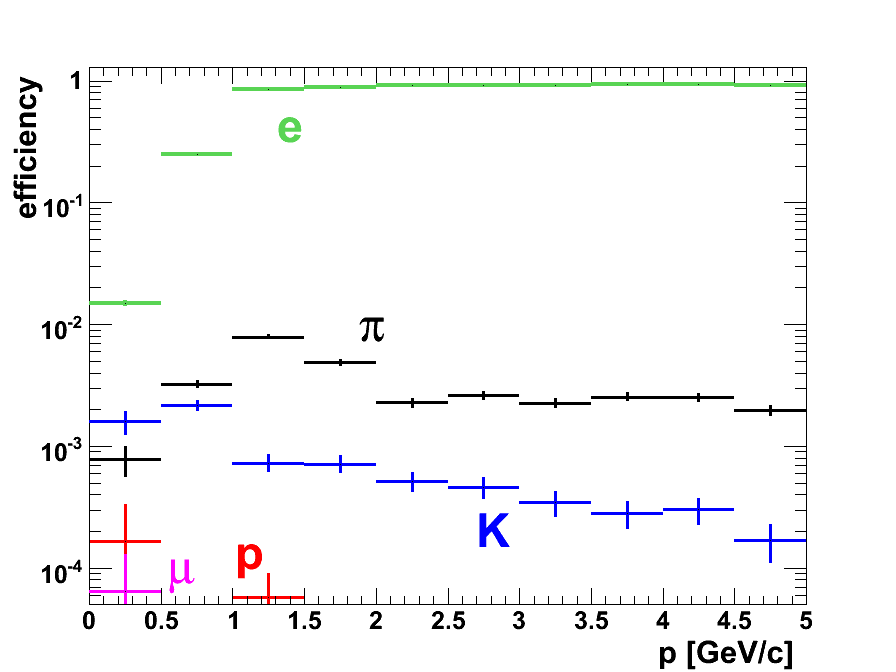}
\caption[Electron efficiency and contamination rate using EMC information.]
{Electron efficiency and contamination rate for muons, pions,
 kaons and protons in different momentum ranges by using the EMC information.}
\label{fig:sim:lhEMC}
\end{center}
\end{figure}

The global PID, which combines the PID informations of the individual sub-detectors, has
been realized with the standard likelihood method. Each sub-detector provides probabilities
for the different particle species, and thus a correlation between the network output and the PID 
likelihood of the EMC has been calculated. \Reffig{fig:sim:lhEMC} shows the electron efficiency 
and contamination rate as a function of momentum achieved  by requiring an 
electron likelihood fraction of the EMC of more than 95\%.
For momenta above 1$\,\gevc$ one can see that the electron efficiency is greater than 98\%
while the contamination by other particles is substantially less than 1\%. 
For momenta below 1$\,\gevc$ instead the electron identification obtained with just the EMC 
is quite bad and not sufficient.
 
\begin{figure}[htb]
\begin{center}
\includegraphics[width=\swidth]{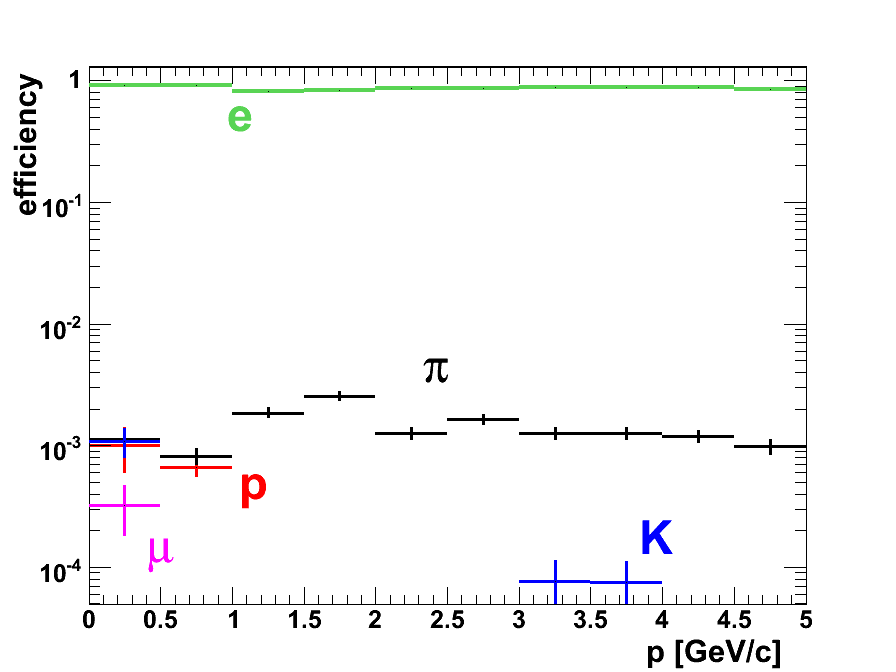}
\caption[Electron efficiency and contamination rate using combined PID information.]
{Electron efficiency and contamination rate for muons, pions,
 kaons and protons in different momentum ranges by using the combined PID informations.}
\label{fig:sim:lhAll}
\end{center}
\end{figure}

A good electron-ID efficiency and small contamination rates can be achieved by taking
into consideration additional sub-detectors (MVD, Cherenkov detectors and muon counters) 
(\Reffig{fig:sim:lhAll}). By applying a 99.5\% cut
on the combined likelihood fraction, the electron efficiency 
becomes better than 98\% while the pion misidentification is just on the $10^{-3}$ level over 
the whole momentum range. The contamination rates for muons, kaons and protons are 
negligible.
     
\section{Material Budget in front of the EMC}
\begin{figure}[htb]
\begin{center}
\includegraphics[width=\swidth]{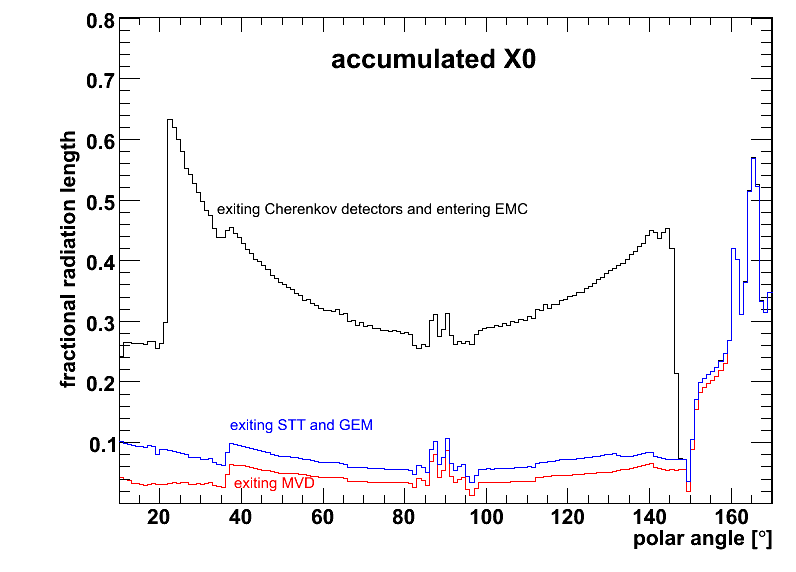}
\caption[Amount of material in front of the EMC.]
{Material in front of the EMC in units of a radiation length $X_0$ as a function of the 
polar angle $\Theta$.}
\label{fig:sim:x0}
\end{center}
\end{figure}

The reconstruction efficiency as well as the energy and spatial resolution of the EMC  
are affected by the interaction of particles with material in front of the calorimeter.
While the dominant interaction process for photons in the energy range of interest
is $e^+\,e^-$ pair production, electrons lose energy mainly via 
Bremsstrahlung ($e \, \to \, e \, \gamma$). \Reffig{fig:sim:x0} illustrates the 
material budget in front of the EMC originating from the individual sub-detectors 
in units of radiation lengths as a function of $\Theta$.
The largest contribution comes from the
Cherenkov detectors, which consist of quartz radiators of 1-2~cm thickness. 
This corresponds to 17\% to 50\% of a radiation length, depending on the polar angle.
     
\subsection{DIRC Preshower}
\begin{figure}[htb]
\begin{center}
\includegraphics[width=\swidth]{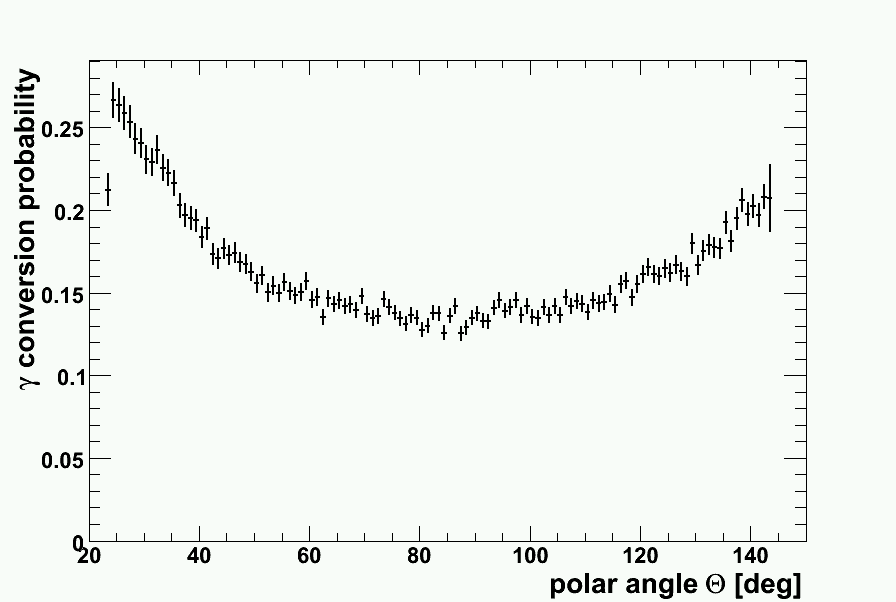}
\caption[$\gamma$ conversion probability in the DIRC as a function of $\Theta$.]
{$\gamma$ conversion probability in the DIRC as a function of $\Theta$.}
\label{fig:sim:gammaConfDIRC}
\end{center}
\end{figure}

As \Reffig{fig:sim:x0} shows, the DIRC detector contributes most to the material budget
in front of the EMC. Therefore detailed Monte Carlo studies have been done
to investigate the impact of this device on the $\gamma$ reconstruction 
efficiency and energy resolution. \Reffig{fig:sim:gammaConfDIRC} represents the 
$\gamma$ conversion probability in the DIRC for
1$\,\gev$ photons generated homogeneously in the $\Theta$ range 
between 22$^\circ$ and 145$^\circ$. While 15\% of the photons convert at the polar angle of 90$^\circ$, the DIRC preshower probability increases up to 27\% at 22$^\circ$. 
        
\begin{figure}[htb]
\begin{center}
\includegraphics[width=\swidth]{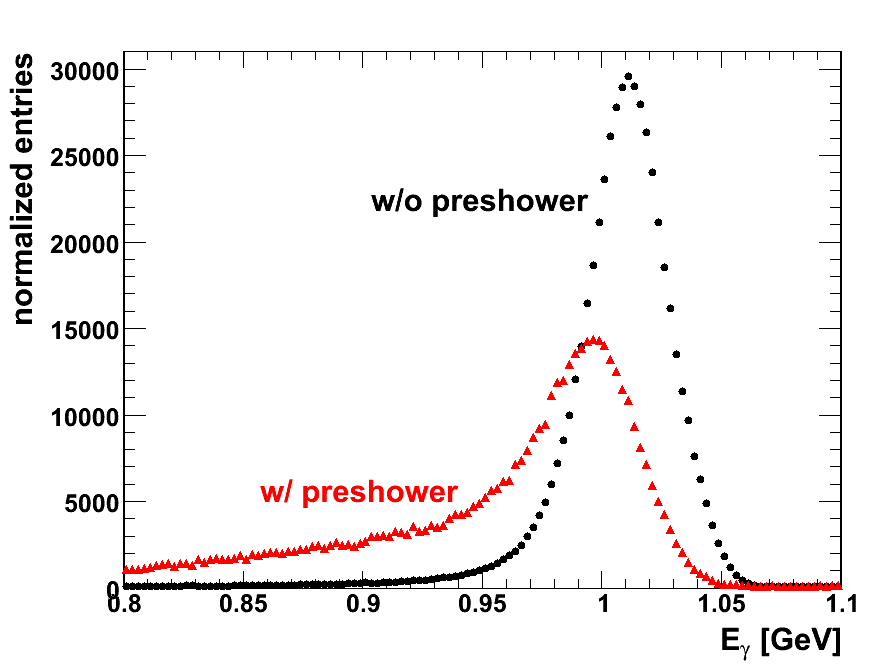}
\caption[Reconstructed photon energy with/without DIRC preshowers.]
{Reconstructed energy of 1$\,\gev$ photons without (black circles) and with (red triangles)
 DIRC preshowers. For a better comparison the plot with DIRC preshowers is scaled by a factor of 4.87.}
\label{fig:sim:resPreshower}
\end{center}
\end{figure}

DIRC preshowers mainly lead to a degradation of the energy 
resolution. A comparison of the reconstructed energies for 1$\,\gev$ photons 
with and without DIRC preshowers is shown in \Reffig{fig:sim:resPreshower}. While
the energy distribution for non-preshower photons can be well described by a 
Gaussian function with $\sigma(\Delta E/E)\,<$~2\%, 
the distribution for DIRC preshower clusters becomes significantly worse. Due to the fact 
that a fraction of the photon 
energy is deposited in the quartz bars, a broad and asymmetric distribution with a huge
low energy tail shows up.
      
\subsubsection{Outlook: Preshower Recognition and Energy Correction.}
If the Cherenkov light originating from the produced $e^+\,e^-$ pairs gets measured, 
the number of detected 
Cherenkov photons provides a measure for the energy loss, and thus an energy correction 
of such clusters could be feasible. A DIRC preshower recognition with an additional 
energy correction would yield in a better performance of the photon reconstruction.

Based upon recent investigations for the BaBar experiment, it is expected to achieve
a DIRC preshower detection efficiency of better than 50\% and an improvement for the photon
energy resolution of more than 1\% \cite{bib:sim:Adametz}.

\section{Benchmark Studies}
As discussed in the introduction of this document, the \Panda physics
program covers many fields and the addressed topics range from light
and charm hadron spectroscopy over the study of charm in nuclear matter
and the measurement of the proton form-factors.  To fulfill the
requirements defined by the different fields, the EMC is essential in
many cases. To demonstrate that the proposed \Panda detector and in
particular the EMC is able to reach the defined goals, detailed Monte
Carlo studies have been performed. In the following the results of
these benchmark studies will be presented, where the EMC plays a
major role. These studies cover charmonium spectroscopy as well as the
measurement of the time-like electromagnetic form-factors of the
proton. In the field of charmonium spectroscopy recent observations of
new states
\cite{bib:sim:SwansonStatusReport} falling into the mass range of the charmonium
system, underline the importance of the ability to detect these states
in as many decay modes as possible to compare the observed decay
patterns with theoretical expectations, and thus understand the nature of
these states. For the envisaged comprehensive study of the charmonium
system at \Panda this implies the detection of hadronic, leptonic
and radiative decay modes, where the final states can consist out of
charged and neutral particles. Therefore photons have to be detected
with precise energy and spatial resolution over a wide energy range
and an excellent coverage of the solid angle. For the reconstruction of
leptonic decays to
\ee (i.e., \jpsi and $\psi(2S)$ decays) the EMC supplements the PID
capabilities of the other detector components and provides together
with these a clean identification of electrons.  This is also crucial
for the measurement of the electromagnetic proton form factors in the
time-like region via the process $\pbarp\,\to\,\ee$, where the rejection
of the dominant $\pbarp \, \to \, \pip \, \pim$ background requires an excellent
separation of electrons and pions.

\subsection{$\mathbf{h_c}$ Detection with the $\mathbf{h_c \, \to \, \eta_{c} \, \gamma}$
Decay and the Role of low Energy $\mathbf{\gamma}$-ray Threshold} \label{sec:sim:h_c}
\subsubsection{Description of the Studied Channel}
$h_c$ is a singlet state of P wave charmonium ($1 ^{1}P_1$) with a mass of M=3526
$\mevcc$. One of its main decay modes is the electromagnetic transition to the
ground state, $\eta_c$, with the emission of a $\gamma$ with an energy of
$E_{\gamma}=503~\mev$. This decay mode was previously observed by E835
\cite{bib:sim:hc_E835} and CLEO \cite{bib:sim:hc_CLEO}. 
$h_c$ can be observed exclusively in many decay modes of
$\eta_c$, neutral ($\eta_c \, \to \, \gamma \, \gamma$) or
hadronic. The given analysis is based on the decay modes $\eta_c \, \to \, \phi \,
\phi$ with a branching fraction of $BR=2.6 \cdot 10^{-3}$, and 
$\phi \, \to \, K^{+} \, K^{-}, BR= 0.49$.

\subsubsection{Background Consideration}
\begin{figure}
\begin{center}
\includegraphics[width=\swidth]{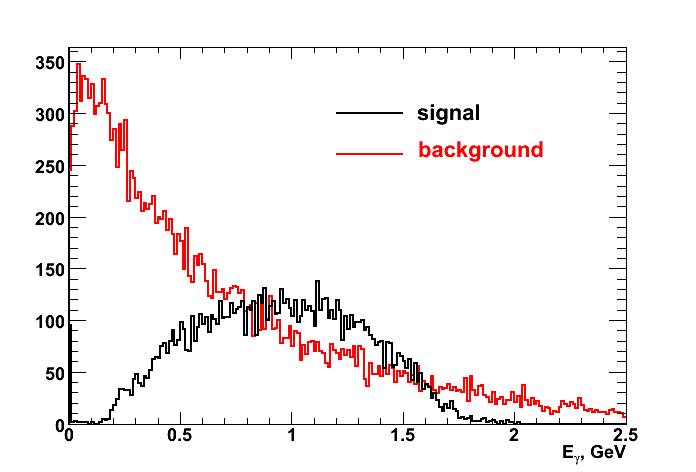}
\caption[Energy distributions of $\gamma$s from  $\pbarp\,\to\,h_c\,\to\,\eta_c\,\gamma$ 
and $\pbarp\,\to\,\phi\,\phi\,\piz$.]
{Energy distribution of $\gamma$s from $\pbarp\,\to\,h_c\,\to\,\eta_c\,\gamma$ 
and $\pbarp\,\to\,\phi\,\phi\,\piz$.}
\label{fig:sim:hc_e_gamma}
\end{center}
\end{figure}

The previous observation of $h_c$ was done in the neutral decay mode of $\eta_c$ in 
\pbarp annihilations \cite{bib:sim:hc_E835}, or in hadronic decay modes of $\eta_c$ 
from $e^{+} \, e^{-} \, \to \, \Psi(2S) \, \to \, \pi^{0} \, h_c, h_c \, \to \, \eta_c \, \gamma$ 
\cite{bib:sim:hc_CLEO}. In PANDA, where 
$h_c$ is produced in $\pbarp$ annihilations and the detector is capable to detect 
hadronic final states, the most important aspect of the analysis is the evaluation of signal to 
background ratios, because the production of the hadronic final states is much
more enhanced in $\pbarp$ annihilation in comparison to $e^{+}e^{-}$. 
For the exclusive decay mode considered in this study:
\begin{center}
$\pbarp \, \to \, h_c \, \to \, \eta_c \, \gamma \, \to \, \phi \, \phi \, \gamma \to K^{+} K^{-} K^{+} K^{-} \gamma$
\end{center}            
the following 3 reactions are considered as the main background contributors:\\
\begin{enumerate}
\item
$\pbarp \to K^{+} K^{-} K^{+} K^{-} \pi^{0}$,
\item
$\pbarp \to \phi K^{+} K^{-} \pi^{0}$,
\item
$\pbarp \to \phi \phi \pi^{0}$.
\end{enumerate}
With one $\gamma$-ray from the $\pi^{0}$ decay left undetected, these reactions have the same 
signature of decay products as the studied $h_c$ decay.\\
In \Reffig{fig:sim:hc_e_gamma} the energy distributions of $\gamma$s are 
presented for the signal and one of the background channels.

$\gamma$s from $h_c$ decays cover an energy range of $[0.15; 2.0]\; \gev$. The 
corresponding distribution for the background also covers all this range, but
increases for small energies. If we want to recover $\pi^{0}$s to separate signal from background, 
the low energy $\gamma$ reconstruction threshold has be as 
low as possible, as it was already discussed in section \ref{sec:sim:recothresh}. Moreover, 
if one $\gamma$ from the $\pi^{0}$ is low energetic, the momenta of the other $\gamma$ 
and the charged hadrons move closer to the total momentum of the initial 
$\pbarp$ system. Such events pass cuts on the fit probability of the 4C-fit, 
and this makes a low energy threshold mandatory for applying anti-cuts on
no $\pi^{0}$ in the event, and correspondingly to suppress the background.

There are no measurements, to our best knowledge, of the cross-sections of the studied background
reactions. The only way to estimate these cross-sections we found was to use the DPM 
(Dual Parton Model) event generator \cite{bib:sim:DPM}. $10^7$ events were generated with DPM at a
beam momentum of $p_z=5.609 \; \gevc$, which corresponds to the studied $h_c$ resonance. 
The corresponding numbers of events are 60 and 6 for the first two background channels. No event
from the $\pbarp\,\to\,\phi\,\phi\,\piz$ reaction was observed. With a
total $\pbarp$ cross-section at this beam momentum of 60~mb, the cross-sections
for the corresponding background channels are estimated to 360~nb, 60~nb and below
6~nb, respectively.

\subsubsection{Event Selection}

The following selection criteria were applied to select the signal:
\begin{enumerate}
\item
constraint on a common vertex of $K^{+}$ and $K^{-}$ with $\phi$, pre-fit selection
$\eta_c$ invariant mass window of [2.6;3.2] $\gevcc$ and $\phi$ mass window of [0.8; 1.2] 
$\gevcc$,
\item
4C-fit to beam momentum, $CL> 0.05$,
\item
$\eta_c$ invariant mass in [2.9;3.06] $\gevcc$,
\item
$E_{\gamma}$ within [0.4;0.6] $\gev$,
\item 
$\phi$ invariant mass in [0.97;1.07] $\gevcc$,
\item
no $\pi^{0}$ candidates in the event, i.e.\ no 2 $\gamma$ invariant mass in the range 
[0.115;0.15] $\gevcc$ with 2 different low energy $\gamma$ threshold options: 30 $\mev$ 
and 10 $\mev$.
\end{enumerate}

\begin{figure}
\begin{center}
\includegraphics[width=\swidth]{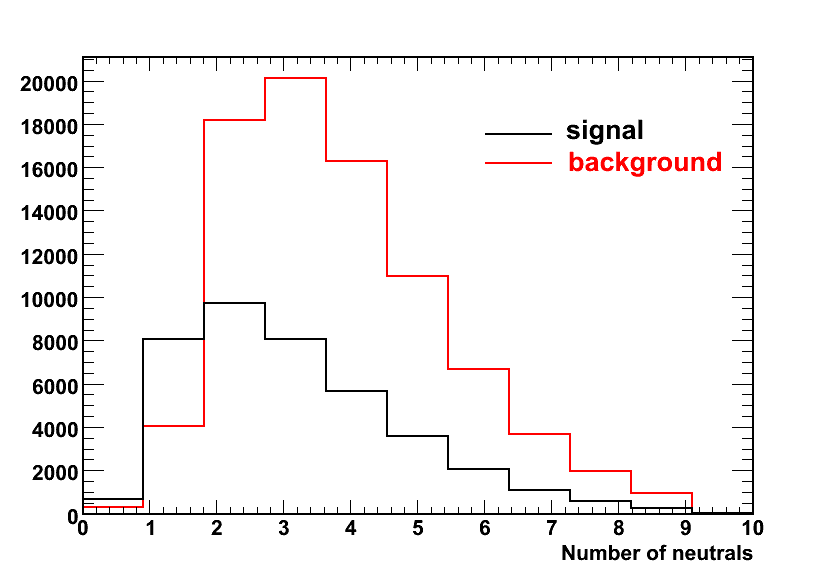}
\caption[The number of reconstructed EMC clusters.]
{The number of reconstructed EMC clusters for the $\pbarp\,\to\,h_c\,\to\,\eta_c\,\gamma$ and $\pbarp\,\to\,\phi\,\phi\,\piz$ reactions.}
\label{fig:sim:hc_n_gamma}
\end{center}
\end{figure}

\Reffig{fig:sim:hc_n_gamma} presents the multiplicity distribution of reconstructed EMC clusters 
for the signal and one of the background channels. One may note that the mean numbers of clusters, 
i.e.\ neutral candidates, significantly exceed one, expected for the signal, or two expected for 
the background from $\pi^{0}$ decay. This is caused by hadronic split-offs and prevents the use of
the number of clusters as a selection criteria. This observation emphasizes the importance of other 
selection criteria, in 
particular of the anti-cut on no $\pi^{0}$ in an event. The latter strongly depends on the assumed 
low energy $\gamma$-ray threshold.

\begin{table*}
\begin{center}
\begin{tabular}{|l|c|c|c|c|}
  \hline
  Selection criteria & signal & $4K \pi^{0}$ & $\phi  K^{+} K^{-} \pi^{0} $ & $\phi \phi \pi^{0}$ \\
    \hline
  pre-selection & 0.51 & $9.8\cdot 10^{-3}$ & $1.3\cdot 10^{-2}$ & $4.9\cdot 10^{-2}$\\
  $CL>0.05$ & 0.36 & $1.6\cdot 10^{-3}$ & $2.0\cdot 10^{-3}$ & $6.8\cdot 10^{-3}$\\
  $m(\eta_c), E_{\gamma}$ & 0.34 & $4.2\cdot 10^{-4}$ & $5.2\cdot 10^{-4}$ & $1.7\cdot 10^{-3}$ \\
  $m(\phi)$ & 0.33 & $1.0\cdot 10^{-5}$ & $1.2\cdot 10^{-4}$ & $1.7\cdot 10^{-3}$ \\
  $no \; \pi^{0} (30 MeV)$ & 0.27 & $1.0\cdot 10^{-5}$ & $4.5\cdot 10^{-5}$ & $8.6\cdot 10^{-4}$ \\
  $no \; \pi^{0} (10 MeV)$ & 0.25 & $5.0\cdot 10^{-6}$ & $3.0\cdot 10^{-5}$ & $7.0\cdot 10^{-4}$ \\

  \hline
\end{tabular}
\caption[Efficiency of different event selection criteria.]{Efficiency of different event selection criteria.}             
\label{tab:perf:hc_eff}
\end{center}
\end{table*}

\begin{table}
\begin{center}
\begin{tabular}{|l|c|}
  \hline
  channel & signal/backgr. ratio\\
    \hline
  $\pbarp \to K^{+} K^{-} K^{+} K^{-} \pi^{0}$ & 3.5 \\
  $\pbarp \to \phi K^{+} K^{-} \pi^{0}$ & 10\\
  $\pbarp \to \phi \phi \pi^{0}$ & $>6$\\
  \hline
\end{tabular}
\caption[Signal to background ratio for different $h_c$ background channels.]{Signal to background ratio for different $h_c$ background channels.}          
\label{tab:perf:hc_sb_ratio}
\end{center}
\end{table}

\subsubsection{Signal to Background Ratio versus low Energy $\gamma$-ray Threshold}
The efficiencies of the chosen selection criteria are given in \Reftbl{tab:perf:hc_eff} 
for the signal and the considered background channels.

Assuming an $h_c$ production cross-section of $40$~nb yields in the signal to background
ratios given in \Reftbl{tab:perf:hc_sb_ratio}.

For the $\pbarp\,\to\,\phi\,\phi\,\piz$ background channel a 
reduction of the $\gamma$-ray threshold from 30 MeV to 10 MeV gives a $20\percent$ improvement 
in the signal to background ratio, for 
$\pbarp\,\to\,\phi\,K^{+}\,K^{-} \pi^{0}$ the corresponding improvement is $40\percent$.

For the final signal selection efficiency of $25\percent$ and the assumed luminosity of
$L=10^{32}\cm^{-2}s^{-1}$ we expect a signal event rate of 55 events/day.

\subsection{$\mathbf{Y(4260)}$ in Formation}
The recently discovered vector-state $Y(4260)$ having a mass and width
of $(4259\pm8^{+2}_{-6})$~\mevcc and $(88\pm23^{+6}_{-4})$~\mev
\cite{bib:sim:PDG2006}, is observed in radiative return events from
$\ee$ collisions \cite{bib:sim:y4260babar} and direct formation in
$\ee$ annihilation \cite{bib:sim:y4260cleo}. Evidence is also
reported in $B\,\to\,\jpsi\,\pip\,\pim K$ decays
\cite{bib:sim:y4260babarBdec}. The $Y(4260)$ is found in the
$\jpsi\pi\pi$ and $\jpsi\Kp\Km$ decay modes. Here the reaction
$Y(4260)\,\to\,\jpsi\,\piz\,\piz$ is studied in \pbarp-formation and the decay
chain is exclusively reconstructed via $\jpsi\,\to\,\ee$ and
$\piz\,\to\,\gamma\,\gamma$.

Electron candidates are identified by the global likelihood selection
algorithm described in \Refsec{sec:sim:electronpid}. For the
reconstruction of $\jpsi\,\to\,\ee$ decays at least one of the lepton
candidates must have a likelihood value $L>0.99$ while the other
candidate should fulfill $L>0.85$. The corresponding tracks of the
$\ee$ candidates are fitted to a common vertex and an accepted
candidate should have a confidence level of $CL>0.1\%$. Photon
candidates are formed from the clusters found in the EMC applying the
reconstruction algorithm disussed in \Refsec{sec:sim:recoalgo}, must have an energy $>20~\mev$ 
and are
combined to $\piz\,\to\,\gamma\,\gamma$ candidates.

\begin{figure*}[tp]
\begin{center}
\includegraphics[width=\dwidth]{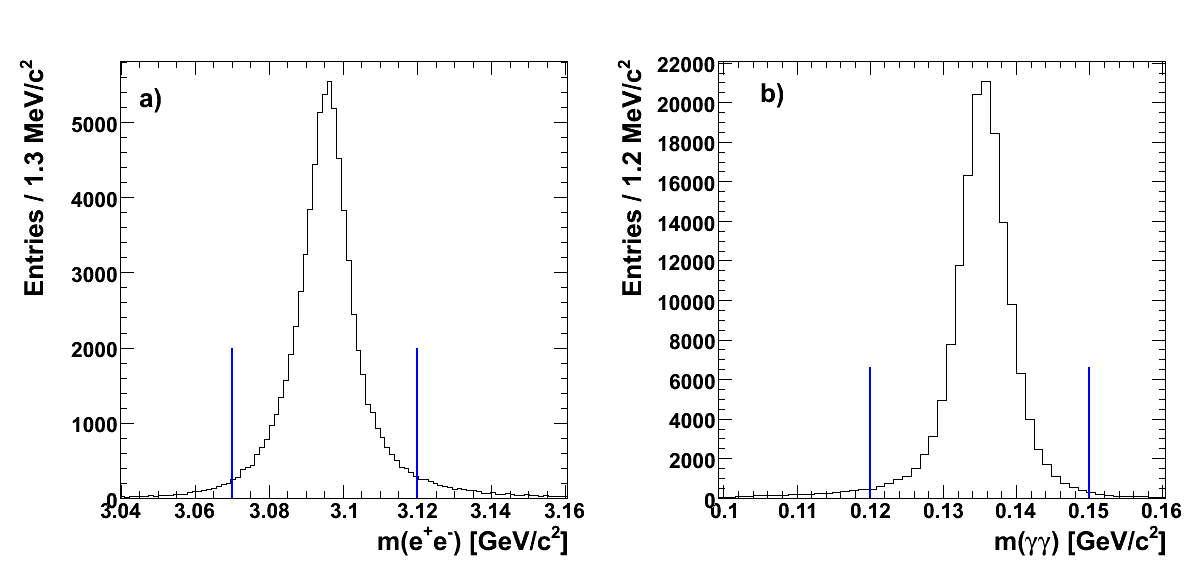}
\caption[Invariant mass for $\jpsi\,\to\,\ee$ and $\piz\,\to\,\gamma\,\gamma$.]
{Invariant a) $\jpsi\,\to\,\ee$ and b) $\piz\,\to\,\gamma\,\gamma$ mass after
the kinematic fit with constraints on the initial \pbarp system as
described in the text. The vertical lines indicate the mass windows
chosen for further selection.}
\label{fig:sim:jpsipipi}
\end{center}
\end{figure*}

Accepted \jpsi and \piz candidates are combined to $\jpsi\piz\piz$
candidates, where the \piz candidates should not share the same photon
candidate. The final state lepton and photon candidates are
kinematically fitted by constraining the sums of their spatial
momentum components and energies to the corresponding values of the
initial \pbarp system. An accepted $\jpsi\piz\piz$ candidate must have
a confidence level of $CL>0.1\%$ and the invariant mass of the \ee and
$\gamma\gamma$ subsystems should be within $[3.07;3.12]\,\gevcc$ and
$[120;150]\,\mevcc$, respectively. A FWHM of $\approx 14\,\mevcc$
($\approx 7\,\mevcc$) is obtained for the \jpsi (\piz) signal after the
kinematic fit (\Reffig{fig:sim:jpsipipi},\Reffig{fig:sim:jpsipipi}). For the final event
selection $\jpsi\piz\piz$ candidates are refitted kinematically by
constraining the invariant
\ee mass to the
\jpsi mass and the invariant $\gamma\gamma$ mass to the \piz mass on
top of the constraints applied to the initial
\pbarp four-vector. Only events where exactly one valid $\jpsi\piz\piz$ candidate
with $CL>0.1\%$ is found are accepted for further analysis.

To reject background from the reactions $\pbarp\,\to\,\jpsi\,\eta\,\eta$
($\pbarp\,\to\,\jpsi\,\eta\,\piz$) with $\eta\,\to\,\gamma\,\gamma$ the events are
reconstructed and kinematically fitted similar to the fit with the
$\jpsi\,\piz\,\piz$ signal hypothesis described above but assuming the
$\jpsi\,\eta\,\eta$ ($\jpsi\eta\piz$) hypothesis, where $\eta$ candidates
should have an invariant mass within $[500;600]\,\mevcc$. Events where
at least one valid $\jpsi\eta\piz$ or $\jpsi\,\eta\,\eta$ candidate with
$CL>0.01\%$ is found are rejected.

The reconstruction efficiency after all selection criteria is found to
be $13.8$\%. For $\pbarp\,\to\,\jpsi\,\eta\,\eta$ and $\pbarp\,\to\,\jpsi\,\eta\,\piz$
events the suppression rate is better than $10^4$ and the
contamination of the $\jpsi\piz\piz$ signal with events from these
reactions is expected to be negligible unless the cross sections for
$\pbarp\,\to\,\jpsi\,\eta\,\eta$ and $\pbarp\,\to\,\jpsi\,\eta\,\piz$ are enhanced by
more than an order of magnitude compared to the signal cross
section. Another source of background which has been investigated are
non-resonant $\pbarp\,\to\,\pip\,\pim\,\piz\,\piz$ events. For this reaction an
upper limit of $132\,\mubarn$ at 90\% CL can be derived from the
measurement at $\sqrt{s}=4.351\,\gevcc$ \cite{bib:sim:y4260BGXsec}, $92\,\mevcc$
above the $Y(4260)$ resonance. The cross section for the signal
reaction is estimated from Ref.~\cite{bib:sim:negrini} to be in the order
of $50\,\pb$. This requires a suppression for
$\pbarp\,\to\,\pip\,\pim\,\piz\,\piz$ of at least $10^7$, whereas with the
currently available amount of MC events a suppression better than
$10^8$ is obtained.

In summary the reconstruction of the $Y(4260)\,\to\,\jpsi\,\piz\,\piz$ decay
mode with one \ee pair and four photons in the final state yields a
good detection efficiency and suppression of the dominant
$\pbarp\,\to\,\pip\,\pim\,\piz\,\piz$ background.

\subsection{Charmonium Hybrid in Production}
Closely connected with charmonium spectroscopy is the search for
predicted exotic hadrons with hidden charm, such as charmonium
molecule, tetra-quark or hybrid states. For the latter the ground
state is generally expected to be a spin-exotic $J^{PC}=1^{-+}$ state
and lattice QCD calculations predict its mass in the range between
$4100$ and $4400\,\mevcc$ \cite{bib:sim:Bernard, bib:sim:Close,
bib:sim:Page}. In \pbarp annihilation this state can be produced only
in association with one or more recoiling particles. Here the results
of a study assuming the production of a state having a mass
$4290\,\mevcc$ and a width of $20\,\mev$ together with a $\eta$ meson are
reported.

Flux-tube model calculations predict for a hybrid state of this mass
suppressed decays to open charm with respect to hidden charm decays
\cite{bib:sim:HybridSuprOpenCharm}. An OZI-allowed decay to hidden
charm would be the transition to $\chi_{c1}$ with emission of light
hadrons, preferable scalar particles \cite{bib:sim:HybridScalars}. The
lightest scalar system is composed out of two neutral pions in a
relative $s$-wave.

In the study the decay of the charmonium hybrid (labeled as $\psi_g$ in
the following) to $\chi_{c1}\,\piz\,\piz$ with the subsequent
radiative $\chi_{c1}\,\to\,\jpsi\,\gamma$ decay with $\jpsi$ decaying to a
lepton pair \ee is considered. The recoiling meson is
reconstructed from the decay $\eta\,\to\,\gamma\,\gamma$. The total
branching fraction for the subsequent $\psi_g$ and $\eta$ decays is
given by $\mathcal{B}(\psi_g\,\to\,\chi_{c1}\,\piz\,\piz)\times 0.81\%$, where
$\mathcal{B}(\psi_g\,\to\,\chi_{c1}\,\piz\,\piz)$ is the unknown branching
fraction of the $\psi_g$ decay.

Photon candidates are selected from the clusters found in the EMC with
the reconstruction algorithm explained in
\Refsec{sec:sim:recoalgo}. Two photon candidates are combined and
accepted as \piz and $\eta$ candidates if their invariant mass is
within the interval [115;150]\,\mevcc and [470;610]\,\mevcc, respectively.

For the reconstruction of \jpsi decays two candidates of opposite
charge, both identified as electrons applying the likelihood selection
algorithm described in \Refsec{sec:sim:electronpid}, where one of the
candidates should have a likelihood value $L>0.2$ and the other a
value $L>0.85$, are combined and accepted as \jpsi candidates if their
invariant mass is within the interval $[2.98;3.16]\,\gevcc$. The
corresponding tracks of the two lepton candidates are kinematically
and geometrically fitted to a common vertex and their
invariant mass is constrained to the nominal \jpsi mass.

\begin{figure*}[tp]
\includegraphics[width=\dwidth]{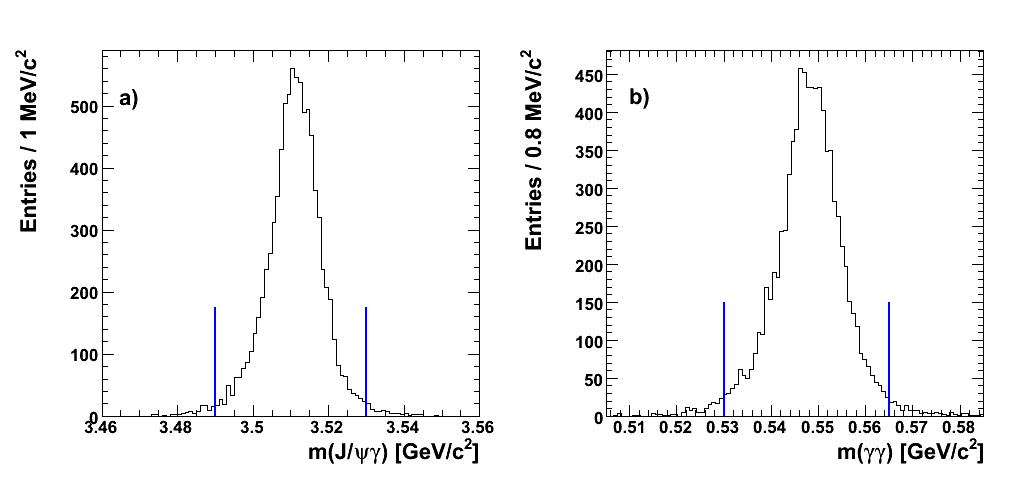}
\caption[Invariant $\chi_{c1}\,\to\,\jpsi\,\gamma$ and $\eta\,\to\,\gamma\,\gamma$ mass for the process $\ppbar\,\to\,\psi_g\,\eta$.]
{Invariant a) $\chi_{c1}\,\to\,\jpsi\,\gamma$ and b) $\eta\,\to\,\gamma\,\gamma$
mass after the kinematic fit with constraints on the initial \pbarp
system as described in the text. The vertical lines indicate the mass
windows chosen for further selection.}
\label{fig:sim:hybridChiEta}
\end{figure*}

Afterwards $\chi_{c1}\,\to\,\jpsi\,\gamma$ candidates are formed by
combining accepted
\jpsi and photon candidates, whose invariant mass is within the range $3.3$-$3.7$\,\gevcc.
From these $\chi_{c1}\,\piz\,\piz\,\eta$ candidates are created, where the
same photon candidate does not occur more than once in the final
state.  The corresponding tracks and photon candidates of the final
state are kinematically fitted by constraining their momentum and
energy sum to the initial \pbarp system and the invariant lepton
candidates mass to the \jpsi mass. Accepted candidates must have a
confidence level of $CL>0.1\%$ and the invariant mass of the
$\jpsi\,\gamma$ subsystem should be within the range
$[3.49;3.53]\,\gevcc$, whereas the invariant mass of the $\eta$
candidates must be within the interval $[530;565]\,\mevcc$. A FWHM of
$\approx 13\mevcc$ is observed for the $\chi_{c1}$ and $\eta$ signals
(\Reffig{fig:sim:hybridChiEta}) after the kinematic fit.

For the final event selection the same kinematic fit is repeated with
additionally constraining the invariant $\chi_{c1}$, \piz and $\eta$
invariant mass to the corresponding nominal mass values. Candidates
having a confidence level less than $0.1\%$ are rejected. If more than
one candidate is found in an event, the candidate with the biggest
confidence level is chosen for further analysis. 

\begin{figure}[htb]
\begin{center}
\includegraphics[width=\swidth]{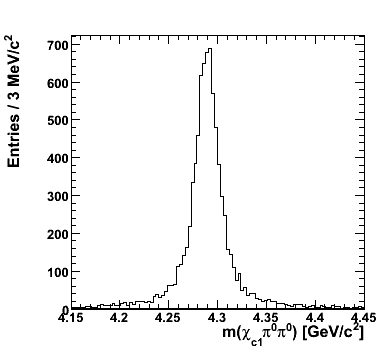}
\caption[Invariant $\chi_{c1}\,\piz\,\piz$ mass for the reaction $\ppbar\,\to\,\psi_g\,\eta$.]
{Invariant $\chi_{c1}\,\piz\,\piz$ mass after the kinematic fit with
constraints on the initial \pbarp system and resonances as described
in the text.}
\label{fig:sim:hybrid}
\end{center}
\end{figure}

The reconstruction efficiency after all selection criteria is
$3.1\%$. The invariant $\chi_{c1}\,\piz\,\piz$ mass is shown in
\Reffig{fig:sim:hybrid}. The $\psi_g$ signal has a FWHM of
$27\,\mevcc$.

The final state with 7 photons and an \ee lepton pair originating from
\jpsi decays has a distinctive signature and the separation from light
hadron background should be feasible.  Another source of background
are events with hidden charm, in particular events including a \jpsi
meson. This type of background has been studied by analyzing
$\pbarp\,\to\,\chi_{c0}\,\piz\,\piz\,\eta$,
$\pbarp\,\to\,\chi_{c1}\,\piz\,\eta\,\eta$,
$\pbarp\,\to\,\chi_{c1}\,\piz\,\piz\,\piz\,\eta$ and
$\pbarp\,\to\,\jpsi\,\piz\,\piz\,\piz\,\eta$. The hypothetical hybrid
state is absent in these reactions, but the $\chi_{c0}$ and
$\chi_{c1}$ mesons decay via the same decay path as for
signal. Therefore these events have a similar topology as signal
events and could potentially pollute the $\psi_g$ signal. The
suppression after application of all selection criteria is found to be
$4\cdot 10^3$ ($\pbarp\,\to\,\chi_{c0}\,\piz\,\piz\,\eta$), $2\cdot
10^4$ ($\pbarp\,\to\,\chi_{c1}\,\piz\,\eta\,\eta$), $>1\cdot 10^5$
($\pbarp\,\to\,\chi_{c1}\,\piz\,\piz\,\piz\,\eta$) and $9\cdot 10^4$
($\pbarp\,\to\,\jpsi\,\piz\,\piz\,\piz\,\eta$). Therefore only low
contamination of the $\psi_g$ signal from this background reactions is
expected.

\subsection{Time-like Electromagnetic Form-Factors}
The electric ($G_E$) and magnetic ($G_M$) form factors of the proton
parameterize the hadronic current in the matrix element for elastic
electron scattering $\ee\to\pbarp$ and in its crossed process
$\pbarp\to\ee$ annihilation. The form-factors are analytic functions
of the four momentum transfer $q^2$ ranging from $q^2=-\infty$ to
$q^2=+\infty$. The annihilation process allows to access the
formfactors in the timelike region ($q^2>0$) starting from the
threshold of $q^2 = 4 m_p^2c^4$. Measurements are concentrated to the
region near threshold and few data points in the $q^2$ range between
$8.8\,(\gevc)^2$ and $14.4\,(\gevc)^2$ \cite{bib:sim:eeE835}. Thus the
determination for low to intermediate momentum transfer remains an
open question. This region can be accessed at \Panda between
$q^2=5\,(\gevc)^2$ and $q^2=22\,(\gevc)^2$. The differential cross section
for the reaction $\pbarp\to\ee$ is given by \cite{bib:sim:eexsec}
\begin{eqnarray}
\frac{\mathrm{d}\sigma}{\mathrm{d}\cos\theta}&=&\frac{\pi\alpha^2 (\hbar c)^2}{8m_p^2\sqrt{\tau\left(\tau-1\right)}} \nonumber  \\
 & &   \Huge(|G_M|^2 \left (1+\cos^2 \theta \right ) \nonumber\\
& + &\frac{|G_E|^2}{\tau} \left( 1-\cos^2 \theta \right ) \Huge) 
\end{eqnarray}
with $\tau=s/4m^2_pc^4$, where $\theta$ is the angle between the electon
and the antiproton beam in the center of mass system. The factors
$|G_E|$ and $|G_M|$ can be determined from the angular distribution of
$\pbarp\to\ee$ events in dependence of the beam momentum. A
measurement of the luminosity will provide a measurement of both
factors $|G_E|$ and $|G_M|$, otherwise the absolute height of the
cross section remains unknown and only the ratio $|G_E|/|G_M|$ is
accessible. However, for a precise determination of the form factors a
suppression better than $10^8$ of the dominant background from
$\pbarp\to\pip\pim$, having a cross section about $10^6$ times higher
than the signal reaction is mandatory. For the simulations, 
a model lagrangian for the process $\pbarp\to\pip\pim$
has been created and fitted to the data \cite{bib:sim:pipluspiminus_eisenhandler,
bib:sim:pipluspiminus_buran,bib:sim:pipluspiminus_berglund,
bib:sim:pipluspiminus_dulude,bib:sim:pipluspiminus_armstrong} 
in order to get cross section predictions at angular regions, where
no data exist.

To achieve the required background suppression a good separation of
electrons and pions over a wide momentum range up to $\approx 15\,\gevc$
is mandatory, where the PID information from the individual detector
components has to be exploited. In particular information from the EMC
is important in the high momentum range where PID measurements from
the tracking, Cherenkov and muon detectors do not allow to distinguish
between the two species.

For the reconstruction of $\pbarp\,\to\,\ee$ two candidates identified as
electrons (see \Refsec{sec:sim:electronpid}) having a likelihood value $>0.998$ are
combined. The two tracks are kinematically fitted by constraining the
sum of the four-momenta to the four-momentum of the initial \pbarp
system and are accepted as a \ee candidate if the fit converges.  
The same fit is repeated but
assuming pion hypothesis for the two tracks.  For events where
more than one valid \ee candidate is found, only the candidate with
the biggest confidence level is considered for further analysis.

\begin{figure}[tb]
\begin{center}
\includegraphics[width=\swidth]{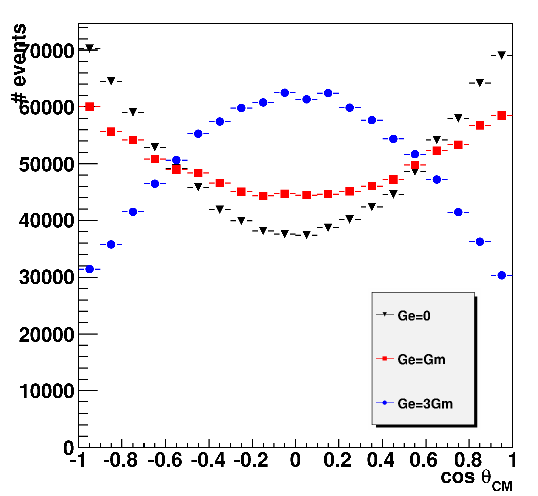}
\caption[Angular distribution of reconstructed $\pbarp\,\to\,\ee$ events.]
{Angular distribution for reconstructed $\pbarp\,\to\,\ee$ events at an
incident beam momentum of $3.3\,\gevc$ after efficiency correction. For
the ratio of the form-factors $|G_E|/|G_M|$ a value of $0$ (black), $1$ (red) and
$3$ (blue) has been assumed in the simulation.}
\label{fig:sim:epemang}
\end{center}
\end{figure}

For the measurement of $|G_E|$ and $|G_M|$ the angle $\theta$ between
the $e^+$ candidate's direction of flight and the beam axis is
computed in the center of mass system. The obtained $\cos(\theta)$
distribution is corrected by the $\cos(\theta)$-dependent
reconstruction efficiency. In lack of an absolute luminosity
measurement for MC data only the ratio $|G_E|/|G_M|$ can be determined
from a fit to the corrected $\cos(\theta)$ distribution. In
\Reffig{fig:sim:epemang} the corrected distribution for
$\sqrt{s}\,=\,3.3\,\gevc$ is shown exemplarily, whereas a ratio
$|G_E|/|G_M|$ of 0, 1 and 3 has been assumed in the simulation. The
values $-0.14\,\pm 0.07$, $1.04\, \pm 0.02$ and $2.98\,\pm 0.03$ derived from a fit to 
these distributions are in good
agreement with the input values of the simulation.

\begin{figure}[tb]
\begin{center}
\includegraphics[width=\swidth]{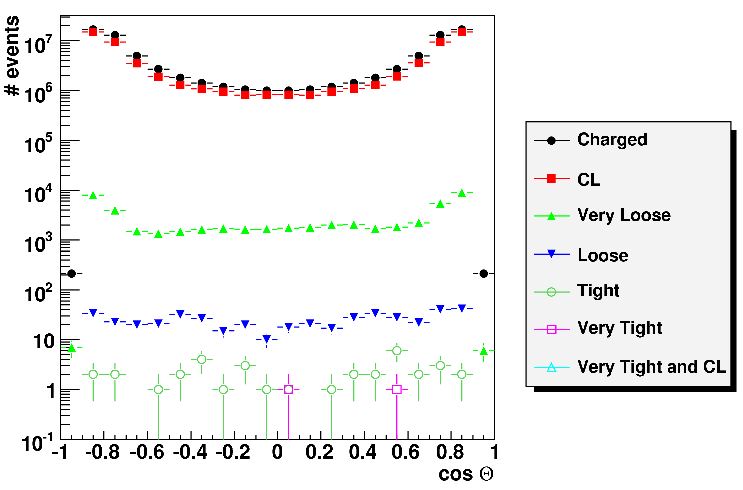}
\caption[Angular distribution of reconstructed $\pbarp\,\to\,\pip\pim$ events.]
{Angular distribution for reconstructed $\pbarp\,\to\,\pip\pim$ events at an
incident beam momentum of $3.3\,\gevc$. Different colors represent different
cuts. When requiring a probability to be an electron greater than 99.8 \% (from a combined PID 
analysis of EMC, DIRC, 
MVD and STT), we have only two out of 10$^8$ simulated and reconstructed 
$\pbarp\,\to\,\pip\pim$ events which are misinterpreted as an electron positron pair. 
If we add the constraints from the kinematical fit, there are no misidentified $\pbarp\,\to\,\pip\pim$
events left over out of the 10$^8$ simulated $\pbarp\,\to\,\pip\pim$ events. 
In the case of $7.9\,\gevc$ beam momentum 8 events out of 10$^8$ $\pbarp\,\to\,\pip\pim$ events
are left over after PID cuts. Those 8 events are also rejected by adding the constraints 
from the kinematical fit.}

\label{fig:sim:pipiang}
\end{center}
\end{figure}

Currently, 10$^8$ MC events are available for the background
process $\pbarp\to\pip\pim$ at a beam momentum of $3.3\,\gevc$ 
and additionally the same amount for the same process at a beam 
momentum of $7.9\,\gevc$. Figure \Reffig{fig:sim:pipiang} shows the 
rejection of $\pbarp\,\to\,\pip\,\pim$ when we apply different cuts based on 
PID and kinemtical fits. We find 2 out of 10$^8$ events misinterpreted
as electron positron pairs at a beam momentum of $3.3\,\gevc$ based 
on PID only and 0 events out of 10$^8$ events misinterpreted if we add 
kinematical constraints. The respective numbers for a beam momentum of $7.9\,\gevc$
are 8 events misinterpreted out of 10$^8$ from PID only, and 0 events 
out of 10$^8$ when adding kinematical constraints.

%
%
\newpage
\bibliographystyle{panda_tdr_lit}
\bibliography{./lit_emc}
%


%
\cleardoublepage
\chapter{Performance}
\label{sec:perf}
%
%
\label{sec:perform:scint}

As documented by the intrinsic performance of individual crystals of
the present quality of \PWOII, such as luminescence yield, decay
kinetics and radiation hardness and the additional gain in light yield
due to cooling down to T=-25$\degC$, there is presently no alternative
scintillator material for the electromagnetic calorimeter of \Panda
besides lead tungstate. The general applicability of PWO for
calorimetry in High-Energy Physics has been promoted and finally
proven by the successful realization of the CMS/ECAL detector as well
as the photon spectrometer \INST{ALICE/PHOS}, both installed at LHC.

The necessary mass production of high-quality crystals has been
achieved primarily at BTCP and North Crystals in Russia and SICCAS in
China. However, driven by the much more stringent requirements to
perform the Physics program of \PANDA, from the beginning it became
mandatory to search for a significantly higher scintillation yield
in order to extend photon detection down to the MeV energy
range. Beyond the realization of a nearly perfect crystal the
operation at temperatures well below room temperature will reduce the
effect of the thermal quenching of the scintillation process and
improve the photon statistics. The spectrometer PHOS is following the
same line. Our requirements are even more stringent, since the \TSEMC
of \Panda has to cover nearly the full solid angle and stresses even
more the concept of thermal insulation and cooling.

The test experiments to prove the concept and applicability of PWO
have concentrated on the response to photons and charged particles at
energies below $1\,\GeV$, since those results are limited by the
photon statistics of the scintillator, the sensitivity and efficiency
of the photo sensor and the noise contributions of the front-end
electronics. The conclusions are drawn based on crystal arrays
comprising up to 25 modules. The individual crystals have a final
length of $200\,\mm$ and a rectangular cross section of
$20\times20\,\mm^{2}$. Only the most recently assembled array PROTO60
consists of 60 crystals in \Panda geometry. The tapered shape will
improve the light collection due to the focusing effect of the
geometry as known from detailed simulations at CMS/ECAL.

The performance tests completed up to now have been aiming at two
complementary aspects. On one hand, the quality of full size PWO-II
crystals has to be verified by in-beam measurements with energy marked
photons covering the most critical energy range up to 1GeV. Therefore,
the scintillator modules were readout with standard photomultiplier
tubes (Philips XP1911) with a bi-alkali photocathode, which covers
$\sim$35$\percent$ of the crystal endface with a typical quantum
efficiency of QE=18$\percent$. The noise contribution of the sensor
can be neglected and the fast response allows an estimate of the time
response. All achieved resolutions, which have been deduced at various
operating temperatures, can be considered as benchmark limits for
further studies including simulations and electronics development.

The second R$\&$D activity was aiming to come close to the final
readout concept with large area avalanche photodiodes (\LAAPD), which
are mandatory for the operation within the magnetic field. The new
developments of the sensor as well as the parallel design of an
extremely low-noise and low-power preamplifier, based on either
discrete components or customized ASIC technology, are described in
detail in separate chapters. All the reported results are performed by
collecting and converting the scintillation light with a single
quadratic \LAAPD of $10\times10\,\mm^{2}$ active area with a quantum
efficiency above 60$\percent$. The final readout considers two \LAAPDs
of identical surface but rectangular shape to fit on the crystal
endface, which is presently prevented by the dead space of the ceramic
support. The reported experiments are using in addition individual
low-noise preamplifiers and commercial electronics for the
digitization. Again, the present data deliver a lower limit of the
final performance, which will be achieved with twice the sensor
surface.

In order to simulate the operation of large arrays, the mechanical
support structures, cooling and temperature stabilization concepts and
long term stabilities, a large prototype comprising 60 tapered
crystals in \Panda-geometry has been designed and brought into
operation. First in-beam tests are scheduled in summer 2008. In a next
iteration, a mechanical prototype for housing 200 crystals is
presently under construction. This device is primarily meant for
studying stable operation and cooling simulating two adjacent detector
slices of the barrel section of the EMC.

In spite of some compromises compared to the final concept, the
achieved resolutions represent excellent lower limits of the
performance to be expected. Operation at T=-25$\degC$ using a
photomultiplier readout delivers an energy resolution of $\sigma / E =
0.95\percent /\sqrt{E} + 0.91\percent$ ($E$ given in $\GeV$) for a
3$\times$3 sub-array accompanied with time resolutions below
$\sigma$=$130\,\ps$. The complementary test using a single \LAAPD for
readout reaches even at T=0$\degC$ a fully sufficient resolution of
$\sigma / E = 1.86\percent /\sqrt{E} + 0.65\percent$ ($E$ given in
$\GeV$). Timing information can be expected with an accuracy well
below $1\,\ns$ for energy depositions above $100\,\MeV$.  Summarizing,
according to the detailed simulations and the selected test results,
operation of the calorimeter at \mbox{T=-25$\degC$} will allow to
perform the very ambitious research program, even if radiation damage
at most forward directions might reduce asymptotically the light
output by up to 30$\percent$.

\section{Results from Prototype Tests}

\subsection{Energy Resolution of \PWO arrays}
\label{chap:perform:energies}

Almost in parallel to the research project for CMS/ECAL,
investigations have been started to explore the applicability of PWO
in the low and medium energy regime. Beside the response to low energy
$\gamma$-rays of radioactive sources the initial pilot experiments
used electrons as well as energy marked photons at the tagging
facility of \INST{MAMI} at Mainz \cite{bib:emc:perf2} between 50 and
850MeV energy \cite{bib:emc:perf3,bib:emc:perf4}. The crystals were
rectangular of $150\,\mm$ length and with a quadratic cross section of
20$\times$20mm$^{2}$ with all sides optically polished. The quality
was similar to CMS samples of the pre-production runs and contained
slow decay components on the percent level due to
Mo-contamination. The readout of the scintillation light was achieved
with photomultiplier tubes and commercial electronics of NIM and CAMAC
standard. In most of the cases the anode signals were carried via RG58
coaxial cables over a distance of $30-50\,\m$ outside the experimental
area to the DAQ system. For practical reasons, energy and time
information were deduced and digitized after long passive delays,
causing a significant signal attenuation and loss of the
high-frequency response. The compact detector arrays were operated at
a stabilized temperature, always above T=0$\degC$ but slightly below
room temperature. Keeping the crystals at a fixed temperature had to
guarantee stable operation, not with the intention to increase the
luminescence yield.

The response to monoenergetic photons showed for the first time an
excellent energy resolution even for photons well below the expected
range for CMS/ECAL. Resolution values of $\sigma / E = 1.54\percent /
\sqrt{E} + 0.30\percent$ ($E$ given in $\GeV$) were achieved for a
matrix of 5$\times$5 elements and in addition time resolutions of
$\sigma$$_{t}$$\leq$130ps for photon energies above $25\,\MeV$
\cite{bib:emc:perf9}.

Besides the studies using low and high-energy photons exploratory
experiments at \INST{KVI}, Groningen, and \INST{COSY}, FZ J\"ulich,
have delivered results for low and high energy protons and pions
\cite{bib:emc:perf1}. However, in all these cases \PWO of standard
quality according to the CMS specifications was used.  For protons,
which are completely stopped in a crystal of 150$\,\mm$ length, an
energy resolution of $\sigma / E = 1.44\percent / \sqrt{E} +
1.97\percent$ ($E$ given in $\gev$) has been extracted as an upper
limit, since the proton energy had to be deduced from the time of
flight measured using a fast plastic start counter. For 90$\,\MeV$
protons typical resolutions between 4$\percent$ and $5\percent$ were
obtained.

In the following chapters, the obtained experimental results are based
on scintillator crystals, which correspond to the performance
parameters of \PWOII \cite{bib:emc:perf7}.

\subsubsection{Energy Response Measured with Photomultiplier Readout}
\label{subchap:perform:energy-pm}

As outlined in the previous chapters on \PWO
(\Refsec{sec:scint:pwo:scintillator}) the specifications of the
calorimeter can be further optimized by the reduction of the thermal
quenching of the relevant scintillation process by cooling the
crystals down to temperatures as low as T=-25$\degC$, which leads to a
significant increase of the light yield by a factor 4 compared to
T=+25$\degC$. For a typical \PWOII crystal 500 photons can be
collected at the endface of a $200\,\mm$ long crystal for an energy
deposition of $1\,\MeV$ at 100\percent quantum efficiency.

The reported experiments \cite{bib:emc:perf5} are based on large size
crystals of $200\,\mm$ length grouped in arrays composed of 3$\times$3
up to 5$\times$5 crystals manufactured at BTCP or comparable in
quality grown at SICCAS. All crystals were rectangular parallelepipeds
to allow in addition a direct comparison to initial pilot experiments
with limited quality as stated in the previous chapter.

\begin{figure}
\begin{center}
\includegraphics[width=\swidth]{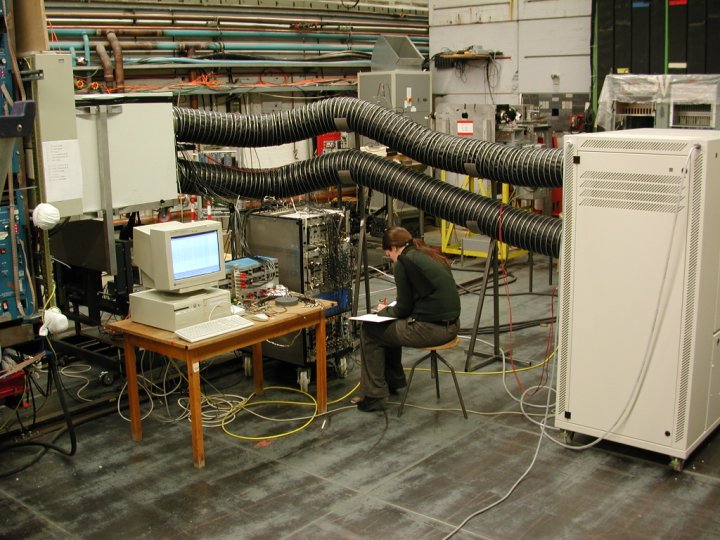}
\caption[The experimental setup at the tagged photon facility at
MAMI.]  {The experimental setup at the tagged photon facility at MAMI
  comprising the cooling unit (right), the connection pipe to
  circulate cold dry air and the insulated container for the test
  detectors (left).}
\label{fig:perf:perf1}
\end{center}
\end{figure}

\begin{figure}
\begin{center}
\includegraphics[width=\swidth]{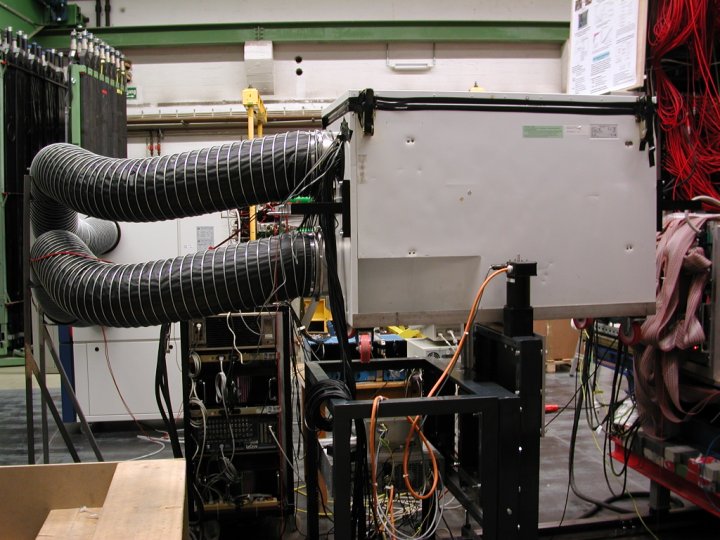}
\caption[The insulated container to house the various test arrays.]
{The insulated container to house the various test arrays. The cold
  air is circulated via the two black pipes.}
\label{fig:perf:perf2}
\end{center}
\end{figure}

\begin{figure}
\begin{center}
\includegraphics[width=\swidth]{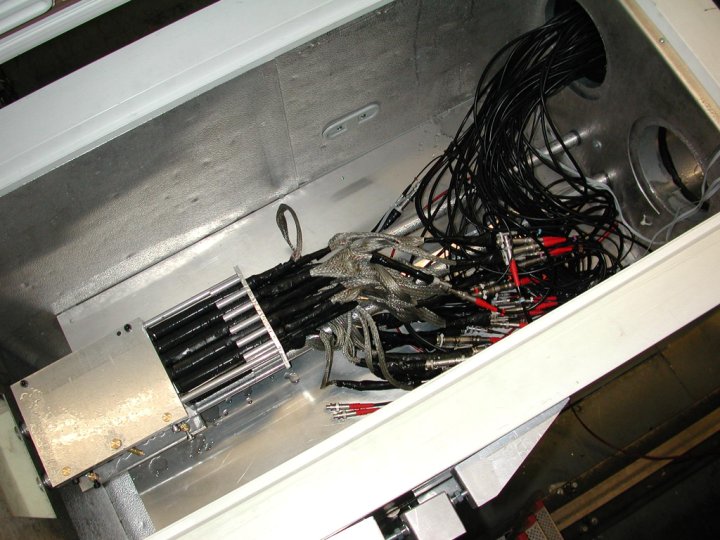}
\caption[A typical 5$\times$5 \PWO matrix.]  {A typical 5$\times$5
  \PWO matrix shown inside the insulated box including photomultiplier
  readout, signal and high voltage cables.}
\label{fig:perf:perf3}
\end{center}
\end{figure}

The temperature stabilization and the operation at significantly lower
temperatures in an atmosphere free of moisture required the
installation of a new experimental setup, which comprises an insulated
container of large volume to house the different detector arrays with
high flexibility. A computer controlled cooling machine circulates
cold and dry air of constant temperature via thermally insulated
$5\,\m$ long pipes of large diameter to the detector
container. \Reffig{fig:perf:perf1}, \Reffig{fig:perf:perf2} and
\Reffig{fig:perf:perf3} show the experimental installation as well as
details of a typical detector matrix mounted in the \INST{A2} hall at
Mainz behind the \INST{CB/TAPS} experiment.

\begin{figure}
\begin{center}
\includegraphics[width=\swidth]{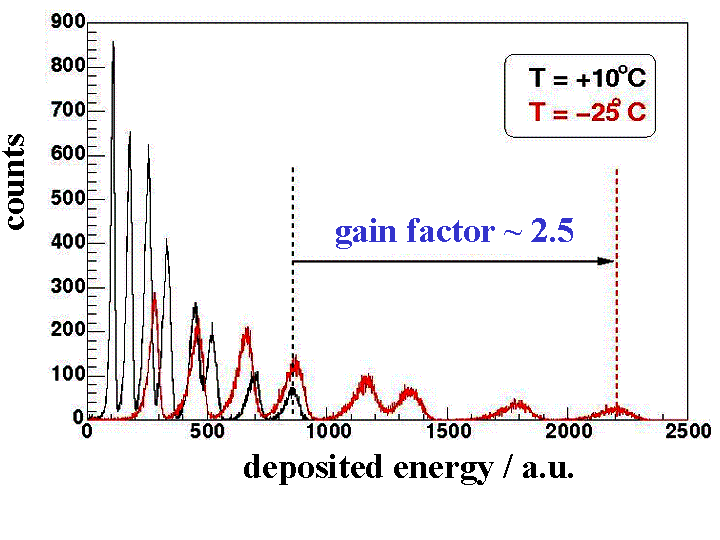}
\caption[Lineshape of the central detector module.]  {Lineshape of the
  central detector module measured at eight different photon energies
  and two operating temperatures. The change of the light yield due to
  the difference in the thermal quenching of the scintillation light
  can be deduced directly.}
\label{fig:perf:perf4}
\end{center}
\end{figure}

The setup shown particularly in \Reffig{fig:perf:perf3} was used to
test the inner 3$\times$3 section of a 5$\times$5 matrix of $200\,\mm$
long crystals from BTCP, individually readout via photomultiplier
tubes (Philips XP 1911). The measurement was performed at T=+10$\degC$
and T=-25$\degC$, respectively, to illustrate the effect of
cooling. The advantage of reduced temperatures becomes immediately
obvious in \Reffig{fig:perf:perf4} in the overlay of the response of
the central detector to eight different photon energies between
$64\,\MeV$ and $520\,\MeV$, respectively, measured at the two
operating temperatures. The light-output was measured under identical
experimental conditions of the photomultiplier bias and electronics
settings and documents a gain factor of 2.6 of the overall
luminescence yield.

\begin{figure}
\begin{center}
\includegraphics[width=\swidth]{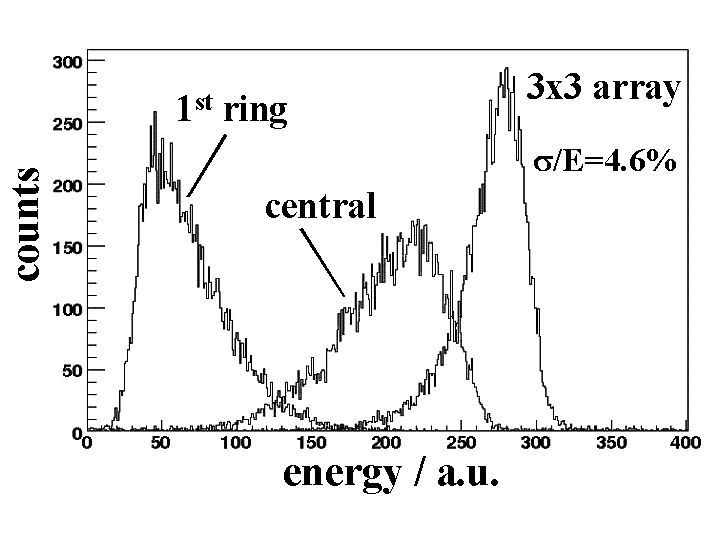}
\caption[Response function of the 3$\times$3 matrix to photons.]
{Response function of the 3$\times$3 matrix to photons of $63.8\,\MeV$
  energy measured at T=-25$\degC$.}
\label{fig:perf:perf5}
\end{center}
\end{figure}

\begin{figure}
\begin{center}
\includegraphics[width=\swidth]{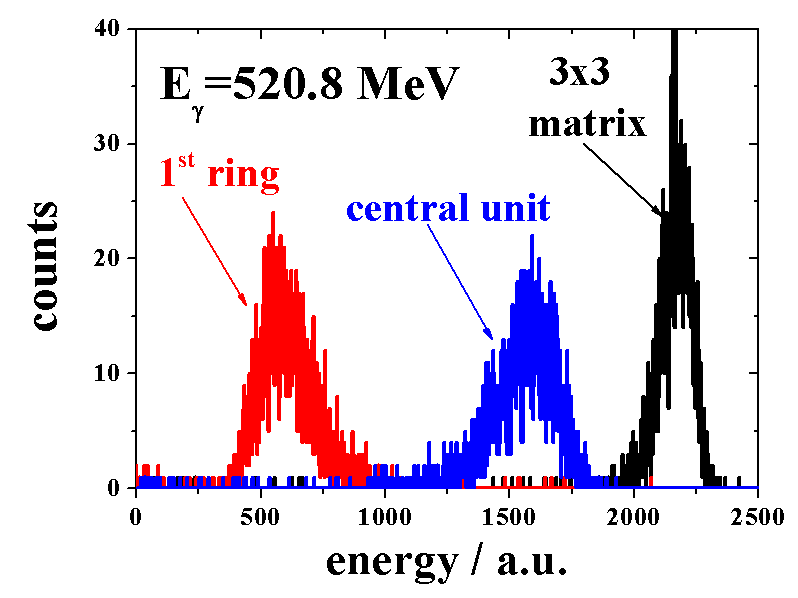}
\caption[Response function of the 3$\times$3 matrix to photons.]
{Response function of the 3$\times$3 matrix to photons of
  $520.8\,\MeV$ energy measured at T=-25$\degC$.}
\label{fig:perf:perf6}
\end{center}
\end{figure}

\begin{figure}
\begin{center}
\includegraphics[width=\swidth]{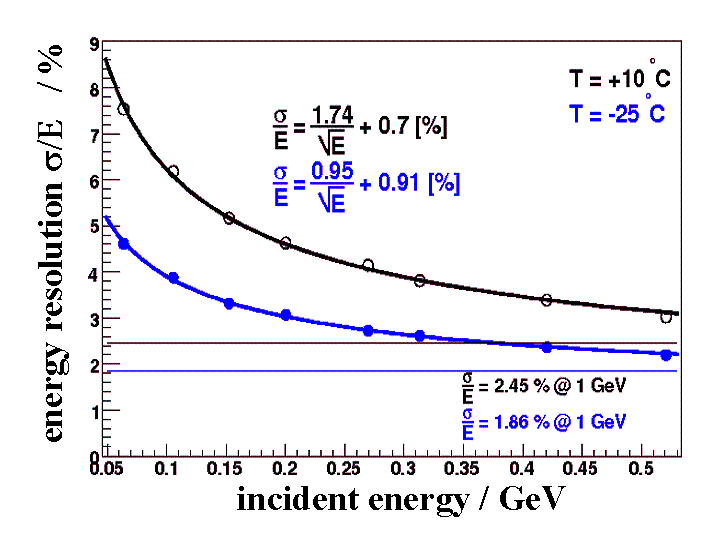}
\caption[Comparison of the energy resolution.]  {Comparison of the
  energy resolution of a 3$\times$3 \PWOII matrix of $200\,\mm$ long
  crystals measured at two different operating temperatures of
  T=-25$\degC$ and T=+10$\degC$, respectively. The crystal responses
  were readout individually with photomultiplier tubes.}
\label{fig:perf:perf7}
\end{center}
\end{figure}

Figure~\ref{fig:perf:perf5} and \Reffig{fig:perf:perf6} show for the
low temperature of T=-25$\degC$ with reduced statistics the lineshape
of the central detector, the total energy deposition in the
surrounding ring of eight crystals and the integral of the full array
for the two extreme photon energies. In particular at the highest
energy, the persisting low energy tail indicates the significant
lateral shower leakage, which limits the obtainable resolution. To
parameterize the achieved energy resolution the reconstructed
lineshapes of the electromagnetic showers have been fitted with an
appropriate function to determine the FWHM. Finally, an excellent
energy resolution of $\sigma / E = 0.95\percent /\sqrt{E} +
0.91\percent$ ($E$ given in $\GeV$) was deduced, which represents the
best resolution ever measured for PWO (see \Reffig{fig:perf:perf7}). A
value well below 2$\percent$ extrapolating to an incident photon
energy of $1\,\gev$ is very similar to the performance of several
operating EM calorimeters, which are based on well known bright
scintillator materials such as CsI(Tl), NaI(Tl) or BGO, respectively.

In comparison, the reduced energy resolution of $\sigma / E =
1.74\percent /\sqrt{E} + 0.70\percent$ ($E$ given in $\GeV$) deduced
at T=+10$\degC$ is consistent with the lower light output due to
thermal quenching and is quantitatively expressed by the higher
statistical term in the parametrization of the resolution
\cite{bib:emc:perf10}.

\begin{figure}
\begin{center}
\includegraphics[width=\swidth]{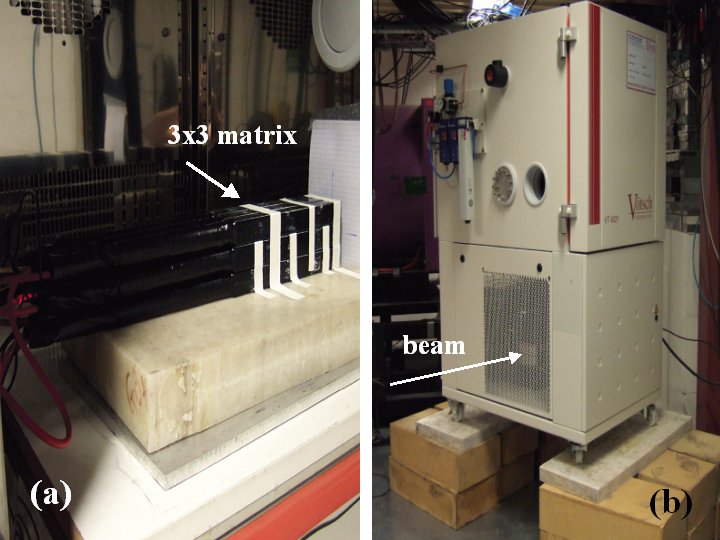}
\caption[Detector operation at low temperatures at MAX-Lab.]  {The
  3$\times$3 \PWOII matrix (left) and the climate chamber (right) for
  detector operation at low temperatures as installed for response
  measurements at MAX-Lab.}
\label{fig:perf:perf8}
\end{center}
\end{figure}

In order to extend and quantify experimentally the response function
to even lower energies, tests were performed at the MAX-Lab laboratory
at Lund, Sweden. The tagging facility delivers low-energy photons in
the energy range between $19\,\MeV$ and $50\,\MeV$ with an intrinsic
photon resolution far below $1\,\MeV$. The rectangular crystals of
20$\times$20$\times$200mm$^{3}$ were individually wrapped in Teflon
and coupled to a photomultiplier tube (Philips XP 1911). The detector
array was mounted in a climate chamber and cooled down to a
temperature of T=-25$\degC$. The assembled detector matrix as well as
the cooling unit accommodating the setup are shown in
\Reffig{fig:perf:perf8}.

\begin{figure}
\begin{center}
\includegraphics[width=\swidth]{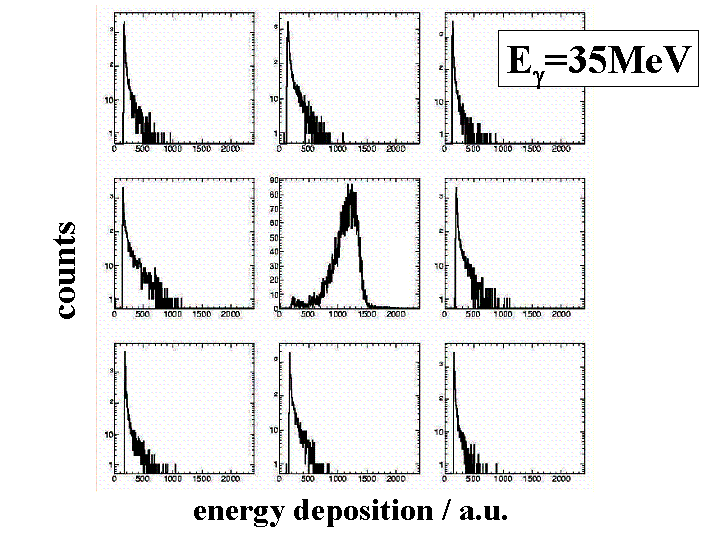}
\caption[Energy deposition of a $35\,\MeV$ incident photon.]
{Energy deposition of a $35\,\MeV$ incident photon in a 3$\times$3 \PWOII matrix measured at a temperature of T=-25$\degC$ at MAX-Lab.}
\label{fig:perf:perf9}
\end{center}
\end{figure}

\begin{figure}
\begin{center}
\includegraphics[width=\swidth]{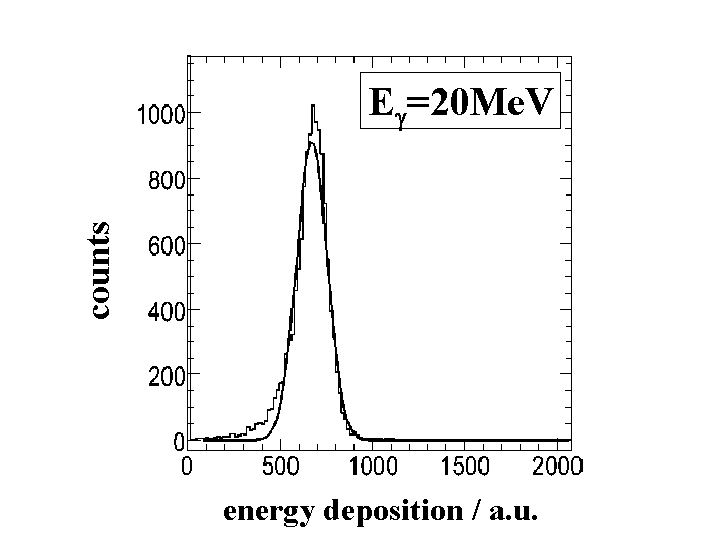}
\caption[Response of a 3$\times$3 \PWOII matrix to incident photons.]
{Response of a 3$\times$3 \PWOII matrix to incident photons of
  $20\,\MeV$ energy measured at a temperature of T=-25$\degC$ at
  MAX-Lab. A Gaussian-like lineshape is shown for comparison.}
\label{fig:perf:perf10}
\end{center}
\end{figure}

The individual modules have been calibrated using the energy loss of
minimum ionizing muons in order to reconstruct the electromagnetic
shower within the matrix. \Reffig{fig:perf:perf9} illustrates the
energy distribution within the array for an incident photon energy of
$35\,\MeV$. Nearly Gaussian-like lineshapes have been deduced for
energies as low as $20\,\MeV$ as illustrated in
\Reffig{fig:perf:perf10}.

\begin{figure}
\begin{center}
\includegraphics[width=\swidth]{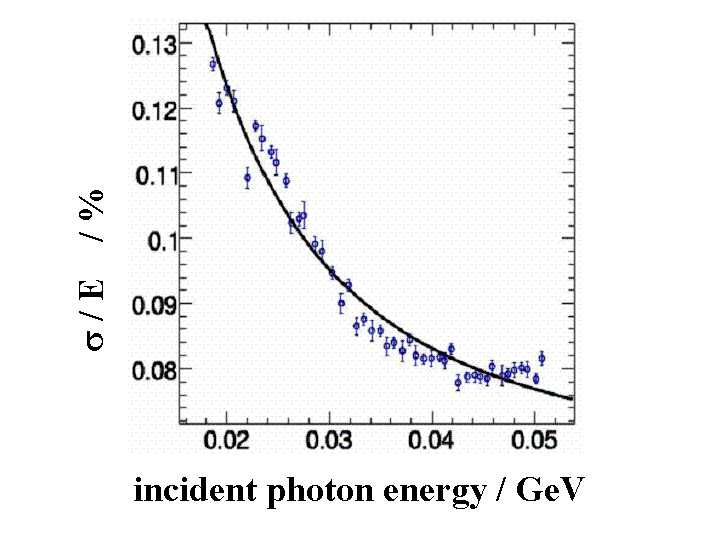}
\caption[Experimental energy resolution of a 3$\times$3 \PWOII
matrix.]  {Experimental energy resolution of a 3$\times$3 \PWOII
  matrix in the energy range between $19\,\MeV$ and $50\,\MeV$,
  respectively, measured at a temperature of T=-25$\degC$ at MAX-Lab.}
\label{fig:perf:perf11}
\end{center}
\end{figure}

\Reffig{fig:perf:perf11} summarizes the obtained energy resolutions
over the investigated range of low incident photon energies and
documents the excellent performance even at the lowest energies to be
studied with the \TSEMC. One should consider that the readout with a
standard photomultiplier represents only a lower limit compared to the
future readout with two \LAAPDs. They will convert a significantly
higher percentage of the scintillation light, which should have a
strong impact on photon detection in particular at very low energies.

\subsubsection{Energy Response Measured with \LAAPD Readout}
\label{subchap:perform:energy-apd}

The operation of the calorimeter inside the solenoid will require a
readout with \LAAPD or equivalent sensors, which are not affected by
strong magnetic fields. As outlined in more detail in the relevant
chapters, a new low-noise and compact charge sensitive preamplifier
was developed and used in prototype tests. As a test of principle,
with low energy $\gamma$-sources and small \PWOII crystals, cooled
down to T=-25$\degC$, a clean identification of the photopeak of
$\gamma$-rays as low as $511\,\keV$ was demonstrated, thus indicating
the low level of noise contributions.

In a first in-beam test \cite{bib:emc:perf6} in 2004 at the tagged
photon facility of MAMI at Mainz, a 3$\times$3 matrix of $150\,\mm$
long crystals of \PWOII quality was operated at T=-25$\degC$. All
modules were read out with 10$\times$10mm$^{2}$ \LAAPD manufactured by
Hamamatsu \cite{bib:emc:perf8}. The charge output was collected in the
new charge sensitive preamplifier and passively split afterwards to
integrate the signals in a spectroscopy amplifier as well as to deduce
a timing information from a constant fraction discriminator after
shaping in a timing filter amplifier (INT=$50\,\ns$, DIFF=$50\,\ns$).
Both signals were transferred via $50\,\m$ coaxial cables to the DAQ
system for digitization.

\begin{figure}
\begin{center}
\includegraphics[width=\swidth]{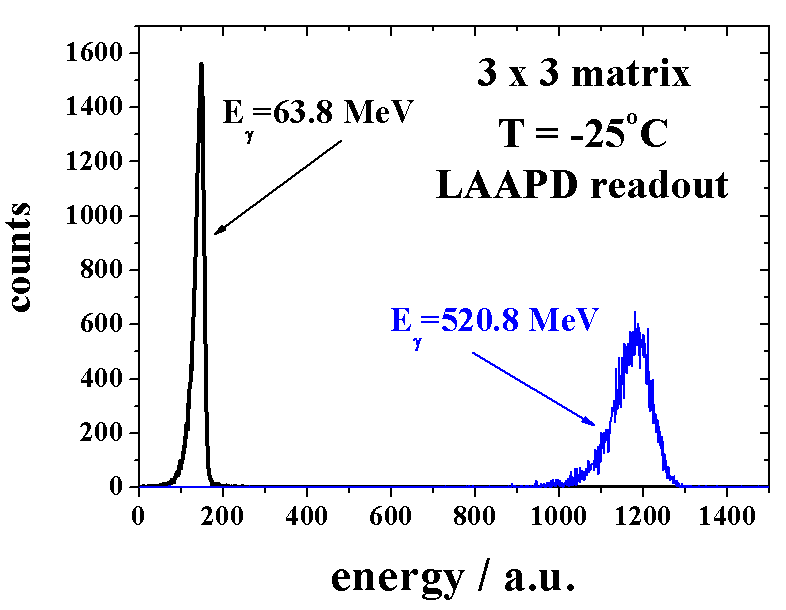}
\caption[Response function of the 3$\times$3 \PWOII matrix to
photons.]  {Response function of the 3$\times$3 \PWOII matrix to
  photons of $63.8\,\MeV$ and $520.8\,\MeV$ energy measured at
  T=-25$\degC$. The $150\,\mm$ long crystals are readout with
  \LAAPDs.}
\label{fig:perf:perf12}
\end{center}
\end{figure}

\begin{figure}
\begin{center}
\includegraphics[width=\swidth]{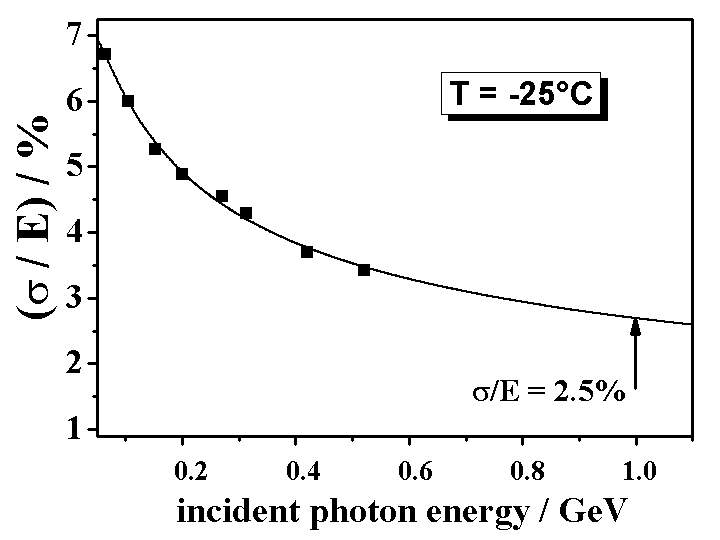}
\caption[Energy resolution of a 3$\times$3 \PWOII matrix.]  {Energy
  resolution of a 3$\times$3 \PWOII matrix of $150\,\mm$ long crystals
  measured at T=-25$\degC$ readout with \LAAPDs.}
\label{fig:perf:perf13}
\end{center}
\end{figure}

In spite of the long transfer lines and the missing option to optimize
the adjustments of the electronics an excellent result has been
achieved. \Reffig{fig:perf:perf12} shows the obtained lineshape at
$64\,\MeV$ and $520\,\MeV$ incident photon energy, respectively, after
reconstruction of the electromagnetic shower deposited within the
array. It should be mentioned that even one of the detector modules
was not functioning at all. The deduced energy resolution is
summarized in \Reffig{fig:perf:perf13} and can be parameterized as
$\sigma / E = 0.90\percent /\sqrt{E} \oplus 2.13\percent$ ($E$ given
in $\GeV$). The low value of the statistical factor reflects the high
electron/hole statistics due to the significantly larger quantum
efficiency of the avalanche diode compared to a standard bialkali
photocathode. The large constant term can be related to the missing
detector, reduced signal/noise ratio due to passive splitting and the
not yet optimized adjustments of the used commercial electronics.

\begin{figure}
\begin{center}
\includegraphics[width=\swidth]{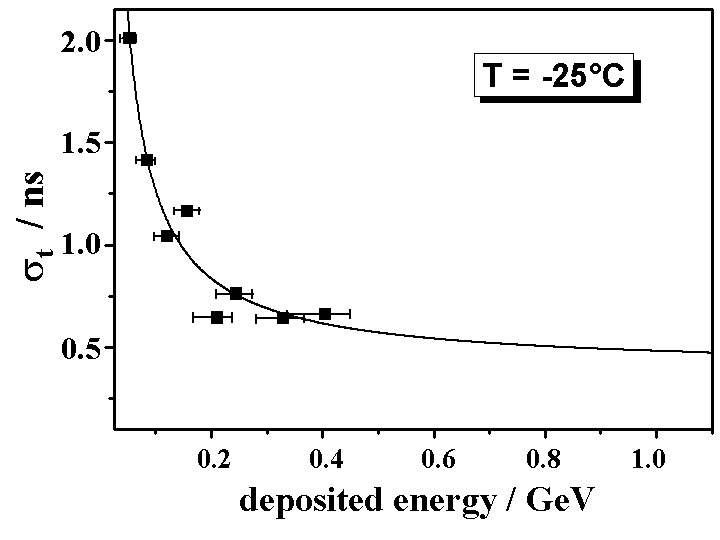}
\caption[Time resolution of a 3$\times$3 \PWOII matrix.]  {Time
  resolution of a 3$\times$3 \PWOII matrix of $150\,\mm$ long crystals
  measured at T=-25$\degC$ readout with \LAAPDs.}
\label{fig:perf:perf14}
\end{center}
\end{figure}

The used readout represents a first attempt to deduce as well timing
information. In order to identify off-line the selected photon
energies the trigger signal of the responding scintillators of the
tagger ladder had to be in coincidence. The recorded time information
serves as a reference with an intrinsic resolution of approximately
$1.2\,\ns$ (FWHM). In spite of the by far not optimized timing
measurement, a resolution of $\sigma$$_{t}$$\ll$$1\,\ns$ can be
reached already above the typical energy deposition in the central
crystal for an incident photon energy of $\geq$$200\,\MeV$. The result
emphasizes the excellent timing capabilities of \PWO even for a
readout via \LAAPDs. The achieved time resolutions are summarized in
\Reffig{fig:perf:perf14}.

In order to investigate a larger range of operational temperature a
3$\times$3 matrix of $200\,\mm$ long \PWOII crystals, produced in 2007
under more stringent specification limits, was tested at a temperature
of T=0$\degC$ \cite{bib:emc:perf14}. In addition, the rectangular
crystals were covered for the first time with one layer of a thermally
molded reflector foil consisting of the 3M radiant mirror film
VM2000$^{TM}$ and mounted in a container made of carbon fibre, similar
to the alveole-structure to be foreseen for the final assembly of the
EMC. Detailed tests with respect to the properties of VM2000 show an
increase of the collected light of 10--15$\percent$ compared to a
wrapping with several layers of Teflon. The matrix was mounted in an
insulated container, which could be cooled down to a low temperature
well stabilized ($\Delta$T$<$0.1$\degC$) by circulating cold and dry
air similar to the setup as shown in \Reffig{fig:perf:perf1} and
\Reffig{fig:perf:perf2}. A thin plastic scintillator covering the
total front of the prototype served as a charged particle veto to
reject electrons and/or positrons due to converted photons upstream of
the detector. The crystal matrix could be moved remote controlled in
two dimensions perpendicular to the axis of the collimated photon beam
by stepping-motors to perform a relative calibration of each detector
element under beam conditions. Each of the optically polished crystals
was coupled to a single \LAAPD using optical grease (BAYSILONE
300.000). The used prototype-2 version (S8664-1010SPL) has an active
area of 10$\times$10mm$^{2}$, a quantum efficiency of 70$\percent$ (at
420nm), a dark current of $10\,\nA$ at a gain of 50 and a capacitance
of $270\,\pF$ at a gain of 50 and at a frequency of $100\,\kHz$. The
charge signal of the \LAAPD was amplified with a low-noise and
low-power charge sensitive preamplifier with a sensitivity of
$0.5\,\V$/pC and a small feedback time constant of $25\,\mus$.

The output signal of the preamplifier was fed into a 16-fold
multi-functional NIM module (MESYTEC, MSCF-16), which provided a
remote controllable multi-channel spectroscopic amplifier ($1\,\mus$
Gaussian shaping), timing filter amplifier (INT=DIFF=$50\,\ns$), and
constant fraction discriminator to serve both the analog and logic
circuit in parallel. The energy response was digitized in a
peak-sensing ADC (CAEN V785N). Time information of the individual
modules was deduced relative to the central module. The coincidence
timing of the central crystal with one of the timing signals of the
chosen tagger channels selected the event type. The energy and time
information of each module together with the timing response of the
relevant tagger channels were recorded event-by-event for an off-line
analysis. The response function of the crystal matrix was measured for
the first time at an intermediate temperature of T=0$\degC$.

\begin{figure}
\begin{center}
\includegraphics[width=\swidth]{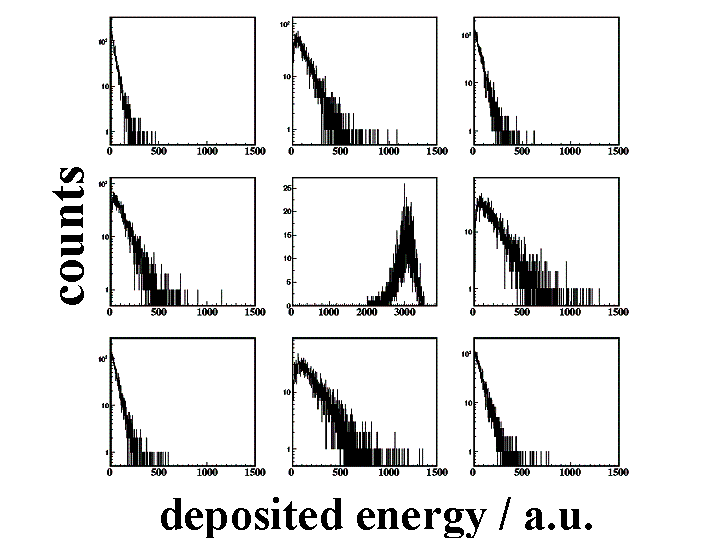}
\caption[Distribution of the electromagnetic shower into the
3$\times$3 \PWOII matrix.]  {Distribution of the electromagnetic
  shower into the 3$\times$3 \PWOII matrix at an incident photon
  energy of $674.5\,\MeV$ readout with single \LAAPDs as photo sensors
  at a temperature of T=0$\degC$.}
\label{fig:perf:perf15}
\end{center}
\end{figure}

The distribution of the electromagnetic shower within the 3$\times$3
crystal matrix has been obtained after a carefully prepared relative
calibration of the individual modules. The readout concept of a common
integration gate did not impose any hardware energy threshold. The
absolute calibration was deduced from the simulation of the total
energy deposition in the array based on a GEANT4 simulation, taking
into account all instrumental details.  \Reffig{fig:perf:perf15}
illustrates the measured energy distribution within the crystal matrix
at an incident photon energy of $674.5\,\MeV$. Since the cross section
of the used crystals is comparable to the Moli\`ere radius,
approximately 70$\percent$ of the shower energy are contained in the
central module.

\begin{figure}
\begin{center}
\includegraphics[width=\swidth]{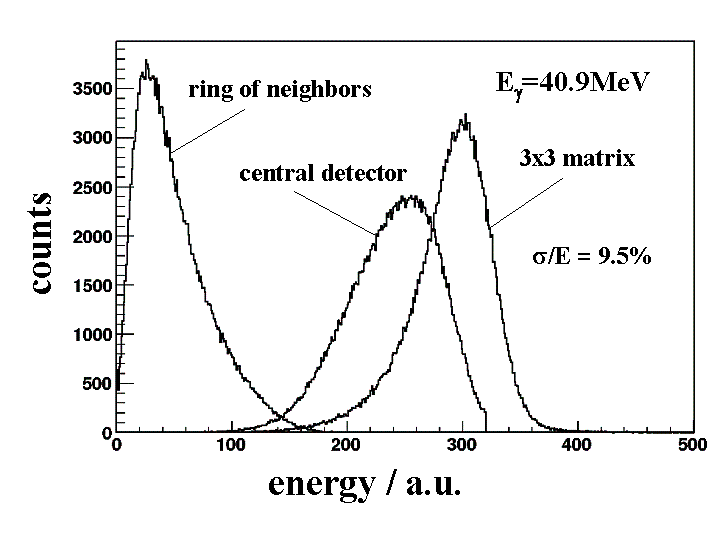}
\caption[Experimental line shape of the 3$\times$3 \PWOII matrix.]
{Experimental line shape of the 3$\times$3 \PWOII matrix measured at
  the lowest photon energy of $40.9\,\MeV$ at a temperature of
  T=0$\degC$. The figure shows the individual energy depositions into
  the central module and the surrounding ring. Each crystal was
  readout with a single \LAAPD.}
\label{fig:perf:perf16}
\end{center}
\end{figure}

\begin{figure}
\begin{center}
\includegraphics[width=\swidth]{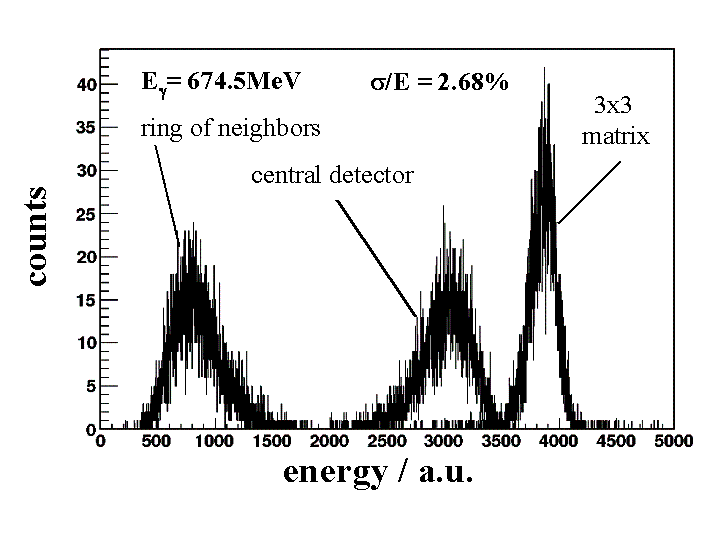}
\caption[Experimental line shape of a 3$\times$3 \PWOII matrix.]
{Experimental line shape of a 3$\times$3 \PWOII matrix measured at the
  highest photon energy of $674.5\,\MeV$ at a temperature of
  T=0$\degC$. The figure shows the individual energy depositions into
  the central module and the surrounding ring. The crystal was readout
  with a single $\LAAPD$.}
\label{fig:perf:perf17}
\end{center}
\end{figure}

The line shape and the resulting energy resolution of the sub-array
are obtained by summing event-wise the responding modules. All
contributions above the pedestal were taken into account corresponding
to a minimum energy threshold of approximately
$1\,\MeV$. \Reffig{fig:perf:perf16} shows the excellent resolution of
$\sigma$/E=9.5$\percent$ for the lowest photon energy measured in such
a configuration. The corresponding distributions for the highest
investigated incident energy are reproduced in
\Reffig{fig:perf:perf17}.

\begin{figure}
\begin{center}
\includegraphics[width=\swidth]{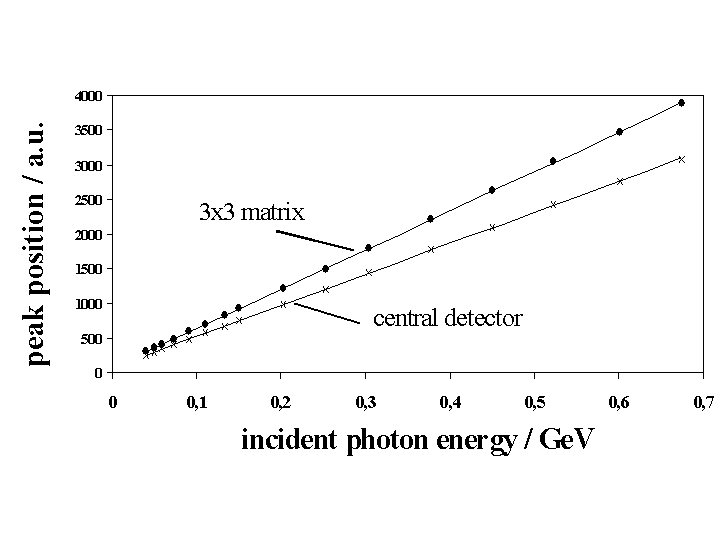}
\caption[The linearity of the response of a 3$\times$3 matrix.]  {The
  linearity of the response of a 3$\times$3 matrix over the entire
  range of photon energies measured at a temperature of
  T=0$\degC$. The most probable energy deposition into the central
  module is shown separately.}
\label{fig:perf:perf18}
\end{center}
\end{figure}

\begin{figure}
\begin{center}
\includegraphics[width=\swidth]{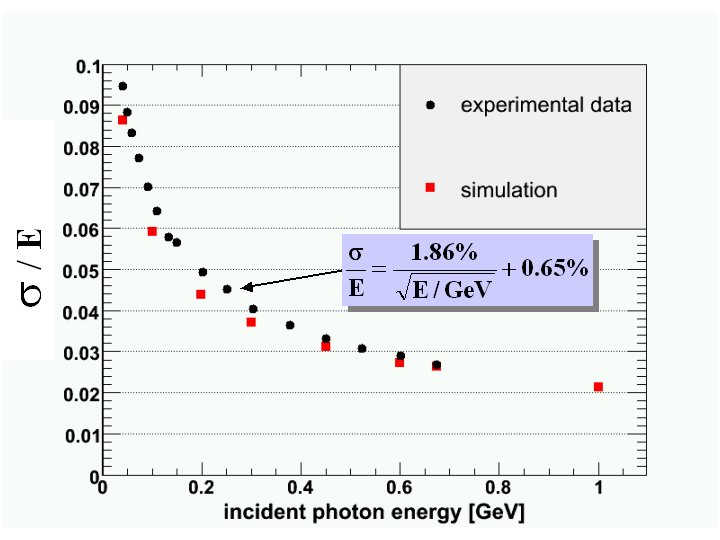}
\caption[Experimental energy resolution as a function of incident
photon energy.]  {Experimental energy resolution as a function of
  incident photon energy measured for a 3$\times$3 \PWOII matrix
  readout with a \LAAPDs at a temperature of T=0$\degC$. The result of
  a GEANT4 simulation is shown for comparison.}
\label{fig:perf:perf19}
\end{center}
\end{figure}

The response of the matrix over the entire energy range follows a
linear relation, both for the central module as well as the whole
detector block. The correlation is shown in
\Reffig{fig:perf:perf18}. The overall performance is summarized in
\Reffig{fig:perf:perf19}. From the experimental data one can
extrapolate an energy resolution of 2.5$\percent$ at $1\,\GeV$ photon
energy in spite of the fact, that the lateral dimensions of the
detector array are not sufficient at that energy. The figure contains
for comparison the expected resolution based on the Monte-Carlo
simulation using the code GEANT4 including the concept of digitization 
as used in the overall simulation of the \PANDA detector. It confirms that the
simulations discussed in chapter 9 are close to the present experimental performance.  

The results show the excellent performance even at operating
conditions well above the envisaged temperature of T=-25$\degC$ and
with a limited collection efficiency of the \LAAPD compared to the
final coverage with two sensors, which will lead to an improvement in
resolution at least in the order of $\geq$1.4 and will be significant
at lower photon energies, when the resolution is dominated by the
photon statistics. However, the final gain factor will strongly depend
on the read-out concept and noise imposed by the front-end electronics
\cite{bib:emc:perf11}.

\subsection{Position Resolution}
\label{chap:perform:position}

The point of impact of the impinging photon can be reconstructed from
the center of gravity of the electromagnetic shower distributed over
the cluster of responding detector modules. Such an analysis can be
performed in a straight forward fashion using data from previously
described measurements, independent of the used readout of the
scintillation response. However, reconstructions and optimizations
adapted to the crystal geometry of the \TSEMC have been performed so
far only based on Monte-Carlo simulations.

Experimental studies in the past with detector arrays of identical
geometry but inferior crystal quality have delivered a position
resolution of $\sigma$$<$$2\,\mm$, slightly depending on the
reconstruction algorithm but typical for crystal cross sections close
to the Moli\`ere radius.

\subsection{Particle Identification}
\label{chap:perform:pid}

The capabilities of particle identification, such as the different
cluster size of detector modules responding to hadron or
photon/electron induced events, have not been investigated so far
experimentally but will be a major task using the PROTO60 device
within the next months. In previous experiments with $2.1\,\gev$
proton induced reactions at \INST{COSY}
\cite{bib:emc:perf1,bib:emc:perf12} a clean separation of hadrons and
photons was possible even in a sub-array of 5$\times$5 units.  More
specific aspects are studied in simulations discussed in the next
section on simulations.

The high quality of the timing information of the calorimeter elements
was documented in several experiments and will be further exploited
strongly depending on the final concept of the front-end electronics
and final digitization. Exploratory measurements were previously
discussed in the context of the electronics, see
\Refsec{sec:elo:Timing}. Time resolutions even below 1ns will allow
selective and precise correlations between different inner detector
components, in particular the tracking as well as the particle
identification in the barrel DIRC. In addition, the calorimeter allows
the extension of muon detection and identification down to low
energies. Finally, the expected fast response can become a selective
tool to discard random coincidences and events initiated by background
such as due to residual gases in the beamline near the target region
or secondary interactions at beamline components up- and downstream
from the central target.

\subsection{Construction and Basic Performance of the Barrel Prototype Comprising 60 Modules}
\label{chap:perform:proto60}

Several prototypes have been constructed in order to validate the
concepts of performance with respect to physics, mechanics, thermics
and integration. The last one and the most complete is the real-size
PROTO60, composed of 60 crystals of \PWOII. The setup is
representative for the center part of a slice, assembled of type 6
crystals in their realistic positions. The design principle is similar
to the final one except that the crystals, \LAAPDs and preamplifiers
are easily accessible for reasons of maintenance or controls as needed
for a prototype. The crystals are resting on a thick support plate
made of aluminum and the inserts into the carbon fiber alveoles are
not permanently glued. The insulation box has a removable back cover.

\begin{figure}
\begin{center}
\includegraphics[width=\swidth]{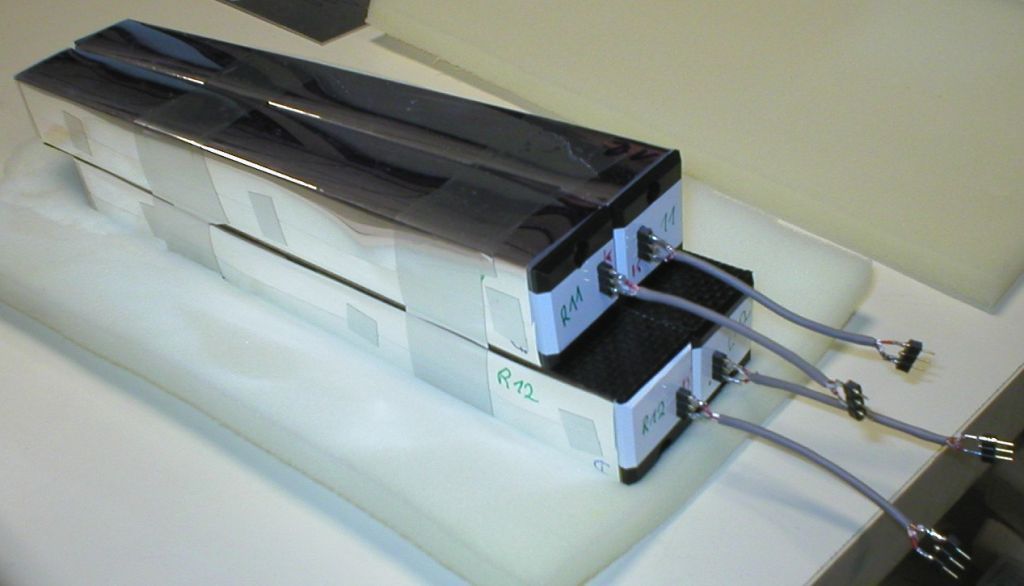}
\caption[Pack of 4 crystals wrapped with reflector.]
{Pack of 4 crystals wrapped with reflector and coupled to a \LAAPD as part of the PROTO60.}
\label{fig:perf:perf20}
\end{center}
\end{figure}

\begin{figure}
\begin{center}
\includegraphics[width=\swidth]{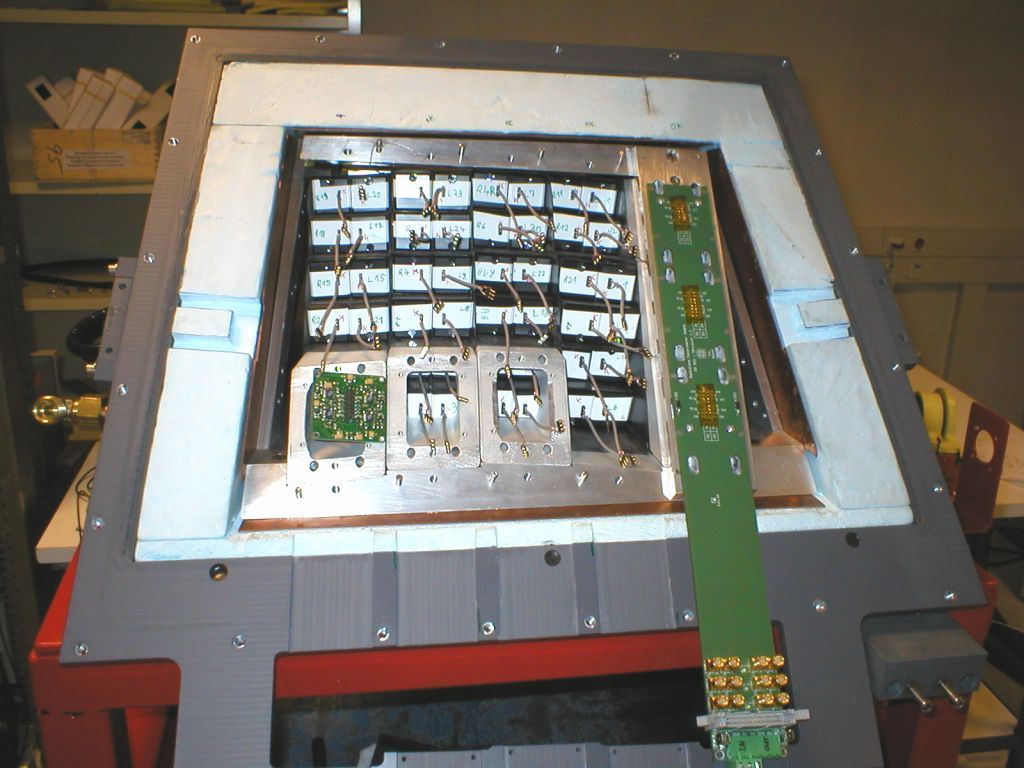}
\caption[Back view of PROTO60.]
{Back view of PROTO60 before installing the fiber system for monitoring.}
\label{fig:perf:perf21}
\end{center}
\end{figure}

\begin{figure}
\begin{center}
\includegraphics[width=\swidth]{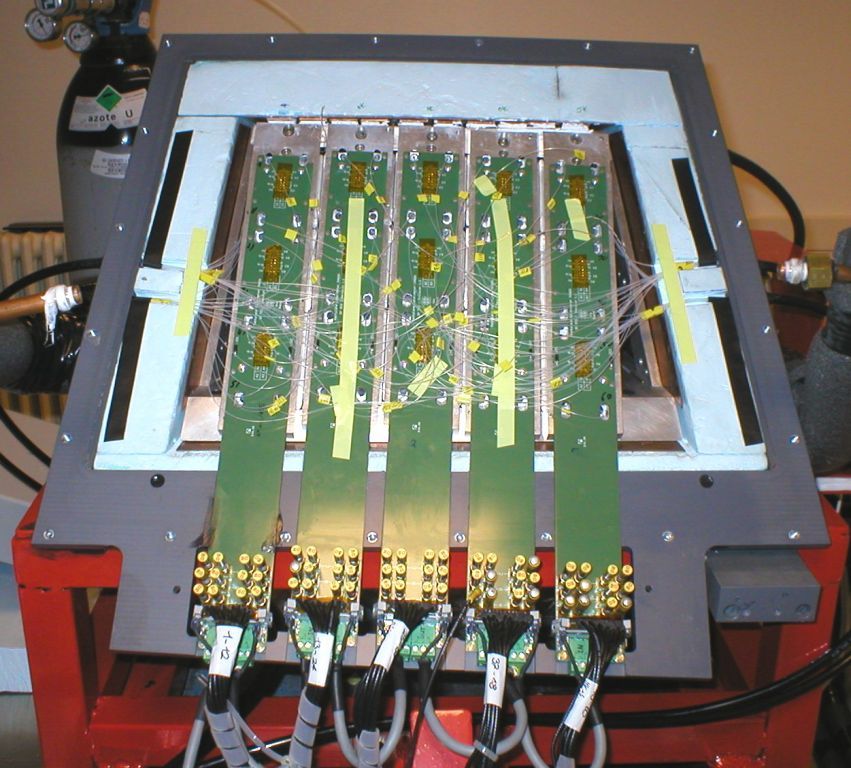}
\caption[Back view of the fully assembled PROTO60.]
{Back view of the fully assembled PROTO60 including the optical fibers.}
\label{fig:perf:perf22}
\end{center}
\end{figure}

\begin{figure}
\begin{center}
\includegraphics[width=\swidth]{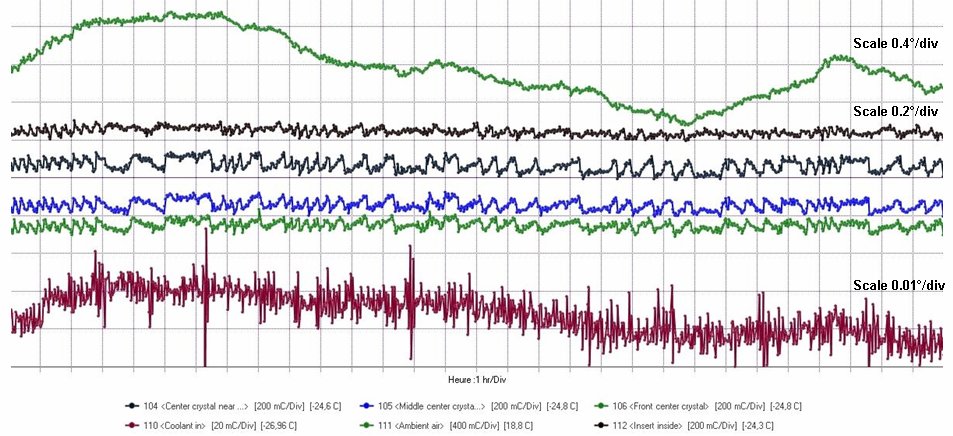}
\caption[Temperature stability of the PROTO60.]  {Temperature
  stability of the PROTO60. The temperature variation of the inlet
  coolant, the crystals and the ambient air stay within
  $\pm$0.01$\degC$, $\pm$0.05$\degC$, and $\pm$0.6$\degC$,
  respectively.}
\label{fig:perf:perf23}
\end{center}
\end{figure}

\begin{figure}
\begin{center}
\includegraphics[width=\swidth]{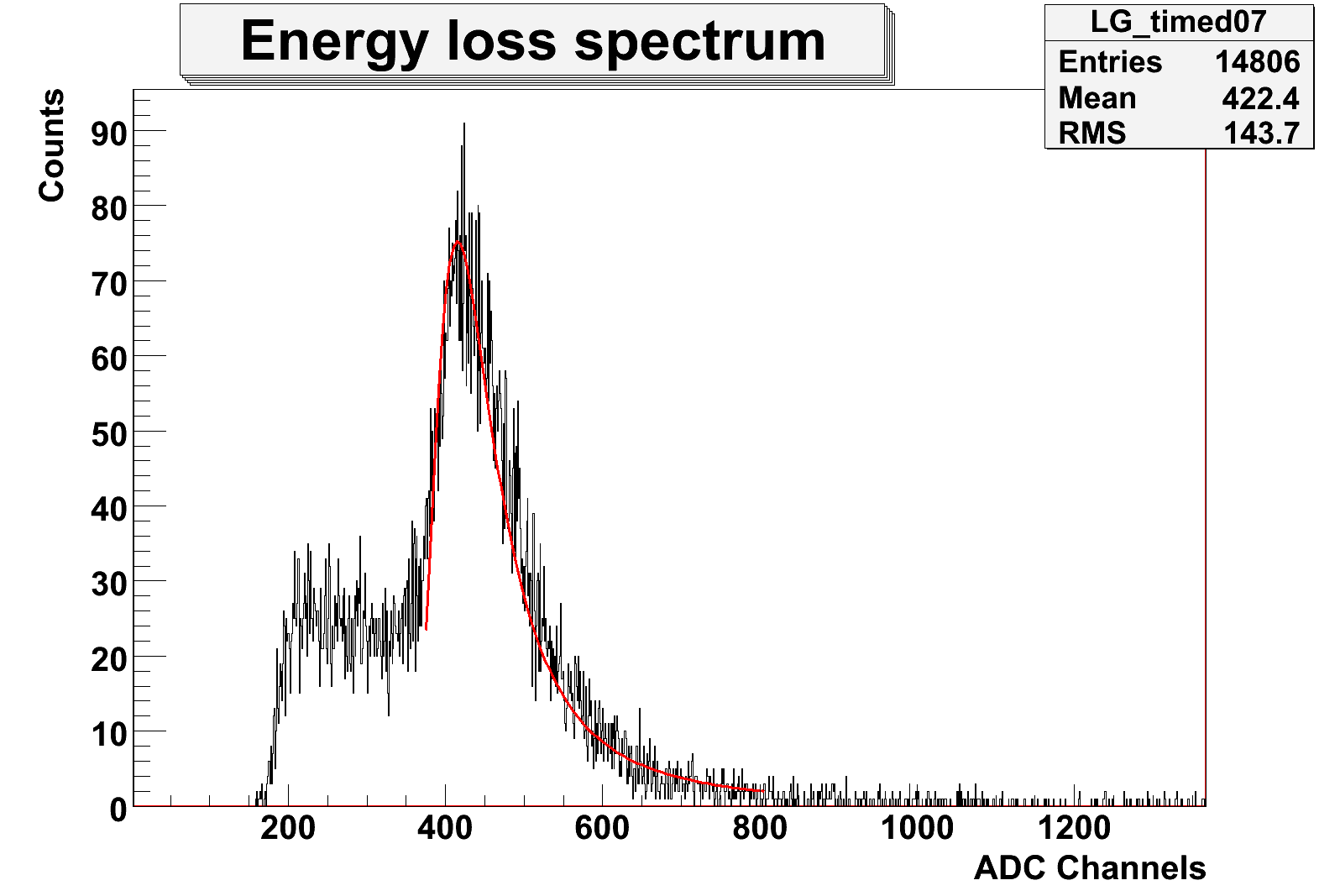}
\caption[Measured response of a single module of the PROTO60.]
{Measured response of a single module of the PROTO60 to the energy
  deposition of penetrating minimum ionizing cosmic muons.}
\label{fig:perf:perf24}
\end{center}
\end{figure}

\Reffig{fig:perf:perf20} shows a group of 2 pairs of left and right
crystals, wrapped with the ESR reflector. The large \LAAPDs (hidden
behind their black capsules) are coupled with optical grease for easy
removal and connected to the preamplifiers by twisted and shielded
wires. They are arranged in groups of 4 precisely positioned with a
thin mylar tape before being inserted into the alveoles.  The
\Reffig{fig:perf:perf21} shows a back view with all alveoles being
already filled, with some inserts and one printed circuit board in
place for signal readout. This multilayer board connects the
preamplifiers and transfers the signals to the data acquisition across
the thermal shield.  The next \Reffig{fig:perf:perf22} shows the back
view before mounting the cover plate. Besides the printed boards one
can notice the distribution of the optical fibers, which inject light
pulses from a LED-pulser individually into each crystal. The fibers
are just contacting the rear surface without grease for better optical
coupling.  The light is provided by a pulsed LED with a variable
frequency up to several kHz and distributed and homogenized via an
optical system to serve the 60 optical fibers made of fused silica.

The insulation is achieved by thermal screens composed of copper
plates with serpentines, confined in an insulated plastic box filled
with dry nitrogen to avoid ice formation. In front of the calorimeter,
a prototype of the vacuum panel with coolant carbon channels is
successfully running with a temperature variation of up to 3$\degC$
between both crystal end faces. It means that the front side is well
above the dew point and avoids any risk of moisture close to the
\Dirc. The set of crystals is operating at a temperature of
T=-25$\degC$ stabilized to $\pm$0.1$\degC$ with a $\emph{Julabo}$
chiller recirculating $\emph{Syltherm}$ coolant at a flow rate of 3.5
liters/minute. A set of 13 thermocouples measures the temperatures at
different locations and is recorded with an $\emph{Agilent}$ data
acquisition. The \Reffig{fig:perf:perf23} shows one cooling cycle over
a period of 30 hours. The temperature variation of the inlet coolant,
selected crystals and the ambient air inside the prototype stay within
$\pm$0.01$\degC$, $\pm$0.05$\degC$, and $\pm$0.6$\degC$, respectively,
and verify the required thermal stability.

In order to prepare the in-beam tests, the functionality of the
calorimeter prototype has been investigated using the response to
cosmic muons. From the known high-voltage settings of the \LAAPDs for
a gain of 50, given by the manufacturer for a temperature at
T=+20$\degC$, the corresponding lower bias voltages at the final
working temperature of T=-25$\degC$ have been adjusted exploiting the
light pulser as a reference. The fine-tuning of the gain including the
amplification settings in the spectroscopy amplifiers have been
performed using the energy deposition of cosmic muons into the
crystals. The individual spectra can be used as a relative calibration
of the calorimeter elements for later reconstruction of an
electromagnetic shower. The linearity of the assembled module
including the electronics chain is checked with the LED light, which
can be diminished in intensity by a set of gray-filters. Presently the
PROTO60 is ready for beam tests in the summer of 2008 including a DAQ
system based on commercial VME-modules for event-wise recording of the
responding detectors. \Reffig{fig:perf:perf24} shows for a single
module a typical spectrum initiated by penetrating muons. The
prominent peak corresponds to the most probable path-length of minimum
ionizing muons, equivalent to an energy deposition of $24\,\MeV$.  The
prototype has been running for weeks under stable conditions.

\section{Expected performance of the EMC}

Summarizing, the experimental data as well as the detailed design
concepts and simulations, respectively, show that the ambitious
physics program can be fully explored with respect to the measurement
of the electromagnetic probes, such as photons, electrons/positrons or
the reconstruction of the invariant mass of neutral mesons.

The calorimeter design in the target region is in full accordance with
the constraints imposed by a fixed target experiment with the strong
focusing of the momenta in forward direction. It is composed of a
barrel and two endcaps comprising in total 15,552 tapered crystal
modules of 11 + 2 different shapes. A nearly full coverage of $\sim$
99\percent solid angle in the center-of-mass system is guaranteed in
combination with the forward electromagnetic calorimeter, which is not
part of the present TDR but is based on well known technology of a
sampling calorimeter and is located downstream beyond the dipole
magnet. The granularity is adapted to the tolerable maximum count rate
of the individual modules and the optimum shower distribution for
energy and position reconstruction by minimizing energy losses due to
dead material. The front faces of the crystal elements cover a nearly
identical solid angle and have lateral dimensions close to the
Moli\`ere radius. All crystals are pointing off the target center.

\PWO has been chosen as the most appropriate scintillator material,
which allows a compact design within the solenoidal magnet, a fast
response and high-rate capability due to the short decay time of the
dominating scintillation process concentrated on the blue
emission. All crystals have a common length of $200\,\mm$
corresponding to 22 radiation lengths, which allows optimum shower
containment up to $15\,\GeV$ photon energy and limits the nuclear
counter effect in the subsequent photo sensor to a tolerable
level. The luminescence yield, which was sufficient for the
high-energy regime at LHC, has been significantly improved by nearly a
factor of two due to the optimization of the production technology,
such as raw material selection, pre-fabrication processing and
optimized co-doping with a minimum concentration of La- and Y-ions. To
cope with the need for low energy photon detection in the tens of MeV
range, the calorimeter will be operated at T=-25$\degC$ in order to
gain another factor of 4 due to the lower thermal quenching of the
scintillation light.

The radiation level will be well below typical LHC values by at
least two orders of magnitude at the most forward direction and even
further below at larger angles. However, the interplay between
radiation damage and recovery processes is almost negligible at the
envisaged temperatures
\cite{bib:emc:perf13,bib:emc:perf15}. Therefore, the change of the
optical transparency of the crystals is degrading asymptotically to a
value, which is given by the absolute concentration of defect centers
in the crystal. As a consequence, the present quality of PWO,
characterized as PWO-II, provides an induced absorption coefficient,
which is almost a factor two below the quality limits of CMS/ECAL. The
light loss, asymptotically reached in a typical running period of 6-8
months, stays below 30\percent, which reduces the optimum gain factor
to a value of 3 compared to an operation at T=+25$\degC$. It should be
noted, that these effects have to be considered only in the very
forward region of the \FWEMC. When the calorimeter is brought up to room temperature
after a typical experimental period of 6-8 months, there will be a full recovery of the optical transmission
within a few weeks. In Y-, La-doped crystals shorter the dominant relaxation times are on the level 20 to 25 hours.
There are only very small contributions due to tunneling processes, with recovery constants near 75 hours.

The operation at low temperatures imposes a technological challenge on
the mechanical design, the cooling concept and thermal insulation
under the constraints of a minimum material budget of dead material in
particular in front and in between substructures. Detailed simulations
and prototyping have confirmed the concept and high accuracy has been
achieved in temperature stabilization taking into account also
realistic scenarios of the power consumption of the front-end
electronics.

The direct readout of the scintillation crystals has been
significantly improved by the development of a large size avalanche
photodiode, similar to the intrinsic structure of the CMS/ECAL
version, in collaboration with Hamamatsu. In total, two \LAAPD{}s of
$100\,\mm^{2}$ active surface each will be used to convert the
scintillation light with high efficiency, since almost 30\percent of
the crystal endface are covered. Irradiation studies with
electromagnetic and hadronic probes have confirmed the radiation
hardness. The operation at low temperatures is even in favor and leads
to a significantly slower increase of the dark current due to
radiation damage. In order to cope with a substantially higher count
rate of the individual crystal modules in the \FWEMC, vacuum photo
triodes are foreseen as photo sensors for the forward
endcap. Besides the design used at CMS, the development and testing of
prototypes continues in parallel at both the manufacturers Hamamatsu
and Photonis, which focus in addition on the implementation of
photocathodes with higher quantum efficiency compared to standard
bialkali versions.

The concept of the readout electronics has been elaborated. As the most
sensitive element, a prototype of a custom designed ASIC implementing
preamplification and shaping stages has been successfully brought into
operation and will provide a large dynamic range of 12,000 with a
typical noise level corresponding to $\sim1\,\MeV$. Again, the
operation at low temperatures reduces further any noise
contributions. Presently most of the experimental tests were based on
low-noise preamplifiers assembled of discrete components. A similar
version might be used for the VPTs after some modifications.

The expected overall performance of the calorimeter is presently based
on a series of test experiments with detector arrays comprising up to
60 full size crystals. The basic quality parameters of \PWOII are
confirmed in many tests including a first pre-production run of 710
crystals which are sufficient in shape and quantity to complete one
out of the 16 slices of the barrel part. The results establish the
readiness for mass production, at least at the manufacturer plant BTCP
in Russia. Parallel developments of full size samples from SICCAS,
China, are expected soon for comparison.

The experimental tests, mostly performed at the tagged photon facility
at MAMI at Mainz and MAXLab at LUND have delivered values for energy
and time resolutions, which should be considered as safety
limits. They have been performed at various temperatures and were
read out via photomultiplier tubes or just with a single \LAAPD
\cite{bib:emc:perf14}. Therefore, only a significantly lower
percentage of the scintillation light was converted into an analogue
signal. From these results one can extrapolate and confirm at an
operating temperature of T=-25$\degC$ an energy resolution, which will
be well below 2.5\percent at $1\,\GeV$ photon energy.  At the lowest
investigated shower energy of $20\,\MeV$ a resolution of 13\percent
has even been obtained, which reflects the excellent statistical term
due to the overall improved and enhanced light yield. Relevant for the
efficient detection and reconstruction of multi-photon events, an
effective energy threshold of $10\,\MeV$ can be considered for the
whole calorimeter as a starting value for cluster identification
along with a single crystal threshold of $3\,\MeV$ for adjacent channels
which will enable us to disentangle physics channels with extremely
low cross section.

The fast timing performance will allow a time resolution close to
$1\,\ns$ even based on the readout with \LAAPD{}s and provide a very
efficient tool to correlate the calorimeter with other devices for
tracking and particle identification or an efficient reduction of
background caused by residual gas in the beam pipe or interactions at
accelerator components acting as secondary targets. In a series of
tests the detection of hadronic probes such as low energy protons or
minimum ionizing particles has been proven and will provide
complementary information to the particle tracking or even
identification of muons based on the energy deposition or different
cluster multiplicity of trespassing hadrons.

The overall performance of the calorimeter will be carefully
controlled by injecting light from LED-sources distributed via optical
fibers via the rear face of the crystal. The change of optical
transmission will be monitored near the emission wavelength at
$420\,\nm$, at a dominant defect center due to Molybdenum at
$530\,\nm$, and far in the red spectral region to control
independently the readout chain including the photo sensor. Radiation
damage is not expected to appear in the red. A first prototype is
already implemented into the PROTO60 array and operating.

Based on the ongoing developments of \PWOII crystals and \LAAPDs, a
detailed program has been elaborated for quality assurance of the
crystals and screening of the photo sensors. The crystal studies
include the preparation of various facilities for irradiation studies
at Protvino and Giessen. In addition, the quality control is planned
to be performed at CERN, taking advantage of the existing
infrastructure and experience developed for CMS. One of the two ACCOS
machines, semi-automatic robots, is presently getting modified for the
different specification limits and geometrical dimensions of the
\Panda crystals.

Similar infrastructure has been developed for the final certification
of the \LAAPD{}s in rectangular shape. The intrinsic layout is
identical to the quadratic version, which has been tested in great
detail. The new geometry allows to fit two sensors on each crystal
irrespective of its individual shape.

The general layout of the mechanical structure is completed including
first estimates of the integration into the \Panda detector. The
concepts for signal- and HV-cables, cooling, slow control, monitoring
as well as the stepwise assembly are worked out to guarantee that the
crystal geometries could be finalized. Prototypes of the individual
crystal containers, based on carbon fiber alveoles, have been
fabricated and tested and are already implemented in the PROTO60
device.

%
%
\newpage
\bibliographystyle{panda_tdr_lit}
\bibliography{./lit_emc}
%

%
\cleardoublepage
\chapter{Organisation}
\label{sec:org}
%
%
\section{Quality Control and Assembly}

\subsection{Production Logistics}
The realization of the two major components of the \TSEMC, the barrel
part and the two endcaps of different size, is split into several
parts, which also require significantly different logistics. The
mechanical parts, such as the basic detector modules including their
housing as well as the support structure, the thermal insulation and
cooling as well as the container for the readout electronics, are
presently under development and first design studies have been
realized in prototype development. The detector components comprise
basically the nearly 16,000 \PWOII crystals and the two types of photo
sensors, \LAAPD{}s and vacuum triodes. Both require a sophisticated
program of production by manufacturers, the quality control and
assurance and screening and pre-radiation in particular of the
avalanche photo diodes. The concept and the procedures of the latter
components are outlined in great detail in the chapter about photo
sensors.

\begin{figure}[htb]
\begin{center}
\includegraphics[width=\swidth]{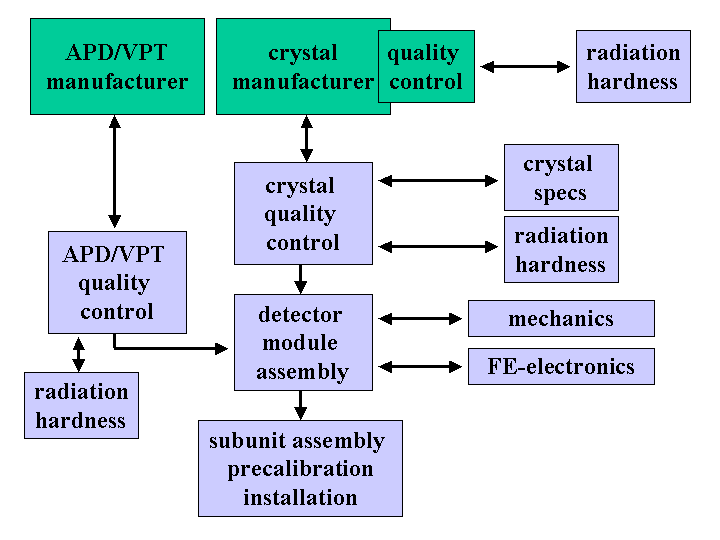}
\caption{Schematic layout for the realization of the \TSEMC of \PANDA.}
\label{fig:org:fig1}
\end{center}
\end{figure}

The major steps in realization are defined on one hand by the optimum
logistics but also by the capabilities and interests of the
participating institutions as well as the supporting funding
profile. The present layout assumes the sufficient funding and a
timing profile to be available when necessary. The logistics is based
on a practicable concept to achieve an optimum realization and to
guarantee the envisaged performance of the calorimeter and is
schematically illustrated in \Reffig{fig:org:fig1}.

\subsection{The \PWOII Crystals}
Presently, the production of high-quality \PWO crystals delivering the
specifications of \PWOII can be expected to be performed only by the
two manufacturers BTCP and SICCAS, respectively, which have been
successfully involved in the realization of the CMS EM
calorimeter. There is foreseen to have prepared at both companies an
infrastructure to measure the specifications primarily to have a fast
response to instabilities of the production line and to deliver
crystals according to the very selective specification
parameters. There will be a major center for the crystal inspection
and assembly of the detector modules. As an option, it could be
located at Giessen, which is in addition advantageous due to the very
close distance to FAIR. Already confirmed by an official offer of the
CMS collaboration, the quality tests will be performed at CERN using
the semi-automatic ACCOS machine. It guarantees a well accepted and
adapted procedure and allows to exploit more than 10 years of
experience. Preparations have been initiated already, to adapt the
installations to the geometry and more selective specifications of the
\PANDA crystals. Afterwards, the crystals will be shipped back to
control for each crystal its radiation hardness and the wavelength
dependence, which is very crucial for the operation at low
temperatures and has to be performed for a very high percentage of the
crystals.  These measurements could be handled at the irradiation
facility at Giessen, as characterized in chapter 4.

\subsection{Detector Module Assembly}
In a second line, to be performed at an appropriate regional center,
the crystals will be equipped with a reflector foil, photo and
temperature sensors, and the front-end electronics. All latter
components are delivered tested and screened from the responsible
laboratories at GSI (\LAAPD, front-end electronics) and Bochum/KVI
(\VPT), most probably.

\subsection{Final Assembly, Pre-calibration and Implementation into \Panda}
The final assembly of the barrel slices or the two endcaps will be
completed by the implementation of the optical fibers for the
monitoring system, the electronics for digitization, the cables for
signal transfer and low- and high voltage support, the sensors and
lines for slow control as well as the installations for cooling. This
task including the intermediate storage have to be performed at or at
least close to FAIR in a dedicated laboratory space. The basic
functionality of the assembled units will be tested using on one hand
the monitoring system to check the electronics chain including the
photo sensors. A first hint of the expected sensitivity of each
detector module is given by the measured light yield of the crystal,
the absolute value of optical transmission at the absorption edge and
the known gain of the photo sensor. These parameters will allow a
first step towards a pre-calibration with an accuracy on the order of
5$\percent$. On the other hand, a more sensitive relative and absolute
calibration of the calorimeter modules will be obtained by using
minimum ionizing cosmic muons. The submodules have to be mounted on a
mechanical support structure, which allows to rotate most of the
detector modules into vertical positions. Requiring anti-coincidence
with all neighboring modules selects the full path length of
$200\,\mm$ of the particle, which corresponds to an energy deposition
of $\sim$$240\,\MeV$. This procedure will deliver calibration
parameters with an intended accuracy of 2.5$\percent$, values which
base on the experience of CMS.  In order to obtain a more accurate
calibration, a cross check with the calibration steps explained above
and to determine finally the response function a substantial part of
the units will be exposed to a direct beam of energy marked photons,
which can be provided at the two tagging facilities at MAMI (Mainz)
and ELSA (Bonn), respectively, covering the complete energy range up
to $\sim$$3.2\,\GeV$. At both places experimental set-ups are under
consideration.

\subsection{Other Calorimeter Components}
Besides the basic calorimeter components, crystals, photo sensors and
pre-amplifier/shaper, all other parts including mechanical structure,
cooling, thermal insulation, electronics and DAQ are under
development. In case of the most critical tasks prototypes have been
designed and realized and confirm the proposed concept. The
responsibility for the final design and realization are distributed
among those groups within the collaboration, which have the
infrastructure and experience with large scale experimental
installations.

\subsection{Integration in PANDA}
The design of the \TSEMC is in full agreement to the overall layout of
the whole \PANDA detector. The components including service
connections and mounting spaces fit within the fiducical volume defined
by the magnet and the other detector components. It is foreseen, that
the fully assembled and tested submodules are stored in an
air-conditioned area with stabilized temperature and low humidity. The
sealed components should be kept in a nitrogen atmosphere.  The
necessary steps for mounting and integrating the calorimeter into
\PANDA have been described already in great detail in the chapter on
mechanics for both major parts and will be the final task performed by
the whole collaboration.

\section{Safety}
The design details and construction of the calorimeter including the
infrastructure for operation will be done according to the safety
requirements of FAIR and the European and German safety regulations.
Design aspects for which the CMS calorimeters are taken as examples
take into consideration the detailed guidelines given by CERN, which
are most appropriate to the case of large scientific installations.
\subsection{Mechanics}
The strength of the EMC support structures has been computed with
physical models in the course of the design process. Details of finite
element analysis are shown in \Refsec{sec:emc:mech:bar:slice_def}.
Each mechanical component will undergo a quality acceptance
examination including stress and loading tests for weight bearing
parts. Spare samples may also be tested up to the breaking point. A
detailed material map of the entire apparatus showing location and
abundance of all materials used in the construction will be
created. For structural components radiation resistance levels will be
taken into account in the selection process and quoted in the material
map. For the calorimeter crystals themselves precautions will be taken
to rule out exposure of persons to toxic material due to exhaustion in
case of fire or due to mechanical operations. In normal conditions
lead tungstate is stable, non-hygroscopic, and hardly soluble in
water, oils, acids and basis.  
\subsection{Electrical Equipment and Cooling}
All electrical equipment in \PANDA will comply to the legally required
safety code and concur to standards for large scientific installations
following guidelines worked out at CERN to ensure the protection of
all personnel working at or close to the components of the \PANDA
system.  Power supplies will have safe mountings independent of large
mechanical loads.  Hazardous voltage supplies and lines will be marked
visibly and protected from damage by near-by forces, like pulling or
squeezing. All supplies will be protected against overcurrent and
overvoltage and have appropriate safety circuits and fuses against
shorts. All cabling and optical fibre connections will be executed
with non-flammable halogen-free materials according to up-to-date
standards and will be dimensioned with proper safety margins to prevent 
overheating. A safe ground scheme will be employed throughout all
electrical installations of the experiment. Smoke detectors will be
mounted in all appropriate locations.

Lasers or high output LEDs will be employed in the monitoring system
and their light is distributed throughout the calorimeter systems.
For these devices all necessary precautions like safe housings, color
coded protection pipes, interlocks, proper warnings and instructions
as well as training of the personnel with access to these components
will be taken.

The operation of the \PWO EMC at temperature down to -25$\degC$
requires a powerful and stable cooling system for the crystals and
sensors. It consists of a system of cooling pipes, pumps and heat
exchangers and employs a liquid silicone polymer as coolant. This
material is highly inert and non-toxic. Its flow will be in the order
of 4 l/s for the full EMC. A control system is employed to detect any
malfunctioning of the cooling system and include interlocks for
electronics and high-voltage supplies. In the case of coolant loss or
in case of abnormal temperature gradients between coolant input and
output the control system enacts the appropriate safety procedures.
A highly redundant system of temperature sensors allows the continuous
monitoring of the effectiveness of the system.

\subsection{Radiation Aspects}
The \PWO crystals can be activated by low energy protons and neutrons
leading to low-energy radioactivity of the activated nuclei. However,
simulations based on the computer code MARS deliver in case of the
maximum luminosity of 2$\times$10$^{32}$/s and for an operation period
of 30 days a dose rate of $20\,\mu$Sv/h at the crystal surface of the
\FWEMC closest to the beam axis, where the rate is expected to be the
highest in the entire EMC by several orders of magnitude. The
estimated rate is one order of magnitude lower than in case of CMS at
LHC. Shielding, operation and maintenance will be planned according to
European and German safety regulations to ensure the proper protection
of all personnel.
\section{Schedule}
\begin{figure*}
\centering
\includegraphics[width=\dwidth]{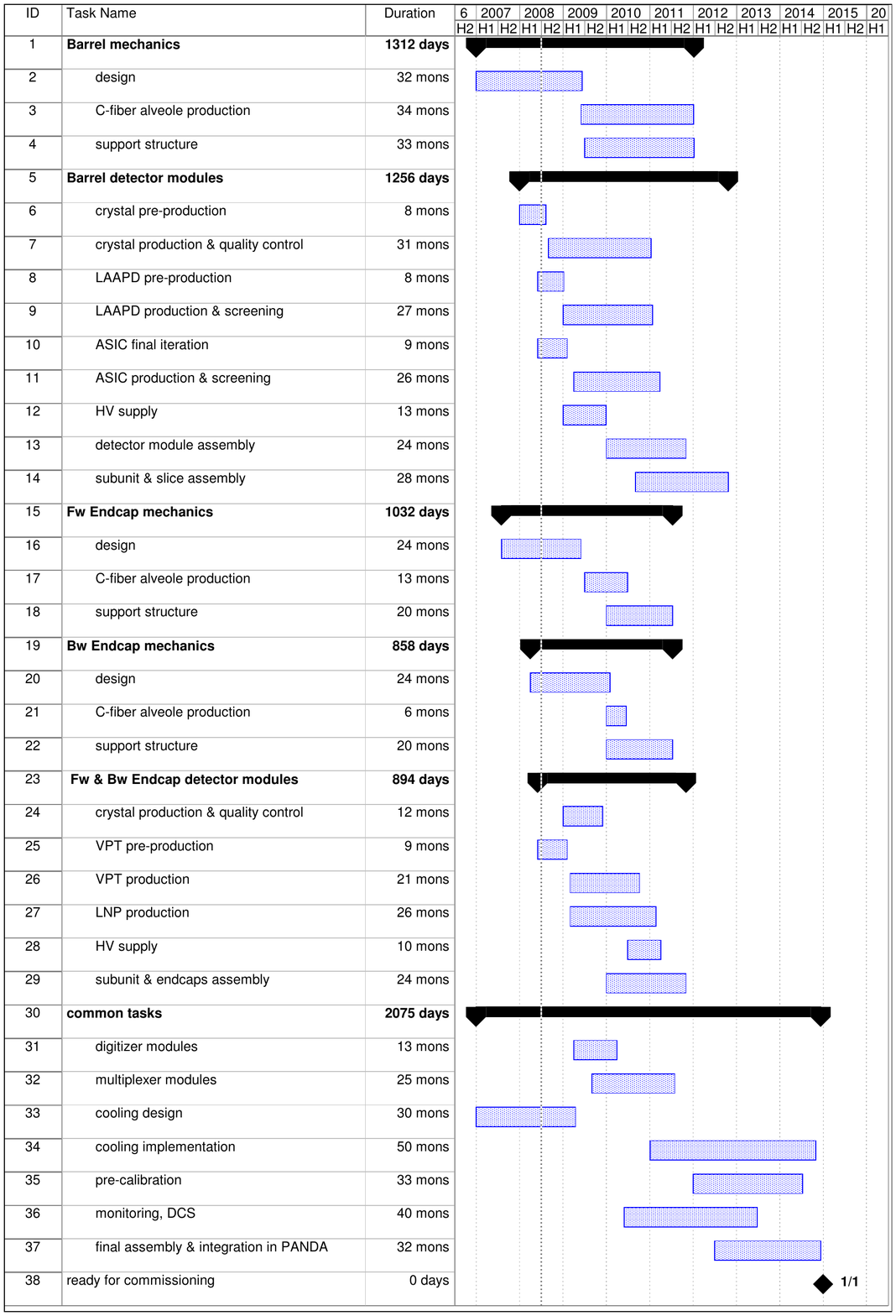}
\caption[Timeline for the realization of the \TSEMC.]{Timeline for the
  realization of the \TSEMC.}
\label{fig:prod:timeline}
\end{figure*}

\Reffig{fig:prod:timeline} shows in some detail the estimated
timelines including the main tasks for the design phase, prototyping,
production, pre-calibration and implementation of the \TSEMC into the
\PANDA detector. The schedule is based on the experience gathered
during the R$\&$D and prototyping phase, the close collaboration with
optional manufacturers and the fruitful exchange of experience in
particular the CMS/ECAL project at CERN. The timelines assume
sufficient funding, which should become available when necessary.

\begin{figure*}
\centering
\includegraphics[width=\dwidth]{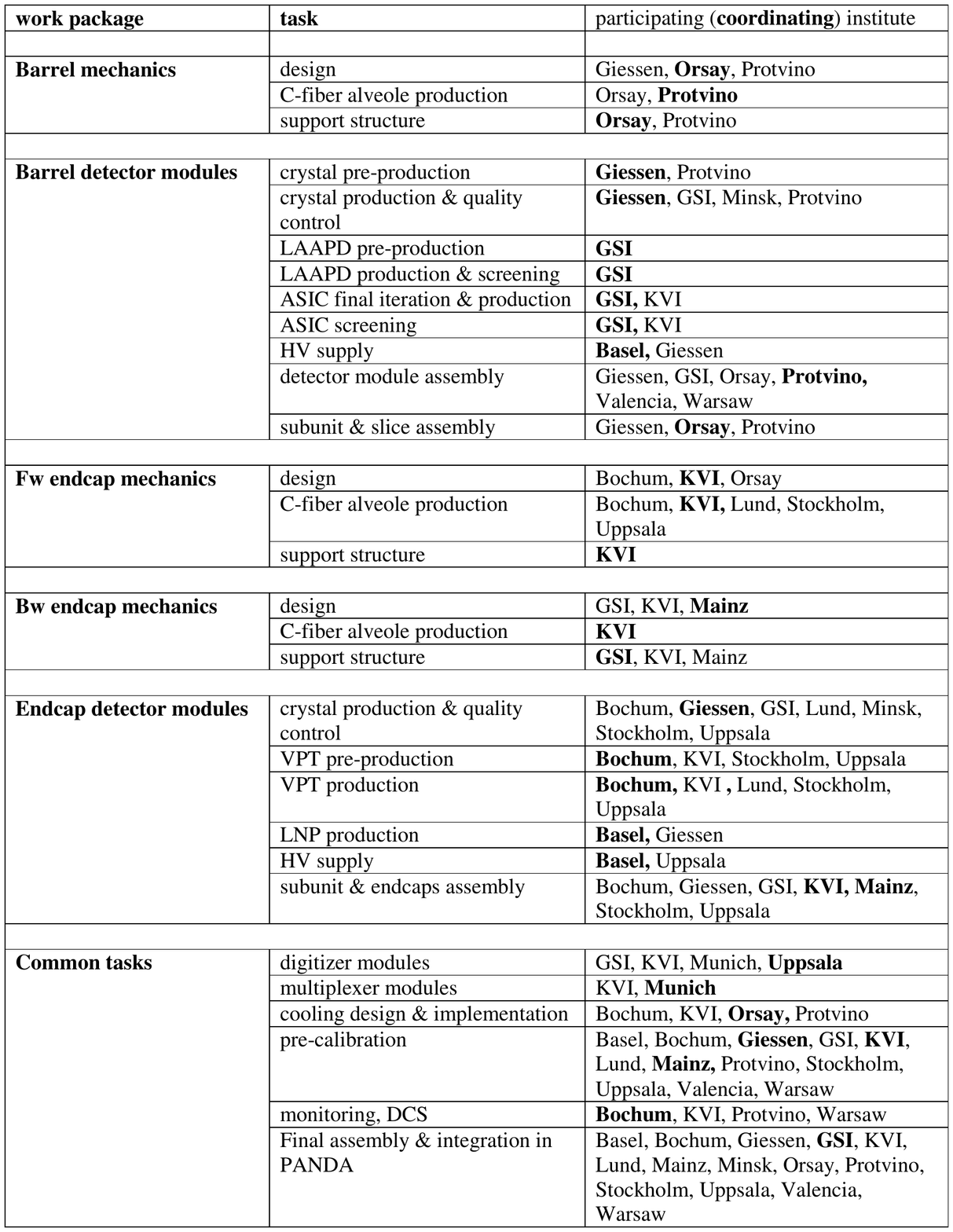}
\caption[Tasks and responsibilities for the realization of the \TSEMC.]{Tasks and responsibilities for the realization of the \TSEMC.}
\label{fig:prod:task}
\end{figure*}

After an initial phase of research and development of the \TSEMC
concentrating on the basic performance of the essential components,
i.e. the optimal scintillator material and the photo sensors of
choice, the collaboration has completed the design phase. The
integration of the mechanical components into the \Panda detector has
been clarified, prototype crystals, sensors, discrete and ASIC
preamplifiers, and digitization modules are at hand, and a prototype
detector has been set up and is ready for beam experiments.  A number
of institutions, who have gained specific expertise in past and
ongoing research programme at large-scale experiments such as
\INST{BaBar}, \INST{TAPS}/\INST{Crystal Ball}/\INST{Crystal Barrel} at
several accelerator facilities, are engaged in the remaining tasks
leading to the completion of the \TSEMC. The responsibilities for the
various work packages and tasks are listed in \Reffig{fig:prod:task}
and show the participating institutions. The coordination of the work
packages is performed by those groups marked in bold-face. However,
the responsibility is not automatically linked to the availability of
future funding.

Since the table documents as well the capabilities of the
collaboration members, the different institutions are shortly
described and characterized by their expertise gathered in previous or
ongoing research activities.

\begin{itemize}
\item \textbf{Basel} (B. Krusche, M. Steinacher): The group is
  conducting experiments at tagged photon facilities and has expertise
  in analogue electronics.
\item \textbf{Bochum}(F.-H. Heinsius, U. Wiedner): The Bochum group is
  engaged in the design and operation of calorimeters for BaBar
  (SLAC), Crystal Barrel (CERN and Bonn/ELSA) and the WASA detector
  (Uppsala and J\"ulich).
\item \textbf{Giessen} (V. Metag, R. Novotny): The Giessen group has a
  longstanding experience in crystal calorimetry. It has been the
  leading institution in building the TAPS detector including
  electronics. The group has played a leading role in the development
  of the \PWOII.
\item \textbf{GSI} (B. Lewandowski, F. Mass, K. Peters, A. Wilms): The GSI group has long
  experience in crystal calorimetry (BaBar (SLAC), Crystal Barrel (CERN), Mami A4 (Mainz))
  and ASIC design.
  Particular fields of expertise include design and construction of sensors
  and electronics as well as calibration, monitoring and front-end software.
  The APD-LAB branch is a renowned R\&{}D center for LAAPDs
  channeling the developments for PANDA and R3B. In addition technical and
  infrastructure coordination for PANDA is performed by GSI.
\item \textbf{KVI} (H. L\"{o}hner, J. Messchendorp): The Groningen
  group participated in the construction and exploitation of the TAPS
  calorimeter. They have major experience in applications of
  scintillation detectors, photo sensors, light pulser systems for
  monitoring crystals, developing FPGA electronics and DSP networks.
\item \textbf{Lund} (B. Schr\"{o}der): The Lund group performs
  research at the MAX-Laboratory with monoenergetic photons in the 15
  to 200 MeV range with an energy resolution of $250\,\keV$ at best.
\item \textbf{Mainz} (F. Maas): The group has a long standing
  experience in fast EM calorimetry, trigger systems and operation of
  a facility for energy marked photons up to $1.5\,\GeV$ energy.
\item \textbf{Minsk} (M. Korzhik, O. Missevitch): The group has long
  year expertise in solid state physics of scintillation materials and
  has been the coordinator for PWO production for CMS at CERN.
\item \textbf{M\"{u}nchen} (I. Konorov, S. Paul): The group has
  expertise in design and development of digital electronics for large
  scale experiments (COMPASS) and has developed ADC systems for physics
  and medical applications.
\item \textbf{Orsay} (T. Hennino, P. Rosier): Long year experience in
  mechanical and electronics engineering and construction of large
  scale detector systems.
\item \textbf{Protvino} (V. A. Kachanov, A. N. Vasiliev): The group
  has been strongly involved in the design and construction of the
  endcap of CMS/ECAL and R\&D for the proposed BTeV experiment.
\item \textbf{Stockholm} (P. E. Tegner): The Stockholm University
  group has a major experience in operating detectors in high magnetic
  fields at the former CELSIUS machine in Uppsala.
\item \textbf{Uppsala} (P. Marciniewski, T. Johansson): The Uppsala
  group has been involved in experiments at CERN, Uppsala and
  J\"ulich. Their experience covers many aspects of calorimeters
  (WASA), electronic signal digitization and readout systems.
\item \textbf{Valencia} (J. Diaz): The group is involved in
  scintillator systems of large scale detectors such as TAPS and
  HADES.
\item \textbf{Warsaw} (B. Zwieglinski): The SINS Warsaw group has
  experience in response studies of scintillators exploiting proton
  accelerators.
\end{itemize}

%
%
%

%
%
\cleardoublepage
\onecolumn
%
\begin{center}
\vspace*{2cm}
{\Large\bf Acknowledgments}
\addcontentsline{toc}{chapter}{Acknowledgments}
\vskip 2cm
\begin{minipage}[t]{8cm}
\sloppy\large
We acknowledge financial support from
the Bundesministerium f\"ur Bildung und Forschung (bmbf),
the Deutsche Forschungsgemeinschaft (DFG),
the University of Groningen, Netherlands,
the Gesellschaft f\"ur Schwerionenforschung mbH (GSI), Darmstadt,
the Helmholtz-Gemeinschaft Deutscher Forschungszentren (HGF),
the Schweizerischer Nationalfonds zur F\"orderung der wissenschaftlichen
Forschung (SNF),
the Russian funding agency ``State Corporation for Atomic Energy Rosatom'',
the CNRS/IN2P3 and the Universit\'e Paris-sud,
the British funding agency ``Science and Technology Facilities
Council'' (STFC),
the Instituto Nazionale di Fisica Nucleare (INFN),
the Swedish Research Council,
the Polish Ministry of Science and Higher Education,
the European Community FP6 FAIR Design Study: DIRACsecondary-Beams,
contract number 515873,
the European Community FP6 Integrated Infrastructure Initiative:
HadronPhysics, contract number RII3-CT-2004-506078,
the INTAS, 
and the Deutscher Akademischer Austauschdienst (DAAD).\par
We also would like to thank the CMS Collaboration at CERN for their support.
\end{minipage}
\end{center}
\vfill
%
%

%
\cleardoublepage
\twocolumn
%
\addcontentsline{toc}{chapter}{List of Acronyms}
\begin{acronym}
\acro{ADC}{Analog to Digital Converter}
\acro{ALICE}{A Large Ion Collider Experiment at CERN LHC}
\acro{ALICE/PHOS}{ALICE Photon Spectrometer}
\acro{APD}{Avalanche Photo Diode}
\acro{APFEL}{ASIC for Panda Front End ELectronics}
\acro{ASIC}{Application Specific Integrated Circuit}
\acro{AWG}{Arbitrary waveform generator}
\acro{BGO}{Bismuth Germanate}
\acro{BTCP}{Bogoroditsk Techno-Chemical Plant}
\acro{C}{Capacitance}
\acro{CAD}{Computer Aided Design}
\acro{CERN}{Conseil European pour la Recherche Nucleaire}
\acro{CLAS}{CEBAF Large Acceptance Spectrometer}
\acro{CMOS}{Complementary Metal Oxide Semiconductor}
\acro{CMS}{Compact Muon Solenoid}
\acro{CMS/ECAL}{CMS electromagnetic calorimeter}
\acro{COMPASS}{Common Muon Proton Apparatus for Structure and Spectroscopy}
\acro{COSY}{Cooler Synchrotron}
\acro{DAQ}{Data Acquisition}
\acro{DCS}{Detector Control System}
\acro{DIRC}{Detector for Internally Reflected Cherenkov Light}
\acro{DPM}{Dual Parton Model}
\acro{DSP}{Digital Signal Processor}
\acro{DVCS}{Deeply Virtual Compton Scattering}
\acro{EMC}{Electromagnetic Calorimeter}
\acro{ENC}{Equivalent Noise Charge}
\acro{EPR}{Electron Paramagnetic Resonance}
\acro{FADC}{Flash ADC}
\acro{FAIR}{Facility for Antiproton and Ion Research}
\acro{FINUDA}{Fisica Nucleare a DAFNE }
\acro{FPGA}{Field Programmable Gate Array}
\acro{GEM}{Gas Electron Multiplier}
\acro{GLAST}{Gamma ray Large Area Space Telescope}
\acro{GSI}{Gesellschaft f\"ur Schwerionenforschnung}
\acro{HEP}{High Energy Physics}
\acro{HESR}{High Energy Storage Ring}
\acro{HV}{High Voltage}
\acro{IHEP}{Institute for High Energy Physics}
\acro{J-FET}{Junction Field-Effect Transistor}
\acro{KVI}{Kernfysisch Versneller Instituut}
\acro{LAAPD}{Large Area APD}
\acro{LEAR}{Low Energy Antiproton Ring}
\acro{LED}{Light Emitting Diode}
\acro{LHC}{Large Hadron Collider}
\acro{LNP}{Low Noise Preamplifier}
\acro{LOI}{Letter of Intent}
\acro{LY}{Light Yield}
\acro{M}{Gain of APD}
\acro{MAMI}{Mainz Microtron}
\acro{MIP}{Minimum Ionizing Particle}
\acro{MLP}{Multi Layer Perceptron}
\acro{MVD}{Micro Vertex Detector}
\acro{NCE}{Nuclear counter effect}
\acro{NIEL}{Non-Ionizing Energy Loss}
\acro{PANDA}{Pbar ANihilation at Darmstadt}
\acro{PCB}{Printed Circuit Board}
\acro{PEEK}{Polyetheretherketone}
\acro{phe}{photo electrons}
\acro{PID}{Particle Identification}
\acro{PMT}{Photomultiplier}
\acro{PSI}{Paul Scherrer Institute}
\acro{PWO}{Lead Tungstate}
\acro{QCD}{Quantum Chromo Dynamics}
\acro{QE}{Quantum efficiency}
\acro{RICH}{Ring Imaging Cherenkov Counter}
\acro{RT}{Room Temperature}
\acro{SCADA}{Supervisory Control And Data Acquisition}
\acro{SICCAS}{Shanghai Institute of Ceramics, Chinese Academy of Sciences}
\acro{SIS}{Heavy ion synchrotron}
\acro{SMD}{Surface Mount Device}
\acro{STT}{Straw Tube Tracker}
\acro{TDC}{Time to Digital Converter}
\acro{TSL}{The Svedberg Laboratory}
\acro{TSL}{Thermo stimulated luminescence}
\acro{UrQMD}{Ultra-relativistic Quantum Molecular Dynamic}
\acro{VME}{Versa Module Eurocard}
\acro{VPT}{Vacuum Photo Triode}
\end{acronym}
\vfill
%
%

\cleardoublepage
\addcontentsline{toc}{chapter}{List of Figures}
\listoffigures
\cleardoublepage
\addcontentsline{toc}{chapter}{List of Tables}
\listoftables
%
%
\end{document}